\documentclass[PhD,binding=0.6cm]{sapthesis}

\usepackage{microtype}
\usepackage[utf8]{inputenx}
\usepackage[T1]{fontenc}
\usepackage[sc,osf]{mathpazo}

\makeatletter
	\let\ps@plain\ps@empty
\makeatother

\usepackage{amssymb,amsthm,amsfonts,braket,mathtools,bm,bbm}
\DeclarePairedDelimiter{\abs}{\lvert}{\rvert}
\DeclarePairedDelimiter{\norm}{\lVert}{\rVert}
\allowdisplaybreaks		

\def\Xint#1{\mathchoice
   {\XXint\displaystyle\textstyle{#1}}%
   {\XXint\textstyle\scriptstyle{#1}}%
   {\XXint\scriptstyle\scriptscriptstyle{#1}}%
   {\XXint\scriptscriptstyle\scriptscriptstyle{#1}}%
   \!\int}
\def\XXint#1#2#3{{\setbox0=\hbox{$#1{#2#3}{\int}$}
     \vcenter{\hbox{$#2#3$}}\kern-.5\wd0}}

\def\dashint{\Xint-}

\usepackage[usenames,dvipsnames,svgnames,table]{xcolor}
\usepackage{graphicx}
\usepackage{pdfpages}
\definecolor{BleuDeFrance}{rgb}{0.19,0.55,0.91}

\usepackage{etoolbox}
\makeatletter
\patchcmd{\@chapter}
  {\chaptermark{#1}}
  {\chaptermark{#1}%
   \addtocontents{loa}{\protect\addvspace{10\p@}}}
  {}{}
\makeatother
\usepackage{algpseudocode}
\usepackage[chapter]{algorithm}
\makeatletter
	\renewcommand{\ALG@name}{Code}
\makeatother

\def\equationautorefname~#1\null{(#1)\null}
\def\chapterautorefname~#1\null{Chapter~#1\null}
\def\sectionautorefname~#1\null{Section~#1\null}
\def\subsectionautorefname~#1\null{Section~#1\null}
\def\appendixautorefname~#1\null{Appendix~#1\null}

\usepackage{enumitem}
\usepackage{xstring}
\usepackage{booktabs,tabularx,multicol,multirow,arydshln,siunitx}
\newcolumntype{Y}{>{\centering\arraybackslash}X}

\usepackage[hyperref=true,
	safeinputenc,
	style=alphabetic,
	citestyle=alphabetic,
	isbn=false,					
	url=false,					
	useprefix=true,				
	backref=true,
	backrefstyle=none,
	backend=biber,
	maxnames=20,					
	maxcitenames=1,				
	sorting=anyt]{biblatex}
\bibliography{myBiblio,bibByHand}
\DeclareLabelalphaTemplate{
  \labelelement{
	\field[strwidth=3,strside=left,names=1]{labelname}
  }
  \labelelement{
    \field[strwidth=2,strside=right]{year}
  }
}

\DeclareSourcemap{
	\maps{
		\map{
			\step[fieldsource=author,
				match=\regexp{\$\\sim\$},
				replace=\regexp{\~}]
		}
	}
}

\DeclareFieldFormat[article]{volume}{\textbf{#1}}

\renewbibmacro{in:}{}

\AtEveryBibitem{%
  \clearfield{day}%
  \clearfield{month}%
  \clearfield{endday}%
  \clearfield{endmonth}%
}

\usepackage{xpatch}\xpatchbibmacro{pageref}{parens}{brackets}{}{}

\DefineBibliographyStrings{english}{%
  backrefpage = {\uppercase{c}ited on page},%
  backrefpages = {\uppercase{c}ited on pages}%
}
\DeclareListFormat{pageref}{%
\ifthenelse{\value{liststop} < 3}
{\ifthenelse{\value{listcount}<\value{liststop}}{\hyperpage{#1} and }{\hyperpage{#1}}} %
{ 
    \ifthenelse{\value{listcount}<\value{liststop}}
      {\hyperpage{#1}%
        \addtocounter{listcount}{1}%
     \ifthenelse{\value{listcount}<\value{liststop}}%
     {\addcomma\addspace}{\addspace}%
    \addtocounter{listcount}{-1}%
      }
      {\ifnumequal{\value{listcount}}{\value{liststop}}
        {and \hyperpage{#1}}
        {}%
      }%
}%
}

\usepackage[]{hyperref}
\hypersetup{
	pdfauthor = {Cosimo Lupo},
	pdftitle = {Critical properties of disordered XY model on sparse random graphs},
	linktocpage = true,
	pdfstartpage = 1,
	pdfstartview = Fit,
	colorlinks = true,
	pdfborder = {0 0 1.0},		
	breaklinks = true,
	anchorcolor = blue, citecolor = Green, citebordercolor = Green, filecolor = RedViolet, filebordercolor = RedViolet, linkcolor = BleuDeFrance, linkbordercolor = BleuDeFrance, urlcolor = Rhodamine, urlbordercolor = Rhodamine
}

\usepackage[nomain,acronyms,toc,nopostdot]{glossaries}
\newglossary[slg]{symbols}{sls}{slo}{\glssymbolsgroupname}
\makeglossaries

\newacronym{1RSB}{1RSB}{1\,-\,step Replica Symmetry Breaking}
\newacronym{2RSB}{2RSB}{2\,-\,step Replica Symmetry Breaking}
\newacronym{BP}{BP}{Belief Propagation}
\newacronym{ChainSuscProp}{ChainSuscProp}{Chain Susceptibility Propagation}
\newacronym{CW}{CW}{Curie\,-\,Weiss}
\newacronym{dAT}{dAT}{de Almeida\,-\,Thouless}
\newacronym{EA}{EA}{Edwards\,-\,Anderson}
\newacronym{ERG}{ERG}{Erd\H{o}s\,-\,R\'enyi Graph}
\newacronym{FEL}{FEL}{Free Energy Landscape}
\newacronym{fRSB}{fRSB}{full\,-\,step Replica Symmetry Breaking}
\newacronym{GCD}{GCD}{Greedy Coordinate Descent}
\newacronym{GSA}{GSA}{Given Sample Algorithm}
\newacronym{GT}{GT}{Gabay\,-\,Toulouse}
\newacronym{IPR}{IPR}{Inverse Participation Ratio}
\newacronym{kRSB}{kRSB}{k\,-\,step Replica Symmetry Breaking}
\newacronym{MF}{MF}{Mean Field}
\newacronym{PDA}{PDA}{Population Dynamics Algorithm}
\newacronym{RFIM}{RFIM}{Random Field Ising Model}
\newacronym{RKKY}{RKKY}{Ruderman\,-\,Kittel\,-\,Kasuya\,-\,Yosida}
\newacronym{RRG}{RRG}{Random Regular Graph}
\newacronym{RS}{RS}{Replica Symmetry}
\newacronym{RSB}{RSB}{Replica Symmetry Breaking}
\newacronym{SK}{SK}{Sherrington\,-\,Kirkpatrick}
\newacronym{SuscProp}{SuscProp}{Susceptibility Propagation}
\newacronym{TAP}{TAP}{Thouless\,-\,Anderson\,-\,Palmer}

\newglossaryentry{thermal_average}{
	sort={001},
	type=symbols,
	name={\ensuremath{\braket{\cdot}}},
	description={Thermal average over $\mathbb{P}_{\textnormal{eq}}$}
}
\newglossaryentry{connected_average}{
	sort={002},
	type=symbols,
	name={\ensuremath{\braket{\cdot}_c}},
	description={Connected average}
}
\newglossaryentry{MF_thermal_average}{
	sort={003},
	type=symbols,
	name={\ensuremath{\braket{\cdot}_{\textnormal{MF}}}},
	description={Thermal average over $\mathbb{P}_{\textnormal{MF}}$}
}
\newglossaryentry{time_average}{
	sort={004},
	type=symbols,
	name={\ensuremath{\braket{\cdot}_t}},
	description={Time average}
}
\newglossaryentry{scala_product}{
	sort={005},
	type=symbols,
	name={\ensuremath{\braket{\cdot|\cdot}}},
	description={Scalar product}
}
\newglossaryentry{disorder_average}{
	sort={006},
	type=symbols,
	name={\ensuremath{\overline{\,\,\cdot\,\,}}},
	description={Average over the quenched disorder distribution}
}
\newglossaryentry{norm_L2}{
	sort={007},
	type=symbols,
	name={\ensuremath{\norm{\cdot}_2}},
	description={$L_2$-norm}
}

\newglossaryentry{a1_a2_a3_dots}{
	sort={a001},
	type=symbols,
	name={\ensuremath{a_1,a_2,a_3,\dots}},
	description={Coefficient in a generic Taylor expansion}
}
\newglossaryentry{a_Fourier}{
	sort={a002},
	type=symbols,
	name={\ensuremath{a^{(i\to j)}_l}},
	description={Cosine coefficient of order $l$ for the expansion of $\eta_{i\to j}(\theta_i)$ in Fourier eigenfunctions}
}
\newglossaryentry{small_abc}{
	sort={a003},
	type=symbols,
	name={\ensuremath{a,b,c,\dots}},
	description={
		$\triangleright$ In~\autoref{chap:tools}: interaction labels, or function/check node labels of the corresponding interaction graph%
		\newline
		$\triangleright$ In~\autoref{chap:sg_replica}: replica indexes%
		\newline
		$\triangleright$ In~\autoref{chap:clock}: direction indexes in the $Q$-state clock model%
		\newline
		$\triangleright$ In~\autoref{sec:beyond_RS}: generic coefficients
	}
}
\newglossaryentry{a_to_i}{
	sort={a004},
	type=symbols,
	name={\ensuremath{a\to i}},
	description={Directed check-to-variable edge on the bipartite interaction graph}
}
\newglossaryentry{capital_A}{
	sort={a005},
	type=symbols,
	name={\ensuremath{A}},
	description={Generic coefficient}
}
\newglossaryentry{Adj_matrix}{
	sort={a006},
	type=symbols,
	name={\ensuremath{\mathbb{A}}},
	description={Adjacency matrix of a graph}
}
\newglossaryentry{site_resolvent_real}{
	sort={a007},
	type=symbols,
	name={\ensuremath{\mathcal{A}_i(\lambda)}},
	description={Real part of $\mathcal{R}_i(\lambda)$}
}
\newglossaryentry{cavity_resolvent_real}{
	sort={a008},
	type=symbols,
	name={\ensuremath{\widetilde{\mathcal{A}}_{i\to j}(\lambda)}},
	description={Real part of $\widetilde{\mathcal{R}}_{i\to j}(\lambda)$}
}
\newglossaryentry{A_k_Fourier}{
	sort={a009},
	type=symbols,
	name={\ensuremath{A_k(l)}},
	description={Growth rate of the disorder-averaged $k$-th moment of the distribution of the Fourier coefficients $\{a^{(i\to j)}_l\}$}
}
\newglossaryentry{alpha}{
	sort={a010},
	type=symbols,
	name={\ensuremath{\alpha}},
	description={
		$\triangleright$ In~\autoref{chap:tools} and~\autoref{chap:XYnoField}: ratio between the number $M$ of function/check nodes and the number $N$ of variable nodes in bipartite graphs%
		\newline
		$\triangleright$ In~\autoref{subsec:conv_discr_Bessel} and~\autoref{sec:GT_dAT_diluted}: generic power-law exponent%
		\newline
		$\triangleright$ In~\autoref{chap:XYinField_zeroTemp}: exponent of the power law of the spectral density $\rho(\lambda)$ close to the lower band edge
	}
}
\newglossaryentry{small_greek_letters}{
	sort={a011},
	type=symbols,
	name={\ensuremath{\alpha,\beta,\gamma,\dots}},
	description={Pure states indexes}
}

\newglossaryentry{small_b}{
	sort={b001},
	type=symbols,
	name={\ensuremath{b}},
	description={In~\autoref{sec:conv_phys_obs}: exponent in the stretched-exponential decay of $\Delta f^{(Q)}$ in the zero-temperature limit}
}
\newglossaryentry{b1_b2_b3_dots}{
	sort={b002},
	type=symbols,
	name={\ensuremath{b_1,b_2,b_3,\dots}},
	description={Coefficient in a generic Taylor expansion}
}
\newglossaryentry{b_Fourier}{
	sort={b003},
	type=symbols,
	name={\ensuremath{b^{(i\to j)}_l}},
	description={Sine coefficient of order $l$ for the expansion of $\eta_{i\to j}(\theta_i)$ in Fourier eigenfunctions}
}
\newglossaryentry{belief_i_BP}{
	sort={b004},
	type=symbols,
	name={\ensuremath{b_i(\boldsymbol{x}_i)}},
	description={Belief about the realization $\boldsymbol{x}_i$ of the $i$-th event}
}
\newglossaryentry{belief_ij_BP}{
	sort={b005},
	type=symbols,
	name={\ensuremath{b_{ij}(\boldsymbol{x}_i,\boldsymbol{x}_j)}},
	description={Belief about the joint realization $(\boldsymbol{x}_i,\boldsymbol{x}_j)$ of the $i$-th and the $j$-th (directly related) events}
}
\newglossaryentry{site_resolvent_imaginary}{
	sort={b006},
	type=symbols,
	name={\ensuremath{\mathcal{B}_i(\lambda)}},
	description={Imaginary part of $\mathcal{R}_i(\lambda)$}
}
\newglossaryentry{cavity_resolvent_imaginary}{
	sort={b007},
	type=symbols,
	name={\ensuremath{\widetilde{\mathcal{B}}_{i\to j}(\lambda)}},
	description={Imaginary part of $\widetilde{\mathcal{R}}_{i\to j}(\lambda)$}
}
\newglossaryentry{inverse_temperature}{
	sort={b008},
	type=symbols,
	name={\ensuremath{\beta}},
	description={
		$\triangleright$ Inverse temperature (with $k_B=1$)%
		\newline
		$\triangleright$ In~\autoref{chap:XYnoField}: magnetization vs temperature critical exponent
	}
}
\newglossaryentry{critical_inverse_temperature}{
	sort={b009},
	type=symbols,
	name={\ensuremath{\beta_c}},
	description={Critical inverse temperature (with $k_B=1$)}
}
\newglossaryentry{critical_inverse_temperature_Qclock}{
	sort={b010},
	type=symbols,
	name={\ensuremath{\beta^{(Q)}_c}},
	description={$Q$-state approximation of $\beta_c$}
}
\newglossaryentry{critical_inverse_temperature_F}{
	sort={b011},
	type=symbols,
	name={\ensuremath{\beta_{\textnormal{F}}}},
	description={Critical inverse temperature (with $k_B=1$) between the paramagnetic and the ferromagnetic phases}
}
\newglossaryentry{critical_inverse_temperature_F_Qclock}{
	sort={b012},
	type=symbols,
	name={\ensuremath{\beta^{(Q)}_{\textnormal{F}}}},
	description={$Q$-state approximation of $\beta_{\text{F}}$}
}
\newglossaryentry{critical_inverse_temperature_SG}{
	sort={b013},
	type=symbols,
	name={\ensuremath{\beta_{\textnormal{SG}}}},
	description={Critical inverse temperature (with $k_B=1$) between the paramagnetic and the spin glass phases}
}
\newglossaryentry{critical_inverse_temperature_SG_Qclock}{
	sort={b014},
	type=symbols,
	name={\ensuremath{\beta^{(Q)}_{\textnormal{SG}}}},
	description={$Q$-state approximation of $\beta_{\text{SG}}$}
}

\newglossaryentry{c_Fourier}{
	sort={c001},
	type=symbols,
	name={\ensuremath{c^{(i\to j)}_l}},
	description={Coefficient of order $l$ for the expansion of $\eta_{i\to j}(\theta_i)$ in Fourier eigenfunctions in the $Q$-state clock model approximation of the XY model}
}
\newglossaryentry{connectivity}{
	sort={c002},
	type=symbols,
	name={\ensuremath{C}},
	description={Fixed connectivity of a RRG or average connectivity of an ERG}
}
\newglossaryentry{Cl_Fourier}{
	sort={c003},
	type=symbols,
	name={\ensuremath{C_l(\theta)}},
	description={Shorthand notation for the $l$-th-order term in a Fourier expansion}
}
\newglossaryentry{cumulative}{
	sort={c004},
	type=symbols,
	name={\ensuremath{\mathcal{C}}},
	description={Cumulative function of a generic probability distribution}
}
\newglossaryentry{specific_heat_v}{
	sort={c005},
	type=symbols,
	name={\ensuremath{C_v}},
	description={Specific heat at constant volume}
}
\newglossaryentry{cumulative_Weib}{
	sort={c006},
	type=symbols,
	name={\ensuremath{\mathcal{C}_{\textnormal{Weib}}}},
	description={Cumulative function of $\mathbb{P}_{\text{Weib}}$}
}
\newglossaryentry{chi}{
	sort={c007},
	type=symbols,
	name={\ensuremath{\chi}},
	description={Zero-field magnetic susceptibility}
}
\newglossaryentry{chi_square}{
	sort={c008},
	type=symbols,
	name={\ensuremath{\chi^2}},
	description={Chi-square}
}
\newglossaryentry{chi_f}{
	sort={c009},
	type=symbols,
	name={\ensuremath{\chi_{\textnormal{F}}}},
	description={Ferromagnetic susceptibility}
}
\newglossaryentry{chi_sg}{
	sort={c010},
	type=symbols,
	name={\ensuremath{\chi_{\textnormal{SG}}}},
	description={Spin glass susceptibility}
}

\newglossaryentry{lattice_dim}{
	sort={d001},
	type=symbols,
	name={\ensuremath{d}},
	description={Dimension of the hypercubic lattice}
}
\newglossaryentry{degree_i}{
	sort={d002},
	type=symbols,
	name={\ensuremath{d_i}},
	description={Degree (or connectivity) of the $i$-th site of a graph}
}
\newglossaryentry{upper_crit_d}{
	sort={d003},
	type=symbols,
	name={\ensuremath{d^u_c}},
	description={Upper critical dimension}
}
\newglossaryentry{distance_alpha_beta}{
	sort={d004},
	type=symbols,
	name={\ensuremath{d_{\alpha\beta}}},
	description={Distance between the pure states $\alpha$ and $\beta$ in the overlap space}
}
\newglossaryentry{delta}{
	sort={d005},
	type=symbols,
	name={\ensuremath{\delta}},
	description={Exponent of the power law of the density of states $g(\omega)$ close to the lower band edge}
}
\newglossaryentry{belief_i_pert}{
	sort={d006},
	type=symbols,
	name={\ensuremath{\delta\eta_i(\boldsymbol{x}_i)}},
	description={Perturbation of $\eta^*_i(\boldsymbol{x}_i)$}
}
\newglossaryentry{cavity_belief_ij_pert}{
	sort={d007},
	type=symbols,
	name={\ensuremath{\delta\eta_{i\to j}(\boldsymbol{x}_i)}},
	description={Perturbation of $\eta^*_{i\to j}(\boldsymbol{x}_i)$}
}
\newglossaryentry{cavity_belief_ij_hat_pert}{
	sort={d008},
	type=symbols,
	name={\ensuremath{\delta\widehat{\eta}_{i\to j}(\boldsymbol{x}_j)}},
	description={Perturbation of $\widehat{\eta}^*_{i\to j}(\boldsymbol{x}_j)$}
}
\newglossaryentry{cavity_field_ij_pert}{
	sort={d009},
	type=symbols,
	name={\ensuremath{\delta h_{i\to j}(\boldsymbol{x}_i)}},
	description={Perturbation of $h^*_{i\to j}(\boldsymbol{x}_i)$}
}
\newglossaryentry{m_i_pert}{
	sort={d010},
	type=symbols,
	name={\ensuremath{\delta\boldsymbol{m}_i}},
	description={Perturbation of $\boldsymbol{m}_i$ at equilibrium}
}
\newglossaryentry{cavity_bias_ij_pert}{
	sort={d011},
	type=symbols,
	name={\ensuremath{\delta u_{i\to j}(\boldsymbol{x}_j)}},
	description={Perturbation of $u^*_{i\to j}(\boldsymbol{x}_j)$}
}
\newglossaryentry{delta_Tc}{
	sort={d012},
	type=symbols,
	name={\ensuremath{\delta T_c}},
	description={Distance $(T_c-T)$ from the critical temperature $T_c$}
}
\newglossaryentry{delta_theta_XY}{
	sort={d013},
	type=symbols,
	name={\ensuremath{\delta\theta_i}},
	description={Perturbation of the angular variable of the $i$-th XY spin}
}
\newglossaryentry{delta_theta_1_largeH}{
	sort={d014},
	type=symbols,
	name={\ensuremath{\delta\theta^{(1)}_i}},
	description={First-order correction to $\theta^*_i$ in the large-field expansion}
}
\newglossaryentry{delta_theta_2_largeH}{
	sort={d015},
	type=symbols,
	name={\ensuremath{\delta\theta^{(2)}_i}},
	description={Second-order correction to $\theta^*_i$ in the large-field expansion}
}
\newglossaryentry{chain_cavity_bias_ij_pert}{
	sort={d016},
	type=symbols,
	name={\ensuremath{\delta w_{i\to j}(\boldsymbol{x}_j)}},
	description={Perturbation of $w^*_{i\to j}(\boldsymbol{x}_j)$}
}
\newglossaryentry{Part_func_Z_i_to_j_pert}{
	sort={d017},
	type=symbols,
	name={\ensuremath{\delta\mathcal{Z}_{i\to j}}},
	description={Normalization constant in the computation of the perturbed cavity message $\delta\eta_{i\to j}(\boldsymbol{x}_i)$}
}
\newglossaryentry{Part_func_Z_i_to_j_hat_pert}{
	sort={d018},
	type=symbols,
	name={\ensuremath{\delta\widehat{\mathcal{Z}}_{i\to j}}},
	description={Normalization constant in the computation of the perturbed cavity message $\delta\widehat{\eta}_{i\to j}(\boldsymbol{x}_j)$}
}
\newglossaryentry{Delta}{
	sort={d019},
	type=symbols,
	name={\ensuremath{\Delta}},
	description={
		$\triangleright$ Interpolation parameter between a pure ferromagnet and an unbiased spin glass in the gauge glass model%
		\newline
		$\triangleright$ In~\autoref{sec:GT_dAT_diluted}: Perturbation parameter in the `Cooling' and `Heating' protocols
	}
}
\newglossaryentry{Delta_Q_above_f}{
	sort={d020},
	type=symbols,
	name={\ensuremath{\Delta^{(Q)}(f)}},
	description={Error committed in the evaluation of $I(f)$ via the $Q$-state approximation}
}
\newglossaryentry{Delta_star_below}{
	sort={d021},
	type=symbols,
	name={\ensuremath{\Delta_*}},
	description={Critical value of $\Delta$ on the $T=0$ axis for the transition line between the mixed and the spin glass phases in the gauge glass model}
}
\newglossaryentry{Delta_dAT}{
	sort={d022},
	type=symbols,
	name={\ensuremath{\Delta_{\textnormal{dAT}}}},
	description={Critical value of $\Delta$ on the $T=0$ axis for the transition line between the ferromagnetic and the mixed phases in the gauge glass model}
}
\newglossaryentry{Delta_GCD}{
	sort={d023},
	type=symbols,
	name={\ensuremath{\Delta_{\textnormal{GCD}}}},
	description={Accuracy in the GCD algorithm}
}
\newglossaryentry{Delta_GSA}{
	sort={d024},
	type=symbols,
	name={\ensuremath{\Delta_{\textnormal{GSA}}}},
	description={Accuracy in the GSA}
}
\newglossaryentry{Delta_mc}{
	sort={d025},
	type=symbols,
	name={\ensuremath{\Delta_{mc}}},
	description={Critical value of $\Delta$ for the multicritical point in the gauge glass model}
}
\newglossaryentry{Delta_E_overline}{
	sort={d026},
	type=symbols,
	name={\ensuremath{\overline{\Delta E}}},
	description={Disorder-averaged percentage energy difference between the combined algorithm GSA plus GCD and the sole algorithm GCD}
}
\newglossaryentry{Delta_f_Q_above}{
	sort={d027},
	type=symbols,
	name={\ensuremath{\Delta f^{(Q)}}},
	description={Error committed in the evaluation of the free energy density $f$ of the XY model via the $Q$-state clock model}
}
\newglossaryentry{Delta_f_i_to_j}{
	sort={d028},
	type=symbols,
	name={\ensuremath{\Delta f_{i\to j}}},
	description={Free energy shift due to the addition of the directed edge $i\to j$ to the original graph}
}
\newglossaryentry{Delta_lambda}{
	sort={d029},
	type=symbols,
	name={\ensuremath{\Delta\lambda}},
	description={Typical spacing between the eigenvalues of the Hessian matrix~$\mathbb{H}$}
}
\newglossaryentry{set_neigh_a}{
	sort={d030},
	type=symbols,
	name={\ensuremath{\partial a}},
	description={Set of the nearest-neighbour variable nodes of the function/check node $a$}
}
\newglossaryentry{set_neigh_a_but_i}{
	sort={d031},
	type=symbols,
	name={\ensuremath{\partial a\setminus i}},
	description={Set of the nearest-neighbour variable nodes of the function/check node $a$, excluded $i$}
}
\newglossaryentry{set_neigh_i}{
	sort={d032},
	type=symbols,
	name={\ensuremath{\partial i}},
	description={Set of the nearest-neighbour variable nodes of $i$}
}
\newglossaryentry{set_neigh_i_but_a}{
	sort={d033},
	type=symbols,
	name={\ensuremath{\partial i\setminus a}},
	description={Set of the nearest-neighbour function/check nodes of the variable node $i$, excluded $a$}
}
\newglossaryentry{set_neigh_i_but_j}{
	sort={d034},
	type=symbols,
	name={\ensuremath{\partial i\setminus j}},
	description={Set of the nearest-neighbour variable nodes of $i$, excluded $j$}
}

\newglossaryentry{basis_vector}{
	sort={e001},
	type=symbols,
	name={\ensuremath{\ket{e_i}}},
	description={$i$-th vector of the canonical basis}
}
\newglossaryentry{expectation_value}{
	sort={e002},
	type=symbols,
	name={\ensuremath{\mathbb{E}[\cdot]}},
	description={Expectation value}
}
\newglossaryentry{epsilon_resolvent}{
	sort={e003},
	type=symbols,
	name={\ensuremath{\epsilon}},
	description={Regularizer in the cavity equations for the resolvent}
}
\newglossaryentry{epsilon_BP}{
	sort={e004},
	type=symbols,
	name={\ensuremath{\varepsilon_i(\{\boldsymbol{x}_k\}_{k\in\partial i}|\boldsymbol{x}_i)}},
	description={Correction term to the Bethe\,-\,Peierls approximation on finite-size sparse random graphs}
}
\newglossaryentry{cavity_belief_ai}{
	sort={e005},
	type=symbols,
	name={\ensuremath{\widehat{\eta}_{a\to i}(\boldsymbol{x}_i)}},
	description={Cavity marginal from function/check node $a$ to variable node $i$ in the factor graph Belief Propagation approach}
}
\newglossaryentry{belief_i}{
	sort={e006},
	type=symbols,
	name={\ensuremath{\eta_i(\boldsymbol{x}_i)}},
	description={Marginal (approximated) probability distribution --- or belief --- of the $i$-th variable at equilibrium}
}
\newglossaryentry{belief_i_star}{
	sort={e007},
	type=symbols,
	name={\ensuremath{\eta^*_i(\boldsymbol{x}_i)}},
	description={BP fixed-point value of $\eta_i(\boldsymbol{x}_i)$}
}
\newglossaryentry{belief_ij}{
	sort={e008},
	type=symbols,
	name={\ensuremath{\eta_{ij}(\boldsymbol{x}_i,\boldsymbol{x}_j)}},
	description={Joint (approximated) probability distribution --- or belief --- of the nearest-neighbour variables $\boldsymbol{x}_i$ and $\boldsymbol{x}_j$ at equilibrium}
}
\newglossaryentry{cavity_belief_ia}{
	sort={e009},
	type=symbols,
	name={\ensuremath{\eta_{i\to a}(\boldsymbol{x}_i)}},
	description={Cavity marginal from variable node $i$ to function/check node~$a$ in the factor graph Belief Propagation approach}
}
\newglossaryentry{cavity_belief_ij}{
	sort={e010},
	type=symbols,
	name={\ensuremath{\eta_{i\to j}(\boldsymbol{x}_i)}},
	description={Cavity marginal from variable node $i$ to variable node $j$ in the pairwise Belief Propagation approach}
}
\newglossaryentry{cavity_belief_ij_star}{
	sort={e011},
	type=symbols,
	name={\ensuremath{\eta^*_{i\to j}(\boldsymbol{x}_i)}},
	description={BP fixed-point value of $\eta_{i\to j}(\boldsymbol{x}_i)$}
}
\newglossaryentry{cavity_belief_ij_hat}{
	sort={e012},
	type=symbols,
	name={\ensuremath{\widehat{\eta}_{i\to j}(\boldsymbol{x}_j)}},
	description={Cavity marginal from variable node $i$ to variable node $j$ in the pairwise Belief Propagation approach}
}
\newglossaryentry{cavity_belief_ij_hat_star}{
	sort={e013},
	type=symbols,
	name={\ensuremath{\widehat{\eta}^*_{i\to j}(\boldsymbol{x}_j)}},
	description={BP fixed-point value of $\widehat{\eta}_{i\to j}(\boldsymbol{x}_j)$}
}
\newglossaryentry{cavity_belief_ij_alpha}{
	sort={e014},
	type=symbols,
	name={\ensuremath{\eta^{(\alpha)}_{i\to j}(\boldsymbol{x}_i)}},
	description={Cavity marginal from variable node $i$ to variable node $j$ in the $\alpha$-th state in the Gibbs measure in the pairwise 1RSB Belief Propagation approach}
}

\newglossaryentry{small_f}{
	sort={f001},
	type=symbols,
	name={\ensuremath{f}},
	description={
		$\triangleright$ Free energy density%
		\newline
		$\triangleright$ In~\autoref{subsec:conv_discr_Bessel}: generic $2\pi$-periodic smooth function
	}
}
\newglossaryentry{small_f_star_above}{
	sort={f002},
	type=symbols,
	name={\ensuremath{f^*}},
	description={Equilibrium value of $f_{\beta}(x)$}
}
\newglossaryentry{small_f_k_above}{
	sort={f003},
	type=symbols,
	name={\ensuremath{f^{(k)}}},
	description={$k$-th derivative of the function $f$}
}
\newglossaryentry{small_f_Q_above}{
	sort={f004},
	type=symbols,
	name={\ensuremath{f^{(Q)}}},
	description={Free energy density of the $Q$-state clock model}
}
\newglossaryentry{free_energy_density_replicated_nto0}{
	sort={f005},
	type=symbols,
	name={\ensuremath{f_0}},
	description={Free energy density of the $n$-times replicated original system evaluated in the $n\to 0$ limit}
}
\newglossaryentry{annealed_free_energy_density}{
	sort={f006},
	type=symbols,
	name={\ensuremath{f_{\textnormal{ann}}}},
	description={Annealed free energy density}
}
\newglossaryentry{free_energy_density_alpha}{
	sort={f007},
	type=symbols,
	name={\ensuremath{f_{\alpha}}},
	description={Free energy density in the $\alpha$-th state of the Gibbs measure}
}
\newglossaryentry{free_energy_density_beta_x}{
	sort={f008},
	type=symbols,
	name={\ensuremath{f_{\beta}(x)}},
	description={Free energy density in the 1RSB framework on diluted graphs}
}
\newglossaryentry{free_energy_site}{
	sort={f009},
	type=symbols,
	name={\ensuremath{f_i}},
	description={Site contribution to the total free energy}
}
\newglossaryentry{free_energy_edge}{
	sort={f010},
	type=symbols,
	name={\ensuremath{f_{ij}}},
	description={Edge contribution to the total free energy}
}
\newglossaryentry{free_energy_density_given_instance}{
	sort={f011},
	type=symbols,
	name={\ensuremath{f_J}},
	description={Free energy density for a given instance of the quenched disorder}
}
\newglossaryentry{free_energy_density_replicated}{
	sort={f012},
	type=symbols,
	name={\ensuremath{f_n}},
	description={Free energy density of the $n$-times replicated original system}
}
\newglossaryentry{f_Parisi_equation}{
	sort={f013},
	type=symbols,
	name={\ensuremath{\mathrm{f}(q,y)}},
	description={Function entering into the Parisi nonlinear antiparabolic differential equation}
}
\newglossaryentry{Helmholtz_free_energy}{
	sort={f014},
	type=symbols,
	name={\ensuremath{F}},
	description={Helmholtz free energy}
}
\newglossaryentry{Bethe_free_energy}{
	sort={f015},
	type=symbols,
	name={\ensuremath{F_{\textnormal{BP}}}},
	description={Bethe free energy}
}
\newglossaryentry{F_BP}{
	sort={f016},
	type=symbols,
	name={\ensuremath{\mathcal{F}[\cdot]}},
	description={Shorthand notation for the finite-temperature pairwise BP equations}
}
\newglossaryentry{F_BP_prime}{
	sort={f017},
	type=symbols,
	name={\ensuremath{\mathcal{F}'[\cdot]}},
	description={Shorthand notation for the linearized finite-temperature pairwise BP equations}
}
\newglossaryentry{F_BP_zeroTemp}{
	sort={f018},
	type=symbols,
	name={\ensuremath{\mathcal{F}_0[\cdot]}},
	description={Shorthand notation for the zero-temperature pairwise BP equations}
}
\newglossaryentry{F_BP_zeroTemp_prime}{
	sort={f019},
	type=symbols,
	name={\ensuremath{\mathcal{F}'_0[\cdot]}},
	description={Shorthand notation for the linearized zero-temperature pairwise BP equations}
}
\newglossaryentry{Reweigh_factor_1RSB_SK}{
	sort={f020},
	type=symbols,
	name={\ensuremath{\mathcal{F}(z_0,z1)}},
	description={Reweigh factor in the fully connected 1RSB framework}
}

\newglossaryentry{density_of_states}{
	sort={g001},
	type=symbols,
	name={\ensuremath{g(\omega)}},
	description={Density of states}
}
\newglossaryentry{Gibbs_free_energy}{
	sort={g002},
	type=symbols,
	name={\ensuremath{G}},
	description={Gibbs free energy}
}
\newglossaryentry{graph}{
	sort={g003},
	type=symbols,
	name={\ensuremath{\mathcal{G}}},
	description={Interaction graph}
}
\newglossaryentry{gauss}{
	sort={g004},
	type=symbols,
	name={\ensuremath{\mathrm{Gauss}(\mu,\sigma^2)}},
	description={Gaussian distribution with mean $\mu$ and variance $\sigma^2$}
}
\newglossaryentry{gamma}{
	sort={g005},
	type=symbols,
	name={\ensuremath{\gamma}},
	description={Damping in the GSA}
}
\newglossaryentry{gamma_i}{
	sort={g006},
	type=symbols,
	name={\ensuremath{\gamma_i}},
	description={Lagrange multiplier for the normalization of the one-point belief $\eta_i(\boldsymbol{x}_i)$}
}
\newglossaryentry{gamma_ij}{
	sort={g007},
	type=symbols,
	name={\ensuremath{\gamma_{ij}}},
	description={Lagrange multiplier for the normalization of the two-point belief $\eta_{ij}(\boldsymbol{x}_i,\boldsymbol{x}_j)$}
}
\newglossaryentry{Gamma_function}{
	sort={g008},
	type=symbols,
	name={\ensuremath{\Gamma(\cdot)}},
	description={Gamma function}
}
\newglossaryentry{Gamma_a}{
	sort={g009},
	type=symbols,
	name={\ensuremath{\Gamma_a}},
	description={$a$-th bin of the $[0,2\pi)$ interval in the $Q$-state approximation}
}
\newglossaryentry{Gamma_coefficients}{
	sort={g010},
	type=symbols,
	name={\ensuremath{\Gamma_1,\Gamma_2,\Gamma_3,\dots}},
	description={Generic coefficients of an expansion}
}

\newglossaryentry{generic_site_term_hi}{
	sort={h001},
	type=symbols,
	name={\ensuremath{h_i(\boldsymbol{x}_i)}},
	description={Site bias coming from a generic external field}
}
\newglossaryentry{cavity_field_ij}{
	sort={h002},
	type=symbols,
	name={\ensuremath{h_{i\to j}(\boldsymbol{x}_i)}},
	description={Cavity field from variable node $i$ to variable node $j$ in the pairwise zero-temperature Belief Propagation approach}
}
\newglossaryentry{cavity_field_ij_star}{
	sort={h003},
	type=symbols,
	name={\ensuremath{h^*_{i\to j}(\boldsymbol{x}_i)}},
	description={BP fixed-point value of $h_{i\to j}(\boldsymbol{x}_i)$}
}
\newglossaryentry{capital_H}{
	sort={h004},
	type=symbols,
	name={\ensuremath{H}},
	description={Intensity of the external magnetic field}
}
\newglossaryentry{capital_H_c}{
	sort={h005},
	type=symbols,
	name={\ensuremath{H_c}},
	description={Critical value of $H$}
}
\newglossaryentry{capital_H_dAT}{
	sort={h006},
	type=symbols,
	name={\ensuremath{H_{\textnormal{dAT}}}},
	description={Critical value of $H$ on the $T=0$ axis for the dAT transition line}
}
\newglossaryentry{capital_H_dAT_Temp}{
	sort={h007},
	type=symbols,
	name={\ensuremath{H_{\textnormal{dAT}}(T)}},
	description={dAT transition line in the $(T,H)$ plane}
}
\newglossaryentry{capital_H_gap}{
	sort={h008},
	type=symbols,
	name={\ensuremath{H_{\textnormal{gap}}}},
	description={Critical value of $H$ on the $T=0$ axis at which the spectral density $\rho(\lambda)$ becomes gapless}
}
\newglossaryentry{capital_H_GT}{
	sort={h009},
	type=symbols,
	name={\ensuremath{H_{\textnormal{GT}}}},
	description={Critical value of $H$ on the $T=0$ axis for the GT transition line}
}
\newglossaryentry{capital_H_GT_Temp}{
	sort={h010},
	type=symbols,
	name={\ensuremath{H_{\textnormal{GT}}(T)}},
	description={GT transition line in the $(T,H)$ plane}
}
\newglossaryentry{capital_H_i}{
	sort={h011},
	type=symbols,
	name={\ensuremath{H_i}},
	description={Intensity of the external magnetic field on the $i$-th site of the graph}
}
\newglossaryentry{vector_field}{
	sort={h012},
	type=symbols,
	name={\ensuremath{\boldsymbol{H}_i}},
	description={External (vector) magnetic field on the $i$-th site of the graph}
}
\newglossaryentry{H_nd}{
	sort={h013},
	type=symbols,
	name={\ensuremath{H_{nd}}},
	description={Value of $H$ in correspondence of a nondifferentiability point in the critical lines}
}
\newglossaryentry{hamiltonian}{
	sort={h014},
	type=symbols,
	name={\ensuremath{\mathcal{H}}},
	description={Hamiltonian}
}
\newglossaryentry{hamiltonian_single}{
	sort={h015},
	type=symbols,
	name={\ensuremath{\mathcal{H}_i}},
	description={Single-particle Hamiltonian}
}
\newglossaryentry{hamiltonian_given_sample}{
	sort={h016},
	type=symbols,
	name={\ensuremath{\mathcal{H}_J}},
	description={Hamiltonian of a given instance of the quenched disorder}
}
\newglossaryentry{hessian}{
	sort={h017},
	type=symbols,
	name={\ensuremath{\mathbb{H}}},
	description={
		$\triangleright$ In~\autoref{chap:sg_replica}: Hessian matrix in the space of replicas%
		\newline
		$\triangleright$ In~\autoref{chap:XYinField_zeroTemp}: Hessian matrix in the space of spin coordinates
	}
}

\newglossaryentry{imaginary_unit}{
	sort={i001},
	type=symbols,
	name={\ensuremath{\mathfrak{i}}},
	description={Imaginary unit}
}
\newglossaryentry{sites}{
	sort={i002},
	type=symbols,
	name={\ensuremath{i,j,k,\dots}},
	description={Variable labels, or node labels of the corresponding interaction graph}
}
\newglossaryentry{couple_ia}{
	sort={i003},
	type=symbols,
	name={\ensuremath{(i,a)}},
	description={Couple of variable and function labels, or the corresponding (undirected) edge on the bipartite interaction graph}
}
\newglossaryentry{couple_ij}{
	sort={i004},
	type=symbols,
	name={\ensuremath{(i,j)}},
	description={Couple of variable labels, or the corresponding (undirected) edge on the interaction graph}
}
\newglossaryentry{i_to_a}{
	sort={i005},
	type=symbols,
	name={\ensuremath{i\to a}},
	description={Directed variable-to-check edge on the bipartite interaction graph}
}
\newglossaryentry{i_to_j}{
	sort={i006},
	type=symbols,
	name={\ensuremath{i\to j}},
	description={Directed edge on the interaction graph}
}
\newglossaryentry{imaginary_part}{
	sort={i007},
	type=symbols,
	name={\ensuremath{\Im[\cdot]}},
	description={Imaginary part of a complex-valued quantity}
}
\newglossaryentry{identity_matrix}{
	sort={i008},
	type=symbols,
	name={\ensuremath{\mathbb{I}}},
	description={Identity matrix}
}
\newglossaryentry{integral_f}{
	sort={i009},
	type=symbols,
	name={\ensuremath{I(f)}},
	description={Integral of a generic function $f$ over the $[0,2\pi)$ interval}
}
\newglossaryentry{integral_f_Qclock}{
	sort={i010},
	type=symbols,
	name={\ensuremath{I^{(Q)}(f)}},
	description={$Q$-state approximation of $I(f)$}
}
\newglossaryentry{Bessel_function}{
	sort={i011},
	type=symbols,
	name={\ensuremath{I_m(\cdot)}},
	description={Modified Bessel function of the first kind of order $m$}
}
\newglossaryentry{Bessel_function_discr}{
	sort={i012},
	type=symbols,
	name={\ensuremath{I^{(Q)}_m(\cdot)}},
	description={$Q$-state approximation of $I_m(\cdot)$}
}
\newglossaryentry{dashint}{
	sort={i013},
	type=symbols,
	name={\ensuremath{\dashint}},
	description={Cauchy principal value of the integral}
}

\newglossaryentry{coupling_J}{
	sort={j001},
	type=symbols,
	name={\ensuremath{J}},
	description={Intensity of the exchange couplings}
}
\newglossaryentry{variance_J}{
	sort={j002},
	type=symbols,
	name={\ensuremath{J^2}},
	description={Variance of the (randomly distributed) exchange couplings}
}
\newglossaryentry{mean_J}{
	sort={j003},
	type=symbols,
	name={\ensuremath{J_0}},
	description={Mean of the (randomly distributed) exchange couplings}
}
\newglossaryentry{generic_interaction_Ja}{
	sort={j004},
	type=symbols,
	name={\ensuremath{J_a(\underline{\boldsymbol{x}}_{\partial a})}},
	description={Generic many-body interaction between variables $\underline{\boldsymbol{x}}_{\partial a}$}
}
\newglossaryentry{coupling_Jij}{
	sort={j005},
	type=symbols,
	name={\ensuremath{J_{ij}}},
	description={Interaction coupling between the $i$-th and the $j$-th degrees of freedom}
}
\newglossaryentry{generic_interaction_Jij}{
	sort={j006},
	type=symbols,
	name={\ensuremath{J_{ij}(\boldsymbol{x}_i,\boldsymbol{x}_j)}},
	description={Generic pairwise interaction between variables $\boldsymbol{x}_i$ and $\boldsymbol{x}_j$}
}
\newglossaryentry{coupling_Jxy}{
	sort={j007},
	type=symbols,
	name={\ensuremath{J_{\boldsymbol{x}\boldsymbol{y}}}},
	description={RKKY interaction coupling between magnetic moments located respectively at $\boldsymbol{x}$ and $\boldsymbol{y}$}
}

\newglossaryentry{Boltzmann_constant}{
	sort={k001},
	type=symbols,
	name={\ensuremath{k_B}},
	description={Boltzmann constant}
}
\newglossaryentry{k_Fermi}{
	sort={k002},
	type=symbols,
	name={\ensuremath{\boldsymbol{k}_F}},
	description={Fermi wave vector}
}
\newglossaryentry{kappa}{
	sort={k003},
	type=symbols,
	name={\ensuremath{\kappa}},
	description={Direction index in the $Q$-state approximation}
}

\newglossaryentry{Fourier_order}{
	sort={l001},
	type=symbols,
	name={\ensuremath{l}},
	description={Order of expansion in Fourier eigenfunctions}
}
\newglossaryentry{work}{
	sort={l002},
	type=symbols,
	name={\ensuremath{L}},
	description={Work}
}
\newglossaryentry{loop_size}{
	sort={l003},
	type=symbols,
	name={\ensuremath{\ell}},
	description={Size of the typical loop on sparse random graphs}
}
\newglossaryentry{lambda}{
	sort={l004},
	type=symbols,
	name={\ensuremath{\lambda}},
	description={
		$\triangleright$ In~\autoref{sec:stat_mech_and_therm}: external parameter acting on the system%
		\newline
		$\triangleright$ In~\autoref{sec:var_appr_Gibbs_free_en}: Lagrange multiplier%
		\newline
		$\triangleright$ In~\autoref{chap:XYinField_zeroTemp}: eigenvalue of the Hessian matrix $\mathbb{H}$
	}
}
\newglossaryentry{lambda_tilde}{
	sort={l005},
	type=symbols,
	name={\ensuremath{\tilde{\lambda}}},
	description={Lower band edge in the spectral density $\rho(\lambda)$}
}
\newglossaryentry{lambda_prime}{
	sort={l006},
	type=symbols,
	name={\ensuremath{\lambda'}},
	description={Parameter of $\mathbb{P}_{\text{Weib}}$}
}
\newglossaryentry{lambda_replicon}{
	sort={l007},
	type=symbols,
	name={\ensuremath{\lambda_1}},
	description={Replicon eigenvalue of the Hessian matrix in the replica space}
}
\newglossaryentry{lambda_BP}{
	sort={l008},
	type=symbols,
	name={\ensuremath{\lambda_{\textnormal{BP}}}},
	description={Stability parameter of the BP iterative algorithm}
}
\newglossaryentry{lambda_i}{
	sort={l009},
	type=symbols,
	name={\ensuremath{\lambda_i}},
	description={$i$-th eigenvalue of the Hessian matrix $\mathbb{H}$, sorted from the smallest to the largest one}
}
\newglossaryentry{lambda_i_to_j}{
	sort={l010},
	type=symbols,
	name={\ensuremath{\lambda_{i\to j}(\boldsymbol{x}_i)}},
	description={Lagrange multiplier for the marginalization over $\boldsymbol{x}_j$ of the two-point belief $\eta_{ij}(\boldsymbol{x}_i,\boldsymbol{x}_j)$}
}
\newglossaryentry{lambda_max}{
	sort={l011},
	type=symbols,
	name={\ensuremath{\lambda_{\textnormal{max}}}},
	description={Largest eigenvalue of the Hessian matrix $\mathbb{H}$}
}
\newglossaryentry{lambda_min}{
	sort={l012},
	type=symbols,
	name={\ensuremath{\lambda_{\textnormal{min}}}},
	description={Smallest eigenvalue of the Hessian matrix $\mathbb{H}$}
}
\newglossaryentry{lambda_i_vector}{
	sort={l013},
	type=symbols,
	name={\ensuremath{\ket{\lambda}_i}},
	description={Eigenvector of the Hessian matrix $\mathbb{H}$ corresponding to the $i$-th eigenvalue $\lambda_i$}
}
\newglossaryentry{lambda_min_vector}{
	sort={l014},
	type=symbols,
	name={\ensuremath{\ket{\lambda}_{\textnormal{min}}}},
	description={Eigenvector of the Hessian matrix $\mathbb{H}$ corresponding to $\lambda_{\textnormal{min}}$}
}

\newglossaryentry{small_m}{
	sort={m001},
	type=symbols,
	name={\ensuremath{m}},
	description={
		$\triangleright$ Number of vector spin components%
		\newline
		$\triangleright$ Norm of the average magnetization%
		\newline
		$\triangleright$ In~\autoref{sec:stat_mech_and_therm}: mass of a gas particle%
		\newline
		$\triangleright$ In~\autoref{sec:beyond_RS}: number of replicas of the original system in the 1RSB framework on diluted graphs%
		\newline
		$\triangleright$ In~\autoref{chap:XYinField_zeroTemp}: number of smallest eigenvalues of the Hessian matrix $\mathbb{H}$ compute via the Arnoldi method
	}
}
\newglossaryentry{block_size_replicas}{
	sort={m002},
	type=symbols,
	name={\ensuremath{m_1,m_2,\dots}},
	description={Sizes of the blocks of the overlap matrix $q_{ab}$ in the RSB framework}
}
\newglossaryentry{m_a}{
	sort={m003},
	type=symbols,
	name={\ensuremath{m_a}},
	description={Magnetization of the $a$-th replica}
}
\newglossaryentry{m_i}{
	sort={m004},
	type=symbols,
	name={\ensuremath{\boldsymbol{m}_i}},
	description={Average magnetization of site $i$}
}
\newglossaryentry{m_i_alpha}{
	sort={m005},
	type=symbols,
	name={\ensuremath{\boldsymbol{m}^{(\alpha)}_i}},
	description={Average magnetization of site $i$ in the $\alpha$-th pure state}
}
\newglossaryentry{m_parallel}{
	sort={m006},
	type=symbols,
	name={\ensuremath{m_{\parallel}}},
	description={Average magnetization along the field direction in the vector spin case}
}
\newglossaryentry{m_perp}{
	sort={m007},
	type=symbols,
	name={\ensuremath{m_{\perp}}},
	description={Average magnetization in the transverse plane with respect to the field direction in the vector spin case}
}
\newglossaryentry{capital_M}{
	sort={m008},
	type=symbols,
	name={\ensuremath{M}},
	description={
		$\triangleright$ Number of interactions (or function/check nodes) in bipartite graphs%
		\newline
		$\triangleright$ In~\autoref{chap:XYinField_zeroTemp}: dimension of the basis of vectors exploited in the Arnoldi method
	}
}
\newglossaryentry{belief_message_M}{
	sort={m009},
	type=symbols,
	name={\ensuremath{M_{i\to j}(\boldsymbol{x}_j)}},
	description={Belief about the effect $\boldsymbol{x}_j$ starting from the cause $\boldsymbol{x}_i$}
}
\newglossaryentry{M_pda_1RSB}{
	sort={m010},
	type=symbols,
	name={\ensuremath{\mathcal{M}}},
	description={Size of the ``inner'' population of cavity messages/fields in the 1RSB PDA}
}
\newglossaryentry{spin_components}{
	sort={m011},
	type=symbols,
	name={\ensuremath{\mu,\nu,\dots}},
	description={Indexes of the vector spin components}
}

\newglossaryentry{n}{
	sort={n001},
	type=symbols,
	name={\ensuremath{n}},
	description={
		$\triangleright$ In~\autoref{sec:stat_mech_and_therm}: density of particles%
		\newline
		$\triangleright$ In~\autoref{chap:sg_replica}: number of replicas of the original system
	}
}
\newglossaryentry{N}{
	sort={n002},
	type=symbols,
	name={\ensuremath{N}},
	description={Total number of particles, or size of a graph}
}
\newglossaryentry{N_r}{
	sort={n003},
	type=symbols,
	name={\ensuremath{N(r)}},
	description={Average number of variable nodes at distance $r$ from the center of a sparse random graph}
}
\newglossaryentry{N_r_tot}{
	sort={n004},
	type=symbols,
	name={\ensuremath{N_{\textnormal{tot}}(r)}},
	description={Average number of variable nodes into the circle of radius $r$ around the center of a sparse random graph}
}
\newglossaryentry{N_pda}{
	sort={n005},
	type=symbols,
	name={\ensuremath{\mathcal{N}}},
	description={
		$\triangleright$ Size of the population of cavity messages/fields in the PDA%
		\newline
		$\triangleright$ Size of the ``outer'' population of cavity messages/fields in the 1RSB PDA
	}
}
\newglossaryentry{N_samples}{
	sort={n006},
	type=symbols,
	name={\ensuremath{\mathcal{N}_s}},
	description={Number of samples for the GSA at each graph size $N$}
}
\newglossaryentry{N_tr}{
	sort={n007},
	type=symbols,
	name={\ensuremath{\mathcal{N}^{(t)}_{tr}}},
	description={Number of tries needed to completely update the population of size $\mathcal{N}$ at the $t$-th time step of the zero-temperature ChainSuscProp algorithm}
}

\newglossaryentry{order_O}{
	sort={o001},
	type=symbols,
	name={\ensuremath{O}},
	description={``Big O'' of the Bachmann\,-\,Landau notation}
}
\newglossaryentry{generic_obs}{
	sort={o002},
	type=symbols,
	name={\ensuremath{\mathcal{O}}},
	description={Generic physical observable}
}
\newglossaryentry{omega}{
	sort={o003},
	type=symbols,
	name={\ensuremath{\omega}},
	description={Frequency}
}
\newglossaryentry{omega_ij}{
	sort={o004},
	type=symbols,
	name={\ensuremath{\omega_{ij}}},
	description={Rotational shift in the interaction between the $i$-th and the $j$-th XY spins}
}
\newglossaryentry{Omega}{
	sort={o004},
	type=symbols,
	name={\ensuremath{\Omega(f)}},
	description={Number of states with free energy density $f$}
}

\newglossaryentry{small_p}{
	sort={p001},
	type=symbols,
	name={\ensuremath{p}},
	description={
		$\triangleright$ Fraction of positive couplings in the $\pm J$ case%
		\newline
		$\triangleright$ In~\autoref{sec:sparse_random_graphs}: probability of drawing an edge between any two nodes on sparse random graphs
	}
}
\newglossaryentry{small_p_star_below}{
	sort={p002},
	type=symbols,
	name={\ensuremath{p_*}},
	description={Critical value of $p$ on the $T=0$ axis for the transition line between the mixed and the spin glass phases in the $\pm J$ case}
}
\newglossaryentry{small_p_dAT}{
	sort={p003},
	type=symbols,
	name={\ensuremath{p_{\textnormal{dAT}}}},
	description={Critical value of $p$ on the $T=0$ axis for the transition line between the ferromagnetic and the mixed phases in the $\pm J$ case}
}
\newglossaryentry{momentum_p_i}{
	sort={p004},
	type=symbols,
	name={\ensuremath{\boldsymbol{p}_i}},
	description={Momentum of a gas particle}
}
\newglossaryentry{small_p_mc}{
	sort={p005},
	type=symbols,
	name={\ensuremath{p_{mc}}},
	description={Critical value of $p$ for the multicritical point in the $\pm J$ case}
}
\newglossaryentry{P_generic_prob_cont}{
	sort={p006},
	type=symbols,
	name={\ensuremath{\mathbb{P}(\cdot)}},
	description={Generic probability distribution}
}
\newglossaryentry{P_generic_prob_discr}{
	sort={p007},
	type=symbols,
	name={\ensuremath{\mathbb{P}[\cdot]}},
	description={The same as $\mathbb{P}(\cdot)$}
}
\newglossaryentry{overlap_distr_zero_n}{
	sort={p008},
	type=symbols,
	name={\ensuremath{\mathbb{P}(q)}},
	description={Probability distribution of the overlap between different replicas in the $n\to 0$ limit}
}
\newglossaryentry{overlap_distr_zero_n_fRSB}{
	sort={p009},
	type=symbols,
	name={\ensuremath{\widetilde{\mathbb{P}}(q)}},
	description={Probability distribution of the overlap between different replicas in the $n\to 0$ limit, restricted to the region $q\in[q_m,q_M]$ in the fRSB framework}
}
\newglossaryentry{approx_eq_distr}{
	sort={p010},
	type=symbols,
	name={\ensuremath{\mathbb{P}_0}},
	description={Approximated probability distribution at equilibrium}
}
\newglossaryentry{P_alpha_state}{
	sort={p011},
	type=symbols,
	name={\ensuremath{P_{\alpha}}},
	description={Statistical weight of the $\alpha$-th pure state of the Gibbs measure}
}
\newglossaryentry{BP_eq_distr}{
	sort={p012},
	type=symbols,
	name={\ensuremath{\mathbb{P}_{\textnormal{BP}}}},
	description={Bethe\,-\,Peierls probability distribution at equilibrium}
}
\newglossaryentry{degree_distr}{
	sort={p013},
	type=symbols,
	name={\ensuremath{\mathbb{P}_{d}}},
	description={Probability distribution of the node degree on sparse random graphs}
}
\newglossaryentry{PDA_prob_distr_zeroTemp}{
	sort={p014},
	type=symbols,
	name={\ensuremath{\mathbb{P}_{h}}},
	description={Probability distribution of the cavity fields in the zero-temperature PDA}
}
\newglossaryentry{PDA_prob_distr_zeroTemp_star}{
	sort={p015},
	type=symbols,
	name={\ensuremath{\mathbb{P}^*_h}},
	description={BP fixed-point value of $\mathbb{P}_h$}
}
\newglossaryentry{Boltzmann_eq_distr}{
	sort={p016},
	type=symbols,
	name={\ensuremath{\mathbb{P}_{\textnormal{eq}}}},
	description={Boltzmann (or canonical) probability distribution at equilibrium}
}
\newglossaryentry{PDA_prob_distr}{
	sort={p017},
	type=symbols,
	name={\ensuremath{\mathbb{P}_{\eta}}},
	description={Probability distribution of the cavity messages in the PDA}
}
\newglossaryentry{PDA_prob_distr_star}{
	sort={p018},
	type=symbols,
	name={\ensuremath{\mathbb{P}^*_{\eta}}},
	description={BP fixed-point value of $\mathbb{P}_{\eta}$}
}
\newglossaryentry{field_intensity_distr}{
	sort={p019},
	type=symbols,
	name={\ensuremath{\mathbb{P}_H}},
	description={Probability distribution of the field intensity $H$}
}
\newglossaryentry{Boltzmann_eq_distr_i}{
	sort={p020},
	type=symbols,
	name={\ensuremath{\mathbb{P}_i}},
	description={Marginal (exact) probability distribution of the $i$-th variable at equilibrium}
}
\newglossaryentry{prob_distr_i_to_j_alpha}{
	sort={p021},
	type=symbols,
	name={\ensuremath{\mathbb{P}_{i\to j}[\eta^{(\alpha)}_{i\to j}]}},
	description={Probability distribution of $\eta^{(\alpha)}_{i\to j}$ over the $\{\alpha\}$ states in the Gibbs measure in the 1RSB framework on diluted graphs}
}
\newglossaryentry{coupling_distr}{
	sort={p022},
	type=symbols,
	name={\ensuremath{\mathbb{P}_J}},
	description={Probability distribution of the exchange coupling $J$}
}
\newglossaryentry{MF_eq_distr}{
	sort={p023},
	type=symbols,
	name={\ensuremath{\mathbb{P}_{\textnormal{MF}}}},
	description={Mean-field probability distribution at equilibrium}
}
\newglossaryentry{overlap_distr_finite_n}{
	sort={p024},
	type=symbols,
	name={\ensuremath{\mathbb{P}_n(q)}},
	description={Probability distribution of the overlap between different replicas at finite $n$}
}
\newglossaryentry{field_direction_distr}{
	sort={p025},
	type=symbols,
	name={\ensuremath{\mathbb{P}_{\phi}}},
	description={Probability distribution of the local direction $\phi$ of the external field}
}
\newglossaryentry{overlap_distr_replicas}{
	sort={p026},
	type=symbols,
	name={\ensuremath{\mathbb{P}_{\textnormal{rep}}(q)}},
	description={Probability distribution of the overlap between different replicas}
}
\newglossaryentry{overlap_distr_states}{
	sort={p027},
	type=symbols,
	name={\ensuremath{\mathbb{P}_{\textnormal{states}}(q)}},
	description={Probability distribution of the overlap between different states}
}
\newglossaryentry{approx_eq_distr_S}{
	sort={p028},
	type=symbols,
	name={\ensuremath{\mathbb{P}_{\mathcal{S}}}},
	description={Approximated probability distribution at equilibrium for the variables in the subset $\mathcal{S}$}
}
\newglossaryentry{PDA_prob_distr_zeroTemp_u}{
	sort={p029},
	type=symbols,
	name={\ensuremath{\mathbb{P}_{u}}},
	description={Probability distribution of the cavity biases in the zero-temperature PDA}
}
\newglossaryentry{PDA_prob_distr_zeroTemp_u_star}{
	sort={p030},
	type=symbols,
	name={\ensuremath{\mathbb{P}^*_u}},
	description={BP fixed-point value of $\mathbb{P}_u$}
}
\newglossaryentry{P_omega}{
	sort={p031},
	type=symbols,
	name={\ensuremath{\mathbb{P}_{\omega}}},
	description={Probability distribution of the angular shift $\omega$}
}
\newglossaryentry{P_Weib}{
	sort={p032},
	type=symbols,
	name={\ensuremath{\mathbb{P}_{\textnormal{Weib}}}},
	description={Weibull probability distribution}
}
\newglossaryentry{P_mathcal}{
	sort={p033},
	type=symbols,
	name={\ensuremath{\mathcal{P}}},
	description={Long-time limit of $\mathcal{P}^{(t)}$}
}
\newglossaryentry{P_mathcal_t}{
	sort={p034},
	type=symbols,
	name={\ensuremath{\mathcal{P}^{(t)}}},
	description={Survival probability of perturbations at the $t$-th time step of the zero-temperature ChainSuscProp algorithm}
}
\newglossaryentry{P_mathcal_1RSB}{
	sort={p035},
	type=symbols,
	name={\ensuremath{\mathcal{P}[\mathbb{P}_{i\to j}]}},
	description={Probability distribution of $\mathbb{P}_{i\to j}$ in the 1RSB PDA}
}
\newglossaryentry{phi_overline}{
	sort={p036},
	type=symbols,
	name={\ensuremath{\overline{\phi}}},
	description={Global direction of the homogeneous external magnetic field for the XY model}
}
\newglossaryentry{phi_beta}{
	sort={p037},
	type=symbols,
	name={\ensuremath{\phi_{\beta}(x)}},
	description={Replicated free energy density in the 1RSB framework on diluted graphs}
}
\newglossaryentry{phi_i}{
	sort={p038},
	type=symbols,
	name={\ensuremath{\phi_i}},
	description={
		$\triangleright$ Local direction of the external magnetic field for the XY model%
		\newline
		$\triangleright$ Site contribution to the replicated free energy density $\phi_{\beta}$
	}
}
\newglossaryentry{phi_i_x_i}{
	sort={p039},
	type=symbols,
	name={\ensuremath{\phi_i(\boldsymbol{x}_i)}},
	description={External bias on the realization $\boldsymbol{x}_i$ of the $i$-th event}
}
\newglossaryentry{phi_ij}{
	sort={p040},
	type=symbols,
	name={\ensuremath{\phi_{ij}}},
	description={Edge contribution to the replicated free energy density $\phi_{\beta}$}
}
\newglossaryentry{psi_r}{
	sort={p041},
	type=symbols,
	name={\ensuremath{\psi(r)}},
	description={``Integrated correlation function'' obtained from a resummation on the components of the eigenvectors of the Hessian matrix~$\mathbb{H}$}
}
\newglossaryentry{psi_a_single_site}{
	sort={p042},
	type=symbols,
	name={\ensuremath{\psi_a(\boldsymbol{x}_i)}},
	description={The same of $\phi_i(\boldsymbol{x}_i)$}
}
\newglossaryentry{psi_a_generic}{
	sort={p043},
	type=symbols,
	name={\ensuremath{\psi_a(\underline{\boldsymbol{x}}_{\partial a})}},
	description={Compatibility function of the $a$-th interaction involving $\underline{\boldsymbol{x}}_{\partial a}$ variables}
}
\newglossaryentry{psi_ij}{
	sort={p044},
	type=symbols,
	name={\ensuremath{\psi_{ij}(\boldsymbol{x}_i,\boldsymbol{x}_j)}},
	description={Compatibility function of the pairwise interaction involving variables $\boldsymbol{x}_i$ and $\boldsymbol{x}_j$}
}

\newglossaryentry{small_q}{
	sort={q001},
	type=symbols,
	name={\ensuremath{q}},
	description={RS overlap between replicas}
}
\newglossaryentry{q_Parisi}{
	sort={q002},
	type=symbols,
	name={\ensuremath{q(x)}},
	description={Parisi overlap function}
}
\newglossaryentry{RSB_overlaps}{
	sort={q003},
	type=symbols,
	name={\ensuremath{q_0,q_1,q_2,\dots}},
	description={Overlaps between replicas in the RSB framework}
}
\newglossaryentry{q_ab}{
	sort={q004},
	type=symbols,
	name={\ensuremath{q_{ab}}},
	description={Overlap between the magnetizations of the $a$-th and the $b$-th replicas}
}
\newglossaryentry{q_ab_star}{
	sort={q005},
	type=symbols,
	name={\ensuremath{q^*_{ab}}},
	description={Saddle-point value of $q_{ab}$}
}
\newglossaryentry{q_alpha_beta}{
	sort={q006},
	type=symbols,
	name={\ensuremath{q_{\alpha\beta}}},
	description={Overlap between the magnetizations of the $\alpha$-th and the $\beta$-th pure states}
}
\newglossaryentry{q_EA}{
	sort={q007},
	type=symbols,
	name={\ensuremath{q_{\textnormal{EA}}}},
	description={Edwards\,-\,Anderson overlap}
}
\newglossaryentry{q_m}{
	sort={q008},
	type=symbols,
	name={\ensuremath{q_m}},
	description={Minimum value taken on by $q(x)$ --- at $x=x_m$ --- in the fRSB framework}
}
\newglossaryentry{q_M}{
	sort={q009},
	type=symbols,
	name={\ensuremath{q_M}},
	description={Maximum value taken on by $q(x)$ --- at $x=x_M$ --- in the fRSB framework}
}
\newglossaryentry{q_parallel}{
	sort={q010},
	type=symbols,
	name={\ensuremath{q_{\parallel}}},
	description={Overlap between the magnetizations along the field direction in the vector spin case}
}
\newglossaryentry{q_parallel_Parisi}{
	sort={q011},
	type=symbols,
	name={\ensuremath{q_{\parallel}(x)}},
	description={Parisi overlap function along the field direction in the vector spin case}
}
\newglossaryentry{q_perp}{
	sort={q012},
	type=symbols,
	name={\ensuremath{q_{\perp}}},
	description={Overlap between the magnetizations in the transverse plane with respect to the field direction in the vector spin case}
}
\newglossaryentry{q_perp_Parisi}{
	sort={q013},
	type=symbols,
	name={\ensuremath{q_{\perp}(x)}},
	description={Parisi overlap function in the transverse plane with respect to the field direction in the vector spin case}
}
\newglossaryentry{capital_Q}{
	sort={q014},
	type=symbols,
	name={\ensuremath{Q}},
	description={
		$\triangleright$ Number of states in the clock model approximation%
		\newline
		$\triangleright$ In~\autoref{sec:stat_mech_and_therm}: heat
	}
}
\newglossaryentry{capital_Q_prime}{
	sort={q015},
	type=symbols,
	name={\ensuremath{Q'}},
	description={Parameter in the probability distribution of the field local direction}
}
\newglossaryentry{capital_Q_star}{
	sort={q016},
	type=symbols,
	name={\ensuremath{Q^*}},
	description={Typical scale of the exponential decay of $\Delta f^{(Q)}$}
}

\newglossaryentry{small_r}{
	sort={r001},
	type=symbols,
	name={\ensuremath{r}},
	description={
		$\triangleright$ In~\autoref{chap:tools} and~\autoref{chap:XYinField_zeroTemp}: Modulus of the distance on the graph%
		\newline
		$\triangleright$ In~\autoref{sec:new_kind_mag}: Modulus of the distance in the real space%
		\newline
		$\triangleright$ In~\autoref{sec:beyond_RS}: redundancy factor in the 1RSB PDA
	}
}
\newglossaryentry{small_r_vector}{
	sort={r002},
	type=symbols,
	name={\ensuremath{\boldsymbol{r}}},
	description={Distance in the real space}
}
\newglossaryentry{position_r_i}{
	sort={r003},
	type=symbols,
	name={\ensuremath{\boldsymbol{r}_i}},
	description={Position of a gas particle}
}
\newglossaryentry{r_ChainSuscProp}{
	sort={r004},
	type=symbols,
	name={\ensuremath{\mathfrak{r}}},
	description={Survival rate of perturbations in the zero-temperature SuscProp algorithm}
}
\newglossaryentry{spectral_density}{
	sort={r005},
	type=symbols,
	name={\ensuremath{\rho(\lambda)}},
	description={Spectral density of the Hessian matrix $\mathbb{H}$}
}
\newglossaryentry{gas_prob_distr}{
	sort={r006},
	type=symbols,
	name={\ensuremath{\rho(\boldsymbol{r},\boldsymbol{p},t)}},
	description={Probability distribution of a gas particle}
}
\newglossaryentry{MB_gas_prob_distr}{
	sort={r007},
	type=symbols,
	name={\ensuremath{\rho^{(\infty)}(p)}},
	description={Equilibrium Maxwell-Boltzmann probability distribution for the momentum of a gas particle}
}
\newglossaryentry{real_part}{
	sort={r008},
	type=symbols,
	name={\ensuremath{\Re[\cdot]}},
	description={Real part of a complex-valued quantity}
}
\newglossaryentry{resolvent}{
	sort={r009},
	type=symbols,
	name={\ensuremath{\mathcal{R}(\lambda)}},
	description={Resolvent of the Hessian matrix $\mathbb{H}$}
}
\newglossaryentry{reduced_resolvent}{
	sort={r010},
	type=symbols,
	name={\ensuremath{\mathcal{R}^{(i)}(\lambda)}},
	description={Resolvent of the Hessian matrix $\mathbb{H}$ without its $i$-th row and column}
}
\newglossaryentry{cavity_resolvent}{
	sort={r011},
	type=symbols,
	name={\ensuremath{\widetilde{\mathcal{R}}_{i\to j}(\lambda)}},
	description={Cavity resolvent from variable node $i$ to variable node $j$}
}
\newglossaryentry{cavity_resolvent_star}{
	sort={r012},
	type=symbols,
	name={\ensuremath{\widetilde{\mathcal{R}}^*_{i\to j}(\lambda)}},
	description={BP fixed-point value of $\widetilde{\mathcal{R}}_{i\to j}(\lambda)$}
}

\newglossaryentry{entropy_density}{
	sort={s001},
	type=symbols,
	name={\ensuremath{s}},
	description={Entropy density}
}
\newglossaryentry{entropy}{
	sort={s002},
	type=symbols,
	name={\ensuremath{S}},
	description={Entropy}
}
\newglossaryentry{entropy_approx}{
	sort={s003},
	type=symbols,
	name={\ensuremath{S_0}},
	description={Entropy averaged over $\mathbb{P}_0$}
}
\newglossaryentry{S_cal}{
	sort={s004},
	type=symbols,
	name={\ensuremath{\mathcal{S}}},
	description={
		$\triangleright$ In~\autoref{chap:tools}: subset of the $N$ degrees of freedom%
		\newline
		$\triangleright$ In~\autoref{chap:sg_replica}: action%
		\newline
		$\triangleright$ In~\autoref{sec:int_GT_dAT}: set of XY spin directions allowed by the $Q$-state clock model
	}
}
\newglossaryentry{vector_spin}{
	sort={s005},
	type=symbols,
	name={\ensuremath{\boldsymbol{\sigma}}},
	description={Vector spin}
}
\newglossaryentry{Ising_spin}{
	sort={s006},
	type=symbols,
	name={\ensuremath{\sigma}},
	description={Ising spin}
}
\newglossaryentry{Ising_spin_replica}{
	sort={s007},
	type=symbols,
	name={\ensuremath{\sigma^{(a)}}},
	description={Ising spin of the $a$-th replica}
}
\newglossaryentry{Sigma_complexity}{
	sort={s008},
	type=symbols,
	name={\ensuremath{\Sigma_{\beta}(f)}},
	description={Configurational entropy or complexity}
}
\newglossaryentry{Sum_all_couples}{
	sort={s009},
	type=symbols,
	name={\ensuremath{\sum_{i,j}}},
	description={Sum over all the couples of indexes}
}
\newglossaryentry{Sum_all_couples_diff}{
	sort={s010},
	type=symbols,
	name={\ensuremath{\sum_{i\neq j}}},
	description={Sum over all the couples of different indexes}
}
\newglossaryentry{Sum_nn_couples}{
	sort={s011},
	type=symbols,
	name={\ensuremath{\sum_{(i,j)}}},
	description={Sum over the couples of indexes of nearest-neighbour variables}
}
\newglossaryentry{Sum_constrained}{
	sort={s012},
	type=symbols,
	name={\ensuremath{\sum_{p_1,p_2,\dots,p_k}^{(l)}}},
	description={Sum over all the choices of the positive integer indexes $\{p_k\}$ that algebraically sum to $l$}
}

\newglossaryentry{time}{
	sort={t001},
	type=symbols,
	name={\ensuremath{t}},
	description={Time, or time step in numerical algorithms}
}
\newglossaryentry{initial_time}{
	sort={t002},
	type=symbols,
	name={\ensuremath{t_0}},
	description={Reference (or initial) time}
}
\newglossaryentry{t_GCD}{
	sort={t003},
	type=symbols,
	name={\ensuremath{t^*_{\textnormal{GCD}}}},
	description={Average number of time steps to reach convergence within $\Delta_{\text{GCD}}$ accuracy in the GCD algorithm}
}
\newglossaryentry{t_GSA}{
	sort={t004},
	type=symbols,
	name={\ensuremath{t^*_{\textnormal{GSA}}}},
	description={Average number of time steps to reach convergence within $\Delta_{\text{GSA}}$ accuracy in the GSA}
}
\newglossaryentry{t_max}{
	sort={t005},
	type=symbols,
	name={\ensuremath{t_{\textnormal{max}}}},
	description={Maximum number of time steps in numerical algorithms}
}
\newglossaryentry{temperature}{
	sort={t006},
	type=symbols,
	name={\ensuremath{T}},
	description={Temperature}
}
\newglossaryentry{critical_temperature}{
	sort={t007},
	type=symbols,
	name={\ensuremath{T_c}},
	description={Critical temperature}
}
\newglossaryentry{critical_temperature_F}{
	sort={t008},
	type=symbols,
	name={\ensuremath{T_{\textnormal{F}}}},
	description={Critical temperature between the paramagnetic and the ferromagnetic phases}
}
\newglossaryentry{critical_temperature_mc}{
	sort={t009},
	type=symbols,
	name={\ensuremath{T_{mc}}},
	description={Critical temperature of the multicritical point}
}
\newglossaryentry{critical_temperature_nd}{
	sort={t010},
	type=symbols,
	name={\ensuremath{T_{nd}}},
	description={Value of $T$ in correspondence of a nondifferentiability point in the critical lines}
}
\newglossaryentry{critical_temperature_SG}{
	sort={t011},
	type=symbols,
	name={\ensuremath{T_{\textnormal{SG}}}},
	description={Critical temperature between the paramagnetic and the spin glass phases}
}
\newglossaryentry{tau}{
	sort={t012},
	type=symbols,
	name={\ensuremath{\tau}},
	description={Reduced temperature $(T_c-T)/T_c$}
}
\newglossaryentry{theta_XY}{
	sort={t013},
	type=symbols,
	name={\ensuremath{\theta_i}},
	description={Angular variable of the $i$-th XY spin}
}
\newglossaryentry{theta_i_star}{
	sort={t014},
	type=symbols,
	name={\ensuremath{\theta^*_i}},
	description={Direction of the angular variable of the $i$-th XY spin in the ground state configuration}
}
\newglossaryentry{vartheta_i}{
	sort={t015},
	type=symbols,
	name={\ensuremath{\vartheta_i}},
	description={Angle between the direction of the perturbed local magnetization and the external field on the $i$-th site of the graph}
}
\newglossaryentry{theta_clock}{
	sort={t016},
	type=symbols,
	name={\ensuremath{\theta_{i,a}}},
	description={Discretized angular variable of the $i$-th XY spin via the $Q$-state clock model}
}

\newglossaryentry{small_u}{
	sort={u001},
	type=symbols,
	name={\ensuremath{u}},
	description={Internal energy density}
}
\newglossaryentry{small_u_Q_above}{
	sort={u002},
	type=symbols,
	name={\ensuremath{u^{(Q)}}},
	description={Internal energy density of the $Q$-state clock model}
}
\newglossaryentry{internal_energy_site}{
	sort={u003},
	type=symbols,
	name={\ensuremath{u_i}},
	description={Site contribution to the total internal energy}
}
\newglossaryentry{internal_energy_edge}{
	sort={u004},
	type=symbols,
	name={\ensuremath{u_{ij}}},
	description={Edge contribution to the total internal energy}
}
\newglossaryentry{cavity_bias_ij}{
	sort={u005},
	type=symbols,
	name={\ensuremath{u_{i\to j}(\boldsymbol{x}_j)}},
	description={Cavity bias from variable node $i$ to variable node $j$ in the pairwise zero-temperature Belief Propagation approach}
}
\newglossaryentry{cavity_bias_ij_star}{
	sort={u006},
	type=symbols,
	name={\ensuremath{u^*_{i\to j}(\boldsymbol{x}_j)}},
	description={BP fixed-point value of $u_{i\to j}(\boldsymbol{x}_j)$}
}
\newglossaryentry{internal_energy}{
	sort={u007},
	type=symbols,
	name={\ensuremath{U}},
	description={Internal energy}
}
\newglossaryentry{generic_U}{
	sort={u008},
	type=symbols,
	name={\ensuremath{U(x)}},
	description={Generic $d=1$ potential}
}
\newglossaryentry{internal_energy_approx}{
	sort={u009},
	type=symbols,
	name={\ensuremath{U_0}},
	description={Internal energy averaged over $\mathbb{P}_0$}
}
\newglossaryentry{random_rotation}{
	sort={u010},
	type=symbols,
	name={\ensuremath{\mathbb{U}}},
	description={Random rotation matrix in the spin space}
}
\newglossaryentry{unif}{
	sort={u011},
	type=symbols,
	name={\ensuremath{\mathrm{Unif}\bigl([a,b]\bigr)}},
	description={Uniform distribution over the $[a,b]$ interval}
}
\newglossaryentry{ipr}{
	sort={u012},
	type=symbols,
	name={\ensuremath{\Upsilon}},
	description={Inverse Participation Ratio (IPR)}
}
\newglossaryentry{ipr_10_averaged}{
	sort={u013},
	type=symbols,
	name={\ensuremath{\overline{\Upsilon}_{10}}},
	description={Sample-averaged IPR of the first eigenvectors $\{\ket{\lambda}_i\}_{i=0,\dots,9}$}
}
\newglossaryentry{ipr_min_averaged}{
	sort={u014},
	type=symbols,
	name={\ensuremath{\overline{\Upsilon}_{\textnormal{min}}}},
	description={Sample-averaged IPR of $\ket{\lambda}_{\text{min}}$}
}

\newglossaryentry{vector}{
	sort={v001},
	type=symbols,
	name={\ensuremath{\boldsymbol{v}}},
	description={Generic vector}
}
\newglossaryentry{volume}{
	sort={v002},
	type=symbols,
	name={\ensuremath{V}},
	description={Volume}
}
\newglossaryentry{potential}{
	sort={v003},
	type=symbols,
	name={\ensuremath{\mathcal{V}(\cdot)}},
	description={Potential of a generic interaction}
}

\newglossaryentry{small_w}{
	sort={w001},
	type=symbols,
	name={\ensuremath{w}},
	description={Parameter in the probability distribution of the field local direction}
}
\newglossaryentry{w_alpha}{
	sort={w002},
	type=symbols,
	name={\ensuremath{w_{\alpha}}},
	description={Statistical weight of the $\alpha$-th state in the Gibbs measure}
}
\newglossaryentry{chain_cavity_bias_ij}{
	sort={w003},
	type=symbols,
	name={\ensuremath{w_{i\to j}(\boldsymbol{x}_j)}},
	description={Cavity bias from variable node $i$ to variable node $j$ in the pairwise zero-temperature Belief Propagation approach propagating along a chain}
}
\newglossaryentry{chain_cavity_bias_ij_star}{
	sort={w004},
	type=symbols,
	name={\ensuremath{w^*_{i\to j}(\boldsymbol{x}_j)}},
	description={BP fixed-point value of $w_{i\to j}(\boldsymbol{x}_j)$}
}

\newglossaryentry{Parisi_x}{
	sort={x001},
	type=symbols,
	name={\ensuremath{x}},
	description={
		$\triangleright$ In~\autoref{chap:sg_replica}: Parisi parameter in the fRSB framework%
		\newline
		$\triangleright$ In~\autoref{sec:beyond_RS}: Parisi parameter in the 1RSB framework on diluted graphs
	}
}
\newglossaryentry{x_star_above}{
	sort={x002},
	type=symbols,
	name={\ensuremath{x^*}},
	description={Equilibrium value of $x$ in the 1RSB framework on diluted graphs}
}
\newglossaryentry{Parisi_x_q}{
	sort={x003},
	type=symbols,
	name={\ensuremath{x(q)}},
	description={Inverse function of $q(x)$}
}
\newglossaryentry{x_zero}{
	sort={x004},
	type=symbols,
	name={\ensuremath{x_0}},
	description={Minimum of $U(x)$}
}
\newglossaryentry{Parisi_x_min}{
	sort={x005},
	type=symbols,
	name={\ensuremath{x_m}},
	description={Minimum value of the Parisi parameter $x$ in the fRSB framework}
}
\newglossaryentry{Parisi_x_max}{
	sort={x006},
	type=symbols,
	name={\ensuremath{x_M}},
	description={Maximum value of the Parisi parameter $x$ in the fRSB framework}
}
\newglossaryentry{generic_dof_x}{
	sort={x007},
	type=symbols,
	name={\ensuremath{\boldsymbol{x}_i}},
	description={Generic degree of freedom}
}
\newglossaryentry{x_S_subset}{
	sort={x008},
	type=symbols,
	name={\ensuremath{\underline{\boldsymbol{x}}_{\mathcal{S}}}},
	description={Set of variables belonging to the subset $\mathcal{S}$ of the graph}
}
\newglossaryentry{x_partial_a}{
	sort={x009},
	type=symbols,
	name={\ensuremath{\underline{\boldsymbol{x}}_{\partial a}}},
	description={Set of variables involved in the $a$-th interaction}
}

\newglossaryentry{small_y}{
	sort={y001},
	type=symbols,
	name={\ensuremath{y}},
	description={Effective local field in the fRSB framework}
}

\newglossaryentry{small_z_dots}{
	sort={z001},
	type=symbols,
	name={\ensuremath{z,z_0,z_1,\dots}},
	description={Auxiliary integration variables}
}
\newglossaryentry{Part_func_Z}{
	sort={z002},
	type=symbols,
	name={\ensuremath{\mathcal{Z}}},
	description={Canonical partition function}
}
\newglossaryentry{Part_func_Z_a}{
	sort={z003},
	type=symbols,
	name={\ensuremath{\mathcal{Z}_a}},
	description={Normalization constant of the function node in the factor graph formalism}
}
\newglossaryentry{Part_func_Z_a_to_i}{
	sort={z004},
	type=symbols,
	name={\ensuremath{\widehat{\mathcal{Z}}_{a\to i}}},
	description={Normalization constant of the cavity message $\widehat{\eta}_{a\to i}(\boldsymbol{x}_i)$ in the factor graph formalism}
}
\newglossaryentry{Part_func_Z_beta}{
	sort={z005},
	type=symbols,
	name={\ensuremath{\mathcal{Z}_{\beta}}},
	description={The same as $\mathcal{Z}$}
}
\newglossaryentry{Part_func_Z_beta_repl}{
	sort={z006},
	type=symbols,
	name={\ensuremath{\mathcal{Z}_{\beta}(m)}},
	description={$m$-times replicated canonical partition function in the 1RSB framework on diluted graphs}
}
\newglossaryentry{Part_func_Z_i}{
	sort={z007},
	type=symbols,
	name={\ensuremath{\mathcal{Z}_i}},
	description={Normalization constant of the one-node belief $\eta_i(\boldsymbol{x}_i)$}
}
\newglossaryentry{Part_func_Z_ia}{
	sort={z008},
	type=symbols,
	name={\ensuremath{\mathcal{Z}_{ia}}},
	description={Normalization constant of the edge between variable and function nodes in the factor graph formalism}
}
\newglossaryentry{Part_func_Z_ij}{
	sort={z009},
	type=symbols,
	name={\ensuremath{\mathcal{Z}_{ij}}},
	description={Normalization constant of the two-node belief $\eta_i(\boldsymbol{x}_i,\boldsymbol{x}_j)$}
}
\newglossaryentry{Part_func_Z_i_to_a}{
	sort={z010},
	type=symbols,
	name={\ensuremath{\mathcal{Z}_{i\to a}}},
	description={Normalization constant of the cavity message $\eta_{i\to a}(\boldsymbol{x}_i)$ in the factor graph formalism}
}
\newglossaryentry{Part_func_Z_i_to_j}{
	sort={z011},
	type=symbols,
	name={\ensuremath{\mathcal{Z}_{i\to j}}},
	description={Normalization constant of the cavity message $\eta_{i\to j}(\boldsymbol{x}_i)$}
}
\newglossaryentry{Part_func_Z_i_to_j_hat}{
	sort={z012},
	type=symbols,
	name={\ensuremath{\widehat{\mathcal{Z}}_{i\to j}}},
	description={Normalization constant of the cavity message $\widehat{\eta}_{i\to j}(\boldsymbol{x}_j)$}
}
\newglossaryentry{Part_func_Z_given_instance}{
	sort={z013},
	type=symbols,
	name={\ensuremath{\mathcal{Z}_J}},
	description={Partition function for a given instance of the quenched disorder}
}
\newglossaryentry{Part_func_Z_replicated}{
	sort={z014},
	type=symbols,
	name={\ensuremath{\mathcal{Z}_n}},
	description={Disorder-averaged partition function of the $n$-times replicated original system}
}

\newglossarystyle{symbolstyle}%
{%
	\glossarystyle{tree}%
	\renewcommand{\glossaryentryfield}[5]{%
		\hangindent0pt\relax 
		\parindent0pt\relax 
		\makebox[\glslistdottedwidth][l]%
		{%
			\glsentryitem{##1}\textbf{\glstarget{##1}{##2}}%
			\unskip\leaders\hbox to 2.9mm{\hss.}\hfill\strut 
		}%
	\parbox[t]{\linewidth-\glslistdottedwidth}{##3}\par\vspace{\baselineskip}}%
}

\usepackage{lipsum}
\usepackage{curve2e}
\definecolor{gray}{gray}{0.4}

\newcommand{\mcH}{\mathcal{H}}

\DeclareMathOperator{\sign}{sign}
\DeclareMathOperator{\sech}{sech}

\DeclareMathOperator*{\argmax}{argmax}
\DeclareMathOperator*{\Tr}{Tr}

\title{Critical properties of disordered XY model \\on sparse random graphs}
\author{Cosimo Lupo}
\IDnumber{1266587}
\course[Physics]{Fisica}
\courseorganizer{Scuola di dottorato Vito Volterra}
\cycle{XXIX}
\submitdate{February 2017}
\copyyear{2017}
\advisor{Prof. Federico Ricci\,-\,Tersenghi}
\authoremail{cosimo.lupo89@gmail.com}
\coadvisor{Prof. Florent Krzakala}
\coadvisor{Prof. Juan Jes\'us Ruiz\,-\,Lorenzo}

\examdate{16 February 2017}
\examiner{Prof. Mauro Carfora}
\examiner{Prof. Sergio Caracciolo}
\examiner{Dr. Antonello Scardicchio}
\versiondate{\today}

\begin{document}

\makeatletter
  \renewcommand{\SAP@ThesisCoAdvisorLabel}{External Referee}
  \renewcommand{\SAP@ThesisCoAdvisorsLabel}{External Referees}
  \renewcommand{\SAP@sapthesisInformationLabel}{The research leading to these results has received funding from the European Research Council (ERC) under the European Union's Horizon 2020 research and innovation programme, grant agreement No.\,694925 (\textbf{LoTGlasSy}).\\ \\This thesis has been typeset by \LaTeX\ and the Sapthesis class}
\makeatother

\frontmatter

\maketitle

\dedication{A Serena,\\\vspace{0.2cm}per quanto cercassi il senso delle cose,\\\vspace{0.2cm}la risposta sei sempre stata tu.}

\dedication{``Il caos \`e un ordine da decifrare.''\\\vspace{0.2cm}\textnormal{Jos\'e Saramago, L'uomo duplicato}\\\quad\\\quad\\\quad\\``\textit{Tutte le isole, anche quelle conosciute, sono sconosciute finch\'e non vi si sbarca.}''\\\vspace{0.2cm}\textnormal{Jos\'e Saramago, Il racconto dell'isola sconosciuta}}

\dedication{``Chaos is order yet undeciphered.''\\\vspace{0.2cm}\textnormal{Jos\'e Saramago, The double}\\\quad\\\quad\\\quad\\``\textit{Even known islands remain unknown until we set foot on them.}''\\\vspace{0.2cm}\textnormal{Jos\'e Saramago, The tale of the unknown island}}

\begingroup
	\makeatletter
	\let\ps@plain\ps@empty
	\begin{abstract}
		This thesis focuses on the XY model, the simplest vector spin model, used for describing numerous physical systems, from random lasers to superconductors, from synchronization problems to superfluids. It is studied for different sources of quenched disorder: random couplings, random fields, or both them. The belief propagation algorithm and the cavity method are exploited to solve the model on the sparse topology provided by Bethe lattices. It is found that the discretized version of the XY model, the so-called $Q$-state clock model, provides a reliable and efficient proxy for the continuous model with an error going to zero exponentially in $Q$, so implying a remarkable speedup in numerical simulations. Interesting results regard the low-temperature solution of the spin glass XY model, which is by far more unstable toward the replica symmetry broken phase with respect to what happens in discrete models. Moreover, even the random field XY~model with ferromagnetic couplings exhibits a replica symmetry broken phase, at variance with both the fully connected version of the same model and the diluted random field Ising model, as a further evidence of a more pronounced glassiness of the diluted XY model. Then, the instabilities of the spin glass XY model in an external field are characterized, recognizing different critical lines according to the different symmetries of the external field. Finally, the inherent structures in the energy landscape of the spin glass XY model in a random field are described, exploiting the capability of the zero-temperature belief propagation algorithm to actually reach the ground state of the system. Remarkably, the density of soft modes in the Hessian matrix shows a non-mean-field behaviour, typical of glasses in finite dimension, while the critical point of replica symmetry instability predicted by the belief propagation algorithm seems to correspond to a delocalization of such soft modes.
	\end{abstract}
	\cleardoublepage
\endgroup

\begingroup
	\makeatletter
	\let\ps@plain\ps@empty
	\begin{abstract}[Sommario]
		Questa tesi si concentra sul modello XY, il pi\`u semplice modello con spin vettoriali, usato per descrivere diversi sistemi fisici, dai random laser ai superconduttori, dal problema della sincronizzazione ai superfluidi. Viene studiato per diverse sorgenti di disordine quenched: accoppiamenti random, campi random, o entrambi. Il~modello XY viene risolto su grafi di Bethe grazie all'algoritmo di belief propagation e al metodo della cavit\`a. Si trova che la versione discreta del modello~XY, il cosiddetto clock model a $Q$ stati, fornisce un'approssimazione affidabile ed efficiente del modello continuo con un errore che va a zero esponenzialmente in~$Q$, fornendo cos\`i un notevole guadagno nelle simulazioni numeriche. La soluzione di bassa temperatura riserva risultati interessanti e inaspettati, essendo di gran lunga pi\`u instabile verso la rottura di simmetria delle repliche rispetto a quanto accade nei modelli discreti. Inoltre, persino il modello XY ferromagnetico in campo random mostra una fase con rottura di simmetria delle repliche, a differenza di quanto accade nell'analogo modello fully connected e nel modello di Ising ferromagnetico in campo random su grafi diluiti, ad ulteriore conferma di una maggiore vetrosit\`a del modello XY diluito. Poi, vengono caratterizzate le instabilit\`a del modello~XY spin glass in campo magnetico esterno, trovando cos\`i diverse linee critiche a seconda delle simmetrie del campo esterno. Infine, vengono studiate le strutture inerenti del panorama energetico del modello XY spin glass in campo random, sfruttando la capacit\`a dell'algoritmo di belief propagation a temperatura nulla di raggiungere esattamente il ground state del sistema. La densit\`a dei modi a pi\`u bassa energia nella matrice Hessiana mostra un comportamento non di campo medio, tipico dei sistemi vetrosi in dimensione finita, mentre il punto critico della rottura di simmetria delle repliche dato dall'algoritmo di belief propagation sembra corrispondere ad una delocalizzazione di tali modi soffici.
	\end{abstract}
	\cleardoublepage
\endgroup

\begingroup
	\makeatletter
	\let\ps@plain\ps@empty
	\begin{acknowledgments}
		This thesis is the final step of a long path started even before the PhD. Hence, there are a lot of people I should say thank you to. First of all, I would like to thank Federico. For me, he has been by far more than just a thesis advisor. I owe him a lot, starting from having taught me how to actually do research. Thanks to him, I learned that there is always a smarter way to look at things, it is just enough to find it. 	Then, I would like to thank Giorgio Parisi. His suggestions have always been unvaluable, providing each time a unique point of view about things. He has been for me an infinite source of new ideas and solutions. I should then thank all the large family of ``Chimera'' group at Sapienza, especially Andrea Maiorano and Luca Leuzzi, both always available for any question; I am very grateful to them. The two external referees, Florent Krzakala and Juan Jes\'us Ruiz-Lorenzo, deserve a special mention, for a sincere interest exhibited for my research and for useful suggestions about it. Last, but not the least, I would like to thank all the people with whom I had several stimulating discussions: Ada Altieri, Fabrizio Antenucci, Marco Baity-Jesi, Francesco Concetti, Carlo Lucibello, Enrico Malatesta, Alessia Marruzzo, Matteo Mori, Gabriele Perugini, Jacopo Rocchi, Francesco Zamponi, as well as all the other people I had the pleasure to met at Sapienza in Rome and during all the conferences, workshops and schools I took place in.
	\end{acknowledgments}
	\cleardoublepage
\endgroup

\begingroup
	\makeatletter
	\let\ps@plain\ps@empty
	\tableofcontents
	\cleardoublepage
\endgroup

\chapter*{\hypertarget{chap:intro}{Introduction}}
\markboth{Introduction}{Introduction}
\addcontentsline{toc}{chapter}{Introduction}
\thispagestyle{empty}

Starting from their very first introduction~\cite{EdwardsAnderson1975} --- more than forty years ago --- spin glasses have now become ubiquitous, being paradigmatic of all those systems where some source of disorder introduces a frustration, namely the impossibility of fully satisfying all the clauses. Then, more broadly, spin glass phenomenology has been recovered every time there is a consistent heterogeneity that can not be effectively described by ``ordered'' statistical mechanics.

Furthermore, in addition to this, the contextually developed tools --- namely the replica method~\cite{Parisi1979b, Parisi1980b} and the cavity method~\cite{MezardParisi2001, MezardParisi2003} --- turned out to be so powerful that they are now applied in an overwhelming quantity of fields apparently unrelated to spin glasses, from combinatorial optimization~\cite{MezardParisi1985, MezardEtAl2002, KrzakalaEtAl2007} to neural networks~\cite{Hopfield1982, AmitEtAl1985, Book_Amit1989, MonassonZecchina1995}, from supercooled liquids and glass forming~\cite{KirkpatrickThirumalai1987, KirkpatrickEtAl1989, Cavagna2009, Book_Gotze2009, ParisiZamponi2010, BerthierBiroli2011} to immunology~\cite{Parisi1990, AgliariEtAl2012}, from inference~\cite{Book_MacKay2003, MorcosEtAl2011, DecelleEtAl2011, ZdeborovaKrzakala2016} to learning~\cite{Book_EngelVanDenBroeck2001}, from finance~\cite{Book_BouchaudPotters2003} to random interfaces~\cite{MezardParisi1991}, from epidemic spreading~\cite{AltarelliEtAl2014, MoroneMakse2015} to game theory~\cite{ChalletEtAl2000, DallAstaEtAl2012}, from photonics~\cite{LeuzziEtAl2009, ContiLeuzzi2011} to random matrices~\cite{Kuhn2008, RogersEtAl2009}, from collective behaviour~\cite{BalleriniEtAl2008, BialekEtAl2014} to computer science~\cite{Book_Nishimori2001, Book_MezardMontanari2009}.

A predominant role in the statistical mechanics has been always played by one of the most simple model one could define, the Ising model~\cite{Ising1925}. Despite the minimal choice for each variable of the system --- on/off, up/down, 0/1 --- it has become the archetypal of any system in which the interaction and the imitation between ``neighbours'' has a key importance in the explanation of the relevant features of the system itself~\cite{Book_Huang1988, Book_Parisi1988}.

Of course, the prototype of a spin glass system could not have been other than the disordered version of the Ising model~\cite{SherringtonKirkpatrick1975, KirkpatrickSherrington1978}, whose exact solution has required a few years to be developed~\cite{Book_MezardEtAl1987} but more than twenty years to be rigorously proven~\cite{GuerraToninelli2002, Book_Talagrand2003}.

Though being effective in a huge number of cases, however, the discrete nature of the Ising model does not allow to model those systems where small fluctuations and smooth changes of configurations have to be taken into account. In this sense, the usage of \textit{soft spins} like the Ginzburg-Landau ones~\cite{Book_Parisi1988} turns out to be helpful.

Moreover, many other physical systems do not show any such strong anisotropy to justify the use of $z$-aligned spins like the Ising and Ginzburg-Landau ones. The generalization of uniaxial spins to vector spins defined in an $m$-dimensional inner space is the necessary --- and immediate --- step forward.

The continuous nature of vector spins allows a plethora of new physical phenomena to occur, from spin waves to the creation of vortexes. Indeed, continuous symmetries can be broken in a number of different ways, corresponding in turn to different instabilities, different properties at the critical point and --- most remarkably --- different phase transitions. Maybe the most famous among them is the Berezinski{\u{\i}}\,-\,Kosterlitz\,-\,Thouless one~\cite{Berezinskii1971, Berezinskii1972, KosterlitzThouless1972, KosterlitzThouless1973} involving the XY model in two-dimensional regular lattices.

It is just the XY model --- the simplest vector spin model, with $m=2$ spin components --- that has acquired a remarkable importance in many fields of statistical physics, due to its capability of correctly reproducing the features of many physical systems, from granular superconductors~\cite{JohnLubensky1985, HuseSeung1990} to superfluid Helium
~\cite{Minnhagen1987, Book_Brezin1989}, from synchronization problems
~\cite{Book_Kuramoto1975, AcebronEtAl2005, SkantzosEtAl2005, BandeiraEtAl2016} to random lasers~\cite{AntenucciEtAl2015, Thesis_Antenucci2016, AntenucciEtAl2016, Thesis_Marruzzo2015}, just for citing a few.

Due to the aforementioned physical applications, most of the works about the XY model refer to the finite dimensional topology. However, when inserting some defects in the system, e.\,g. by diluting the lattice, long-range order provided by Berezinski{\u{\i}}\,-\,Kosterlitz\,-\,Thouless vortexes is no longer possible, continuous symmetries are broken and some kind of arrangement with respect to the local topology takes place. In this way, the resulting physical picture can be better described through the mean-field framework provided by sparse random graphs.

This kind of mean-field approach is by far more ``physical'' than the one provided by the fully connected topology, since the key concepts of heterogeneity and distance find here a natural interpretation, getting closer to the finite dimensional case. Moreover, an effective approach to solve systems defined on sparse graphs is provided by the cavity method and the related belief propagation technique, developed in the context of inference networks~\cite{Book_Pearl1988, Book_JensenNielsen2007}.

Notwithstanding the importance of continuous-variable models in the physics of disordered systems, there are very few results about them on sparse random graphs, even despite the flourishing literature about discrete models on the same topology. This scarcity is strictly related to the difficulty to handle continuous variables both analytically, due to more cumbersome computations with respect to the case of discrete variables, and numerically, due to very demanding simulations and to the problem of efficiently discretizing such variables.

A further motivation that recently become to boost the interest in continuous-variable models is the very deep difference in the nature of continuous-variable models with respect to discrete-variable models, especially at the critical point. Indeed, the possibility of having low-energy excitations opens the doors to soft long-range correlations even in the zero-temperature limit, with a whole different phenomenology with respect to discrete models. The most direct way to appreciate these strong differences is to look at the (free) energy landscape in the space of configurations of the degrees of freedom of a system, which provides several precious information about its equilibrium properties and even beyond the equilibrium. Indeed, the characterization of the energy landscape~\cite{Hastings2000, DebenedettiStillinger2001, CharbonneauEtAl2014a, Thesis_BaityJesi2016, JinYoshino2017} is considerably capturing the researchers' attention in maybe one of the most appealing though challenging problem of the nowadays physics: the glass transition in structural glasses~\cite{Book_LeuzziNieuwenhuizen2008, Cavagna2009, BerthierBiroli2011, Book_WolynesLubchenko2012, CharbonneauEtAl2017}.

Given all these reasons, it would be a great achievement to fully understand the physics of a simple disordered model with continuous variable, among which the aforementioned XY model plays a predominant role. Hence, this thesis is devoted to shed light on the comprehension of the properties of the disordered XY model, focusing on a sparse random topology so to get more insights on the finite dimensional case and to get rid of some unphysical, misleading predictions coming from the mean field approach of fully connected graphs.

The topics dealt with in this thesis are organized according to a ``logic'' sequence, starting from simpler cases up to the most cumbersome situations. Actually, it turns out to be also the same order how they have been developed during the Ph.\,D. program, with the first topics being propaedeutic for the following ones. There is a total number of five parts, including some preliminary material and the conclusive remarks and perspectives about the thesis results.

\autoref{part:Preliminaries} is meant to provide a general introduction to the field of disordered systems and to the related tools typically exploited in it. Our aim is two-fold: on one hand to make comfortable those readers not very well experienced in spin glasses, and on the other hand to remind main features and results about vector spin glasses on fully connected graphs. In more detail, in~\autoref{chap:tools} we introduce the reader to the fascinating world of statistical mechanics, recalling the fundamental goal of finding the Boltzmann measure of a generic statistical system, describing its thermodynamic properties at equilibrium. Apart from some particular cases, in most of times it is a hard problem and some further assumptions or simplifications have to be taken into account. We start from the most common approach, the ``na\"ive'' mean field, showing that if on one hand it is enlightening about the physics of the model studied, on the other hand the results are often misleading with respect to what actually happens in the ``real world''. Hence, we move to a more refined approach, getting closer to the topology of real systems by means of sparse random graphs and the \textit{belief propagation} approach, which can be effectively exploited on them. In~\autoref{chap:sg_replica} then we actually introduce spin glasses, a special class of \textit{disordered systems} in which the \textit{quenched} disorder is introduced directly in the Hamiltonian ``from the outside'' through random couplings, random fields or random topologies. In particular, for historical reasons the fully connected topology with Gaussian distributed couplings is analyzed. Even though far from actual magnetic alloys --- for which they were introduced --- they constitute the basis above which all the replica theory has been built, starting from the seminal works of Sherrington and Kirkpatrick to the outstanding exact solution provided by Parisi. We briefly review these results --- focusing in particular on vector spin glasses --- in order to set the starting point for the analysis of the sparse topology.

First original results, appeared in Ref.~\cite{LupoRicciTersenghi2017a}, are discussed in~\autoref{part:XYmodelNoField}, where we actually introduce the XY model. The main goal of this part is to characterize the spin glass version of this model on sparse random graphs in absence of any external field. In~\autoref{chap:XYnoField} we exploit the belief propagation approach to analytically solve it. Then, in order to describe the low-temperature region, we also introduce the corresponding numeric algorithms. Similar algorithms are used throughout the thesis, so in this sense this chapter is indeed preparatory for what follows. The typical phases of the temperature versus ferromagnetic bias phase diagram are recognized, together with a peculiar glassy behaviour distinguishing vector spin glasses from scalar ones in the low-temperature limit. Given the need for an efficient discretization of continuous variables in numeric simulations, in~\autoref{chap:clock} we analyze the discretized version of the XY model, the so-called \textit{$Q$-state clock model}, focusing on the effects of such discretization. By looking at the phase diagram, a fast convergence of the critical lines is found to occur for a rather small number $Q$ of states of the clock model. The same happens for physical observables, that show an exponential convergence in $Q$. Finally, also the universality class is examined through a replica symmetry broken ansatz, again expliciting its dependence on~$Q$. This chapter provides also the chance to highlight the key difference between discrete and continuous disordered models, in particular from the algorithmic point of view.

In~\autoref{part:XYmodelInField} we insert a magnetic field, so to analyze the response of the XY model in presence of external perturbations. Being the XY spins modelized as $m=2$-component vectors, also the external field is characterized as a (site-dependent) two-dimensional vector as well. \autoref{chap:RFXY} is meant to reverse the point of view with respect to previous chapters, with the quenched disorder provided via a suitable random field, while interactions between spins are purely ferromagnetic. The resulting model, better known as the~\textit{random field XY model}, is hence analyzed via the previously developed belief propagation algorithms. A particular attention is devoted to the search of a replica symmetric unstable phase, whose presence can not be excluded by analytic arguments as it instead occurs for the random field Ising model. Again, as a combined effect of the sparsity of the underlying topology and of the presence of continuous degrees of freedom, a strong glassy behaviour is found to take place in the region of very low temperatures. Corresponding reference, at the stage of a working paper, is~\cite{LupoRicciTersenghi2017c}. In~\autoref{chap:XYinField} we finally join the two previous scenari, facing the \textit{spin glass XY model in a field}. According to the direction of the magnetic field --- uniform over the whole system or randomly chosen for each site --- different phase transitions occur, corresponding in turn to the breaking of different symmetries. So de Almeida\,-\,Thouless and Gabay\,-\,Toulouse instability lines known from the fully connected case can be recovered also in the sparse case. However, due to the heterogeneity allowed by the topology used, a deeper characterization of the spin symmetries can be performed by looking at perturbations around the belief propagation fixed point. Finally, intermediate distributions of the field direction are analyzed, so to study the crossover between the two types of phase transitions. All the results of this Chapter can be found in Ref.~\cite{LupoRicciTersenghi2017b}.

Once having fully characterized the behaviour of the disordered XY model with and without an external magnetic field, in~\autoref{part:EnergyLandscape} we focus on a specific situation, the zero-temperature limit of the spin glass XY model in a random field. Our main aim is to regard it as a simple, exactly solvable disordered model in which low-energy excitations can take place, focusing in particular on the features of its energy landscape when approaching the critical point. Hence, in~\autoref{chap:XYinField_zeroTemp} we exactly compute the ground state of several instances of such system when lowering the strength of the external field via the zero-temperature belief propagation algorithm. Then, we explore the energy landscape via the Hessian matrix of the energy function, looking at first at its spectral density. Indeed, the rugged, disorder-dependent nature of the energy landscape is believed to be strongly related to the occurrence of an anomalous excess of low-frequency modes in the spectral density --- known as the \textit{boson peak}~\cite{MalinovskySokolov1986, BeltukovParshin2016} --- in turn related to peculiar features of structural glasses~\cite{GrigeraEtAl2003, XuEtAl2007}. Moreover, we try to relate the occurrence of replica symmetry breaking for a low enough value of the field strength with some kind of delocalization in the lowest eigenvectors of the Hessian, likely corresponding to flat extended directions in the energy landscape. These results are also exposed into the working paper~\cite{LupoEtAl2017}.

Before starting with the first chapter of this thesis, the author claims that all the results presented in this thesis are original, if not otherwise contextually stated. In this latter case, suitable bibliographic references are provided, as well as when comparing new results with those already obtained by other authors. So all the references listed in this thesis come from a careful bibliographic research. However, it may occur to the reader to notice that a particular result has already been obtained in a different work the author was not aware of, or that some representative reference is missing. In that unpleasant occurrence --- as well as in case of any misprint or other kind of error --- the reader is kindly asked to inform the author about it, so that it will be fixed in a possible future version of the thesis.

\clearpage{\pagestyle{empty}\cleardoublepage}

\mainmatter

\begingroup
	\makeatletter
	\let\ps@plain\ps@empty
	\part{Preliminaries}
	\label{part:Preliminaries}
	\cleardoublepage
\endgroup

\chapter{Statistical mechanics and the mean field}
\label{chap:tools}
\thispagestyle{empty}

We start this Chapter providing a very brief introduction to statistical mechanics and its tools, showing its connection with the thermodynamics and applying it to the study of a paradigmatic model for magnetic systems, the Ising model. Then, we show how the Boltzmann law of statistical mechanics at equilibrium can be derived as a variational principle, so allowing the possibility of a mean field description of the system. A first, na\"{i}ve mean field approach is described, corresponding to a fully connected topology, namely weak and long-range interactions. Consequently, starting from the Bethe approximation, a more refined approach is presented, the belief propagation one. In this way it is possible to get closer to the physics of real, finite-dimensional systems, where short-range interactions typically take place. Finally, we review some key features of sparse random graphs, on which the belief propagation approach is shown to be correct.

\section{Statistical mechanics and thermodynamics}
\label{sec:stat_mech_and_therm}

The foundation of statistical mechanics grounds on the attempt to describe a perfect gas of particles of mass $m$ in thermal equilibrium at temperature $T$. Each particle has its own position $\boldsymbol{r}_i$ and momentum $\boldsymbol{p}_i$, while $\mathcal{H}[\{\boldsymbol{p}_i\}]=\sum_i p_i^2/2m$ is the Hamiltonian describing the whole system. Given the set of initial conditions, the trajectories of each particle can be obtained by integrating the Hamilton equations:
\begin{equation}
	\dot{\boldsymbol{r}}_i = \pder{\mathcal{H}}{\boldsymbol{p}_i} \qquad , \qquad \dot{\boldsymbol{p}}_i = -\pder{\mathcal{H}}{\boldsymbol{r}_i}
\end{equation}
where derivatives on the left hand sides are meant as total derivatives with respect to the time $t$. Actually, even gases with very low densities are made up of a huge number of particles, hence direct integration of Hamilton equations is unfeasible. A probabilistic approach is so needed.

Keeping in mind the Liouville theorem and the fact that in a perfect gas the unique interactions between particles are provided by pairwise collisions, the Boltzmann equation for the single-particle probability distribution $\rho(\boldsymbol{r},\boldsymbol{p},t)$ can be easily obtained~\cite{Book_Huang1988}. A stationary solution of this equation is nothing but the well known Maxwell\,-\,Boltzmann distribution of single-particle momentum:
\begin{equation}
	\rho^{(\infty)}(p)=\frac{n}{\left(2\pi m k_B T\right)^{3/2}}\,e^{-p^2/2m k_B T}
	\label{eq:Maxweel_Boltzmann_distr}
\end{equation}
with $n=N/V$ being the density of the gas, and $k_B$ the Boltzmann constant.

This result turns out to be by far more general than how it could appear. Indeed, the exponential just contains the single-particle Hamiltonian, $\exp{\{-\beta\mathcal{H}_i\}}$, so even when the ideal gas hypothesis is relaxed by taking into account a generic pairwise interaction $\mathcal{V}(\boldsymbol{r}_i-\boldsymbol{r}_j)$ between the particles, then the equilibrium probability distribution can still be written in this form. The unique difference is that now the whole Hamiltonian is no longer \textit{factorizable} into single-particle contributions, due to the pairwise interactions between the gas particles, and hence also the whole probability distribution can not factorize any longer as in~\autoref{eq:Maxweel_Boltzmann_distr}:
\begin{equation}
	\mathbb{P}_{\text{eq}}(\{\boldsymbol{r}_i,\boldsymbol{p}_i\}) \equiv \frac{e^{-\beta\mathcal{H}[\{\boldsymbol{r}_i,\boldsymbol{p}_i\}]}}{\mathcal{Z}}
\end{equation}
with the \textit{partition function} $\mathcal{Z}=\Tr_{\{\boldsymbol{r}_i,\boldsymbol{p}_i\}}e^{-\beta\mathcal{H}[\{\boldsymbol{r}_i,\boldsymbol{p}_i\}]}$ enforcing the correct normalization of $\mathbb{P}_{\text{eq}}$ as a probability distribution. This is nothing but the canonical ensemble description of equilibrium statistical mechanics, also known as the Boltzmann law. Its derivation can also be performed starting from the microcanonical ensemble and the ergodic hypothesis, as for example shown in~\cite{Book_Huang1988}.

More generally, the Boltzmann law is so powerful that it represents a standard tool for providing the equilibrium properties of every Hamiltonian system with $N \gg 1$ generic degrees of freedom~$\{\boldsymbol{x}_i\}$
\begin{equation}
	\mathbb{P}_{\text{eq}}(\{\boldsymbol{x}_i\}) \equiv \frac{e^{-\beta\mathcal{H}[\{\boldsymbol{x}_i\}]}}{\mathcal{Z}}
	\label{eq:Boltzmann_law}
\end{equation}
irrespective of their nature: they could be positions and momenta of gas particles, vibrational and rotational degrees of freedom of molecules, magnetic spins of a regular lattice of atoms, and so on.

The equilibrium description provided in the canonical ensemble by Boltzmann distribution is directly related with the thermodynamics of the system. Indeed, from the knowledge of the partition function~$\mathcal{Z}$ we can define the Helmholtz free energy~$F$:
\begin{equation}
	F \equiv -\frac{1}{\beta}\ln{\mathcal{Z}}
	\label{eq:Helmholtz_F_stat}
\end{equation}
which is completely equivalent to the one defined in the thermodynamics:
\begin{equation}
	F \equiv U-TS
	\label{eq:Helmholtz_F_thermo}
\end{equation}
The demonstration is straightforward, once assumed that $\beta$ is nothing but the \textit{inverse absolute temperature} $T$ of thermodynamics
\begin{equation}
	\beta = \frac{1}{k_B T}
\end{equation}
with Boltzmann constant $k_B$ set equal to $1$ from now on, and recognizing that the macroscopic energy~$U$ and the macroscopic entropy~$S$ can be obtained by averaging over all the possible configurations $\{\boldsymbol{x}_i\}$ according to their Boltzmann weight $\mathbb{P}_{\text{eq}}(\{\boldsymbol{x}_i\})$:
\begin{equation}
	U=\braket{\mathcal{H}} \qquad , \qquad S=\braket{-\ln{\mathbb{P}}}
\end{equation}
with $\braket{\cdot}$ also known as \textit{thermal average}:
\begin{equation}
	\braket{\mathcal{O}(\{\boldsymbol{x}_i\})} \equiv \Tr_{\{\boldsymbol{x}_i\}}\mathbb{P}_{\text{eq}}(\{\boldsymbol{x}_i\})\,\mathcal{O}(\{\boldsymbol{x}_i\})
\end{equation}
Being $\lambda$ an external parameter whose variation implies an energy change in the system, we have that:
\begin{equation}
\begin{split}
	\di\left(-\beta F\right) &= \pder{\ln{\mathcal{Z}}}{\beta}\di\beta + \pder{\ln{\mathcal{Z}}}{\lambda}\di\lambda\\
	&= -\braket{\mathcal{H}}\di\beta - \beta\braket{\der{\mathcal{H}}{\lambda}}\di\lambda\\
	&= -U\di\beta - \beta\,\delta L
\end{split}
\label{eq:differential_F}
\end{equation}
since the average value of the derivative of $\mathcal{H}$ with respect to $\lambda$ gives the infinitesimal work $\delta L$ applied on the system.

In turn, work is linked to the energy and to heat exchanges through the first principle of thermodynamics:
\begin{equation}
	\delta Q \equiv T\di S = \di U - \delta L
	\label{eq:1st_thermo_princ}
\end{equation}
from which:
\begin{equation}
	\beta\,\delta L = \beta\di U - \di S
\end{equation}
Substituting back into~\autoref{eq:differential_F} we get:
\begin{equation}
	\di\left(-\beta F\right) = -U\di\beta - \beta\di U + \di S = \di\left(-\beta U + S\right)
\end{equation}
so recovering the thermodynamic definition of the Helmholtz free energy~\autoref{eq:Helmholtz_F_thermo}.

Then, usual thermodynamic functions can be obtained from the derivatives of~$F$ instead of computing them directly from $\mathbb{P}_{\text{eq}}$:
\begin{equation}
	U=\pder{(\beta F)}{\beta} \qquad , \qquad S=-\pder{F}{T}
\end{equation}
and so on for the other physical observables as the average position and momentum of particles, the average magnetization of spins, etc.

\section{The Ising model}
\label{sec:Ising_ferro}

If one of the earliest tasks of statistical mechanics has been the description of a gas of particles, here we describe another milestone of this theory, concerning magnetic interactions.

Even though the exact description of the interaction between electrons and nuclei at the microscopic level --- which is at the very basis of magnetic phenomena~--- requires a quantum mechanical treatment, a classical description is still possible by focusing on few interesting degrees of freedom, such as the magnetic moments $\boldsymbol{\sigma}_i$'s generated by the motion of electrons around atoms or molecules inside the medium, also known as \textit{spins}.

They can be described as $m$-dimensional unit vectors, whose orientation is randomized by thermal excitation. Moreover, they are subjected to the aligning effects of the external field $\boldsymbol{H}$ --- if any --- and of their mutual interaction. All these features can be embedded into a statistical mechanics model, whose basic Hamiltonian reads:
\begin{equation}
	\mathcal{H}[\{\boldsymbol{\sigma}_i\}]=-\sum_{i \neq j}J_{ij}\,\boldsymbol{\sigma}_i\cdot\boldsymbol{\sigma}_j-\sum_i\boldsymbol{H}\cdot\boldsymbol{\sigma}_i
\end{equation}
where exchange couplings $J_{ij}$'s take into account the mutual interaction in each  couple $(\boldsymbol{\sigma}_i,\boldsymbol{\sigma}_j)$ of magnetic moments.

In most of materials, the interaction between close spins is such that the exchange energy lowers when they align, so that their interaction is said to be \textit{ferromagnetic}. This is obtained by setting all couplings equal to positive values, $J_{ij}>0$. Moreover, the strength of the magnetic interaction depends on the distance between the two atomic (or molecular) sites and rapidly decreases with it. Hence, the interactions are typically restricted to those couples of spins that are nearest neighbours on the considered topology, e.\,g. a $d$-dimensional hypercubic lattice $\mathcal{G}$. In this way one gets $J_{ij}=J$ when the couple $i,j$ is an \textit{edge} of $\mathcal{G}$ --- and it will be referred to as $(i,j)$ --- and $J_{ij}=0$ elsewhere.

Finally, a further simplification occurs when the spins are projected onto a global direction, e.\,g. the $z$ axis, so that the resulting degrees of freedom are scalar spins, $\sigma_i=\pm 1$. From the classical point of view, such projection can be justified by some strong anisotropy present in the system, while it automatically comes out when moving to the quantum mechanical description of the spin angular moment. The corresponding model is universally known as the \textit{Ising model}:
\begin{equation}
	\mathcal{H}[\{\sigma_i\}]=-J\sum_{(i,j)}\sigma_i\,\sigma_j-H\sum_i\sigma_i
	\label{eq:Hamiltonian_ferro_Ising}
\end{equation}

Invented by Lenz in 1920~\cite{Lenz1920}, who gave it to his student Ising as a topic for his doctoral thesis~\cite{Ising1925}, it has become one of the most studied theoretical models in statistical mechanics. Indeed, once written its equilibrium probability distribution
\begin{equation}
	\mathbb{P}_{\text{eq}}(\{\sigma_i\})=\frac{1}{\mathcal{Z}}\prod_{(i,j)}e^{\,\beta J\sigma_i\,\sigma_j}\prod_i e^{\,\beta H\sigma_i}
\end{equation}
and then computed $\mathcal{Z}$ and $F$, all the key features of typical ferromagnets can be caught, despite its apparent simplicity: the increasing alignment to the external field~$H$ when lowering the temperature, or when $H=0$ the presence of a ferromagnetic ordering at low temperatures and the absence of any ordering at high temperatures, with a second-order phase transition separating the two regimes. Moreover, it has become the paradigm of all those systems in which some ``degrees of freedom'' interact in an imitative way, while ``thermal excitations'' from the outside act against this ordering.

The $d$-dimensional hypercubic lattice is the most interesting topology on which solve the Ising model, due to the close connection with typical geometries of solid state physics. Even though the behaviour sketched above can be checked numerically quite easily also in this case, its analytic solution represents one of the most challenging task. Indeed, for the $d=1$ case the solution can be obtained quite easily via the transfer matrix approach~\cite{Book_Huang1988, Book_Parisi1988} and already Ising himself showed that no phase transitions occur in this case. Instead, the solution for the $d=2$ case requires a quite sophisticated and involving computation, firstly performed by Onsager in 1944~\cite{Onsager1944}, who showed that a phase transition actually takes place at a certain critical temperature $T_c=2/\ln{(1+\sqrt{2})}$. Finally, the $d \geqslant 3$ case still lacks of an analytic solution, even though the presence of a phase transition is certain and the values of the corresponding critical exponents have been analytically and numerically computed with a remarkable precision.

\section{Variational approaches on the Gibbs free energy}
\label{sec:var_appr_Gibbs_free_en}

The Ising model is an archetypal of the way of proceeding in statistical mechanics. Indeed, starting from a quite involving problem --- such as the description of magnetic interactions between electrons and nuclei --- one has to identify the relevant degrees of freedom~$\{\boldsymbol{x_i}\}$ and write an effective Hamiltonian $\mathcal{H}[\{\boldsymbol{x_i}\}]$ that catches all the key features of the original model. Then, the partition function~$\mathcal{Z}$ has to be computed and from it the Helmholtz free energy $F$. Finally, the derivatives of $F$ give the other interesting equilibrium physical observables.

However, though appearing quite straightforward, in most cases the bottleneck of this procedure is represented by the computation of partition function $\mathcal{Z}$. Indeed, when the equilibrium probability distribution $\mathbb{P}_{\text{eq}}$ does not factorize into single-particle contributions, the trace has to be performed over all the possible states --- discrete or even continuous --- of each single degree of freedom, so implying an overwhelming computational effort.

This is for example the main difficulty in the search of the analytic solution for the Ising model on $d \geqslant 3$-dimensional hypercubic lattices. Even worse, it occurs when some kind of \textit{heterogeneity} is present in the model, as it will be throughout this dissertation. So in most cases of interest a ``brute force'' computation of $\mathcal{Z}$ is unfeasible, compelling one to search for some approximation.

The kind of approximation to be used can be suggested from those cases in which the computation of $\mathcal{Z}$ is easy. Indeed, as already stated before, the probability distribution of the whole system factorizes when degrees of freedom are actually independent. In this way, the computation can be accomplished by tracing over the states of a single degree of freedom. In those cases when degrees of freedom are no longer actually independent, it could still make some sense to consider some kind of factorization over the whole set of degrees of freedom. For example, when interactions are short-range, then it could be reasonable to consider as independent those variables being \textit{spacially} well separated.

The usual trick exploited in these cases is to try to ``simplify'' the actual equilibrium probability distribution $\mathbb{P}_{\text{eq}}$ by involving only some small subset $\mathcal{S}$ of variables at each time:
\begin{equation}
	\mathbb{P}_{\text{eq}}(\{\boldsymbol{x}_i\}) \qquad \to \qquad \mathbb{P}_0(\{\boldsymbol{x}_i\}) \sim \prod_{\mathcal{S}}\mathbb{P}_{\mathcal{S}}(\{\boldsymbol{x}'_i\}_{\boldsymbol{x}'_i\in\mathcal{S}})
\end{equation}
instead of involving all the degrees of freedom at the same time. In this way, there is some chance to make feasible the computation of the partition function $\mathcal{Z}$.

Before going on in analyzing this kind of approximation, let us try to understand its meaning. Indeed, let us suppose to use a generic probability distribution~$\mathbb{P}_0$ instead of the Boltzmann one~\autoref{eq:Boltzmann_law}. In order to deal with well defined probability distributions, it has to hold:
\begin{equation}
	\mathbb{P}_0 \succeq 0 \qquad , \qquad \Tr{\mathbb{P}_0}=1
\end{equation}
where the trace is meant to run again over all the degrees of freedom. Notwithstanding $\mathbb{P}_0$ being a generic probability distribution and not the actual equilibrium one, we can still define a \textit{functional} free energy having exactly the same structure of the Helmholtz one in~\autoref{eq:Helmholtz_F_thermo}. This is known as \textit{Gibbs free energy}:
\begin{equation}
	G[\mathbb{P}_0] \equiv \Tr{\left(\mathbb{P}_0\,\mathcal{H}\right)} - \frac{1}{\beta}\Tr{\left(-\mathbb{P}_0\,\ln{\mathbb{P}_0}\right)}
	\label{eq:Gibbs_F}
\end{equation}
so that the first term reproduces the energy $U_0$ averaged over $\mathbb{P}_0$ while the second term reproduces the entropy $S_0$ of $\mathbb{P}_0$ itself.

Gibbs free energy in~\autoref{eq:Gibbs_F} can be seen as a \textit{variational} construction that provides the Boltzmann probability distribution $\mathbb{P}_{\text{eq}}$ as the one that yields its global minimum. In this spirit, let us compute the first-order variation of~$G$ with respect to~$\mathbb{P}_0$, still enforcing the normalization constraint $\Tr\left(\mathbb{P}_0+\delta\mathbb{P}_0\right)=1$ through a suitable Lagrange multiplier $\lambda$:
\begin{equation}
	\delta G[\mathbb{P}_0;\,\lambda] = \Tr{\left[\left(\mathcal{H} + \frac{1}{\beta}\ln{\mathbb{P}_0} + \frac{1}{\beta} + \lambda\right)\delta\mathbb{P}_0\right]}
\end{equation}
So the stationary condition $\delta G=0$ provides the explicit expression of the extremal point $\mathbb{P}^*_0$:
\begin{equation}
	\mathbb{P}^*_0 = \frac{e^{-\beta\mathcal{H}}}{e^{\,1+\beta\lambda}}
\end{equation}
where $\lambda$ can be found by imposing the normalization constraint $\Tr{\mathbb{P}_0}=1$, from which in the end the identification of this extremal point with the Boltzmann probability distribution:
\begin{equation}
	\mathbb{P}^*_0 = \mathbb{P}_{\text{eq}} = \frac{e^{-\beta\mathcal{H}}}{\Tr{e^{-\beta\mathcal{H}}}}
	\label{eq:Gibbs_extremal}
\end{equation}

Furthermore, a second-order variation of $G$ provides its convexity, so that the previous extremal point is actually found to be a global minimum:
\begin{equation}
	\delta^2 G[\mathbb{P}^*_0] = \Tr{\frac{\left(\delta\mathbb{P}_0\right)^2}{\beta\mathbb{P}_0}}\biggr{|}_{\mathbb{P}^*_0} > 0
\end{equation}

In the end, substituting back~\autoref{eq:Gibbs_extremal} into $G$, we can finally confirm that the global minimum of the Gibbs free energy provides the Helmholtz free energy~$F$:
\begin{equation}
	G[\mathbb{P}_{\text{eq}}]=F
\end{equation}

The importance of this variational construction relies on the fact that a generic error $\delta\mathbb{P}$ in evaluating the equilibrium distribution of the system implies an error of the same order of magnitude in physical observables, when they are computed directly from $\mathbb{P}$. Instead, when minimizing a suitable Gibbs free energy, then an error $\delta\mathbb{P}$ in the equilibrium distribution causes an error $O[(\delta\mathbb{P})^2]$ in physical observables when computed as derivatives of $G$, so obtaining a valuable enhancement. This is the fundamental principle at the basis of every approximation made on the equilibrium probability distribution of the system.

\section{The na\"{i}ve mean field}
\label{sec:naive_mean_field}

So the most common simplification of the actual equilibrium probability distribution $\mathbb{P}_{\text{eq}}$ is to somehow factorize it. In particular, given that the simplest case is when degrees of freedom are independent, then the roughest approximation is to consider all degrees of freedom as actually independent from each other even when they are not, so completely factorizing the probability distribution of the whole system:
\begin{equation}
	\mathbb{P}_{\text{eq}}(\{\boldsymbol{x}_i\}) \qquad \to \qquad \mathbb{P}_{\text{MF}}(\{\boldsymbol{x}_i\}) = \prod_i\eta_i(\boldsymbol{x}_i)
	\label{eq:naive_mean_field}
\end{equation}
with the normalization constraint $\Tr_{\boldsymbol{x}_i}\eta_i(\boldsymbol{x}_i)=1$. This factorization is typically referred to as the~\acrfull{MF} one.

Notice that $\eta_i(\boldsymbol{x}_i)$ just represents the probability distribution of the degree of freedom~$\boldsymbol{x}_i$ in this approximation, and in this sense it is also called the \textit{belief} of the variable $\boldsymbol{x}_i$. So in general it can be quite different from the actual marginal probability distribution~$\mathbb{P}_i(\boldsymbol{x}_i)$:
\begin{equation}
	\mathbb{P}_i(\boldsymbol{x}_i) \equiv \Tr_{\{\boldsymbol{x}_j\}_{j \neq i}}\mathbb{P}_{\text{eq}}(\{\boldsymbol{x}_i\})
\end{equation}
For this reason, we use a different notation when referring to the former or to the latter.

Let us consider the following Hamiltonian
\begin{equation}
	\mathcal{H}[\{\boldsymbol{x}_i\}] = - \sum_{(i,j)}J_{ij}(\boldsymbol{x}_i,\boldsymbol{x}_j) - \sum_i h_i(\boldsymbol{x}_i)
	\label{eq:generic_magnetic_H}
\end{equation}
where $J_{ij}(\boldsymbol{x}_i,\boldsymbol{x}_j)$ represents the pairwise interaction between $\boldsymbol{x}_i$ and $\boldsymbol{x}_j$, while $h_i(\boldsymbol{x}_i)$ refers to the local bias given by the external field on $\boldsymbol{x}_i$. By using the mean-field factorization~\autoref{eq:naive_mean_field}, Gibbs free energy~\autoref{eq:Gibbs_F} becomes:
\begin{equation}
\begin{split}
	G[\{\eta_i\}] = &- \sum_{(i,j)}\Tr_{\boldsymbol{x}_i,\boldsymbol{x}_j}\Bigl[J_{ij}(\boldsymbol{x}_i,\boldsymbol{x}_j)\eta_i(\boldsymbol{x}_i)\eta_j(\boldsymbol{x}_j)\Bigr] - \sum_i\Tr_{\boldsymbol{x}_i}\Bigl[h_i(\boldsymbol{x}_i)\eta_i(\boldsymbol{x}_i)\Bigr]\\
	&\qquad+\frac{1}{\beta}\sum_i\Tr_{\boldsymbol{x}_i}\Bigl[\eta_i(\boldsymbol{x}_i)\ln{\eta_i(\boldsymbol{x}_i)}\Bigr]
\end{split}
\end{equation}
in which beliefs $\eta_i$'s appear as variational parameters. Indeed, when minimizing~$G$ with respect to each one of them, one gets their equilibrium expressions according to the factorized probability distribution $\mathbb{P}_{\text{MF}}$:
\begin{equation}
	\frac{\delta G}{\delta\eta_i} = 0 \qquad \forall i
\end{equation}

In order to provide an explicit example of how this mean-field approximation works, let us specialize to the case of the Ising model presented in~\autoref{sec:Ising_ferro}, where each variable $\boldsymbol{x}_i$ takes on only the two values $\sigma_i=\pm 1$:
\begin{equation}
	\mathcal{H}[\{\sigma_i\}] = -\sum_{(i,j)}J_{ij}\,\sigma_i\sigma_j - H\sum_i\sigma_i
\end{equation}
So $\eta_i(\sigma_i)$ is a discrete probability distribution over two states $\sigma_i=\pm 1$ and hence it is more suitably described by a unique parameter $m_i$
\begin{equation}
	m_i \equiv \braket{\sigma_i}_{\text{MF}} = \eta_i(\sigma_i=+1) - \eta_i(\sigma_i=-1)
\end{equation}
which represents the \textit{local magnetization} of spin $\sigma_i$, while $\eta_i(\sigma_i=+1)+\eta_i(\sigma_i=-1)=1$ due to normalization.

In terms of $m_i$'s, Gibbs free energy becomes:
\begin{equation}
\begin{split}
	G[\{m_i\}] = &-\sum_{(i,j)}J_{ij} m_i m_j -H\sum_i m_i\\
	&+ \frac{1}{\beta}\sum_i\left[\frac{1+m_i}{2}\ln{\left(\frac{1+m_i}{2}\right)}+\frac{1-m_i}{2}\ln{\left(\frac{1-m_i}{2}\right)}\right]
\end{split}
\end{equation}
so that, when minimizing with respect to the set of local magnetizations, we get the well known mean-field equation:
\begin{equation}
	m_i = \tanh{\left[\beta\left(\sum_{j\in\partial i}J_{ij}m_j+H\right)\right]}
	\label{eq:Ising_mean_field_eq}
\end{equation}

If the underlying graph $\mathcal{G}$ is a $d$-dimensional hypercubic lattice and all the interactions are purely ferromagnetic ($J_{ij}\equiv J>0$ for each edge) with no external field acting on the system ($H=0$), then we get the mean-field equation of the $d$-dimensional ferromagnetic Ising model:
\begin{equation}
	m = \tanh{\left(2d\beta J m\right)}
\end{equation}
which yields the paramagnetic solution $m=0$ for any value of inverse temperature~$\beta$, and also a magnetized solution $m \neq 0$ for $\beta>\beta_c\equiv 1/2dJ$. Indeed, such $\beta_c$ is found to be the inverse critical temperature of a second-order phase transition, in correspondence of which the paramagnetic solution becomes unstable as a consequence of the \textit{breaking} of the inversion symmetry $\sigma_i \to -\sigma_i$.

The prediction of a phase transition turns out to be correct when $d \geqslant 2$, as stated in~\autoref{sec:Ising_ferro}, even though this does not mean that the equilibrium probability distribution $\mathbb{P}_{\text{eq}}$ is actually given by~\autoref{eq:naive_mean_field}. Indeed, other physical quantities like the specific heat $C_v \equiv \partial U/\partial T$ and the magnetic susceptibility $\chi \equiv \partial m/\partial H$ are predicted to be divergent for $d<4$, while they actually do not.

Eventually, it can be shown that all the qualitative predictions of this mean-field description are correct for $d>4$, namely $d=4$ is the upper critical dimension $d^u_c$ of this model. However, it becomes exact only in the $d\to\infty$ limit, i.\,e. in the fully connected topology. Indeed, in this limit the mean-field Ising model is better known as the \acrfull{CW} model, with interactions that are long-range but weak enough to have a total energy that scales as the size~$N$:
\begin{equation}
	\mathcal{H}[\{\sigma_i\}]=-\frac{J}{2N}\sum_{i,j}\sigma_i\,\sigma_j-H\sum_i\sigma_i
	\label{eq:Hamiltonian_ferro_CW}
\end{equation}
with the explicit $O(1/N)$ scaling of the exchange coupling in front of the sum over each couple of spins.

The reliability of mean-field approximation given in~\autoref{eq:naive_mean_field} so depends on the details of the system under study. In particular, from the analysis of the previous example it should be clear that the larger the fluctuations, the less reliable the mean-field approximation. Ultimately, this is due to the fact that we are neglecting any type of correlations, even the pairwise ones:
\begin{equation}
	\braket{\sigma_i\sigma_j}_{\text{MF}} = \braket{\sigma_i}_{\text{MF}}\braket{\sigma_j}_{\text{MF}}
\end{equation}
and this turns out to be a too rough approximation in most cases, especially when each degree of freedom interacts with very few nearest neighbours.

\section{The Bethe\,-\,Peierls approximation}
\label{sec:BP_approx}

The logical consequence of the previous analysis is that most of times one-point beliefs $\eta_i(\boldsymbol{x}_i)$'s are not enough. Indeed, each site is affected at least by the presence of its nearest neighbours, hence also the corresponding two-point beliefs $\eta_{ij}(\boldsymbol{x}_i,\boldsymbol{x}_j)$'s should be taken into account, as firstly suggested by Bethe~\cite{Bethe1935} and Peierls~\cite{Peierls1936}. This results in the so called Bethe\,-\,Peierls (BP) approximation, according to which the factorized probability distribution~\autoref{eq:naive_mean_field} now becomes~\cite{YedidiaEtAl2000, Yedidia2001}:
\begin{equation}
	\mathbb{P}_{\text{BP}}(\{\boldsymbol{x}_i\}) = \prod_{(i,j)}\frac{\eta_{ij}(\boldsymbol{x}_i,\boldsymbol{x}_j)}{\eta_i(\boldsymbol{x}_i)\eta_j(\boldsymbol{x}_j)}\prod_i\eta_i(\boldsymbol{x}_i) = \prod_{(i,j)}\eta_{ij}(\boldsymbol{x}_i,\boldsymbol{x}_j)\prod_i\eta^{1-d_i}_i(\boldsymbol{x}_i)
	\label{eq:BP_mean_field}
\end{equation}
where $d_i$ is the \textit{degree} of node $i$ in the graph $\mathcal{G}$, namely the number of nearest-neighbour variables of $\boldsymbol{x}_i$. Normalization constraints have to be taken into account:
\begin{equation}
	\Tr_{\boldsymbol{x}_i}\eta_i(\boldsymbol{x}_i) = 1 \qquad , \qquad \Tr_{\boldsymbol{x}_i,\boldsymbol{x}_j}\eta_{ij}(\boldsymbol{x}_i,\boldsymbol{x}_j) = 1
\end{equation}
along with the marginalization conditions:
\begin{equation}
	\Tr_{\boldsymbol{x}_j}\eta_{ij}(\boldsymbol{x}_i,\boldsymbol{x}_j) = \eta_i(\boldsymbol{x}_i) \qquad , \qquad \Tr_{\boldsymbol{x}_i}\eta_{ij}(\boldsymbol{x}_i,\boldsymbol{x}_j) = \eta_j(\boldsymbol{x}_j)
\end{equation}

If we consider again the generic Hamiltonian~\autoref{eq:generic_magnetic_H}, factorized probability distribution~\autoref{eq:BP_mean_field} gives the following expressions for the energy:
\begin{equation}
\begin{split}
	U = & -\sum_{(i,j)}\Tr_{\boldsymbol{x}_i,\boldsymbol{x}_j}\Bigl\{\eta_{ij}(\boldsymbol{x}_i,\boldsymbol{x}_j)\Bigl[J_{ij}(\boldsymbol{x}_i,\boldsymbol{x}_j)+h_i(\boldsymbol{x}_i)+h_j(\boldsymbol{x}_j)\Bigr]\Bigr\}\\
	&\qquad -\sum_i(1-d_i)\Tr_{\boldsymbol{x}_i}\Bigl[\eta_i(\boldsymbol{x}_i)h_i(\boldsymbol{x}_i)\Bigr]
\end{split}
\end{equation}
and the entropy:
\begin{equation}
\begin{split}
	S = & -\sum_{(i,j)}\Tr_{\boldsymbol{x}_i,\boldsymbol{x}_j}\Bigl[\eta_{ij}(\boldsymbol{x}_i,\boldsymbol{x}_j)\ln{\eta_{ij}(\boldsymbol{x}_i,\boldsymbol{x}_j)}\Bigr]\\
	&\qquad -\sum_i(1-d_i)\Tr_{\boldsymbol{x}_i}\Bigl[\eta_i(\boldsymbol{x}_i)\ln{\eta_i(\boldsymbol{x}_i)}\Bigr]
\end{split}
\end{equation}
Putting the two contributions together, we get the expression for the Gibbs free energy as a functional of one-node and two-node beliefs:
\begin{equation}
\begin{split}
	G[\{\eta_i,\eta_{ij}\}] = & -\sum_{(i,j)}\Tr_{\boldsymbol{x}_i,\boldsymbol{x}_j}\Bigl\{\eta_{ij}(\boldsymbol{x}_i,\boldsymbol{x}_j)\Bigl[J_{ij}(\boldsymbol{x}_i,\boldsymbol{x}_j)+h_i(\boldsymbol{x}_i)+h_j(\boldsymbol{x}_j)-\frac{1}{\beta}\ln{\eta_{ij}(\boldsymbol{x}_i,\boldsymbol{x}_j)}\Bigr]\Bigr\}\\
	&\qquad -\sum_i(1-d_i)\Tr_{\boldsymbol{x}_i}\Bigl\{\eta_i(\boldsymbol{x}_i)\Bigl[h_i(\boldsymbol{x}_i)-\frac{1}{\beta}\ln{\eta_i(\boldsymbol{x}_i)}\Bigr]\Bigr\}
	\label{eq:Gibbs_free_energy_Bethe}
\end{split}
\end{equation}

At this point, we want to minimize this expression with respect to variational parameters $\{\eta_i\}$ and $\{\eta_{ij}\}$, by also taking into account their normalizations as well as marginalization conditions. This can be achieved in the usual Lagrangian formalism:
\begin{equation}
\begin{split}
	&\mathcal{L}[\{\eta_i,\eta_{ij}\};\{\gamma_i,\gamma_{ij},\lambda_{i\to j},\lambda_{j\to i}\}] \equiv G[\{\eta_i,\eta_{ij}\}] + \sum_i\gamma_i\Bigl[1-\Tr_{\boldsymbol{x}_i}{\eta_i(\boldsymbol{x}_i)}\Bigr]\\
	&\qquad + \sum_{(i,j)}\gamma_{ij}\Bigl[1-\Tr_{\boldsymbol{x}_i,\boldsymbol{x}_j}{\eta_{ij}(\boldsymbol{x}_i,\boldsymbol{x}_j)}\Bigr] + \sum_{(i,j)}\Tr_{\boldsymbol{x}_j}\Bigl\{\lambda_{j\to i}(\boldsymbol{x}_j)\Bigl[\eta_j(\boldsymbol{x}_j)-\Tr_{\boldsymbol{x}_i}{\eta_{ij}(\boldsymbol{x}_i,\boldsymbol{x}_j)}\Bigr]\Bigr\}\\
	&\qquad + \sum_{(i,j)}\Tr_{\boldsymbol{x}_i}\Bigl\{\lambda_{i\to j}(\boldsymbol{x}_i)\Bigl[\eta_i(\boldsymbol{x}_i)-\Tr_{\boldsymbol{x}_j}{\eta_{ij}(\boldsymbol{x}_i,\boldsymbol{x}_j)}\Bigr]\Bigr\}
\end{split}
\end{equation}
with $\gamma_i$'s, $\gamma_{ij}$'s, $\lambda_{i\to j}(\boldsymbol{x}_i)$'s and $\lambda_{j\to i}(\boldsymbol{x}_j)$'s being the corresponding Lagrange multipliers.

So let us perform functional derivatives with respect to one-node beliefs and put them equal to zero, obtaining:
\begin{equation}
\begin{split}
	\eta_i(\boldsymbol{x}_i) &= \frac{\exp{\left\{\beta\left[h_i(\boldsymbol{x}_i)+\frac{1}{d_i-1}\sum_{j\in\partial i}\lambda_{i\to j}(\boldsymbol{x}_i)\right]\right\}}}{\exp{\left\{1+\beta\frac{\gamma_i}{d_i-1}\right\}}}\\
	&\equiv\frac{1}{\mathcal{Z}_i}\exp{\left\{\beta\left[h_i(\boldsymbol{x}_i)+\frac{1}{d_i-1}\sum_{j\in\partial i}\lambda_{i\to j}(\boldsymbol{x}_i)\right]\right\}}
	\label{eq:one_node_belief_selfCons}
\end{split}
\end{equation}
where we also enforced the normalization of beliefs:
\begin{equation}
	\mathcal{Z}_i \equiv \exp{\left\{1+\beta\frac{\gamma_i}{d_i-1}\right\}} = \Tr_{\boldsymbol{x}_i}\exp{\left\{\beta\left[h_i(\boldsymbol{x}_i)+\frac{1}{d_i-1}\sum_{j\in\partial i}\lambda_{i\to j}(\boldsymbol{x}_i)\right]\right\}}
\end{equation}
The same has to be done for two-node beliefs:
\begin{equation}
\begin{split}
	\eta_{ij}(\boldsymbol{x}_i,\boldsymbol{x}_j) &= \frac{\exp{\Bigl\{\beta\left[J_{ij}(\boldsymbol{x}_i,\boldsymbol{x}_j)+h_i(\boldsymbol{x}_i)+h_j(\boldsymbol{x}_j)+\lambda_{j\to i}(\boldsymbol{x}_j)+\lambda_{i\to j}(\boldsymbol{x}_i)\right]\Bigr\}}}{\exp{\left\{1-\beta\gamma_{ij}\right\}}}\\
	&\equiv\frac{1}{\mathcal{Z}_{ij}}\exp{\Bigl\{\beta\left[J_{ij}(\boldsymbol{x}_i,\boldsymbol{x}_j)+h_i(\boldsymbol{x}_i)+h_j(\boldsymbol{x}_j)+\lambda_{j\to i}(\boldsymbol{x}_j)+\lambda_{i\to j}(\boldsymbol{x}_i)\right]\Bigr\}}
	\label{eq:two_node_belief_selfCons}
\end{split}
\end{equation}
where:
\begin{equation}
\begin{split}
	\mathcal{Z}_{ij} &\equiv \exp{\left\{1-\beta\gamma_{ij}\right\}}\\
	&= \Tr_{\boldsymbol{x}_i,\boldsymbol{x}_j}\exp{\Bigl\{\beta\left[J_{ij}(\boldsymbol{x}_i,\boldsymbol{x}_j)+h_i(\boldsymbol{x}_i)+h_j(\boldsymbol{x}_j)+\lambda_{j\to i}(\boldsymbol{x}_j)+\lambda_{i\to j}(\boldsymbol{x}_i)\right]\Bigr\}}
\end{split}
\end{equation}

At this point, we are left with the marginalization constraints, which provide a set of self-consistency equations for the Lagrange multipliers $\lambda_{i\to j}$'s. So let us plug expressions~\autoref{eq:one_node_belief_selfCons} and~\autoref{eq:two_node_belief_selfCons} into marginalization condition $\eta_i(\boldsymbol{x}_i)=\Tr_{\boldsymbol{x}_j}{\eta_{ij}(\boldsymbol{x}_i,\boldsymbol{x}_j)}$:
\begin{equation}
\begin{split}
	&\frac{1}{\mathcal{Z}_i}\exp{\left\{\beta\left[h_i(\boldsymbol{x}_i)+\frac{1}{d_i-1}\sum_{j\in\partial i}\lambda_{i\to j}(\boldsymbol{x}_i)\right]\right\}}\\
	&\qquad = \Tr_{\boldsymbol{x}_j}\left\{\frac{1}{\mathcal{Z}_{ij}}\exp{\Bigl\{\beta\left[J_{ij}(\boldsymbol{x}_i,\boldsymbol{x}_j)+h_i(\boldsymbol{x}_i)+h_j(\boldsymbol{x}_j)+\lambda_{j\to i}(\boldsymbol{x}_j)+\lambda_{i\to j}(\boldsymbol{x}_i)\right]\Bigr\}}\right\}
\end{split}
\end{equation}
from which, after some manipulations:
\begin{equation}
\begin{split}
	\lambda_{i\to j}(\boldsymbol{x}_i) = &\frac{1}{d_i-2}\sum_{k\in\partial i\setminus j}\lambda_{i\to k}(\boldsymbol{x}_i) - \frac{1}{\beta}\,\frac{d_i-1}{d_i-2}\ln{\frac{\mathcal{Z}_i}{\mathcal{Z}_{ij}}}\\
	&\qquad - \frac{1}{\beta}\,\frac{d_i-1}{d_i-2}\ln{\Tr_{\boldsymbol{x}_j}{\biggl\{\exp{\Bigl\{\beta\left[J_{ij}(\boldsymbol{x}_i,\boldsymbol{x}_j)+h_j(\boldsymbol{x}_j)+\lambda_{j\to i}(\boldsymbol{x}_j)\right]\Bigr\}}\biggr\}}}
	\label{eq:Lagrange_mult_selfCons}
\end{split}
\end{equation}

So expressions~\autoref{eq:one_node_belief_selfCons} and~\autoref{eq:two_node_belief_selfCons}, where Lagrange multipliers $\lambda_{i\to j}$'s satisfy self-consistency equations~\autoref{eq:Lagrange_mult_selfCons}, provide the one-node and two-nodes beliefs that actually minimize the Gibbs free energy~\autoref{eq:Gibbs_free_energy_Bethe} in the Bethe\,-\,Peierls approximation. It is clear from these expressions that, apart from interactions $J$'s and fields $h$'s which are usually known \textit{a priori}, the most relevant correlations between variables $\boldsymbol{x}_i$'s are hidden inside Lagrange multipliers $\lambda_{i\to j}$'s. Hence, they should also possess some physical interpretation. However, self-consistency equations~\autoref{eq:Lagrange_mult_selfCons} can not be solved efficiently as they are, but they have to be rewritten in a smarter way. A~hint on how to do it has been provided by a completely different field.

\section{Belief Propagation}
\label{sec:bp}

Suppose we have a set of events $\{i\}$, with each one of them having several realizations labeled by the states of the variable $\boldsymbol{x}_i$. The \textit{degree of belief} we have about each realization of the $i$-th event can be described by the probability distribution $b_i(\boldsymbol{x}_i)$, which can be also referred to as the~\textit{belief} about $\boldsymbol{x}_i$.

Such belief must depend on the realizations of the other events in a direct causal relation with it and furthermore on the evidences about it that can be provided from the outside. For each event $i$ we can identify the set $\partial i$ of events that are in a direct cause-and-effect relation with it, drawing \textit{directed edges} from the ``cause'' nodes to the ``effect'' nodes. The resulting network is a \textit{Bayesian network}, where in general the directed edges are also associated with a weight depending on both the realization $\boldsymbol{x}_i$ of the cause and the output $\boldsymbol{x}_j$ of the effect. Indeed, along each directed edge $i\to j$ it flows a \textit{message} $M_{i\to j}(\boldsymbol{x}_j)$, representing the belief about the effect $\boldsymbol{x}_j$ starting from the cause $\boldsymbol{x}_i$, independently from other events that can affect~$\boldsymbol{x}_j$. Also external evidences can be represented in such network, as external biases~$\phi_i(\boldsymbol{x}_i)$ acting on each node.

These networks can be generalized to \textit{Markov networks}, where the cause-and-effect relation is substituted by a more generic dependency $\psi_{ij}(\boldsymbol{x}_i,\boldsymbol{x}_j)$, namely the corresponding edge is no longer directed. The reason of the `Markov' name is straightforward: the probability of a certain realization of event $i$ conditionally depends only on the outcomes of events $j\in\partial i$ in direct connection with it, and on no other event. Indeed, once fixed the outcomes $\{\boldsymbol{x}_j\}_{j\in\partial i}$ of the events which are nearest neighbours of~$i$, then $b_i(\boldsymbol{x}_i)$ is independent from all the other events in the network
\begin{equation}
	\mathbb{P}(\boldsymbol{x}_i|\{\boldsymbol{x}_j\}_{j \neq i}) = \mathbb{P}(\boldsymbol{x}_i|\{\boldsymbol{x}_j\}_{j\in\partial i})
	\label{eq:conditional_independence}
\end{equation}
just as it happens in Markov chains.

It is clear that Markov networks actually constitute a simple though effective model to represent probabilistic knowledge about a set of interrelated events, while Bayesian networks can be seen as the paradigm of inference reasoning~\cite{Book_Pearl1988, Book_MacKay2003, Book_JensenNielsen2007, WainwrightJordan2008}. Notice that in both cases the update of beliefs and evidences should propagate through the network via local operations, just due to the Markovianity of the networks. A quantitative description of this updating process has been firstly provided by Pearl~\cite{Book_Pearl1988} and goes under the name of~\acrfull{BP}.

The key observation is that, due to the conditional independence~\autoref{eq:conditional_independence}, the belief about $\boldsymbol{x}_i$ can be computed as a product of the messages $M_{k\to i}$'s coming from ``cause'' events and of the external evidence $\phi_i$:
\begin{equation}
	b_i(\boldsymbol{x}_i) \propto \phi_i(\boldsymbol{x}_i)\prod_{k\in\partial i}M_{k\to i}(\boldsymbol{x}_i)
	\label{eq:one_node_belief_selfCons_Pearl}
\end{equation}

Moreover, also two-point beliefs $b_{ij}(\boldsymbol{x}_i,\boldsymbol{x}_j)$ for directly interrelated events can be computed following the same strategy. Indeed, they depend on all the messages $M_{k\to i}$'s and $M_{l\to j}$'s entering in both the nodes (included the ones flowing through the inbetween edge), on their external evidences $\phi_i$ and $\phi_j$ and eventually on the weight $\psi_{ij}$ characterizing their cause-and-effect relation:
\begin{equation}
	b_{ij}(\boldsymbol{x}_i,\boldsymbol{x}_j) \propto \psi_{ij}(\boldsymbol{x}_i,\boldsymbol{x}_j)\phi_i(\boldsymbol{x}_i)\phi_j(\boldsymbol{x}_j)\prod_{k\in\partial i}M_{k\to i}(\boldsymbol{x}_i)\prod_{l\in\partial j}M_{l\to j}(\boldsymbol{x}_j)
	\label{eq:two_node_belief_selfCons_Pearl}
\end{equation}

Since one-point and two-point beliefs are probability distributions as well, the following marginality condition has to be fulfilled:
\begin{equation}
	b_i(\boldsymbol{x}_i) = \Tr_{\boldsymbol{x}_j}b_{ij}(\boldsymbol{x}_i,\boldsymbol{x}_j)
	\label{eq:marginality_condition_Pearl}
\end{equation}
So when plugging expressions~\autoref{eq:one_node_belief_selfCons_Pearl} and~\autoref{eq:two_node_belief_selfCons_Pearl} in it, we get the following self-consistency equation for the messages:
\begin{equation}
	M_{i\to j}(\boldsymbol{x}_i) \propto \Tr_{\boldsymbol{x}_i}\Biggl[\psi_{ij}(\boldsymbol{x}_i,\boldsymbol{x}_j)\phi_i(\boldsymbol{x}_i)\prod_{k\in\partial i\setminus j}M_{k\to i}(\boldsymbol{x}_i)\Biggr]
	\label{eq:selfCons_messages_Pearl}
\end{equation}

This set of relations provides quantitative rules for the \textit{spreading of knowledge} about the probability distribution of each event during the inference reasoning, so looking at it on a ``dynamical'' setting. Starting from the earliest causes, beliefs flow through directed edges and reach all the nodes, updating the corresponding set of probabilities for the outcomes. This propagation mechanism has a two-fold advantage. First of all, it actually allows a \textit{local} update for the belief of each node, just due to the Markovianity of the networks. This implies a linear growth of computational resources and hence an exponential speedup in the computation of equilibrium beliefs. Secondly, each update of beliefs is quite transparent and can be given a meaningful interpretation, as we just saw.

At this point, we can relate the self-consistency equations~\autoref{eq:Lagrange_mult_selfCons} for Lagrange multipliers $\lambda_{i\to j}$'s of the Bethe\,-\,Peierls variational approach with the~\autoref{eq:selfCons_messages_Pearl} ones for messages $M_{i\to j}$'s of the~\acrshort{BP} approach. Indeed, once identified the edge weight~$\psi_{ij}$ with the exchange interaction and the external evidence with the external field
\begin{equation}
	\psi_{ij}(\boldsymbol{x}_i,\boldsymbol{x}_j) \leftrightarrow \exp{\left\{\beta J_{ij}(\boldsymbol{x}_i,\boldsymbol{x}_j)\right\}} \qquad , \qquad \phi_i(\boldsymbol{x}_i) \leftrightarrow \exp{\left\{\beta h_i(\boldsymbol{x}_i)\right\}}
\end{equation}
then the correspondence is completed by setting
\begin{equation}
	\lambda_{i\to j}(\boldsymbol{x}_j) \equiv \frac{1}{\beta}\ln{\prod_{k\in\partial i\setminus j}}M_{k\to i}(\boldsymbol{x}_i)
	\label{eq:BP_equivalence}
\end{equation}
Consequently, self-consistency equations~\autoref{eq:Lagrange_mult_selfCons} become:
\begin{equation}
	M_{i\to j}(\boldsymbol{x}_j) = \frac{\mathcal{Z}_j}{\mathcal{Z}_{ij}}\Tr_{\boldsymbol{x}_i}\Biggl\{\exp{\Bigl\{\beta J_{ij}(\boldsymbol{x}_i,\boldsymbol{x}_j)+\beta h_i(\boldsymbol{x}_i)\Bigr\}}\prod_{k\in\partial i\setminus j}M_{k\to i}(\boldsymbol{x}_i)\Biggr\}
	\label{eq:BP_selfCons_EtaHatMess}
\end{equation}

Finally, a more useful expression is obtained when referring to the messages \textit{outgoing} from each node:
\begin{equation}
	\eta_{i\to j}(\boldsymbol{x}_i) \equiv \exp{\Bigl\{\beta h_i(\boldsymbol{x}_i)\Bigr\}}\prod_{k\in\partial i\setminus j}M_{k\to i}(\boldsymbol{x}_i)
\end{equation}
from which:
\begin{equation}
	\eta_{i\to j}(\boldsymbol{x}_i) = \frac{1}{\mathcal{Z}_{i\to j}}\exp{\Bigl\{\beta h_i(\boldsymbol{x}_i)\Bigr\}}\prod_{k\in\partial i\setminus j}\Tr_{\boldsymbol{x}_k}\biggl[\exp{\Bigl\{\beta J_{ik}(\boldsymbol{x}_i,\boldsymbol{x}_k)\Bigr\}}\eta_{k\to i}(\boldsymbol{x}_k)\biggr]
	\label{eq:BP_selfCons_EtaMess}
\end{equation}
with $\mathcal{Z}_{i\to j}$ enforcing the normalization constraint $\Tr_{\boldsymbol{x}_i}\eta_{i\to j}(\boldsymbol{x}_i)=1$:
\begin{equation}
	\mathcal{Z}_{i\to j} = \Tr_{\boldsymbol{x}_i}\Biggl\{\exp{\Bigl\{\beta h_i(\boldsymbol{x}_i)\Bigr\}}\prod_{k\in\partial i\setminus j}\Tr_{\boldsymbol{x}_k}\biggl[\exp{\Bigl\{\beta J_{ik}(\boldsymbol{x}_i,\boldsymbol{x}_k)\Bigr\}}\eta_{k\to i}(\boldsymbol{x}_k)\biggr]\Biggr\}
	\label{eq:Zeta_EtaMess}
\end{equation}

\begin{figure}[!t]
	\centering
	\includegraphics[scale=1]{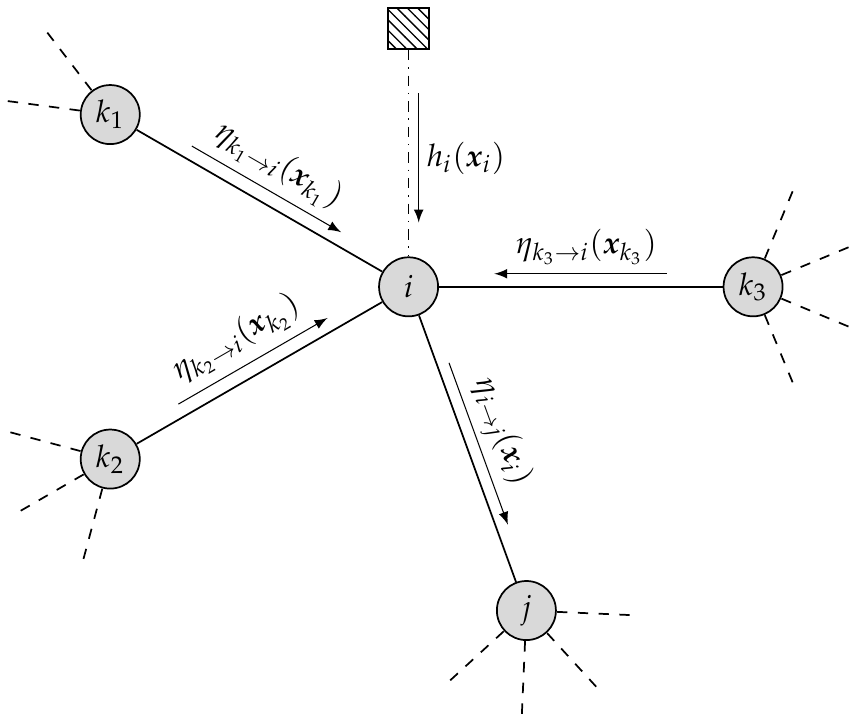}
	\caption[Pairwise Belief Propagation]{Graphical representation of the pairwise~\acrshort{BP} equations. The message $\eta_{i\to j}(\boldsymbol{x}_i)$ going out from site $i$ toward site $j$ only depends on the set of messages $\{\eta_{k\to i}(\boldsymbol{x}_k)\}$ going out from sites $\{k\in\partial i\setminus j\}$ toward site $i$, plus the external evidence $h_i$ about $\boldsymbol{x}_i$, here represented as a square node.}
	\label{fig:BP_tree}
\end{figure}

This set of equations --- known as pairwise \acrshort{BP} equations --- can be efficiently solved, since involving only local operations: the message $\eta_{i\to j}(\boldsymbol{x}_i)$ going out from site $i$ toward site $j$ only depends on the set of messages $\{\eta_{k\to i}(\boldsymbol{x}_k)\}$ going out from sites $\{k\in\partial i\setminus j\}$ toward site $i$, plus the external evidence $h_i$ about $\boldsymbol{x}_i$, as also effectively represented in~\autoref{fig:BP_tree}. Moreover, these messages have a precise physical meaning: $\eta_{i\to j}(\boldsymbol{x}_i)$ is the probability distribution of variable $\boldsymbol{x}_i$ when its dependence on $\boldsymbol{x}_j$ has been neglected, namely when the corresponding edge has been removed from the system.

Finally, by means of self-consistency equations~\autoref{eq:BP_selfCons_EtaMess}, one-point~\autoref{eq:one_node_belief_selfCons} and two-point~\autoref{eq:two_node_belief_selfCons} beliefs read:
\begin{subequations}
	\begin{equation}
		\eta_i(\boldsymbol{x}_i) = \frac{1}{\mathcal{Z}_i}\exp{\Bigl\{\beta h_i(\boldsymbol{x}_i)\Bigr\}}\prod_{k\in\partial i}\Tr_{\boldsymbol{x}_k}\biggl[\exp{\Bigl\{\beta J_{ik}(\boldsymbol{x}_i,\boldsymbol{x}_k)\Bigr\}}\eta_{k\to i}(\boldsymbol{x}_k)\biggr]
		\label{eq:one_node_belief_BP}
	\end{equation}
	\begin{equation}
		\eta_{ij}(\boldsymbol{x}_i,\boldsymbol{x}_j) = \frac{1}{\mathcal{Z}_{ij}}\exp{\Bigl\{\beta J_{ij}(\boldsymbol{x}_i,\boldsymbol{x}_j)\Bigr\}}\eta_{i\to j}(\boldsymbol{x}_i)\eta_{j\to i}(\boldsymbol{x}_j)
		\label{eq:two_node_belief_BP}
	\end{equation}
	\label{eq:beliefs_BP}%
\end{subequations}
where the explicit expressions of the normalization constants are given by:
\begin{subequations}
	\begin{equation}
		\mathcal{Z}_i = \Tr_{\boldsymbol{x}_i}\Biggl\{\exp{\Bigl\{\beta h_i(\boldsymbol{x}_i)\Bigr\}}\prod_{k\in\partial i}\Tr_{\boldsymbol{x}_k}\biggl[\exp{\Bigl\{\beta J_{ik}(\boldsymbol{x}_i,\boldsymbol{x}_k)\Bigr\}}\eta_{k\to i}(\boldsymbol{x}_k)\biggr]\Biggr\}
		\label{eq:Zeta_one_node_belief_BP}
	\end{equation}
	\begin{equation}
		\mathcal{Z}_{ij} = \Tr_{\boldsymbol{x}_i,\boldsymbol{x}_j}\biggl\{\exp{\Bigl\{\beta J_{ij}(\boldsymbol{x}_i,\boldsymbol{x}_j)\Bigr\}}\eta_{i\to j}(\boldsymbol{x}_i)\eta_{j\to i}(\boldsymbol{x}_j)\biggr\}
		\label{eq:Zeta_two_node_belief_BP}
	\end{equation}
	\label{eq:Zeta_beliefs_BP}%
\end{subequations}
in turn related to the normalization constant of the \acrshort{BP} messages $\eta_{i\to j}$'s as:
\begin{equation}
	\mathcal{Z}_{ij} = \frac{\mathcal{Z}_i}{\mathcal{Z}_{i\to j}} = \frac{\mathcal{Z}_j}{\mathcal{Z}_{j\to i}}
\end{equation}

Finally, the Gibbs free energy~\autoref{eq:Gibbs_free_energy_Bethe} in the Bethe\,-\,Peierls approximation can be evaluated in the extremal point provided by~\acrshort{BP} self-consistency equations~\autoref{eq:BP_selfCons_EtaMess}, obtaining the so-called \textit{Bethe free energy}:
\begin{equation}
	F_{\text{BP}} = -\frac{1}{\beta}\biggl(\sum_i\ln{\mathcal{Z}_i}-\sum_{(i,j)}\ln{\mathcal{Z}_{ij}}\biggr)
	\label{eq:Bethe_free_energy}
\end{equation}
whose physical interpretation is again quite straightforward: the total free energy is made up of local contributions, respectively corresponding to each node and each edge appearing in the graph. The same holds for the internal energy $U$, whose expression can be easily derived via the derivative of $\beta F_{\text{BP}}$ with respect to~$\beta$:
\begin{equation}
\begin{split}
	U&=-\sum_i\frac{1}{\mathcal{Z}_i}\Tr_{\boldsymbol{x}_i}{\Biggl\{h_i(\boldsymbol{x}_i)\exp{\Bigl\{\beta h_i(\boldsymbol{x}_i)\Bigr\}}\prod_{k\in\partial i}\Tr_{\boldsymbol{x}_k}\biggl[\exp{\Bigl\{\beta J_{ik}(\boldsymbol{x}_i,\boldsymbol{x}_k)\Bigr\}}\eta_{k\to i}(\boldsymbol{x}_k)\biggr]\Biggr\}}\\
	&\qquad -\sum_{(i,j)}\frac{1}{\mathcal{Z}_{ij}}\Tr_{\boldsymbol{x}_i,\boldsymbol{x}_j}{\biggl[J_{ij}(\boldsymbol{x}_i,\boldsymbol{x}_j)\exp{\Bigl\{\beta J_{ij}(\boldsymbol{x}_i,\boldsymbol{x}_j)\Bigr\}}\eta_{i\to j}(\boldsymbol{x}_i)\eta_{j\to i}(\boldsymbol{x}_j)\biggr]}
	\label{eq:Bethe_internal_energy}
\end{split}
\end{equation}
and which in turn corresponds to the sum of the local contributions to the total energy respectively from the external field~$h_i$ on each node --- say $u_i$ --- and from the interaction~$J_{ij}$ on each edge --- say $u_{ij}$.

Hence the advantage of the~\acrshort{BP} approach is clear: it allows the computation of physical observables with a computational effort that grows linearly in the size $N$ of the system, by summing over node and edge contributions just as seen before for the free energy and the internal energy. It is a clear enhancement with respect to their ``brute-force'' computation in the canonical ensemble via the exact equilibrium probability distribution $\mathbb{P}_{\text{eq}}$, whose computational complexity grows exponentially in the size $N$ of the system.

As usual, we can now wonder when the Bethe\,-\,Peierls approximation actually provides reliable results. We surely expect it to be more accurate with respect to the na\"{i}ve mean field, but it is anyway a mean-field approach --- having discarded correlation functions with more than two variables --- and hence it has some limits of applicability. This question will be addressed in~\autoref{sec:sparse_random_graphs}.

\section{Factor graph formalism}
\label{sec:fact_graph_form}

In physical systems involving a huge number of variables $\{\boldsymbol{x}_i\}$, it turns out that each of them interacts only with a finite number of other variables at each time, and this number is usually not extensive, i.\,e. it remains finite even in the thermodynamic limit. This is due to the fact that most of physical interactions are very \textit{local}: e.\,g. collisions between particles, magnetic interactions between dipoles, Lennard-Jones-like interactions between molecules, and so on.

A direct consequence of this is that the whole set of interactions can be \textit{factorized} in a non trivial way, recognizing that each interaction only involves a finite subset of variables. At the same time, each variable takes part only in a small number of interactions among all those taking place in the system. This factorized structure can be suitably represented in the \textit{factor graph formalism}~\cite{Book_Jordan1999, KschischangEtAl2001, Book_MezardMontanari2009}.

In this formalism, the joint probability distribution $\mathbb{P}(\{\boldsymbol{x}_i\})$ can be decomposed in the product of each ``elementary'' interaction
\begin{equation}
	\mathbb{P}(\{\boldsymbol{x}_i\}) = \frac{1}{\mathcal{Z}}\prod_{a=1}^{M}\psi_a(\underline{\boldsymbol{x}}_{\partial a})
	\label{eq:factor_graph_definition}
\end{equation}
with $M$ representing the total number of interactions. $\underline{\boldsymbol{x}}_{\partial a}$ is just a shorthand notation for $\{\boldsymbol{x}_i\}_{i\in\partial a}$, namely for the subset of variables involved in the $a$-th interaction. \textit{Compatibility function} $\psi_a$ is a nonnegative function describing the interaction between variables $\underline{\boldsymbol{x}}_{\partial a}$ and it can be of two types: \textit{i)} it can involve just a variable $\boldsymbol{x}_i$ at each time, so providing an external bias or evidence for such variable, as an external field would do:
\begin{equation}
	\psi_a(\boldsymbol{x}_i) \quad \to \quad \phi_i(\boldsymbol{x}_i) = \exp{\{\beta h_i(\boldsymbol{x}_i)\}}
\end{equation}
or \textit{ii)} it can involve more than one variable at each time, so representing a proper interaction in the statistical mechanics sense:
\begin{equation}
	\psi_a(\underline{\boldsymbol{x}}_{\partial a}) = \exp{\{\beta J_a(\underline{\boldsymbol{x}}_{\partial a})\}}
\end{equation}
with $J_a$ containing the explicit expression of the many-body interaction between variables $\underline{\boldsymbol{x}}_{\partial a}$. Hence a suitable many-body Hamiltonian --- still quite generic --- can be written down:
\begin{equation}
	\mathcal{H}[\{\boldsymbol{x}_i\}] = - \sum_a J_a(\underline{\boldsymbol{x}}_{\partial a}) - \sum_i h_i(\boldsymbol{x}_i)
	\label{eq:generic_magnetic_H_factor_graph}
\end{equation}
such that the joint probability distribution $\mathbb{P}(\{\boldsymbol{x}_i\})$ can be actually written as in~\autoref{eq:factor_graph_definition}. Finally, in the case of pairwise interactions, it holds $\underline{\boldsymbol{x}}_{\partial a}=(\boldsymbol{x}_i,\boldsymbol{x}_j)$ and hence we can recover the pairwise formalism exploited so far:
\begin{equation}
	\mathbb{P}(\{\boldsymbol{x}_i\}) = \frac{1}{\mathcal{Z}}\prod_{(i,j)}\psi_{ij}(\boldsymbol{x}_i,\boldsymbol{x}_j)\prod_i\phi_i(\boldsymbol{x}_i)
	\label{eq:factor_graph_definition_pairwise}
\end{equation}
in turn corresponding to the pairwise-interaction Hamiltonian~\autoref{eq:generic_magnetic_H}.

\begin{figure}[!t]
	\centering
	\includegraphics[scale=1]{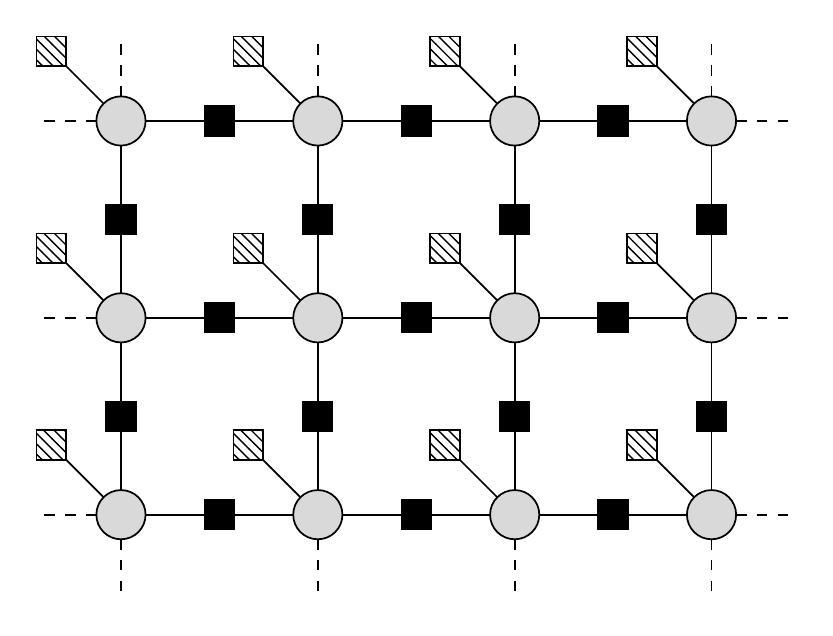}
	\caption[Factor graph of the $d=2$-dimensional Ising model]{Graphical model associated with the Ising model on a $d=2$-dimensional hypercubic lattice. Gray circles represent Ising variables, black squares represent their pairwise interaction and finally striped squares represent the external field acting on each variable.}
	\label{fig:factor_graph_Ising_2D}
\end{figure}

Factor graph formalism allows to translate nearly every statistical mechanics model for which factorization~\autoref{eq:factor_graph_definition} holds into a graphical model, namely a \textit{bipartite graph} with undirected edges joining two main kinds of nodes: the $N$ \textit{variable nodes}, each associated with a variable $\boldsymbol{x}_i$ and typically represented by a circle, and the $M$ \textit{check nodes} or \textit{function nodes}, each associated with an interaction $J_a$ and represented by a square. Also the external field can be represented as a function node, even though involving only a site each time; for this reason, we still represent it by a square, though with a different filling pattern. For example, in~\autoref{fig:factor_graph_Ising_2D} there is the graphical model associated with the Ising model on a $d=2$-dimensional hypercubic lattice.

\begin{figure}[!t]
	\centering
	\includegraphics[width=\columnwidth]{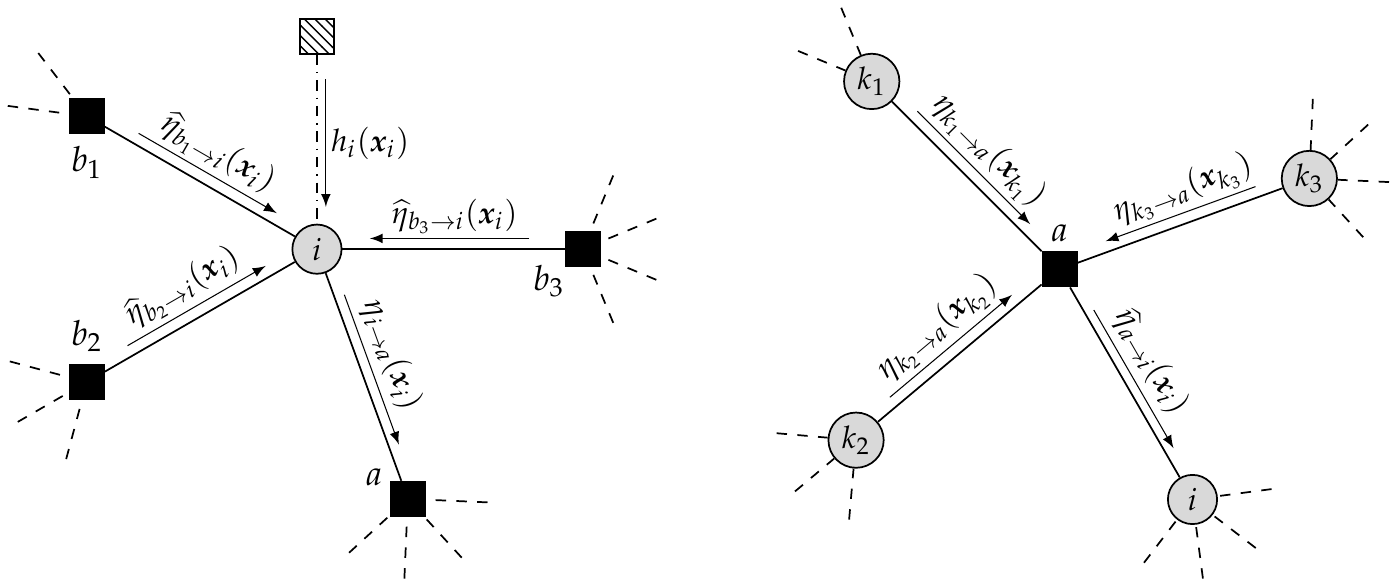}
	\caption[Factor graph Belief Propagation]{Graphical representation of the~\acrshort{BP} equations~\autoref{eq:BP_selfCons_Eta_EtaHat} in the factor graph case. \textit{Left panel}: the outgoing message $\eta_{i\to a}(\boldsymbol{x}_i)$ from site $i$ toward function node $a$ depends on the set of incoming messages $\{\widehat{\eta}_{b\to i}(\boldsymbol{x}_i)\}$ from function nodes $\{b\in\partial i\setminus a\}$ toward site $i$, plus the external evidence $h_i$ about $\boldsymbol{x}_i$. \textit{Right panel}: the incoming message $\widehat{\eta}_{a\to i}(\boldsymbol{x}_i)$ from function node $a$ to site $i$ only depends on the set of outgoing messages $\{\eta_{k\to a}(\boldsymbol{x}_k)\}$ from sites $\{k\in\partial a\setminus i\}$ toward function node $a$.}
	\label{fig:BP_factor_graph}
\end{figure}

The~\acrshort{BP} formalism developed in~\autoref{sec:bp} can be easily generalized also to bipartite graphs, which actually become necessary when dealing with many-body interactions. The key point is that in the factor graph case there are two types of ``messages'' flowing through edges: the \textit{check-to-variable} message $\widehat{\eta}_{a\to i}(\boldsymbol{x}_i)$, which is meant to leave function node $a$ and enter site $i$, and the \textit{variable-to-check} message $\eta_{i\to a}(\boldsymbol{x}_i)$, which flows away from site $i$ toward the $a$-th function node. \acrshort{BP} self-consistency equations~\autoref{eq:BP_selfCons_EtaMess} can be suddenly generalized~\cite{YedidiaEtAl2005, Book_MezardMontanari2009}; referring to~\autoref{fig:BP_factor_graph}, we have:
\begin{equation}
	\left\{
	\begin{aligned}
		\eta_{i\to a}(\boldsymbol{x}_i) &= \frac{1}{\mathcal{Z}_{i\to a}}\exp{\Bigl\{\beta h_i(\boldsymbol{x}_i)\Bigr\}}\prod_{b\in\partial i\setminus a}\widehat{\eta}_{b\to i}(\boldsymbol{x}_i)\\
	\widehat{\eta}_{a\to j}(\boldsymbol{x}_j) &= \frac{1}{\widehat{\mathcal{Z}}_{a\to j}}\Tr_{\underline{\boldsymbol{x}}_{\partial a\setminus j}}\biggl[\exp{\Bigl\{\beta J_a(\underline{\boldsymbol{x}}_{\partial a})\Bigr\}}\prod_{k\in\partial a\setminus j}\eta_{k\to a}(\boldsymbol{x}_k)\biggr]\\
	\end{aligned}
	\right.
	\label{eq:BP_selfCons_Eta_EtaHat}
\end{equation}
with $a,b,c,\dots$ labeling check nodes and $i,j,k,\dots$ labeling variable nodes. So $\partial i\setminus a$ refers to the check nodes directly linked to variable node~$i$ but~$a$, while $\partial a\setminus j$ stands for the set of variables that take part into the $a$-th interaction but~$j$. Normalization constants finally read:
\begin{subequations}
	\begin{equation}
		\mathcal{Z}_{i\to a} \equiv \Tr_{\boldsymbol{x}_i}{\biggl[\exp{\Bigl\{\beta h_i(\boldsymbol{x}_i)\Bigr\}}\prod_{b\in\partial i\setminus a}\widehat{\eta}_{b\to i}(\boldsymbol{x}_i)\biggr]}
		\label{eq:BP_selfCons_Zeta_Eta}
	\end{equation}
	\begin{equation}
		\widehat{\mathcal{Z}}_{a\to j} \equiv \Tr_{\boldsymbol{x}_j}{\Biggl\{\Tr_{\underline{\boldsymbol{x}}_{\partial a\setminus j}}\biggl[\exp{\Bigl\{\beta J_a(\underline{\boldsymbol{x}}_{\partial a})\Bigr\}}\prod_{k\in\partial a\setminus j}\eta_{k\to a}(\boldsymbol{x}_k)\biggr]\Biggr\}}
		\label{eq:BP_selfCons_Zeta_EtaHat}
	\end{equation}
	\label{eq:BP_selfCons_Zeta_Eta_EtaHat}%
\end{subequations}

Once noticed that the local contributions to the Bethe free energy~\autoref{eq:Bethe_free_energy} in the pairwise case come from the two objects present in the graphical model --- namely the variable nodes and the edges between them ---, then it is easy to generalize its expression to the factor graph formalism, where the basic objects are three: the variable nodes, the check nodes and the edges joining them. Hence, considering the three types of local contributions, it holds~\cite{Book_MezardMontanari2009}:
\begin{equation}
	F_{\text{BP}} = -\frac{1}{\beta}\biggl(\sum_i\ln{\mathcal{Z}_i}+\sum_a\ln{\mathcal{Z}_a}-\sum_{(i,a)}\ln{\mathcal{Z}_{ia}}\biggr)
	\label{eq:Bethe_free_energy_factor_graph}
\end{equation}
where:
\begin{subequations}
	\begin{equation}
		\mathcal{Z}_i = \Tr_{\boldsymbol{x}_i}{\biggl[\exp{\Bigl\{\beta h_i(\boldsymbol{x}_i)\Bigr\}}\prod_{a\in\partial i}\widehat{\eta}_{a\to i}(\boldsymbol{x}_i)\biggr]}
	\end{equation}
	\begin{equation}
		\mathcal{Z}_a = \Tr_{\underline{\boldsymbol{x}}_{\partial a}}{\biggl[\exp{\Bigl\{\beta J_a(\underline{\boldsymbol{x}}_{\partial a})\Bigr\}}\prod_{k\in\partial a}\eta_{k\to a}(\boldsymbol{x}_k)\biggr]}
	\end{equation}
	\begin{equation}
		\mathcal{Z}_{ia} = \Tr_{\boldsymbol{x}_i}{\Bigl[\eta_{i\to a}(\boldsymbol{x}_i)\,\widehat{\eta}_{a\to i}(\boldsymbol{x}_i)\Bigr]}
	\end{equation}
\end{subequations}
Of course, this expression can be rigorously recovered when generalizing the Gibbs free energy~\autoref{eq:Gibbs_free_energy_Bethe} in the Bethe\,-\,Peierls approximation to the case of many-body interactions and then minimizing it with respect to the new Lagrange multipliers, which are in turn related to the messages $\eta_{i\to a}$'s and $\widehat{\eta}_{a\to i}$'s of the factor graph formalism~\cite{YedidiaEtAl2005}.

Finally, also the internal energy $U$ can be written in terms of the two types of messages, so generalizing the pairwise expression~\autoref{eq:Bethe_internal_energy}:
\begin{equation}
\begin{split}
	U&=-\sum_i\frac{1}{\mathcal{Z}_i}\Tr_{\boldsymbol{x}_i}{\biggl[h_i(\boldsymbol{x}_i)\exp{\Bigl\{\beta h_i(\boldsymbol{x}_i)\Bigr\}}\prod_{a\in\partial i}\widehat{\eta}_{a\to i}(\boldsymbol{x}_i)\biggr]}\\
	&\qquad -\sum_a\frac{1}{\mathcal{Z}_a}\Tr_{\underline{\boldsymbol{x}}_{\partial a}}{\biggl[J_a(\underline{\boldsymbol{x}}_{\partial a})\exp{\Bigl\{\beta J_a(\underline{\boldsymbol{x}}_{\partial a})\Bigr\}}\prod_{k\in\partial a}\eta_{k\to a}(\boldsymbol{x}_k)\biggr]}
	\label{eq:Bethe_internal_energy_factor_graph}
\end{split}
\end{equation}
while keeping the straightforward physical interpretation of sum of the local contributions from the field and the proper many-body interactions.

\section{Sparse random graphs and applicability of Belief Propagation}
\label{sec:sparse_random_graphs}

Factorization~\autoref{eq:factor_graph_definition}, which defines a factor graph representation, has actually a highly nontrivial consequence. Indeed, every time it holds, then any two variables $\boldsymbol{x}_i$ and $\boldsymbol{x}_j$ separated by a set $\mathcal{S}$ of other variables $\underline{\boldsymbol{x}}_{\mathcal{S}}\equiv\{\boldsymbol{x}_k\}_{k\in\mathcal{S}}$ --- namely there is no path joining $\boldsymbol{x}_i$ and $\boldsymbol{x}_j$ without involving also some variable in $\mathcal{S}$ --- are \textit{conditionally independent}:
\begin{equation}
	\mathbb{P}(\boldsymbol{x}_i,\boldsymbol{x}_j|\underline{\boldsymbol{x}}_{\mathcal{S}}) = \mathbb{P}(\boldsymbol{x}_i|\underline{\boldsymbol{x}}_{\mathcal{S}})\,\mathbb{P}(\boldsymbol{x}_j|\underline{\boldsymbol{x}}_{\mathcal{S}})
\end{equation}
This property is nothing but the Markovianity anticipated in~\autoref{sec:bp}, indeed it goes under the name of~\textit{global Markov property}.

At this point, it is clear that the Markovianity is necessary for the Bethe\,-\,Peierls approximation to be meaningful, as already seen. However, it is not enough: the key ingredient that allows to compute one-point and two-point beliefs in an efficient way is the factorization~\autoref{eq:BP_equivalence}, that is not harmless at all. Indeed, it requires that the nearest-neighbour variables $\{\boldsymbol{x}_k\}$ of $\boldsymbol{x}_i$ are conditionally independent only with respect to $\boldsymbol{x}_i$ itself
\begin{equation}
	\mathbb{P}(\{\boldsymbol{x}_k\}_{k\in\partial i}|\boldsymbol{x}_i) = \prod_{k\in\partial i}\mathbb{P}(\boldsymbol{x}_k|\boldsymbol{x}_i)
	\label{eq:BP_condition}
\end{equation}
namely along $\mathcal{G}$ there is a unique path joining them and it necessarily passes through the site $i$. So if such site is removed from the graph, then they actually become independent.

This property is stronger than the global Markov property implied by factorization~\autoref{eq:factor_graph_definition}, as for example it can seen by looking at the graphical model in~\autoref{fig:factor_graph_Ising_2D}: once removed a certain site $i$, its nearest-neighbour variables do not become independent, due to the presence of \textit{very short loops} joining them. This observation actually translates into the following rigorous statement: the Bethe\,-\,Peierls approximation is exact on \textit{treelike} graphs, with~\acrshort{BP} fixed point in one-to-one correspondence with the stationary points of Bethe free energy~\autoref{eq:Gibbs_free_energy_Bethe}~\cite{YedidiaEtAl2000, YedidiaEtAl2005}. Indeed, treelike graphs have no loops and hence there is always a unique path joining any two variables, so that condition~\autoref{eq:BP_condition} is always verified. However, if correlations decay ``fast enough'' with the distance, the previous factorization would be nearly exact also in presence of large loops, namely on the so-called \textit{locally treelike} graphs.

An important class of graphs of this kind is that of \textit{sparse random graphs}~\cite{Book_JansonEtAl2000, Book_Bollobas2001}. These graphs are such that the total number $M$ of interactions scales linearly with the number $N$ of variables, so that the ratio $\alpha\equiv M/N$ approaches a finite value in the thermodynamic limit ($N\to\infty$). Moreover, the randomness is given by the fact that each edge between any two nodes is drawn with a fixed probability $p=O(1/N)$, independently from other edges. In this way, each node has a finite average number $C$ of neighbours, from which also their sparsity property.

A distinctive feature of sparse random graphs is represented by their degree profile, namely by the probability distribution of the degree $d_i$ of their variable nodes. If it has been fixed to be equal to $C$ for all the sites
\begin{equation}
	\mathbb{P}_d(d_i) = \delta(d_i-C)
	\label{eq:deg_distr_RRG}
\end{equation}
then the resulting ensemble of graphs is the~\acrfull{RRG} one. Instead, if the probability $p$ of drawing each edge is set equal to $C/(N-1)$, then the degree turns out to be distributed according to a Poisson distribution of first moment $C$
\begin{equation}
	\mathbb{P}_d(d_i) = C^{d_i}\,\frac{e^{-C}}{d_i!}
	\label{eq:deg_distr_ERG}
\end{equation}
and the corresponding ensemble is the~\acrfull{ERG} one~\cite{ErdosRenyi1959}.

The two properties of these classes of graphs, the sparsity and the randomness, jointly provide the local treelike structure necessary for the Bethe\,-\,Peierls approximation to work. Indeed, the factorization~\autoref{eq:BP_condition} is not strictly true on sparse random graphs, but it is violated by a term $\varepsilon$ that contains the correlations between $\{\boldsymbol{x}_k\}_{k\in\partial i}$ variables along paths that do not contain the site $i$, namely along the loops that eventually join them at finite $N$:
\begin{equation}
	\mathbb{P}(\{\boldsymbol{x}_k\}_{k\in\partial i}|\boldsymbol{x}_i) = \prod_{k\in\partial i}\mathbb{P}(\boldsymbol{x}_k|\boldsymbol{x}_i) + \varepsilon_i(\{\boldsymbol{x}_k\}_{k\in\partial i}|\boldsymbol{x}_i)
	\label{eq:BP_condition_violated}
\end{equation}
In order to quantify such error, let us evaluate the loop typical size. Since these graphs are locally treelike, the average number of sites $N(r)$ at distance $r$ from the center of the graph grows as $C^r$. However, when $N(r)$ becomes comparable with the size $N$, loops start to appear. So the typical size $\ell$ of the loops can be estimated as
\begin{equation}
	N(r) \sim N \quad \Rightarrow \quad r \sim \frac{\ln{N}}{\ln{C}} \quad \Rightarrow \quad \ell \sim \frac{\ln{N}}{\ln{C}}
\end{equation}
so that in the thermodynamic limit they can actually be neglected. This will allow us to actually exploit the~\acrshort{BP} approach on large enough sparse random graphs.

Notice, however, that a further assumption is needed, namely that the correlations must decay fast enough with the distance. This is certainly true when there is a \textit{single pure state}, with correlations decaying exponentially with the distance. But when the Gibbs measure breaks into more pure states, then clustering property does not hold any longer and hence correlations may not decay fast enough to neglect the correction term $\varepsilon$ in~\autoref{eq:BP_condition_violated}, even in presence of very large graphs. This is a crucial point when dealing with disordered systems, as we will see throughout this thesis.

Finally, one last comment about the~\acrshort{BP} method. The removal of a site from graph $\mathcal{G}$ --- together with its incoming edges --- so to make its nearest-neighbour variables uncorrelated can be effectively depicted as the creation of a ``cavity'' in the original graph. This operation --- or equivalently the addition of further nodes and links --- is at the basis of the \textit{cavity method}, introduced in Ref.~\cite{MezardParisi1987} and then refined and revised in Refs.~\cite{MezardParisi2001, MezardParisi2003}. According to this method, the total free energy $F$ can be computed as the sum of the local terms coming from the elementary operations of link and site additions, and again this strategy provides exact results on (at least locally) treelike graphs. Such method is indeed completely equivalent to the~\acrshort{BP} approach, once noticed that the free energy shifts occurring at the addition of nodes and links are strictly related to the~\acrshort{BP} normalization constants $\mathcal{Z}_i$ and $\mathcal{Z}_{ij}$. For this reason, the~\acrshort{BP} probability distributions $\{\eta_{i\to j}\}$ are often referred to as \textit{cavity messages}, as well as their counterparts in the factor graph formalism.

\clearpage{\pagestyle{empty}\cleardoublepage}

\chapter{Spin glasses: the replica approach}
\label{chap:sg_replica}
\thispagestyle{empty}

This Chapter is meant to provide a brief recap of the results obtained for spin glasses on fully connected graphs --- namely within the na{\"i}ve mean-field framework~---, with a particular focus on the features of the different phase diagrams. Essential references are hence represented by~\cite{BinderYoung1986, Book_MezardEtAl1987, Book_FischerHertz1991, Book_Nishimori2001}. The starting point is represented by the experimental evidences that boosted the theoretical research on quenched disordered systems. Then, we move to the disordered version of the Ising model, reviewing its mean-field solution via the replica method. Finally, we face also the vector case, highligthing the similarities and the novelties with respect to the scalar case. Even though the results in this Chapter are already known in the literature, so that the experienced reader could safely skip the first two Sections, we highly recommend to read the Section about vector models, since we display several results that will be referred to throughout the thesis and that unfortunately seem not to be well recognized as for the ubiquitous Ising case.

\section{A new kind of magnetism}
\label{sec:new_kind_mag}

In~\autoref{chap:tools} we introduced the finite dimensional Ising model and its mean-field version, the~\acrshort{CW} model, showing how they reproduce quite well some key features of ordered ferromagnets. The attempts to solve them basically rely on two fundamental properties: the invariance under translation and the equivalence of all the sites. Unfortunately, in several \textit{real} cases they are not verified because some kind of heterogeneity occurs, due e.\,g. to vacancies in the arrangement of atoms or to replacements with atoms of another species. This particular kind of disorder is named \textit{quenched}, in order to stress the fact that it is fixed and does not enter in the Hamiltonian as a further degree of freedom, namely it does not change in the time.

There are two key features that identify these peculiar magnets with respect to ordered ones:
\begin{itemize}
	\item \textit{frustration}, namely the competition between the interactions of the various spins such that there does not exist a configuration of the spins that satisfies all the interactions;
	\item \textit{randomness} in the exchange couplings between spins, so to mix both ferromagnetic and antiferromagnetic interactions.
\end{itemize}

From the historical point of view, the first widely studied examples of quenched disorder in condensed matter have been the diluted solutions of transition metal impurities in a substrate of noble metals, such as manganese ($\mathrm{Mn}$) on gold ($\mathrm{Au}$), on silver ($\mathrm{Ag}$) or --- more often --- on copper ($\mathrm{Cu}$)~\cite{OwenEtAl1956, DeNobelDuChatenier1959, ZimmermanHoare1960}. It is referring to these magnetic alloys that the terms \textit{random magnets} and later \textit{spin glasses}\footnote{It appeared for the first time in a paper~\cite{Anderson1970} by Anderson in 1970.} have been introduced.

In these alloys, frustration and randomness are provided by the $d$-shell electrons of transition metals that polarize the $s$-shell conduction electrons of noble host metals. Indeed, the induced polarization has a characteristic oscillatory behaviour with the distance from the impurity~\cite{Book_AshcroftMermin1976}, so that also the resulting effective interaction between magnetic moments of impurities can be both positive or negative depending on their mutual distance.

The mechanism of the $s-d$ interaction was firstly proposed by Zener in~1951 \cite{Zener1951a, Zener1951b, Zener1951c}, though not taking into account antiferromagnetic effects. The oscillatory interaction was instead already known from studies on nuclear magnetism by Ruderman and Kittel~\cite{RudermanKittel1954}. So slightly later it was applied to the $s-d$ coupling in diluted magnetic alloys by Kasuya~\cite{Kasuya1956} and Yosida~\cite{Yosida1957}, leading to the famous \acrfull{RKKY} interaction:
\begin{equation}
	J_{\boldsymbol{x}\boldsymbol{y}} \sim \frac{\cos(2\boldsymbol{k}_F\cdot\boldsymbol{r})}{r^3}
\end{equation}
with $\boldsymbol{r}$ being the distance between $\boldsymbol{x}$ and $\boldsymbol{y}$, while $\boldsymbol{k}_F$ being the Fermi wave vector. So due to the random location of impurities, both strength and sign of exchange couplings are randomly distributed, providing the two key ingredients for a random magnet. Indeed, the first theoretical work on spin glasses actually considered \acrshort{RKKY} interactions~\cite{KleinBrout1963}.

The peculiar difference between ordinary ferromagnets and these random magnets, or spin glasses, lies in the features of the low-temperature phase. Indeed, in the latter case it is characterized by a set of ``frozen'' local spontaneous magnetizations giving a vanishing global magnetization, while in the former case the ferromagnetic long-range order is highlighted by the nonzero value of the global magnetization. As a direct consequence of the random orientation of local magnetizations, magnetic susceptibility in the low-temperature phase has a lower value with respect to its extrapolation from the high temperature phase.

Moreover, the two regimes are separated by a pronounced cusp at a well defined temperature, as also measured in several experiments on diluted magnetic alloys~\cite{CannellaEtAl1971, CannellaMydosh1972, Book_Mydosh1977}, so suggesting the presence of a second-order phase transition. The location of this phase transition is also experimentally found out to depend on the frequency $\omega$ of the applied magnetic field~\cite{HuserEtAl1973}, so that the ``true'' critical temperature is the \textit{static} one, obtained in the $\omega\to 0$ limit.

This crucial dependence on the \textit{protocol} through which the external field is applied is a signature of the presence of many metastable states, or \textit{valleys} --- i.\,e. stable spin configurations --- roughly equivalent from the energetic point of view but separated by energy barriers of different heights. This picture is analogous to what happens in Ising ferromagnets, where below the critical point there are two stable states --- the one with positive magnetization and the one with negative magnetization --- separated by an energy barrier that is exponentially large in the system size. However, in this case the number of these stable states is huge, with each one of them characterized by a certain set of local magnetizations $\{m_i\}$.

Other peculiar features of the so-called \textit{spin glass phase}, directly related to the rugged landscape of valleys and barriers sketched above, are the dramatic slowdown of the dynamics, remanence and memory effects, rejuvenation, chaos in temperature.

Finally, as for ordinary ferromagnets spin correlations extend to the whole system at the critical point, so making the ferromagnetic susceptibility $\chi_{\text{F}}$
\begin{equation}
	\chi_{\text{F}} \equiv \frac{\beta}{N}\sum_{i,j}\braket{\sigma_i\sigma_j}_c
\end{equation}
diverge, also for spin glasses there is a suitable susceptibility that diverges at the critical point, the spin glass one:
\begin{equation}
	\chi_{\text{SG}} \equiv \frac{\beta^2}{N}\sum_{i,j}\Bigl(\braket{\sigma_i\sigma_j}_c\Bigr)^2
\end{equation}
where it is the square of spin correlations that extends over the whole system, while due to the randomness of exchange couplings the spin correlations themselves decay quite fast.

If from one hand the presence of a diverging correlation length suggests to study the spin glass transition within the framework of second-order phase transitions, on the other hand the large number of metastable states and the other peculiar features of the low-temperature phase need for more refined and involved approaches, which we will briefly summarize in what follows.

\section{The Ising spin glass model}

As the archetypal of ferromagnets is represented by the Ising model, the same has happened for spin glasses. So we start our review about fully connected models of spin glass from the Ising case.

\subsection{The Edwards\,-\,Anderson model}

In 1975 Edwards and Anderson introduced a simple though effective spin glass version of the ferromagnetic Ising model on a $d$-dimensional hypercubic lattice~$\mathcal{G}$ --- with Hamiltonian~\autoref{eq:Hamiltonian_ferro_Ising} ---, that has been actually named as the~\acrfull{EA} model after them:
\begin{equation}
	\mathcal{H}_J[\{\sigma_i\}] = -\sum_{(i,j)}J_{ij}\,\sigma_i\sigma_j
\end{equation}
with the subscript `$J$' indicating the particular set of couplings $J_{ij}$'s identically and independently distributed according to a Gaussian distribution, so to mimic the randomness of the \acrshort{RKKY} distribution:
\begin{equation}
	\mathbb{P}_{J}(J) \sim \mathrm{Gauss}(J_0,J^2)
\end{equation}
The set of couplings $J_{ij}$'s for a lattice $\mathcal{G}$ represents a \textit{sample}, i.\,e. a fixed configuration of the disorder\footnote{In the following we will also refer to the samples as \textit{instances} of the problem. Moreover, we will see that the randomness can also involve the topology and/or the external field, so in general ``sample'' and ``instance'' refer to the whole realization of the quenched disorder.}.

Also the corresponding partition function has to depend on the set of couplings, as well as the free energy density:
\begin{equation}
	f_J = -\frac{1}{\beta N}\ln{\mathcal{Z}_J} = -\frac{1}{\beta N}\ln{\biggl[\sum_{\{\sigma_i\}}e^{-\beta\mathcal{H}_J[\{\sigma_i\}]}\biggr]}
\end{equation}
Being an extensive physical observable, it is meaningful to look at its average value over the disorder distribution:
\begin{equation}
	f \equiv \overline{f_J} = \int \di J \, \mathbb{P}_J(J) \, f_J
\end{equation}
with $\overline{\,\,\cdot\,\,}$ labeling the average over the disorder, namely over the samples. However, standard arguments of statistical mechanics claim that in the thermodynamic limit all the different realizations of the disorder yield the same value for the free energy density:
\begin{equation}
	\overline{f^2_J} - \left(\overline{f_J}\right)^2 = O\left(\frac{1}{N}\right)
\end{equation}
namely $f$ is said to be \textit{self averaging}. This property, possessed by several other physical observables, has a quite straightforward interpretation: indeed the larger the sample, the weaker the dependence on the realization of the quenched disorder.

However, not all physical observables are self averaging, but some of them must depend on the realization of the disorder. Indeed, the crucial idea of Edwards and Anderson is that --- on a given sample --- there exists an ensemble of local preferred directions for each spin such that, below the critical temperature $T_c$, the spin relaxes toward one of them. However, from the macroscopic point of view, the arrangement of such directions is completely random and it also changes according to different initial conditions. In other words, when decreasing the temperature, the system chooses a valley --- depending on the initial condition --- and then relaxes toward its bottom.

For the previous reason, the order parameter of this transition can not be the global magnetization as for ordinary ferromagnets, being zero both in the high- as well as in the low-temperature region. Instead, they define an \textit{overlap} --- later named after them --- between the direction of each spin at some reference time $t_0$ and the direction of the same spin at later times:
\begin{equation}
	q_{\text{EA}} \equiv \lim_{N\to\infty}\lim_{t\to\infty}\frac{1}{N}\sum_i\braket{\sigma_i(t_0)\,\sigma_i(t_0+t)}_t
\end{equation}
where $\braket{\cdot}_t$ is just the time average:
\begin{equation}
	\braket{\mathcal{O}(t)}_t \equiv \frac{1}{t}\int_0^t\di t'\mathcal{O}(t')
\end{equation}
So if the system remains trapped in a certain valley at $T<T_c$, then the resulting overlap $q_{\text{EA}}$ must differ from zero, while the restoration of the ergodicity --- namely the possibility of exploring \textit{all} the valleys --- would yield a vanishing overlap, as it actually happens above the critical point. Notice that the order of the two limits, $N\to\infty$ and $t\to\infty$, is crucial. If $N$ is finite, then in the $t\to\infty$ limit the system has actually the possibility to explore all the valleys and hence it does not remain trapped into a certain valley: the resulting overlap $q_{\text{EA}}$ must hence be zero. Conversely, if the infinite-time limit is performed before than the infinite-size one, the system actually gets trapped, so providing a non vanishing overlap $q_{\text{EA}}$.

Two main difficulties regard the solution of the~\acrshort{EA} model, and in particular the computation of the average free energy density $f$. Firstly, as for the ferromagnetic Ising model, an analytic computation of $f$ is not straightforward to obtain in the generic $d$-dimensional hypercubic case.

The second, and most important, issue regards the average over the disorder. Indeed, the \textit{quenched} average over the disorder mentioned above
\begin{equation}
	f = -\frac{1}{\beta N}\,\overline{\ln{\mathcal{Z}_J}}
	\label{eq:f_quenched}
\end{equation}
is not feasible, due to the non linearity of the logarithm. Conversely, if the average over the disorder is performed directly on the partition function $\mathcal{Z}$ and no longer on its logarithm, then the computation is by far simpler~\cite{Book_MezardEtAl1987, Book_FischerHertz1991}, leading to the definition of the \textit{annealed} free energy density:
\begin{equation}
	f_{ann} \equiv -\frac{1}{\beta N}\ln{\overline{\mathcal{Z}_J}}
	\label{eq:f_annealed}
\end{equation}

The physical interpretation of the two kinds of averages is well explained by the two terms borrowed by metallurgy and used to label them. Indeed, `annealing' means performing a very slow cooling of the sample during its preparation, so that the disorder degrees of freedom are allowed to change on the same time scale of spins. Hence, they actually enter in the Hamiltonian as further degrees of freedom and it is meaningful to average the partition function on both spin and disorder degrees of freedom at the same time. On the other side, `quenching' suggests a very rapid cooling of the sample, meaning that spins have to arrange toward their best configuration provided the frozen disorder realization. In this case, the disorder average has to be performed only on extensive physical observables, such as free energy density.

Even though the~\acrshort{EA} model contains quenched disorder and hence $f$ has to be computed as in~\autoref{eq:f_quenched}, the other crucial idea of Edwards and Anderson consists in the computation of the quenched average of $f_J$ via an analytic continuation on the annealed average of the free energy density of $n$ uncoupled \textit{replicas} of the initial system:
\begin{equation}
	 \mathcal{Z}_n \equiv \overline{(\mathcal{Z}_J)^n} = \overline{\sum_{\{\sigma^{(1)}_i\}}\sum_{\{\sigma^{(2)}_i\}}\dots\sum_{\{\sigma^{(n)}_i\}}e^{-\beta\sum_{a=1}^n\mathcal{H}_J[\{\sigma^{(a)}_i\}]}}
\end{equation}
so that:
\begin{equation}
	f \equiv f_0 = \lim_{n\to 0}\,f_n \qquad , \qquad f_n \equiv -\frac{1}{\beta Nn}\,\ln{\mathcal{Z}_n}
\end{equation}
This is nothing but the \textit{replica method}, based on the elementary identity:
\begin{equation}
	\ln{\mathcal{Z}} = \lim_{n\to 0}\,\frac{\mathcal{Z}^n-1}{n}
\end{equation}
which in turn comes from the expansion $\mathcal{Z}^n \simeq 1 + n\ln{\mathcal{Z}}$ for $n$ close to zero.

As long as $n$ is integer, the annealed average of the partition function can be computed quite easily, being the $n$ replicas uncoupled. So in the end the replica trick provides a way to actually compute the quenched free energy density:
\begin{equation}
	f = -\lim_{N\to\infty}\,\lim_{n\to 0}\,\frac{1}{\beta N}\frac{\overline{(\mathcal{Z}_J)^n}-1}{n}
\end{equation}

\subsection{The Sherrington\,-\,Kirkpatrick model}

A first successful application of the replica method developed by Edwards and Anderson has been performed by Sherrington and Kirkpatrick in the mean-field fully connected version  of the~\acrshort{EA} model: the~\acrfull{SK} model~\cite{SherringtonKirkpatrick1975, KirkpatrickSherrington1978}
\begin{equation}
	\mathcal{H}_J[\{\sigma_i\}] = -\frac{1}{2}\sum_{i \neq j}J_{ij}\,\sigma_i\sigma_j
	\label{eq:H_SK}
\end{equation}
with $J_{ij}$'s drawn from a Gaussian distribution of mean $J_0/N$ and variance $J^2/N$, where $J_0$ and $J^2$ are both of order one.

The first step of the replica method consists in the average over the disorder, which yields a surprising result:
\begin{equation}
\begin{split}
	\overline{(\mathcal{Z}_J)^n} &= \overline{\sum_{\{\sigma^{(1)}_i\}}\sum_{\{\sigma^{(2)}_i\}}\dots\sum_{\{\sigma^{(n)}_i\}}\exp{\biggl\{-\beta\sum_{a=1}^n\mathcal{H}_J[\{\sigma^{(a)}_i\}]\biggr\}}}\\
	&= \sum_{\{\sigma^{(1)}_i\}}\sum_{\{\sigma^{(2)}_i\}}\dots\sum_{\{\sigma^{(n)}_i\}}\exp{\Biggl\{\sum_{i \neq j}\biggl[\frac{\beta J_0}{2N}\sum_a\sigma^{(a)}_i\sigma^{(a)}_j+\frac{\beta^2J^2}{4N}\sum_{a,b}\sigma^{(a)}_i\sigma^{(b)}_i\sigma^{(a)}_j\sigma^{(b)}_j\biggr]\Biggr\}}
\end{split}
\end{equation}
namely the $n$ replicas of the original system have become coupled via a $4$-spin interaction, while disorder has disappeared from the system.

After some manipulations with the sums over the site and the replica indexes, these multi-spin interactions can be then linearized via the Hubbard\,-\,Stratonovich method, so inserting the auxiliary variable $m_a$ for terms with a single replica index
\begin{equation}
	\exp{\Biggl\{\frac{\beta J_0}{2N}\biggl(\sum_i\sigma^{(a)}_i\biggr)^2\Biggr\}} = \sqrt{\frac{\beta N}{2\pi}}\int \di m_a \, \exp{\biggl\{-\frac{\beta N}{2}m^2_a+\beta\sqrt{J_0} \, m_a\sum_i\sigma^{(a)}_i\biggr\}}
\end{equation}
and the auxiliary variable $q_{ab}$ for terms with two replica indexes
\begin{equation}
	\exp{\Biggl\{\frac{\beta^2 J^2}{4N}\biggl(\sum_i\sigma^{(a)}_i\sigma^{(b)}_i\biggr)^2\Biggr\}} = \sqrt{\frac{\beta^2 N}{\pi}}\int \di q_{ab} \, \exp{\biggl\{-\beta^2 Nq^2_{ab}+\beta^2 J \, q_{ab}\sum_i\sigma^{(a)}_i\sigma^{(b)}_i\biggr\}}
\end{equation}
So $m_a$ is a $n$-dimensional vector, while $q_{ab}$ is a $n \times n$ symmetric matrix, with the diagonal entries $q_{aa}$'s referring to subleading terms in $N$ with respect to the others and hence set equal to zero.

After some further manipulations, all the leading terms in $N$ and $n$ can be put into an effective action~$\mathcal{S}$
\begin{equation}
	\mathcal{Z}_n = \int \prod_a\left(\di m_a\right) \prod_{a,b}\left(\di q_{ab}\right) \exp{\Bigl\{-N\mathcal{S}\bigl[\{m_a\},\{q_{ab}\}\bigr]\Bigr\}}
\end{equation}
so that in the thermodynamic limit it is possible to obtain --- via a saddle-point evaluation --- the expression of the replicated free energy density $f_n$ averaged over the disorder:
\begin{equation}
	f_n = \frac{1}{\beta n}\min_{\{m_a\},\{q_{ab}\}}{\mathcal{S}\bigl[\{m_a\},\{q_{ab}\}\bigr]}
\end{equation}
from which in turn the quenched free energy density can be obtained:
\begin{equation}
	f = \lim_{n\to 0} \, f_n
\end{equation}

At this point, in order to go further with the evaluation of the saddle-point values for $m_a$'s and $q_{ab}$'s, Sherrington and Kirkpatrick made the crucial assumption of the \textit{symmetry between replicas}, which seemed to be quite reasonable. Indeed, the action $\mathcal{S}$ is symmetric under the permutation of the $n$ replicas. The~\acrfull{RS} ansatz so yields:
\begin{equation}
	m_a = m \qquad \forall\,a \quad, \qquad q_{ab}=q(1-\delta_{ab}) \qquad \forall\,a,b
\end{equation}
greatly simplifying the saddle-point equations, who finally read:
\begin{subequations}
\begin{equation}
	m = \int \di z \, \frac{e^{-z^2/2}}{\sqrt{2\pi}}\tanh{\bigl[\beta(J z \sqrt{q}+J_0 m)\bigr]}
	\label{eq:SK_RS_m}
\end{equation}
\begin{equation}
	q = \int \di z \, \frac{e^{-z^2/2}}{\sqrt{2\pi}}\tanh^2{\bigl[\beta(J z \sqrt{q}+J_0 m)\bigr]}
	\label{eq:SK_RS_q}
\end{equation}
	\label{eq:SK_RS}%
\end{subequations}

The physical interpretation of $m$ and $q$ so naturally comes out. Indeed, the former turns out to be related to the average magnetization of each replica of the system, related to the ferromagnetic bias $J_0$ of couplings, while the latter represents the correlations between spins from different replicas, related to the variance $J^2$ of the couplings:
\begin{equation}
	m = \overline{\braket{\sigma_i}} \qquad , \qquad q = \overline{\braket{\sigma_i}^2}
\end{equation}
So a phase with both $m$ and $q$ vanishing is paramagnetic, while when both them are different from zero there is a ferromagnetic ordering. Finally, when $q$ is different from zero while $m$ is equal to zero, a spin glass phase with a breaking of ergodicity is taking place, following the idea by Edwards and Anderson.

\begin{figure}[t]
	\centering
	\includegraphics[scale=0.8]{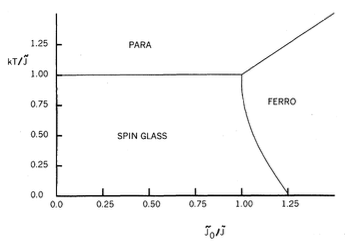}
	\caption[Phase diagram of the SK model with the RS ansatz]{Phase diagram of the~\acrshort{SK} model with the~\acrshort{RS} assumption. Temperatures $T$ on the vertical axis are rescaled by $J$, as well as $J_0$ values on the horizontal axis. Reprinted from~\cite{SherringtonKirkpatrick1975}.}
	\label{fig:SK_phase_diagram}
\end{figure}

The critical lines between these three different phases have been then computed by Sherrington and Kirkpatrick from the solutions of self-consistency equations~\autoref{eq:SK_RS}, as shown in the $J_0$ vs $T$ phase diagram of~\autoref{fig:SK_phase_diagram}. Paramagnetic solution becomes unstable in correspondence of the largest temperature between $J_0$ and~$J$, $T_c=\max{\{J_0,J\}}$, while the critical line between the two different ordered states is \textit{reentrant} toward the ferromagnetic phase, going from $J_0=J,T=J$ to $J_0=\sqrt{\pi/2}\,J,T=0$.

Unfortunately, Sherrington and Kirkpatrick found some issues in the low-temperature region. In particular, the entropy density was negative at low enough temperatures
\begin{equation}
	\lim_{T\to 0}\,s(T) = -\frac{1}{2\pi}
\end{equation}
clearly providing an unphysical result for a discrete model as the~\acrshort{SK} one. Even if it was initially blamed the interchange between the $N\to\infty$ limit and the $n\to 0$ limit, quite suddenly it became clear that the issue was actually rooted in the~\acrshort{RS} assumption.

\subsection{The breaking of replica symmetry}

Indeed, the saddle point condition
\begin{equation}
	\frac{\partial \mathcal{S}}{\partial q_{ab}} = 0 \qquad \forall\,a,b
	\label{eq:replica_stat_cond}
\end{equation}
is not on its own a guarantee of the stability of the extremal point $\{q^*_{ab}\}$. Moreover, given that $q_{ab}$ matrix is symmetric and has zero entries on the diagonal, the number of its independent variable is $n(n-1)/2$ and hence when performing the analytic continuation toward $n=0$, such number becomes negative below $n=1$. Direct consequences of this are, for example, the change of minima into maxima and the appearance of negative determinants~\cite{Book_MezardEtAl1987}.

In order to check the stability of the extremal points coming from~\autoref{eq:replica_stat_cond}, the standard study on the second derivatives of $\mathcal{S}$ has to be performed out, so defining its Hessian matrix $\mathbb{H}$ in the space of replicas:
\begin{equation}
	\mathbb{H}_{ab,cd} \equiv \frac{\partial^2 \mathcal{S}}{\partial q_{ab}\,\partial q_{cd}}\Biggr{|}_{\{q^*_{ab}\}}
\end{equation}
In this way, the stability of the stationary point $\{q^*_{ab}\}$ is ensured when $\mathbb{H}$ is positive definite.

The stability of the~\acrshort{RS} solution $q^*_{ab}=q$ has been finally checked by de Almeida and Thouless in 1978~\cite{deAlmeidaThouless1978}, who realized that among the three distinct eigenvalues of $\mathbb{H}$ for $n>1$, one of them --- named the \textit{replicon}~\cite{BrayMoore1978} --- actually becomes negative for $T<T_c$ and $J_0=0$ when $n\to 0$:
\begin{equation}
	\lambda_1 = 1 - \beta^2 \int \di z \, \frac{e^{-z^2/2}}{\sqrt{2\pi}}\biggl[1-\tanh^2{(\beta J z \sqrt{q})}\biggr]^2
\end{equation}
So the~\acrshort{RS} ansatz is definitely wrong in the spin glass phase, hence a different solution must be considered, which takes into account the~\textit{breaking of the symmetry} between replicas, so leading to the~\acrfull{RSB} scenario.

\begin{figure}[t]
	\centering
	\includegraphics[scale=0.6]{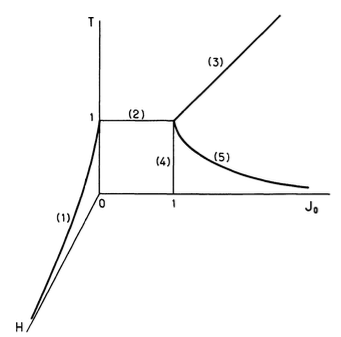}
	\caption[Phase diagram of the SK model with the RSB ansatz]{Phase diagram of the~\acrshort{SK} model when taking into account~\acrshort{RS} instability. Values of $T$, $H$ and $J_0$ are again meant to be rescaled by $J$. Line (1) corresponds to the~\acrshort{RS} instability in presence of a field $H$, while lines (2) and (5) refer to the case with $J_0 \neq 0$. So line (3), separating paramagnetic from ferromagnetic phase, entirely lies in the~\acrshort{RS} stable region. Vertical line (4) is where global magnetization $m$ actually vanishes, according to the Toulouse argument. Reprinted from~\cite{Toulouse1980}.}
	\label{fig:dAT_phase_diagram}
\end{figure}

An analogous instability has been found by de Almeida and Thouless in the case of an external homogeneous field $H$. Indeed, if on one hand there are no phase transitions in this case within the~\acrshort{RS} ansatz --- there always being a globally magnetized phase --- on the other hand also in this case the replicon $\lambda_1$ becomes negative on a well defined line in the $T$ vs $H$ plane, identified by the condition
\begin{equation}
	\left(\frac{1}{\beta J}\right)^2 = \int \di z \, \frac{e^{-z^2/2}}{\sqrt{2\pi}} \sech^4{\bigl[\beta(J z \sqrt{q}+H)\bigr]}
\end{equation}
and named~\acrfull{dAT} line, with the~\acrshort{RS} values of $m$ and $q$ given by
\begin{subequations}
\begin{equation}
	m = \int \di z \, \frac{e^{-z^2/2}}{\sqrt{2\pi}}\tanh{\bigl[\beta(J z \sqrt{q}+H)\bigr]}
	\label{eq:SK_RS_m_H}
\end{equation}
\begin{equation}
	q = \int \di z \, \frac{e^{-z^2/2}}{\sqrt{2\pi}}\tanh^2{\bigl[\beta(J z \sqrt{q}+H)\bigr]}
	\label{eq:SK_RS_q_H}
\end{equation}
	\label{eq:SK_RS_H}%
\end{subequations}
It can be observed in the $T$ vs $H$ plane of the phase diagram in~\autoref{fig:dAT_phase_diagram}.

It is worth highlighting two key features of the~\acrshort{dAT} line, since we will recall them in the course of the thesis. Firstly, a power series expansion for small fields yields:
\begin{equation}
	\left(\frac{H}{J}\right)^2 \simeq \frac{4}{3}\left(\frac{T_c-T}{T_c}\right)^3
	\label{eq:scaling_dAT_line_near_zeroH_fullyConnected}
\end{equation}
from which the $3/2$ value of the~\acrshort{dAT} line critical exponent in the~\acrshort{SK} model. Secondly, a power series expansion for large fields gives:
\begin{equation}
	\frac{1}{\beta J} \simeq \frac{4}{3\sqrt{2\pi}}\exp{\left\{-\frac{H^2}{2J^2}\right\}}
	\label{eq:scaling_dAT_line_near_zeroT_fullyConnected_H}
\end{equation}
showing that for any value of the field $H$, there is always a small enough temperature such that the breaking of replica symmetry occurs.

Finally, an analogous~\acrshort{RS} instability line also appears when a ferromagnetic bias $J_0$ is present, instead of the external field $H$. Indeed, for $J_0<J$ it corresponds to the second-order critical line between the paramagnetic and the spin glass phase. Then, for $J_0>J$ the~\acrshort{RS} instability line replaces the transition line between the ferromagnetic and the spin glass phase computed by Sherrington and Kirkpatrick, since occurring at higher critical temperatures, given by:
\begin{equation}
	\left(\frac{1}{\beta J}\right)^2 = \int \di z \, \frac{e^{-z^2/2}}{\sqrt{2\pi}} \sech^4{\bigl[\beta(J z \sqrt{q}+J_0 m)\bigr]}
\end{equation}
The exponent of this line close to the multicritical point $J_0=J,T=J$ can be computed by looking at the Toulouse argument~\cite{Toulouse1980, Book_FischerHertz1991} for the equivalence between the $H=0,J_0\neq 0$ case and the $H\neq 0,J_0=0$ case. Indeed, keeping in mind the correspondence between the effective field $J_0 m$ of the former case and the actual field $H$ of the latter case, and moreover recalling the mean-field prediction of the growing of the global magnetization below the critical temperature, $m \propto (T_c-T)^{1/2}$, the~\acrshort{dAT} line decreases as a square root for $J_0 \gtrsim J$:
\begin{equation}
	\left(\frac{J_0-J}{J}\right) \propto \left(\frac{T_c-T}{T_c}\right)^2
	\label{eq:scaling_dAT_line_near_mcPoint_fullyConnected}
\end{equation}
Instead, on the other end of the~\acrshort{dAT} line, it turns out that in fact there is no endpoint on the $T=0$ axis, since for any value of $J_0$ there always exists a critical temperature such that $\lambda_1$ becomes negative:
\begin{equation}
	\frac{1}{\beta J} \simeq \frac{4}{3\sqrt{2\pi}}\exp{\left\{-\frac{J^2_0}{2J^2}\right\}}
	\label{eq:scaling_dAT_line_near_zeroT_fullyConnected_J0}
\end{equation}
Notice that this scaling has exactly the same expression of~\autoref{eq:scaling_dAT_line_near_zeroT_fullyConnected_H}, again due to the correspondence $H \leftrightarrow J_0 m$ and to the fact that $m\to 1$ when $T\to 0$. Indeed, they actually belong to the same~\acrshort{dAT} hypersurface in the $T$ vs $H$ vs $J_0$ space, below which the~\acrshort{RS} ansatz is definitely unstable~\cite{Toulouse1980}.

Another striking difference between the $T$ vs $J_0$ phase diagram depicted by Sherrington and Kirkpatrick in the~\acrshort{RS} ansatz and the one obtained by de Almeida and Thouless taking into account~\acrshort{RS} instability is the presence of a new phase lying between the ferromagnetic one and the spin glass one. Indeed, for $J_0>J$ the global magnetization $m$ is correctly given by the~\acrshort{RS} expression~\autoref{eq:SK_RS_m} until reaching the~\acrshort{dAT} line, hence it is still different from zero when crossing it. So there should be a further line where $m$ actually goes to zero, but such line can not be represented by the reentrant one found by Sherrington and Kirkpatrick in the~\acrshort{RS} approach. Instead, it has been predicted by Toulouse to be located at $J_0=J$ for any $T<T_c$, according to an argument regarding the zero-field susceptibility~$\chi$~\cite{Toulouse1980}. So the region of~\acrshort{RS} instability actually splits into two different phases: the proper spin glass one with $m=0$ and a \textit{mixed} one with $m \neq 0$ but still~\acrshort{RS} unstable. The correct phase diagram of the~\acrshort{SK} model, taking into account the breaking of replica symmetry, is eventually reported in~\autoref{fig:dAT_phase_diagram}.

Even though it was established that the~\acrshort{RS} assumption is incorrect in the low-temperature region, it was not yet clear how to break such symmetry. An attempt was made by Thouless, Anderson and Palmer~\cite{ThoulessEtAl1977}, who constructed the mean-field theory of the~\acrshort{SK} model \textit{before} averaging over the disorder and hence with no need of introducing replicas. Indeed, starting from the Bethe formulation of the cavity method~\cite{Bethe1935}, they tried to extend the mean-field theory of ordered ferromagnets to the case of spin glasses, finding the~\acrfull{TAP} equation:
\begin{equation}
	m_i = \tanh{\Bigl[\beta H + \beta\sum_j J_{ij}m_j - \beta^2 m_i (1-q_{\text{EA}})\Bigr]}
	\label{eq:TAP_eq}
\end{equation}
that actually recalls the ferromagnetic one~\autoref{eq:Ising_mean_field_eq}, apart from the \textit{Onsager reaction term} $\beta^2 m_i (1-q_{\text{EA}})$~\cite{Onsager1936, BarkerWatts1973} that takes into account the feedback from changes in $m_i$ itself. Unfortunately, the~\acrshort{TAP} approach has been proved to be correct only where the~\acrshort{RS} solution is stable as well~\cite{KirkpatrickSherrington1978, Plefka1982, Plefka2002}.

The puzzle has been finally solved slightly later by Parisi in a series of papers, finding the right direction along which the breaking of replica symmetry actually takes place. Indeed, he firstly observed~\cite{Parisi1979a} that the addition of a further parameter in the matrix overlap $q_{ab}$ improves the~\acrshort{RS} solution given by Sherrington and Kirkpatrick, still providing a negative entropy in the $T\to 0$ limit, even though by far closer to zero. More precisely, the~\acrfull{1RSB} method consists in allowing the overlap between different replicas to take on two different values, so that each replica has an overlap $q_1$ with other $m_1-1$ replicas, while it has a smaller overlap $q_0$ with the remaining $n-m_1$ replicas. This ansatz can be concisely expressed by means of the probability distribution of the overlaps in the replica space:
\begin{equation}
	\mathbb{P}_n(q) = \frac{m_1-1}{n-1}\,\delta(q-q_1) + \frac{n-m_1}{n-1}\,\delta(q-q_0)
	\label{eq:1RSB_q_distrib_n}
\end{equation}

More visually, the~\acrshort{1RSB} ansatz gives the following \textit{block representation} for the overlap matrix $q_{ab}$:
\begingroup
\renewcommand{\arraystretch}{1.25}
\[
	q = 
	\left(
		\begin{array}{ccc:ccc:ccc}
			0 & q_1 & q_1 & \multicolumn{3}{c}{\multirow{3}{*}{}} & \multicolumn{3}{c}{\multirow{3}{*}{$q_0$}}\\
			q_1 & 0 & q_1 & \multicolumn{3}{c}{} & \multicolumn{3}{c}{}\\
			q_1 & q_1 & 0 & \multicolumn{3}{c}{} & \multicolumn{3}{c}{}\\
			\cdashline{1-6}
			\multicolumn{3}{c:}{\multirow{3}{*}{}} & 0 & q_1 & q_1 & \multicolumn{3}{c}{\multirow{3}{*}{}}\\
			\multicolumn{3}{c:}{} & q_1 & 0 & q_1 & \multicolumn{3}{c}{}\\
			\multicolumn{3}{c:}{} & q_1 & q_1 & 0 & \multicolumn{3}{c}{}\\
			\cdashline{4-9}
			\multicolumn{3}{c}{\multirow{3}{*}{$q_0$}} & \multicolumn{3}{c:}{\multirow{3}{*}{}} & 0 & q_1 & q_1\\
			\multicolumn{3}{c}{} & \multicolumn{3}{c:}{} & q_1 & 0 & q_1\\
			\multicolumn{3}{c}{} & \multicolumn{3}{c:}{} & q_1 & q_1 & 0
		\end{array}
	\right)
\]
\endgroup
that is divided into $m_1/n \times m_1/n$ square blocks of linear size $m_1$, such that diagonal blocks have the larger overlap $q_1$, with the diagonal entries $q_{aa}$'s still set equal to zero, while off-diagonal blocks have the smaller overlap $q_0$.

In the $n\to 0$ limit, the overlap probability distribution~\autoref{eq:1RSB_q_distrib_n} remains well defined if $m_1$ belongs to the $[0,1]$ interval (so acquiring the meaning of a \textit{fraction} of replicas):
\begin{equation}
	\mathbb{P}(q) = (1-m_1)\delta(q-q_1) + m_1\delta(q-q_0)
	\label{eq:1RSB_q_distrib_nTo0}
\end{equation}
Its value is the one given by the saddle-point evaluation, $\partial\mathcal{S}/\partial m_1=0$, together with the saddle-point equations for $q_0$ and $q_1$, that in the isotropic case $J_0=H=0$ read:
\begin{subequations}
	\begin{equation}
		q_0 = \int \mathcal{D}z_0 \, \left\{\frac{\int \mathcal{D}z_1 \, \bigl[\mathcal{F}(z_0,z_1)\bigr]^{m_1}\tanh{\bigl[\beta J(z_1\sqrt{q_1-q_0}+z_0\sqrt{q_0})\bigr]}}{\int \mathcal{D}z_1 \, \bigl[\mathcal{F}(z_0,z_1)\bigr]^{m_1}} \right\}^2
		\label{eq:SK_1RSB_q0}
	\end{equation}
	\begin{equation}
		q_1 = \int \mathcal{D}z_0 \, \frac{\int \mathcal{D}z_1 \, \bigl[\mathcal{F}(z_0,z_1)\bigr]^{m_1}\tanh^2{\bigl[\beta J(z_1\sqrt{q_1-q_0}+z_0\sqrt{q_0})\bigr]}}{\int \mathcal{D}z_1 \, \bigl[\mathcal{F}(z_0,z_1)\bigr]^{m_1}}
		\label{eq:SK_1RSB_q1}
	\end{equation}
	\label{eq:SK_1RSB}%
\end{subequations}
where $\mathcal{D}z_{0,1} \equiv \di z_{0,1} \exp{\{-z^2_{0,1}/2\}}/\sqrt{2\pi}$ is the Gaussian measure already encountered in the~\acrshort{RS} ansatz, while $\mathcal{F}(z_0,z_1)$ contains the usual contribution from Ising spins:
\begin{equation}
	\mathcal{F}(z_0,z_1) \equiv \exp{\biggl\{-\frac{\beta^2 q_1}{2}\biggr\}}\,2\cosh{\Bigl[\beta J(z_1\sqrt{q_1-q_0}+z_0\sqrt{q_0})\Bigr]}
\end{equation}
It is evident the similarity between these saddle-point equations and the~\acrshort{RS} ones~\autoref{eq:SK_RS}, with the further level of average over the Gaussian measure $\mathcal{D}z_1$ and the \textit{reweighing} through the factor $\bigl[\mathcal{F}(z_0,z_1)\bigr]^{m_1}$.

The generalization to further steps of~\acrshort{RSB} followed quite suddenly~\cite{Parisi1979b, Parisi1980a, Parisi1980b}. Indeed, in the~\acrfull{2RSB} ansatz the diagonal blocks of linear size $m_1$ are again divided into sub-blocks of linear size $m_2$, while off-diagonal blocks are left unchanged, so providing the following block representation of the overlap matrix:
\begingroup
\renewcommand{\arraystretch}{1.25}
\[
	q = 
	\left(
		\begin{array}{cc:cc:cc:cc}
			0 & q_2 & \multicolumn{2}{c:}{\multirow{2}{*}{$q_1$}} & \multicolumn{4}{c}{\multirow{4}{*}{$q_0$}}\\
			q_2 & 0 & \multicolumn{2}{c:}{} & \multicolumn{4}{c}{}\\
			\cdashline{1-4}
			\multicolumn{2}{c:}{\multirow{2}{*}{$q_1$}} & 0 & q_2 & \multicolumn{4}{c}{}\\
			\multicolumn{2}{c:}{} & q_2 & 0 & \multicolumn{4}{c}{}\\
			\cdashline{1-8}
			\multicolumn{4}{c:}{\multirow{4}{*}{$q_0$}} & 0 & q_2 & \multicolumn{2}{c}{\multirow{2}{*}{$q_1$}}\\
			\multicolumn{4}{c:}{} & q_2 & 0 & \multicolumn{2}{c}{}\\
			\cdashline{5-8}
			\multicolumn{4}{c:}{} & \multicolumn{2}{c:}{\multirow{2}{*}{$q_1$}} & 0 & q_2\\
			\multicolumn{4}{c:}{} & \multicolumn{2}{c:}{} & q_2 & 0
		\end{array}
	\right)
\]
\endgroup
with the corresponding overlap distribution in the $n\to 0$ limit:
\begin{equation}
	\mathbb{P}(q) = (1-m_2)\delta(q-q_2) + (m_2-m_1)\delta(q-q_1) + m_1\delta(q-q_0)
	\label{eq:2RSB_q_distrib_nTo0}
\end{equation}
The~\acrfull{kRSB} ansatz acts along the same way, so producing an overlap matrix with a \textit{hierarchy} of sub-blocks along the diagonal, characterized by $k+1$ different values of the replica overlap, while the overlap distribution reads in the $n\to 0$ limit:
\begin{equation}
\begin{split}
	\mathbb{P}(q) =& \,(1-m_k)\delta(q-q_k) + (m_k-m_{k-1})\delta(q-q_{k-1})\\
		&\qquad + \dots + (m_2-m_1)\delta(q-q_1) + m_1\delta(q-q_0)
	\label{eq:kRSB_q_distrib_nTo0}
\end{split}
\end{equation}
with $0<m_1<m_2<\dots<m_{k-1}<m_k<1$ so to have a properly defined probability distribution.

Two key observations mainly suggest that the correct solution should be the one obtained in the $k\to\infty$ limit. First, the negative value acquired by both the entropy density at zero temperature and the replicon $\lambda_1$ get closer and closer to zero when increasing the number $k$ of~\acrshort{RSB} steps. Second, the increasing sequence of overlap values $\{q_i\}$ can be arranged as a step-wise function
\begin{equation}
	q(x) \equiv q_i \qquad \text{for} \quad m_i < x < m_{i+1}
\end{equation}
that for large values of $k$ gets closer to a smooth function~\cite{Parisi1980a}. So Parisi eventually sent $k$ to infinity, moving to the~\acrfull{fRSB} ansatz~\cite{Parisi1980b}. In this frame, the order parameter $q(x)$ of spin glasses is finally recognized to be a continuous function on the $[0,1]$ interval:
\begin{equation}
	q(x) = 
	\left\{
	\begin{aligned}
		&q_m \qquad && x < x_m\\
		&q_m < q(x) < q_M \qquad && x_m < x < x_M\\
		&q_M \qquad && x > x_M\\
	\end{aligned}
	\right.
\end{equation}
with $q_m$ and $x_m$ that are different from zero only in presence of an external field, and that go to zero with $H$ as $H^{2/3}$. The explicit expression of $q(x)$ can be obtained as usual from the stationarity condition $\delta \mathcal{S}[q]/\delta q(x)=0$, which now translates into the Parisi nonlinear antiparabolic differential equation~\cite{Parisi1980b} for $\mathrm{f}(q,y)$:
\begin{equation}
	\frac{\partial \mathrm{f}}{\partial q} = -\frac{J^2}{2}\left[\frac{\partial^2 \mathrm{f}}{\partial y^2}+x(q)\left(\frac{\partial \mathrm{f}}{\partial y}\right)^2\right]
\end{equation}
to be solved in the interval $[q_m,q_M]$ with the boundary condition $\mathrm{f}(q_M,y)=\ln{(2\cosh{\beta y})}$, where $x(q)$ is given by
\begin{equation}
	x(q) = \int_0^q \di q' \, \mathbb{P}(q')
\end{equation}
and where $y$ is the local effective field
\begin{equation}
	y = \beta\Bigl[H + J_0 m + J\bigl(z_0\sqrt{q_0}+z_1\sqrt{q_1-q_0}+z_2\sqrt{q_2-q_1}+\dots\bigr)\Bigr]
\end{equation}

An equivalent description is given by the overlap distribution $\mathbb{P}(q)$. Indeed, in the~\acrshort{fRSB} ansatz it has two Dirac delta functions respectively at $q=q_m$ and $q=q_M$, while on the inbetween values it is a smooth function $\widetilde{\mathbb{P}}(q)$ with support in the interval $(q_m,q_M)$:
\begin{equation}
	\mathbb{P}(q) = x_m\,\delta(q-q_m) + (1-x_m-x_M)\widetilde{\mathbb{P}}(q) + x_M\,\delta(q-q_M)
	\label{eq:fRSB_q_distrib_nTo0}
\end{equation}
This translates into the following picture: with probability $x_m$ two replicas have an overlap $q_m$, with probability $x_M$ they have an overlap $q_M$ and finally with probability $1-x_m-x_M$ they have an overlap intermediate between $q_m$ and $q_M$.

\subsection{The nature of the spin glass phase}

A first confirmation of the correctness of the~\acrshort{fRSB} solution provided by Parisi is given by the value of the entropy density at zero temperature. Indeed, if when increasing the number $k$ of~\acrshort{RSB} steps it gets closer and closer to zero, it turns out to actually vanish in the $k\to\infty$ limit, as expected for a discrete model at zero temperature. Moreover, also the replicon $\lambda_1$ ceases to be negative below $T_c$ only in such limit, so highlighting a peculiar feature of the~\acrshort{fRSB} solution of the~\acrshort{SK} model: it is \textit{marginally stable}~\cite{deDominicisKondor1983}.

However, a formal proof of the correctness of the solution provided by Parisi had to wait for a long time. Indeed, if on one hand the possibility of the interchange of the two limits $n\to 0$ and the $N\to\infty$ has been assured in 1979~\cite{vanHemmenPalmer1979}, some other mathematical subtleties --- e.\,g. the existence and the uniqueness of the solution in the thermodynamic limit --- had to wait two decades to be rigorously proven~\cite{GuerraToninelli2002, Guerra2003, Book_Talagrand2003}.

Once solved the~\acrshort{SK} model, the next step was to assign a physical interpretation to replicas. This goal has been achieved slightly later by Parisi~\cite{Parisi1983}, who showed that the probability distribution $\mathbb{P}_{\text{rep}}(q)$ of the overlaps between replicas
\begin{equation}
	\mathbb{P}_{\text{rep}}(q) \equiv \lim_{n\to 0}\frac{1}{n(n-1)}\sum_{a \neq b}\delta(q-q_{ab}) \qquad , \qquad q_{ab} \equiv \frac{1}{N}\sum_i \sigma^{(a)}_i \sigma^{(b)}_i
\end{equation}
actually coincides with the probability distribution $\mathbb{P}_{\text{states}}(q)$ of the overlaps between magnetizations in different states of the Gibbs measure
\begin{equation}
	\mathbb{P}_{\text{states}}(q) \equiv \sum_{\alpha,\beta}P_{\alpha}P_{\beta}\,\delta(q-q_{\alpha\beta}) \qquad , \qquad q_{\alpha\beta} \equiv \frac{1}{N}\sum_i m^{(\alpha)}_i m^{(\beta)}_i
\end{equation}
with $P_{\alpha}$ being the statistical weight of the pure state $\alpha$ coming from the decomposition of the Gibbs measure. So the replicas actually acquire a physical meaning: if different replicas of the systems behave in the same way, i.\,e. they all have the same mutual overlap $q=q_{\text{EA}}$, then there is a unique pure state and hence no replica symmetry breaking does occur. Instead, if the replicas differ from each other, $q$ is described by a broad probability distribution, replica symmetry breaking has occurred and hence Gibbs measure has broken into several pure states. Keeping this in mind, the~\acrshort{fRSB} probability distribution~\autoref{eq:fRSB_q_distrib_nTo0} can be also thought to refer to the overlap $q_{\alpha\beta}$ between couples of states $\alpha$ and $\beta$, rather than between couples of replicas $a$ and $b$.

At this point, it is clear that the hierarchical structure of replica overlaps has to reflect on the topological structure of the states. Indeed, states are found to be arranged according to their mutual overlap in a ultrametric structure~\cite{MezardEtAl1984a, MezardVirasoro1985, RammalEtAl1986}. So given any three states $\alpha$, $\beta$ and $\gamma$, it turns out to hold for their mutual distances in the overlap space that
\begin{equation}
	d_{\alpha\beta} \leqslant \max{\bigl\{d_{\alpha\gamma},d_{\beta\gamma}\bigr\}}
\end{equation}
instead of the usual (weaker) triangular inequality $d_{\alpha\beta} \leqslant d_{\alpha\gamma} + d_{\beta\gamma}$. In terms of overlaps, the ultrametric property can also be written as
\begin{equation}
	q_{\alpha\beta} \geqslant \min{\bigl\{q_{\alpha\gamma},q_{\beta\gamma}\bigr\}}
\end{equation}
More pictorially, the three states can be thought as located at the vertices of a triangle that is either equilateral or isosceles, with the sides measured in terms of mutual overlap.

Other striking features of the spin glass phase can be interpreted in terms of the breaking of the Gibbs measure in a large number of~\textit{metastable} states. Indeed, the space of spin configurations can be imagined as a \textit{rugged} landscape with a huge number of valleys --- i.\,e. the minima --- separated by high mountains --- i.\,e. the free energy barriers separating different stable configurations. The number of such minima turns out to be exponentially large in the size $N$~\cite{BrayMoore1980, DeDominicisEtAl1980, Young1981}. Moreover, also the height of the free energy barriers diverge with some power of $N$, so that the time spent by the system in a certain valley is exponentially large in $N$ as well, resulting in an extremely slow dynamics~\cite{MackenzieYoung1982, MackenzieYoung1983}. In the thermodynamic limit the ergodicity eventually breaks and the system remains trapped in a given \textit{metastable} state.

A first consequence of this is the lack of self averaging~\cite{YoungEtAl1984}, already mentioned at the beginning of this Section. In this picture, non self-averaging observables are just the ones that contain \textit{intervalley} correlations, which change according to the different realizations of the disorder. So when ergodicity breaks, the system is no longer able to visit all the configurations and a strong dependence on the sample occurs. On the other hand, self-averaging quantities --- e.\,g. the free energy and the global magnetization --- are just computed by means of single valley contributions, hence they turn out to be self averaging.

\section{Vector spin glasses}
\label{sec:vector_sg_fully}

The choice of Ising spins for the~\acrshort{SK} model mainly rests on the simplification induced by dealing with scalar spins instead of vector spins, namely with discrete instead of continuous degrees of freedom. However, several physical situations do not show any such particular anisotropy that justifies the projection of magnetic moments on the $z$ axis.

Moreover, from experiments performed on spin glass materials with different degrees of anysotropy~\cite{BertEtAl2004}, it is known that some peculiar features of the spin glass phase --- e.\,g. memory effects, aging, rejuvenation, and so on --- are continuously reduced when decreasing the spin anysotropy. On the other hand, no clear evidence about the aforementioned signatures of the complex and hierarchical organization of the spin glass long-range order has been found in numerical simulations on Ising spins~\cite{PiccoEtAl2001, MaioranoEtAl2005}. So it is crucial to take into account the vectorial nature of spins in the previous theoretical description.

The natural generalization is hence to allow spins $\boldsymbol{\sigma}_i$'s to rotate in a $m$-dimensional space, with their norm $\sigma_i=\sum_{\mu=1}^m\sigma^2_{i,\mu}$ fixed to $m$. In this way, we will be able to compare more easily the resulting phase diagrams with the Ising ones, since the corresponding critical temperatures will be $m$-independent.

\subsection{The isotropic case}

Let us start the analysis from the isotropic case, namely $J_0=H=0$. The~\acrshort{SK} Hamiltonian~\autoref{eq:H_SK} preserves its formal structure when moving to the vector case
\begin{equation}
	\mathcal{H}_J[\{\boldsymbol{\sigma}_i\}] = -\frac{1}{2}\sum_{i \neq j}J_{ij}\,\boldsymbol{\sigma}_i\cdot\boldsymbol{\sigma}_j
	\label{eq:H_SK_m}
\end{equation}
so that the model can be again solved via the replica method. Obviously, the overlap $q_{ab}$ between two replicas now becomes an $m \times m$ tensor
\begin{equation}
	q_{ab,\mu\nu} \equiv \frac{1}{N}\sum_i\sigma^{(a)}_{i,\mu}\sigma^{(b)}_{i,\nu}
\end{equation}
with $\mu,\nu=1,\dots,m$ labeling spatial components. However, being in the isotropic case, it reduces to a multiple of the identity
\begin{equation}
	q_{ab,\mu\nu} = q_{ab}\delta_{\mu\nu}
\end{equation}
due to the invariance of the model under rotations of the spins.

The~\acrshort{RS} ansatz $q_{ab}=q(1-\delta_{ab})$ --- firstly exploited by Sherrington and Kirkpatrick themselves for the XY model~\cite{KirkpatrickSherrington1978} and later generalized by de Almeida \textit{et al.} for the generic-$m$ case~\cite{deAlmeidaEtAl1978} --- leads to a self-consistency equation for~$q$ that is formally equivalent to the one~\autoref{eq:SK_RS_q} of the Ising case\footnote{Notice that in this Equation $m$ represents the number of spin components and not the magnetization.}
\begin{equation}
	q = \frac{2^{\,(2-m)/2}}{\Gamma(m/2)}\int_0^{\infty} \di z \, z^{m-1} \, e^{-z^2/2} \left[\frac{I_{m/2}(\beta J z \sqrt{mq})}{I_{(m-2)/2}(\beta J z \sqrt{mq})}\right]^2
	\label{eq:SK_RS_q_m}
\end{equation}
where $\Gamma(\cdot)$ is the usual gamma function and $I_m(\cdot)$ is the modified Bessel function of the first kind of order $m$~\cite{Book_AbramowitzStegun1964}. Special cases are the $m=1$ one
\begin{equation}
	\frac{I_{1/2}(x)}{I_{-1/2}(x)} = \tanh{x}
\end{equation}
so that it is actually possible to recover~\autoref{eq:SK_RS_q}, and the $m=3$ one
\begin{equation}
	\frac{I_{3/2}(x)}{I_{1/2}(x)} = \coth{x} - \frac{1}{x}
\end{equation}
that gives back the well known Langevin function.

For any finite value of $m$, Equation~\autoref{eq:SK_RS_q_m} admits the paramagnetic solution $q=0$ in the high-temperature region and the spin glass solution $q>0$ in the low-temperature region, separated by a second-order phase transition at $T_c=J$. However, exactly as it happens for Ising spins, the~\acrshort{RS} solution is unstable in the low-temperature region~\cite{deAlmeidaEtAl1978}, since in the $n\to 0$ limit there are nine distinct eigenvalues of the Hessian $\mathbb{H}$ and some of them are negative below $T_c$. So again a~\acrshort{RSB} ansatz has to be used, following the same steps of the Parisi solution in the Ising case. Consequently, peculiar features of the~\acrshort{fRSB} ansatz --- e.\,g. lack of self averaging, marginal stability, ultrametricity, and so on --- will occur as well.

A peculiar case is instead represented by the $m\to\infty$ limit, that is found to be~\acrshort{RS} stable~\cite{deAlmeidaEtAl1978}, with
\begin{equation}
	q =
	\left\{
	\begin{aligned}
		& 0 \qquad && T>T_c\\
		& 1-T/T_c \qquad && T<T_c
	\end{aligned}
	\right.
\end{equation}
Indeed, it can be mapped into the spin glass \textit{spherical model} --- where scalar spins satisfy the global constraint $\sum_i \sigma^2_i=N$ ---, which has been shown to be~\acrshort{RS} stable as well~\cite{KosterlitzEtAl1976}. However, it still represents an interesting model, since some analytical computations can be performed in the $m\to\infty$ limit on different graph topologies~\cite{AspelmeierMoore2004, BeyerEtAl2012, JavanmardEtAl2016}. Moreover, an appealing question regards the way the physics of the low-temperature phase changes when increasing the number of spin components, so focusing on the case of large but finite values of $m$~\cite{BaityJesiParisi2015, RicciTersenghiEtAl2016}.

\subsection{The anisotropic case}

If the behaviour of isotropic vector spin glasses is quite similar to the Ising case, important differences arise when considering some source of anisotropy in the model. Indeed, due to the fact that there is more than one possible direction for each spin, different phase transitions can in principle take place, involving different spin degrees of freedom.

The simplest case of anisotropy is represented by the insertion of a uniform field, say along the direction $\mu=1$:
\begin{equation}
	\mathcal{H}_J[\{\boldsymbol{\sigma}_i\}] = -\frac{1}{2}\sum_{i \neq j}J_{ij}\,\boldsymbol{\sigma}_i\cdot\boldsymbol{\sigma}_j - H\sum_i\sigma_{i,1}
	\label{eq:H_SK_m_Huniform}
\end{equation}
The presence of the uniform field breaks the invariance under rotation, leading to two different kinds of overlaps, the longitudinal one and the transverse one:
\begin{equation}
	q_{ab,\mu\nu} =
	\left\{
	\begin{aligned}
		&q_{\parallel} \equiv \frac{1}{N}\sum_i\sigma^{(a)}_{i,1}\sigma^{(b)}_{i,1}\\
		&q_{\perp} \equiv \frac{1}{N}\sum_i\sigma^{(a)}_{i,\mu}\sigma^{(b)}_{i,\mu} \quad , \quad \mu=2,\dots,m
	\end{aligned}
	\right.
\end{equation}
For any value of the field strength $H$ and of the temperature $T$, direction $\mu=1$ is marked by both a non vanishing magnetization and a non vanishing overlap:
\begin{equation}
	m_{\parallel} \neq 0 \qquad , \qquad q_{\parallel} \neq 0
\end{equation}
while the transverse direction with respect to the field does never show a global magnetization different from zero, $m_{\perp} = 0$.

\begin{figure}[t]
	\centering
	\includegraphics[scale=0.6]{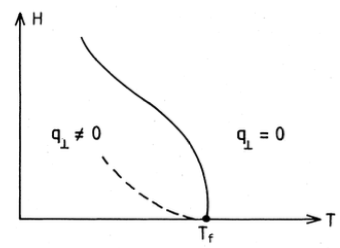}
	\caption[GT and dAT lines for vector spin glasses in a field]{Phase diagram $H$ vs $T$ of the $m$-component spin glass in a homogeneous external field. Full curve refers to the~\acrshort{GT} line~\autoref{eq:scaling_GT_line_near_zeroH_fullyConnected_Vector} and identifies the ordering of transverse components, while dashed curve refers to the~\acrshort{dAT} line~\autoref{eq:scaling_dAT_line_near_zeroH_fullyConnected_Vector} and it is related to the ordering of longitudinal degrees of freedom. Reprinted from~\cite{BinderYoung1986}.}
	\label{fig:GT_phase_diagram_H}
\end{figure}

The transverse overlap, instead, may or may not vanish according to if a freezing of the transverse degrees of freedom does occur or it does not. Indeed, as firstly studied by Gabay and Toulouse~\cite{GabayToulouse1981}, it vanishes in the region of large $H$ and $T$, but it becomes different from zero on a well defined critical line in the $H$ vs $T$ plane, whose expansion close to the zero-field axis --- computed\footnote{In Ref.~\cite{GabayToulouse1981} this expansion reads differently, due to an algebraic error.} in Refs.~\cite{CraggEtAl1982, GabayEtAl1982} --- reads as
\begin{equation}
	\left(\frac{H}{J}\right)^2 \simeq \frac{4(m+2)^2}{m^2+4m+2}\left(\frac{T_c-T}{T_c}\right)
	\label{eq:scaling_GT_line_near_zeroH_fullyConnected_Vector}
\end{equation}
so yielding a $1/2$ critical exponent, different from the $3/2$ one of the~\acrshort{dAT} line. This new critical line has been then named as the~\acrfull{GT} one and it is represented by the full curve in~\autoref{fig:GT_phase_diagram_H}.

According to first computations of Gabay and Toulouse~\cite{GabayToulouse1981}, the~\acrshort{GT} line does not imply any~\acrshort{RSB}, but just an ordering in the transverse degrees of freedom. Instead, the~\acrshort{RS} solution becomes unstable at lower critical temperatures, corresponding to the~\acrshort{dAT} line
\begin{equation}
	\left(\frac{H}{J}\right)^2 \simeq \frac{4}{m+2}\left(\frac{T_c-T}{T_c}\right)^3
	\label{eq:scaling_dAT_line_near_zeroH_fullyConnected_Vector}
\end{equation}
with exactly the same features of the~\acrshort{RS} instability line~\autoref{eq:scaling_dAT_line_near_zeroH_fullyConnected} of the $m=1$ case, included the coefficient of the expansion.

In fact, a careful stability analysis of the~\acrshort{RS} solution~\cite{CraggEtAl1982} subsequently showed that~\acrshort{RSB} already occurs on the~\acrshort{GT} line and involves both longitudinal and transverse degrees of freedom at the same time. Hence, the~\acrshort{fRSB} solution has been applied~\cite{GabayEtAl1982} as a generalization of the Ising case, showing that the longitudinal overlap $q_{\parallel}(x)$ depends only ``weakly'' on the Parisi parameter $x$ in the region slightly below the~\acrshort{GT} line. Instead, a ``strong''~\acrshort{RSB} suddenly affects the transverse overlap $q_{\perp}(x)$ as soon as the~\acrshort{GT} line is crossed.

Moreover, since the field does not couple to the transverse spin components and since the small-$x$ expansion of $q_{\perp}(x)$ just below the~\acrshort{GT} line recalls that of $q(x)$ in the $(m-1)$-dimensional isotropic case, it can be claimed that the~\acrshort{RSB} occurring on the~\acrshort{GT} line for a $m$-dimensional anisotropic vector spin glass is exactly of the same kind of the one occurring for a $(m-1)$-dimensional isotropic vector spin glass~\cite{MooreBray1982, GabayEtAl1982}. So the presence of the field just rescales the critical temperature at which the~\acrshort{RS} instability occurs. Furthermore, due to this $(m-1)$ dependence, in the $m \to 1$ limit --- where there are no transverse spin components --- the~\acrshort{GT} line eventually disappears, as expected to be.

Since the~\acrshort{RSB} does already occur on the~\acrshort{GT} line, then the~\acrshort{dAT} line as a sharp transition given by~\acrshort{RS} instability can not exist any longer. However, for values of the field $H$ scaling as in~\autoref{eq:scaling_dAT_line_near_zeroH_fullyConnected_Vector}, it has been observed a change of regime in the shape of the longitudinal overlap $q_{\parallel}(x)$, that acquires a relevant dependence on the Parisi parameter $x$~\cite{GabayEtAl1982}. So the picture is that the~\acrshort{dAT} line disappears as a proper transition line, but leaves some traces of itself as a \textit{crossover} between a weak and a strong~\acrshort{RSB} along the direction of the field. Further evidences of this change of regime have been provided also by Elderfield and Sherrington in a series of works~\cite{ElderfieldSherrington1982b, ElderfieldSherrington1982c, ElderfieldSherrington1984}.

Proofs for the scenario depicted above --- namely the freezing of transverse degrees of freedom for $H \sim (\delta T_c)^{1/2}$ and the change of behaviour of longitudinal degrees of freedom for $H \sim (\delta T_c)^{3/2}$, with $\delta T_c \equiv (T_c-T)$ --- have been also provided by several experiments~\cite{MonodBouchiat1982, LauerKeune1982, FogleEtAl1983, CampbellEtAl1983, CampbellEtAl1984}, so further proving the effectiveness and the validity of the~\acrshort{fRSB} scheme developed by Parisi.

A further interesting case of anisotropy in vector spin glasses is --- in analogy with the scalar model $m=1$ --- when a ferromagnetic bias $J_0$ is present in the coupling distribution. The Hamiltonian reads the same as~\autoref{eq:H_SK_m}, but again a transverse overlap and a longitudinal overlap have to be distinguished, referring to the direction along which the $\mathrm{O}(2)$ symmetry is spontaneously broken.

\begin{figure}[t]
	\centering
	\includegraphics[scale=0.7]{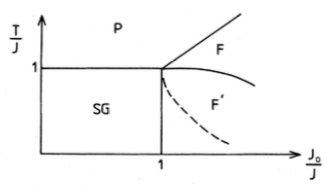}
	\caption[Phase diagram of vector spin glasses with ferromagnetic bias]{Phase diagram $T$ vs $J_0$ of the $m$~component spin glass with a ferromagnetic bias in coupling distribution. Reprinted from~\cite{BinderYoung1986}.}
	\label{fig:GT_phase_diagram_J0}
\end{figure}

The high-temperature region is correctly described by the~\acrshort{RS} saddle-point equations, with a second-order phase transition toward a spin glass phase at $T_c=J$ for $J_0<J$ and toward a ferromagnetic phase at $T_c=J_0$ for $J_0>J$, just as in the Ising case. Then, the~\acrshort{RS} instability line again coincides with the horizontal line $T=T_c$ for $J_0<J$ and with a line lower than the $T=J_0$ for $J_0>J$. Finally, Toulouse argument~\cite{Toulouse1980} also applies here, so providing a vertical critical line at $J_0=J$ separating the magnetized from the unmagnetized~\acrshort{RSB} regions.

The unique differences, even though not negligible, regard the mixed phase and the line separating it from the~\acrshort{RS} ferromagnetic phase. Indeed, Gabay and Toulouse~\cite{GabayToulouse1981} found that the~\acrshort{RS} instability line for $J_0 \gtrsim J$ does not go as a square root as in the $m=1$ case, but it goes as~\cite{DubielEtAl1987, Book_FischerHertz1991}
\begin{equation}
	\left(\frac{T_c-T}{T_c}\right) \simeq 2\,\frac{m^2+4m+2}{4(m+2)^2}\left(\frac{J_0-J}{J}\right)^2
	\label{eq:scaling_GT_line_near_mcPoint_fullyConnected_Vector}
\end{equation}
with a coefficient\footnote{Again the coefficient of the expansion provided by Gabay and Toulouse~\cite{GabayToulouse1981} is different from the one reported here, due to an algebraic error.} that is just the same one of the~\acrshort{GT} line~\autoref{eq:scaling_GT_line_near_zeroH_fullyConnected_Vector} in the $H$ vs $T$ plane, apart from a factor $2$. The reason for such change in the exponent must be searched in the vector nature of the model. Indeed, along this line there is again a freezing in the transverse degrees of freedom, with $q_{\perp}$ becoming different from zero and also depending on the Parisi parameter $x$. That is why also this line is commonly referred to as~\acrshort{GT}. Moreover, at the same time also the overlap $q_{\parallel}$ acquires a dependence on $x$, though weak as for the~\acrshort{GT} line in a homogeneous field, so coupling together the longitudinal and the transverse components. This results in an anomalous growth of the spontaneous magnetization very close to the multicritical point, namely $m \propto (T_c-T)$ instead of $m \propto (T_c-T)^{1/2}$, in turn causing the change of the exponent from $1/2$ to $2$ for the~\acrshort{GT} line~\cite{DubielEtAl1987}.

Then, when further lowering the temperature, also in this case there can be detected a crossover between a weak and a strong~\acrshort{RSB} for the longitudinal degrees of freedom --- initially claimed to be a sharp phase transition by Gabay and Toulouse~\cite{GabayToulouse1981} --- in correspondence of a \textit{vestige} of the~\acrshort{dAT} line
\begin{equation}
	\left(\frac{T_c-T}{T_c}\right)^2 \simeq \frac{m+2}{3}\left(\frac{J_0-J}{J}\right)
	\label{eq:scaling_dAT_line_near_mcPoint_fullyConnected_Vector}
\end{equation}
whose exponent is the same of the~\acrshort{dAT} line of the $m=1$ case~\autoref{eq:scaling_dAT_line_near_mcPoint_fullyConnected} close to the multicritical point.

We conclude noting that there seems to be a general fact that the~\acrshort{dAT} line of the scalar $m=1$ case --- both for $H \neq 0$ and for $J_0 \neq 0$ --- survives in the vector case just as a crossover, though retaining the same exponent, while the sharp phase transition from~\acrshort{RS} to~\acrshort{RSB} moves at higher temperatures, due to the freezing of the transverse degrees of freedom.

\clearpage{\pagestyle{empty}\cleardoublepage}

\begingroup
	\makeatletter
	\let\ps@plain\ps@empty
	\part{The XY model and the clock model}
	\label{part:XYmodelNoField}
	\cleardoublepage
\endgroup

\chapter{The XY model in absence of a field}
\label{chap:XYnoField}
\thispagestyle{empty}

In this Chapter we finally introduce the simplest spin model with continuous variables, namely the XY model. We analyze its behaviour according to different degrees of the quenched disorder, provided via suitable probability distributions for the exchange couplings. For the moment, the presence of an external field is not taken into account. The use of sparse random graphs allows us to exploit the~\acrshort{BP} techniques introduced in~\autoref{chap:tools}, both analytically (in the high-temperature regime and slightly below the critical point) and numerically (in the deep low-temperature phase, as well as in the zero-temperature limit). Finally, the temperature versus disorder phase diagrams are obtained, according to the different probability distributions of the quenched disorder.

\section{The model}

As we saw in~\autoref{chap:sg_replica}, the Hamiltonian of a vector spin model has formally the same structure of the Ising case, namely
\begin{equation}
	\mathcal{H}[\{\boldsymbol{\sigma}_i\}] = -\sum_{(i,j)}J_{ij}\boldsymbol{\sigma}_i\cdot\boldsymbol{\sigma}_j
	\label{eq:H_vector_noField}
\end{equation}
having restricted the interactions only to nearest-neighbour spins and having excluded the presence of any external field. Since we are interesting in the XY model, spins are represented by unit vectors with $m=2$ components:
\begin{equation}
	\boldsymbol{\sigma}_i = \bigl(\sigma_{i,x}\,,\,\sigma_{i,y}\bigr)
\end{equation}
subjected to the normalization constraint
\begin{equation}
	\norm{\boldsymbol{\sigma}_i} = \sum_{\mu=x,y}\sigma^2_{i,\mu} = 1
\end{equation}
Hence, each spin can be efficiently represented via a unique real-valued parameter, e.\,g. the angle $\theta$ it forms counterclockwise with the $\boldsymbol{x}$ axis:
\begin{equation}
	\boldsymbol{\sigma}_i = e^{\mathfrak{i}\theta_i} \quad , \qquad \theta_i\in[0,2\pi)
\end{equation}
With this notation, Hamiltonian~\autoref{eq:H_vector_noField} becomes for the XY model:
\begin{equation}
	\mathcal{H}[\{\theta_i\}]= -\sum_{(i,j)}J_{ij}\cos{(\theta_i-\theta_j)}
	\label{eq:Hamiltonian_XY_sparse}
\end{equation}

For the moment we do not make any assumption on the coupling distribution~$\mathbb{P}_J$, neither on the degree distribution~$\mathbb{P}_d$ of the underlying graph $\mathcal{G}$; the unique assumption regards its sparsity, so that we are allowed to exploit the~\acrshort{BP} approach.

The importance of the XY model, as already remarked in the~\hyperlink{chap:intro}{Introduction}, rests on the fact that it is the simplest spin model with continuous variables, so that analytic computations are the least involved possible and numerical simulations are the least demanding possible, though preserving all the features that distinguish vector models from scalar ones, as lengthy discussed in~\autoref{chap:sg_replica}. Moreover, it is just the two-dimensional arrangement of the spins that allows the XY model to correctly reproduce a plethora of physical phenomena, ranging e.\,g. from granular superconductors~\cite{JohnLubensky1985, HuseSeung1990} to superfluid Helium
~\cite{Minnhagen1987, Book_Brezin1989}, from synchronization problems
~\cite{Book_Kuramoto1975, AcebronEtAl2005, SkantzosEtAl2005, BandeiraEtAl2016} to random lasers~\cite{AntenucciEtAl2015, Thesis_Antenucci2016, AntenucciEtAl2016, Thesis_Marruzzo2015}.

\section{The BP equations}
\label{sec:BP_eqs_XY}

In~\autoref{chap:tools} we presented the~\acrshort{BP} approach as an efficient tool for solving models on sparse topologies. The key object of this method --- focusing on the pairwise case --- is the set of cavity messages $\{\eta_{i\to j}(\boldsymbol{x}_i)\}$, that satisfies the self-consistency \acrshort{BP}~equations~\autoref{eq:BP_selfCons_EtaMess}.

In the case of the XY model, $\boldsymbol{x}_i$ is nothing but $\theta_i$, so that the cavity messages become probability distributions defined over the $[0,2\pi)$ interval with periodic boundary conditions. It is easy then to rewrite the~\acrshort{BP}~equations~\autoref{eq:BP_selfCons_EtaMess} for the XY model, as shown in~\autoref{app:BPeqs_XYmodel} for both the factor graph formalism and the pairwise case. For the Hamiltonian~\autoref{eq:Hamiltonian_XY_sparse}, where no external field is present and interactions are just pairwise, they read:
\begin{equation}
	\eta_{i\to j}(\theta_i) = \frac{1}{\mathcal{Z}_{i\to j}}\prod_{k\in\partial i\setminus j}\int \di\theta_k\,e^{\,\beta J_{ik}\cos{(\theta_i-\theta_k)}}\,\eta_{k\to i}(\theta_k)
	\label{eq:BP_eqs_XY}
\end{equation}
where normalization constant $\mathcal{Z}_{i\to j}$ is given by:
\begin{equation}
	\mathcal{Z}_{i\to j} = \int\di\theta_i\,\prod_{k\in\partial i\setminus j}\int \di\theta_k\,e^{\,\beta J_{ik}\cos{(\theta_i-\theta_k)}}\,\eta_{k\to i}(\theta_k)
\end{equation}
Notice that all the integrals over angular variables are meant to be over the $[0,2\pi)$ interval, if not otherwise stated.

Once solved the set of~\acrshort{BP}~equations~\autoref{eq:BP_eqs_XY} for the XY model, then it is possible to compute the Bethe free energy $f$ and all the other physical observables, as explained in~\autoref{chap:tools}. However, let us first of all try to guess how the fixed-point cavity messages $\{\eta^*_{i\to j}\}$ should look like.

\subsection{Paramagnetic solution}

Since we have no external field, then in the high-temperature regime it is reasonable to have a \textit{symmetric} paramagnetic phase in which all the directions on the $xy$ plane are exactly equivalent for each spin, as it happens for ordinary magnetic models~\cite{Book_Huang1988,Book_Parisi1988}. In other words, the one-point marginals $\eta_i(\theta_i)$'s should acquire the following expression:
\begin{equation}
	\eta_i(\theta_i) = \frac{1}{2\pi} \quad , \qquad \forall i
	\label{eq:para_sol_XY}
\end{equation}
The corresponding expression for the cavity messages is the uniform distribution over~$\theta_i$'s as well:
\begin{equation}
	\eta_{i\to j}(\theta_i) = \frac{1}{2\pi} \quad , \qquad \forall i\to j
	\label{eq:para_sol_XY_cavityMarg}
\end{equation}
Indeed, it can be easily checked that this set of cavity messages automatically yields the paramagnetic solution~\autoref{eq:para_sol_XY} once plugged into~\autoref{eq:one_node_belief_BP}.

Also the expression for the Bethe free energy density in the paramagnetic phase can be easily computed. Indeed, by recalling its general expression for pairwise models, Eq.~\autoref{eq:Bethe_free_energy}, and by plugging the paramagnetic solution~\autoref{eq:para_sol_XY} into it, we get the following expressions for $\mathcal{Z}_i$ and $\mathcal{Z}_{ij}$:
\begin{subequations}
	\begin{equation}
		\mathcal{Z}_i = \int \di \theta_i\prod_{k\in\partial i}\int \di \theta_k \, e^{\,\beta J_{ik}\cos{(\theta_i-\theta_k)}}\eta_{k\to i}(\theta_k) = 2\pi \prod_{k\in\partial i} I_0(\beta J_{ik})
		\label{eq:Zeta_i_XY}
	\end{equation}
	\begin{equation}
		\mathcal{Z}_{ij} = \int \di \theta_i \di \theta_j \, e^{\,\beta J_{ij}\cos{(\theta_i-\theta_j)}}\eta_{i\to j}(\theta_i)\eta_{j\to i}(\theta_j) = I_0(\beta J_{ij})
		\label{eq:Zeta_ij_XY}
	\end{equation}
	\label{eq:Zeta_i_and_ij_XY}%
\end{subequations}
having recalled the definition of the modified Bessel functions of the first kind \cite{Book_AbramowitzStegun1964}:
\begin{equation}
	I_n(x) \equiv \frac{1}{2\pi}\int_0^{2\pi} \di\theta\,e^{\,x\cos{\theta}}\,\cos{(n\theta)} \quad , \qquad n\in\mathbb{Z}
	\label{eq:modif_Bessel_I_XY}
\end{equation}
At this point, one should take into account both the topology of the underlying graph $\mathcal{G}$ and the probability distribution of the random couplings $J_{ij}$'s, and then average over them\footnote{As anticipated in~\autoref{chap:tools}, quenched disorder can be given not only by random couplings, but also by the random topology of the underlying graph.}. We choose $\mathcal{G}$ to be a~\acrshort{RRG} of fixed connectivity $C$, namely each site has exactly $C$ nearest neighbours. Moreover, we let $J_{ij}$'s be i.\,i.\,d. variables drawn from the following bimodal probability distribution
\begin{equation}
	\mathbb{P}_J(J_{ij}) = p\,\delta(J_{ij}-J) + (1-p)\,\delta(J_{ij}+J)
	\label{eq:disorder_distribution_pmJ}
\end{equation}
with $J>0$ and $p\in[0.5,1]$, so using also for the XY model one of the most exploited coupling probability distributions for the Ising case. Being the parity of Bessel function $I_n(\cdot)$ with respect to its argument the same of its order $n$, we can finally obtain the disorder-averaged expression for the free energy density $f$ of this model for the paramagnetic solution:
\begin{equation}
	f(\beta) = -\frac{1}{\beta}\ln{2\pi} - \frac{C}{2\beta}\ln{I_0(\beta J)}
	\label{eq:f_para_XY_RRG}
\end{equation}
given that the ratio $\alpha$ between edges and nodes on the $C$-\acrshort{RRG} ensemble is exactly equal to $C/2$.

Analogously, also the disorder-averaged internal energy density $u$ can be evaluated for the paramagnetic solution, recalling the generic pairwise expression~\autoref{eq:Bethe_internal_energy} of $U$. Since there is no external field, the unique contributions to the total internal energy are the $\{u_{ij}\}$ ones coming from the pairwise interactions between nearest-neighbour spins:
\begin{equation}
	u_{ij} = -\frac{J_{ij}}{\mathcal{Z}_{ij}}\int \di\theta_i \di\theta_j \cos{(\theta_i-\theta_j)} \, e^{\,\beta J_{ij}\cos{(\theta_i-\theta_j)}}\eta_{i\to j}(\theta_i)\eta_{j\to i}(\theta_j) = -J_{ij}\frac{I_1(\beta J_{ij})}{I_0(\beta J_{ij})}
\end{equation}
from which, averaging over the bimodal $\mathbb{P}_J$ and over the graph ensemble:
\begin{equation}
	u(\beta) = -\frac{JC}{2}\frac{I_1(\beta J)}{I_0(\beta J)}
	\label{eq:u_para_XY_RRG}
\end{equation}
Notice that in the high-temperature phase there is no dependence on the fraction $p$ of positive couplings in both $f(\beta)$ and $u(\beta)$, due to the fact that each site marginal $\eta_i(\theta_i)$ is completely flat over the $[0,2\pi)$ interval. So there is no difference between a purely ferromagnetic model and a disordered one in the paramagnetic region.

At this point, we notice that the paramagnetic solution~\autoref{eq:para_sol_XY_cavityMarg} satisfies the self-consistency~\acrshort{BP} equations~\autoref{eq:BP_eqs_XY} for any value of the inverse temperature~$\beta$. However, we remind that the usual picture for ordered magnetic systems with pairwise interactions and no external field is that the paramagnetic solution becomes unstable below a certain \textit{critical temperature}~$T_c$ through a second-order phase transition~\cite{Book_Parisi1988}. Then, the low-temperature phase is characterized by a nonvanishing order parameter --- typically the global magnetization $m$ --- which grows continuously from zero below the critical point. Since we are now dealing with a disordered ferromagnet instead of a pure one, its low-temperature solution should in principle depend on the quantity of quenched disorder inserted in the system, namely on the probability distribution $\mathbb{P}_J$ from which couplings $J_{ij}$'s are drawn.

A hint about the low-temperature physics for the XY model on sparse random graphs comes from the solution of vector spin glass models in the fully connected case. Indeed, we saw in~\autoref{sec:vector_sg_fully} that when there is a strong bias toward positive couplings, then the nonvanishing local magnetizations~$\boldsymbol{m}_i$'s are mostly coherent and yield a ``usual'' ferromagnetic phase with a nonzero average magnetization~$\boldsymbol{m}$. Instead, if positive and negative couplings are drawn with almost the same frequency, then the nonvanishing local magnetizations $\boldsymbol{m}_i$'s incoherently sum to zero, so producing a spin glass phase. In order to verify the validity of these claims also in the sparse case, we need first of all to check the stability of the paramagnetic solution, and then to inspect the resulting directions of instability. In this sense, it should turn out to be useful to build a perturbative expansion around the paramagnetic solution.

\subsection{Expanding around the paramagnetic solution}
\label{subsec:exp_around_para}

The goal of a perturbative expansion around the paramagnetic solution~\autoref{eq:para_sol_XY_cavityMarg} is hence twofold: \textit{i)} to estimate the critical temperature $T_c$ where it becomes unstable, and \textit{ii)} to detect the features of the low-temperature phase toward which the system moves.

Since the cavity messages are $2\pi$-periodic functions, then it is natural to expand them around the paramagnetic solution in the Fourier basis of eigenfunctions:
\begin{equation}
	\eta_{i\to j}(\theta_i)=\frac{1}{2\pi}\left[1+\sum_{l=1}^{\infty}a^{(i\to j)}_l\cos{(l\theta_i)}+\sum_{l=1}^{\infty}b^{(i\to j)}_l\sin{(l\theta_i)}\right]
	\label{eq:Fourier_expansion_XY}
\end{equation}
where the Fourier coefficients are given by:
\begin{subequations}
	\begin{equation}
		a^{(i\to j)}_l = 2\int\di\theta_i\,\eta_{i\to j}(\theta_i)\,\cos{(l\theta_i)}
		\label{eq:Fourier_coeff_XY_a}
	\end{equation}
	\begin{equation}
		b^{(i\to j)}_l = 2\int\di\theta_i\,\eta_{i\to j}(\theta_i)\,\sin{(l\theta_i)}
		\label{eq:Fourier_coeff_XY_b}
	\end{equation}
	\label{eq:Fourier_coeff_XY}%
\end{subequations}
The paramagnetic solution \autoref{eq:para_sol_XY_cavityMarg} obviously corresponds to all Fourier coefficients being equal to zero, while a (locally) magnetized solution has at least some of them which do not vanish. So the check of the stability of the paramagnetic solution turns into the check of the stability of the null solution for all the Fourier coefficients of the expansion~\autoref{eq:Fourier_expansion_XY}.

In order to do this, let us plug the Fourier expansion~\autoref{eq:Fourier_expansion_XY} into the right hand side of~\acrshort{BP} equations~\autoref{eq:BP_eqs_XY}, obtaining:
\begin{equation}
\begin{split}
	\eta_{i\to j}(\theta_i) = \frac{1}{\mathcal{Z}_{i\to j}}\prod_{k\in\partial k\setminus j}\Biggl\{&\frac{1}{2\pi} \int\di\theta_k\,e^{\,\beta J_{ik}\cos{(\theta_i-\theta_k)}}\\
	& \times\biggl[1+\sum_{l=1}^{\infty}a^{(k\to i)}_l\cos{(l\theta_k)} + \sum_{l=1}^{\infty}b^{(k\to i)}_l\sin{(l\theta_k)}\biggr] \Biggr\}
\end{split}
\end{equation}
Integrals over $\theta_k$'s can be evaluated by using again the modified Bessel functions $I_n(\cdot)$'s defined in~\autoref{eq:modif_Bessel_I_XY}, together with the related properties and identities~\cite{Book_AbramowitzStegun1964}, so getting:
\begin{equation}
	\eta_{i\to j}(\theta_i) = \frac{1}{\mathcal{Z}_{i\to j}}\prod_{k\in\partial i\setminus j}\biggl\{I_0(\beta J_{ik})+\sum_{l=1}^{\infty}I_l(\beta J_{ik})\left[a^{(k\to i)}_l\cos{(l\theta_i)}+b^{(k\to i)}_l\sin{(l\theta_i)}\right]\biggr\}
\end{equation}
where also $\mathcal{Z}_{i\to j}$ has to be rewritten in terms of the modified Bessel functions:
\begin{equation}
	\mathcal{Z}_{i\to j} = \int\di\theta_i\prod_{k\in\partial i\setminus j}\biggl\{I_0(\beta J_{ik})+\sum_{l=1}^{\infty}I_l(\beta J_{ik})\left[a^{(k\to i)}_l\cos{(l\theta_i)}+b^{(k\to i)}_l\sin{(l\theta_i)}\right]\biggr\}
	\label{eq:Fourier_coeff_XY_Z}
\end{equation}

At this point, we can plug this expansion back into relations~\autoref{eq:Fourier_coeff_XY} for the $a_l$'s and the $b_l$'s coefficients, obtaining a set of self-consistency equations for them:
\begin{subequations}
	\begin{equation}
		\begin{split}
			a^{(i\to j)}_l &= \frac{2}{\mathcal{Z}_{i\to j}}\int\di\theta\,\cos{(l\theta)}\prod_{k\in\partial i\setminus j}\Biggl\{I_0(\beta J_{ik})\\
	& \qquad +\sum_{p=1}^{\infty}I_p(\beta J_{ik})\left[a^{(k\to i)}_p\cos{(p\theta)}+b^{(k\to i)}_p\sin{(p\theta)}\right]\Biggr\}
		\end{split}
		\label{eq:Fourier_coeff_XY_selfcons_a}
	\end{equation}
	\begin{equation}
		\begin{split}
			b^{(i\to j)}_l &= \frac{2}{\mathcal{Z}_{i\to j}}\int\di\theta\,\sin{(l\theta)}\prod_{k\in\partial i\setminus j}\Biggl\{I_0(\beta J_{ik})\\
			& \qquad +\sum_{p=1}^{\infty}I_p(\beta J_{ik})\left[a^{(k\to i)}_p\cos{(p\theta)}+b^{(k\to i)}_p\sin{(p\theta)}\right]\Biggr\}
		\end{split}
		\label{eq:Fourier_coeff_XY_selfcons_b}
	\end{equation}
	\label{eq:Fourier_coeff_XY_selfcons}%
\end{subequations}
with $\mathcal{Z}_{i\to j}$ still given by~\autoref{eq:Fourier_coeff_XY_Z}. In this way we got another set of~\acrshort{BP} equations, now in terms of the ``cavity'' Fourier coefficients rather than in terms of the cavity messages.

Now it is possible to easily study the stability of the paramagnetic solution. Indeed, let us focus on the equation for $a^{(i\to j)}_l$ and let us truncate the Fourier expansion in the right hand side at the linear term in each coefficient. For the numerator we get:
\begin{equation}
\begin{split}
	& 2\int\di\theta\,\cos{(l\theta)}\prod_{k\in\partial i\setminus j}\biggl\{I_0(\beta J_{ik})+\sum_{p=1}^{\infty}I_p(\beta J_{ik})\Bigl[a^{(k\to i)}_p\cos{(p\theta)}+b^{(k\to i)}_p\sin{(p\theta)}\Bigr]\biggr\}\\
	& \qquad = 2\prod_{k\in\partial i\setminus j}I_0(\beta J_{ik})\int\di\theta\cos{(l\theta)}\\
	& \qquad \qquad \times \prod_{k\in\partial i\setminus j}\biggl\{1+\sum_{p=1}^{\infty}\frac{I_p(\beta J_{ik})}{I_0(\beta J_{ik})}\Bigl[a^{(k\to i)}_p\cos{(p\theta)}+b^{(k\to i)}_p\sin{(p\theta)}\Bigr]\biggr\}\\
	& \qquad \simeq 2\prod_{k\in\partial i\setminus j}I_0(\beta J_{ik})\int\di\theta\cos{(l\theta)}\\
	& \qquad \qquad \times \biggl\{1+\sum_{k\in\partial i\setminus j}\,\sum_{p=1}^{\infty}\frac{I_p(\beta J_{ik})}{I_0(\beta J_{ik})}\Bigl[a^{(k\to i)}_p\cos{(p\theta)}+b^{(k\to i)}_p\sin{(p\theta)}\Bigr]\biggr\}\\
	& \qquad = 2\pi\prod_{k\in\partial i\setminus j}I_0(\beta J_{ik})\,\sum_{k\in\partial i\setminus j}\,\frac{I_l(\beta J_{ik})}{I_0(\beta J_{ik})}\,a^{(k\to i)}_l
\end{split}
\label{eq:al_linear_numerator}
\end{equation}
while denominator can be evaluated in the same manner:
\begin{equation}
\begin{split}
	& \int\di\theta\prod_{k\in\partial i\setminus j}\biggl\{I_0(\beta J_{ik})+\sum_{p=1}^{\infty}I_p(\beta J_{ik})\Bigl[a^{(k\to i)}_p\cos{(p\theta)}+b^{(k\to i)}_p\sin{(p\theta)}\Bigr]\biggr\}\\
	& \quad = \prod_{k\in\partial i\setminus j}I_0(\beta J_{ik})\int\di\theta\,\prod_{k\in\partial i\setminus j}\biggl\{1+\sum_{p=1}^{\infty}\frac{I_p(\beta J_{ik})}{I_0(\beta J_{ik})}\Bigl[a^{(k\to i)}_p\cos{(p\theta)}+b^{(k\to i)}_p\sin{(p\theta)}\Bigr]\biggr\}\\
	& \quad \simeq \prod_{k\in\partial i\setminus j}I_0(\beta J_{ik})\int\di\theta\,\biggl\{1+\sum_{k\in\partial i\setminus j}\,\sum_{p=1}^{\infty}\frac{I_p(\beta J_{ik})}{I_0(\beta J_{ik})}\Bigl[a^{(k\to i)}_p\cos{(p\theta)}+b^{(k\to i)}_p\sin{(p\theta)}\Bigr]\biggr\}\\
	& \quad = 2\pi\prod_{k\in\partial i\setminus j}I_0(\beta J_{ik})
\end{split}
\label{eq:al_linear_denominator}
\end{equation}
By following the same steps for the equation for $b^{(i\to j)}_l$, the linear expansion in Fourier coefficients finally provides a set of recursive relations:
\begin{subequations}
	\begin{equation}
		a^{(i\to j)}_l = \sum_{k\in\partial i\setminus j}\frac{I_l(\beta J_{ik})}{I_0(\beta J_{ik})}\,a^{(k\to i)}_l
		\label{eq:al_linear}
	\end{equation}
	\begin{equation}
		b^{(i\to j)}_l = \sum_{k\in\partial i\setminus j}\frac{I_l(\beta J_{ik})}{I_0(\beta J_{ik})}\,b^{(k\to i)}_l
		\label{eq:bl_linear}
	\end{equation}
	\label{eq:al_and_bl_linear}%
\end{subequations}

It is just this set of linear equations that directly provides the stability condition for the paramagnetic solution. Indeed, we already know that physical observables have to be averaged over the coupling distribution and the graph ensemble. In particular, if we focus on the first two moments of the distribution of $a_l$'s Fourier coefficients (the same will hold for $b_l$'s):
\begin{equation}
	\overline{a_l} \equiv \mathbb{E}_{\mathcal{G},J}\left[a^{(i\to j)}_l\right] \qquad , \qquad \overline{a^2_l} \equiv \mathbb{E}_{\mathcal{G},J}\left[\left(a^{(i\to j)}_l\right)^2\right]
\end{equation}
then from~\autoref{eq:al_linear} we get:
\begin{subequations}
	\begin{equation}
		\overline{a_l} = \overline{\sum_{k\in\partial i\setminus j}\frac{I_l(\beta J_{ik})}{I_0(\beta J_{ik})}\,a^{(k\to i)}_l} \equiv A_1(l)\,\overline{a_l}
	\end{equation}
	\begin{equation}
		\overline{a^2_l} = \overline{\left[\sum_{k\in\partial i\setminus j}\frac{I_l(\beta J_{ik})}{I_0(\beta J_{ik})}\,a^{(k\to i)}_l\right]^2} \equiv A_2(l)\,\overline{a^2_l}
	\end{equation}
\end{subequations}
The paramagnetic solution, namely $a^{(i\to j)}_l=0$ for each directed edge $i\to j$ and for each order $l$, is stable as long as it holds
\begin{equation}
	A_{1,2}(l) < 1 \quad , \qquad \forall l
\end{equation}
where $A_1(l)$ and $A_2(l)$ obviously depend on the particular coupling distribution~$\mathbb{P}_J$ chosen, as well as on the ensemble of random graphs considered. Still referring to the bimodal coupling distribution~\autoref{eq:disorder_distribution_pmJ} and to the $C$-\acrshort{RRG} ensemble, for the first moment we get:
\begin{equation}
\begin{split}
	\overline{a_l} &= \mathbb{E}_{\mathcal{G},J}\left[\sum_{k\in\partial i\setminus j}\frac{I_l(\beta J_{ik})}{I_0(\beta J_{ik})}\,a^{(k\to i)}_l\right]\\
	&= (C-1)\,\mathbb{E}_J\left[\frac{I_l(\beta J_{ik})}{I_0(\beta J_{ik})}\right]\overline{a_l}
\end{split}
\end{equation}
Now, being the Bessel functions $I_n(x)$ even or odd depending on their index $n$, then the result of the average over the bimodal distribution has to depend on the parity of $l$ as well. So in the end the $A_1(l)$ factor reads:
\begin{equation}
	A_1(l) = 
	\left\{
	\begin{aligned}
		&(C-1)(2p-1)\frac{I_l(\beta J)}{I_0(\beta J)} \qquad && \text{for $l$ odd}\\
		&(C-1)\frac{I_l(\beta J)}{I_0(\beta J)} \qquad && \text{for $l$ even}
	\end{aligned}
	\right.
\end{equation}
In order to have a stable paramagnetic phase, then $A_1(l)$ has to be smaller than one for any order $l$. But first of all, let us analyze the behaviour of the modified Bessel functions. They are monotonically increasing with respect to their argument and monotonically decreasing with respect to their order, so that once fixed~$C$ and~$p$, also the~$\{A_1(2l)\}$ and the~$\{A_1(2l+1)\}$ successions are separately decreasing with respect to their index $l$. So, depending on the value of $p$, when lowering the temperature, the first coefficient that becomes equal to one is either~$A_1(1)$ or~$A_1(2)$.

At this point, we have to focus also on $A_2(l)$:
\begin{equation}
\begin{split}
	\overline{a^2_l} &= \mathbb{E}_{\mathcal{G},J}\Biggl[\biggl(\sum_{k\in\partial i\setminus j}\frac{I_l(\beta J_{ik})}{I_0(\beta J_{ik})}\,a^{(k\to i)}_l\biggr)^2\Biggr]\\
	&= \mathbb{E}_{\mathcal{G},J}\Biggl[\sum_{k_1,k_2\in\partial i\setminus j}\frac{I_l(\beta J_{ik_1})I_l(\beta J_{ik_2})}{I_0(\beta J_{ik_1})I_0(\beta J_{ik_2})}\,a^{(k_1\to i)}_l\,a^{(k_2\to i)}_l\Biggr]\\
	&= \mathbb{E}_{\mathcal{G},J}\Biggl[\sum_{k\in\partial i\setminus j}\frac{I^2_l(\beta J_{ik})}{I^2_0(\beta J_{ik})}\,\Bigl(a^{(k\to i)}_l\Bigr)^2 + \sum_{\substack{k_1,k_2\in\partial i\setminus j \\ k_1\neq k_2}}\frac{I_l(\beta J_{ik_1})I_l(\beta J_{ik_2})}{I_0(\beta J_{ik_1})I_0(\beta J_{ik_2})}\,a^{(k_1\to i)}_l\,a^{(k_2\to i)}_l\Biggr]\\
	&= (C-1)\frac{I^2_l(\beta J)}{I^2_0(\beta J)}\,\overline{a^2_l}+
	\left\{
	\begin{aligned}
		&(C-1)(C-2)\biggl[(2p-1)\frac{I_l(\beta J)}{I_0(\beta J)}\,\overline{a_l}\biggr]^2 \qquad && \text{for $l$ odd}\\
		&(C-1)(C-2)\biggl[\frac{I_l(\beta J)}{I_0(\beta J)}\,\overline{a_l}\biggr]^2 \qquad && \text{for $l$ even}
	\end{aligned}
	\right.
\end{split}
\end{equation}
Hence, $A_2(l)$ coefficient is given by:
\begin{equation}
	A_2(l) = (C-1)\frac{I^2_l(\beta J)}{I^2_0(\beta J)}
\end{equation}
and again the related succession $\{A_2(l)\}$ is monotonically decreasing with respect to the index $l$. So in the end the stability condition for a vanishing second moment of the distribution of Fourier coefficients $a_l$'s is given by $A_2(1)$.

Finally, we have to compare the three stability conditions given by the three coefficients $A_1(1)$, $A_1(2)$ and $A_2(1)$:
\begin{equation}
	\left\{
	\begin{aligned}
		&A_1(1)<1 \qquad &&\Rightarrow \qquad &&(C-1)(2p-1)\frac{I_1(\beta J)}{I_0(\beta J)}<1\\
		&A_1(2)<1 \qquad &&\Rightarrow \qquad &&(C-1)\frac{I_2(\beta J)}{I_0(\beta J)}<1\\
		&A_2(1)<1 \qquad &&\Rightarrow \qquad &&(C-1)\frac{I^2_1(\beta J)}{I^2_0(\beta J)}<1
	\end{aligned}
	\right.
\end{equation}
It turns out from the properties of the modified Bessel functions that $I_2(\beta J)/I_0(\beta J)$ is strictly smaller than $I^2_1(\beta J)/I^2_0(\beta J)$ for any positive value of their argument. So eventually the stability of the paramagnetic solution is given by the first two moments of the distribution of first-order Fourier coefficients $a_1$'s: a nonzero mean value $\overline{a_1}$ implies a ferromagnetic long-range ordering, with a global nonvanishing magnetization $\boldsymbol{m}$, while a zero mean value $\overline{a_1}$ together with a nonzero second moment $\overline{a^2_1}$ signals a spin glass phase, with a set of incoherent local magnetizations~$\boldsymbol{m}_i$'s giving a vanishing global magnetization $\boldsymbol{m}$. In turn, the appearance of either the ferromagnetic phase or the spin glass phase depends on the fraction~$p$ of positive couplings, just as we saw in~\autoref{chap:sg_replica} for the fully connected case. Indeed, the phase toward which the transition away from the paramagnetic phase takes place when lowering the temperature is given by the lowest among the two following critical inverse temperatures:
\begin{equation}
	\beta_c(p) \equiv \min{\bigl\{\beta_{\text{F}}(p),\beta_{\text{SG}}\bigr\}}
	\label{eq:beta_c_bimodal_XY}
\end{equation}
which are implicitly defined by the two marginality conditions given by $A_1(1)$ and $A_2(1)$:
\begin{equation}
	(C-1)(2p-1)\frac{I_1(\beta_{\text{F}} J)}{I_0(\beta_{\text{F}} J)}=1 \qquad , \qquad (C-1)\frac{I^2_1(\beta_{\text{SG}} J)}{I^2_0(\beta_{\text{SG}} J)}=1
	\label{eq:XY_para_stability_cRRG}
\end{equation}
Notice that also in the sparse case $\beta_{\text{SG}}$ does not actually depend on $p$, while $\beta_{\text{F}}$ does. In particular, by comparing the two left hand sides of the above equations, we get the exact location of the \textit{multicritical} point $(p_{mc},T_{mc})$ separating the two regimes:
\begin{equation}
	p_{mc} = \frac{1+(C-1)^{-1/2}}{2} \quad , \quad T_{mc} = 1/\beta_{\text{SG}}
	\label{eq:mc_point_XY}
\end{equation}

Finally, these results can be straightforwardly generalized to the case of the~\acrshort{ERG} ensemble with average degree $C$. Indeed, when averaging over the Poissonian degree distribution~\autoref{eq:deg_distr_ERG}, the two stability conditions for the paramagnetic solution become:
\begin{equation}
	C(2p-1)\frac{I_1(\beta_{\text{F}} J)}{I_0(\beta_{\text{F}} J)}=1 \qquad , \qquad C\frac{I^2_1(\beta_{\text{SG}} J)}{I^2_0(\beta_{\text{SG}} J)}=1
	\label{eq:XY_para_stability_cERG}
\end{equation}
which are exactly the same ones found in Refs.~\cite{SkantzosEtAl2005, CoolenEtAl2005}, where the $m$-dimensional vector spin glass model is analyzed for a generic disorder distribution through a functional expansion around the paramagnetic solution. Notice that the replacement of $C-1$ by $C$ when passing from the~\acrshort{RRG} ensemble to the~\acrshort{ERG} ensemble is recurrent in cavity calculations, due to the corresponding average branching ratios $1/(C-1)$ and $1/C$, respectively.

\subsection{Scaling of the Fourier coefficients below \ensuremath{T_c}}
\label{subsec:scaling_Four_coeff}

In the study of the stability of the paramagnetic solution, we discovered that at $T=T_c$ all the first-order Fourier coefficients $a_1$'s become different from zero, since the vanishing solution $\overline{a_1}=\overline{a^2_1}=0$ becomes unstable. Furthermore, the same would seem to happen for higher-order Fourier coefficients, but in correspondence of lower critical temperatures, so that in the end they do not correspond to further physical phase transitions.

Actually, this is true only if we truncate the right hand sides of \acrshort{BP} self-consistency equations~\autoref{eq:Fourier_coeff_XY_selfcons} for the cavity coefficients just at the linear term in each coefficient. Instead, if we go further retaining also nonlinear terms --- as lengthy shown in~\autoref{app:Fourier_coeff_expansion} ---, it is clear that all higher-order coefficients become different from zero as soon as first-order coefficients do, namely in correspondence of the critical temperature $T=T_c(p)$ given by Eqs.~\autoref{eq:beta_c_bimodal_XY} and~\autoref{eq:XY_para_stability_cRRG}. Indeed, from the third-order expansion it turns out that nonlinear terms have the following shape:
\begin{equation}
	a^{(i\to j)}_l \propto \prod_{k\in\partial i\setminus j}a^{(k\to i)}_{p_k}
\end{equation}
with $\{p_k\}$ positive integer indexes algebraically summing to $l$. In particular, if we choose $p_k=1$ $\forall k$, then we easily realize that slightly below the critical point the larger Fourier order $l$, the smaller coefficients $a_l$'s and $b_l$'s:
\begin{equation}
	a^{(i\to j)}_l \propto \left(a^{(i\to j)}_1\right)^l
	\label{eq:al_scaling_wrt_a1}
\end{equation}

This remark eventually allows us to find the scaling of the coefficients with the distance from the critical point. Indeed, noticing that the largest nonlinear term in the expansion of self-consistency equations for $a_1$'s is cubic, then a square-root behaviour has to arise. For $p>p_{mc}$, namely for the transition from paramagnetic to ferromagnetic phase, second moments are negligible with respect to first ones, so we get:
\begin{equation}
\begin{split}
	\overline{a_1} & \simeq \mathbb{E}_{\mathcal{G},J}\Biggl[\sum_{k\in\partial i\setminus j}\frac{I_1(\beta J_{ik})}{I_0(\beta J_{ik})}\,a_1 + \Gamma_1 a_1^3 + \Gamma_2 a_1 a_2\Biggr]\\
	& \simeq (C-1)(2p-1)\frac{I_1(\beta J)}{I_0(\beta J)}\,\overline{a_1} + \Gamma_3 \overline{a_1}^3\\
	& \simeq \biggl[(C-1)(2p-1)\frac{I_1(\beta_c J)}{I_0(\beta_c J)} + \Gamma_4(\beta-\beta_c)\biggr]\overline{a_1} + \Gamma_3 \overline{a_1}^3\\
	& = \Bigl[1+\Gamma_4(\beta-\beta_c)\Bigr]\overline{a_1} + \Gamma_3 \overline{a_1}^3
\end{split}
\end{equation}
from which:
\begin{equation}
	\overline{a_1} \simeq \sqrt{-\frac{\Gamma_4}{\Gamma_3}}\,(\beta-\beta_c)^{1/2} \propto (T_c-T)^{1/2}
	\label{eq:a1_scaling}
\end{equation}
Then, by exploiting~\autoref{eq:al_scaling_wrt_a1}, also the growth of the first moment of higher-order Fourier coefficients can be predicted:
\begin{equation}
	\overline{a_l} \propto (T_c-T)^{l/2}
	\label{eq:al_scaling}
\end{equation}

Following exactly the same steps for $p<p_{mc}$ --- i.\,e. for the transition from paramagnetic to spin glass phase, where only second moments have to be taken into account --- we get for the first-order coefficients:
\begin{equation}
	\overline{a^2_1} \propto T_c-T
	\label{eq:a2_scaling}
\end{equation}
and analogously for the higher-order ones:
\begin{equation}
	\overline{a^2_l} \propto (T_c-T)^l
	\label{eq:al_square_scaling}
\end{equation}

At this point, by recalling their definition~\autoref{eq:Fourier_coeff_XY}, it is also clear that $\overline{a_1}$ is directly related to the global magnetization along the $\boldsymbol{x}$ axis, while~$\overline{b_1}$ is linked to the global magnetization along the $\boldsymbol{y}$ axis, so allowing us to recover the well known mean-field value of the critical exponent $\beta$
\begin{equation}
	m \propto (T_c-T)^{\beta} \quad , \qquad \beta = \frac{1}{2}
	\label{eq:mag_growth}
\end{equation}
in the ferromagnetic phase. Analogously, also the mean-field linear behaviour of the average overlap $q$ slightly below $T_c$ can be recovered
\begin{equation}
	q \equiv \frac{1}{N}\sum_i|\boldsymbol{m}_i|^2 \quad , \qquad q \propto (T_c-T)
	\label{eq:overlap_growth}
\end{equation}
for both the ferromagnetic and the spin glass phases.

\section{The RS cavity method}
\label{sec:RS_cavity_method}

So far, we have analytically obtained the critical lines in the $T$ vs $p$ phase diagram of the XY model through an expansion around the paramagnetic solution. Our guess was that, in absence of any field and slightly below the critical temperature $T_c$, the Fourier decomposition~\autoref{eq:Fourier_expansion_XY} of cavity marginals would have allowed us to keep only very few coefficients, and it is what we actually found, since scaling~\autoref{eq:al_scaling_wrt_a1} holds, together with~\autoref{eq:a1_scaling} and~\autoref{eq:a2_scaling}. However, deeply in the low-temperature phase, all Fourier coefficients become of order one and hence it is unfeasible to use them to describe cavity distributions~$\eta_{i\to j}$'s. Hence, we have to move to a different approach, which bases on the \textit{numerical solution} of the~\acrshort{BP} equations~\autoref{eq:BP_eqs_XY}: the \textit{cavity method}.

But first of all, we have to clarify a key point: how to numerically deal with cavity messages $\eta_{i\to j}$'s? Indeed, being continuous functions defined over the $[0,2\pi)$ interval and hence belonging to an infinite-dimensional space, any projection onto a finite-dimensional space would necessarily imply a loss of information. One of the possible ways out is given by the na\"ivest --- though effective --- approach one can figure out: discretize the $[0,2\pi)$ interval into a number $Q$ of equal bins, each one of which having size $2\pi/Q$. This proxy of the XY model is well known in the literature as the \textit{$Q$-state clock model}~\cite{NobreSherrington1986, NobreSherrington1989, IlkerBerker2013, IlkerBerker2014}. Of course, in the $Q\to\infty$ limit the XY model can be exactly recovered. 

Such kind of discretization could seem very rough, because it would get rid of the continuous nature of the model for any finite value of $Q$. However, it actually works very well in most cases. Throughout this chapter we will use $Q=64$ in our numerical simulations, allowing both a reliable and efficient approximation of the XY model. Then, in~\autoref{chap:clock} we will discuss the physics of the $Q$-state clock model when changing $Q$ and in particular its convergence toward the XY model, fully justifying the apparently small value of $Q$ used in this Chapter.

\subsection{Population Dynamics Algorithm}

Let us now focus on the cavity method. We would like to numerically solve the~\acrshort{BP} equations~\autoref{eq:BP_eqs_XY}, e.\,g. by discretizing each cavity marginal $\eta_{i\to j}(\theta_i)$ into a~$Q$-component array via the aforementioned clock model proxy. Then, starting from a certain initial condition, cavity marginals can be computed iteratively according to the \acrshort{BP} equations~\autoref{eq:BP_eqs_XY} until reaching the fixed point $\{\eta^*_{i\to j}\}$. Of~course, it has to depend on the particular realization of the underlying graph~$\mathcal{G}$ and of the set of exchange couplings $\{J_{ij}\}$, and the same holds for any physical observable, e.\,g. $f$, $u$, and so on. If on one hand the sample-to-sample fluctuations in self-averaging observables would be reduced by increasing as much as possible the size of the graph, on the other hand performing the average over $\mathbb{P}_{\mathcal{G}}$ and $\mathbb{P}_J$ is still necessary in order to obtain a meaningful disorder-averaged description of the model. This implies the need of averaging over a large enough number of different instances of the system.

However, as long as our task is to characterize the low-temperature region, recognizing the different thermodynamic phases that take place and evaluating the relevant physical observables, we can turn to a simpler and more effective approach: the \textit{sampled density evolution} technique or, as it is better known in statistical physics, the~\acrfull{PDA}. Firstly introduced in Ref.~\cite{AbouChacraEtAl1973} and then perfectioned and revised in Refs.~\cite{MezardParisi2001,MezardParisi2003}, it bases on a very effective idea: on each sample, the set of fixed-point cavity marginals $\{\eta^*_{i\to j}\}$ can be regarded as ``random variables'' whose realization in turn depends on that of the quenched disorder; hence, provided the probability distributions from which the disorder is drawn, it should be possible to compute the corresponding probability distribution $\mathbb{P}^*_{\eta}$ of the \acrshort{BP} fixed-point cavity marginals over the space of positive semi-definite functions.

At this point, the question of how to actually compute such probability distribution $\mathbb{P}^*_{\eta}$ can be easily addressed. Indeed, it can be seen as the fixed point of an iterative process for an empirical distribution $\mathbb{P}_{\eta}$, exactly in the same spirit of~\acrshort{BP}. If we introduce the following shorthand notation for the~\acrshort{BP} equations~\autoref{eq:BP_eqs_XY}
\begin{equation}
	\eta_{i\to j}(\theta_i) \equiv \mathcal{F}\bigl[\{\eta_{k\to i}(\theta_k)\},\{J_{ik}\}\bigr]
\end{equation}
then the probability of a certain realization of the cavity message can be iteratively computed as
\begin{equation}
	\mathbb{P}_{\eta}[\eta_{i\to j}] = \mathbb{E}_{\mathcal{G},J}\int\prod_{k=1}^{d_i-1}\Bigl(\mathcal{D}\eta_{k\to i}\,\mathbb{P}_{\eta}[\eta_{k\to i}]\Bigr)\delta\Bigl[\eta_{i\to j}-\mathcal{F}\bigr[\{\eta_{k\to i}\},\{J_{ik}\}\bigr]\Bigr]
	\label{eq:def_PDA}
\end{equation}
with the functional delta enforcing the validity of \acrshort{BP} equations and with $\mathcal{D}\eta$ being the measure over the space of positive semi-definite functions. In this way, the resulting fixed-point probability distribution is just the seeked $\mathbb{P}^*_{\eta}$~\cite{Book_MezardMontanari2009}. In other words, within this method the~\acrshort{BP} equations~\autoref{eq:BP_eqs_XY} are meant to be solved via a stochastic approach, automatically accomplishing the average over the quenched disorder.

Of course, when numerically implementing this method, a further discretization has to be taken into account. Indeed, the probability distribution $\mathbb{P}_{\eta}$ is approximated through a \textit{population} of $\mathcal{N}$ cavity messages $\eta_i$'s\footnote{Notice that here the single index $i$ refers to the $i$-th cavity marginal in the population and not the one-point marginal for the $i$-th spin.}. So after a (random) initialization of the population at the time step $t=0$, we can calculate it at the time step~$t+1$ by drawing the node degree $d_i$ from $\mathbb{P}_d$ --- being always equal to $C$ for the \acrshort{RRG} ensemble, while changing for the \acrshort{ERG} ensemble --- and then by picking at random $d_i-1$ couplings from $\mathbb{P}_J$ and $d_i-1$ cavity messages from the population at the time step~$t$. In the pseudocode~\ref{alg:RS_PDA} we list the key steps to implement~\acrshort{PDA} by numerically solving Eq.~\autoref{eq:def_PDA}.

\begin{algorithm}[t]
	\caption{RS Population Dynamics Algorithm ($T>0$)}
	\label{alg:RS_PDA}
	\begin{algorithmic}[1]
		\For {$i=1,\dots,\mathcal{N}$}
			\State Initialize $\eta^{(0)}_i$ \Comment{We use a random initialization}
		\EndFor
		\For {$t=1,\dots,t_{\text{max}}$}
			\For {$i=1,\dots,\mathcal{N}$}
				\State Draw an integer $d_i$ from the degree distribution $\mathbb{P}_d$
				\State Draw $d_i-1$ integers $\{k\}$ uniformly in the range $[1,\mathcal{N}]$
				\State Draw $d_i-1$ couplings $\{J_k\}$ from the coupling distribution $\mathbb{P}_J$
				\State $\eta^{(t)}_i \gets \mathcal{F}[\{\eta^{(t-1)}_{k}\},\{J_k\}]$
			\EndFor
		\EndFor
		\State \textbf{return} $\{\eta^{(t_{\text{max}})}_i\}$
	\end{algorithmic}
\end{algorithm}

Notice that the updates stop after $t_{\text{max}}$ iterations, where $t_{\text{max}}$ has to be large enough to ensure the convergence toward $\mathbb{P}^*_{\eta}$. What it is typically done is hence to compute at each time step some extensive physical observable  --- e.\,g. the free energy density $f$ --- and then to look at its time series. When it reaches a stationary regime, with fluctuations of order $O(1/\sqrt{\mathcal{N}})$ that can be hence interpreted as proper statistical fluctuations, then it can be safely claimed that the convergence has been reached. Also notice that $\mathcal{N}$ should be chosen large enough to reduce as much as possible the finite-size effects and hence to let $\mathbb{P}^*_{\eta}$ reproduce the actual probability distribution of the \acrshort{BP} fixed-point cavity marginals in the thermodynamic limit $N\to\infty$.

The computation of $f$ via the~\acrshort{PDA} follows the same idea of the ``translation'' of~\acrshort{BP} equations~\autoref{eq:BP_eqs_XY} on a given instance into their distributional version~\autoref{eq:def_PDA}. Indeed, the disorder-averaged value of $f$ can be obtained via a suitable average of the node contribution $f_i$ and of the edge contribution $f_{ij}$ over $\mathbb{P}^*_{\eta}$:
\begin{equation}
	f = \mathbb{E}_{\mathcal{G},J,\eta}\bigl[f_i\bigr]-\alpha\mathbb{E}_{\mathcal{G},J,\eta}\bigl[f_{ij}\bigr]
	\label{eq:def_PDA_f}
\end{equation}
where $\alpha$ is the average number of edges per node --- exactly equal to $C/2$ for the $C$-\acrshort{RRG} ensemble --- and where $f_i$ and $f_{ij}$ are given by
\begin{equation}
	f_i \equiv -\frac{1}{\beta}\ln{\mathcal{Z}_i} \qquad , \qquad f_{ij} \equiv -\frac{1}{\beta}\ln{\mathcal{Z}_{ij}}
	\label{eq:def_fi_fij}
\end{equation}
The key steps of the corresponding numerical implementation are listed in the pseudocode~\ref{alg:RS_PDA_free_energy_density}.

\begin{algorithm}[t]
	\caption{RS free energy density in the~\acrshort{PDA}}
	\label{alg:RS_PDA_free_energy_density}
	\begin{algorithmic}[1]
		\State Reach the fixed point $\mathbb{P}^*_{\eta}$ as in pseudocode~\ref{alg:RS_PDA}
		\State $\triangleright$ Node contribution $f_i$:
		\For {$i=1,\dots,\mathcal{N}$}
			\State Draw an integer $d_i$ from the degree distribution $\mathbb{P}_d$
			\State Draw $d_i-1$ integers $\{k\}$ uniformly in the range $[1,\mathcal{N}]$
			\State Draw $d_i-1$ couplings $\{J_k\}$ from the coupling distribution $\mathbb{P}_J$
			\State Compute $f_i$ as in Eq.~\autoref{eq:def_fi_fij}
		\EndFor
		\State $\overline{f_i} \gets (\sum_i f_i)/\mathcal{N}$
		\State $\triangleright$ Edge contribution $f_{ij}$:
		\For {$i=1,\dots,\mathcal{N}$}
			\State Draw an integer $j$ uniformly in the range $[1,\mathcal{N}]\setminus i$
			\State Draw a coupling $J$ from the coupling distribution $\mathbb{P}_J$
			\State Compute $f_{ij}$ as in Eq.~\autoref{eq:def_fi_fij}
		\EndFor
		\State $\overline{f_{ij}} \gets (\sum_i f_{ij})/\mathcal{N}$
		\State \textbf{return} $f \gets \overline{f_i}-\alpha\overline{f_{ij}}$
	\end{algorithmic}
\end{algorithm}

The computation of the other physical observables in the~\acrshort{PDA} can be performed exactly in the same way as for $f$, generalizing their expression on a given instance of the system into an averaged version over $\mathbb{P}^*_{\eta}$ and over all the sources of randomness.

The~\acrshort{PDA} is a widely exploited method, due to its very simple numerical implementation and to the advantage of automatically yielding a disorder-averaged description of the model. Moreover, it turns out to be very effective since it is found to provide a fixed-point probability distribution $\mathbb{P}^*_{\eta}$ almost irrespective of the initial conditions, namely of the initialization of the $\mathcal{N}$ cavity messages in the population at the time step $t=0$ (from which our random initialization). However, the~\acrshort{PDA} suffers the same limits of validity of the \acrshort{BP} approach on a given instance. Indeed, in Eq.~\autoref{eq:def_PDA} it is clear that the incoming messages $\eta_{k\to i}$'s are taken as independent from each other, exactly in the same spirit of the Bethe\,-\,Peierls approximation on a given instance of the problem. As anticipated in~\autoref{sec:sparse_random_graphs}, the latter ceases to converge to a fixed point $\{\eta^*_{i\to j}\}$ when the Gibbs measure breaks into several pure states, and this surely occurs in the low-temperature phases characterized by the~\acrshort{RSB}.

A formal proof of the conditions under which the~\acrshort{RS} \acrshort{BP} approach provides correct results does not exists yet. However, the requirement of a unique pure state in the Gibbs measure seems to be too strict, since it often turns out to work properly even when the Gibbs measure is just extremal\footnote{Roughly speaking, the extremality of the Gibbs measure means that the behaviour of a spin in the bulk of the system depends only on a set of boundary conditions with null measure.} rather than unique~\cite{KrzakalaEtAl2007}. Indeed, in such case the unique relevant solution in the thermodynamic limit is just the~\acrshort{RS} one.

Fortunately, being a stochastic approach, the~\acrshort{PDA} still provides a fixed point~$\mathbb{P}^*_{\eta}$ even in the \acrshort{RSB} region, though it has to be considered just as a~\acrshort{RS} proxy of the exact \acrshort{RSB} solution. In this way, we will be able to get much precious information about the low-temperature physics of the model under investigation. Moreover, we will see in~\autoref{chap:clock} that it is also possible to take into account the breaking of replica symmetry within the~\acrshort{PDA}, obtaining an algorithm that is equivalent to the~\acrshort{1RSB} ansatz of the fully connected case~\cite{MezardParisi2001, MezardParisi2003}. Unfortunately, the~\acrshort{fRSB} stage for the~\acrshort{BP} approach --- or equivalently for the cavity method --- has not been developed yet~\cite{Parisi2017}, due to the extreme richness and heterogeneity provided by the sparse topology with respect to the fully connected case.

\subsection{Numerical analysis of the Fourier expansion}

\begin{figure}[!t]
	\centering
	\includegraphics[scale=1]{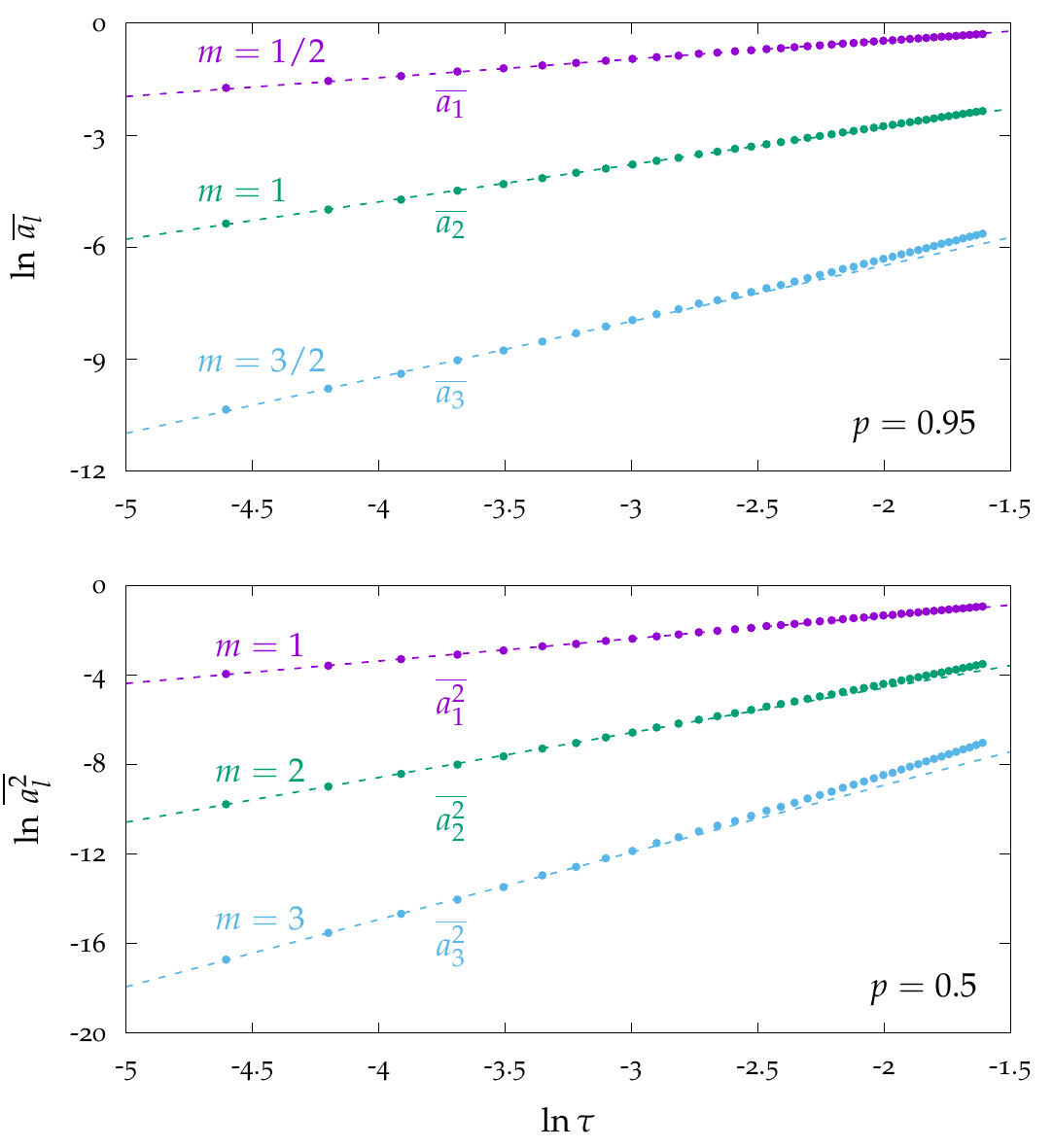}
	\caption[Fourier coefficients below $T_c$]{Scaling of the first moments of the cavity Fourier coefficients $a$'s slightly below the critical temperature $T_c$, measured via the~\acrshort{PDA} with a population of $\mathcal{N}=10^5$ cavity marginals on the $C=3$~\acrshort{RRG} ensemble. In the upper panel, we use $p=0.95$ so to focus on the paramagnetic\,-\,ferromagnetic transition, plotting the first moment. In the lower panel, we use $p=0.5$ so to focus on the paramagnetic\,-\,spin glass transition, plotting the second moment. The analytic predictions, Eqs.~\autoref{eq:al_scaling} and~\autoref{eq:al_square_scaling} respectively, are well reproduced in a wide range of values for the reduced temperature $\tau\equiv(T_c-T)/T_c$, as shown by the straight line of slope $m$ superimposed on each dataset. The corresponding statistical errors are smaller than the symbol size.}
	\label{fig:fourier}
\end{figure}

The first task we can accomplish by using~\acrshort{PDA} is to verify the scaling of first and second moments of the Fourier coefficients slightly below the critical temperature~$T_c$, Eqs.~\autoref{eq:al_scaling} and~\autoref{eq:al_square_scaling}. In order to do it, we randomly initialize a population of $\mathcal{N}=10^5$ cavity messages and then we let them evolve according to the~\acrshort{PDA} until reaching the fixed-point probability distribution $\mathbb{P}^*_{\eta}$; then, we compute the corresponding probability distributions of the Fourier coefficients of order $l=1,2,3$. Their first and second moments are reported in~\autoref{fig:fourier}, respectively corresponding to the paramagnetic\,-\,ferromagnetic transition ($p=0.95$, upper panel) and to the paramagnetic\,-\,spin glass transition ($p=0.5$, lower panel). For small values of the reduced temperature $\tau\equiv(T_c-T)/T_c$, the mean-field scaling
\begin{equation*}
	\overline{a_l} \propto \tau^{\,l/2} \qquad , \qquad \overline{a^2_l} \propto \tau^{\,l}
\end{equation*}
is nicely confirmed, while when getting deep into the low-temperature region this scaling starts to be violated, due to the growth of nonlinear terms in the self-consistency equation for each coefficient.

\subsection{Exploring the low-temperature region}

As anticipated at the beginning of~\autoref{subsec:exp_around_para}, one of the chased goals of the Fourier expansion around the paramagnetic solution was to get some insights about the ordering of spins in the low-temperature region. In this sense, we obtained strong evidences of the presence of either a ferromagnetic phase or a spin glass phase depending on the ferromagnetic bias of the coupling probability distribution $\mathbb{P}_J$. So far, there seems to occur nothing new with respect to the fully connected predictions of~\autoref{sec:vector_sg_fully}. Moreover, these results are in agreement with other works about disordered vector models defined on sparse random graphs~\cite{SkantzosEtAl2005, CoolenEtAl2005}.

Indeed, for a fraction $p$ of ferromagnetic couplings larger than the value $p_{mc}$ given in~\autoref{eq:mc_point_XY}, a global magnetization is found to appear:
\begin{equation}
	\boldsymbol{m} = \frac{1}{N}\sum_i\boldsymbol{m}_i
\end{equation}
where each $\boldsymbol{m}_i$ is directly related to first-order Fourier coefficients $a_1$'s and $b_1$'s, while $\boldsymbol{m}$ is linked to their average value. Moreover, we also succeeded in recovering the mean-field value $\beta=1/2$ of the critical exponent of the norm $m$ of the global magnetization versus the reduced temperature below the critical point
\begin{equation}
	m \propto \tau^{1/2}
\end{equation}
which hence becomes different with a square-root-like behaviour and eventually approaches the unity in the zero-temperature limit (with, however, a behaviour that is no longer square-root-like).

On the other side, when $p$ is smaller than $p_{mc}$, the disorder exhibits in an incoherent arrangement of spin directions, in turn resulting in a vanishing global magnetization. However, most of local magnetizations $\boldsymbol{m}_i$'s continue to be largely different from zero, so that in the spin glass phase the order parameter to look at in the~\acrshort{RS} approximation is the square magnetization or overlap $q$:
\begin{equation}
	q = \frac{1}{N}\sum_i|\boldsymbol{m}_i|^2
\end{equation}
The growth of $q$ with the distance from the critical point has already been predicted, Eq.~\autoref{eq:overlap_growth}, since it is linked to the average value of the square of the first-order Fourier coefficients $a_1$'s and $b_1$'s:
\begin{equation}
	q \propto \tau
\end{equation}
Furthermore, from the triangular inequality of norms, it holds for any $p$ that
\begin{equation*}
	m^2 \leqslant q
\end{equation*}
where the equality exactly holds only in the pure ferromagnetic case ($p=1$).

Order parameters $m$ and $q$ are hence enough to identify and separate the ferromagnetic phase from the ``unbiased'' (namely globally unmagnetized) spin glass phase --- and both them from the paramagnetic phase --- at least in the~\acrshort{RS} ansatz, as summarized in~\autoref{tab:RS_order_parameters}.

\begin{table}[t]
	\setlength{\tabcolsep}{8pt}		
	\centering
	\caption[RS order parameters]{Order parameters that allow us to identify paramagnetic, ferromagnetic, mixed and spin glass phases in the~\acrshort{RS} ansatz.}
	\label{tab:RS_order_parameters}
	\begin{tabular}{cccc}
		\toprule
		Phase & $m$ & $q$ & $\lambda_{\text{BP}}$\\
		\midrule
		Paramagnetic & $=0$ & $=0$ & $<0$\\
		Ferromagnetic & $>0$ & $>0$ & $<0$\\
		Mixed & $>0$ & $>0$ & $>0$\\
		Spin glass & $=0$ & $>0$ & $>0$\\
		\bottomrule
	\end{tabular}
\end{table}

However, the presence of a nonvanishing global magnetization is not on its own a guarantee of a \acrshort{RS} ferromagnetic phase. Indeed, we already saw in~\autoref{chap:sg_replica} that in the fully connected case --- for both scalar and vector spins --- a \textit{mixed} phase lies inbetween the ferromagnetic and the unbiased spin glass phases such that a global magnetization $m$ is still present, though the assumption of replica symmetry is no longer valid. Evidences of the presence of a mixed phase also in the diluted case have been alredy provided for the Ising model~\cite{VianaBray1985, Kabashima2003, CastellaniEtAl2005, MatsudaEtAl2010} as well as for vector models~\cite{SkantzosEtAl2005}, so it could be the same also for our model.

At this point, a further order parameter should be defined in order to distinguish the \acrshort{RS} ferromagnetic phase from the \acrshort{RSB} mixed phase, detecting the \acrshort{dAT} line --- in analogy with the replica computation --- that separates them. So let us introduce a stability parameter $\lambda_{\text{BP}}$ such that it is negative when the \acrshort{RS} \acrshort{BP} algorithm converges to a stable fixed point, while it is positive when an instability appears. In this way, the list of order parameters in the \acrshort{RS} ansatz, \autoref{tab:RS_order_parameters}, can be completed. We postpone the operative definition of $\lambda_{\text{BP}}$ to the next Subsection.

\begin{figure}[p]
	\centering
	\includegraphics[scale=1]{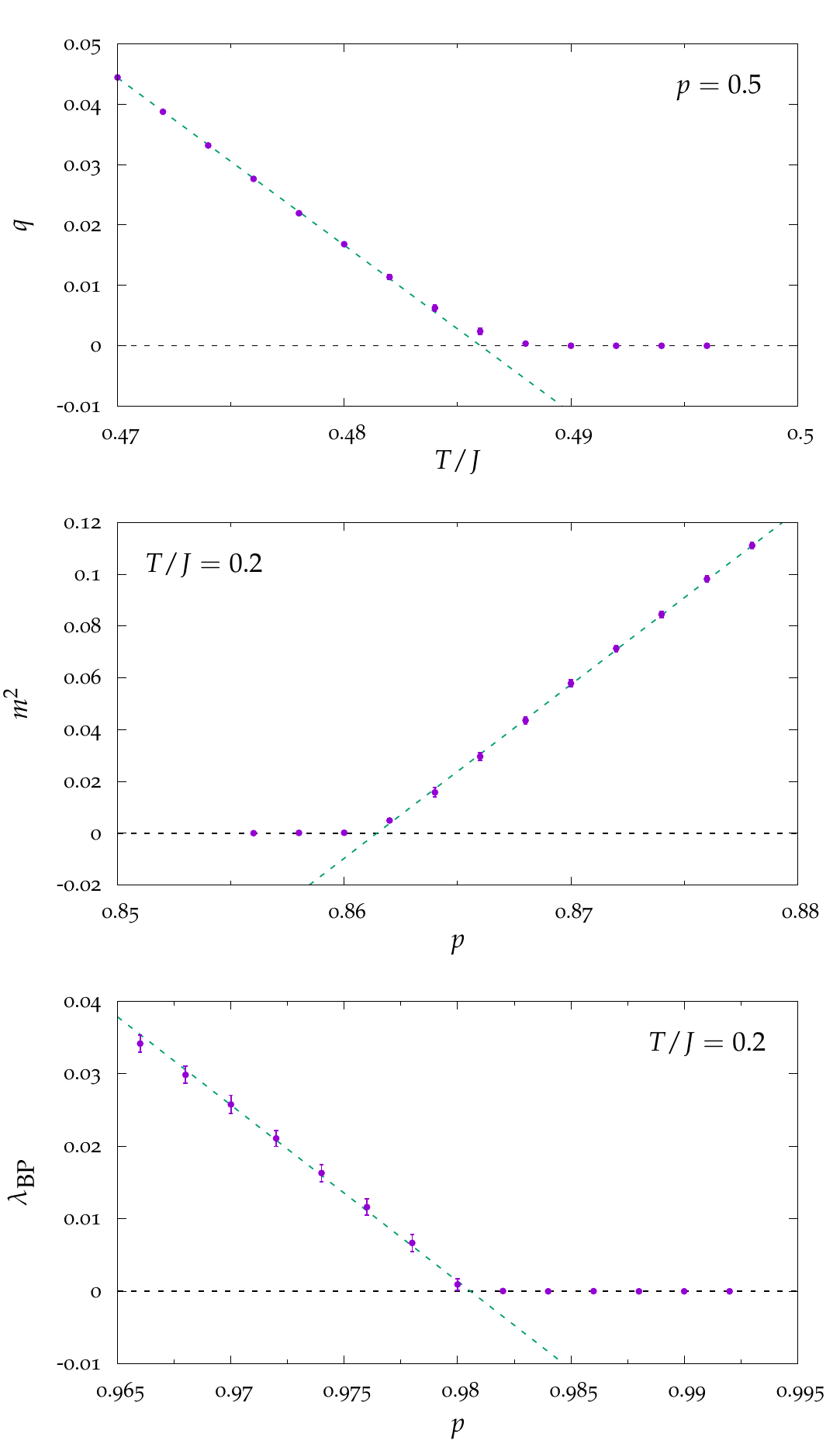}
	\caption[Detection of critical lines in the $T$ vs $p$ plane]{Detection of the critical lines between paramagnetic and ferromagnetic phases (upper panel, $p=0.5$), mixed and spin glass phases (central panel, $T/J=0.2$), ferromagnetic and mixed phases (lower panel, $T/J=0.2$) through~\acrshort{PDA} and~\acrshort{SuscProp}, with a population of $\mathcal{N}=10^6$ on the $C=3$~\acrshort{RRG} ensemble. In the first two plots error bars are too small with respect to the symbol size for almost each data point, while in the third case the larger error bars are due to a wide heterogeneity of perturbations $\delta\eta$'s, resulting in larger fluctuations of $\lambda_{\text{BP}}$ measurement.}
	\label{fig:critical_lines}
\end{figure}

The critical line between the paramagnetic phase and low-temperature region has already been evaluated analytically in~\autoref{subsec:exp_around_para}, so here we just compute the $q(T)$ curve for a fixed value of $p$ in order to check our numerical algorithm. E.\,g., when dealing with the~\acrshort{RRG} ensemble with connectivity $C=3$, the phase transition between the high- and the low-temperature regions locates at $T/J \simeq 0.4859$ for the unbiased spin glass ($p=0.5$), which is well confirmed by the linear fit over $q(T)$ dataset in the upper panel of~\autoref{fig:critical_lines}. Notice the usual finite-size effects that smooth the nondifferentiable point corresponding to the exact location of the phase transition.

The order parameter $m$, instead, can be used to detect the critical line between globally magnetized phase (that will be recongnized by looking at the $\lambda_{\text{BP}}$ stability parameter) and the unbiased spin glass phase in the~\acrshort{RS} approximation, which otherwise can not be analytically evaluated in the sparse case. As expected for these models, this line starts from the multicritical point $(p_{mc},T_{mc})$ and goes down almost vertically to the zero-temperature axis, recalling ---~still as a~\acrshort{RS} approximation~--- the Toulouse argument from the fully connected case~\cite{Toulouse1980}. In the central panel of~\autoref{fig:critical_lines} we again analyze the $C=3$~\acrshort{RRG} case for the temperature value $T/J=0.2$, finding the expected linear behaviour for $m^2$ when getting closer to the phase transition. Analogously, the whole transition line between the globally magnetized phase and the unbiased spin glass phase can be numerically computed.

\subsection{Susceptibility Propagation}

In order to actually distinguish the long-range order of the~\acrshort{RS} ferromagnetic phase from the one of the~\acrshort{RSB} mixed phase, we introduced the order parameter~$\lambda_{\text{BP}}$. If the~\acrshort{RS} stability in the fully connected case can be studied by looking at the eigenvalues of the Hessian matrix in the replica space (\autoref{chap:sg_replica} and Refs.~\cite{deAlmeidaThouless1978, deAlmeidaEtAl1978, Thesis_deAlmeida1980}), in the sparse case it is quite easier. Indeed, the stability parameter $\lambda_{\text{BP}}$ can be defined in several ways, which of course turn out to be equivalent, as shown in Ref.~\cite{Thesis_Zdeborova2009}. For example, since the breaking of replica symmetry corresponds to the breaking of the Gibbs measure in a large number of pure states, the related failure of the clustering property can be measured as a dependence of the \acrshort{BP} fixed point on the initial condition for the cavity messages on a given instance of the model~\cite{PagnaniEtAl2003, NewmanStein2003}.

Instead, since we are interested in the stochastic approach of the~\acrshort{PDA}, we can define a different stability parameter. Let us focus on the distributional fixed point~$\mathbb{P}^*_{\eta}$ found through~\acrshort{PDA}. In the \acrshort{RS} region it is surely stable, while in the~\acrshort{RSB} region it is unstable, even though it can still be reached, at variance with the \acrshort{BP} approach on a given instance. So an analysis of the linear stability of the fixed point $\mathbb{P}^*_{\eta}$ would seem to be enough.

For each cavity message~$\eta_{i\to j}$, let us consider its perturbation $\delta\eta_{i\to j}$, which is a function over the $[0,2\pi)$ interval as well. On a given instance of the model, it has to evolve according to the linearized version of the \acrshort{BP} equations~\autoref{eq:BP_eqs_XY}, which we explicitly compute in~\autoref{app:BPeqs_XYmodel} and which we rewrite here in a shorthand notation:
\begin{equation}
	\begin{split}
		\delta\eta_{i\to j} &= \sum_{k\in\partial i\setminus j}\biggl|\frac{\delta \mathcal{F}[\{\eta_{k\to i}\},\{J_{ik}\}]}{\delta \eta_{k\to i}}\biggr|_{\eta^*_{k\to i}}\delta\eta_{k\to i}\\
		&\equiv \mathcal{F}'[\{\eta^*_{k\to i}\},\{\delta\eta_{k\to i}\},\{J_{ik}\}]
	\end{split}
	\label{eq:BP_eqs_XY_linear}
\end{equation}
Then, if we move to the~\acrshort{PDA}, the population of $\mathcal{N}$ cavity messages $\{\eta_{i\to j}\}$ becomes a population of $\mathcal{N}$ couples $(\eta_{i\to j},\delta\eta_{i\to j})$. So after having reached the fixed point~$\mathbb{P}^*_{\eta}$, at each time step we implement the distributional version of~\autoref{eq:BP_eqs_XY_linear}:
\begin{equation}
\begin{split}
	\mathbb{P}[(\eta^*_{i\to j},\delta\eta_{i\to j})] &= \mathbb{E}_{\mathcal{G},J}\int\prod_{k=1}^{d_i-1}\Bigl(\mathcal{D}\eta_{k\to i}\,\mathbb{P}[(\eta^*_{k\to i},\delta\eta_{k\to i})]\Bigr)\\
	&\qquad\qquad\times\delta\Bigl[\delta\eta_{i\to j}-\mathcal{F}'[\{\eta^*_{k\to i}\},\{\delta\eta_{k\to i}\},\{J_{ik}\}]\Bigr]
	\label{eq:def_PDA_SuscProp}
\end{split}
\end{equation}
in analogy with~\autoref{eq:def_PDA}, where now we have a \textit{joint} probability distribution of fixed-point cavity marginals and the corresponding perturbations.

We can evaluate the $L_2$ norm of each perturbation at each time step, summing over the $Q$ values of the corresponding clock model proxy
\begin{equation}
	\norm{\delta\eta^{(t)}_i}_2^2 = \sum_{a=0}^{Q-1}\abs{\delta\eta^{(t)}_i(\theta_{i,a})}^2
\end{equation}
and then we can average over the population, in order to get the global norm of the perturbations at the $t$-th time step. However, as an effect of the sparsity of the underlying graph, there appears a strong heterogeneity in the population. A striking consequence of this is that perturbations span several orders of magnitude. So it turns out to be more robust and reliable to average the \textit{logarithm} of the norms rather than the norms themselves, so defining the following global norm:
\begin{equation}
	\norm{\delta\eta^{(t)}} \equiv \frac{1}{\mathcal{N}}\sum_i\,\ln{\norm{\delta\eta^{(t)}_i}^2_2}
	\label{eq:delta_eta_norm}
\end{equation}
The corresponding growth rate at each time step can be then computed as the \textit{difference} with respect the previous time step:
\begin{equation}
	\lambda^{(t)}_{\text{BP}} \equiv \norm{\delta\eta^{(t)}} - \norm{\delta\eta^{(t-1)}}
	\label{eq:lambdaBP_XY}
\end{equation}
and its time series asymptotically approaches the corresponding Lyapunov exponent:
\begin{equation}
	\lambda_{\text{BP}} \equiv \lim_{t\to\infty}\frac{1}{t}\norm{\delta\eta^{(t)}}
	\label{eq:lambdaBP_XY_def}
\end{equation}
This approach is known as \acrfull{SuscProp}~\cite{Book_MezardMontanari2009}, and the key steps of its numerical implementation are shown in pseudocode~\ref{alg:RS_PDA_SuscProp}.

\begin{algorithm}[t]
\caption{Susceptibility Propagation in the~\acrshort{PDA} ($T>0$)}
\label{alg:RS_PDA_SuscProp}
\begin{algorithmic}[1]
\State Reach the fixed point $\mathbb{P}^*_{\eta}$ as in pseudocode~\ref{alg:RS_PDA}
\For {$i=1,\dots,\mathcal{N}$}
	\State Initialize $\delta\eta^{(0)}_i$ \Comment{We use a random initialization}
\EndFor
\For {$t=1,\dots,t_{\text{max}}$}
	\For {$i=1,\dots,\mathcal{N}$}
		\State Draw an integer $d_i$ from the degree distribution $\mathbb{P}_d$
		\State Draw $d_i-1$ integers $\{k\}$ uniformly in the range $[1,\mathcal{N}]$
		\State Draw $d_i-1$ couplings $\{J_k\}$ from the coupling distribution $\mathbb{P}_J$
		\State $\eta^*_i \gets \mathcal{F}[\{\eta^*_{k}\},\{J_k\}]$ \Comment{Just a ``refresh'' of the population}
		\State $\delta\eta^{(t)}_i \gets \mathcal{F}'[\{\eta^*_k\},\{\delta\eta^{(t-1)}_{k}\},\{J_k\}]$
	\EndFor
	\State Compute $\norm{\delta\eta^{(t)}}$ as in~\autoref{eq:delta_eta_norm}
	\State $\lambda^{(t)}_{\text{BP}} \gets \norm{\delta\eta^{(t)}} - \norm{\delta\eta^{(t-1)}}$
\EndFor
\State Average $\lambda^{(t)}_{\text{BP}}$ over the $t_{\text{max}}$ iterations \Comment{Pay attention to thermalization}
\State \textbf{return} $\lambda_{\text{BP}}$
\end{algorithmic}
\end{algorithm}

At this point, a negative Lyapunov exponent $\lambda_{\text{BP}}$ actually implies stability, while a positive one signals an instability. Finally, the~\acrshort{dAT} line can be detected through the marginality condition
\begin{equation*}
	\lambda_{\text{BP}}=0
\end{equation*}
Still referring to the $C=3$ \acrshort{RRG} topology, in the lower panel of~\autoref{fig:critical_lines} we can see how $\lambda_{\text{BP}}$ is almost linear close to the critical point, while changing the slope when trespassing it, and hence a linear interpolation easily gives the estimation of the critical point itself.

Being rigorous, the marginality condition $\lambda_{\text{BP}}=0$ does not univocally implies \acrshort{RS} instability and so the occurrence of the~\acrshort{dAT} line; indeed, according to its definition, $\lambda_{\text{BP}}$ vanishes every time the fixed point $\mathbb{P}^*_{\eta}$ becomes unstable along some direction, namely when the system undergoes a generic (second-order) phase transition. However, if the ``new'' fixed point is \acrshort{RS} stable as well, then $\lambda_{\text{BP}}$ would again reach a negative value, provided the convergence to the new fixed point has been actually attained. Instead, in presence of a \acrshort{RSB} solution, the \acrshort{BP} fixed point would always be unstable, namely $\lambda_{\text{BP}}>0$, since the \acrshort{RSB} solutions are not reachable within the \acrshort{RS} ansatz.

In this way, the whole \acrshort{dAT} line can be finally detected. It corresponds to the critical line between the paramagnetic and the spin glass phases as long as $p<p_{mc}$, while for $p>p_{mc}$ it is lower than the transition line between the paramagnetic and the ferromagnetic phases --- just as in the fully connected case --- so allowing us to properly identify the boundaries of the mixed phase.

\section{The limit of zero temperature}
\label{sec:XY_zeroTemp}

The previous numerical techniques hold as long as the system stays at a finite temperature, so that in principle each configuration of the system has a finite --- even if sometimes very small --- probability to be realized. But when $T$ goes to zero, probability distributions acquire a singular behaviour, as well as compatibility functions and evidences (see~\autoref{chap:tools}).

However, it is still possible to exploit the message-passing technique provided by the~\acrshort{BP} approach, though in a different form. Indeed, by rewriting the cavity probability distributions as large-deviation functions
\begin{equation}
	\eta_{i\to j}(\theta_i) \equiv e^{\,\beta h_{i\to j}(\theta_i)}
\end{equation}
and then evaluating the integrals over the angular variables via the saddle-point method with $\beta\to\infty$, the zero-temperature~\acrshort{BP} equations can be obtained, as thoroughly explained in~\autoref{app:BPeqs_XYmodel}:
\begin{equation}
	\begin{split}
		h_{i\to j}(\theta_i) &\cong \sum_{k\in\partial i\setminus j}\max_{\theta_k}{\bigl[h_{k\to i}(\theta_k)+J_{ik}\cos{(\theta_i-\theta_k)}\bigr]}\\
		&\equiv \mathcal{F}_0[\{h_{k\to i}\},\{J_{ik}\}]
	\end{split}
	\label{eq:BP_eqs_zeroTemp}
\end{equation}
where $\cong$ takes into account the additive constant for the normalization, so that the \textit{cavity fields} $h_{i\to j}$'s become negative semidefinite functions:
\begin{equation}
	\max_{\theta_i}h_{i\to j}(\theta_i) = 0
\end{equation}

The interesting physical observables as the free energy density~$f$ (that in the $T\to 0$ limit matches with the internal energy density~$u$), the average magnetization~$\boldsymbol{m}$, and so on, can be computed in the same spirit, namely by evaluating the integrals over the angular variables via the saddle-point method in the $\beta\to\infty$ limit. For example, node and edge contributions $f_i$ and $f_{ij}$ to the free energy, defined in Eq.~\autoref{eq:def_fi_fij}, become:
\begin{subequations}
	\begin{equation}
		f_i = -\max_{\theta_i}\Biggl[\sum_{k\in\partial i}\max_{\theta_k}\bigl[h_{k\to i}(\theta_k)+J_{ik}\cos{(\theta_i-\theta_k)}\bigr]\Biggr]
	\end{equation}
	\begin{equation}
		f_{ij} = -\max_{\theta_i,\theta_j}\bigl[h_{i\to j}(\theta_i)+h_{j\to i}(\theta_j)+J_{ij}\cos{(\theta_i-\theta_j)}\bigr]
	\end{equation}
\end{subequations}

Of course, also in the zero-temperature limit it is possible to solve the~\acrshort{BP} equations via the~\acrshort{PDA}, focusing on the probability distribution $\mathbb{P}_h[h_{i\to j}]$ of the cavity fields and then searching for its fixed point by implementing the following distributional equation
\begin{equation}
	\mathbb{P}_{h}[h_{i\to j}] = \mathbb{E}_{\mathcal{G},J}\int\prod_{k=1}^{d_i-1}\Bigl(\mathcal{D}h_{k\to i}\,\mathbb{P}_{h}[h_{k\to i}]\Bigr)\delta\Bigl[h_{i\to j}-\mathcal{F}_0\bigr[\{h_{k\to i}\},\{J_{ik}\}\bigr]\Bigr]
	\label{eq:def_PDA_zeroTemp}
\end{equation}
as listed in the pseudocode~\ref{alg:RS_PDA_zeroTemp}.

The stability of the fixed point $\mathbb{P}^*_h$ reached in this way can be then checked in an analogous manner as we did in the finite-temperature case. Indeed, we can either perturb each finite-temperature cavity message and then rewrite the resulting perturbation as a large-deviation function, or directly perturb each zero-temperature cavity field. Both computations are equivalent (see~\autoref{app:BPeqs_XYmodel}) and lead to the following zero-temperature linearized~\acrshort{BP} equations:
\begin{equation}
\begin{split}
	\delta h_{i\to j}(\theta_i) &\cong \sum_{k\in\partial i\setminus j}\delta h_{k\to i}(\theta^*_k(\theta_i))\\
	&\equiv \mathcal{F}'_0\bigl[\{h^*_{k\to i}\},\{\delta h_{k\to i}\},\{J_{ik}\}\bigr]
	\label{eq:BP_eqs_zeroTemp_linear}
\end{split}
\end{equation}
with $\theta^*_k(\theta_i)$ given by:
\begin{equation}
	\theta^*_k(\theta_i) = \argmax_{\theta_k}{\bigl[h_{k\to i}(\theta_k)+J_{ik}\cos{(\theta_i-\theta_k)}\bigr]}
	\label{eq:BP_eqs_zeroTemp_linear_thetaStar}
\end{equation}
So for each directed edge, the outgoing perturbation is equal to the sum of the incoming perturbations respectively evaluated in the direction that maximizes the incoming message. As a consequence of this, the zero-temperature perturbations can be very singular. Moreover, notice that these perturbations are again defined up to an additive constant, which actually possesses a meaningful physical interpretation, as discussed in~\autoref{app:BPeqs_XYmodel}: each properly normalized $\delta h_{i\to j}$ crosses the zero in correspondence of the maximum of the related cavity field $h_{i\to j}$, namely in correspondence of its most probable value in the $\beta\to\infty$ limit
\begin{equation}
	\delta h_{i\to j}(\theta^*_i)=0 \quad , \qquad \theta^*_i \equiv \argmax_{\theta_i}{\Biggl[\sum_{k\in\partial i\setminus j}\Bigl[h_{k\to i}(\theta^*_k(\theta_i)) + J_{ik}\cos{(\theta^*_k(\theta_i)-\theta_k)}\Bigr]\Biggr]}
\end{equation}
In this way, the property of being negative semidefinite is preserved for the perturbed cavity fields, provided perturbations are small enough.

\begin{algorithm}[t]
	\caption{RS Population Dynamics Algorithm ($T=0$)}
	\label{alg:RS_PDA_zeroTemp}
	\begin{algorithmic}[1]
		\For {$i=1,\dots,\mathcal{N}$}
			\State Initialize $h^{(0)}_i$ \Comment{We use a random initialization}
		\EndFor
		\For {$t=1,\dots,t_{\text{max}}$}
			\For {$i=1,\dots,\mathcal{N}$}
				\State Draw an integer $d_i$ from the degree distribution $\mathbb{P}_d$
				\State Draw $d_i-1$ integers $\{k\}$ uniformly in the range $[1,\mathcal{N}]$
				\State Draw $d_i-1$ couplings $\{J_k\}$ from the coupling distribution $\mathbb{P}_J$
				\State $h^{(t)}_i \gets \mathcal{F}_0[\{h^{(t-1)}_{k}\},\{J_k\}]$
			\EndFor
		\EndFor
		\State \textbf{return} $\{h^{(t_{\text{max}})}_i\}$
	\end{algorithmic}
\end{algorithm}

Finally, we can write down the stochastic equation for the joint probability distribution of the fixed-point cavity fields and of their perturbations:
\begin{equation}
\begin{split}
	\mathbb{P}[(h^*_{i\to j},\delta h_{i\to j})] &= \mathbb{E}_{\mathcal{G},J}\int\prod_{k=1}^{d_i-1}\Bigl(\mathcal{D}h_{k\to i}\,\mathbb{P}[(h^*_{k\to i},\delta h_{k\to i})]\Bigr)\\
	&\qquad\qquad\times\delta\Bigl[\delta h_{i\to j}-\mathcal{F}'_0[\{h^*_{k\to i}\},\{\delta h_{k\to i}\},\{J_{ik}\}]\Bigr]
	\label{eq:def_PDA_SuscProp_zeroTemp}
\end{split}
\end{equation}
which can be solved by implementing~\acrshort{SuscProp} as outlined in the pseudocode~\ref{alg:RS_PDA_SuscProp_zeroTemp}. However, in the zero-temperature limit some further precautions have to be taken in the~\acrshort{SuscProp} algorithm with respect to the finite-temperature case, as we discuss below.

Indeed, as long as angular variables are actually evaluated as continuous, the correct behaviour of the XY model is reproduced even in the zero-temperature limit. But we have to remember that in the whole low-temperature region the~\acrshort{BP}~equations for the XY model have to be solved numerically by means of a $Q$-state clock model with $Q$ large enough, but still finite, and this is true also on the $T=0$ axis. This discretization does not qualitatively change the solution of the~\acrshort{BP} equations~\autoref{eq:BP_eqs_XY} at finite temperature, and the same holds for their linearized version, as we will check in~\autoref{chap:clock}. $Q$-dependence still remains smooth enough in the zero-temperature limit for what regards the~\acrshort{BP} equations~\autoref{eq:BP_eqs_zeroTemp}. But what happens for their linearized version~\autoref{eq:BP_eqs_zeroTemp_linear}?

The main issue of the zero-temperature linearized~\acrshort{BP} equations is the evaluation of angles $\theta^*_k(\theta_i)$'s which maximize the right hand side of~\autoref{eq:BP_eqs_zeroTemp}. Provided $h$'s are still smooth functions when $Q$ is large but finite, it is no longer true for $\delta h$'s. As long as $Q$ is finite, it could happen that for all the $\theta_i$ directions allowed by the clock model, the argmax is given by the same value $\theta^*_k$, so that the resulting perturbation $\delta h_{i\to j}(\theta_i)$ is a constant function. Then, when normalizing it through the suitable additive constant, it becomes an identically vanishing function. In the end, this results in a cascade of exactly null perturbations when iterating~\acrshort{BP} on both a given instance or via the~\acrshort{PDA}.

In this way, the global norm of perturbations~$\norm{\delta h^{(t)}}$, computed as in the finite-temperature case:
\begin{equation}
	\norm{\delta h^{(t)}} \equiv \frac{1}{\mathcal{N}}\sum_i\ln{\norm{\delta h^{(t)}_i}^2_2}
	\label{eq:delta_eta_norm_zeroTemp}
\end{equation}
becomes meaningless, since each $\norm{\delta h^{(t)}_i}$ eventually collapses to zero. This is exactly the mechanism that has been proved to hold at zero temperature for the Ising model on Bethe lattice~\cite{CastellaniEtAl2005, MoroneEtAl2014} and for several other discrete models, such as the Potts model~\cite{KrzakalaZdeborova2008} and the colouring problem~\cite{ZdeborovaKrzakala2007, Thesis_Zdeborova2009}. We will come back to this issue in~\autoref{chap:clock}, when we will show how to define $\lambda_{\text{BP}}$ also for discrete models and in particular for the $Q$-state clock model, taking into account the ratio at which the fraction of nonvanishing perturbation shrinks to zero with $t$.

The situation drastically changes when the system under analysis is described by continuous variables. Due to the possibility of having infinitesimal perturbations even in the zero-temperature limit, the fraction of nonvanishing perturbations does not shrink any longer with $t$. This is due to the fact that any infinitesimal shift in $\theta_i$ on the left hand side of~\autoref{eq:BP_eqs_zeroTemp_linear} always causes an infinitesimal change in the argmax $\theta^*_k$ as well, and hence a nonconstant incoming perturbation $\delta h_{k\to i}(\theta_k)$ typically does not turn into a vanishing outgoing perturbation $\delta h_{i\to j}(\theta_i)$.

At this point, the dramatic difference in the evolution of perturbations at zero temperature between discrete and continuous models may suggest that the clock model could not be able to correctly reproduce the physics of the XY model in such limit for any finite number $Q$ of states. Luckily, it is not the case. Indeed, the correct computation of perturbations at zero temperature can still be performed by using the $Q$-state clock model, with the caveat of evaluating the ``continuous'' argmax $\theta^*_k(\theta_i)$ by interpolating around the ``discrete'' one $\widetilde{\theta^*_k}(\theta_i)$ provided by the clock model, and then evaluating the incoming perturbation $\delta h_{k\to i}$ just in correspondence of the actual value of the argmax. This trick is crucial in order to match the zero-temperature limit of the data collected at finite temperature with the data directly collected through the zero-temperature algorithm, as we will see also in~\autoref{chap:RFXY}. The whole procedure is outlined in the pseudocode~\ref{alg:RS_PDA_SuscProp_zeroTemp}.

\begin{algorithm}[t]
\caption{Susceptibility Propagation in the~\acrshort{PDA} ($T=0$)}
\label{alg:RS_PDA_SuscProp_zeroTemp}
\begin{algorithmic}[1]
\State Reach the fixed point $\mathbb{P}^*_h$ as in pseudocode~\ref{alg:RS_PDA_SuscProp_zeroTemp}
\For {$i=1,\dots,\mathcal{N}$}
	\State Initialize $\delta h^{(0)}_i$ \Comment{We use a random initialization}
\EndFor
\For {$t=1,\dots,t_{\text{max}}$}
	\For {$i=1,\dots,\mathcal{N}$}
		\State Draw an integer $d_i$ from the degree distribution $\mathbb{P}_d$
		\State Draw $d_i-1$ integers $\{k\}$ uniformly in the range $[1,\mathcal{N}]$
		\State Draw $d_i-1$ couplings $\{J_k\}$ from the coupling distribution $\mathbb{P}_J$
		\State $h^*_i \gets \mathcal{F}_0[\{h^*_{k}\},\{J_k\}]$ \Comment{Just a ``refresh'' of the population}
		\State Compute the discrete argmax $\widetilde{\theta^*_k}$ over the $Q$ clock model states
		\State Interpolate the continuous argmax $\theta^*_k$ around $\widetilde{\theta^*_k}$
		\State Interpolate the actual value of $\delta h^{(t-1)}_{k}$ for $\theta_k=\theta^*_k$
		\State $\delta h^{(t)}_i \gets \mathcal{F}'_0[\{h^*_k\},\{\delta h^{(t-1)}_k\},\{J_k\}]$
	\EndFor
	\State Compute $\norm{\delta h^{(t)}}$ as in~\autoref{eq:delta_eta_norm_zeroTemp}
	\State $\lambda^{(t)}_{\text{BP}} \gets \norm{\delta h^{(t)}} - \norm{\delta h^{(t-1)}}$
\EndFor
\State Average $\lambda^{(t)}_{\text{BP}}$ over the $t_{\text{max}}$ iterations \Comment{Pay attention to thermalization}
\State \textbf{return} $\lambda_{\text{BP}}$
\end{algorithmic}
\end{algorithm}

An evidence in favour of the reasoning outlined above is given by the fact that --- without applying the previous trick --- the fraction of nonvanishing perturbations decays with a time rate which is smaller the larger $Q$, and hence it is reasonable to claim that in the $Q\to\infty$ limit it does not decay any longer. So we can finally claim that, when dealing with continuous variable models, the stability of the~\acrshort{RS} \acrshort{BP} fixed point is ruled by the same mechanism seen at finite temperature, namely by the growth of the norm of the perturbations. It is in the case of discrete models that it changes, involving a fraction of identically vanishing perturbations that grows up with $t$, as we mentioned above and as we will see more extensively in~\autoref{chap:clock}.

So the~\acrshort{BP} fixed-point stability at $T=0$ for the XY model can be checked again by looking at a Lyapunov exponent $\lambda_{\text{BP}}$ defined in perfect analogy with the finite-temperature case:
\begin{equation}
	\lambda_{\text{BP}} \equiv \lim_{t\to\infty}\frac{1}{t}\norm{\delta h^{(t)}}
	\label{eq:lambdaBP_XY_zeroTemp_def}
\end{equation}
so that $\lambda_{\text{BP}}<0$ refers to a stable \acrshort{BP} fixed point and $\lambda_{\text{BP}}>0$ to an unstable \acrshort{BP} fixed point. In this way, we will finally succeed in matching the extrapolation of the \acrshort{dAT} line toward the $T=0$ axis with the location of such endpoint directly computed in the zero-temperature framework.

\section{Phase diagram for the bimodal distribution}

\begin{figure}[!b]
	\centering
	\includegraphics[scale=1]{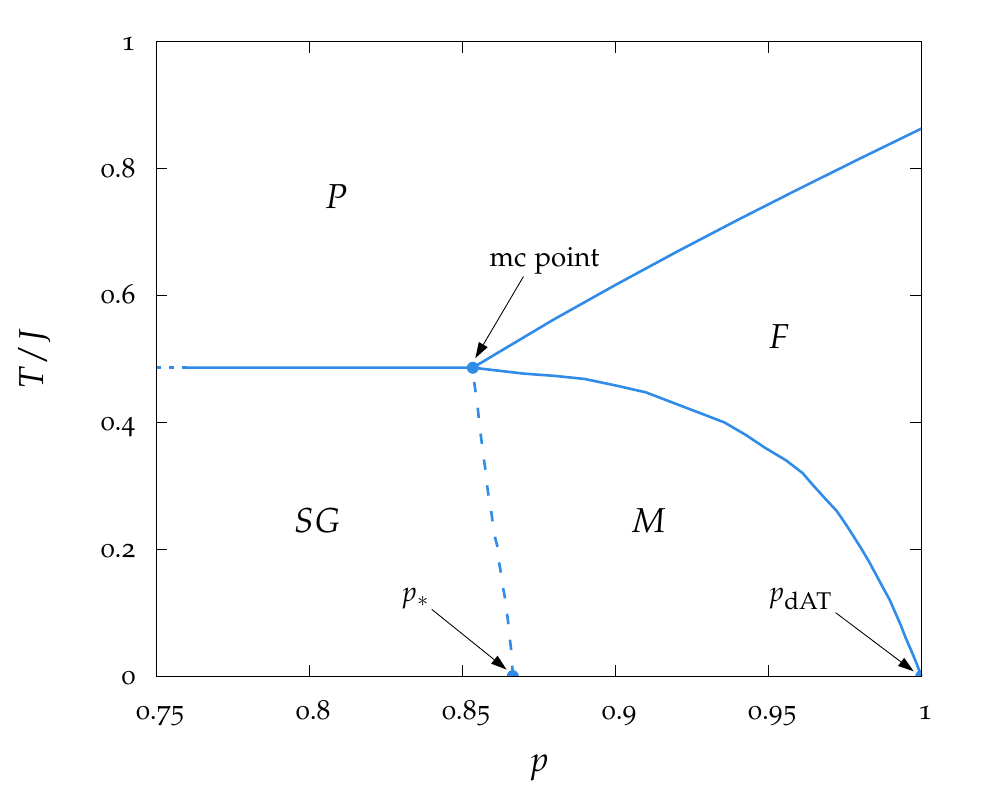}
	\caption[Phase diagram of the XY model with bimodal couplings]{Phase diagram of the XY model on the $C=3$~\acrshort{RRG} ensemble with random couplings $J_{ij}$'s drawn from the bimodal disorder distribution $\mathbb{P}_J$ of Eq.~\autoref{eq:disorder_distribution_pmJ}.}
	\label{fig:XY_phase_diag_pmJ}
\end{figure}

At this point, the complete phase diagram for the XY model on sparse random graphs with a bimodal coupling distribution $\mathbb{P}_J$ can be drawn. In particular, \autoref{fig:XY_phase_diag_pmJ} refers to the $C=3$~\acrshort{RRG} ensemble, with the following values for some particular points on the critical lines:
\begin{gather*}
	p_{mc} = 0.854(1) \qquad , \qquad T_{mc}/J=0.486(1) \qquad , \qquad T_c(p=1)/J=0.863(1)\\
	p_*=0.867(1) \qquad , \qquad p_{\text{dAT}}=1.000(1)
\end{gather*}
with the location of the multicritical point that remarkably agrees with the analytical prediction~\autoref{eq:mc_point_XY}, as well as the critical temperature away from paramagnetic phase in the pure ferromagnetic model, as predicted in Eq.~\autoref{eq:XY_para_stability_cRRG}. $p_*$, instead, stands for the value on the zero-temperature axis where $m$ goes to zero in the~\acrshort{RS} ansatz, namely the endpoint of the dashed line in the phase diagram, marking the transition between the mixed and the spin glass phases. Finally, $p_{\text{dAT}}$ is the endpoint of the~\acrshort{dAT} line on the $T=0$ axis.

The arrangement of the different thermodynamic phases in the $T$ vs $p$ plane is compatible with the results obtained on the same topology for the Ising model at zero temperature~\cite{Kabashima2003, CastellaniEtAl2005, MatsudaEtAl2010} and in the fully connected case for both Ising and vector spins (\autoref{chap:sg_replica}), included the presence of the mixed phase inbetween the ferromagnetic and the spin glass phases.

However, an important difference with respect to the Ising case rises up: the~\acrshort{dAT} line does not approach the $T=0$ axis for a fraction $p_{\text{dAT}}$ of positive couplings smaller than one, but it exactly reaches the right lower corner of the phase diagram, corresponding to the zero-temperature ferromagnetic XY model. This is a very important feature, not yet known in the literature (to the best of our knowledge), remarking a peculiar behaviour of vector models with respect to scalar models: as soon as an infinitesimal quantity of disorder is introduced in the model, the zero-temperature solution ceases to be \acrshort{RS}~stable and becomes \textit{glassy}. Indeed, a weak disorder may not be enough to force discrete variables to align in different directions, while continuous variables can arrange more easily to several different orientations, favouring the appearance of many states in the Gibbs measure. This feature may provide a strong connection with the theory of structural glasses and the glass transition~\cite{ParisiZamponi2010, BerthierBiroli2011, CharbonneauEtAl2017}, and in~\autoref{chap:XYinField_zeroTemp} we will deeply study the zero-temperature energy landscape of the XY model in a random field in order to better characterize this peculiar behaviour.

Finally, the main features of the phase diagram in~\autoref{fig:XY_phase_diag_pmJ} are expected not to change qualitatively when increasing the connectivity~$C$ --- still for finite values of $C$ --- or when moving to the~\acrshort{ERG} ensemble.

\section{The gauge glass model}
\label{sec:gauge_glass_XY}

In the Ising case, the unique ways to insert disorder in the exchange couplings are to change their sign, e.\,g. through the bimodal distribution used so far, or to change their magnitude, e.\,g. by drawing it from a Gaussian distribution, as typically done in the fully connected case. Instead, in the vector case, there is a huge variety of ways to insert randomness in the exchange constants, involving modification in magnitudes, phase shifts, or both.

So far, we just exploited the same coupling distribution used in the Ising case, namely a complete flip of the coupling so to imply a purely antiferromagnetic interaction. However, we could also leave the magnitude $J$ unchanged and then act on the \textit{direction} of the interaction
\begin{equation}
	-J_{ij}\,\boldsymbol{\sigma}_i\cdot\boldsymbol{\sigma}_j \quad \to \quad -J\,\boldsymbol{\sigma}_i\mathbb{U}(\{\omega_{ij}\})\boldsymbol{\sigma}_j
\end{equation}
through a suitable random rotation $\mathbb{U}$ in the spin space, characterized by the set of angles~$\{\omega_{ij}\}$.

In particular, for the XY model the rotation matrix $\mathbb{U}$ reduces to a random rotation in the $xy$ plane, that can be identified through a single angle $\omega_{ij}\in[0,2\pi)$. Hence, the corresponding Hamiltonian reads:
\begin{equation}
	\mcH[\{\theta\}]= -J\sum_{(i,j)}\cos{(\theta_i-\theta_j-\omega_{ij})}
	\label{eq:Hamiltonian_XY_sparse_gg}
\end{equation}
with $\omega_{ij}$'s drawn from a suitable probability distribution $\mathbb{P}_{\omega}$~\cite{CoolenEtAl2005}. Notice that if we choose $\omega_{ij}\in\{0,\pi\}$, the unbiased ($p=0.5$) bimodal case can be recovered.

The model with Hamiltonian~\autoref{eq:Hamiltonian_XY_sparse_gg} is known as \textit{gauge glass}, and it is typically used to describe granular disordered type-II superconductors~\cite{ShihEtAl1984, JohnLubensky1985, HuseSeung1990}, with quenched random phase shifts $\omega_{ij}$'s linked to the spacial distribution of the external field.

In order to get complementary results with respect to the bimodal case --- which is very anisotropic --- we could choose $\mathbb{P}_{\omega}$ as the flat distribution over the $[0,2\pi)$ interval, so allowing a completely isotropic choice of the couplings between nearest-neighbour spins:
\begin{equation}
	\mathbb{P}_{\omega}(\omega_{ij}) = \mathrm{Unif}\bigl([0,2\pi)\bigr)
\end{equation}
However, in order to get even more general results, we choose a suitable probability distribution that allows us to interpolate between the pure ferromagnet ($\omega=0$ for each link) and the unbiased spin glass ($\omega$ uniformly drawn from the flat distribution for each link):
\begin{equation}
	\mathbb{P}_{\omega}(\omega_{ij}) = (1-\Delta)\,\delta(\omega_{ij})+\Delta\,\mathrm{Unif}\bigl([0,2\pi)\bigr)
	\label{eq:disorder_distribution_gg}
\end{equation}
so that for $\Delta=0$ we recover the former case and for $\Delta=1$ we recover the latter case. So $\Delta\in[0,1]$ plays the same role of \textit{ferromagnetic bias} as $p$.

The~\acrshort{BP} equations for the gauge glass XY model can be easily written down, referring to the ones for the bimodal case and substituting the interaction term:
\begin{equation}
	\eta_{i\to j}(\theta_i) = \frac{1}{\mathcal{Z}_{i\to j}}\prod_{k\in\partial i\setminus j}\int \di\theta_k\,e^{\,\beta J\cos{(\theta_i-\theta_k-\omega_{ik})}}\,\eta_{k\to i}(\theta_k)
	\label{eq:BP_eqs_XY_gg}
\end{equation}
Analogously, at $T=0$ we have:
\begin{equation}
	h_{i\to j}(\theta_i) \cong \sum_{k\in\partial i\setminus j}\max_{\theta_k}{\bigl[h_{k\to i}(\theta_k)+J\cos{(\theta_i-\theta_k-\omega_{ik})}\bigr]}
	\label{eq:BP_eqs_zeroTemp_gg}
\end{equation}
The corresponding linearized versions are then straightforward to be obtained (see~\autoref{app:BPeqs_XYmodel}).

\begin{figure}[t]
	\centering
	\includegraphics[scale=1]{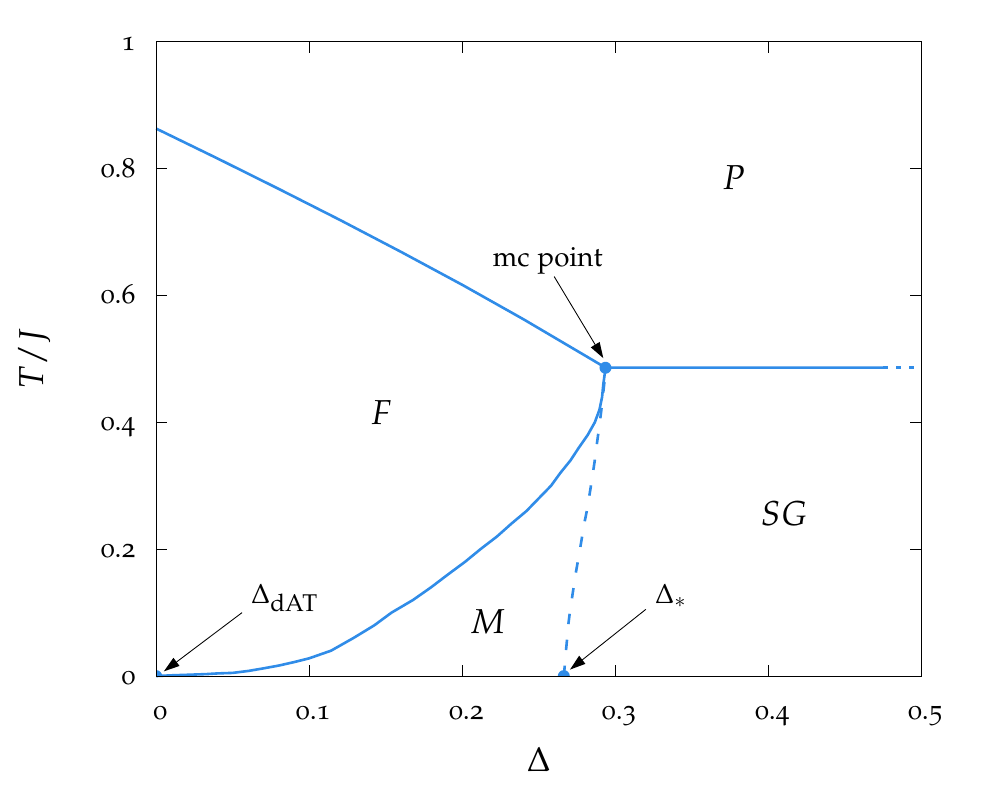}
	\caption[Phase diagram of the gauge glass XY model]{Phase diagram of the XY model on the $C=3$~\acrshort{RRG} ensemble with random angular shifts $\omega_{ij}$'s drawn from the gauge glass disorder distribution $\mathbb{P}_{\omega}$ of Eq.~\autoref{eq:disorder_distribution_gg}.}
	\label{fig:XY_phase_diag_gg}
\end{figure}

By exploiting again~\acrshort{PDA} and~\acrshort{SuscProp}, the~\acrshort{BP} equations~\autoref{eq:BP_eqs_XY_gg} can be numerically solved --- e.\,g. via the $Q=64$-state clock model --- and all the thermodynamic phases of the $T$ vs $\Delta$ plane can be characterized, with the corresponding critical lines between them depicted in~\autoref{fig:XY_phase_diag_gg} for the $C=3$ \acrshort{RRG} ensemble. The corresponding coordinates for some relevant points in the phase diagram are the following:
\begin{gather*}
	\Delta_{mc} = 0.294(1) \qquad , \qquad T_{mc}/J=0.486(1) \qquad , \qquad T_c(\Delta=0)/J=0.863(1)\\
	\Delta_*=0.266(1) \qquad , \qquad \Delta_{\text{dAT}}=0.000(1)
\end{gather*}

Notice that, if properly mapping the parameter $\Delta$ of the gauge glass distribution~\autoref{eq:disorder_distribution_gg} into the parameter $p$ of the bimodal distribution~\autoref{eq:disorder_distribution_pmJ}
\begin{equation}
	p \, \leftrightarrow \, 1-\frac{\Delta}{2}
\end{equation}
then it is easy to see that the two corresponding phase diagrams exactly match --- apart from the~\acrshort{dAT} line between the ferromagnetic and the mixed phases --- so highlighting one of the most remarkable features of spin glass models: a universal behaviour of the instability line of the paramagnetic solution and of the critical line separating the unbiased spin glass phase from the mixed phase (here computed in the \acrshort{RS} ansatz, even though a~\acrshort{RSB} ansatz would be necessary).

At variance, the~\acrshort{dAT} line is dramatically different between the two cases: the~\acrshort{RS} ferromagnetic phase of the gauge glass seems to be more stable when lowering the temperature with respect to the bimodal case when a comparable quantity of disorder is inserted. Furthermore, the critical exponents of the two lines close to the multicritical point are largely different.

The reason of this very different behaviour seems to be hidden in the type of spin symmetry involved in the~\acrshort{RS} instability. Indeed, the strong anisotropy of the bimodal case identifies a preferred direction with respect to which it is possible to break the inversion symmetry of the transverse spin components. Instead, in the gauge glass case, each interaction is shifted of an angle randomly distributed on the unit circle and hence, due to the isotropy in the spin space, no transverse spin symmetry can be globally broken. In fact, in the latter case it could be that the longitudinal\footnote{In absence of any global preferred direction, we refer to the \textit{local} direction identified by the effective field given by the neighbours.} degrees of freedom undergo a freezing, bringing to the~\acrshort{RS} instability.

Some evidences in this sense come from the comparison with the behaviour of the anisotropic fully connected vector model~(\autoref{sec:vector_sg_fully}), where the freezing of the transverse degrees of freedom yields the~\acrshort{GT} line, $(\delta T_c) \sim (\delta J_0)^2$. Then, if further decreasing the temperature, the freezing of the longitudinal degrees of freedom yields a crossover line --- rather than a sharp phase transition --- recognizable as the~\acrshort{dAT} line of the scalar case, $(\delta T_c) \sim (\delta J_0)^{1/2}$. Compatible exponents can be found in the diluted case respectively for the gauge glass and the bimodal XY model:
\begin{subequations}
	\begin{equation}
		\delta T_c(p) \simeq (p-p_{mc})^{\alpha} \quad , \qquad \alpha \simeq 2
	\end{equation}
	\begin{equation}
		\delta T_c(\Delta) \simeq (\Delta_{mc}-\Delta)^{\alpha} \quad , \qquad \alpha \simeq \frac{1}{2}
	\end{equation}
\end{subequations}
even though their precise evaluation is not easily feasible due to the very proximity of the multicritical point.

Notice that, up to our knowledge, we are not aware of any other similar result for the sparse case in the literature. A full characterization of this ``double'' behaviour according to the presence (or not) of a global direction of anisotropy --- and the corresponding interpretation as transverse or longitudinal perturbations, respectively --- will be provided in~\autoref{chap:XYinField}, where the directional bias of the couplings seen here will be substituted by the directional bias of the external field.

\clearpage{\pagestyle{empty}\cleardoublepage}

\chapter{Discretizing the XY model: the clock model}
\label{chap:clock}
\thispagestyle{empty}

The need for a discretization arises every time we want to set up a numerical simulation involving continuous variables, e.\,g. the ubiquitous problem of numerically solving an ordinary differential equation.

If in most cases the na{\"i}ve strategy of dividing intervals in a very large number of bins works quite well, however there can be some nonnegligible issues. What if the features of the system are such that any finite number of intervals provides wrong results? What if the required number of bins is so large to make practically unfeasible numerical simulations? So the success of this approach generally depends on the given problem and on the optimal values for the parameters used (e.\,g. the number of the bins).

In our case, the problem of discretizing the $[0,2\pi)$ interval in order to provide a numerical solution to the~\acrshort{BP} equations for the XY model arises a few fundamental questions: \textit{i)} Does the solution qualitatively change when passing from the continuous to the discrete model? \textit{ii)} How does the error committed in the discretization decay with the number of bins? \textit{iii)} Is there any strong dependence on the temperature and/or the quantity of disorder? \textit{iv)} Does the universality class change?

We will provide the answers to all these questions throughout this Chapter, eventually justifying some statements claimed in~\autoref{chap:XYnoField} about the clock model approximation.

\section{From the XY model to the clock model}
\label{sec:clock_model}

In~\autoref{sec:RS_cavity_method} we already introduced the $Q$-state clock model, since we wanted to numerically solve the \acrshort{BP} equations for the XY model in the low-temperature region. Let us recap its definition: the $[0,2\pi)$ interval is divided into $Q$ equal bins of elementary width $2\pi/Q$. Consequently, each spin can align only along one of these $Q$ allowed directions, labeled by index $a$:
\begin{equation}
	\theta \in [0,2\pi) \qquad \Rightarrow \qquad \theta_a \in \left\{0,\,\frac{1}{Q}\,2\pi,\,\frac{2}{Q}\,2\pi,\,\dots,\,\frac{Q-1}{Q}\,2\pi\right\}
\end{equation}
$Q=2$ case is nothing but the Ising model, and indeed we will exploit this peculiar value in order to test our algorithms. Instead, the opposite limit $Q\to\infty$ allows us to exactly recover the XY model.

As a consequence of this discretization, each integral over the $[0,2\pi)$ interval becomes a sum over the $Q$ states:
\begin{equation}
	\frac{1}{2\pi}\int_0^{2\pi}\di\theta f(\theta) \qquad \Rightarrow \qquad \frac{1}{Q}\sum_{a=0}^{Q-1} f(\theta_a)
\end{equation}
Notice that the ``conversion'' factor $2\pi/Q$ will appear throughout this Chapter, since it allows to exactly recover the XY results in the $Q\to\infty$ limit.

The Hamiltonian~\autoref{eq:Hamiltonian_XY_sparse} of the XY model remains formally unchanged when passing to the $Q$-state clock model:
\begin{equation}
	\mathcal{H}[\{\theta\}]= -\sum_{(i,j)}J_{ij}\cos{(\theta_{i,a}-\theta_{j,b})}
	\label{eq:Hamiltonian_Qclock_sparse}
\end{equation}
with the unique prescription that angles $\theta$'s take on only the $Q$ allowed values. Consequently, also the~\acrshort{BP} equations read formally the same, apart from the necessary substitution of the integrals with the sums. For example, at finite temperature we get:
\begin{equation}
	\eta_{i\to j}(\theta_{i,a}) = \frac{1}{\mathcal{Z}_{i\to j}}\prod_{k\in\partial i\setminus j}\,\sum_{b_k=0}^{Q-1}\,e^{\,\beta J_{ik}\cos{(\theta_{i,a}-\theta_{k,b_{\scriptsize{k}}})}}\,\eta_{k\to i}(\theta_{k,b_k})
	\label{eq:BP_eqs_Qclock}
\end{equation}
with normalization constant $\mathcal{Z}_{i\to j}$ which is now given by:
\begin{equation}
	\mathcal{Z}_{i\to j} = \frac{2\pi}{Q}\sum_{a=0}^{Q-1}\biggl[\,\prod_{k\in\partial i\setminus j}\,\sum_{b_k=0}^{Q-1}\,e^{\,\beta J_{ik}\cos{(\theta_{i,a}-\theta_{k,b_{\scriptsize{k}}})}}\,\eta_{k\to i}(\theta_{k,b_k})\biggr]
\end{equation}
Notice the presence of the $2\pi/Q$ factor in front of $\mathcal{Z}$, which actually ensures a well defined behaviour in the $Q\to\infty$ limit. We will further discuss this normalization factor in the next Section.

\section{RS solution of the BP equations in the bimodal case}
\label{sec:RS_solution_Qclock}

Since the~\acrshort{BP} equations remained formally unchanged through the discretization, we expect that also the features of their solution should remain qualitatively unchanged, at least for large values of $Q$. In this Section we will actually prove this statement.

However, before going on, a caveat about the parity of $Q$ has been remarked. Since we are going to use the bimodal distribution $\mathbb{P}_J$, it implies that a certain link must be fully satisfied in both cases of a ferromagnetic and an antiferromagnetic coupling. If $J_{ij}$ is positive, then the perfect alignment between $\boldsymbol{\sigma}_i$ and $\boldsymbol{\sigma}_j$ always ensures it. But when $J_{ij}$ is negative, hence $\boldsymbol{\sigma}_i$ and $\boldsymbol{\sigma}_j$ have to differ by $\pi$ and this is not possible if we are using an odd $Q$ number of bins. For this reason, in this Section we will use only even values for $Q$.

\subsection{Paramagnetic solution and expansion around it}

Let us start from the analysis of the paramagnetic solution. In absence of any external field, it is the uniform distribution over the $Q$ allowed directions:
\begin{equation}
	\eta_{i\to j}(\theta_{i,a}) = \frac{1}{Q} \quad , \qquad \forall a \, , \, \forall i\to j
	\label{eq:para_sol_Qclock}
\end{equation}
and it obviously satisfies the~\acrshort{BP} equations~\autoref{eq:BP_eqs_Qclock} at any temperature. Notice that we are using normalization of cavity marginals up to unity:
\begin{equation}
	\sum_a \eta_{i\to j}(\theta_{i,a}) = 1 \quad , \qquad \forall i\to j
	\label{eq:Qclock_norm_1}
\end{equation}
as it is usual for models with a finite number of states. Instead, if we want to recover the XY limit, we have to use a different normalization:
\begin{equation}
	\sum_a \eta_{i\to j}(\theta_{i,a}) = \frac{Q}{2\pi} \quad , \qquad \forall i\to j
	\label{eq:Qclock_norm_Qover2pi}
\end{equation}
so that e.\,g. paramagnetic solution reads $1/2\pi$. Depending on what is our task, we will use the former (say the `clock model' or `discrete' one) or the latter (say the `XY' or `continuous' one) throughout this Chapter.

Free energy density and internal energy density in the paramagnetic phase acquire analogous expressions with respect to the XY model ones, Eqs.~\autoref{eq:f_para_XY_RRG} and~\autoref{eq:u_para_XY_RRG}:
\begin{equation}
	f(\beta) = -\frac{1}{\beta}\ln{2\pi} - \frac{C}{2\beta}\ln{I^{(Q)}_0(\beta J)} \qquad , \qquad u(\beta) = -\frac{JC}{2}\frac{I^{(Q)}_1(\beta J)}{I^{(Q)}_0(\beta J)}
\end{equation}
where \textit{discrete} modified Bessel functions of the first kind has been defined:
\begin{equation}
	I^{(Q)}_n(x) \equiv \frac{1}{Q}\sum_{a=0}^{Q-1}\,e^{\,x\cos{(2\pi a / Q)}}\,\cos{\left(n\frac{2\pi a}{Q}\right)} \quad , \qquad n\in\mathbb{Z}
	\label{eq:modif_Bessel_I_Qclock}
\end{equation}
in perfect analogy with their continuous version defined in~\autoref{eq:modif_Bessel_I_XY}, which can be recovered in the $Q\to\infty$ limit:
\begin{equation}
	\lim_{Q\to\infty}{I^{(Q)}_n(x)} = I_n(x)
\end{equation}
Consequently, it is easy to see that in the definition of free energy density, the normalization just affects the entropy term: the clock model normalization~\autoref{eq:Qclock_norm_1} ensures a positive definite entropy for all the temperatures and for any finite value of $Q$, while the XY normalization~\autoref{eq:Qclock_norm_Qover2pi} allows to exactly recover the XY entropy in the $Q\to\infty$ limit even when the latter is negative.

We already know from the analysis performed on the XY model in~\autoref{chap:XYnoField} that the paramagnetic solution is expected to be stable only down to a certain critical temperature $T_c$. So in analogy with analytical computations performed for the XY model, the (discrete) periodicity of $\theta$'s variables suggests to study the stability of paramagnetic solution~\autoref{eq:para_sol_Qclock} by using a Fourier expansion for cavity marginals, which has now to be discrete as well:
\begin{equation}
	\eta_{i\to j}(\theta_{i,a}) = \frac{1}{Q}\sum_{b=0}^{Q-1}\,c^{(i\to j)}_b\,e^{\,2\pi \mathfrak{i} a b / Q}
	\label{eq:discrete_Fourier_expansion_1}
\end{equation}
with Fourier coefficients $c$'s satisfying the following relation:
\begin{equation}
	c^{(i\to j)}_b = \sum_{a=0}^{Q-1}\,\eta_{i\to j}(\theta_{i,a})\,e^{-2\pi \mathfrak{i} a b / Q}
	\label{eq:definition_Fourier_c_coeff}
\end{equation}

Following this definition, the zeroth-order coefficient represents the sum of values taken by each cavity marginal on the $Q$ directions:
\begin{equation}
	c^{(i\to j)}_0 = \sum_{a=0}^{Q-1}\,\eta_{i\to j}(\theta_{i,a})
\end{equation}
namely it is directly related to the normalization of cavity messages. So if we choose the discrete normalization~\autoref{eq:Qclock_norm_1}, then we have:
\begin{equation*}
	c^{(i\to j)}_0 = 1
\end{equation*}
while if we choose the continuous normalization~\autoref{eq:Qclock_norm_Qover2pi} we get:
\begin{equation*}
	c^{(i\to j)}_0 = \frac{Q}{2\pi}
\end{equation*}
In what follows, we will use the continuous normalization, so to rewrite the Fourier expansion~\autoref{eq:discrete_Fourier_expansion_1} in a more useful shape:
\begin{equation}
	\eta_{i\to j}(\theta_{i,a}) = \frac{1}{2\pi}\left[1+\frac{2\pi}{Q}\sum_{b=1}^{Q-1}\,c^{(i\to j)}_b\,e^{\,2\pi \mathfrak{i} a b / Q}\right]
	\label{eq:discrete_Fourier_expansion_2}
\end{equation}

A further remark regards the nature of these $c$'s coefficients. Indeed, since we are using the exponential form instead of the cosine and sine form, they are expected to be complex instead of real. However, due to periodicity of cavity marginals over the $[0,2\pi)$ interval, it holds:
\begin{equation*}
	c^{(i\to j)}_l = \left(c^{(i\to j)}_{Q-l}\right)^*
\end{equation*}
and, when using even values for $Q$ as in this case, this implies that $c_{Q/2}$ is real.

At this point, we can proceed exactly as done for the XY model, expanding the right hand side of~\autoref{eq:BP_eqs_Qclock} as in~\autoref{eq:discrete_Fourier_expansion_2} and then substituting it into the right hand side of~\autoref{eq:definition_Fourier_c_coeff}, so getting a set of self-consistency equations for the discrete Fourier coefficients $c$'s:
\begin{equation}
	c^{(i\to j)}_l = \frac{1}{\mathcal{Z}_{i\to j}}\sum_{a=0}^{Q-1}\,e^{-2\pi \mathfrak{i} a l / Q}\prod_{k\in\partial i\setminus j}\,\sum_{b_k=0}^{Q-1}\,c^{(k\to i)}_{b_k}I^{(Q)}_{b_k}(\beta J_{ik})\,e^{\,2\pi \mathfrak{i} a b_k / Q}
	\label{eq:Fourier_coeff_Qclock_selfcons}
\end{equation}
where also $\mathcal{Z}_{i\to j}$ has to be expanded in terms of $c$'s.

A perturbative expansion of the right hand side can be performed as for the XY model, obtaining at the linear order in each coefficient:
\begin{equation}
	c^{(i\to j)}_l = \sum_{k\in\partial i\setminus j}\frac{I^{(Q)}_l(\beta J_{ik})}{I^{(Q)}_0(\beta J_{ik})}\,c^{(k\to i)}_l
	\label{eq:cl_linear}
\end{equation}
in perfect analogy with the result obtained for the XY model, Eq.~\autoref{eq:al_and_bl_linear}. Then, when averaging over the bimodal disorder distribution $\mathbb{P}_J$ and over the realization of the $C$-\acrshort{RRG}, from the first and the second moment of $c$'s coefficients we get the marginality conditions between the paramagnetic and the low-temperature solutions:
\begin{equation}
	(C-1)(2p-1)\frac{I^{(Q)}_1\bigl(\beta^{(Q)}_{\text{F}} J\bigr)}{I^{(Q)}_0\bigl(\beta^{(Q)}_{\text{F}} J\bigr)}=1 \qquad , \qquad (C-1)\left[\frac{I^{(Q)}_1\bigl(\beta^{(Q)}_{\text{SG}} J\bigr)}{I^{(Q)}_0\bigl(\beta^{(Q)}_{\text{SG}} J\bigr)}\right]^2=1
	\label{eq:Qclock_para_stability_cRRG}
\end{equation}
again formally equal to the ones for the XY model. The critical inverse temperature away from the paramagnetic phase as a function of the fraction $p$ of positive couplings is finally given by:
\begin{equation}
	\beta^{(Q)}_c(p) \equiv \min{\left\{\beta^{(Q)}_{\text{F}}(p),\beta^{(Q)}_{\text{SG}}\right\}}
	\label{eq:beta_c_bimodal_Qclock}
\end{equation}

Not so surprisingly, the features of the phase transition away from the paramagnetic phase are the same found for the XY model, i.\,e. a transition toward a ferromagnetic phase for values of $p$ close to $1$ and a transition toward a spin glass phase for values of $p$ close to $1/2$. Furthermore, the expression of the corresponding critical line is formally the same, and even the abscissa of multicritical point coincides with the one found for the XY model, Eq.~\autoref{eq:mc_point_XY}, since it does not depend on~$Q$:
\begin{equation}
	p_{mc} = \frac{1+(C-1)^{-1/2}}{2}
\end{equation}
What actually changes when varying $Q$ is the height of this critical line, which in turn depends on the discretized Bessel functions $I^{(Q)}_n$'s. So in order to understand how fast this line approaches the XY model one in the large-$Q$ limit, we firstly have to study the rate of convergence of the discretized Bessel functions toward the continuous ones.

\subsection{Convergence of the discretized Bessel functions}
\label{subsec:conv_discr_Bessel}

\begin{figure}[t]
	\centering
	\includegraphics[scale=1]{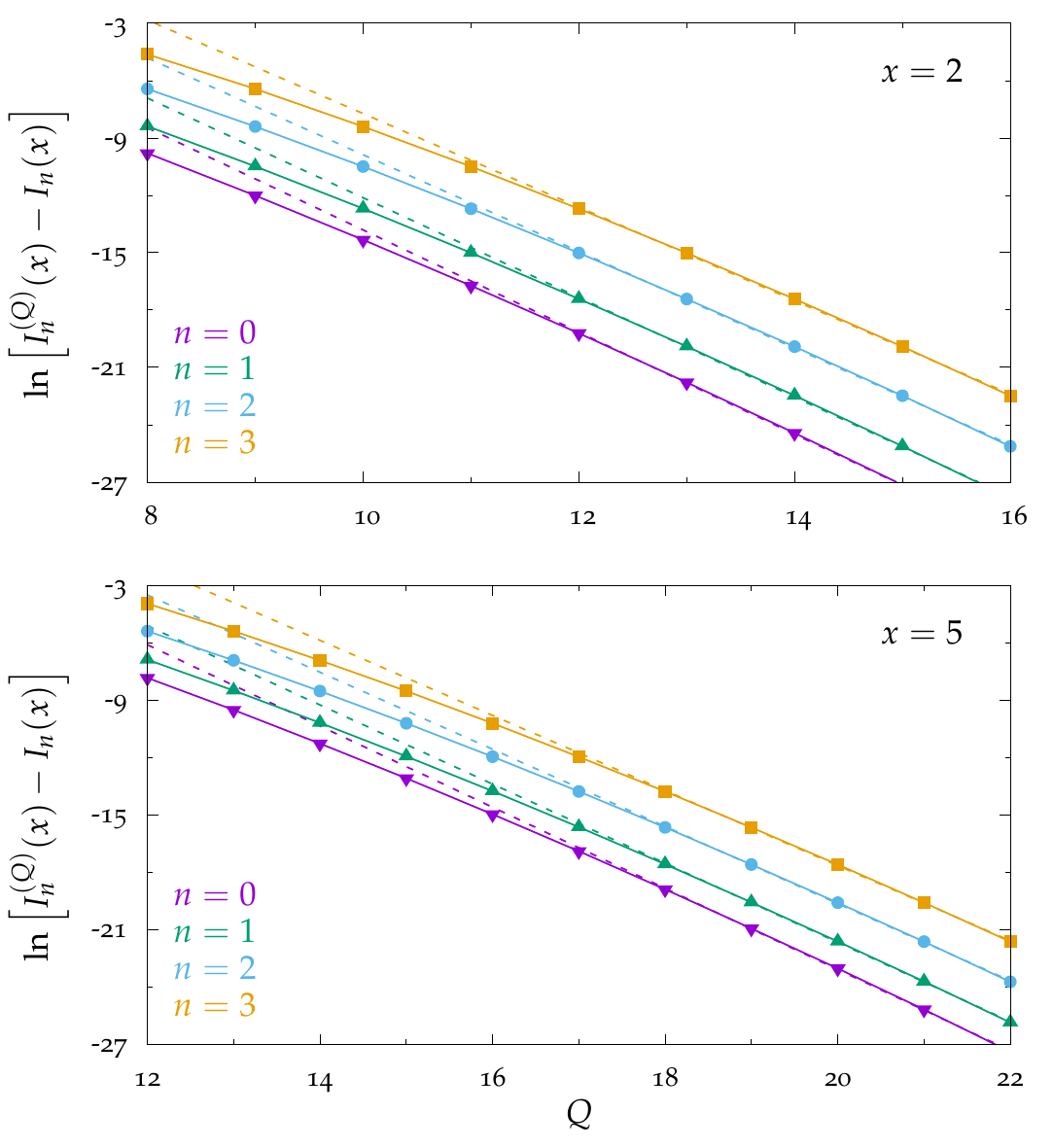}
	\caption[Convergence of discretized Bessel functions]{Convergence of the discretized Bessel functions $I^{(Q)}_n(x)$ toward their $Q\to\infty$ value. The convergence is exponential, apart from an initial transient, as it can be seen from the linear behaviour of the logarithm of the difference $I^{(Q)}_n(x)-I_n(x)$ for two different values of their argument $x$.}
	\label{fig:Bessel}
\end{figure}

A first numerical study shows an exponential convergence of discretized Bessel functions toward the continuous ones:
\begin{equation}
	I^{(Q)}_n(x) - I_n(x) \sim A\exp{(-Q/Q^*)}
\end{equation}
as shown in~\autoref{fig:Bessel}, with a characteristic scale $Q^*$ increasing with the argument~$x$ of the Bessel functions --- namely with the inverse temperature~$\beta$ in our case:
\begin{equation*}
\begin{split}
	&Q^*(x=2) \simeq 2.0\\
	&Q^*(x=5) \simeq 2.5
\end{split}
\end{equation*}
Since Bessel functions are explicitly contained in both physical observables and marginality conditions, an exponential convergence of the discretized Bessel functions toward their continuous limit would provide strong evidences in favour of an exponential convergence in physical observables and in critical lines as well, when considering the large-$Q$ limit of the clock model toward the XY model.

Even though we have no fully analytic proof of this exponential convergence for the Bessel functions, we can produce an argument which should convince the reader that a power-law decay in $Q$ can not take place whenever the integral of a periodic function is approximated with a discrete sum of $Q$ terms. Let us take a $2\pi$-periodic and infinitely differentiable function $f(x)$ and let us focus on the approximation of its integral over the $[0,2\pi)$ interval:
\begin{equation}
	I(f) \equiv \frac{1}{2\pi}\int_0^{2\pi}\di x\,f(x)
\end{equation}
with the following sum:
\begin{equation}
	I^{(Q)}(f) \equiv \frac{1}{Q}\sum_{a=0}^{Q-1}\,f(2\pi a / Q)
\end{equation}
This sum can be rewritten as an integral of a stepwise function:
\begin{equation}
	I^{(Q)}(f) = \frac{1}{Q}\sum_{a=0}^{Q-1}\,\left[\frac{Q}{2\pi}\int_{\Gamma_a}\di x\,f(2\pi a / Q)\right] = \frac{1}{2\pi}\sum_{a=0}^{Q-1}\,\int_{\Gamma_a}\di x\,f(2\pi a / Q)
\end{equation}
where $\Gamma_a$ is the $a$-th subinterval of size $2\pi/Q$ centered around $2\pi a / Q$ in which the~$[0,2\pi)$ interval has been divided:
\begin{equation*}
	\Gamma_a \equiv \biggl[\frac{2\pi}{Q}\Bigl(a-\frac{1}{2}\Bigr),\frac{2\pi}{Q}\Bigl(a+\frac{1}{2}\Bigr)\biggr)
\end{equation*}

Hence, the error $\Delta^{(Q)}$ committed when approximating the integral with the sum of~$Q$ terms reads:
\begin{equation}
	\Delta^{(Q)}(f) = \frac{1}{2\pi}\sum_{a=0}^{Q-1}\,\int_{\Gamma_a}\di x\,\bigl[f(x)-f(2\pi a / Q)\bigr]
	\label{eq:error_DeltaQ_1}
\end{equation}
When $Q$ is large, a Taylor expansion of $f(x)$ around the central point of each $\Gamma_a$ interval gives:
\begin{equation}
	f(x) = f(2\pi a / Q) + \sum_{k=1}^{\infty} f^{(k)}(2\pi a / Q)\,\frac{(x-2\pi a / Q)^k}{k!}
\end{equation}
where $f^{(k)}$ is the $k$-th derivative of $f$. Substituting this expansion into the right hand side of~\autoref{eq:error_DeltaQ_1} and then integrating, we get:
\begin{equation}
	\Delta^{(Q)}(f) = \sum_{\substack{\text{$k$ even}\\k>0}}\frac{\pi^k}{Q^{k}\,(k+1)!}\,\frac{1}{Q}\sum_{a=0}^{Q-1}\,f^{(k)}(2\pi a / Q)
	\label{eq:error_DeltaQ_2}
\end{equation}
At this point, also the internal sum over the $k$-th derivative of $f$ can be substituted by the corresponding integral in the $Q\to\infty$ limit, plus the correction term:
\begin{equation}
\begin{split}
	\frac{1}{Q}\sum_{a=0}^{Q-1}\,f^{(k)}(2\pi a / Q) &= \frac{1}{2\pi}\int_0^{2\pi}\di x\,f^{(k)}(x) + \Delta^{(Q)}(f^{(k)})\\
	&= \frac{f^{(k-1)}(2\pi)-f^{(k-1)}(0)}{2\pi} + \Delta^{(Q)}(f^{(k)})\\
	&= \Delta^{(Q)}(f^{(k)})
\end{split}
\end{equation}
due to the $2\pi$-periodicity of $f$ and of its derivatives. Going back to Eq.~\autoref{eq:error_DeltaQ_2}, we finally have:
\begin{equation}
	\Delta^{(Q)}(f) = \sum_{\substack{\text{$k$ even}\\k>0}}\frac{\pi^k}{Q^{k}\,(k+1)!}\,\Delta^{(Q)}(f^{(k)})
	\label{eq:error_DeltaQ_3}
\end{equation}
so relating the error committed in the approximation of the integral of $f$ with the error in the approximation of the integral of its derivatives.

For a function $f$ smooth enough --- as $\exp{(x\cos{\theta})}\cos{\theta}$ in modified Bessel functions $I_n$'s --- the error $\Delta^{(Q)}(f^{(k)})$ on derivatives is expected to decay in the same manner as the error $\Delta^{(Q)}(f)$ on function itself. Indeed, $I_1$ is the derivative of~$I_0$ and in~\autoref{fig:Bessel} it is quite clear that their errors decay both in the same way with $Q$. Since a power-law decay
\begin{equation*}
	\Delta^{(Q)}(f) \sim \Delta^{(Q)}(f^{(k)}) \sim Q^{-\alpha}
\end{equation*}
is not compatible with relation~\autoref{eq:error_DeltaQ_3} for any value of $\alpha$, then we can conclude that these errors should decay faster than any power law.

As stated at the beginning of this reasoning, this is not a proof of a clear exponential convergence, since for example it could be the case of a stretched-exponential convergence. In fact, in~\autoref{sec:conv_phys_obs} we will actually see an exponential convergence of physical observables as long as $T>0$, which then turns into a stretched exponential convergence when~$T=0$.

\subsection{Low- and zero-temperature solutions}

Below the critical temperature signaling the paramagnetic instability, the clock model \acrshort{BP} equations~\autoref{eq:BP_eqs_Qclock} have to be solved numerically, as done for the XY model.

For finite temperatures, we can exploit exactly the same~\acrshort{PDA} outlined in the pseudocode~\ref{alg:RS_PDA}, now choosing any even value for $Q$ (while we chose $Q=64$ for the analysis of the XY model). Furthermore, also~\acrshort{SuscProp} can be used for the detection of the~\acrshort{dAT} line just as explained in the pseudocode~\ref{alg:RS_PDA_SuscProp}, since the mechanism of the propagation of perturbations is the same: their global norm grows or decreases according to the \acrshort{RS} stability of the \acrshort{BP} fixed point.

In the zero-temperature limit we have to be more careful, as anticipated in~\autoref{sec:XY_zeroTemp}. Indeed, full~\acrshort{BP} equations at zero temperature for the $Q$-state clock model
\begin{equation}
	h_{i\to j}(\theta_{i,a}) \cong \sum_{k\in\partial i\setminus j}\max_{b}{\bigl[h_{k\to i}(\theta_{k,b})+J_{ik}\cos{(\theta_{i,a}-\theta_{k,b})}\bigr]}
	\label{eq:BP_eqs_zeroTemp_Qclock}
\end{equation}
can be solved without any particular difference with respect to the continuous case, showing a smooth convergence of the solution in the large-$Q$ limit. Instead, their linearized version:
\begin{subequations}
	\begin{equation}
		\delta h_{i\to j}(\theta_{i,a}) \cong \sum_{k\in\partial i\setminus j}\delta h_{k\to i}(\theta^*_{k,b}(\theta_{i,a}))
		\label{eq:BP_eqs_zeroTemp_linear_Qclock}
	\end{equation}
	\begin{equation}
		\theta^*_{k,b}(\theta_{i,a}) = \argmax_{b}{\bigl[h_{k\to i}(\theta_{k,b})+J_{ik}\cos{(\theta_{i,a}-\theta_{k,b})}\bigr]}
		\label{eq:BP_eqs_zeroTemp_linear_Qclock_thetaStar}
	\end{equation}
	\label{eq:BP_eqs_zeroTemp_linear_Qclock_ALL}%
\end{subequations}
has to be solved in a different manner. Indeed, the maximum in the right hand side of~\autoref{eq:BP_eqs_zeroTemp_Qclock} can take on different values if evaluated on the $Q$ states of the clock model or on the whole unit circle, and hence also the corresponding argmax of Eq.~\autoref{eq:BP_eqs_zeroTemp_linear_Qclock_thetaStar} has to change consequently. So, as already anticipated in~\autoref{sec:XY_zeroTemp}, it may happen that for all $\theta_{i,a}$'s directions the argmax is given by the same angle $\theta^*_{k,b}$, so yielding a constant incoming perturbation~$\delta h_{k\to i}$ in the right hand side of~\autoref{eq:BP_eqs_zeroTemp_linear_Qclock}. After normalization, it actually becomes null, and if this occurs simultaneously for all the neighbours of $i$ but $j$, then also the outgoing perturbation $\delta h_{i\to j}$ is constant and hence identically null as well.

This phenomenon is the more probable, the smaller $Q$, but anyway it has a finite probability to occur as long as $Q$ is finite. Indeed, it can be observed in any model with discrete variables (Ising~\cite{CastellaniEtAl2005, MoroneEtAl2014}, Potts~\cite{KrzakalaZdeborova2008}, colouring~\cite{ZdeborovaKrzakala2007, Thesis_Zdeborova2009}, and so on). So even if the initialiation of perturbations in the $T=0$ \acrshort{SuscProp} is such that none of them is null at $t=0$, at each time step new identically vanishing perturbations appear in the population and so, due to a cascade effect, in the long-time limit all the perturbations become vanishing as well.

\begin{figure}[!t]
	\centering
	\includegraphics[scale=1]{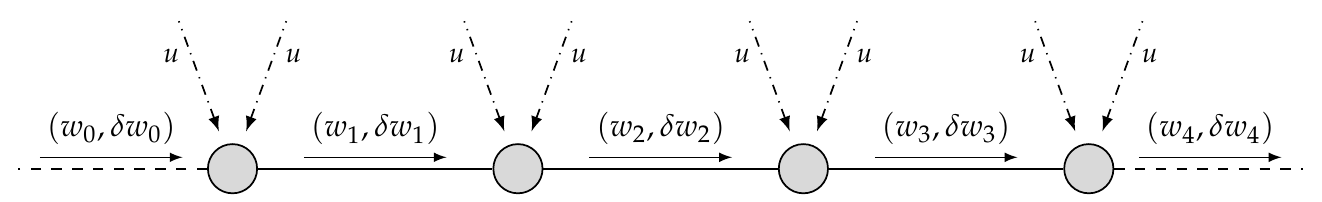}
	\caption[Susceptibility Propagation along a chain at $T=0$]{Sketch of the propagation of cavity perturbations along a chain at $T=0$. After each node $i$, the ``new'' cavity bias $w_{i+1}$ takes the contribution from the ``old'' one $w_i$ along the chain and from the $d_i-2$ other biases $u$'s with identically null perturbations from the outside of the chain, with $d_i$ drawn from the degree distribution~$\mathbb{P}_d$. Instead, the new perturbation $\delta w_{i+1}$ is due uniquely to the old one $\delta w_i$ along the chain, evaluated as in Eq.~\autoref{eq:BP_eqs_zeroTemp_linear_Qclock_u}. The survival probability of these perturbations is directly linked to the \acrshort{RS} stability of the \acrshort{BP} fixed point. Further details are provided in the main text.}
	\label{fig:BP_chain}
\end{figure}

Even though this could seem a serious issue for the evaluation of the \acrshort{RS} stability of the~\acrshort{BP} fixed point as outlined in~\autoref{chap:XYnoField}, namely by looking at the global growth rate of perturbations, in fact it is actually just the decrease of the fraction of nonvanishing perturbations at each time step that provides a measurement of the \acrshort{RS} stability. Indeed, we can generalize the stochastic approach of~\acrshort{PDA} and~\acrshort{SuscProp} from a generic sparse random graph to a chain, so that we can focus on the survival probability of a nonvanishing perturbation propagating along such chain (\autoref{fig:BP_chain} and Ref.~\cite{MoroneEtAl2014}). Let us refer to it as the~\acrfull{ChainSuscProp} approach.

First of all, it is more convenient to focus on the message that enters in each node, namely the cavity bias $u$, and no longer on the message that leaves it, namely the cavity field $h$. Remembering their definition within the factor graph formalism (\autoref{sec:fact_graph_form}) and the corresponding \acrshort{BP} equations for the XY model (\autoref{app:BPeqs_XYmodel}), it is easy to rewrite the pairwise zero-temperature \acrshort{BP} equations~\autoref{eq:BP_eqs_zeroTemp_Qclock} for the $u$'s:
\begin{equation}
	u_{i\to j}(\theta_{j,b}) \cong \max_{a}{\Biggl[\sum_{k\in\partial i\setminus j}u_{k\to i}(\theta_{i,a})+J_{ij}\cos{(\theta_{i,a}-\theta_{j,b})}\Biggr]}
	\label{eq:BP_eqs_zeroTemp_Qclock_u}
\end{equation}
as well as their linearized version:
\begin{equation}
		\delta u_{i\to j}(\theta_{j,b}) = \sum_{k\in\partial i\setminus j}\delta u_{k\to i}(\theta^*_{i,a}(\theta_{j,b})) - \sum_{k\in\partial i\setminus j}\delta u_{k\to i}(\theta^*_{j,b})
	\label{eq:BP_eqs_zeroTemp_linear_Qclock_u}
\end{equation}
where:
\begin{subequations}
	\begin{equation}
		\theta^*_{i,a}(\theta_{j,b}) \equiv \argmax_{a}{\Biggl[\sum_{k\in\partial i\setminus j}u_{k\to i}(\theta_{i,a})+J_{ij}\cos{(\theta_{i,a}-\theta_{j,b})}\Biggr]}
	\end{equation}
	\begin{equation}
		\theta^*_{j,b} \equiv \argmax_{b}{\Biggl[\sum_{k\in\partial i\setminus j}u_{k\to i}(\theta^*_{i,a}(\theta_{j,b}))+J_{ij}\cos{(\theta^*_{i,a}(\theta_{j,b})-\theta_{j,b})}\Biggr]}
	\end{equation}
	\label{eq:BP_eqs_zeroTemp_linear_Qclock_thetastar_u}%
\end{subequations}
Notice also in this case the normalization of $\delta u$'s such that they vanish in correspondence of the maximum of the related (negative semidefinite) cavity bias.

Once reached the fixed-point probability distribution $\mathbb{P}^*_u$ --- e.\,g. for a \acrshort{RRG} with connectivity $C$ --- by implementing the zero-temperature \acrshort{PDA}, we can as usual add some small perturbations $\delta u$'s. However, we can no longer let them evolve directly on the graph, since all them will eventually become identically null. Rather, we would actually like to evaluate the rate at which they vanish. In order to achieve such goal, the~\acrshort{ChainSuscProp} algorithm prescribes to let $u$'s propagate along a chain rather than on a proper sparse graph. The population of the \textit{chain cavity biases}, being in principle different from the $u$'s propagating on the graph, will be then referred to as $w$'s.

So at this point we have two sets of cavity biases: the $u$'s, propagating on the graph and having identically vanishing perturbations, and the $w$'s, propagating on the chain and endowed with nonvanishing perturbations. Notice that the latter population can be initialized by directly sampling from $\mathbb{P}^*_u$. Then, at each time step~$t$, we pick at random an incoming cavity bias $w$ with its perturbation $\delta w$ and $d_i-2$ incoming cavity biases $u$'s from the ``sides'' of the chain --- where $d_i$ is drawn from $\mathbb{P}_d$ as usual --- and then we compute the outgoing couple $(w,\delta w)$ as:
\begin{subequations}
	\begin{equation}
		w_{i\to j}(\theta_{j,b}) \cong \max_{a}{\Biggl[w_{k\to i}(\theta_{i,a})+\sum_{k'=1}^{d_i-2}u_{k'\to i}(\theta_{i,a})+J_{ij}\cos{(\theta_{i,a}-\theta_{j,b})}\Biggr]}
		\label{eq:BP_eqs_zeroTemp_Qclock_w_bias}
	\end{equation}
	\begin{equation}
		\delta w_{i\to j}(\theta_{j,b}) \cong \delta w_{k\to i}(\theta^*_{i,a}(\theta_{j,b}))
		\label{eq:BP_eqs_zeroTemp_linear_Qclock_w_pert}
	\end{equation}
	\label{eq:BP_eqs_zeroTemp_linear_Qclock_w}%
\end{subequations}
taking also into account the proper normalizations. At the same time, it is useful to ``refresh'' the population of $u$'s satisfying the proper \acrshort{BP} equations~\autoref{eq:BP_eqs_zeroTemp_Qclock_u} on the graph. Only if the outgoing $\delta w$ is actually nonvanishing as well as the incoming one, then we can store both the message and the perturbation in the population of~$w$'s; otherwise, we discard it. The procedure is also illustrated in~\autoref{fig:BP_chain}.

Finally, since we are interested in the survival probability of the $\{w\}$ population, we can compute the number $\mathcal{N}^{(t)}_{tr}$ of tries needed to completely update the population of $\mathcal{N}$~$w$'s at the $t$-th time step. In the long-time limit, then, the probability $\mathcal{P}$ of generating a nonvanishing $\delta w$ starting from another nonvanishing $\delta w$ can be computed as:
\begin{equation}
	\mathcal{P} \equiv \lim_{t\to\infty}\mathcal{P}^{(t)} \qquad , \qquad \mathcal{P}^{(t)} \equiv \frac{\mathcal{N}}{\mathcal{N}^{(t)}_{tr}}
\end{equation}
In order to come back from the chain to the graph, we multiply $\mathcal{P}$ by the inverse of the \textit{branching ratio} of the graph, actually estimating the rate $\mathfrak{r}$ at which nonvanishing perturbations survive on the graph:
\begin{equation}
	\mathfrak{r} \equiv
	\left\{
	\begin{aligned}
	&(C-1)\,\mathcal{P} \qquad &&\text{for $C$-\acrshort{RRG}}\\
	&C\mathcal{P} \qquad &&\text{for $C$-\acrshort{ERG}}\\
	\end{aligned}
	\right.
	\label{eq:fracNonNullPert}
\end{equation}
Finally, given the (numerically verified) remark that the norm of each $\delta w$ either collapses to zero as shown above or stays quite constant, we can define the proper zero-temperature stability parameter of the~\acrshort{BP} fixed point also for discrete models:
\begin{equation}
	\lambda_{\text{BP}} \equiv \ln{\mathfrak{r}}
\end{equation}
where the~\acrshort{dAT} line corresponds again to $\lambda_{\text{BP}}=0$, as usual. In the pseudocode~\ref{alg:RS_CPDA_SuscProp_zeroTemp} we list the key steps of the~\acrshort{ChainSuscProp} algorithm.

\begin{algorithm}[t]
\caption{Chain Susceptibility Propagation in the~\acrshort{PDA} ($T=0$)}
\label{alg:RS_CPDA_SuscProp_zeroTemp}
\begin{algorithmic}[1]
\State Reach the fixed point $\mathbb{P}^*_u$
\For {$i=1,\dots,\mathcal{N}$}
	\State Initialize $w^{(0)}_i$ \Comment{We sample from $\mathbb{P}^*_u$}
	\State Initialize $\delta w^{(0)}_i$ \Comment{We use a random initialization}
\EndFor
\For {$t=1,\dots,t_{max}$}
	\For {$i=1,\dots,\mathcal{N}$}
		\State $u^{(t)} \gets \mathcal{F}_0[\{u^{(t-1)}_k\},J]$ \Comment{Just a ``refresh'' of the population}
		\State $\mathcal{N}^{(t)}_{tr}=0$
		\Repeat
			\State Draw an integer $d_i$ from the degree distribution $\mathbb{P}_d$
			\State Draw a $w^{(t-1)}$ from $w$'s population
			\State Draw $d_i-2$ $\{u^{(t-1)}_k\}$ from $h$'s population
			\State Draw a $J$ from the coupling distribution $\mathbb{P}_J$
			\State $\widetilde{w}^{(t)} \gets \mathcal{F}_0[w^{(t-1)},\{u^{(t-1)}_k\},J]$ \Comment{Need to thermalize $w$'s}
			\State $\delta\widetilde{w}^{(t)} \gets \mathcal{F}'_0[(w^{(t-1)},\delta w^{(t-1)}),\{u^{(t-1)}_{k}\},J]$
			\If{$\norm{\delta\widetilde{w}^{(t)}}>0$}
				\State Put $(\widetilde{w}^{(t)},\delta\widetilde{w}^{(t)})$ into $w$'s population
			\EndIf
			\State $\mathcal{N}^{(t)}_{tr} \gets \mathcal{N}^{(t)}_{tr}+1$
		\Until{$\norm{\delta\widetilde{w}^{(t)}}>0$}
	\EndFor
	\State $\mathcal{P}^{(t)} \gets \mathcal{N}/\mathcal{N}^{(t)}_{tr}$
	\State Compute $\mathfrak{r}^{(t)}$ as in Eq.~\autoref{eq:fracNonNullPert}
	\State $\lambda^{(t)}_{\text{BP}} \gets \ln{\mathfrak{r}^{(t)}}$
\EndFor
\State Average $\lambda^{(t)}_{\text{BP}}$ over the $t_{max}$ iterations \Comment{Pay attention to thermalization}
\State \textbf{return} $\lambda_{\text{BP}}$
\end{algorithmic}
\end{algorithm}

At this point we can make a brief recap of the effects introduced by the discretization. As long as $T$ stays finite, no qualitative deviations with respect to the continuous model are introduced, so the same algorithms can be used and we are also confident that the values of $Q$ required for a reliable approximation should not be too large. When $T$ approaches zero, dependence on $Q$ is still smooth for what regards the cavity messages, but not their perturbations. The basic mechanism of their growth is drastically different for the two models: in the continuous one, they are all different from zero and it is their global norm that shrinks if the fixed point is stable and grows up if unstable; instead, in the discrete model there is an extensive fraction of identically null perturbations, and it is the decay rate of the remaining fraction of nonvanishing perturbations that rules the stability of the fixed point, while their global norm stays quite constant.

As a preliminar test of this algorithm, we set $Q=2$ --- namely the Ising model --- in order to recover the values of $p$ on the $T=0$ axis at which: \textit{i)} global magnetization $m$ goes to zero, and \textit{ii)} \acrshort{RSB} occurs when coming from higher values of $p$, i.\,e. the $p_*$ and $p_{\text{dAT}}$ values already defined for the XY model, respectively. For both them, we find a high agreement with the corresponding values in the \acrshort{RS} ansatz already known from the literature~\cite{CastellaniEtAl2005}.

\section{Phase diagrams of the clock model}
\label{sec:Qclock_phase_diagrams}

At this point, for any even value of $Q$ it is possible to solve the~\acrshort{BP} equations for the clock model with bimodal couplings $\pm J$ and then find all the critical lines between the different phases. From the literature and from~\autoref{chap:XYnoField}, we expect the presence of four phases: the paramagnetic one in the high-temperature region, and the ferromagnetic, the mixed and the unbiased spin glass ones in the low-temperature region. Indeed, we already know that they are present in the two opposite limits $Q=2$ and $Q\to\infty$, so it is reasonable to expect their presence also for intermediate values of~$Q$.

\begin{figure}[!t]
	\centering
	\includegraphics[scale=1]{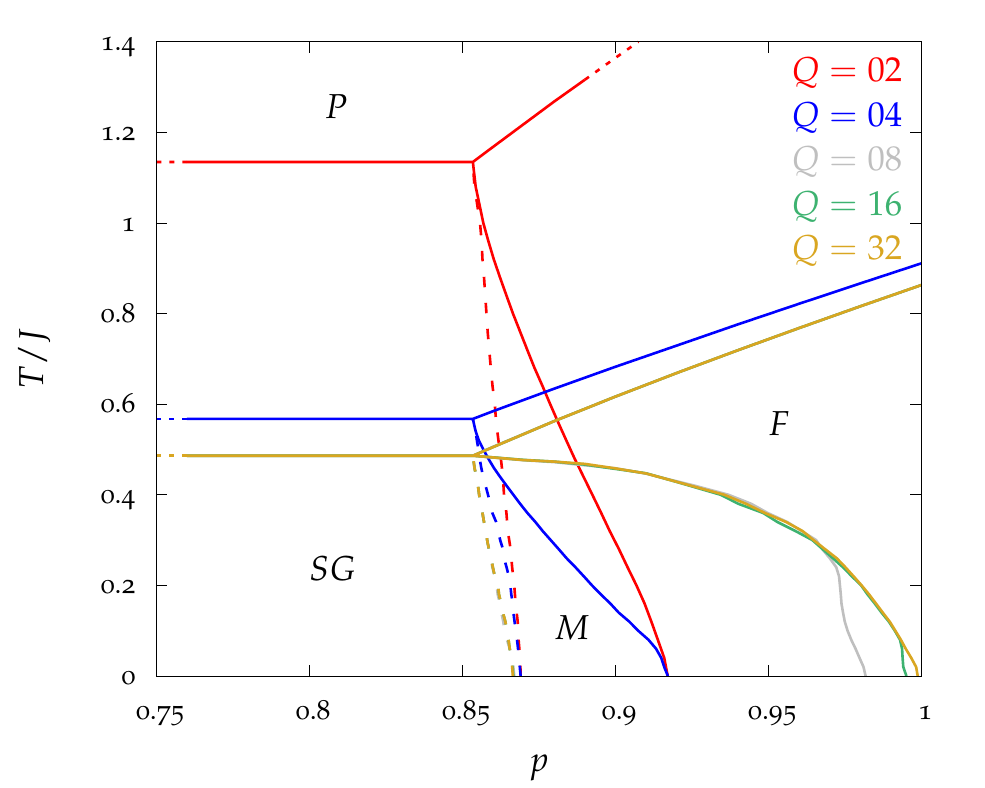}
	\caption[Clock model phase diagram with the bimodal disorder distribution]{Phase diagram of the $Q$-state clock model on the $C=3$~\acrshort{RRG} ensemble with couplings drawn from the bimodal disorder distribution $\mathbb{P}_J$ of Eq.~\autoref{eq:disorder_distribution_pmJ}, for several values of $Q$. The value $Q=32$ does no longer show appreciable differences with respect to the XY limit.}
	\label{fig:Qclock_phase_diag_pmJ}
\end{figure}

In~\autoref{fig:Qclock_phase_diag_pmJ} we report the $T$ vs $p$ phase diagram for $Q=2,4,8,16,32$. The most striking feature is the very fast convergence in~$Q$ for most of critical lines, since for $Q=8$ they are already quite indistinguishable, with the arrangement of the different phases that remains unchanged when increasing $Q$. This is a direct consequence of the exponential convergence of the discretized Bessel function, as seen in~\autoref{subsec:conv_discr_Bessel}. Furthermore, as already pointed out in~\autoref{chap:XYnoField}, the abscissa $p_{mc}$ of the multicritical point actually turns out to be independent from $Q$.

The unique region where a stronger dependence of critical lines from $Q$ is visible is the one in the right lower corner, namely the region of low temperatures and weak disorder. In particular, it is the~\acrshort{dAT} line that moves toward larger values of~$p$ when increasing $Q$, recovering in the XY limit the value $p_{\text{dAT}}=1$ on the $T=0$ axis. The reason is as simple as profound: when the temperature is low and the system is close to a pure ferromagnet, the perfect alignment of spins is fundamental. So if $Q$ is large enough, a slightly misalignment can be activated by a small energy cost, proportional to the cosine of the elementary angle $2\pi/Q$ of the clock model. The large number of resulting configurations corresponds to several states appearing in the Gibbs measure, in turn favouring the breaking of replica symmetry for smaller quantities of disorder. At variance, if $Q$ is rather small, for a low enough temperature no fluctuations are allowed and hence the system arranges in a single configuration as in the case of a pure ferromagnet, even though it is not. \acrshort{RSB} is hence realized at larger quantities of disorder. So it is the value of~$Q$ that \textit{activates} some \textit{soft modes} or does not --- depending on the temperature and on the quantity of disorder --- radically changing the behaviour of the system. We will come back to this point in~\autoref{chap:XYinField_zeroTemp}, where we will deeply study the zero-temperature physics of the spin glass XY model. So even in correspondence of the same values of temperature~$T$ and ferromagnetic bias~$p$, two values of $Q$ rather different can give rise to very different \acrshort{BP} fixed points, even lying on opposite sides of the~\acrshort{dAT} line.

An analogous behaviour regarding the motion in $Q$ of critical lines can be found by analyzing the $Q$-state clock model on a $d=3$ cubic lattice by means of the Migdal - Kadanoff approximated renormalization group~\cite{IlkerBerker2013}. Also in this case the convergence of paramagnetic - ferromagnetic transition line is very fast in $Q$, while a stronger dependence in $Q$ is observed in the convergence of ferromagnetic - spin glass transition line. Furthermore, this latter critical line moves toward larger values of the ferromagnetic bias, again in analogy with our results. However, unlike our case, the critical line between paramagnetic and spin glass phases approaches the $T=0$ axis in the $Q\to\infty$, so making the spin glass phase to disappear in such limit.

Furthermore, notice the change of concavity of the \acrshort{dAT} line close to the multicritical point when increasing $Q$, moving from the Ising-like mechanism of~\acrshort{RS} instability to the vector one discussed at the end of~\autoref{sec:gauge_glass_XY} when in presence of a strong directional anisotropy.

Another important feature, already known from the literature~\cite{NobreSherrington1986}, is the perfect superposition of the critical lines for $Q=2$ and $Q=4$ when in the latter case the strength $J$ of the couplings is properly rescaled by a factor $2$, so doubling the corresponding critical temperatures. Indeed, since nearest directions are orthogonal in the $Q=4$ clock model, it actually behaves as two independent Ising models, with a halved effective coupling strength and a corresponding doubled critical temperature.

Also the continuous gauge glass model, introduced in~\autoref{chap:XYnoField}, can be translated into the clock model version. Actually, it is just how we solved it in~\autoref{sec:gauge_glass_XY}, by setting $Q=64$. For a generic value of $Q$, then, the Hamiltonian remains formally the same:
\begin{equation}
	\mcH[\{\theta\}]= -J\sum_{(i,j)}\cos{(\theta_{i,a}-\theta_{j,b}-\omega_{ij})}
	\label{eq:Hamiltonian_Qclock_sparse_gg}
\end{equation}
with the angular shifts $\omega_{ij}$'s that have now to be chosen as integer multiples of the elementary angle $2\pi/Q$:
\begin{equation}
	\mathbb{P}_{\omega}(\omega_{ij}) = (1-\Delta)\,\delta(\omega_{ij})+\frac{\Delta}{Q}\sum_{a=0}^{Q-1}\delta\Bigl(\omega_{ij}-\frac{2\pi a}{Q}\Bigr)
	\label{eq:disorder_distribution_gg_Qclock}
\end{equation}
and no longer uniformly on the $[0,2\pi)$ interval. Notice that in this case also odd values of $Q$ can be used, since there always exists a suitable arrangement of angles $\theta_{i,a}$ and $\theta_{j,b}$ that perfectly satisfies the interaction $-J\cos{(\theta_{i,a}-\theta_{j,b}-\omega_{ij})}$ for any value of the angular shift.

The corresponding \acrshort{BP} equations have exactly the same expression of the continuous case, Eq.~\autoref{eq:BP_eqs_XY_gg} at finite $T$ and Eq.~\autoref{eq:BP_eqs_zeroTemp_gg} at zero $T$, apart from the substitution of integrals with sums at $T>0$ and the evaluation of the max and the argmax on the $Q$ discrete directions at $T=0$. Of course, the reasoning developed about the spreading of perturbations at zero temperature holds also in this case.

\begin{figure}[!t]
	\centering
	\includegraphics[scale=1]{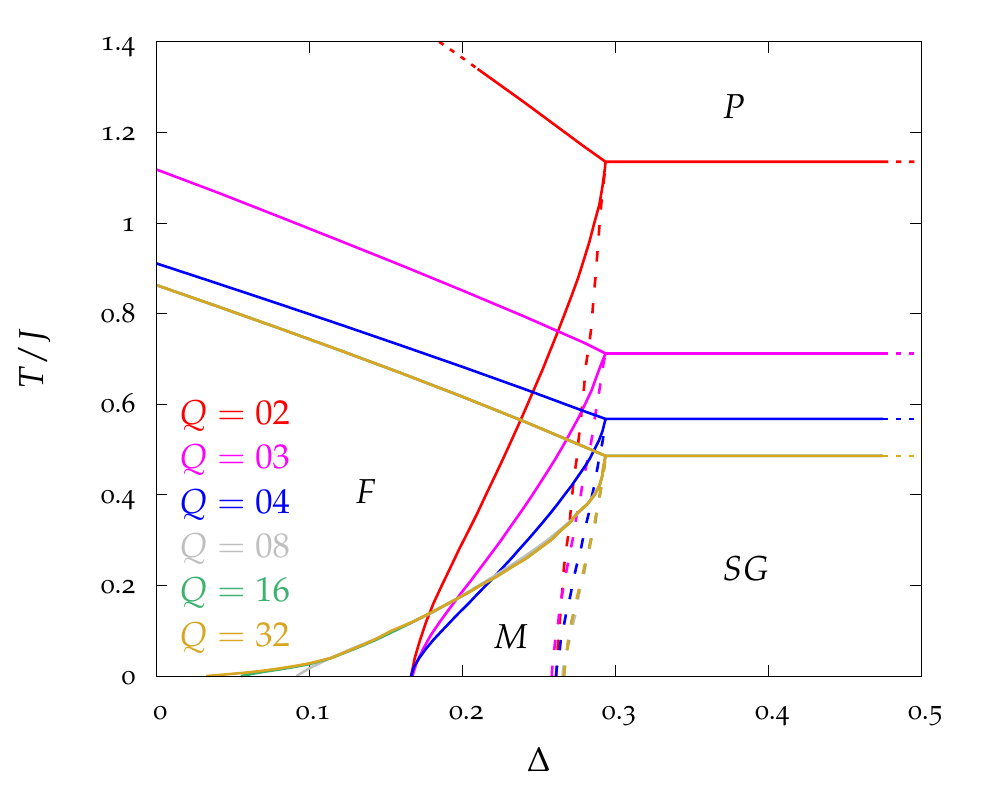}
	\caption[Clock model phase diagram with the gauge glass disorder distribution]{Phase diagram of the $Q$-state clock model on the $C=3$~\acrshort{RRG} ensemble with random angular shifts $\omega_{ij}$'s drawn from the gauge glass disorder distribution $\mathbb{P}_{\omega}$ of Eq.~\autoref{eq:disorder_distribution_gg}, for several values of $Q$. The value $Q=32$ does no longer show appreciable differences with respect to the XY limit.}
	\label{fig:Qclock_phase_diag_gg}
\end{figure}

The phase diagram in the temperature vs ferromagnetic bias plane is depicted in~\autoref{fig:Qclock_phase_diag_gg} for $Q=2,3,4,8,16,32$. Also in this case, the qualitative structure of the phase diagram remains unchanged when increasing $Q$, with most of critical lines converging very fast in $Q$, so that they are practically indistinguishable when $Q \gtrsim 8$. The slowest convergence is again exhibited by the \acrshort{dAT} line in the region of low temperatures and weak disorder, due to the same mechanism holding in the bimodal case. However, as discussed in~\autoref{sec:gauge_glass_XY}, the gauge glass disorder distribution does not provide any strong directional anisotropy, implying no changes in the concavity of the \acrshort{dAT} line close to the multicritical point when moving from very small to larger values of $Q$.

It is interesting to notice that in the Ising case the phase diagrams of the two classes of disorder distributions --- the bimodal coupling one and the gauge glass --- are perfectly mappable one into each other through the transformation $p \leftrightarrow 1-\Delta/2$, as it can be checked by direct comparison. The mapping continues to hold also for larger values of $Q$, if neglecting the \acrshort{dAT} line, in agreement with the analogous observation for the XY model. Finally, also in this case it can be observed the superposition of $Q=2$ and $Q=4$ critical lines when properly rescaling the coupling strength $J$.

\section{Convergence of physical observables}
\label{sec:conv_phys_obs}

The above numerical evidences of a very fast convergence of the critical lines when increasing $Q$, together with the heuristic argument about Bessel functions provided in~\autoref{sec:RS_solution_Qclock}, strongly suggest an exponential convergence of the clock model physical observables toward the XY model ones for large values of $Q$.

Main physical observable we focus on is the Bethe free energy density $f^{(Q)}(\beta)$, computed by using the $2\pi/Q$ normalization for the cavity messages of the clock model. Indeed, as already pointed out at the beginning of~\autoref{sec:RS_solution_Qclock}, this choice correctly reproduces the XY behaviour in the $Q\to\infty$ limit. Furthermore, notice that in this Section we use a very large size of the population in the~\acrshort{PDA}, namely $\mathcal{N}=10^7$, since we want to recognize the effects introduced through the discretization for quite large values of $Q$, and it is not possible if statistical fluctuations are too wide.

\subsection{The finite-temperature regime}

As a first step, let us focus on the expression of the free energy density derived in~\autoref{sec:RS_solution_Qclock} for the $Q$-state clock model in the paramagnetic phase:
\begin{equation}
	f^{(Q)}(\beta) = -\frac{1}{\beta}\ln{2\pi} - \frac{C}{2\beta}\ln{I^{(Q)}_0(\beta J)}
\end{equation}
with $I^{(Q)}_0(\beta J)$ that exponentially converges to $I_0(\beta J)$ when $Q$ is large enough, as numerically checked in~\autoref{subsec:conv_discr_Bessel}. Hence, we can perform the following expansion:
\begin{equation}
\begin{split}
	f^{(Q)}(\beta) &= -\frac{1}{\beta}\ln{2\pi} - \frac{C}{2\beta}\ln{I^{(Q)}_0(\beta J)}\\
	&\simeq -\frac{1}{\beta}\ln{2\pi} - \frac{C}{2\beta}\ln{\bigl[I_0(\beta J)+A\exp{(-Q/Q^*)}\bigr]}\\
	&= -\frac{1}{\beta}\ln{2\pi} - \frac{C}{2\beta}\ln{I_0(\beta J)}+\ln{\left[1+\frac{A}{I_0(\beta J)}\exp{(-Q/Q^*)}\right]}\\
	&\simeq f(\beta) + \frac{A}{I_0(\beta J)}\exp{(-Q/Q^*)}
\end{split}
\end{equation}
which yields an exponentially small difference between the free energy density $f^{(Q)}(\beta)$ of the clock model and the one $f(\beta)$ of the XY model when $Q$ is large enough (compared to the typical scale $Q^*$):
\begin{equation}
	\Delta f^{(Q)}(\beta) \equiv f^{(Q)}(\beta) - f(\beta) \propto \exp{(-Q/Q^*)}
\end{equation}
The convergence scale $Q^*$ depends on $\beta J$ and has been numerically evaluated in~\autoref{sec:RS_solution_Qclock} --- when looking at the convergence of discretized Bessel functions --- for two representative values:
\begin{equation}
	Q^*(\beta J=2) \simeq 2.0 \qquad , \qquad Q^*(\beta J=5) \simeq 2.5
\end{equation}
respectively corresponding to a point very close to the paramagnetic to spin glass transition in the XY model with unbiased bimodal couplings and to a point deeply into the spin glass phase. For smaller values of $\beta J$, then, convergence is even faster, with smaller values of $Q^*$.

So we can conclude that the error committed when approximating the XY model free energy density with the $Q$-state clock model one is exponentially small in $Q$ in the paramagnetic phase, with a scale of convergence $Q^*$ which can be safely assumed to be at most of order one. Furthermore, the same behaviour is expected to be observed in the paramagnetic phase of the gauge glass model, due to the many analogies pointed out between the two classes of disorder in previous Chapters.

\begin{figure}[p]
	\centering
	\includegraphics[scale=0.99]{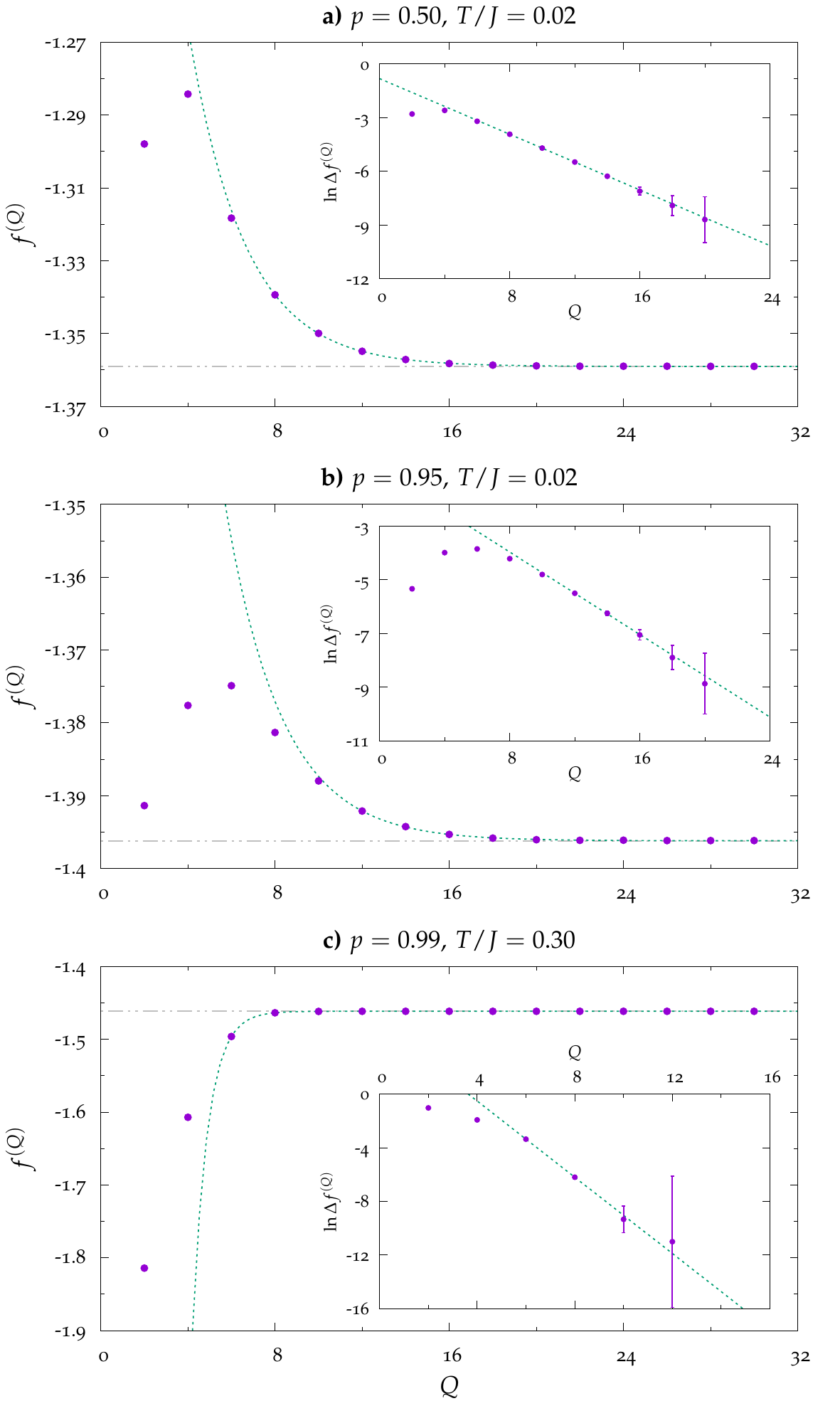}
	\caption[Convergence of clock model free energy density at finite $T$ for the bimodal disorder distribution]{Convergence in $Q$ of the Bethe free energy density $f^{(Q)}$ of the clock model with bimodal couplings for different points in the phase diagram: \textbf{a)}~unbiased spin glass phase, \textbf{b)}~mixed phase, \textbf{c)}~ferromagnetic phase. Main plots show an exponential convergence of $f^{(Q)}$ after an initial transient, while the insets show a linear behaviour for the logarithm of the difference with the corresponding value for the XY model.}
	\label{fig:conv_physObs_pmJ}
\end{figure}

At this point, let us move to the low-temperature region. Our guess is that, given the very fast convergence of critical lines even for small temperatures and the exponential convergence of $f^{(Q)}$ in the paramagnetic phase, also in the low-temperature region the convergence of clock model physical observables is exponentially fast in $Q$. So let us compute the Bethe free energy density $f^{(Q)}(\beta)$ for some representative points in the unbiased spin glass, mixed and ferromagnetic phases, showing its $Q$-dependence in the main plots of~\autoref{fig:conv_physObs_pmJ}.

After an initial transient involving very small values of $Q$, the convergence is actually exponential toward the corresponding XY value for all the analyzed points in the $(p,T)$ plane. So we can perform a linear fit
\begin{equation}
	\ln{\Delta f^{(Q)}(\beta)} = A - Q/Q^*
\end{equation}
and hence accurately evaluate the convergence scale $Q^*$ for each $(p,T)$ point (insets of~\autoref{fig:Qclock_phase_diag_pmJ}), reporting the resulting values in~\autoref{tab:conv_physObs_pmJ}. Each fit has an acceptable $\chi^2$ per degree of freedom, as shown by the last column of the table.

\begin{table}[!t]
	\setlength{\tabcolsep}{8pt}		
	\centering
	\caption[Convergence scale of the clock model free energy density at finite $T$ for the bimodal disorder distribution]{Values of the convergence scale $Q^*$ for some points in the three low-temperature phases of the $Q$-state clock model with bimodal couplings. The last column reports the values of the total $\chi^2$ over the total number of degrees of freedom for each fit.}
	\label{tab:conv_physObs_pmJ}
	\begin{tabular}{ccccc}
		\toprule
		Phase & $p$ & $T/J$ & $Q^*$ & $\chi^2/dof$\\
		\midrule
		Spin glass & $0.50$ & $0.02$ & $2.57(1)$ & $0.20/5$\\
		Mixed & $0.95$ & $0.02$ & $2.60(5)$ & $0.19/3$\\
		Ferromagnetic & $0.99$ & $0.30$ & $0.70(1)$ & $0.12/2$\\
		\bottomrule
	\end{tabular}
\end{table}

Some important features come out from this analysis, performed also on several other points of the low-temperature region. First of all, there occurs an increasing of the convergence scale $Q^*$ when lowering the temperature, consistently with the results found in the high-temperature region. Secondly, the characteristic scale of convergence $Q^*$ remains quite unchanged when moving inside the \acrshort{RSB} region at a fixed temperature, as it usually happens for physical observables. Instead, when disorder is weak enough to yield a~\acrshort{RS} ferromagnetic solution, the scale $Q^*$ slightly increases: again, the effects of the discretization are found to be more evident when getting closer to the point ($p=1,T=0$) of the phase diagram.

The quite small values of the convergence scale $Q^*$ found in the~\acrshort{RSB} region suggest that disorder and frustration provide an \textit{enhancement} in the convergence of the discretized model toward the continuous one. This can be understood by the following reasoning. If the system is a pure ferromagnet, then the smallest misalignment in the discretized model implies a strong breaking in the symmetry of the system and a corresponding large energy cost. Instead, if the system is already misaligned and frustrated due to the presence of the quenched disorder, then a further slight misalignment due to the discretization gives just a minor effect, without any further strong breaking of symmetry or energy cost.

Notice that it was already known in the literature that pure ferromagnetic models with continuous variables are badly approximated by the corresponding discretized models in the low-temperature regime, since small fluctuations are highly suppressed for any not large enough number of states of the discretized model. E.\,g., this is the reason why the discretization of $\mathrm{SU}(2)$ and $\mathrm{SU}(3)$ symmetries in the lattice gauge theory works quite badly in the low-temperature or strong-coupling regime~\cite{Rebbi1980, PetcherWeingarten1980, GrosseKuhnelt1982}.

\begin{figure}[p]
	\centering
	\includegraphics[scale=0.99]{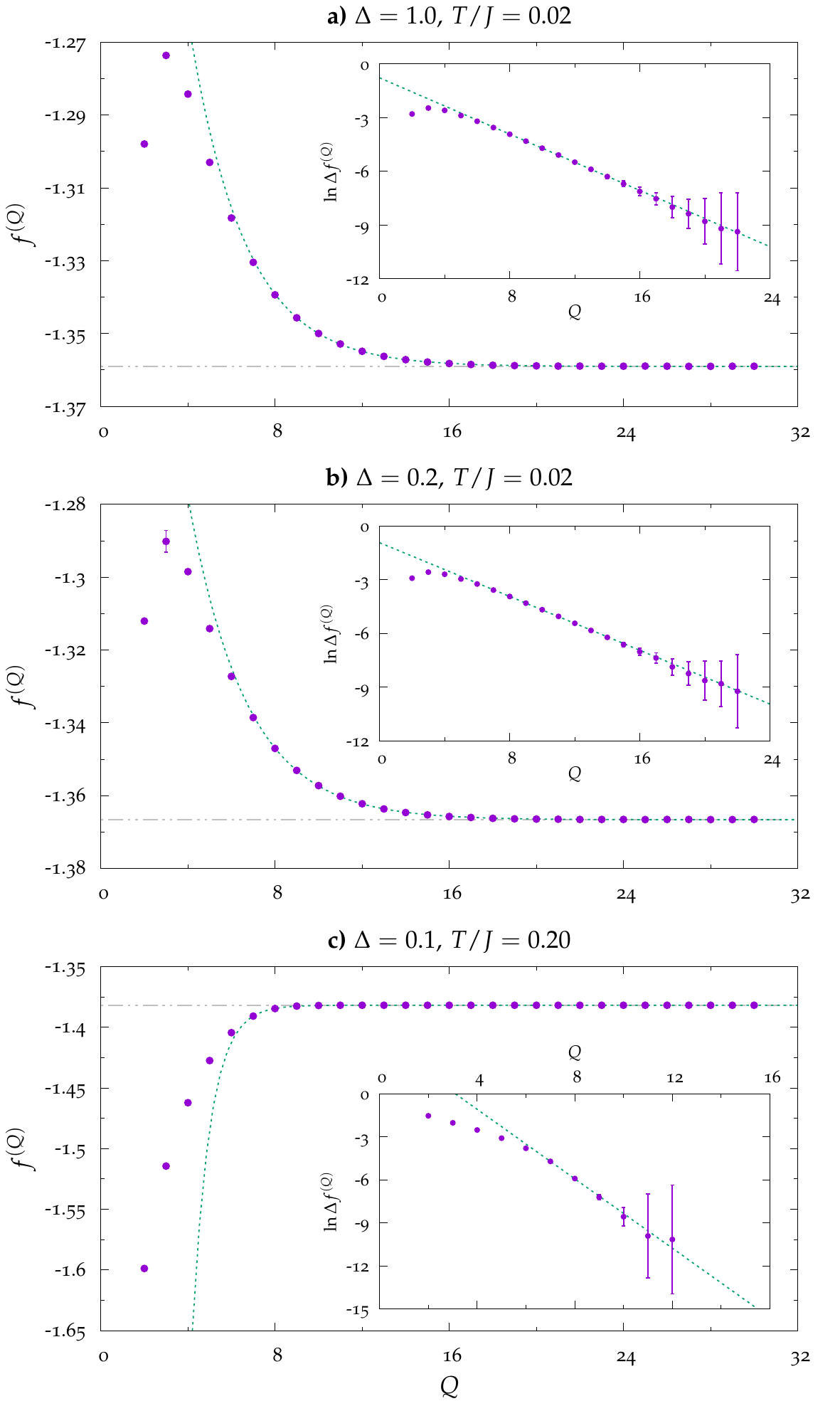}
	\caption[Convergence of clock model free energy density at finite $T$ for the gauge glass disorder distribution]{Convergence in $Q$ of the Bethe free energy density $f^{(Q)}$ of the clock model in the gauge glass case for different points in the phase diagram: \textbf{a)}~unbiased spin glass phase, \textbf{b)}~mixed phase, \textbf{c)}~ferromagnetic phase. Main plots show an exponential convergence of $f^{(Q)}$ after an initial transient, while the insets show a linear behaviour for the logarithm of the difference with the corresponding value for the XY model.}
	\label{fig:conv_physObs_gg}
\end{figure}

Analogous results are shown in~\autoref{fig:conv_physObs_gg} for the gauge glass disorder distribution, again considering a point ($\Delta,T$) for each low-temperature phase. In this case the exponential convergence is even more clear, due to the fact that also odd values of $Q$ can be used. Using the same exponential shape for $\Delta f^{(Q)}(\beta)$ as before, we can fit the data and estimate the convergence scale $Q^*$ corresponding to each point, then reporting them in~\autoref{tab:conv_physObs_gg}.

The resulting values are highly compatible with the ones found for the bimodal disorder distribution, highlighting a sort of universality also in the convergence features. In particular, the dataset for the bimodal disorder distribution at $p=0.5$ and $T/J=0.02$ can be perfectly superimposed on the dataset with $Q$ even for the gauge glass disorder distribution at $\Delta=1$ and $T/J=0.02$. This is not surprising, since both cases refer to the same temperature and to a completely symmetric choice of the disorder, and this is enough to observe this universality.

\begin{table}[!t]
	\setlength{\tabcolsep}{8pt}		
	\centering
	\caption[Convergence scale of clock model free energy density at finite $T$ for the gauge glass disorder distribution]{Values of the convergence scale $Q^*$ for some points in the three low-temperature phases of the $Q$-state clock model with the gauge glass disorder. The last column reports the values of the total $\chi^2$ over the total number of degrees of freedom for each fit.}
	\label{tab:conv_physObs_gg}
	\begin{tabular}{ccccc}
		\toprule
		Phase & $\Delta$ & $T/J$ & $Q^*$ & $\chi^2/dof$\\
		\midrule
		Spin glass & $1.00$ & $0.02$ & $2.55(1)$ & $0.48/12$\\
		Mixed & $0.20$ & $0.02$ & $2.66(5)$ & $1.35/13$\\
		Ferromagnetic & $0.10$ & $0.20$ & $0.83(1)$ & $0.43/4$\\
		\bottomrule
	\end{tabular}
\end{table}

So we can conclude that at any finite temperature and for any quantity of disorder introduced in the system, the $Q$-state clock model provides an efficient and reliable approximation of the XY model, with an error in the evaluation of physical observables that is exponentially small in $Q$. Furthermore, the presence of the disorder strongly enhances such convergence, giving a convergence scale $Q^*$ of order one down to the very low-temperature region. These results fully justify \textit{a posteriori} the value $Q=64$ adopted when numerically studying the XY model in~\autoref{chap:XYnoField} via the $Q$-state clock model.

Remarkably, the convergence features seem also to be independent from the full shape of the probability distribution of the quenched disorder, recovering the sort of universality already seen for the arrangement of the different phases and the inbetween critical lines.

\subsection{The zero-temperature regime}

The picture described so far could in principle dramatically change when moving to the $T=0$ axis. Indeed, any finite value of $Q$ implies a finite energy cost for the smallest possible fluctuation around the ground state configuration, and it is only in the $Q\to\infty$ limit that this energy cost goes to zero, so corresponding to a soft mode.

In particular, as also pointed out for the finite-temperature regime, there are two main mechanisms that can excite the system from the perfectly aligned ground state, so making the difference between the discretized and the continuous model less evident: thermic fluctuations induced by a finite temperature $T$, and misalignment induced by the quenched disorder. Indeed, it is their combined action that provides the exponential convergence seen above. But when $T$ goes to zero, it is only the disorder that can excite the system with respect to the fully ordered ground state configuration of the pure ferromagnet.

So it reasonable to expect a fast convergence of physical observables even at zero temperature, provided there is a large enough quantity of disorder to ``unfreeze'' the system. This ``minimum'' quantity of disorder is likely related to location of the endpoint $p_{\text{dAT}}$ of the~\acrshort{dAT} on the $T=0$ axis: since it moves towards $p=1$ when increasing $Q$, then for large quantities of disorder the required ``threshold'' is quite small, while for a very weak disorder it is required a quite large number of states $Q$ to activate the zero-temperature soft modes of the system.

\begin{figure}[!t]
	\centering
	\includegraphics[scale=0.99]{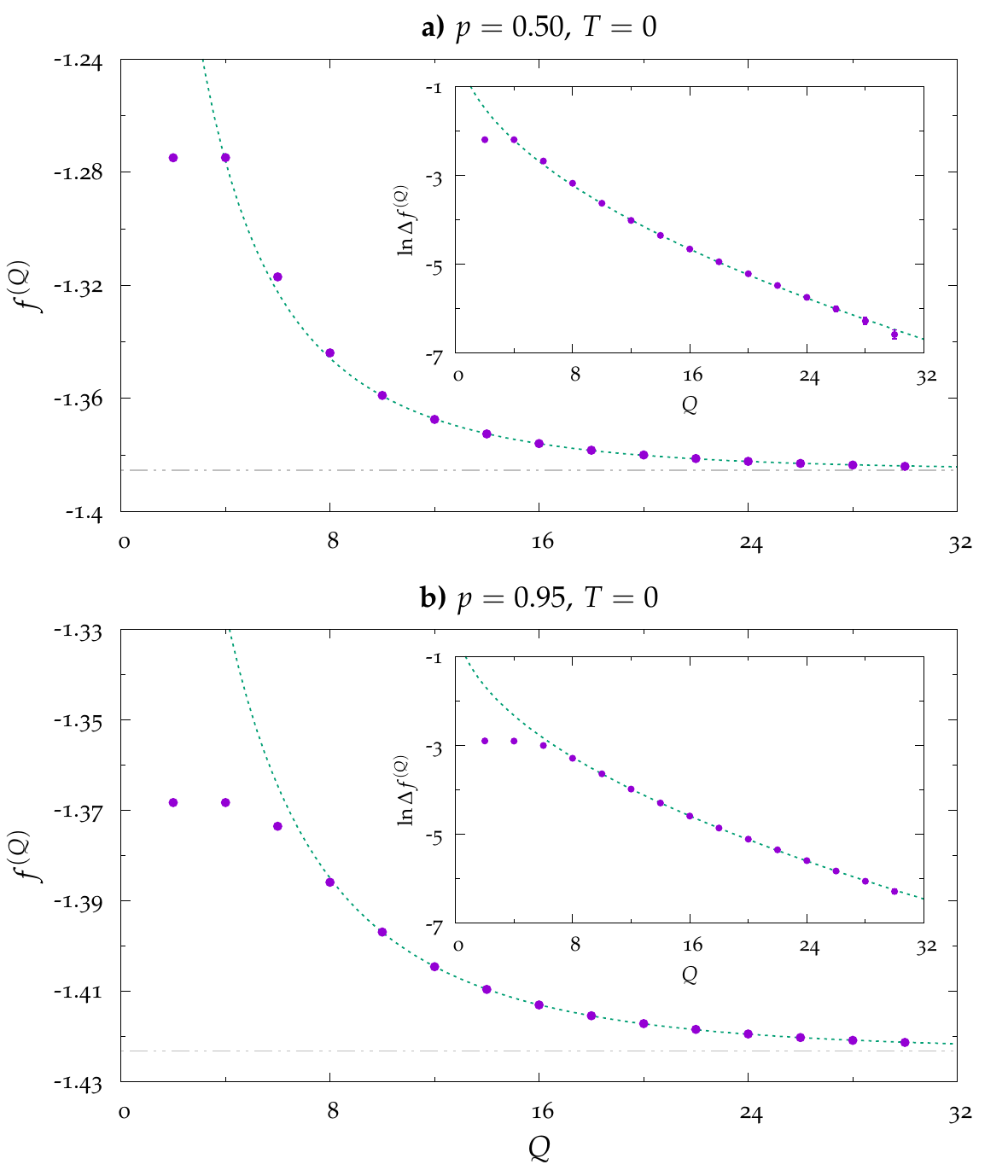}
	\caption[Convergence of clock model free energy density at zero $T$ for the bimodal disorder distribution]{Convergence in $Q$ of the Bethe free energy density $f^{(Q)}$ of the clock model with bimodal couplings at $T=0$ for different points in the phase diagram: \textbf{a)}~unbiased spin glass phase, \textbf{b)}~mixed phase. Main plots show a stretched-exponential convergence of $f^{(Q)}$ after an initial transient, while the insets show a nonlinear behaviour for the logarithm of the difference with the corresponding value for the XY model, compatible with a $b=0.5$ exponent.}
	\label{fig:conv_physObs_pmJ_zeroTemp}
\end{figure}

We study again the convergence of the Bethe free energy density $f^{(Q)}$ --- which actually reduces to the energy density $u^{(Q)}$ for $\beta\to\infty$ --- when increasing $Q$, focusing on two significative points on the $T=0$ axis, respectively in the unbiased spin glass phase and in the mixed phase. The two datasets are then reported in~\autoref{fig:conv_physObs_pmJ_zeroTemp}.

Since the decay of $f^{(Q)}$ toward its limiting value seems still to be close to an exponential, we try to fit the data with a stretched exponential:
\begin{equation}
	\ln{\Delta f^{(Q)}(\beta=\infty)} = A - (Q/Q^*)^{b}
\end{equation}
finding that this functional form actually reproduces very well the data, apart from the initial transient, with a $b$~exponent compatible with $1/2$. Hence, in order to avoid the overfitting, we fix $b=0.5$ and then we fit over $A$ and $Q^*$ parameters, as in the $T>0$ case. The corresponding values for the convergence scale are listed in~\autoref{tab:conv_physObs_pmJ_zeroTemp}. The acceptable $\chi^2$ values reported in the last column of the table strengthens our confidence in the $1/2$ value for the $b$ exponent of the stretched exponential. Unfortunately, we have no analytic arguments to explain the physical meaning of $b=1/2$ in the zero-temperature limit. It could be just related to the entropic contributions that vanish in such limit, so causing a slight slowdown in the convergence with respect to the finite-temperature case.

\begin{table}[!t]
	\setlength{\tabcolsep}{8pt}		
	\centering
	\caption[Convergence scale of clock model free energy density at zero $T$ for the bimodal disorder distribution]{Values of the convergence scale $Q^*$ for some points in the two zero-temperature phases of the $Q$-state clock model with bimodal couplings. The last column reports the values of the total $\chi^2$ over the total number of degrees of freedom for each fit.}
	\label{tab:conv_physObs_pmJ_zeroTemp}
	\begin{tabular}{cccc}
		\toprule
		Phase & $p$ & $Q^*$ & $\chi^2/dof$\\
		\midrule
		Spin glass & $0.50$ & $0.67(1)$ & $6.4/9$\\
		Mixed & $0.95$ & $0.79(1)$ & $2.4/9$\\
		\bottomrule
	\end{tabular}
\end{table}

Again, when looking at the convergence for further points on the $T=0$ axis, there seems to be independence from the quantity of disorder, as long as it is enough to break the replica symmetry. However, when $p$~goes to one, the system becomes a pure ferromagnet and the convergence at zero temperature dramatically slows down. Indeed, it is the worst case for what regards the discretization of a continuous model, as already pointed out before. In this case, the scale $Q^*$ is believed to diverge, so modifying the decay of $f^{(Q)}(\beta=\infty)$ from an exponential to a power law.

However, the ($p=1,T=0$) point is the least interesting for us and hence we can safely claim that everywhere in the phase diagram, both at finite and zero temperature, the physical observables of the XY model can be reliably and efficiently computed through a $Q$-state clock model with moderate values of $Q$, committing an error which is exponentially small in $Q$ itself. Since the numerical effort for solving~\acrshort{BP} equations scales as $Q^2$ for pairwise interactions --- and as $Q^k$ for $k$-spin interactions ---, the proof of an exponential convergence represents a remarkable result, since the use of smaller values of $Q$ yields a strong speedup in numerical simulations.

\begin{figure}[!t]
	\centering
	\includegraphics[scale=0.99]{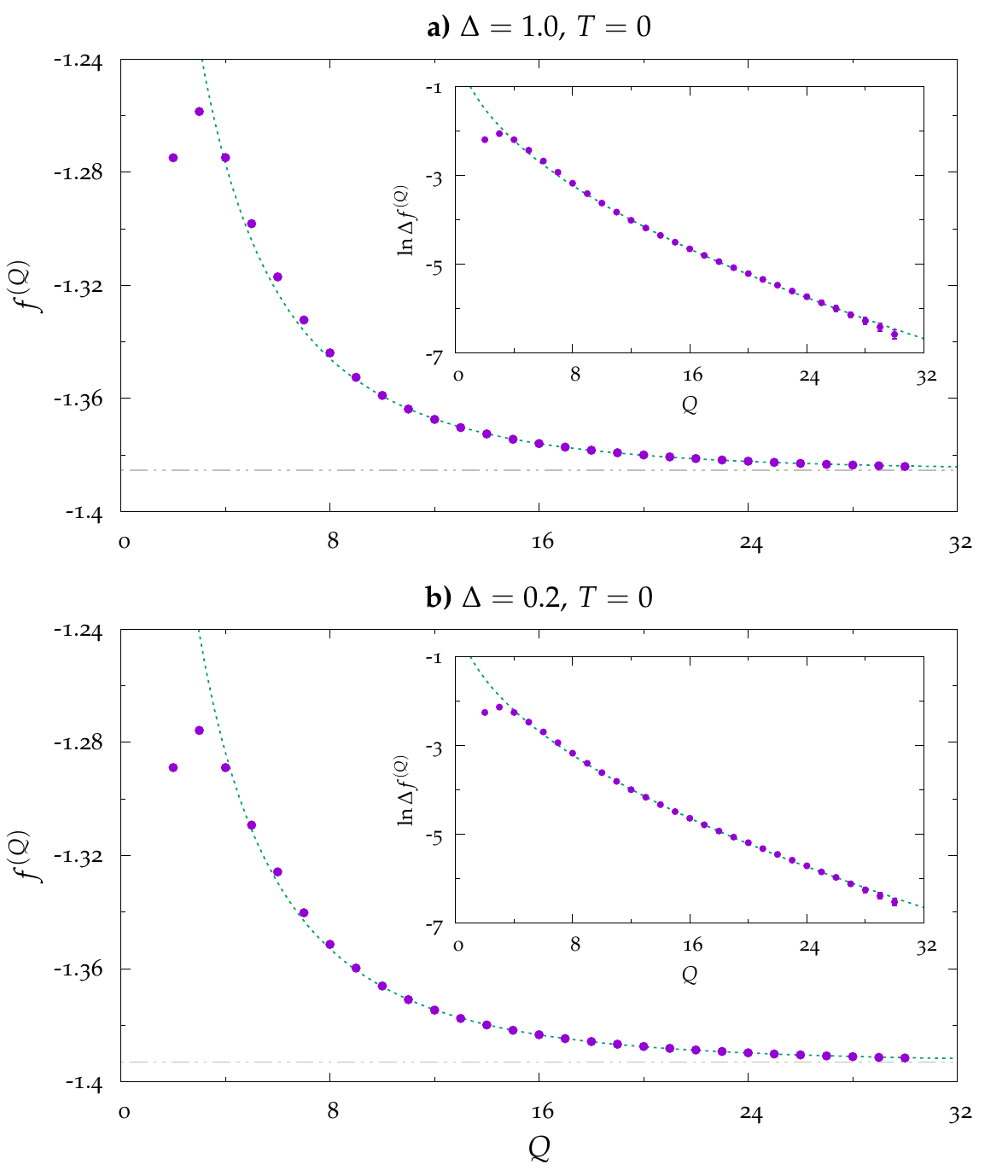}
	\caption[Convergence of clock model free energy density at zero $T$ for the gauge glass disorder distribution]{Convergence in $Q$ of the Bethe free energy density $f^{(Q)}$ of the clock model in the gauge glass case at $T=0$ for different points in the phase diagram: \textbf{a)}~unbiased spin glass phase, \textbf{b)}~mixed phase. Main plots show a stretched-exponential convergence of $f^{(Q)}$ after an initial transient, while the insets show a nonlinear behaviour for the logarithm of the difference with the corresponding value for the XY model, compatible with a $b=0.5$ exponent.}
	\label{fig:conv_physObs_gg_zeroTemp}
\end{figure}

Finally, no qualitative changes are found to occur when moving to the gauge glass clock model, just as in the finite-temperature regime. In~\autoref{fig:conv_physObs_gg_zeroTemp} we report the decay in $Q$ of the Bethe free energy density for two points in the unbiased spin glass phase and in the mixed phase, respectively, finding the values of $Q^*$ listed in~\autoref{tab:conv_physObs_gg_zeroTemp}. Notice that the perfect superposition of the data at $p=0.5$ for the bimodal XY model and those at $\Delta=1.0$ for the gauge glass XY model --- for even values of $Q$ --- still survives in the zero-temperature limit, as it can be seen by comparing the upper panels of~\autoref{fig:conv_physObs_pmJ_zeroTemp} and~\autoref{fig:conv_physObs_gg_zeroTemp}, respectively. Finally, also in this case the $b$ exponent has been reliably set equal to $1/2$, with acceptable values of the resulting $\chi^2$.

\begin{table}[!b]
	\setlength{\tabcolsep}{8pt}		
	\centering
	\caption[Convergence scale of clock model free energy density at zero $T$ for the gauge glass disorder distribution]{Values of the convergence scale $Q^*$ for some points in the two zero-temperature phases of the $Q$-state clock model with the gauge glass disorder. The last column reports the values of the total $\chi^2$ over the total number of degrees of freedom for each fit.}
	\label{tab:conv_physObs_gg_zeroTemp}
	\begin{tabular}{cccc}
		\toprule
		Phase & $\Delta$ & $Q^*$ & $\chi^2/dof$\\
		\midrule
		Spin glass & $1.00$ & $0.67(1)$ & $12.5/19$\\
		Mixed & $0.20$ & $0.68(1)$ & $12.8/19$\\
		\bottomrule
	\end{tabular}
\end{table}

\section{Beyond the RS solution}
\label{sec:beyond_RS}

So far, we have provided several evidences of a very fast convergence of the $Q$-state clock model toward the XY model, by firstly looking at the convergence of the discretized Bessel functions, then at the convergence of critical lines in the phase diagrams and finally at the convergence of physical observables in both cases of finite temperatures and zero temperature.

However, all the computations have been performed out in the~\acrshort{RS} ansatz, that we know to be incorrect once trespassed the~\acrshort{dAT} line, and specifically in the unbiased spin glass phase and in the mixed phase. Even though the~\acrshort{RS} ansatz typically provides a rather good approximation of the exact~\acrshort{RSB} solution, in particular for what regards the evaluation of self-averaging physical observables, at this point we wonder if the results about the convergence are robust when getting closer to the exact solution. Actually, this question is connected to another one, which is even more fundamental: how does the universality class of the $Q$-state clock model change when increasing~$Q$?

A hint about the answer to this question comes from the results on the fully connected topology. In particular, in the second half of the eighties Nobre and Sherrington focused just on the $Q$-state clock model with small values of $Q$, showing that for $Q \geqslant 5$ the clock model belongs to the same universality class of the XY model~\cite{NobreSherrington1986}. Furthermore, they also showed that the absence of the~$\mathrm{Z}_2$ inversion symmetry for odd values of $Q$ --- when considering Gaussian distributed couplings --- becomes irrelevant for $Q \geqslant 5$~\cite{NobreSherrington1989}, adding a further evidence for their previous statement.

If on one hand the exact solution on the fully connected topology requires a~\acrshort{fRSB} ansatz for all the values of $Q$ different from~$3$~\cite{NobreSherrington1986, NobreSherrington1989}, on the other hand it is not known what happens on the sparse topology. Indeed, even for the Ising model it has not been proved yet if the exact solution actually requires a~\acrshort{fRSB} scheme or e.\,g. just a~\acrshort{kRSB} scheme with a finite number $k$ of~\acrshort{RSB} steps, even though it is generally believed that a~\acrshort{fRSB} ansatz is required in the diluted case as well~\cite{Parisi2017}. So we guess that also for the $Q$-state clock model the transition is \textit{continuous} --- hence involving a smooth Parisi order parameter $q(x)$ --- for all the values of~$Q$ but $3$. This expectation relies on the following clues:
\begin{enumerate}[label=\textit{\roman*)}]
	\item $Q=2$ corresponds to the well known Ising model, whose exact solution is believed to be~\acrshort{fRSB} as well on diluted graphs;
	\item $Q=4$ is a ``double Ising'', as already pointed out in this Chapter, and hence also in this case the transition should be continuous;
	\item $Q=3$ is a particular case, since it corresponds to a $3$-state Potts model or to a $3$-colouring problem, both which having no thermodynamic phase transitions on the $C=3$~\acrshort{RRG} ensemble, but only a dynamic phase transition well described in the~\acrshort{1RSB} ansatz~\cite{KrzakalaEtAl2004, ZdeborovaKrzakala2007};
	\item for $Q \geqslant 5$ the \acrshort{RS} overlap $q$ always becomes different from zero in a continuous way below the critical temperature away from the paramagnetic phase.
\end{enumerate}
Notice, instead, that the $Q$-state Potts model on sparse graphs always shows a discontinuous transition, well described by the~\acrshort{1RSB} ansatz and stable under further steps of~\acrshort{RSB}. This is not in contrast with our guess, since $Q$-Potts interactions --- as well as $Q$-colouring ones --- are not ferromagnetic and hence nothing particular is expected to happen in the $Q\to\infty$ limit, while in our case such limit allows to recover a well known model, endowed with a specific ordering of the states that results in the continuous $\mathrm{O}(2)$ symmetry.

The development of a~\acrshort{fRSB} algorithm for the solution on diluted graphs is highly nontrivial, due to the spatial heterogeneity, and hence it is yet to come~\cite{Parisi2017}. Nonetheless, the change from the~\acrshort{RS} frame to the~\acrshort{1RSB} one is a strong qualitative improvement in the search of the exact solution, since we would be able to keep track of some~\acrshort{RSB} effects as e.\,g. the appearance of many states in the Gibbs measure, and hence further steps of~\acrshort{RSB} are believed not to add nothing in this sense. Moreover, the evaluation of the~\acrshort{1RSB} parameters when increasing $Q$ would provide substantial information about the universality class of the $Q$-state clock model --- that again could be exactly studied only in the~\acrshort{fRSB} ansatz ---, which is our ultimate goal within this Section.

\subsection{The 1RSB picture}

The failure of the~\acrshort{RS} ansatz occurs when factorization~\autoref{eq:BP_condition} is no longer valid, due to the breaking of the Gibbs measure in a large number of states, in each one of which the exact solution is still~\acrshort{RS} stable. However, each of these states produces a different free energy shift when nodes and links are added to the graph by following the prescriptions of the \acrshort{RS} cavity method~\cite{MezardParisi2001}, and hence a \textit{weighed average} over the states is needed.

This observation is at the basis of the~\acrshort{1RSB} cavity method, derived by M\'ezard and Parisi~\cite{MezardParisi2001, MezardParisi2003}. We redirect the reader to Refs.~\cite{Zamponi2008, Book_MezardMontanari2009}, in addition to the original papers by M\'ezard and Parisi, for an exhaustive description of the~\acrshort{1RSB} cavity method for solving disordered models on diluted graphs, while in the following we just provide a sketch of the key concepts.

As usual in statistical mechanics, entropy counts the number of configurations in which the system can stay. In the field of disordered systems, when counting the states instead of the configurations, the corresponding entropy $\Sigma$ assumes a key role, being known as \textit{configurational entropy} for structural glasses and \textit{complexity} for spin glasses (to which we will refer). Its definition is straightforward:
\begin{equation}
	\Omega(f) \equiv e^{\,N\Sigma_{\beta}(f)}
\end{equation}
once given the number of states $\Omega(f)$ as a function of the free energy density~$f$. Since $\exp{(-\beta N f_{\alpha})}$ is the Gibbs weight of the $\alpha$-th state --- apart from a multiplicative constant ---, it is then possible to rewrite the total partition function of the system by summing over all the possible states:
\begin{equation}
	\mathcal{Z}_{\beta} \equiv \sum_{\alpha}e^{-\beta N f_{\alpha}} \simeq \int\di f\,\Omega(f)\,e^{-\beta N f} = \int\di f\,e^{\,N[\Sigma_{\beta}(f)-\beta f]}
\end{equation}
finally evaluating the integral through the saddle-point method in the large-$N$ limit. However, this computation relies on the previous knowledge of the complexity~$\Sigma_{\beta}$, which is actually not possible. In order to overcome this issue, Monasson introduced the replicated partition function $\mathcal{Z}_{\beta}(m)$~\cite{Monasson1995}:
\begin{equation}
	\mathcal{Z}_{\beta}(m) \equiv \sum_{\alpha}e^{-\beta m N f_{\alpha}} \simeq \int\di f\,e^{\,N[\Sigma_{\beta}(f)-\beta m f]}
\end{equation}
with $m$ integer. Then, if $m$ is analytically continuated over real values and in particular below the unity --- in analogy with the $m_1$ parameter of the \acrshort{1RSB} ansatz seen in~\autoref{chap:sg_replica} --- we can redefine it as $x$, so that $x\in[0,1]$ provides a well behaved probability distribution for the Gibbs measure over the states. Then, a \textit{replicated} free energy density $\phi_{\beta}(x)$ can be defined:
\begin{equation}
	\phi_{\beta}(x) \equiv -\frac{1}{\beta x N}\ln{\mathcal{Z}_{\beta}(x)}
\end{equation}
in perfect analogy with the definition of $f_{\beta}$ inside each state.

At this point, the saddle point can be evaluated as before:
\begin{equation}
	\mathcal{Z}_{\beta}(x) \simeq e^{\,N[\Sigma_{\beta}(f^*)-\beta x f^*]}
\end{equation}
with $f^*=f^*_{\beta}(x)$ being the maximizer of the exponent, depending on both~$\beta$ and~$x$:
\begin{equation}
	\frac{\partial}{\partial f}\left[\Sigma_{\beta}(f)-\beta x f\right]\biggr{|}_{f^*}=0
\end{equation}
Now, just thanks to the presence of $x$, the complexity $\Sigma_{\beta}(f)$ can be recognized as the Legendre transform of the replicated free energy density $\phi_{\beta}(x)$:
\begin{equation}
	\phi_{\beta}(x) = f^*_{\beta}(x) - \frac{1}{\beta x}\Sigma_{\beta}(f^*)
\end{equation}
Consequently, in this way we are able to actually compute $f^*_{\beta}(x)$:
\begin{equation}
\begin{split}
	f^*_{\beta}(x) &= \frac{\partial}{\partial x}\left(x\phi_{\beta}(x)\right)\\
	&= \phi_{\beta}(x) - x\frac{\partial}{\partial x}\phi_{\beta}(x)
	\label{eq:f_1RSB}
\end{split}
\end{equation}
and $\Sigma_{\beta}(x)$:
\begin{equation}
\begin{split}
	\Sigma_{\beta}(x) &= \Sigma_{\beta}(f^*_{\beta}(x))\\
	&= \beta x^2\frac{\partial}{\partial x}\phi_{\beta}(x)\\
	&= \beta x \bigl[f^*_{\beta}(x) - \phi_{\beta}(x)\bigr]
	\label{eq:Sigma_1RSB}
\end{split}
\end{equation}
both which being functions of $\beta$ and $x$. Finally, $\Sigma_{\beta}(f)$ can be obtained by fixing $\beta$ and then parametrically plotting $\Sigma_{\beta}(x)$ versus $f^*_{\beta}(x)$ for $x\in[0,1]$.

Thermodynamic observables can be then recovered by setting $x=1$ if the corresponding complexity is nonnegative, namely $\Sigma_{\beta}(x=1) \geqslant 0$, so that corresponding states are not subdominant in the thermodynamic limit. Instead, if $\Sigma_{\beta}(x=1) < 0$, the partition function is dominated by the largest value of $x$ --- say $x^*$ --- that makes $\Sigma_{\beta}(x)$ vanish:
\begin{equation}
	x^* \equiv \max{\bigl\{x\,|\,\Sigma_{\beta}(x)=0\bigr\}}
\end{equation}
which is also known as the \acrshort{1RSB}~\textit{Parisi parameter}.

\subsection{The 1RSB BP equations}

The computation of the replicated free energy density $\phi_{\beta}$ --- and from it the other relevant observables --- can be performed after having generalized the~\acrshort{RS}~\acrshort{BP} equations to the \acrshort{1RSB} ones and then solved them.

The starting point is that the Gibbs measure breaks into many states~$\{\alpha\}$, each having a weight $w_{\alpha}\propto\exp{(-\beta N f_{\alpha})}$. So, for each directed edge $i\to j$ of the original graph, we have no longer a single cavity marginal $\eta_{i\to j}$, but a set of them, $\{\eta^{(\alpha)}_{i\to j}\}$, one for each state. In other words, the $\alpha$-th state is completely characterized by the set $\{\eta^{(\alpha)}_{i\to j}\}$ with the lower index running over all the directed edges of the graph. Since in each state the \acrshort{RS} \acrshort{BP} equations are still valid, the corresponding fixed point can be then easily computed. Actually, each state just corresponds to one of the (many) \acrshort{BP} fixed points attainable in the \acrshort{RS} ansatz when \acrshort{RSB} occurs.

However, also the reweigh should be taken into account. To employ it, further check nodes are added to each variable node and to each edge of the original graph, whose function is just that of inserting the suitable reweigh factor (exactly as a usual compatibility function in a factor graph). The resulting modified graph is still sparse and is known as the \textit{auxiliary model}~\cite{Book_MezardMontanari2009}, on which the \acrshort{BP} approach can be exploited as well. The unique difference is that the corresponding cavity messages are no longer the $\eta_{i\to j}$'s, but their probability distributions $\mathbb{P}_{i\to j}[\eta^{(\alpha)}_{i\to j}]$'s over the states.

The resulting \acrshort{BP} self-consistency equations for the $\mathbb{P}_{i\to j}$'s are nothing but the \acrshort{1RSB} \acrshort{BP} equations for the $\eta_{i\to j}$'s:
\begin{equation}
\begin{split}
	\mathbb{P}_{i\to j}[\eta^{(\alpha)}_{i\to j}] &= \widetilde{\mathcal{F}}\bigl[\{\mathbb{P}_{k\to i}[\eta^{(\alpha_k)}_{k\to i}]\},\{J_{ik}\},x\bigr]\\
	&\equiv \int\prod_{k=1}^{d_i-1}\Bigl(\mathcal{D}\eta^{(\alpha_k)}_{k\to i}\,\mathbb{P}_{k\to i}[\eta^{(\alpha_k)}_{k\to i}]\Bigr)\,\delta\Bigl[\eta^{(\alpha)}_{i\to j}-\mathcal{F}[\{\eta^{(\alpha_k)}_{k\to i}\},\{J_{ik}\}]\Bigr]\\
	&\qquad\times\Bigl(\mathcal{Z}_{i\to j}[\{\eta^{(\alpha_k)}_{k\to i}\}]\Bigr)^x
	\label{eq:1RSB_BP_eqs}
\end{split}
\end{equation}
where $\mathcal{Z}_{i\to j}$ is the normalization constant that comes from the computation of the cavity message $\eta^{(\alpha)}_{i\to j}$ via the \acrshort{RS} \acrshort{BP} equations. In this way we are actually implementing the reweigh by $\exp{(-\beta x N f_{\alpha})}$, since $\mathcal{Z}_{i\to j}$ is directly related to the corresponding free energy shift $\Delta f_{i\to j}$:
\begin{equation}
	\mathcal{Z}_{i\to j}=\exp{(-\beta N\Delta f_{i\to j})}
\end{equation}

On a given instance of the problem, it is enough to solve Eqs.~\autoref{eq:1RSB_BP_eqs}, where the probability distribution $\mathbb{P}_{i\to j}[\eta^{(\alpha)}_{i\to j}]$ for each directed edge $i\to j$ can be numerically evaluated via a population of $\mathcal{M}$ cavity messages $\{\eta^{(\alpha)}_{i\to j}\}$, with $\mathcal{M}$ eventually going to infinity. So we can drop the state labels and introduce the following shorthand notation for the \acrshort{1RSB} \acrshort{BP} equations on a given instance of the model:
\begin{equation}
	\mathbb{P}_{i\to j}[\eta_{i\to j}] \equiv \mathbb{E}_{\alpha}\biggl[\delta\Bigl[\eta_{i\to j}-\mathcal{F}[\{\eta_{k\to i}\},\{J_{ik}\}]\Bigr]\Bigl(\mathcal{Z}_{i\to j}[\{\eta_{k\to i}\}]\Bigr)^x\biggr]
\end{equation}
with $\mathbb{E}_{\alpha}$ highlighting the average over the states through the (explicitly written) measure $w_{\alpha}$.

The corresponding fixed point, say $\mathbb{P}^*_{i\to j}$, allows to compute the replicated free energy density $\phi_{\beta}$ on the auxiliary model in the same spirit of the \acrshort{RS} one on the original graph:
\begin{equation}
	\phi_{\beta}(x) = \sum_i\phi_i(x)-\sum_{(i,j)}\phi_{ij}(x)
	\label{eq:1RSB_phi}
\end{equation}
with the replicated node and edge contributions that are a \textit{reweighed} version of their \acrshort{RS} counterparts:
\begin{subequations}
	\begin{equation}
		\phi_i(x) \equiv -\frac{1}{\beta x}\ln{\mathbb{E}_{\alpha}\bigl[\mathcal{Z}^x_i\bigr]}
	\end{equation}
	\begin{equation}
		\phi_{ij}(x) \equiv -\frac{1}{\beta x}\ln{\mathbb{E}_{\alpha}\bigl[\mathcal{Z}^x_{ij}\bigr]}
	\end{equation}
	\label{eq:1RSB_phi_i_phi_ij}
\end{subequations}

An exhaustive discussion about the \acrshort{1RSB} \acrshort{BP} equations on a given instance can be found in Ref.~\cite{Book_MezardMontanari2009}. Instead, here we are more interested in the disorder-averaged description, namely we would like to generalize the \acrshort{RS} \acrshort{PDA} to the \acrshort{1RSB} ansatz. This task can be easily accomplished by considering also the probability distributions $\mathbb{P}_{i\to j}$'s as random variables, distributed according to a suitable $\mathcal{P}[\mathbb{P}_{i\to j}]$. In other words, $\mathcal{P}[\mathbb{P}_{i\to j}]$ in the \acrshort{1RSB} \acrshort{PDA} plays exactly the same role of $\mathbb{P}[\eta_{i\to j}]$ in the \acrshort{RS} \acrshort{PDA}. It is the probability distribution $\mathbb{P}[\eta_{i\to j}]$ over the states that was absent in the \acrshort{RS} ansatz and that has been induced in the \acrshort{1RSB} ansatz by the breaking of the Gibbs measure into several states. Then, when numerically implementing this algorithm, $\mathcal{P}[\mathbb{P}]$ can be substituted by a population of $\mathcal{N}$ elements $\mathbb{P}$'s --- in turn being populations of $\mathcal{M}$ cavity messages each --- that satisfy the following stochastic equation:
\begin{equation}
	\mathcal{P}[\mathbb{P}_{i\to j}] = \mathbb{E}_{\mathcal{G},J}\int\prod_{k=1}^{d_i-1}\Bigl(\mathcal{D}\mathbb{P}_{k\to i}\,\mathcal{P}[\mathbb{P}_{k\to i}]\Bigr)\,\delta\Bigl[\mathbb{P}_{i\to j}-\widetilde{\mathcal{F}}[\{\mathbb{P}_{k\to i}\},\{J_{ik}\},x]\Bigr]
	\label{eq:def_PDA_1RSB}
\end{equation}

At this point, it is clear that a two-level hierarchy of populations comes out in the \acrshort{1RSB} \acrshort{PDA}: \textit{i)} an ``inner'' population of $\mathcal{M}$ cavity messages $\eta_{i\to j}$, describing the probability distribution $\mathbb{P}[\eta]$ over the states for a given directed edge; \textit{ii)} an ``outer'' population of $\mathcal{N}$ elements $\mathbb{P}_{i\to j}$, describing the probability distribution $\mathcal{P}[\mathbb{P}]$ over the random realization of the disorder (i.\,e. the graph, the couplings, the field --- if any ---, and so on).

These two levels of populations actually reflect the two-level hierarchy of the~\acrshort{1RSB} solution and correspond to two different averages to be performed in this approach: a first average $\mathbb{E}_{\alpha}$ over the states, which corresponds to the average over $\mathbb{P}[\eta]$
\begin{equation}
	\mathbb{E}_{\alpha} \leftrightarrow \mathbb{E}_{\eta}
\end{equation}
and a second average $\mathbb{E}_{\mathcal{G},J}$ over the disorder, which corresponds to the average over $\mathcal{P}[\mathbb{P}]$
\begin{equation}
	\mathbb{E}_{\mathcal{G},J} \leftrightarrow \mathbb{E}_{\mathbb{P}}
\end{equation}
with the Parisi parameter $x$ playing the role of modifying the weight of each state --- namely of \textit{reweighing} them --- in order to get access to the complexity~$\Sigma_{\beta}$ and eventually to its equilibrium value $x^*$.

The numerical implementation of the \acrshort{1RSB} \acrshort{PDA} is sketched in the pseudocode~\ref{alg:1RSB_PDA}. After having randomly initialized the population of $\mathcal{N}$ subpopulations of $\mathcal{M}$ cavity messages each, each subpopulation $\mathbb{P}_i$ is iteratively updated by choosing $d_i-1$ ``neighbour'' subpopulations~$\{\mathbb{P}_k\}$ and $d_i-1$ couplings~$\{J_k\}$. The new cavity message $\eta$ is then computed via the \acrshort{RS} \acrshort{BP} equations by picking at random an incoming message from each of the $d_i-1$ previously chosen subpopulations. Notice, indeed, that the ``incoming'' $\mathbb{P}_k$'s and the related $J_k$'s are chosen just once for each subpopulation $\mathbb{P}_i$ (and each time step $t$), since they refer to a well defined node of the auxiliary model.

\begin{algorithm}[t]
	\caption{1RSB Population Dynamics Algorithm ($T>0$)}
	\label{alg:1RSB_PDA}
	\begin{algorithmic}[1]
		\For {$i=1,\dots,\mathcal{N}$} \Comment{The ``outer'' population}
			\For {$\alpha=1,\dots,\mathcal{M}$} \Comment{The ``inner'' population}
				\State Initialize $\eta^{(\alpha)}_i(t=0)$ \Comment{We use a random initialization}
			\EndFor
		\EndFor
		\For {$t=1,\dots,t_{max}$}
			\For {$i=1,\dots,\mathcal{N}$}
				\State Draw an integer $d_i$ from the degree distribution $\mathbb{P}_d$
				\State Draw $d_i-1$ integers $\{k\}$ uniformly in the range $[1,\mathcal{N}]$
				\State Draw $d_i-1$ couplings $\{J_k\}$ from the coupling distribution $\mathbb{P}_J$
				\For {$\alpha=1,\dots,r\mathcal{M}$}
					\State Draw $d_i-1$ integers $\{\alpha_k\}$ uniformly in the range $[1,\mathcal{M}]$
					\State $\eta^{(\alpha)}_{i}(\text{temp}) \gets \mathcal{F}[\{\eta^{(\alpha_k)}_k(t-1)\},\{J_k\}]$
					\State $w_{\alpha} \gets \bigl(\mathcal{Z}_k[\{\eta^{(\alpha_k)}_k(t-1)\},\{J_k\}]\bigr)^x$ \Comment{Needed for the reweigh}
				\EndFor
				\State Reweigh the $i$-th inner population according to $\{w_{\alpha}\}$:
				\State $\{\eta^{(\alpha)}_{i}(t)\}_{\alpha=1,\dots,\mathcal{M}} \gets \{\eta^{(\alpha)}_{i}(\text{temp})\}_{\alpha=1,\dots,r\mathcal{M}}$
			\EndFor
		\EndFor
		\State \textbf{return} $\{\eta^{(\alpha)}_i(t_{max})\}$
	\end{algorithmic}
\end{algorithm}

We do not compute exactly $\mathcal{M}$ new cavity messages for each subpopulation, but $r\mathcal{M}$ with $r>1$. This redundancy is crucial for the reweigh according to the Gibbs weight of the states, whose numerical implementation is outlined in the pseudocode~\ref{alg:1RSB_PDA_reweigh}, and to avoid ``twins'' in the updated subpopulation. Each of the (temporary) $r\mathcal{M}$ cavity marginals $\eta^{(\alpha)}_i$ just computed has a weight $w_{\alpha}$ equal to its normalization constant $\mathcal{Z}_{i\to j}$ to the power $x$:
\begin{equation}
	w_{\alpha} = \bigl(\mathcal{Z}_k[\{\eta^{(\alpha_k)}_k(t-1)\},\{J_k\}]\bigr)^x
\end{equation}
Hence, in order to reweigh them, we normalize the weights, $p_{\alpha}\equiv w_{\alpha}/\sum_{\alpha'=1}^{r\mathcal{M}}w_{\alpha'}$, and then we choose $\mathcal{M}$ cavity messages out of them proportionally to $p_{\alpha}$, e.\,g. by uniformly sampling from the cumulative distribution of the weights. These $\mathcal{M}$ cavity messages finally update the $i$-th population at the time step $t$.

\begin{algorithm}[t]
	\caption{Reweigh in the 1RSB Population Dynamics Algorithm ($T>0$)}
	\label{alg:1RSB_PDA_reweigh}
	\begin{algorithmic}[1]
		\For {$\alpha=1,\dots,r\mathcal{M}$}
			\State $p_{\alpha} \equiv w_{\alpha}/\sum_{\alpha'=1}^{r\mathcal{M}}w_{\alpha'}$ \Comment{Just a normalization of the weights}
		\EndFor
		\For {$\alpha=1,\dots,\mathcal{M}$}
			\State Draw an integer $s$ in the range $[1,r\mathcal{M}]$ according to probabilities $\{p_{s}\}$
			\State $\eta^{(\alpha)}_i(t) \gets \eta^{(s)}_i(\text{temp})$
		\EndFor
		\State \textbf{return} $\{\eta^{(\alpha)}_i(t)\}_{\alpha=1,\dots,\mathcal{M}}$
	\end{algorithmic}
\end{algorithm}

In order to have a proper reweighing, $r$ should go to infinity together with $\mathcal{M}$ and $\mathcal{N}$. However, we dinamically choose $r$ according to the statistical significance of the cavity messages computed up to that point, and the typical values of $r$ chosen in this way are in the range $[2,5]$.

Once reached the fixed point $\mathcal{P}^*[\mathbb{P}^*]$ over both the levels of population, physical observables can be computed eventually computed, following the same stochastic approach of the \acrshort{RS} \acrshort{PDA}. For example, the replicated free energy density $\phi_{\beta}$ can be evaluated via the double average over the populations, giving for the $C$-\acrshort{RRG} ensemble:
\begin{equation}
	\phi_{\beta}(x) = -\frac{1}{\beta x}\mathbb{E}_{\mathbb{P}}\bigl[\ln{\mathbb{E}_{\eta}[\mathcal{Z}^x_i]}\bigr]+\frac{C}{2\beta x}\mathbb{E}_{\mathbb{P}}\bigl[\ln{\mathbb{E}_{\eta}[\mathcal{Z}^x_{ij}]}\bigr]
	\label{eq:1RSB_phi_PDA}
\end{equation}
Then, also the free energy density can be evaluated, recalling Eq.~\autoref{eq:f_1RSB}:
\begin{equation}
	f_{\beta}(x) = -\frac{1}{\beta}\mathbb{E}_{\mathbb{P}}\left[\frac{\mathbb{E}_{\eta}[\mathcal{Z}^x_i\ln{\mathcal{Z}_i}]}{\mathbb{E}_{\eta}[\mathcal{Z}^x_i]}\right] + \frac{C}{2\beta}\mathbb{E}_{\mathbb{P}}\left[\frac{\mathbb{E}_{\eta}[\mathcal{Z}^x_{ij}\ln{\mathcal{Z}_{ij}}]}{\mathbb{E}_{\eta}[\mathcal{Z}^x_{ij}]}\right]
\end{equation}
whose equilibrium value is given by setting $x=x^*$. Finally, when knowing $\phi_{\beta}(x)$ and $f_{\beta}(x)$, the whole curve of the complexity $\Sigma_{\beta}(x)$ for $x\in[0,1]$ can be obtained, Eq.~\autoref{eq:Sigma_1RSB}:
\begin{equation}
	\Sigma_{\beta}(x) = \beta x \bigl[f_{\beta}(x) - \phi_{\beta}(x)\bigr]
\end{equation}

Furthermore, the two-level hierarchy of populations yields two different values of the overlap $q$. Indeed, we have an inner overlap $q_1$, related to the similarity of local magnetizations inside each state:
\begin{equation}
	q_1 = \mathbb{E}_{\mathbb{P}}\left[\frac{\mathbb{E}_{\eta}[\mathcal{Z}^x_i(m^2_{i,x}+m^2_{i,y})]}{\mathbb{E}_{\eta}[\mathcal{Z}^x_i]}\right]
\end{equation}
and an outer overlap $q_0$, describing the similarity of magnetizations between different states:
\begin{equation}
	q_0 = \mathbb{E}_{\mathbb{P}}\left[\left(\frac{\mathbb{E}_{\eta}[\mathcal{Z}^x_i m_{i,x}]}{\mathbb{E}_{\eta}[\mathcal{Z}^x_i]}\right)^2\right] + \mathbb{E}_{\mathbb{P}}\left[\left(\frac{\mathbb{E}_{\eta}[\mathcal{Z}^x_i m_{i,y}]}{\mathbb{E}_{\eta}[\mathcal{Z}^x_i]}\right)^2\right]
\end{equation}
From their definition, it is easy to show that $q_1$ is always equal to or larger than $q_0$. In particular, the equality holds only if the \acrshort{1RSB} algorithm gives back the \acrshort{RS} solution, namely no \acrshort{RSB} has occurred. Notice that the two overlaps $q_1$ and $q_0$ are analogous to those defined within the \acrshort{1RSB} ansatz for fully connected models (\autoref{chap:sg_replica} and Refs.~\cite{Parisi1979b, Parisi1980a}). We will use them to approximate the true order parameter $q(x)$ --- which is supposed to be continuos, so requiring a \acrshort{fRSB} ansatz, not yet developed for sparse models --- in the \acrshort{1RSB} scheme.

Before going on with the study of the $Q$-state clock model within the \acrshort{1RSB} framework, we firstly check our numerical implementation of the \acrshort{1RSB} \acrshort{PDA}, by solving the spin glass Ising model (namely we set $Q=2$) with unbiased bimodal couplings $J_{ij}=\pm 1$ on the $C=6$ \acrshort{RRG} ensemble. The corresponding results are then compared with the ones provided in Ref.~\cite{MezardParisi2001}, finding a remarkable agreement.

\subsection{1RSB solution of the $Q$-state clock model}

The \acrshort{1RSB} ansatz has been widely exploited so far on diluted models, from $p$-spin models~\cite{MezardEtAl2003} to random $k$-SAT problems~\cite{MontanariEtAl2004}, from random colouring problems~\cite{KrzakalaEtAl2004, ZdeborovaKrzakala2007} to Potts models~\cite{KrzakalaZdeborova2008}, just for citing a few. In these models --- at least in a certain range for their parameters --- the \acrshort{1RSB} solution turns out to be stable toward further steps of \acrshort{RSB}, namely it is not just an approximation of a \acrshort{fRSB} solution, but actually the exact one.

This typically occurs when dealing with many-body interactions --- e.\,g. $p$-spin models or $k$-SAT problems --- or when discrete spins take on more than two values --- as it occurs for Potts models and colouring problems. In particular, it is the case for a many-body version of the model we are studying, namely the $Q$-state clock model with $4$-spin interactions~\cite{MarruzzoLeuzzi2016}, whose dynamical transition can be studied by taking $x^*=1$ for the \acrshort{1RSB} Parisi parameter, so highly simplifying the related \acrshort{BP} equations~\cite{Book_MezardMontanari2009}. At variance, the choice $x=0$ avoids the reweigh, making \acrshort{1RSB} \acrshort{BP} equations reduce to the \acrshort{RS} ones.

The main goal we want to reach through the study of the \acrshort{1RSB} solution is the check of the universality class of the $Q$-state clock model when varying $Q$. In the fully connected case, it is enough an expansion of the Parisi order parameter $q(x)$ close to the critical point, i.\,e. at at the reduced temperature $\tau \equiv (T_c-T)/T_c \ll 1$, where it is well approximated by a linear function $q(x)=ax$ for $x<b\tau$ and a constant function $q(x)=ab\tau$ for $x>b\tau$. The universality class is then identified by these two numbers $a$ and $b$. Nobre and Sherrington used exactly this approach to show that for $Q \geqslant 5$ the universality class of the clock model is the same of the XY model~\cite{NobreSherrington1986}.

Unfortunately, for sparse models with a continuous transition, it is not feasible to expand the Parisi function very close to the critical point, because in this case the replicated free energy $\phi_{\beta}(x)$ has a too mild dependence on $x$:
\begin{equation}
	\phi_{\beta}(x) - \phi_{\beta}(0) \propto \tau \quad , \qquad \tau \ll 1
\end{equation}
and hence a numerical estimation of $\phi_{\beta}(x)$ would be too noisy when very close to the critical point. For this reason, we choose $\tau = 1/2$, so computing the replicated free energy density, the free energy density, the complexity and the two overlaps in the middle of the spin glass phase.

We focus on values of $Q$ from $2$ to $8$, apart from the special case $Q=3$, that we already know to be characterized by a dynamic phase transition instead of a static one, hence being \acrshort{1RSB} stable. Since \acrshort{1RSB} \acrshort{PDA} requires a storage of $\mathcal{N}\times\mathcal{M}$ cavity marginals --- each one of which contains $Q$ values --- the optimal values for $\mathcal{N}$ and $\mathcal{M}$ have to be chosen carefully. We observe that, if using the reweigh procedure previously exposed, the complexity $\Sigma_{\beta}(x)$ suffers larger finite size effects in $\mathcal{N}$ than in $\mathcal{M}$, so we opt for an unbalanced choice:
\begin{equation}
	\mathcal{N} = 262\,144 \qquad , \qquad \mathcal{M} = 512
\end{equation}
However, different algorithms for the reweigh would in general yield different finite-size effects in $\mathcal{N}$ and $\mathcal{M}$~\cite{MezardParisi2001}, so the necessity of taking $\mathcal{N}\gg\mathcal{M}$ is not a general feature. Then, the reweighing factor $r$ is dynamically chosen in the range $[2,10]$ in order to avoid ``twins'' in the populations and hence reduce their statistical significance.

\begin{figure}[!t]
	\centering
	\includegraphics[scale=1]{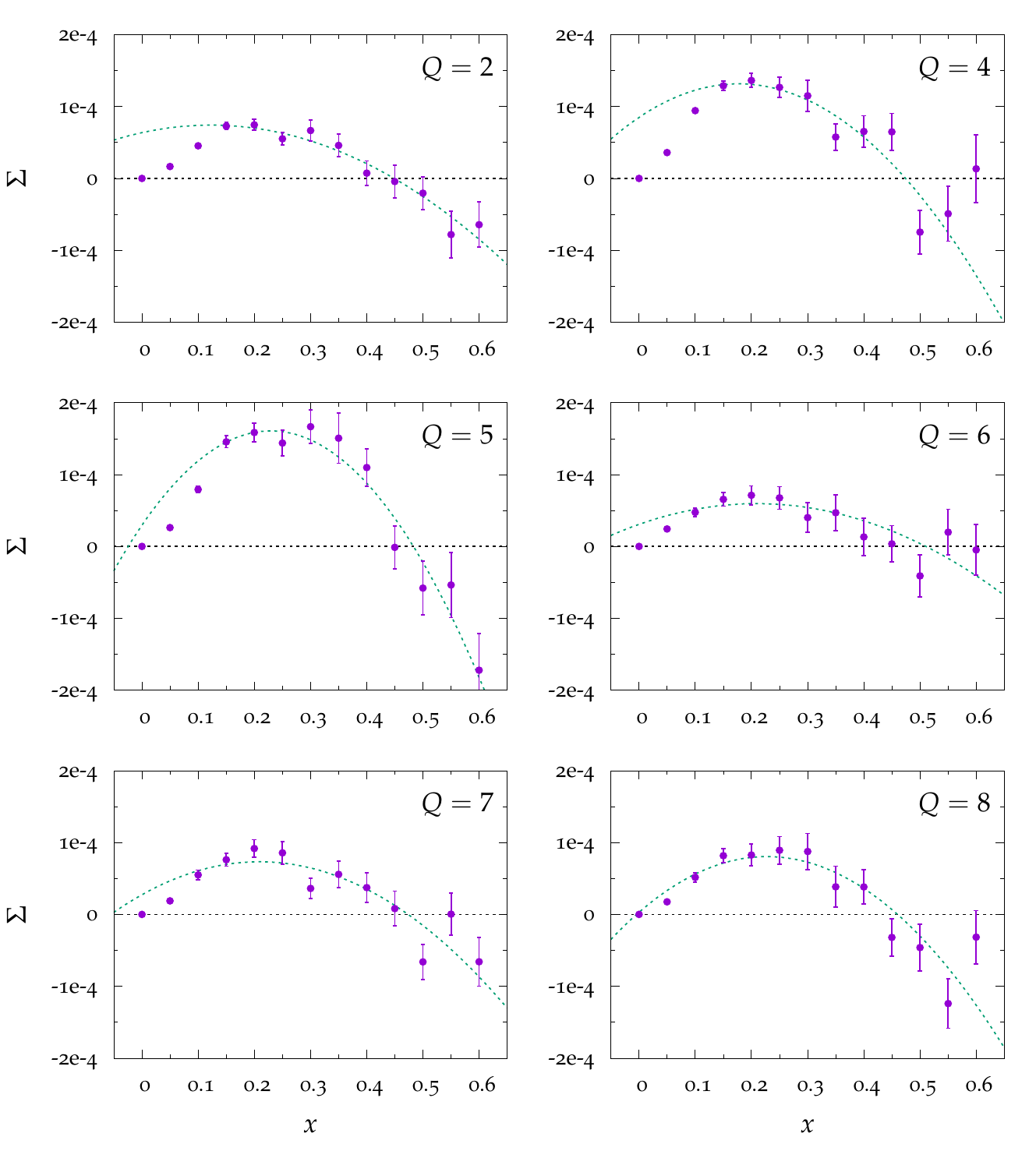}
	\caption[1RSB complexity of the $Q$-state clock model]{Plot of the complexity $\Sigma_{\beta}(x)$ for the $Q$-state clock model with unbiased bimodal couplings ($p=1/2$) at the reduced temperature $\tau=1/2$ on the $C=3$~\acrshort{RRG} ensemble. Plot ranges are the same for all panels, allowing the comparison between the different values of $Q$. Green dashed lines correspond to the fitting quadratic curve used for the estimation of the Parisi parameter $x^*$ such that $\Sigma_{\beta}(x^*)=0$.}
	\label{fig:Sigma_1RSB}
\end{figure}

Resulting complexities $\Sigma_{\beta}(x)$ for $\beta=2\beta_c$ are plotted in~\autoref{fig:Sigma_1RSB} for all the values of $Q$ analyzed. The $x$ and $y$ ranges are the same for each panel, so to allow a direct comparison between the different values of $Q$. Large errors are due to the fact that we are measuring a very small complexity, being of order $10^{-4}$. Together with $\Sigma_{\beta}(x)$, in each panel we also report the quadratic function used for estimating the value of the Parisi parameter $x^*$ where complexity vanishes. The value of $x^*$ corresponding to each choice of $Q$ is reported in~\autoref{tab:1RSB_values}.

\begin{table}[!b]
	\setlength{\tabcolsep}{8pt}		
	\centering
	\caption[1RSB parameters of the $Q$-state clock model]{\acrshort{1RSB} parameters of the $Q$-state clock model with unbiased bimodal couplings ($p=1/2$) at the reduced temperature $\tau=1/2$ on the $C=3$~\acrshort{RRG} ensemble.}
	\label{tab:1RSB_values}
	\begin{tabular}{cccc}
		\toprule
		$Q$ & $x^*$ & $q_0$ & $q_1$\\
		\midrule
		2 & 0.45(1) & 0.497(4) & 0.748(1)\\
		4 & 0.47(2) & 0.506(9) & 0.750(2)\\
		5 & 0.48(1) & 0.427(5) & 0.700(1)\\
		6 & 0.51(3) & 0.499(9) & 0.685(3)\\
		7 & 0.47(2) & 0.447(8) & 0.666(2)\\
		8 & 0.46(2) & 0.449(7) & 0.664(2)\\
		\bottomrule
	\end{tabular}
\end{table}

\begin{figure}[!t]
	\centering
	\includegraphics[scale=1]{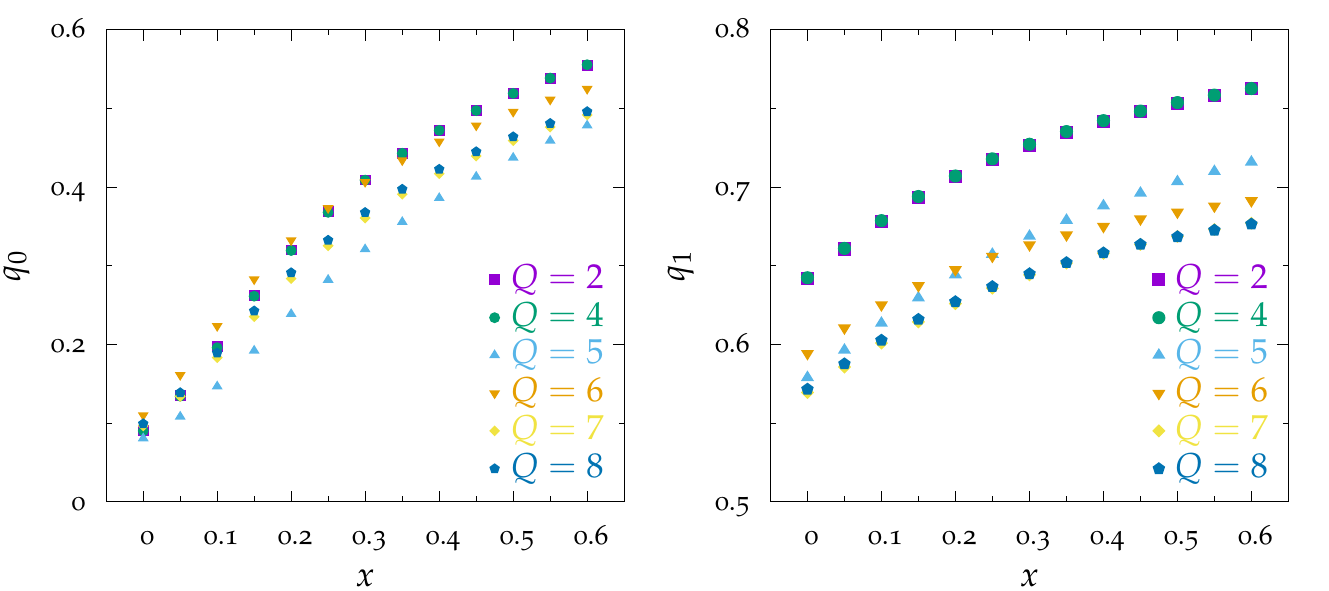}
	\caption[1RSB overlaps of the $Q$-state clock model]{Plot of the overlaps $q_0(x)$ (left panel) and $q_1(x)$ (right panel) for the $Q$-state clock model with unbiased bimodal couplings ($p=1/2$) at the reduced temperature $\tau=1/2$ on the $C=3$~\acrshort{RRG} ensemble. Please notice the different ranges on the $y$ axis. Error bars are always smaller than the symbol size.}
	\label{fig:Overlap_1RSB}
\end{figure}

In the same table we also report the values of the two overlaps $q_0$ and $q_1$ at $x=x^*$, while the entire curves $q_0(x)$ and $q_1(x)$ are plotted in~\autoref{fig:Overlap_1RSB}. In this case error bars given by statistical fluctuations are by far smaller than the symbol size, and hence the uncertainties reported in~\autoref{tab:1RSB_values} are completely due to the error in the location of $x^*$. We notice that both overlaps for $Q=2$ and $Q=4$ coincide, due to the fact that the $4$-state clock model is nothing but a ``double'' Ising, as already explained above. More remarkably, also the data for $Q=7$ and $Q=8$ already coincide, suggesting a very strong convergence of the clock model toward the XY model also in the \acrshort{1RSB} ansatz, after a short transient for $Q=5$ and $Q=6$.

A direct comparison with the corresponding fully connected results can be done only for $Q=2$, namely the \acrshort{SK} model. At the reduced temperature $\tau=1/2$, the \acrshort{1RSB} solution returns the following parameters~\cite{Parisi1980a}:
\begin{gather*}
	x^* \simeq 0.28 \qquad , \qquad q_0(x^*) \simeq 0.213 \qquad , \qquad q_1(x^*) \simeq 0.619
\end{gather*}
which are rather different from the ones listed in~\autoref{tab:1RSB_values}. So the~\acrshort{1RSB} solution in the sparse case is quite far from the one in the fully connected case. Nonetheless, similarly to what has been observed on the fully connected topology by Nobre and Sherrington~\cite{NobreSherrington1986}, also in the sparse case the \acrshort{1RSB} parameters vary little for $Q \geqslant 5$, being compatible with $Q$-independent values within the error bars. Only the $Q=6$ case shows a peculiar trend, maybe due to some reminiscence of the $Q=3$ case.

So we can finally conclude that the very fast convergence of the $Q$-state clock model toward the XY model is actually preserved even deeply in the~\acrshort{RSB} region when taking into account the breaking into many states, providing the same values for the \acrshort{1RSB} parameters $x^*$, $q_0$ and $q_1$ and hence presumably the same universality class of the XY model already from $Q=8$ --- willing to have a precautionary attitude.

Similar results have been also obtained in finite dimension, where the $Q$-state clock model on a $d=3$ cubic lattice has been studied by means of the Migdal - Kadanoff renormalization group approximation~\cite{IlkerBerker2014}: $Q=2$ and $Q=4$ have the same critical exponents, $Q=3$ has a peculiar behaviour and finally $Q \geqslant 5$ behaves like the XY model, providing the same critical exponents.

\clearpage{\pagestyle{empty}\cleardoublepage}

\begingroup
	\makeatletter
	\let\ps@plain\ps@empty
	\part{The XY model in a field}
	\label{part:XYmodelInField}
	\cleardoublepage
\endgroup

\chapter{The random field XY model}
\label{chap:RFXY}
\thispagestyle{empty}

At this point, it is clear that the behaviour of vector spin glasses is very different from the one of scalar spin glasses, especially the Ising ones. This difference is particularly evident in the very low-temperature and weak-disorder region, where the possibility for continuous models to exhibit small fluctuations and hence to easily adapt along several directions is not present for discrete models. As a direct consequence of this, we found that smaller quantities of disorder are enough to break the replica symmetry with respect to the scalar case, and the smaller $T$, the closer such threshold to the pure ferromagnet. In this sense, continuous models are more glassy than discrete models.

In this Chapter, we change the point of view used so far, inserting the quenched disorder no longer via the couplings, but via an external field that randomly varies from site to site. The resulting model, namely the random field XY model, is again studied on diluted graphs via the \acrshort{BP} approach, then computing the transition line in the field versus temperature plane between the disordered high-temperature phase and the long-range ordered low-temperature phase. Even though the analogous model with scalar spins, i.\,e. the random field Ising model, has been proved to always provide a \acrshort{RS} stable solution via a rigorous analytic argument, the same does not forbid --- at least in principle --- continuous-variable models to exhibit \acrshort{RSB} solutions. To this aim, the low-temperature -- low-field regime for the random field XY model is carefully analyzed, eventually finding remarkable results about the critical properties of this model, especially in the very low-temperature limit.

\section{From random couplings to random fields}

In order to better characterize the critical properties of continuous models in presence of quenched disorder, we change the point of view used so far. Indeed, in previous chapters the quenched disorder has been introduced by acting on the interaction between nearest-neighbour variables, focusing in particular on the bimodal distribution $\mathbb{P}_J$ and on the gauge glass distribution $\mathbb{P}_{\omega}$.

Now, we use a ferromagnetic exchange coupling $J>0$ for each edge of the underlying graph and then we insert a random quenched external field acting on each site. In this way, we can refer to our model as the \textit{random field XY model}, described by the following Hamiltonian:
\begin{equation}
	\mcH[\{\theta\}]= - J\sum_{(i,j)}\cos{(\theta_i-\theta_j)} - \sum_i H_i\cos{(\theta_i-\phi_i)}
\end{equation}
with $H_i$ and $\phi_i$ labeling respectively the field strength and the field direction on each site:
\begin{equation}
	\boldsymbol{H}_i = H_i\,e^{\mathfrak{i}\phi_i} \quad, \qquad \phi_i\in[0,2\pi)
\end{equation}

In order to fully exploit the continuous symmetry possessed by the XY model, the field should not possess a unique orientation $\phi_i=\overline{\phi}$ for all the sites, with the strength $H_i$ e.\,g. Gaussian distributed. Indeed, this choice would imply a too strong anisotropy, making the XY model behaviour too close to the Ising one. Instead, it is the direction of the field that has to be randomly chosen for each site. Then, once the field direction is random, its strength can be taken constant without significantly changing the physics, since gauge invariance holds~\cite{Book_Nishimori2001}. In this way, the more pronounced glassy behaviour of vector spin glasses with respect to scalar ones should be the most evident possible.

The need of a field which must possess a direction randomly changing from site to site in order to recover some peculiar features of the continuous symmetry --- while random magnitude is not essential in this sense --- has been firstly pointed out by Sharma and Young~\cite{SharmaYoung2010}, referring to fully connected vector spin glasses in a field. We will deeply investigate the analogous scenario for the XY model on diluted graphs in~\autoref{chap:XYinField}, taking into account different probability distributions for the field direction.

So we eventually set the field strength to be equal to $H$ for all sites, obtaining the following Hamiltonian for the random field XY model:
\begin{equation}
	\mcH[\{\theta\}]= - J\sum_{(i,j)}\cos{(\theta_i-\theta_j)} - H\sum_i\cos{(\theta_i-\phi_i)}
	\label{eq:Hamiltonian_XY_sparse_in_Field_Jferro}
\end{equation}
with $\phi_i$'s randomly drawn from the uniform probability distribution over the unit circle:
\begin{equation}
	\mathbb{P}_{\phi}(\phi_i) = \frac{1}{2\pi} \quad , \qquad \phi_i\in[0,2\pi)
\end{equation}
in order to make the most random possible the field arrangement on the system.

As usual, we can now write down the corresponding~\acrshort{BP} equations and then solve them, so characterizing the behaviour of this model on sparse random graphs. However, before going on along this direction, let us briefly remind some key results on the random field Ising model, so to later make a comparison with our model.

\section{A brief overview of the random field Ising model}
\label{sec:overview_RFIM}

The~\acrfull{RFIM} maybe represents the simplest way to introduce some disorder in a purely ferromagnetic system, so to simulate the presence of impurities within an ordered substrate. Initially introduced by Larkin in order to model the pinning of vortices in superconductors~\cite{Larkin1970}, then it has been exploited in several and apparently unrelated fields, such as diluted antiferromagnets in a homogeneous external field~\cite{FishmanAharony1979}, binary liquids in porous media~\cite{deGennes1984, VinkEtAl2006}, strongly correlated electronic systems~\cite{EfrosShklovskii1975, KirkpatrickBelitz1994, Dagotto2005}, hysteresis and avalanches~\cite{SethnaEtAl1993}, opinion dynamics and social interactions~\cite{Galam1997, MichardBouchaud2005}, just listing a few applications.

Unfortunately, despite its simplicity, a full understanding of its static and dynamical properties is still lacking. In finite dimension, e.\,g., a few points have been finally clarified only after a long debate. Among them, the existence of a long-range order for $d>2$~\cite{ImryMa1975, BricmontKupiainen1987}, and the settlement of its upper critical dimension $d^u_c$ to $6$, as firstly claimed by the famous \textit{dimensional reduction} argument~\cite{ParisiSourlas1979}.

However, for a long time it has not been clear if below $d=6$ --- and in particular in the $d=3$ case --- a spin glass phase occurs inbetween the paramagnetic and the ferromagnetic phases. Large $m$-expansion~\cite{MezardParisi1990, MezardYoung1992, MezardMonasson1994} as well as perturbative~\cite{DeDominicisEtAl1995, BrezinDeDominicis1998} and nonperturbative~\cite{ParisiDotsenko1992} field theory approaches seem to validate the presence of this~\acrshort{RSB} phase, while no evidences of it have been found in numerical simulations~\cite{NewmanBarkema1996, MiddletonFisher2002, ParisiEtAl1999}. Moreover, no hints come from the fully connected case --- where spin glass phase does not occur~\cite{SchneiderPytte1977} --- as well as from the sparse case --- where some works claim its existence~\cite{Shapir1986, Bruinsma1986, deAlmeidaBruinsma1987, PastorEtAl2002}, other do not~\cite{Thouless1986a, Thouless1986b, HaseEtAl2005} and further ones just remain inconclusive~\cite{BleherEtAl1998, NowotnyEtAl2001}, though providing some bounds for its location.

A full point has been finally set only quite recently by Krzakala, Ricci-Tersenghi and Zdeborov\'a~\cite{KrzakalaEtAl2010}, which showed that the spin glass susceptibility $\chi_{\text{SG}}$
\begin{equation}
	\chi_{\text{SG}} \equiv \frac{\beta^2}{N}\sum_{i,j}\Bigl(\braket{\sigma_i\sigma_j}_c\Bigr)^2
	\label{eq:chi_SG}
\end{equation}
is always upper bounded by the ferromagnetic susceptibility $\chi_{\text{F}}$
\begin{equation}
	\chi_{\text{F}} \equiv \frac{\beta}{N}\sum_{i,j}\braket{\sigma_i\sigma_j}_c
	\label{eq:chi_F}
\end{equation}
for any topology, any dimension of the lattice and any arrangement of the external field. This automatically implies that \textit{at thermal equilibrium} no spin glass phase can actually take place in the~\acrshort{RFIM} out of those points where $\chi_{\text{F}}$ itself diverges. Slightly later, this result has been extended also to the Ginzburg-Landau model~\cite{KrzakalaEtAl2011}.

The key point of their argument is the demonstration of the positive semidefiniteness of the two-point connected correlation functions for scalar spins:
\begin{equation}
	\braket{\sigma_i\sigma_j}_c \equiv \braket{\sigma_i\sigma_j} - \braket{\sigma_i}\braket{\sigma_j} \geqslant 0
\end{equation}
provided all the couplings are nonnegative, where $\braket{\cdot}$ refers as usual to the thermal average. This is nothing but a special case of the well known Fortuin, Kasteleyn and Ginibre inequality~\cite{FortuinEtAl1971}.

Unfortunately, the previous argument is valid only for scalar spins, so that nothing can be inferred about the \acrshort{RS} stability in vector spins models, which could at variance exploit transverse fluctuations to yield negative connected correlation functions even in presence of positive couplings.

\section{Back to the XY model}

Let us now come back to the random field XY model, finally analyzing its properties on sparse random graphs. To this aim, we can as usual exploit the~\acrshort{BP} algorithm in the \acrshort{RS} ansatz, whose self-consistency equations on a given instance read:
\begin{equation}
	\eta_{i\to j}(\theta_i) = \frac{1}{\mathcal{Z}_{i\to j}}\,e^{\,\beta H\cos{(\theta_i-\phi_i)}}\prod_{k\in\partial i\setminus j}\int \di\theta_k\,e^{\,\beta J\cos{(\theta_i-\theta_k)}}\,\eta_{k\to i}(\theta_k)
	\label{eq:BP_eqs_XY_Field_Jferro}
\end{equation}
while the disorder-averaged description can be again attained via the \acrshort{RS} \acrshort{PDA}.

Two extremal behaviours for the solutions of these equations can be suddenly analyzed. Indeed, in the high-field limit, each marginal is aligned along the direction of the local field $\boldsymbol{H}_i$, since the messages from the neighbours become quite negligible. Instead, in the high-temperature limit, each marginal is nearly flat, since thermal fluctuations overcome the ordering effect due to both the local field and the couplings with the neighbours. These two extremal cases, as well as the intermediate ones, belong to the high-$H$ -- high-$T$ region that is said to be paramagnetic.

\subsection{The paramagnetic phase}

Unfortunately, it is clear that the paramagnetic solution is not the uniform one over the unit circle --- unless we are exactly on the $H=0$ axis --- and hence it is not possible to exploit a Fourier expansion as we did in~\autoref{chap:XYnoField} in order to detect the instability line of this solution. So the~\acrshort{BP} equations~\autoref{eq:BP_eqs_XY_Field_Jferro} have to be solved numerically also in this region.

The numerical tools to accomplish this task have been largely described in~\autoref{chap:XYnoField}. In particular, the~\acrshort{PDA} allows us to find the~\acrshort{BP} fixed-point probability distribution $\mathbb{P}^*_{\eta}$ over the cavity messages, while via the~\acrshort{SuscProp} algorithm we can check its stability. Again, during numerical simulations, the $[0,2\pi)$ interval has to be discretized; by exploiting the results obtained in~\autoref{chap:clock}, we can efficiently approximate the XY model via a $Q=64$-state clock model, committing a negligible error.

\begin{figure}[!t]
	\centering
	\includegraphics[scale=1]{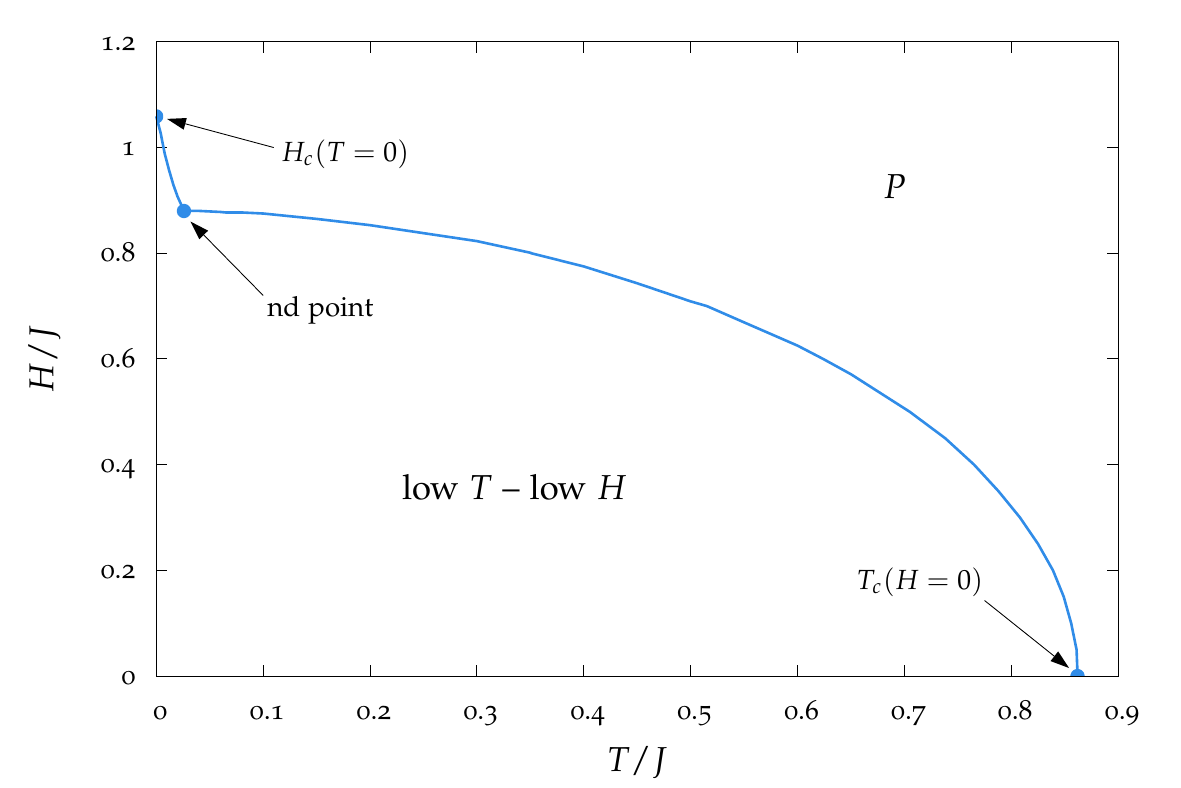}
	\caption[Instability line of the paramagnetic solution in the RFXY model]{Instability line of the paramagnetic solution in the random field XY model, computed on the $C=3$~\acrshort{RRG} ensemble. A nondifferentiability (nd) point can be clearly seen close to the $T=0$ axis, which may represent a multicritical point with further instability lines departing from it.}
	\label{fig:stability_line_RFXY}
\end{figure}

The resulting instability line for the paramagnetic solution is reported in~\autoref{fig:stability_line_RFXY} for the $C=3$~\acrshort{RRG} case. The following values are the two corresponding endpoints, respectively on the zero-temperature axis and on the zero-field axis:
\begin{gather*}
	T_c(H=0)/J = 0.863(1) \qquad , \qquad H_c(T=0)/J = 1.058(2)
\end{gather*}
where the value given here for the zero-field critical temperature is in agreement with its analytic expression for a pure ferromagnet, $(C-1)I_1(\beta_c J)/I_0(\beta_c J)=1$, found in~\autoref{chap:XYnoField}. In addition, a quite unexpected singularity point (which we refer to as `nd', namely `nondifferentiability') for such instability line has been found close to the zero-temperature axis, with exact location:
\begin{gather*}
	T_{nd}/J = 0.026(2) \qquad , \qquad H_{nd}/J = 0.880(1)
\end{gather*}
It could be the signature of a bifurcation of the critical line: a multicritical point may hence be present, with further critical lines departing from it. So the picture of the low-$T$ -- low-$H$ region could not be so plain as it happens for the~\acrshort{RFIM}, where the only ferromagnetic phase takes place at the equilibrium~\cite{KrzakalaEtAl2010}.

\begin{figure}[p]
	\centering
	\includegraphics[scale=1]{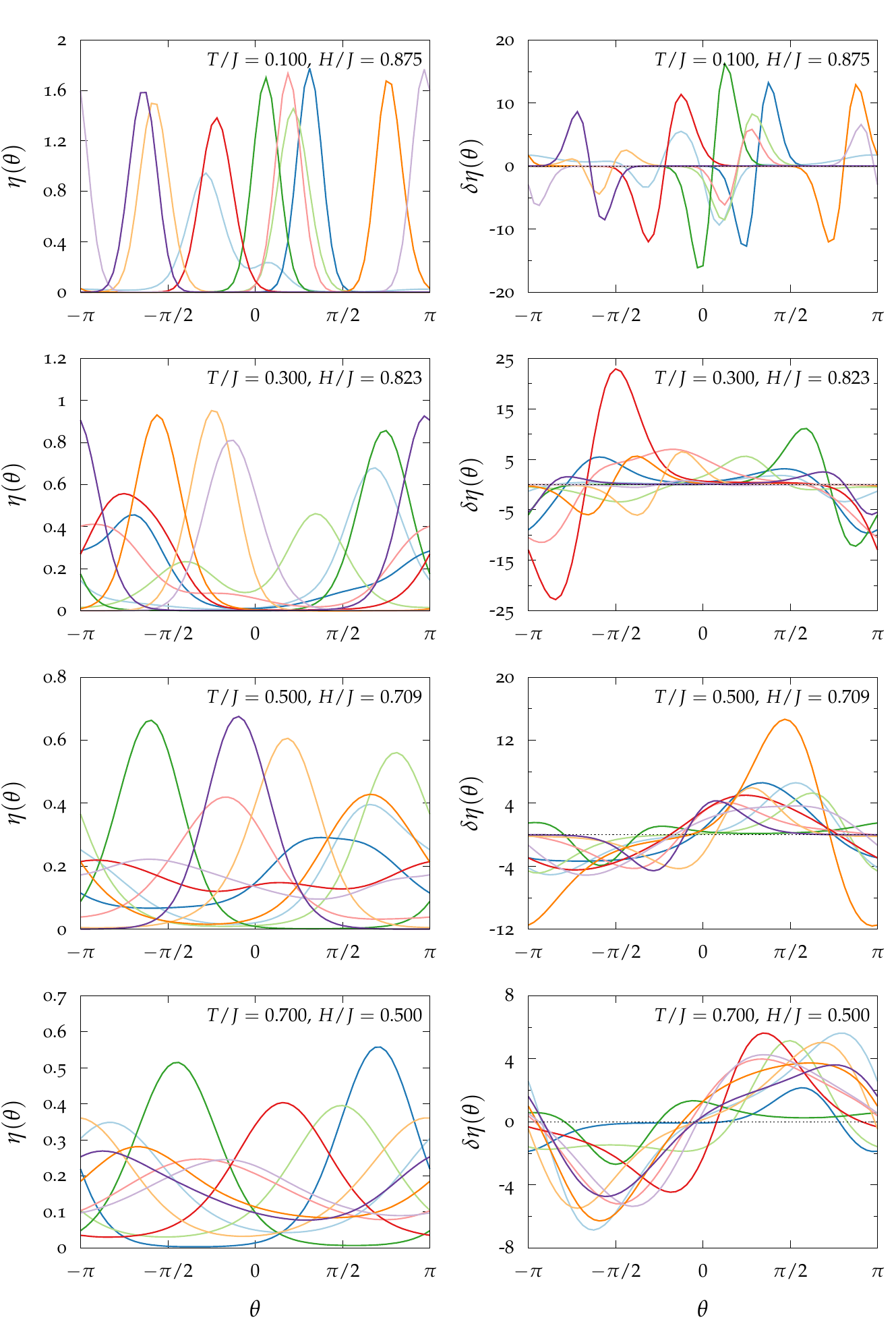}
	\caption[Marginals and perturbations along the critical line of the random field XY model]{Several full site marginals (left panels) and the corresponding perturbations (right panels) for some points along the critical line between the paramagnetic phase and the low-$T$ -- low-$H$ region for the random field XY model on the $C=3$ \acrshort{RRG} ensemble.}
	\label{fig:someMargPert_along_criticalLine}
\end{figure}

In~\autoref{fig:someMargPert_along_criticalLine} we then report several site marginals $\eta_i(\theta_i)$'s computed by sampling $C=3$ cavity marginals from the fixed-point probability distribution $\mathbb{P}^*_{\eta}$ --- together with the corresponding perturbations --- for some points along the critical line computed above. It is evident the growing polarization of marginals around the local direction of the field when moving toward the $T=0$ axis --- and hence toward larger values of $H$ ---, while they broaden when increasing the temperature and lowering the field strength. This is a quite general feature, that can be observed e.\,g. along the \acrshort{RS} instability line of the spin glass Ising model in a uniform field~\cite{ParisiEtAl2014}, where however a unique parameter $u_i$ --- instead of a probability distribution $\eta_i(\theta_i)$ --- is enough to describe each site marginal.

It is also interesting to notice that when the temperature is low enough, most of perturbations are well localized around the peaks of the corresponding marginals, linearly crossing the zero axis almost in correspondence of such peaks. This is a signature of perturbations that are \textit{transverse} with respect to the local external field, since they would just shift the peaks of the marginals almost without modifying their shape. At variance, a \textit{longitudinal} perturbation would fatten or shrink the peak without shifting it, and hence it should exhibit a quadratic shape in correspondence of the peak of the related marginal --- as actually occurring e.\,g. for the light-orange one at $T/J=0.1$. In the low-temperature limit, indeed, purely transverse or longitudinal perturbations are the most energetically favourable ones --- depending on the local effective field --- and hence they can be easily observed in the joint probability distribution $\mathbb{P}^*[(\eta,\delta\eta)]$. Instead, when temperature increases, perturbations become quite extended over the $[0,2\pi)$ interval, with no direct relation between the peaks of the related marginals and the points where they --- or their first derivatives --- vanish, hence providing no direct interpretation as purely transverse or longitudinal perturbations.

The transverse and longitudinal behaviour of perturbations will be lengthy discussed in~\autoref{chap:XYinField}, where the spin glass XY model in an external field will be analyzed. Notice that the low-temperature transverse perturbations mentioned above are not coherent with each other, due to the presence of the randomly oriented external field, so no global transverse symmetry is going to be broken when trespassing the critical line of~\autoref{fig:stability_line_RFXY}. At variance, for the spin glass XY model in a \textit{homogeneous} external field, a breaking of the global transverse symmetry with respect to field direction will take place, at the same time implying a breaking of the replica symmetry.

\subsection{The low-temperature -- low-field region}

\begin{figure}[!b]
	\centering
	\includegraphics[scale=1]{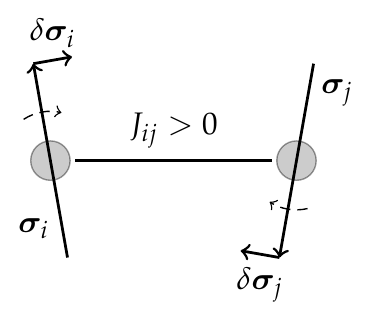}
	\caption[Negative correlations in the random field XY model]{Sketch of a configuration of two nearest-neighbour spins that yields a negative connected correlation between them. Indeed, a tiny clockwise rotation of $\boldsymbol{\sigma}_i$ produces a corresponding tiny clockwise rotation of $\boldsymbol{\sigma}_j$, provided the coupling $J_{ij}$ between them is positive. The corresponding spin fluctuations are then almost oppositely oriented, yielding a negative scalar product between them.}
	\label{fig:negative_corr_XY}
\end{figure}

The possible nontriviality of the low-$T$ -- low-$H$ region is also suggested by a second observation, already grasped in~\autoref{sec:overview_RFIM}. Indeed, if the~\acrshort{RFIM} has nonnegative connected correlation functions, $\braket{\sigma_i\sigma_j}_c \geqslant 0$, it is no longer true for the random field XY model. This claim can be easily understood considering the following situation, also sketched in~\autoref{fig:negative_corr_XY}. Suppose we have a spin $\boldsymbol{\sigma}_i$ oriented along the direction marked by a $\theta_i$ slightly larger than $\pi/2$. Then, let its neighbour $\boldsymbol{\sigma}_j$ be oriented along the direction given by a $\theta_j$ slightly smaller than $3\pi/2$. This is an equilibrium configuration for~$\boldsymbol{\sigma}_i$ and~$\boldsymbol{\sigma}_j$ for suitable values of the field intensity $H/J$ and the temperature $T/J$ and for a suitable orientation of the two fields $\boldsymbol{H}_i$ and $\boldsymbol{H}_j$. Then, if we perturb the orientation of the first spin by a small angle $\delta\theta_i<0$, then the positive coupling~$J_{ij}$ makes the second spin rotate as well by a small angle $\delta\theta_j<0$. However, the corresponding spin fluctuations $\delta\boldsymbol{\sigma}_i$ and $\delta\boldsymbol{\sigma}_j$ are oriented almost the opposite, so eventually producing a negative connected correlation between the two spins:
\begin{equation}
	\braket{\boldsymbol{\sigma}_i\cdot\boldsymbol{\sigma}_j}_c \equiv \braket{\delta\boldsymbol{\sigma}_i\cdot\delta\boldsymbol{\sigma}_j} < 0
\end{equation}
This is a direct consequence of the continuous nature of vector spins, that are allowed not only to flip along a certain direction, but more easily to rotate by a generic angle. Indeed, the generic connected correlation function actually becomes a $m\times m$ matrix for $m$-component vector models:
\begin{equation}
	\braket{\sigma_{i,\mu}\sigma_{j,\nu}}_c \equiv \braket{\delta\sigma_{i,\mu}\delta\sigma_{j,\nu}} \quad , \qquad \mu,\nu=1,\dots,m
\end{equation}
with the $m$ corresponding eigenvalues that can be both positive or negative as well. This observation implies that the argument of Ref.~\cite{KrzakalaEtAl2010} is no longer valid for vector spins. Hence, the presence of a spin glass phase can not be ruled out for the random field XY model.

So let us now solve the~\acrshort{BP} equations in the low-$T$ -- low-$H$ region, keeping in mind the two evidences found before: a candidate multicritical point ($T_{nd},H_{nd}$) and the possible presence of a spin glass phase, likely for $T\leqslant T_{nd}$.

From~\autoref{chap:XYnoField} we already expect a pure ferromagnetic phase on the $H=0$ axis for $T<T_c(H=0)$, and the same should hold in the low-temperature region where $H \ll T$. This is confirmed by the analysis of the~\acrshort{BP} fixed point $\mathbb{P}^*[\eta]$ reached via the~\acrshort{PDA}, which shows a global magnetization $m$ substantially larger than zero.

\begin{figure}[!t]
	\centering
	\includegraphics[scale=1]{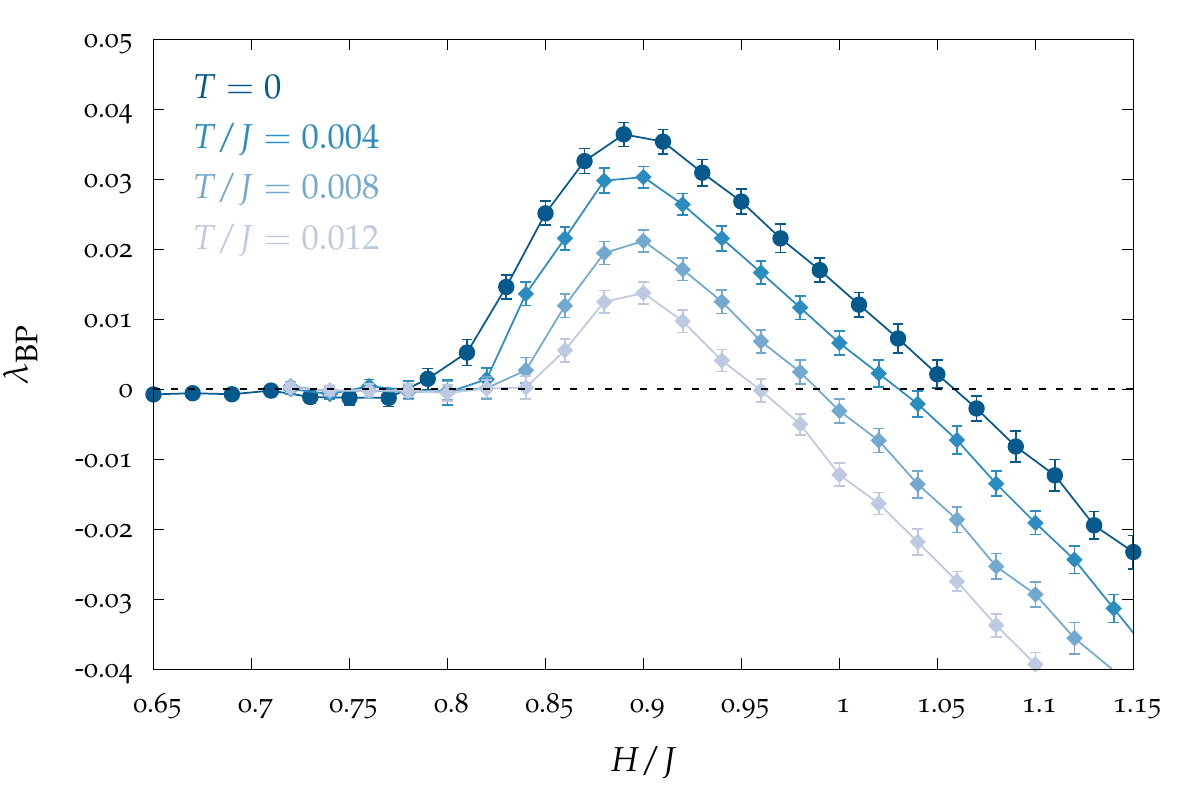}
	\caption[Instability parameter of the RFXY model close to zero temperature]{Instability parameter $\lambda_{\text{BP}}$ as a function of the field strength $H$ for several values of the temperature $T$ in the random field XY model on the $C=3$ \acrshort{RRG} ensemble. The presence of a \acrshort{RSB} phase in a small range of field values centered on $H \simeq 0.9$ for very low values of $T$ can be clearly detected.}
	\label{fig:lambda_BP_RFXY}
\end{figure}

Instead, when $H$ becomes much larger than $T$, it is not obvious if the ferromagnetic phase is still stable, though it is so for the \acrshort{RFIM}. In particular, we have to be very careful in the region close to the $T=0$ axis suggested by the location of the nondifferentiability point found before. So we fix the temperature $T<T_{nd}$ and study the stability parameter~$\lambda_{\text{BP}}$ via an annealing protocol in $H$ via~\acrshort{SuscProp}, obtaining the curves $\lambda_{\text{BP}}(H)$ in~\autoref{fig:lambda_BP_RFXY} for several values of~$T$. It is clear that for very low values of~$T$ there is a whole range of values of~$H$ for which the \acrshort{RS} \acrshort{BP} fixed point is unstable, strongly suggesting the presence of a \acrshort{RSB} phase. However, this spin glass phase is very tiny, and shrinks to zero as soon as the temperature becomes larger than $T_{nd}$.

The evaluation of the curves $\lambda_{\text{BP}}(H)$ in the very low-temperature limit, eventually down to the $T=0$ axis, is a good opportunity to further check the zero-temperature \acrshort{SuscProp} algorithms developed in~\autoref{chap:XYnoField} for the XY model and in~\autoref{chap:clock} for the $Q$-state clock model. Indeed, when comparing the evolution of perturbations in the two cases, we already highlighted as the underlying mechanism is different, according to the continuous or discrete nature of spin variables. It is worth reminding it here: in the former case, no perturbation is identically equal to zero, while it is the global norm $\norm{\delta h}$ that grows up or shrinks according to the stability of the \acrshort{RS} fixed point $\mathbb{P}^*[h]$; in the latter case, at variance, perturbations divide in two well defined groups, the identically vanishing ones and the ones different from zero, with the global norm of the latter that stays approximately constant under \acrshort{BP} iterations, while it is their fraction that decays --- or not --- according to the \acrshort{RS} stability. Finally, in order to mimic the evolution of perturbations of the zero-temperature XY model via its discrete approximation --- the $Q$-state clock model ---, we realized that the way out was to evaluate the real-valued maximizers in the zero-temperature \acrshort{BP} equations by interpolating among the discrete directions allowed by the $Q$-state clock model.

The zero-temperature curve $\lambda_{\text{BP}}(H)$ in~\autoref{fig:lambda_BP_RFXY} has been computed just in this way, and it perfectly matches with the extrapolation over the finite-temperature curves in the $T\to 0$ limit. Conversely, if we had evaluated the maximizers just over the discrete set of directions, the corresponding curve $\lambda_{\text{BP}}(H)$ at $T=0$ would not have matched with the previous extrapolation, wrongly providing a much larger region of \acrshort{RS} instability for our model.

\begin{figure}[!t]
	\centering
	\includegraphics[scale=1]{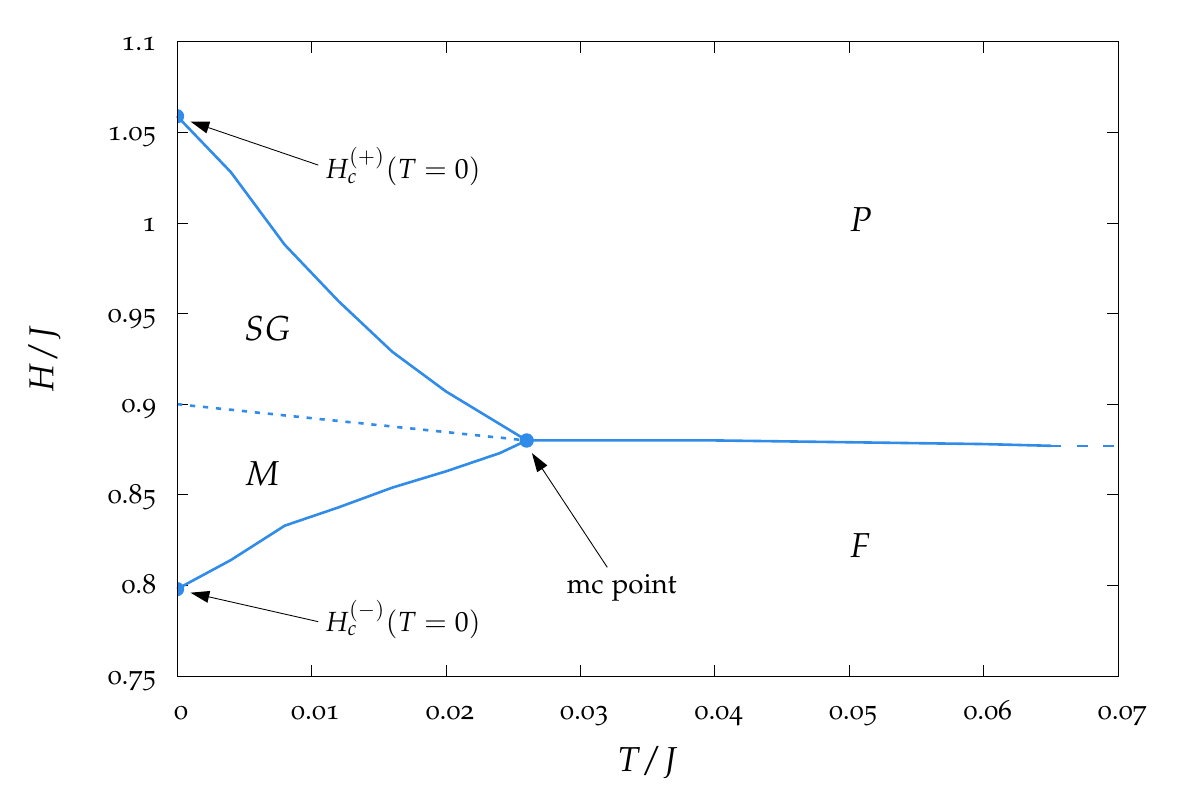}
	\caption[Phase diagram of the RFXY model]{Complete phase diagram of the random field XY model close to the zero-temperature axis. A nontrivial structure in the low-$T$ -- low-$H$ region is present, with a \acrshort{RS} ferromagnetic phase ($F$), an unbiased \acrshort{RSB} spin glass phase ($SG$) and a magnetized \acrshort{RSB} mixed phase ($M$). All numerical values refer to the $C=3$ \acrshort{RRG} ensemble.}
	\label{fig:phase_diagram_RFXY}
\end{figure}

At this point, we can actually recognize the nondifferentiability point $(T_{nd},H_{nd})$ found before as a multicritical point, where the second instability line departing from it separates the \acrshort{RSB} spin glass phase from the \acrshort{RS} ferromagnetic phase. The resulting phase diagram, restricted to the nontrivial region around the multicritical point, is finally depicted in~\autoref{fig:phase_diagram_RFXY}. The two critical values of the field strength on the zero-temperature axis are given for a $C=3$ \acrshort{RRG} by:
\begin{gather*}
	H^{(-)}_c(T=0)/J = 0.80(1) \qquad , \qquad H^{(+)}_c(T=0)/J = 1.058(2)
\end{gather*}
On the right side, the $\lambda_{\text{BP}}(H)$ curves linearly cross the zero axis --- as it typically occurs when coming from the paramagnetic fixed point --- so resulting in a precise evaluation of $H^{(+)}_c(T=0)$. At variance, on the left side, they slowly approach the zero axis rather than crossing it, implying a more noisy evaluation of $H^{(-)}_c(T=0)$. This behaviour is due to the marginal stability of the \acrshort{RS} ferromagnetic phase --- which will be discussed in a while --- hence implying a critical slowing down in the thermalization toward the \acrshort{BP} fixed point during the annealing protocol in the field strength $H$. The smooth behaviour of the $\lambda_{\text{BP}}(H)$ curves close to the critical point is indeed due to a not perfect thermalization in that region. Finally, $H^{(-)}_c(T=0)$ can be evaluated via a linear extrapolation over the data points in the region $H>H^{(-)}_c(T=0)$, discarding those points mostly affected by the proximity to the critical point.

The presence of a spin glass phase in the random field XY model with ferromagnetic couplings was rather unexpected, maybe due to an improper extension of the argument developed for the \acrshort{RFIM} also to this model. However, we showed that the continuous nature of vector spins allows the appearance of \textit{negative effective couplings} between the spin fluctuations around their equilibrium configuration, so making incorrect the claim $\chi_{\text{F}} \geqslant \chi_{\text{SG}}$. On the other hand, in previous Chapters we already realized that the XY model is more glassy than the Ising model for very low temperatures, just due to the continuous nature of the vector spins, and now we provide a further evidence of this strongly different behaviour.

Quite remarkably, the curve of the global magnetization $m(H)$ at a fixed temperature~$T \leqslant T_{nd}$ goes to zero exactly at the center of the spin glass phase, namely in correspondence of the maximum instability of the \acrshort{BP} fixed point. So in principle the tiny spin glass phase occurring for the random field XY model should be split in an unbiased spin glass phase, with $m=0$, and a mixed phase, with $m>0$, exactly as done for the spin glass XY model and the spin glass clock model in absence of a field. However, also in this case the line dividing these two~\acrshort{RSB} phases is just an approximation, since in order to compute it we are using the \acrshort{RS} ansatz though in presence of the \acrshort{RSB}: that is why this line is dashed in the phase diagram of~\autoref{fig:phase_diagram_RFXY}.

\begin{figure}[!t]
	\centering
	\includegraphics[scale=1]{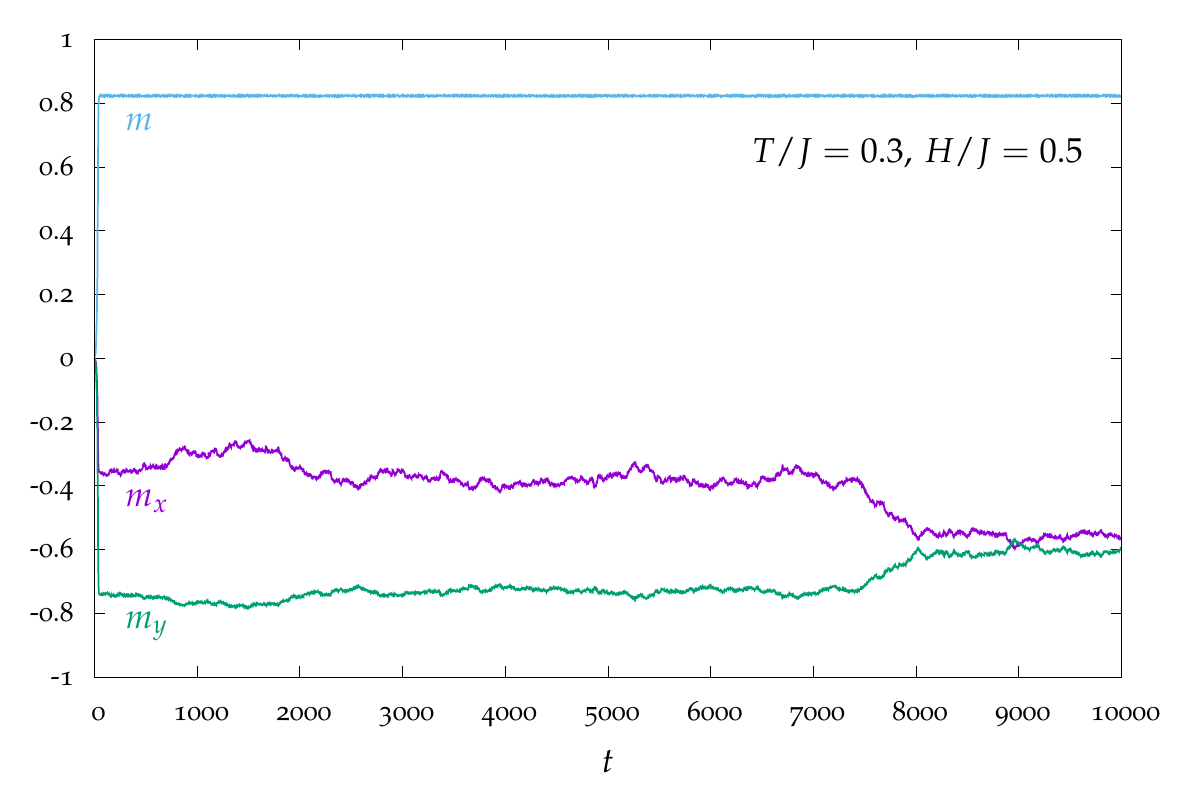}
	\caption[Magnetization of the RFXY model in the ferromagnetic phase]{Global magnetization of the random field XY model in the deep ferromagnetic phase ($T/J=0.3,\,H/J=0.5$) during \acrshort{BP} iterations in the \acrshort{PDA} over the $C=3$ \acrshort{RRG} ensemble. The fixed point is reached after few tens of iterations, hence the modulus $m$ reaches its stationary value and stays constant within statistical fluctuations. Instead, due to the marginality of the XY model in this phase, the direction of the global magnetization continues to change even at large times.}
	\label{fig:mag_RFXY}
\end{figure}

Another important feature of the random field XY model is the aforementioned marginality of the whole ferromagnetic phase, as shown by $\lambda_{\text{BP}} \simeq 0$ (to the best of our numerical evidences) in~\autoref{fig:lambda_BP_RFXY} at very low temperatures and for low values of the field strength $H$. It could be interpreted as a signature of the survival of the $\mathrm{O}(2)$ symmetry possessed by the XY model even in presence of the random field. Indeed, when~$H$ is larger than zero, one would expect the breaking of such symmetry; instead, it does not happen here, since the field direction is randomly distributed on each site and hence the coupling between the global magnetization~$\boldsymbol{m}$ and the field is weak. This implies that, during \acrshort{BP} iterations in the \acrshort{PDA}, the global magnetization $\boldsymbol{m}$ can rigidly rotate in the $xy$ plane, as shown in~\autoref{fig:mag_RFXY}, just due to the marginality highlighted above. As a further evidence, the marginally stable ferromagnetic phase suddenly becomes strictly stable as soon as too small values of $Q$ are used, so actually suppressing the soft modes provided by the $\mathrm{O}(2)$ symmetry.

A conclusive remark will finally suggest further investigations. Indeed, we found a \acrshort{RSB} region for the random field XY model on the \acrshort{RRG} ensemble, and the same would likely yield for the other classes of sparse random graphs, as well as if taking into account other vector models with $m \geqslant 3$ spin components (provided $m$ is still finite). At variance, in the fully connected limit, any ferromagnetic model --- neither with Ising spins nor with $m$-dimensional vector spins --- would not yield a \acrshort{RSB} solution, even if in presence of a random field. That is due to the fact that the $4$-spin interaction $\boldsymbol{\sigma}^{(a)}_i\boldsymbol{\sigma}^{(a)}_j\boldsymbol{\sigma}^{(b)}_i\boldsymbol{\sigma}^{(b)}_j$ --- which then couples different replicas via the overlap $q_{ab}$ --- just comes out when averaging over the Gaussian distribution of the exchange couplings, while the average over the Gaussian distribution of the fields just provides a pairwise interaction $\boldsymbol{\sigma}^{(a)}_i\boldsymbol{\sigma}^{(b)}_i$ --- eventually leading to the single-replica term $m_a$, with no coupling between different replicas. It is hence clear that the \acrshort{RSB} occurring in our model is a direct consequence of the combined action of \textit{both} the continuous nature of vector spins and the sparsity of the underlying network. It would be interesting, then, to study how the \acrshort{RSB} region shrinks and eventually disappears when increasing enough the (average) connectivity~$C$, approaching the fully connected limit. In addition, the enhancement of \acrshort{RSB} by the joint action of continuous variables and sparse networks could suggest interesting connections with some features of real glasses, as it will be further analyzed in~\autoref{chap:XYinField_zeroTemp}.

\clearpage{\pagestyle{empty}\cleardoublepage}

\chapter{The spin glass XY model in a field}
\label{chap:XYinField}
\thispagestyle{empty}

As we saw in the previous Chapters, when moving from Ising spins to vector spins, the resulting behaviour of the system can dramatically change, due to the basic observation that in the latter case the system has more degrees of freedom that can be excited, and in particular small fluctuations are allowed at very low temperatures. Hence, even in presence of a very weak disorder, vector spins can adapt more easily to several different directions, enhancing the instability toward the \acrshort{RSB}. The main consequence of this is a more pronounced glassiness of vector spin glasses with respect to discrete spin glasses, as seen at the end of~\autoref{chap:XYnoField}.

The scenario becomes richer when an external field is switched on, as seen in~\autoref{chap:RFXY} for the random field XY model with ferromagnetic couplings. The difference with respect to the scalar case is even more striking, since we detected a~\acrshort{RSB} phase in a tiny region of the field versus temperature plane, despite its absence in the corresponding Ising model.

Here we perform a further step forward, considering the presence of both spin glass couplings and an external field. The picture is now supposed to be even more cumbersome, due to the fact that different types of continuous symmetries can be broken, depending on the distribution of the external field, as already seen in the fully connected case (\autoref{chap:sg_replica}).

In the vector case, indeed, randomness can involve the field strength, its direction, or both them. We hence analyze two different kinds of external field, homogeneous or randomly varying in direction from site to site, analytically computing the resulting critical lines in the field versus temperature plane. Then, we also provide a physical interpretation of such instabilities in terms of transverse and longitudinal perturbations with respect to the local direction of the field. Finally, we study the crossover between the two regimes when changing the degree of randomness in the direction of the external field.

\section{Instabilities in the fully connected case}
\label{sec:XYinField_fully_conn}

The behaviour of vector spin glass models in a field has already been well characterized in the fully connected case via the replica method. Indeed, as we saw in~\autoref{sec:vector_sg_fully}, according to the distribution of the external field, different instabilities can be detected: the \acrfull{GT} one in presence of a homogeneous field and the \acrfull{dAT} one in presence of a random field. Here we briefly recap their features, referring to~\autoref{sec:vector_sg_fully} --- and to the original references therein --- for a more detailed discussion.

\subsection{A homogeneous field}

De Almeida and Thouless~\cite{deAlmeidaThouless1978} have been the first to study the effect of the presence of an external field on a spin glass, focusing on the Ising case ($m=1$). They found that the \acrshort{RS} solution is unstable in the low-temperature region not only when no field is present, but even for any value of the external field, given the temperature is low enough. In this way, the corresponding critical line of \acrshort{RS} instability can be detected in the $(T,H)$ plane, which has been later named~\acrfull{dAT} line.

The \acrshort{RS} instability also occurs in the vector spin glass case, but the first proof has been given in absence of a field~\cite{Thesis_deAlmeida1980, deAlmeidaEtAl1978}. It is the works by Gabay and Toulouse that faced this problem~\cite{ToulouseGabay1981, GabayToulouse1981}, finding that the presence of a uniform field identifies a global direction with respect to which identify a transverse and a longitudinal order parameter. Coming from the high-$H$ -- high-$T$ region, namely from the paramagnetic phase, the first instability occurs when the transverse overlap $q_{\perp}$ becomes different from zero, implying a breaking of the inversion symmetry for the transverse components of the spins. This line has been named~\acrfull{GT} after them. Its small-field expansion yields a characteristic square-root behaviour:
\begin{equation}
	H_{\text{GT}} \propto \tau^{1/2}
\end{equation}
with $\tau\equiv(T_c-T)$, where $T_c$ is the zero-field critical temperature between the paramagnetic and the spin glass phase. Then, by further lowering $T$ and/or $H$ in the 	\acrshort{RS} ansatz, a second critical line can be met, where it is the longitudinal overlap $q_{\parallel}$ that acquires a nontrivial probability distribution. This line was initially associated with the~\acrshort{dAT} line found in the Ising case, since also in the vector case it was claimed to correspond to the onset of the~\acrshort{RS} instability, even showing the same exponent for the small-field expansion:
\begin{equation}
	H_{\text{dAT}} \propto \tau^{3/2}
\end{equation}
In fact, further computations~\cite{CraggEtAl1982, GabayEtAl1982} showed that in the \acrshort{fRSB} ansatz this second line is just a crossover between a ``weak'' and a ``strong'' \acrshort{RSB} in the longitudinal overlap $q_{\parallel}$, namely in its dependence on the Parisi parameter $x$, while the~\acrshort{RSB} already occurs in correspondence of the \acrshort{GT} line via a ``strong'' dependence in~$x$ of $q_{\perp}$.

\subsection{A random field}

The picture dramatically changes when considering a random field. Indeed, already in the Ising case, the field strength can be independently chosen from site to site, so considering a further source of quenched disorder inside the system. In particular, the sign of the field is not crucial at all when couplings are both positive and negative, since it can be removed through a gauge transformation~\cite{Book_Nishimori2001}. Instead, its modulus determines the position of the instability line in the field -- temperature plane, namely the~\acrshort{dAT} line. According to the probability distribution $\mathbb{P}_{H}$ from which we draw the moduli, the~\acrshort{dAT} line moves, changing the exponent in its small-field expansion. If a homogeneous field yields a $3/2$ exponent, as well as a Gaussian distributed field with zero mean and $H^2$ variance~\cite{deAlmeidaThouless1978}, when $H_i$ is chosen proportionally to the $i$-th eigenvector component corresponding to the largest eigenvalue of the interaction matrix, then the \acrshort{dAT} exponent increases up to~$9/2$, so strongly suppressing the~\acrshort{dAT} line though without removing it~\cite{Bray1982}.

However, even if the value of the critical exponent changes, the nature of the~\acrshort{RS} instability along the~\acrshort{dAT} line does not change. Indeed, randomness in the sign or in the magnitude of the field affects only the \textit{longitudinal} order parameter --- referring to the unique direction, e.\,g. the $z$ axis, identified by the spins as well as by the field --- while no transverse degrees of freedom are present, whose freezing would eventually lead to the \acrshort{GT} critical line.

The situation remarkably changes when moving from the scalar to the vector case, since the randomness can also involve the direction of the external field, so enriching the picture of the possible instabilities. Indeed, if a homogeneous field allows the breaking of the transverse symmetry, causing the \acrshort{GT} instability, for a randomly oriented field there is no longer a strong directional anisotropy, so presumably causing the disappearance of the~\acrshort{GT} line. In such case, \acrshort{RSB} would eventually occur on the line previously corresponding to the crossover in~$q_{\parallel}$, namely the \acrshort{dAT} line, whose presence as a sharp transition could hence be recovered also in the vector case.

This crucial observation has been pointed out for the first time by Sharma and Young in 2010, still referring to the fully connected topology~\cite{SharmaYoung2010}. They explicitly recovered the small-field expansion of the~\acrshort{dAT} line performed by Gabay and Toulouse~\cite{GabayToulouse1981}\footnote{Notice that here the normalization $|\boldsymbol{\sigma}_i|^2=m$ have been used. By using the unitary norm --- as we do --- one gets a rescaling of both $J$ and $H$, so that the zero-field transition occurs at $T_c=J/m$.}:
\begin{equation}
	\left(\frac{H}{J}\right)^2 \simeq \frac{4}{m+2}\left(\frac{T_c-T}{T_c}\right)^3
	\label{eq:scaling_dAT_line_near_zeroH_fullyConnected_2}
\end{equation}
by using a vector field whose components $H_{i,\mu}$ are independently Gaussian distributed. Notwithstanding the same scaling, the two results are fundamentally different. The~\acrshort{dAT} line found by Sharma and Young by using a random field actually corresponds to the onset of the~\acrshort{RSB} instability, but no change in spin symmetry occurs on it. Instead, the candidate for the \acrshort{dAT} line found by Gabay and Toulouse by using the homogeneous field was rather a crossover in $q_{\parallel}$, as discussed above. So Ref.~\cite{SharmaYoung2010} is the first one to recognize the need for a field which is random in direction rather than in magnitude when looking for the~\acrshort{dAT} line of a vector spin glass. Notice, moreover, that it has exactly the same exponent and the same features of the \acrshort{dAT} line in the Ising case.

Another important remark regards the zero-temperature limit. When $m \leqslant 2$, the~\acrshort{dAT} line diverges, while for $m>2$ it actually touches the $T=0$ axis at a finite value of the critical field~\cite{SharmaYoung2010}, so getting closer to the finite dimensional case.

\section{The RS solution in the diluted case}
\label{sec:GT_dAT_diluted}

Let us now move back to the diluted case. The basic Hamiltonian of the XY model on a sparse random graph with spin glass couplings~$J_{ij}$'s and with an external field~$\boldsymbol{H}_i$ acting on each site reads:
\begin{equation}
	\mcH[\{\theta\}]= - \sum_{(i,j)}J_{ij}\cos{(\theta_i-\theta_j)} - \sum_i H_i\cos{(\theta_i-\phi_i)}
	\label{eq:Hamiltonian_XY_sparse_in_Field}
\end{equation}

Then, in~\autoref{sec:XYinField_fully_conn} we saw that the randomness of the field for vector spin models has to necessarily involve its direction in order to highlight the peculiarities with respect to the scalar case, while the randomness in magnitude is not essential in this sense. So we fix the strength of the field to $H$ for each site, without any loss of generality, letting the field randomness be entirely provided by the probability distribution $\mathbb{P}_{\phi}(\phi_i)$ of the field direction $\phi_i$ on each site. The Hamiltonian~\autoref{eq:Hamiltonian_XY_sparse_in_Field} so becomes:
\begin{equation}
	\mcH[\{\theta\}]= - \sum_{(i,j)}J_{ij}\cos{(\theta_i-\theta_j)} - H\sum_i\cos{(\theta_i-\phi_i)}
	\label{eq:Hamiltonian_XY_sparse_in_Field_Hfixed}
\end{equation}
with the quenched couplings $J_{ij}$'s distributed according to a certain $\mathbb{P}_J$, e.\,g. the unbiased bimodal one:
\begin{equation}
	\mathbb{P}_{J}(J_{ij})=\frac{1}{2}\delta(J_{ij}-J)+\frac{1}{2}\delta(J_{ij}+J)
\end{equation}

Let us rewrite here the pairwise~\acrshort{BP} equations for the XY model in presence of an external field:
\begin{equation}
	\eta_{i\to j}(\theta_i) = \frac{1}{\mathcal{Z}_{i\to j}}\,e^{\,\beta H\cos{(\theta_i-\phi_i)}}\prod_{k\in\partial i\setminus j}\int \di\theta_k\,e^{\,\beta J_{ik}\cos{(\theta_i-\theta_k)}}\,\eta_{k\to i}(\theta_k)
	\label{eq:BP_eqs_XY_Field}
\end{equation}
In the high-$H$ -- high-$T$ region, these equations admit a paramagnetic solution, which is however no longer given by the uniform distribution over the $[0,2\pi)$ interval. Indeed, it depends on the direction of the \textit{local effective field} given by the external field $\boldsymbol{H}_i$ acting on site $i$ and by the contributions coming from the neighbours $\partial i$ of $i$, exactly as seen in~\autoref{chap:RFXY} for the random field XY model. This does not allow us to exploit the analytic approach of an expansion around the paramagnetic solution, as done in~\autoref{chap:XYnoField}, so we have to rely only on numerical tools.

\acrshort{BP} equations~\autoref{eq:BP_eqs_XY_Field} are hence solved via the~\acrshort{PDA}, passing to the following distributional equation:
\begin{equation}
	\mathbb{P}_{\eta}[\eta_{i\to j}] = \mathbb{E}_{\mathcal{G},J,\phi}\int\prod_{k=1}^{d_i-1}\mathcal{D}\eta_{k\to i}\,\mathbb{P}_{\eta}[\eta_{k\to i}]\,\delta\Bigl[\eta_{i\to j}-\mathcal{F}[\{\eta_{k\to i}\},\{J_{ik}\},\phi_i]\Bigr]
	\label{eq:def_PDA_inField}
\end{equation}
Notice that, as in the previous Chapters, we are actually solving these equations via a $Q$-state clock model with $Q=64$ states.

Since we are dealing with vector spins and hence there are several directions in the phase space along which instabilities can evolve, the location of the instability line of the paramagnetic solution could strongly depend on the probability distribution $\mathbb{P}_{\phi}$ from which field directions~$\phi_i$'s are drawn, as it actually occurs in the fully connected case. So let us firstly discuss the most efficient and reliable way to detect the \acrshort{RS} instability, then analyzing the features of the corresponding critical line according to different distributions of the field direction.

\subsection{The detection of instability lines}

The stability of the \acrshort{BP} fixed point $\mathbb{P}^*[\eta]$ can be as usual analyzed via the~\acrshort{SuscProp} algorithm (\autoref{chap:XYnoField}), where each cavity marginal $\eta_{i\to j}(\theta_i)$ is accompanied by a small perturbation $\delta\eta_{i\to j}(\theta_i)$, which evolves according to the linearized version of the~\acrshort{BP} equations~\autoref{eq:BP_eqs_XY_Field}:
\begin{equation}
	\begin{split}
		\delta\eta_{i\to j} &= \sum_{k\in\partial i\setminus j}\Biggl|\frac{\delta \mathcal{F}[\{\eta_{k\to i}\},\{J_{ik}\},\phi_i]}{\delta \eta_{k\to i}}\Biggr|_{\eta^*_{k\to i}}\delta\eta_{k\to i}
	\end{split}
\end{equation}
In the \acrshort{PDA} approach, then, we store a population of $\mathcal{N}$ couples $(\eta_i,\delta\eta_i)$ of cavity marginals and related perturbations --- with index $i$ just referring to the population element --- and let them evolve according to the full and the linearized \acrshort{BP} equations, respectively. Then, we evaluate the growth rate $\lambda_{\text{BP}}$ of the global norm of the perturbations, Eq.~\autoref{eq:lambdaBP_XY_def}, that we rewrite here for the reader's convenience:
\begin{equation}
	\lambda_{\text{BP}} \equiv \lim_{t\to\infty}\frac{1}{t\mathcal{N}}\sum_{i=1}^{\mathcal{N}}\ln{\norm{\delta\eta^{(t)}_i}}
	\label{eq:lambdaBP_XY_def_Chap6}
\end{equation}
so that $\lambda_{\text{BP}}<0$ eventually refers to a \acrshort{RS} stable fixed point and $\lambda_{\text{BP}}>0$ otherwise.

In~\autoref{chap:XYnoField} we exploited the~\acrshort{SuscProp} algorithm just to detect the \acrshort{RS} instability line between the ferromagnetic and the mixed phases, while the one between the paramagnetic and the unbiased spin glass phases had already been computed analytically. Unfortunately, here we have to numerically evaluate it, since no perturbative expansion around the paramagnetic solution is available.

Due to the strong heterogeneity of perturbations, whose norms span several orders of magnitude, the measurement of $\lambda_{\text{BP}}$ have to be carefully performed. Moreover, when approaching the critical point, the slowing down in the thermalization affects $\lambda_{\text{BP}}$ more than other physical observables, resulting in a difficult determination of the critical point. So here we expose a way out of it. Let us firstly consider the zero-field case. In order to compute the critical temperature~$T_c$, avoiding the critical slowing close to $T_c$, we can solve the \acrshort{BP} equations at a certain temperature using as initial condition the fixed point previously reached at a nearby temperature. According to the decreasing or the increasing of the temperature during this protocol, we will refer to it as \textit{cooling} or \textit{heating}. In this way, since the fixed point $\mathbb{P}^*[\eta]$ should not be so different for two close enough temperatures, the critical slowing down is typically largely reduced.

\begin{figure}[t]
	\centering
	\includegraphics[scale=1]{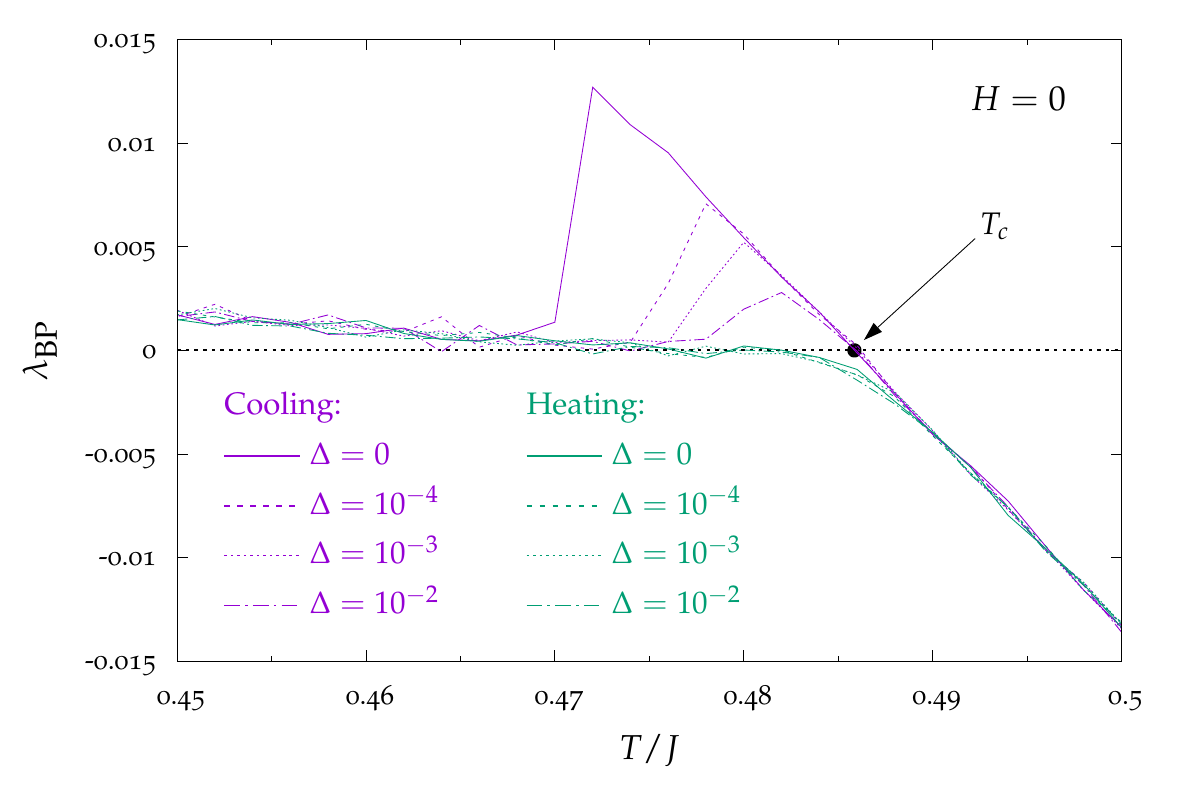}
	\caption[Cooling and heating protocols for the critical-point detection]{Stability parameter $\lambda_{\text{BP}}(T)$ at zero field for the spin glass XY model on a $C=3$~\acrshort{RRG}. The two colours respectively correspond to the two protocols, `cooling' and `heating', used to reach more rapidly the correct \acrshort{BP} fixed point. The analytical value of the zero-field critical temperature $T_c$ is represented by the black dot.}
	\label{fig:lambdaBP_zeroField_roundTrip}
\end{figure}

Unfortunately, this approach has some disadvantages. In particular, since the previously reached fixed point is used as initial condition for the next round of the protocol, it may happen that the system remains stuck into a particular fixed point even when it has already become unstable, so providing the wrong solution at that temperature. This issue is very relevant when the ``starting'' fixed point is endowed with particular symmetries, e.\,g. the paramagnetic solution in absence of a field or in presence of a homogeneous field. The former case corresponds to~\autoref{fig:lambdaBP_zeroField_roundTrip}, where the metastability of the paramagnetic solution can be well observed on the $\lambda_{\text{BP}}(T)$ curve obtained during a cooling protocol with the parameter $\Delta=0$ (which will be defined in a while). In order to avoid such metastabilities, the trick is hence to perturb the previous fixed point by a tiny amount before using it as initial condition of the new step of the protocol. In particular, the $\eta_{i\to j}$'s can be perturbed component-wise by adding a random number $\Delta\abs{z}$ with $z \sim \mathrm{Gauss}(0,1)$ and $\Delta \ll 1$.

The resulting $\lambda_{\text{BP}}(T)$ curves, for both the protocols and for several values of the perturbation parameter $\Delta$, are reported in~\autoref{fig:lambdaBP_zeroField_roundTrip}, together with the exact location of the zero-field critical temperature $T_c$, analytically computed in~\autoref{chap:XYnoField}:
\begin{equation}
	(C-1)\,\biggl[\frac{I_1(\beta_c J)}{I_0(\beta_c J)}\biggr]^2=1
	\label{eq:zeroField_Tc}
\end{equation}
Notice that for each temperature, we waited $t=150$ time steps for the thermalization and then we averaged $\lambda_{\text{BP}}$ over the next $t=150$ time steps.

First of all, we note that the metastability of the paramagnetic solution below the critical point --- due to the use of the cooling protocol --- is strongly reduced when increasing~$\Delta$, just as we guessed. Then, it is rather clear that the cooling protocol is the best way to evaluate the position of the critical point --- whatever the value of~$\Delta$, provided it is small ---, via a linear extrapolation from the $\lambda_{\text{BP}}<0$ data points. On the contrary, the heating protocol is quite useless for this purpose, mainly due to two issues: \textit{i)} a very slow power-law growth of the stability parameter $\lambda_{\text{BP}}$ below the critical point, that induces a large statistical error on the estimate of $T_c$, and \textit{ii)} systematic errors given by a very slow convergence of the population toward the paramagnetic fixed point when $T$ is close to $T_c$ and whatever the value of $\Delta$, so yielding a value of $\lambda_{\text{BP}}$ smaller than expected. In summary, if on one hand a random perturbation is useful for leaving the trivial fixed point, it becomes useless when trying to reach it again from a nontrivial fixed point.

\begin{figure}[t]
	\centering
	\includegraphics[scale=1]{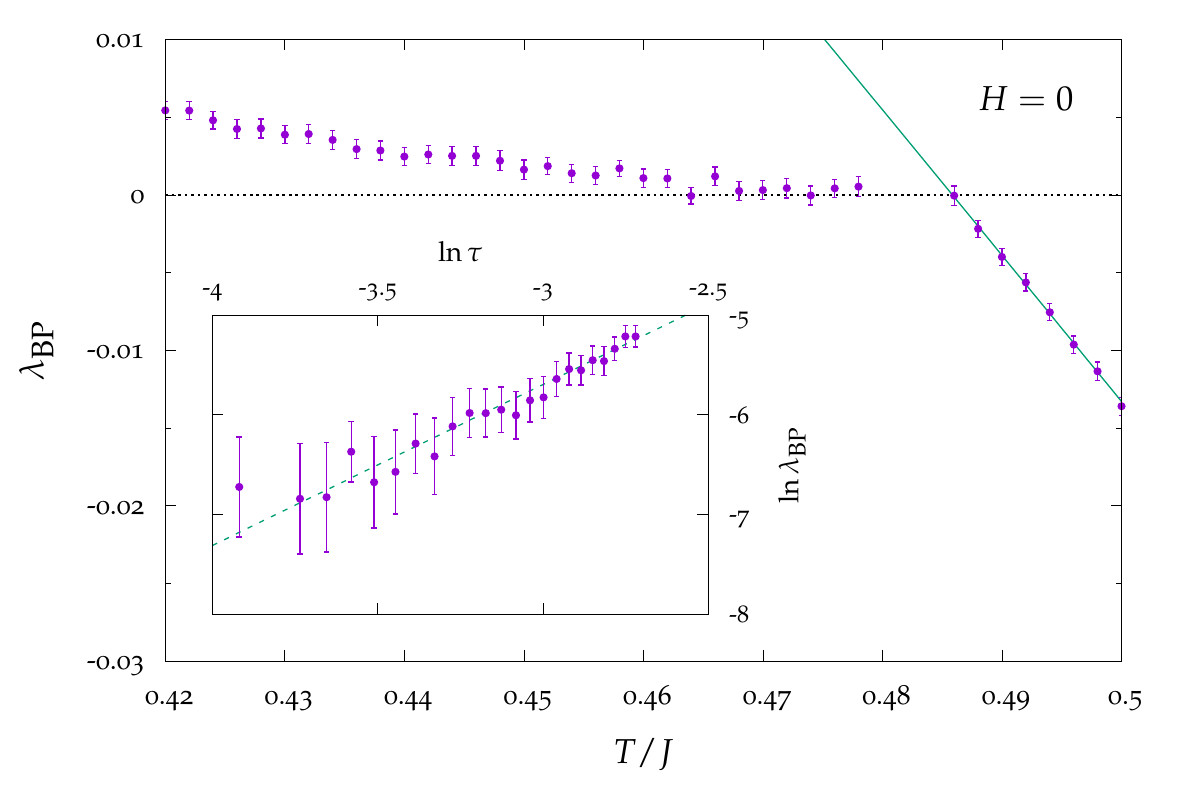}
	\caption[Stability parameter for the zero-field spin glass XY model under cooling protocol]{Stability parameter $\lambda_{\text{BP}}(T)$ at zero field for the spin glass XY model on a $C=3$~\acrshort{RRG}. All the data points have been collected in the stationary regime via the cooling protocol at $\Delta=10^{-2}$, with the analytic prediction of $\lambda_{\text{BP}}(T)$ for the paramagnetic solution represented in the main plot by the green full line. In the inset, instead, we evaluate the $\alpha$ exponent of the power-law behaviour $\lambda_{\text{BP}}\propto\tau^{\alpha}$ below the critical point, obtaining $\alpha=1.6(1)$.}
	\label{fig:lambdaBP_zeroField_Cooling_Delta1e-2}
\end{figure}

From now on we can safely exploit the cooling protocol with e.\,g. $\Delta=10^{-2}$, reporting in~\autoref{fig:lambdaBP_zeroField_Cooling_Delta1e-2} the corresponding data for $\lambda_{\text{BP}}$ in the stationary regime (hence some not thermalized points are missing, though being irrelevant for the determination of $T_c$). For $T>T_c$ they are in agreement with the analytic prediction of $\lambda_{\text{BP}}(T)$ for the paramagnetic solution, namely the logarithm of the left hand side of Eq.~\autoref{eq:zeroField_Tc}. Instead for $T<T_c$ they follow the aforementioned power law, with an exponent that has been computed in the inset of~\autoref{fig:lambdaBP_zeroField_Cooling_Delta1e-2}:
\begin{equation}
	\lambda_{\text{BP}} \propto \tau^{\alpha} \quad , \qquad \alpha=1.6(1)
\end{equation}

\subsection{The case of a homogeneous field}

\begin{figure}[!t]
	\centering
	\includegraphics[scale=0.99]{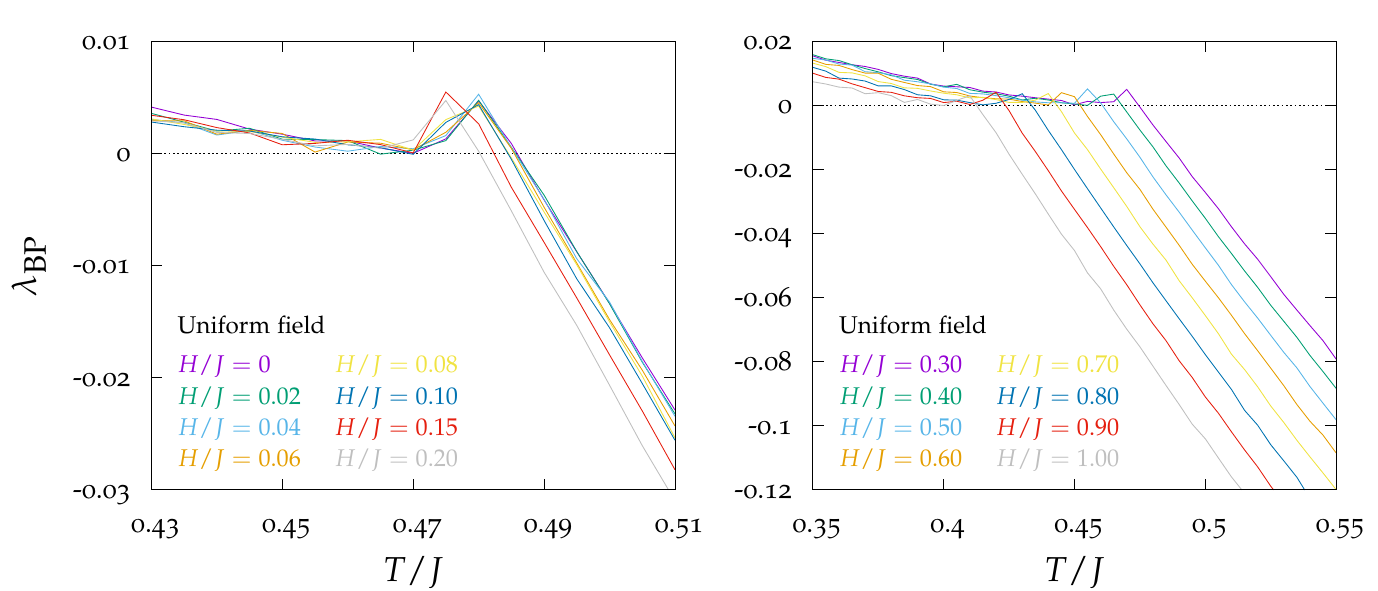}
	\caption[Stability parameter for the spin glass XY model in a uniform field]{Stability parameter $\lambda_{\text{BP}}(T)$ for the spin glass XY model on a $C=3$~\acrshort{RRG} with a uniform field of intensity $H$, with the two panels referring to different ranges of $H$. It is rather clear the leftward shift of the entire curve $\lambda_{\text{BP}}(T)$ when increasing $H$.}
	\label{fig:lambdaBP_UF_severalH_Cooling_Delta1e-2}
\end{figure}

In order to check if a~\acrshort{GT}-like instability is present also in the sparse case, we choose a homogeneous field over the whole system, namely
\begin{equation}
	\mathbb{P}_{\phi}(\phi_i)=\delta(\phi_i-\overline{\phi})
\end{equation}
with $\overline{\phi}=0$ without any loss of generality, and then we solve the~\acrshort{BP} equations~\autoref{eq:BP_eqs_XY_Field} via the~\acrshort{PDA}. Also in this case it turns out to be useful to exploit the cooling protocol (we set $\Delta=10^{-2}$), since the paramagnetic solution still suffers from metastability, as it can be appreciated in~\autoref{fig:lambdaBP_UF_severalH_Cooling_Delta1e-2}, where the resulting curve $\lambda_{\text{BP}}(T)$ is plotted for several values of the field strength $H$. The data slightly below the critical points are clearly not thermalized within the $t_{max}=300$ iterations we chose, but anyway they do not enter into the computation of the critical temperature when varying $H$. Both panels --- and mainly the right one --- clearly show how the main effect of the increasing of the intensity $H$ of the uniform field is that of shifting leftward the entire curve $\lambda_{\text{BP}}(T)$, so that the same instability parameter is attained at lower temperatures for larger values of $H$.

\begin{figure}[!t]
	\centering
	\includegraphics[scale=1]{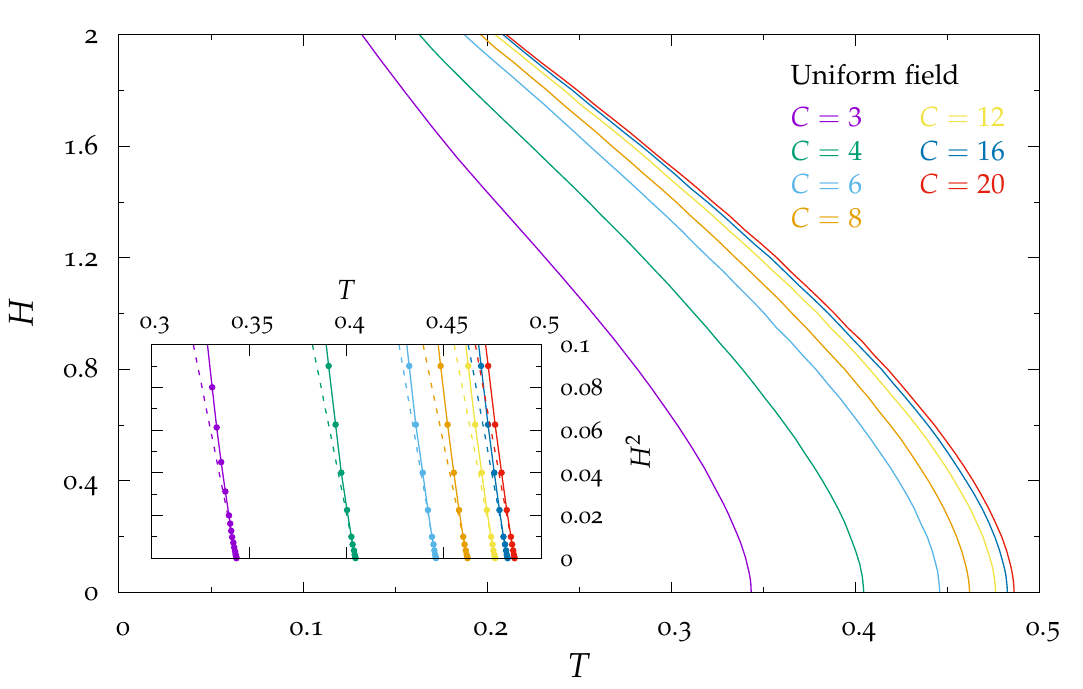}
	\caption[Instability lines in a uniform field for several connectivities of the RRG]{Critical lines $H_c(T)$ in a uniform field for the spin glass XY model on a $C$-\acrshort{RRG} for several values of the connectivity $C$. The inset shows evidences for the typical small-field behaviour $H_c\propto\tau^{1/2}$ of the~\acrshort{GT} instability line.}
	\label{fig:GT_line_severalC_Jrescaled_H2_vs_T_THESIS}
\end{figure}

We see in~\autoref{fig:lambdaBP_UF_severalH_Cooling_Delta1e-2} that the critical temperature as a function of $H$ can be again computed via a linear fit over the $T>T_c$ region, so obtaining the entire curve of \acrshort{RS} instability of the paramagnetic solution on a $C=3$ \acrshort{RRG}. It is reported in purple in~\autoref{fig:GT_line_severalC_Jrescaled_H2_vs_T_THESIS}, together with the analogous curves $H_c(T)$ for several other values of the connectivity $C$ of the \acrshort{RRG}. In order to make the instability curves in the $(T,H)$ plane to collapse onto each other in the large-$C$ limit, so to recover the fully connected behaviour, we have to properly rescale the strength $J$ of the couplings. Indeed, from~\autoref{tab:Tc_severalC} we can see how the zero-field critical temperature $T_c/J$ diverges with $C$; then, if setting $J=1/\sqrt{C-1}$, $T_c$ eventually approaches the fully connected prediction $T_c=1/m=1/2$ in the $C\to\infty$ limit.

\begin{table}[t]
	\setlength{\tabcolsep}{8pt}		
	\centering
	\caption[Zero-field critical temperatures on RRG]{The zero-field critical temperature $T_c$ for the spin glass XY model with (equally distributed) bimodal couplings $J_{ij}=\pm J$ on a \acrshort{RRG} for different values of the connectivity $C$. If choosing $J=1/\sqrt{C-1}$, $T_c$ approaches the fully connected value $1/2$ in the large-$C$ limit.}
	\label{tab:Tc_severalC}
	\begin{tabular}{ccc}
		\toprule
		$C$ & $T_c/J$ & $T_c$\\
		\midrule
		$3$ & $0.4859$ & $0.3436$\\
		$4$ & $0.7012$ & $0.4048$\\
		$6$ & $0.9977$ & $0.4462$\\
		$8$ & $1.2234$ & $0.4624$\\
		$12$ & $1.5805$ & $0.4765$\\
		$16$ & $1.8704$ & $0.4829$\\
		$20$ & $2.1211$ & $0.4866$\\
		\bottomrule
	\end{tabular}
\end{table}

All the instability curves $H_c(T)$ seem to have the same small-field behaviour, compatible with the \acrshort{GT} scaling $H_{\text{GT}}\propto\tau^{1/2}$. Indeed, in the inset of~\autoref{fig:GT_line_severalC_Jrescaled_H2_vs_T_THESIS} we plotted the same curves in the $(T,H^2)$ plane: a clear linear behaviour in $\tau$ can be seen in the region of very small fields, namely the ones where the small-field expansion $H_{\text{GT}}\propto\tau^{1/2}$ actually holds. Then, such scaling is soon lost for larger values of $H$, since the $H_c(T)$ curves change their concavity.

Finally, being $\lambda_{\text{BP}}$ larger than zero in the whole region below the critical line $H_c(T)$ for all the analyzed values of $C$, we can confirm the occurrence of the~\acrshort{RSB} once trespassed the \acrshort{GT} line also in the diluted case.

\subsection{The case of a randomly oriented field}

Once accomplished the first task, namely the check of the occurrence of a \acrshort{GT}-like instability also in the diluted case and the correspondent breaking of replica symmetry, we now want to study the onset of the \acrshort{dAT} instability in our model. To this aim, we exploit the suggestion of Ref.~\cite{SharmaYoung2010}, considering the case of a randomly oriented field, with each direction $\phi_i$ uniformly drawn from the $[0,2\pi)$ interval:
\begin{equation}
	\mathbb{P}_{\phi}(\phi_i)=\mathrm{Unif}\Bigl([0,2\pi)\Bigr)
\end{equation}
As discussed in~\autoref{sec:XYinField_fully_conn}, due to the randomness in the direction of the field, the global order parameters $q_{\parallel}$ and $q_{\perp}$ can not be any longer defined, due to the absence of the directional anisotropy earlier provided by the uniform field. In fact, longitudinal and transverse instabilities could be eventually defined only locally, as we will see in~\autoref{sec:diff_ways_RSB}.

\begin{figure}[!t]
	\centering
	\includegraphics[scale=1]{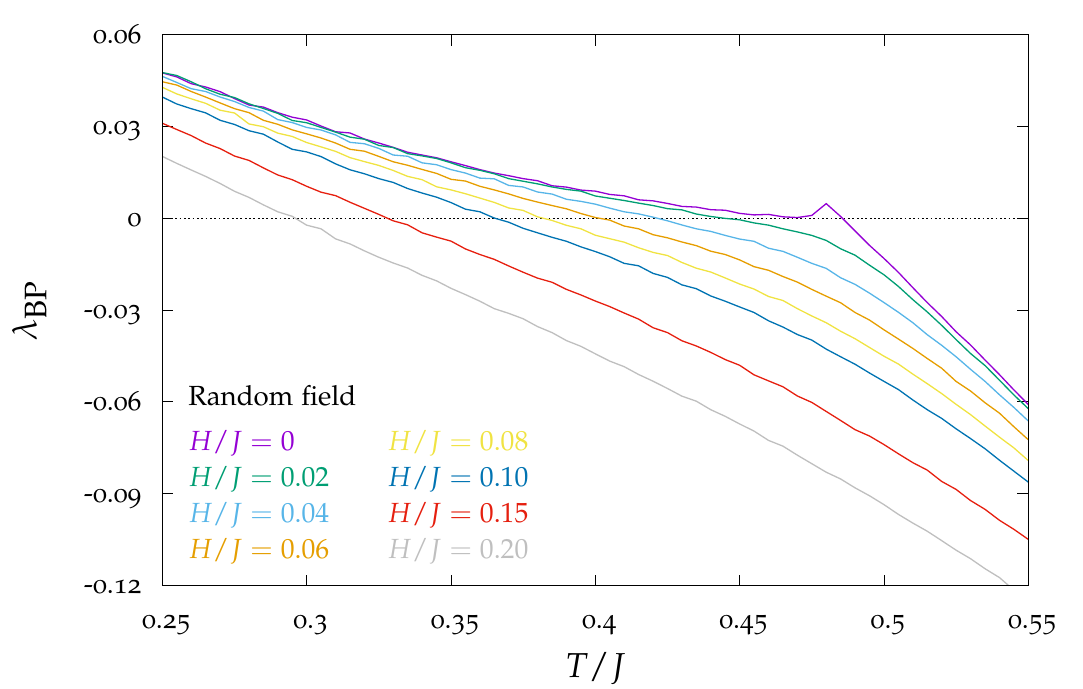}
	\caption[Stability parameter for the spin glass XY model in a random field]{Stability parameter $\lambda_{\text{BP}}(T)$ for the spin glass XY model on a $C=3$~\acrshort{RRG} with a randomly oriented field of intensity $H$. Contrarily to the uniform-field case, the curve $\lambda_{\text{BP}}(T)$ moves downward when increasing $H$, while smoothing away the zero-field singularity.}
	\label{fig:lambdaBP_RF_severalH_Cooling_Delta1e-2}
\end{figure}

\begin{figure}[!t]
	\centering
	\includegraphics[scale=1]{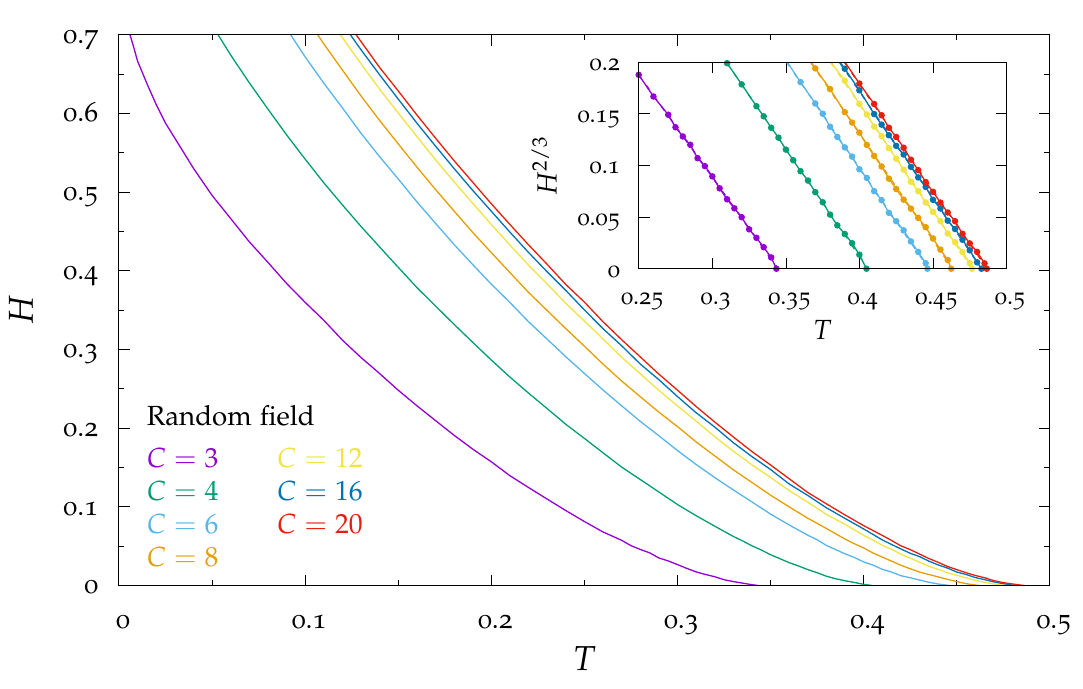}
	\caption[Instability lines in a random field for several connectivities of the RRG]{Critical lines $H_c(T)$ in a randomly oriented field for the spin glass XY model on a $C$-\acrshort{RRG} for several values of the connectivity $C$. The inset shows evidences for the typical small-field behaviour $H_c\propto\tau^{3/2}$ of the~\acrshort{dAT} instability line.}
	\label{fig:dAT_line_severalC_Jrescaled_H2_3_vs_T_THESIS}
\end{figure}

As in the case of a uniform field, we solve the~\acrshort{BP} equations~\autoref{eq:BP_eqs_XY_Field} via the \acrshort{PDA}, by exploiting the cooling protocol with $\Delta=10^{-2}$. The stability parameter $\lambda_{\text{BP}}(T)$ for several intensities $H$ of the field is reported in~\autoref{fig:lambdaBP_RF_severalH_Cooling_Delta1e-2}.

First of all, we notice that the previous instability line has actually vanished, namely no trace of a \acrshort{GT}-like transition has left. Then, contrarily to the uniform-field case, the $\lambda_{\text{BP}}(T)$ curve mainly shifts downward when increasing $H$, at the same time smoothing away the zero-field singularity. Indeed, remarkable changes occur in the stability parameter $\lambda_{\text{BP}}$ even for very small values of the external field $H$ --- while it was not so for the uniform-field case ---, as it can be seen by comparing~\autoref{fig:lambdaBP_RF_severalH_Cooling_Delta1e-2} with the left panel of~\autoref{fig:lambdaBP_UF_severalH_Cooling_Delta1e-2}.

The computation of the whole instability curve $H_c(T)$ in the $(T,H)$ plane proceeds as before, with the result for the $C=3$ \acrshort{RRG} represented by the purple curve in~\autoref{fig:dAT_line_severalC_Jrescaled_H2_3_vs_T_THESIS}. Again, a direct comparison between the $H_c(T)$ curves for several values of $C$ is provided, with the fully connected limit easily recognizable from the remarkable superposition of the instability curves for the largest values of $C$ analyzed. The \acrshort{dAT}-like behaviour $H_c\propto\tau^{3/2}$ can be eventually appreciated for all of them, with the inset showing a linear behaviour for a rather wide range of field values, when plotting the $H_c(T)$ curves in the $(T,H^{2/3})$ plane. Indeed, at variance with respect to the previous case, they do not change their concavity at higher values of $H$.

\section{Different ways of breaking the replica symmetry}
\label{sec:diff_ways_RSB}

\begin{figure}[t]
	\centering
	\includegraphics[scale=1]{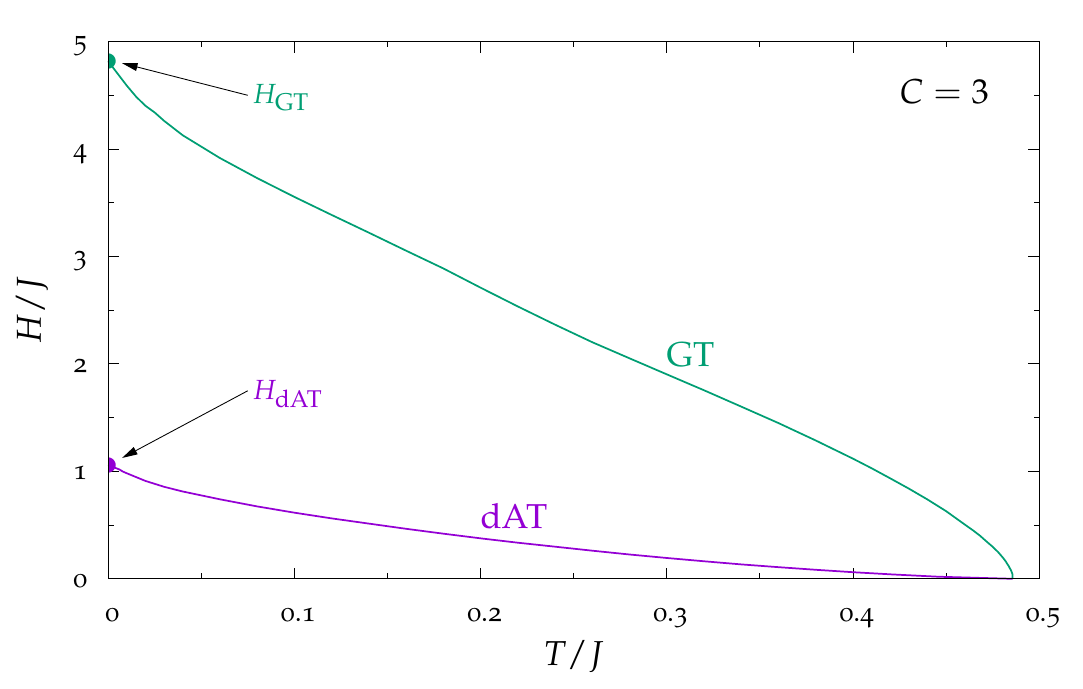}
	\caption[dAT and GT lines for a $C=3$ RRG]{Comparison of the \acrshort{GT} and \acrshort{dAT} critical lines for the spin glass XY model with unbiased bimodal couplings on a $C=3$~\acrshort{RRG}.}
	\label{fig:dAT_plus_GT_line_C3_THESIS}
\end{figure}

At this point, it is clear that several features of the \acrshort{GT} and the \acrshort{dAT} lines in the fully connected case can be recovered also in the diluted case: first of all, their occurrence according to the distribution of the \textit{direction} of the external field; their critical exponents for the small-field expansion; the corresponding breaking of replica symmetry when trespassing them. However, the well known unphysical prediction of diverging critical fields in the zero-temperature limit is lost when moving from the fully connected to the diluted topology, as we would have expected. In particular, for a $C=3$ \acrshort{RRG} we get the following values for the endpoints of the two lines:
\begin{equation}
	H_{\text{GT}}/J = 1.06(1) \qquad , \qquad H_{\text{dAT}}/J = 4.82(1)
\end{equation}
that can be evaluated by extrapolating the finite-temperature data or directly at zero temperature by exploiting the suitable algorithm defined in the previous Chapters. Finally, the direct comparison between the two full curves can be appreciated in~\autoref{fig:dAT_plus_GT_line_C3_THESIS}.

At this point, we are interested in understanding the physical meaning of these two kinds of instabilities, both leading to the \acrshort{RSB}. From the analysis in the fully connected case, we already know that the~\acrshort{GT} line corresponds to a breaking of the inversion symmetry for the spins in the transverse direction with respect to the external field, with a corresponding freezing of the transverse degrees of freedom. Instead, the~\acrshort{dAT} line does not involve any breaking in the spin symmetry, due to the presence of the randomly oriented field, but is linked to the freezing of longitudinal degrees of freedom.

In the sparse case, the heterogeneity naturally coming out provides us a precious tool to better characterize such instabilities. Indeed, once reached the \acrshort{BP} fixed point, we can perform a local analysis, looking for each spin at the direction along which the most probable fluctuation may take place, then relating it to the~\textit{local} external field. This task can be accomplished by looking at the site marginals $\eta_i(\theta_i)$'s and their perturbations $\delta\eta_i(\theta_i)$'s, relying on the following observation:
\begin{itemize}
	\item on the~\acrshort{GT} line, the freezing of the transverse degrees of freedom should imply that perturbations $\delta\eta_i$'s are preferably perpendicular to the local field~$\boldsymbol{H}_i$;
	\item on the~\acrshort{dAT} line, the freezing of the longitudinal degrees of freedom should show itself as a longitudinal arrangement of perturbations $\delta\eta_i$'s with respect to the local field~$\boldsymbol{H}_i$.
\end{itemize}
So, once reached the fixed point for the joint ``cavity'' population $\{(\eta,\delta\eta)\}$, the joint population of site marginals and related perturbations can be computed; then, for each element of such population, the following two local vectors can be defined:
\begin{subequations}
	\begin{equation}
		\boldsymbol{m}_i \equiv \int \di\theta_i\,\eta_i(\theta_i)\bigl(\cos{\theta_i},\sin{\theta_i}\bigr)
	\end{equation}
		\begin{equation}
		\delta\boldsymbol{m}_i \equiv \int \di\theta_i\,\delta\eta_i(\theta_i)\bigl(\cos{\theta_i},\sin{\theta_i}\bigr)
	\end{equation}
\end{subequations}
They respectively correspond to the average magnetization of the $i$-th site, i.\,e. the direction along which the spin most probably aligns, and the perturbed magnetization of the $i$-th site, i.\,e. the direction of the most probable fluctuation.

At this point, the mutual orientation of $\delta\boldsymbol{m}_i$ and $\boldsymbol{H}_i$ on each site should clarify the kind of perturbation to the \acrshort{BP} fixed point. It can be quantified by the cosine of the inbetween angle:
\begin{equation}
	\cos{\vartheta_i} \equiv \frac{\delta\boldsymbol{m}_i\cdot\boldsymbol{H}_i}{\norm{\delta\boldsymbol{m}_i}\norm{\boldsymbol{H}_i}} = \frac{\delta\boldsymbol{m}_i\cdot\boldsymbol{H}_i}{\delta m_i\,H}
	\label{eq:def_cos_varTheta}
\end{equation}
so that a transverse perturbation would yield a value of $\cos{\vartheta_i}$ close to zero, while a longitudinal perturbation would show a strong bias toward the $\pm 1$ values.

\begin{figure}[t]
	\centering
	\includegraphics[scale=1]{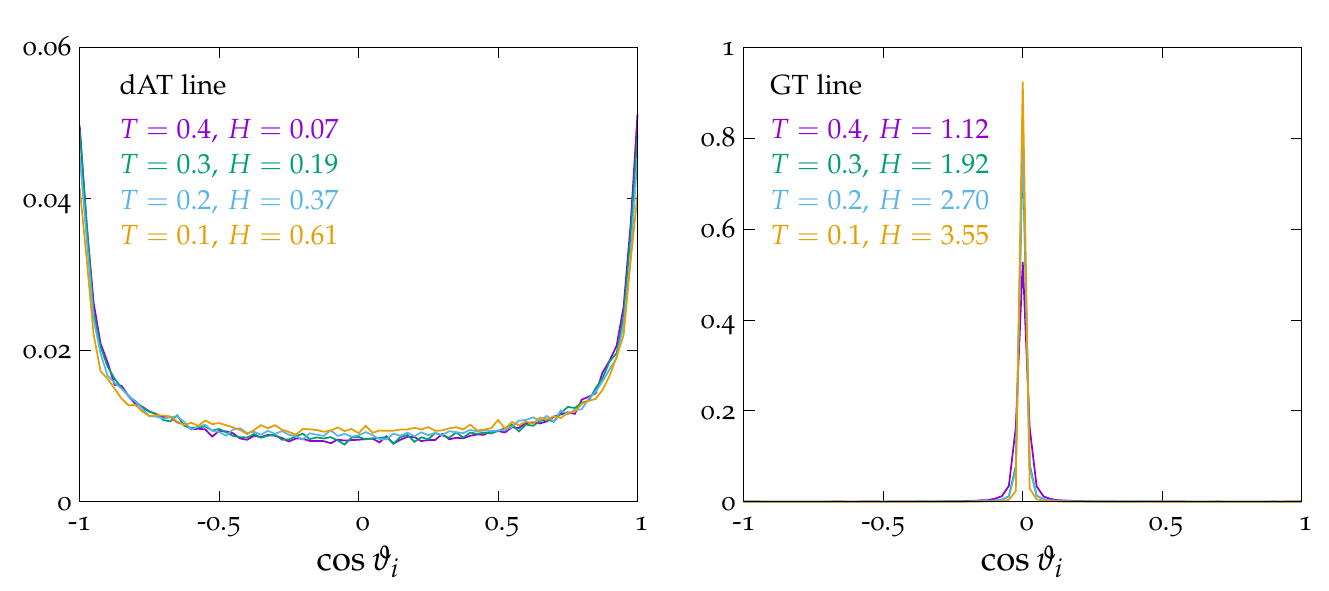}
	\caption[Transverse and longitudinal behaviour of GT and dAT lines on RRG (1D histogram)]{Probability distribution of the cosine of $\vartheta_i$'s angles between the perturbed magnetization $\delta\boldsymbol{m}_i$ and the external field $\boldsymbol{H}_i$ on each site, for several points on the~\acrshort{dAT} and on the~\acrshort{GT} lines. The respective longitudinal and transverse behaviour is made evident by the location of the peaks of the probability distribution in the two cases. Here $T$ and $H$ values refer to the choice $J=1$.}
	\label{fig:scalProd_GT_vs_dAT}
\end{figure}

We compute the probability distribution of $\cos{\vartheta_i}$ for several points along the \acrshort{GT} and the \acrshort{dAT} lines for a $C=3$~\acrshort{RRG}, showing the results in~\autoref{fig:scalProd_GT_vs_dAT}. As expected, on the~\acrshort{dAT} line a clear bias toward the $\pm 1$ values is shown, while on the~\acrshort{GT} line the $0$ value is by far the most preferred, so confirming the interpretation of the two critical lines as instabilities in the longitudinal direction or in the transverse direction, respectively, referred to the local direction of the external field.

Notice that the longitudinal behaviour of perturbations along the \acrshort{dAT} line does not show up at the same manner on all the sites. Indeed, since $H$ is not so large along such line, also transverse perturbations are allowed, even though with a smaller probability: their energy cost is surely larger than that of a longitudinal perturbation, but not enough to dramatically suppress them. It is for this reason that there is apparently no dependence on the specific point of the \acrshort{dAT} line in the left panel of~\autoref{fig:scalProd_GT_vs_dAT}. At variance, along the \acrshort{GT} line, $H$ reaches rather larger values: hence the transverse behaviour of perturbations is by far more pronounced with respect to the longitudinal one, being further enhanced when increasing $H$.

\begin{figure}[p]
	\centering
	\includegraphics[width=0.96\columnwidth]{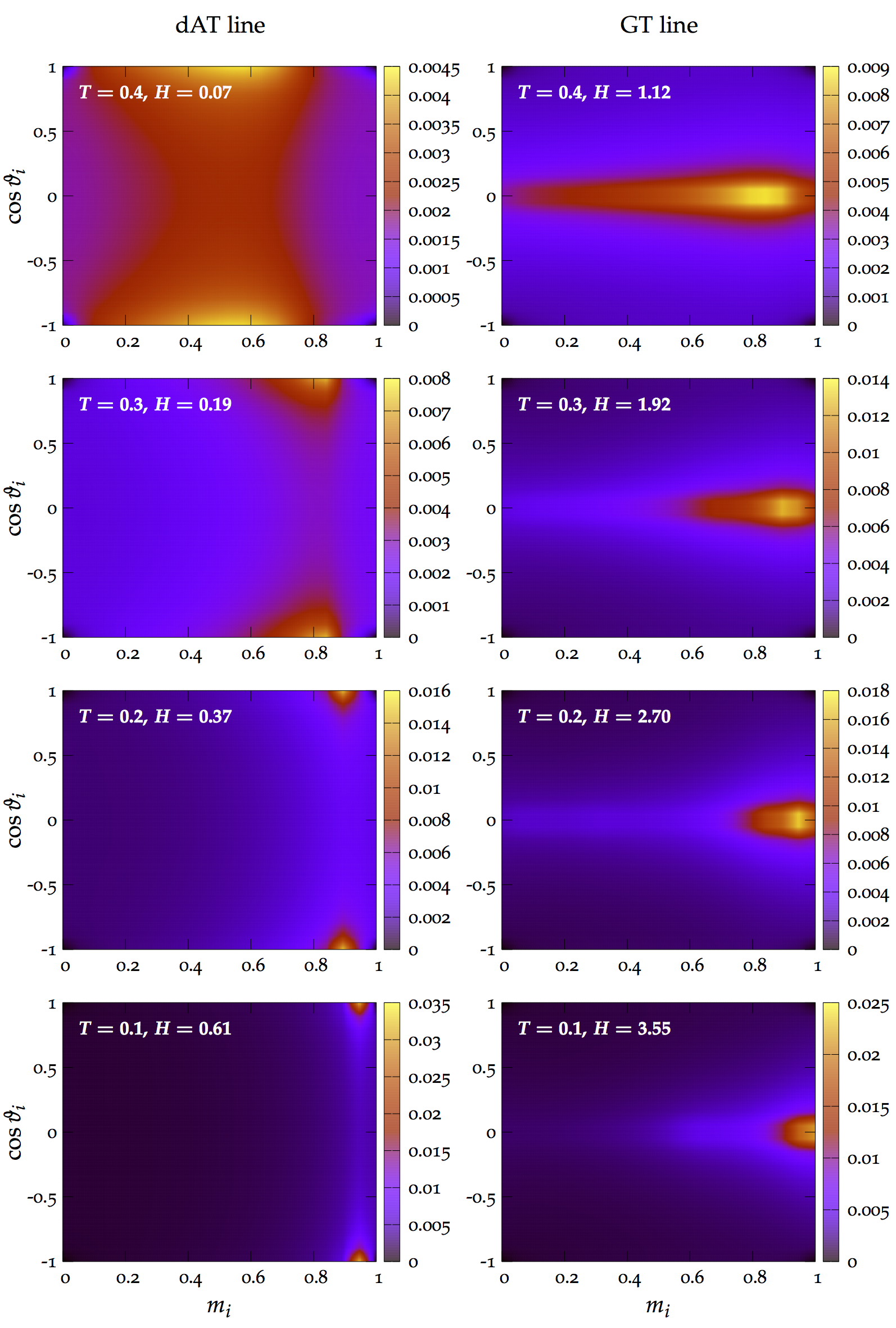}
	\caption[Transverse and longitudinal behaviour of GT and dAT lines (2D histogram)]{Joint probability distribution of $m_i$ and $\cos{\vartheta_i}$ for the same points along the two critical lines in~\autoref{fig:scalProd_GT_vs_dAT}. The longitudinal behaviour of the~\acrshort{dAT} line and the transverse behaviour of the~\acrshort{GT} line are again very sharp, together with a narrowing of the peaks and their shift toward the $m_i=1$ region when lowering $T$. Here $T$ and $H$ values refer to the choice $J=1$.}
	\label{fig:scalProd_GT_vs_dAT_joint}
\end{figure}

The dependence on the specific point along the two critical lines can be more effectively studied when also taking into account the local effective field. It can be done by looking at the joint probability distribution of $(m_i,\cos{\vartheta_i})$ for the same points of~\autoref{fig:scalProd_GT_vs_dAT} along the two lines, reported in~\autoref{fig:scalProd_GT_vs_dAT_joint}. Again, the longitudinal and transverse behaviours are highlighted by a preference for $\cos{\vartheta_i}=\pm 1$ and $\cos{\vartheta_i}=0$, respectively. Then, a sharp narrowing of the peaks is observed when lowering the temperature, together with a shift toward larger values of $m_i$. Indeed, when the temperature is large and hence the local effective field is weak $(m_i \simeq 0)$, the energetic cost of the two kinds of perturbations is rather comparable, resulting in broad peaks. So on the~\acrshort{GT} line a transverse perturbation could also show a longitudinal component by paying a relatively small energetic cost, and an analogous reasoning holds on the~\acrshort{dAT} line. Instead, when lowering the temperature, the local effective field becomes rather strong, the site marginals highly polarize $(m_i\to 1)$ and hence the likely perturbations --- e.\,g. the transverse ones on the \acrshort{GT} line --- become more and more energetically favourable with respect to the unlikely ones.

\section{Intermediate behaviours}
\label{sec:int_GT_dAT}

Once studied the behaviour of the spin glass XY model in a field with constant intensity $H_i=H$ and the local direction $\phi_i$ that can be the same for the all the sites $(\phi_i=\overline{\phi})$ or randomly drawn from the flat distribution over the $[0,2\pi)$ interval, one may wonder what happens if considering intermediate cases of randomness, namely a $\mathbb{P}_{\phi}$ probability distribution which is neither a delta function nor a constant. Our goal is to check if a crossover occurs in the shape of the resulting critical line, or if one of two behaviours is the dominant one in a more general case.

To this aim, we choose two representative classes of interpolating probability distributions. It is convenient to directly define them on the $Q$-state clock model rather than on the XY model, since that is the way we actually solve the \acrshort{BP} equations. So let first of all denote as $\mathcal{S}$ the set of the directions allowed by the clock model:
\begin{equation}
	\mathcal{S} \equiv \kappa\frac{2\pi}{Q}
\end{equation}
with $\kappa$ integer belonging to the range $0 \leqslant \kappa < Q$.

The first class of interpolating probability distributions is the one that, for the XY model, uniformly samples in a subset of the unit circle; hence, for the $Q$-state clock model it reads:
\begin{equation}
	\mathbb{P}[\kappa_i=n] = \frac{1}{Q'}\sum_{a=0}^{Q'-1}\delta_{n,a} \quad , \qquad Q' \text{ integer} \in \{1,\dots,Q\}
\end{equation}
The second class, instead, let us to sample $\phi_i$ from the whole unit circle via a linear combination of the two extremal probability distributions seen in~\autoref{sec:GT_dAT_diluted}; in more detail, for the $Q$-state clock model it reads:
\begin{equation}
	\mathbb{P}[\kappa_i=n] = \frac{w}{Q}\sum_{a=0}^{Q-1}\delta_{n,a} + (1-w)\delta_{n,0} \quad , \qquad w \text{ real-valued} \in [0,1]
\end{equation}

The two extremal cases of a uniform field (giving the \acrshort{GT} line) and a randomly oriented field with a flat measure over the $[0,2\pi)$ interval (giving the \acrshort{dAT} line) can be easily recovered by a suitable choice of the extremal values for the two parameters $Q'$ and $w$. Then, by varying them in the allowed ranges, we can insert a different degree of directional bias, so studying the crossover between the \acrshort{GT}-like regime and the \acrshort{dAT}-like regime.

\begin{figure}[!t]
	\centering
	\includegraphics[scale=1]{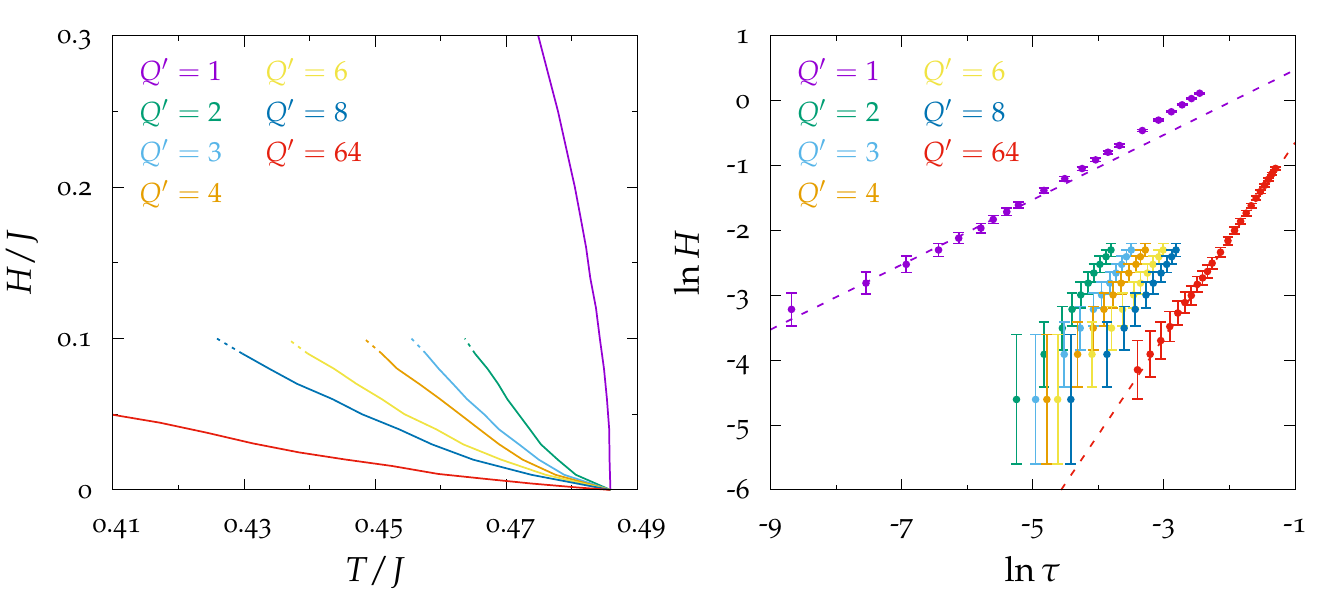}
	\caption[Instability lines in a random field with parameter $Q'$]{Critical lines $H_c(T)$ for the spin glass XY model on a $C$-\acrshort{RRG} with field directions $\phi_i=2\pi\kappa_i/Q$ with $\kappa_i$ uniformly drawn from the set $\{0,\,\dots,Q'-1\}$. The \acrshort{GT} line can be recovered with $Q'=1$, while $Q'=Q=64$ gives back the \acrshort{dAT} line. As soon as $Q'>1$, the data suggest a \acrshort{dAT}-like critical behaviour. In the right panel, the dashed lines have slope $1/2$ and $3/2$, respectively.}
	\label{fig:GT_plus_dAT_plus_varGT_lines_C03}
\end{figure}

In~\autoref{fig:GT_plus_dAT_plus_varGT_lines_C03} we analyze the behaviour of the first class of interpolating distributions on a $C=3$ \acrshort{RRG} for several values of the parameter $Q'$. It is quite clear that even the smallest nontrivial choice for $Q'$, namely $Q'=2$, causes a dramatic change in the instability line with respect to the \acrshort{GT} case $Q'=1$. So the loss of a perfect alignment of the local directions of the field implies a considerable change in the critical properties of the model. In particular, a careful study on the small-field expansion is performed in the right panel of~\autoref{fig:GT_plus_dAT_plus_varGT_lines_C03}: as soon as $Q'$ changes from~$1$ to larger values, the exponent seems to suddenly change from the~\acrshort{GT} value~$1/2$ to the~\acrshort{dAT} value~$3/2$. So, to the best of our numerical evidences, this class of intermediate distributions for the local directions of the field yields an abrupt change in the critical exponent, rather than a smooth crossover in it.

\begin{figure}[!t]
	\centering
	\includegraphics[scale=1]{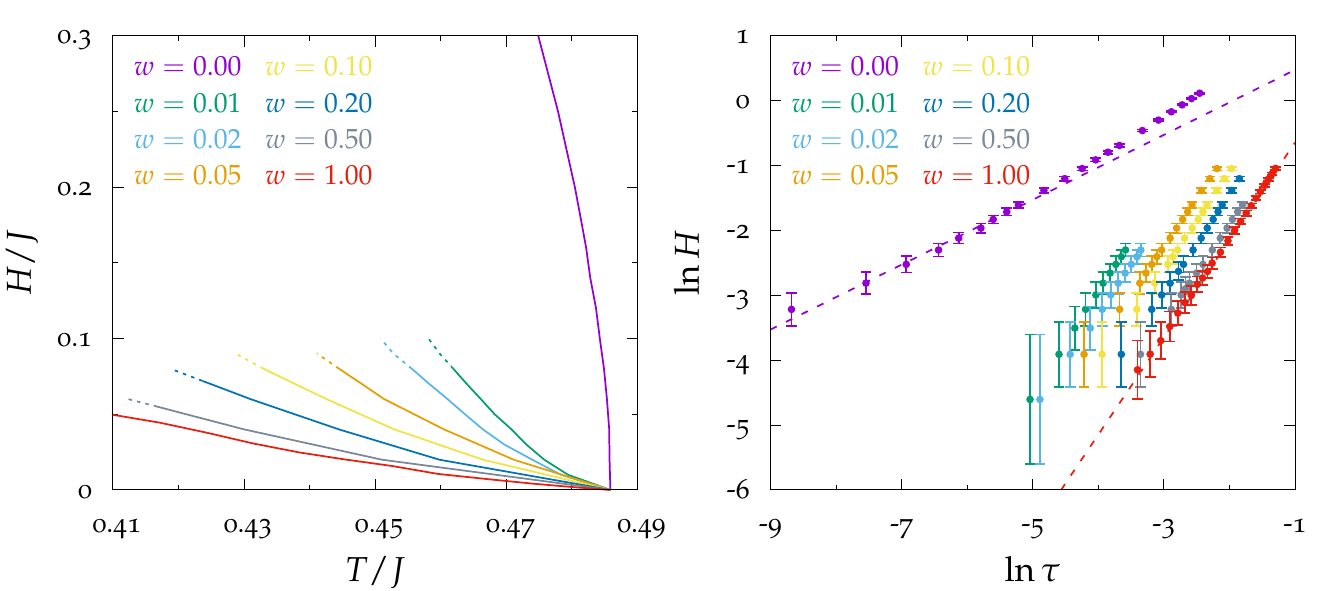}
	\caption[Instability lines in a random field with parameter $w$]{Critical lines $H_c(T)$ for the spin glass XY model on a $C$-\acrshort{RRG} with field directions $\phi_i=2\pi\kappa_i/Q$ with $\kappa_i$ randomly drawn according to $\mathbb{P}[\kappa_i=n]=w/Q\sum_{a=0}^{Q-1}\delta_{n,a}+(1-w)\delta_{n,0}$. The \acrshort{GT} line can be recovered with $w=0$, while $w=1$ gives back the \acrshort{dAT} line. As soon as $w>0$, the data suggest a \acrshort{dAT}-like critical behaviour. In the right panel, the dashed lines have slope $1/2$ and $3/2$, respectively.}
	\label{fig:GT_plus_dAT_plus_intGT_lines_C03}
\end{figure}

An analogous situation is found to occur for the second class of distributions. Moreover, here the real-valued parameter $w$ can be tuned even more finely, allowing to perturb in a very tiny way the uniform-field case. In~\autoref{fig:GT_plus_dAT_plus_intGT_lines_C03} we report the resulting critical lines on a $C=3$ \acrshort{RRG} for several values of $w$. In the left panel we can again appreciate a dramatic change in the instability line as soon as the uniform-field case is perturbed: the value $w=0.01$ already shows a \acrshort{dAT}-like behaviour instead of the \acrshort{GT}-like one, confirmed by the small-field analysis reported in the right panel. So it seems that --- to the best of our numerical evidences --- the~\acrshort{GT}-like behaviour is lost as soon as $w$ becomes strictly larger than zero.

The resulting picture is hence that even a very tiny perturbation of $O(10^{-2})$ in the perfect alignment of the local directions of the field --- no matter how it is specifically implemented --- makes the critical behaviour to change from \acrshort{GT}-like to \acrshort{dAT}-like, with the latter hence being much more generic and robust with respect to the former, which at variance seems to be yielded only in the case of a perfect alignment of both the couplings (remember the discussion at the end of~\autoref{chap:XYnoField}) and the fields.

\clearpage{\pagestyle{empty}\cleardoublepage}

\begingroup
	\makeatletter
	\let\ps@plain\ps@empty
	\part{The energy landscape}
	\label{part:EnergyLandscape}
	\cleardoublepage
\endgroup

\chapter{The zero-temperature spin glass XY model in a field}
\label{chap:XYinField_zeroTemp}
\thispagestyle{empty}

In~\autoref{chap:XYinField} we saw how the probability distribution of the external field has important consequences on the features of the system: the instability line of the paramagnetic solution moves, the kind of instability changes, different perturbations take place, the critical exponents change too, and so on.

This difference becomes more and more striking when lowering the temperature and eventually reaching the $T=0$ axis, as we saw in~\autoref{fig:scalProd_GT_vs_dAT_joint}. In this limit, some spin configurations become more and more energetically favourable than others, and the same occurs for the typical perturbations to such solutions.

The most common picture exploited to represent this flourishing of states, due to the breaking of the replica symmetry, is the so-called \textit{free energy landscape} or simply \textit{energy landscape}, since in the $T \to 0$ limit the two observables do coincide. The space of spin configurations is so depicted as an ensemble of valleys --- or minima, namely the states --- separated by high (free) energy barriers --- eventually diverging in the thermodynamic limit ---, among which the system wanders during its relaxation toward the equilibrium. The deeper a valley, the most energetic favourable it is~\cite{Parisi1983}.

If in ordered systems the energy landscape is typically rather trivial, with just very few minima, in disordered systems it acquires a very rugged aspect, with a large number of metastable configurations almost equivalent from the energetic point of view. Due to this, the relaxation toward the equilibrium configuration is rather slow, and the system can sooner or later get stuck in one of the (many) metastable minima.

The structure of the energy landscape provides very useful insights on the physics of the system. E.\,g., when the solution is found to be~\acrshort{fRSB}, the corresponding energy landscape exhibits a hierarchy of minima organized in a fractal arrangement~\cite{CharbonneauEtAl2014a}. Moreover, several physical phenomena can be explained referring to the structure of the correspoding energy landscape. For example, avalanches in the Ising model bring the system to a different local minimum through the flip of a large number of spins, corresponding to the overcome of the inbetween energy barrier. This is linked to the so-called marginal stability of the Ising model~\cite{MullerWyart2015}.

At variance, models with continuous spins can also exhibit small fluctuations around the energy local minima, implying a whole different set of physical phenomena. Indeed, phonon excitations and the phenomenon of the \textit{boson peak} in structural glasses~\cite{WyartEtAl2005, CharbonneauEtAl2016, LernerEtAl2016, CharbonneauEtAl2017} can be theoretically interpreted by means of low-energy excitations in vector spin glasses~\cite{BaityJesiEtAl2015}. Even more interestingly, the \acrshort{fRSB} low-$T$ -- low-$H$ phase found for vector spin glasses in a field (as we saw for the XY model in~\autoref{chap:XYinField}) is conjectured~\cite{SharmaEtAl2016} to be of the same type of the Gardner phase~\cite{Gardner1985} predicted for high-dimensional hard spheres~\cite{CharbonneauEtAl2017}.

In this Chapter we try to provide a self-consistent description of the energy landscape of the spin glass XY model in a random field, starting from the analysis of the Hessian matrix of the Bethe free energy in the $T \to 0$ limit. The ultimate task is to find a connection between the features of the energy landscape studied via the Hessian matrix and the results provided by the~\acrshort{BP} algorithm, in particular at the critical point $H_{\text{dAT}}$. Computations are performed on given instances of the model, due to the need of preserving all the long-range correlations, that at variance would be lost via the~\acrshort{PDA}. Interesting properties regarding the spectral density of the Hessian matrix arise, that strongly suggest a connection with the vibrational spectrum of real glasses. Hence, this model can be taken as a simple, \textit{exactly solvable} model for reproducing the soft vibrational modes of glasses.

\section{The Hessian matrix}

Let us consider again the spin glass XY model with unbiased bimodal couplings $J_{ij}=\pm J$ and an external field with constant intensity $H$ and random local directions~$\phi_i$'s drawn from the flat distribution over the $[0,2\pi)$ interval:
\begin{equation}
	\mcH[\{\theta\}]= - \sum_{(i,j)}J_{ij}\cos{(\theta_i-\theta_j)} - H\sum_i\cos{(\theta_i-\phi_i)}
	\label{eq:Hamiltonian_XY_sparse_in_Field_chap7}
\end{equation}
defined on a \acrshort{RRG} of connectivity $C=3$ and size $N$. As usual, such topology will allow us to solve the model by means of the \acrshort{BP} algorithm.

The energy landscape of this model can be characterized by looking at the \textit{Hessian matrix} $\mathbb{H}$ --- namely the matrix of the second derivatives --- of the total Bethe free energy $F$ with respect to angles~$\theta_i$'s in the zero-temperature limit:
\begin{equation}
	\mathbb{H}_{ij} \equiv \frac{\partial^2 F}{\partial\theta_i\,\partial\theta_j}\Biggr|_{T=0,\{\theta^*_i\}} = \frac{\partial^2 \mathcal{H}}{\partial\theta_i\,\partial\theta_j}\Biggr|_{\{\theta^*_i\}}
\end{equation}
computed on the configuration $\{\theta^*_i\}$ of the minimum where the system has relaxed (hopefully the true ground state), and keeping in mind that in the $T \to 0$ limit the free energy reduces to the internal energy, i.\,e. to the Hamiltonian $\mathcal{H}$. More explicitly:
\begin{equation}
	\mathbb{H}_{ij} = 
	\left\{
	\begin{aligned}
		&\sum_{k\in\partial i}J_{ik}\cos(\theta^*_i-\theta^*_k) + H\cos(\theta^*_i-\phi_i) && \quad i=j\\
		&-J_{ij}\cos(\theta^*_i-\theta^*_j) && \quad (i,j) \text{ edge of } \mathcal{G}\\
		&0 && \quad \text{otherwise}
	\end{aligned}
	\right.
	\label{eq:Hessian_XY}
\end{equation}
From this definition, it is quite clear that~$\mathbb{H}$ is a \textit{sparse} $N \times N$ matrix, with exactly $C$ nonvanishing offdiagonal entries per row and per column. More in general, for a $m$-dimensional vector spin glass it is a $(m-1)N \times (m-1)N$ matrix, due to the constraint over the norm of each spin. E.\,g., for Heisenberg spins ($m=3$) the linear size of $\mathbb{H}$ is $2N$. Notice that the description of XY spins in terms of $\theta$'s angles automatically takes into account the small fluctuations around the minimum in the transverse direction with respect to the spin orientation, while for $m>2$-component spins --- where a description in terms of polar coordinates is not easily attainable --- the \textit{local} transverse direction has to be explicitely computed each time~\cite{Thesis_BaityJesi2016}.

The physical meaning of $\mathbb{H}$ is well known: its eigenvalues $\{\lambda_i\}$ give the energetic cost of excitations in the energy landscape, and hence they rule the stability of the configuration $\{\theta^*_i\}$, while the corresponding eigenvectors $\{\ket{\lambda_i}\}$ describe the directions along which such excitations extend. However, since some properties are well defined only in the thermodynamic limit, in which $\mathbb{H}$ becomes an infinite matrix, it is more useful to deal with the eigenvalue spectral density~$\rho(\lambda)$:
\begin{equation}
	\rho(\lambda) \equiv \lim_{N\to\infty}\frac{1}{N}\sum_{i=1}^{N}\delta(\lambda-\lambda_i)
	\label{eq:rhoLambda_def}
\end{equation}

As usual in statistical mechanics, a positive definite Hessian matrix implies the stability of the state in which the system is; in turn, this corresponds to the presence of a~\textit{gap} in the spectral density, namely any fluctuation around the minimum has a finite energetic cost, and hence exactly at zero temperature it does not occur. Instead, the closure of the gap in the Hessian matrix usually corresponds to a second-order phase transition: the previously stable minimum becomes now unstable and the system moves toward a new minimum, that is stable and characterized by a lower free energy~\cite{Book_Huang1988, Book_Parisi1988}.

However, this is not the case of spin glasses. In particular, the stability of the~\acrshort{RS} solution is not given by the Hessian of the free energy in the space of ``real'' configurations --- i.\,e. in the space of spin configurations --- but in the space of \textit{replica}, as briefly shown in~\autoref{chap:sg_replica}. The closure of the gap in this case corresponds to the well known~\acrshort{RSB}~\cite{deAlmeidaThouless1978, Parisi1980b}. Unfortunately, so far it is not yet clear if the two Hessians are related, and in particular if an instability in one of them does correspond to an instability also in the other one. We will try to answer this question in the following of this Chapter.

\subsection{How to compute the Hessian}

In order to analyze the Hessian matrix $\mathbb{H}$, the first task is to compute it. Indeed, it is a nontrivial operation and a careful discussion has to be done. So far, we have exploited the~\acrshort{PDA} in order to get an average over the ensemble of random graphs and over the disorder distribution of couplings and fields, so actually solving the distributional version of the~\acrshort{BP} equations at zero temperature:
\begin{equation}
	\mathbb{P}_{h}[h_{i\to j}] = \mathbb{E}_{\mathcal{G},J,\phi}\int\prod_{k=1}^{d_i-1}\mathcal{D}h_{k\to i}\,\mathbb{P}_{h}[h_{k\to i}]\,\delta\Bigl[h_{i\to j}-\mathcal{F}_0[\{h_{k\to i}\},\{J_{ik}\},\phi_i]\Bigr]
	\label{eq:def_PDA_zeroTemp_inField}
\end{equation}
The main advantage of this approach is the possibility of computing physical observables by summing local terms, and hence by picking at random the cavity fields $h$'s directly from the fixed-point probability distribution $\mathbb{P}^*_{h}$. This turns out to be directly related to the extensive nature of observables like the free energy and the internal energy, so that on (large enough) treelike topologies they can be actually computed as a sum of local terms and hence no long-range correlations enter in their computation (as long as the validity of the~\acrshort{BP} assumptions is ensured).

However, objects like the Hessian contain long-range correlations even on very large sparse random graphs, that can not be reproduced through the previous approach. Indeed, the likely presence of extended eigenvectors of $\mathbb{H}$ is intimately related to such long-correlations, and the diagonalization of $\mathbb{H}$ can be performed only through nonlocal operations, as e.\,g. the computation of the determinant.

The unique way to recover the aforementioned long-range correlations is to actually provide a given instance of the sample, namely a given sparse random graph with a quenched set of couplings $\{J_{ij}\}$ and field directions $\{\phi_i\}$, and then solve the zero-temperature~\acrshort{BP} equations on it (easily obtained by generalizing the zero-field case, as shown in~\autoref{app:BPeqs_XYmodel}):
\begin{equation}
	h_{i\to j}(\theta_i) \cong H\cos{(\theta_i-\phi_i)} + \sum_{k\in\partial i\setminus j}\max_{\theta_k}\bigl[h_{k\to i}(\theta_k)+J_{ik}\cos{(\theta_i-\theta_k)}\bigr]
	\label{eq:BP_eqs_XY_zeroTemp_inField}
\end{equation}
The resulting fixed-point $\{h^*_{i\to j}\}$ then yields --- at least in principle, as we will discuss later --- the ground state $\{\theta^*_i\}$. We will refer to this approach as the~\acrfull{GSA}, whose key steps are listed in the pseudocode~\ref{alg:RS_GI_zeroTemp}. Finally, we can evaluate the Hessian matrix~\autoref{eq:Hessian_XY} on it, so that in this way it actually contains the relevant features of the energy landscape for a given instance of the model close to its ground state.

\begin{algorithm}[!t]
	\caption{RS Given Sample Algorithm ($T=0$)}
	\label{alg:RS_GI_zeroTemp}
	\begin{algorithmic}[1]
		\State Fix an accuracy $\Delta_{\text{GSA}}$
		\State Generate the sparse random graph $\mathcal{G}$
		\For {$i=1,\dots,N$}
			\State Draw a field direction $\phi_i$ from the probability distribution $\mathbb{P}_{\phi}$
		\EndFor
		\For {$i=1,\dots,N$}
			\For {$j\in\partial i,\,j>i$}
				\State Draw a coupling $J_{ij}$ from the probability distribution $\mathbb{P}_J$
			\EndFor
		\EndFor
		\For {$i=1,\dots,N$}
			\For {$j\in\partial i$}
				\State Initialize $h^{(0)}_{i\to j}$ \Comment{We use a random initialization}
			\EndFor
		\EndFor
		\For {$t=1,\dots,t_{\text{max}}$}
			\For {$i=1,\dots,N$}
				\For {$j\in\partial i$}
					\State $h^{(t)}_{i\to j} \gets \mathcal{F}_0[\{h^{(t-1)}_{k\to i}\},\{J_{ik}\},\phi_i]$ \Comment{Use a damping $\gamma$ if necessary}
				\EndFor
			\EndFor
			\State $\Delta^{(t)} \gets \max_{i\to j}\norm{h^{(t)}_{i\to j}-h^{(t-1)}_{i\to j}}$
		\EndFor
		\If{$\Delta^{(t_{\text{max}})}<\Delta_{\text{GSA}}$}
			\State \textbf{return} $\{h^{(t_{\text{max}})}_{i\to j}\}$
		\Else
			\State \textbf{stop}: no convergence within accuracy $\Delta_{\text{GSA}}$
		\EndIf
	\end{algorithmic}
\end{algorithm}

\subsection{Reaching the ground state on a given sample}

Unfortunately, at variance with respect to the~\acrshort{PDA} --- which actually solves the distributional version~\autoref{eq:def_PDA_zeroTemp_inField} of the~\acrshort{BP} equations --- it may happen that the~\acrshort{GSA} not only does not reach the true ground state, but even it may not reach any fixed point of the~\acrshort{BP} equations at all, keeping wandering forever. This is a very common problem of~\acrshort{BP} approaches on given instances of a problem, strictly related to the presence of an enormous number of states --- i.\,e. \acrshort{BP} fixed points for $T<T_c$ ---, that in turn makes very hard to find the actual ground state.

In order to help the convergence, a trick typically exploited in these cases is the insertion of a damping $\gamma$ in the~\acrshort{BP} equations, so to actually perform just a fraction of a \acrshort{BP} iteration at each time step of the algorithm:
\begin{equation}
	h^{(t)}_{i\to j}(\theta_i) \quad \leftarrow \quad (1-\gamma)\,h^{(t)}_{i\to j}(\theta_i) + \gamma\,h^{(t-1)}_{i\to j}(\theta_i)
\end{equation}
The value of $\gamma$ is typically chosen according to the ``difficulty'' of reaching the convergence, e.\,g. according to the gradient around the fixed point: larger values correspond to smaller \acrshort{BP} steps, so increasing the possibility of actually reaching the fixed point. On the other hand, larger values of $\gamma$ of course imply a slower algorithm. In all the~\acrshort{GSA} simulations of this Chapter we use $\gamma=0.1$, which in our case represents a good compromise between efficiency and reliability.

Moreover, due to the continuous nature of XY spins, the space of configurations is continuous as well and hence infinitesimal displacements are always well defined. In particular, due to the presence of the random field, any degeneration of the minima is broken and hence wherever the system is, a gradient descent would allow us to reach the bottom of one of the nearest valleys with probability one. According to this observation, when we realize that the~\acrshort{GSA} has performed enough steps without reaching the convergence within the assigned accuracy $\Delta_{\text{GSA}}$, we can stop it and try to reach one of the nearest minima, in order to obtain a properly positive definite Hessian matrix.

Before explaining how to implement this further trick, let us focus on a detail that so far has remained hidden in this Chapter. Every time we perform a numerical simulation on the XY model, we are actually using a $Q$-state clock model with a reasonably large $Q$ value. In particular, since~\autoref{chap:XYnoField} we used $Q=64$, then justifying this choice in~\autoref{chap:clock}. Hence, the discretization of the XY model via the $Q$-state clock model does not show any negative effects in the~\acrshort{PDA}.

When using the~\acrshort{GSA}, however, discretization becomes crucial. The system is moving in a well defined (quenched) energy landscape, following a $N$-dimensional grid of lattice spacing $a=2\pi/Q$. Any minimum of the ``true'' landscape is with probability one out of the nodes of the grid for any finite value of $Q$, and hence the~\acrshort{GSA} never stops in an actual minimum of the XY model energy landscape. The consequence of this is straightforward: even if the actual fixed point of the XY model landscape were a minimum, the Hessian matrix computed through such~$\{\theta^*_i\}$ given by the $Q$-state clock model could not be positive definite, so providing negative eigenvalues.

In this case, the prescription is to stop in the lowest configuration $\{\theta^*_i\}$ reached via the~\acrshort{GSA} at $t=t_{\text{max}}$ --- whether the algorithm has reached the convergence or not --- and then forget about the $Q$-state clock model. Each discrete angle $\theta^*_i$ is then associated with the corresponding magnetization vector $\boldsymbol{m}_i$ on the unit circle (remembering that $T=0$ and that degeneracies have been removed by the randomly oriented field):
\begin{equation}
	\boldsymbol{\sigma}_i = \boldsymbol{m}_i \equiv \bigl(\cos{\theta^*_i},\sin{\theta^*_i}\bigr)
\end{equation}
Finally, we implement the~\acrfull{GCD}, namely a zero-temperature dynamics which aligns each spin to the effective local field, also known as the Gauss-Seidel procedure:
\begin{equation}
	\boldsymbol{\sigma}^{(t)}_i \cong \boldsymbol{H}_i + \sum_{k\in\partial i}J_{ik}\boldsymbol{\sigma}^{(t-1)}_k
	\label{eq:GCD}
\end{equation}
enforcing again the unit normalization for each spin at the end of each time step. It is described in more detail in the pseudocode~\ref{alg:zeroTemp_Dynamics}. This allows the system to actually reach the bottom of the valley where the~\acrshort{GSA} stopped and hence obtain a positive definite Hessian matrix. Notice that this procedure \textit{always} reaches a stable minimum, provided $t_{\text{max}}$ is large enough to stay within the given accuracy $\Delta_{\text{GCD}}$ and the degeneracies have been removed by the presence of the randomly oriented field.

\begin{algorithm}[!t]
	\caption{Greedy Coordinate Descent}
	\label{alg:zeroTemp_Dynamics}
	\begin{algorithmic}[1]
		\State Fix an accuracy $\Delta_{\text{GCD}}$
		\State Take the final configuration $\{h^{(t_{\text{max}})}_{i\to j}\}$ of~\acrshort{GSA} \Comment{Or random, if necessary}
		\For {$i=1,\dots,N$}
			\State Compute the site marginal $h^{(t_{\text{max}})}_i$
			\State Compute the most probable direction $\theta^*_i$ over the allowed $Q$ ones
			\State $\boldsymbol{\sigma}^{(0)}_i \gets (\cos{\theta^*_i},\sin{\theta^*_i})$
		\EndFor
		\For {$t=1,\dots,t_{\text{max}}$}
			\For {$i=1,\dots,N$}
				\State $\boldsymbol{\sigma}^{(t)}_i \gets \boldsymbol{H}_i + \sum_{k\in\partial i}J_{ik}\boldsymbol{\sigma}^{(t-1)}_k$
				\State $\boldsymbol{\sigma}^{(t)}_i \gets \boldsymbol{\sigma}^{(t)}_i/\norm{\boldsymbol{\sigma}^{(t)}_i}$
			\EndFor
			\State $\Delta^{(t)} \gets \max_i\norm{\boldsymbol{\sigma}^{(t)}_i-\boldsymbol{\sigma}^{(t-1)}_i}$
		\EndFor
		\If{$\Delta^{(t_{\text{max}})}<\Delta_{\text{GCD}}$}
			\State Compute the real-valued $\{\theta^*_i\}$ from $\boldsymbol{\sigma}^{(t_{\text{max}})}_i$
			\State \textbf{return} $\{\theta^*_i\}$
		\Else
			\State \textbf{stop}: no convergence within accuracy $\Delta_{\text{GCD}}$, increase $t_{\text{max}}$
		\EndIf
	\end{algorithmic}
\end{algorithm}

At this point, we can go back to the issue illustrated above. When the~\acrshort{GSA} does not converge within a reasonable time, we can stop the algorithm at $t=t_{\text{max}}$, move from the angle $\theta^*_i$ --- namely the best approximation of the ground state provided by the~\acrshort{GSA} --- to the magnetization vector $\boldsymbol{\sigma}_i$ for each site and then exploit the zero-temperature dynamics given by the~\acrshort{GCD}, reaching the bottom of the closest valley in the energy landscape. This allows us to obtain a meaningful Hessian matrix to analyze even in those cases in which a \acrshort{RS} solution of the~\acrshort{BP} equations is very hard to find --- or even impossible. The careful reader could, however, object that these particular instances introduce some bias in our analysis. We will show in the following that this is not the case; on the other hand, their inclusion will actually enlarge the statistics in the region of hard convergence for the~\acrshort{GSA}.

Once explained how to compute $\mathbb{H}$, we can finally do it by choosing $Q=64$ for the clock model in the~\acrshort{GSA} and then by implementing the zero-temperature dynamics through the~\acrshort{GCD}. We study several instances of the system for different sizes of the graph --- $N=10^3$, $10^4$, $10^5$ and $10^6$ --- starting from high values of the field intensity $H/J$ and then performing an annealing in it. In~\autoref{tab:GSA_statistics} we report some statistics about these simulations, including the values of $H$ taken into account, the number of samples $\mathcal{N}_s$ analyzed for each size $N$, the number of samples $\mathcal{N}^*_s$ on which the~\acrshort{GSA} actually converged within $t_{\text{max}}=500$ iterations, the corresponding average number $t^*_{\text{GSA}}$ of time steps spent by the~\acrshort{GSA}, the average number $t^*_{\text{GCD}}$ of time steps spent by the~\acrshort{GCD} (also taking into account the samples where the~\acrshort{GSA} did not converge within the $t_{\text{max}}$ iterations). The convergence thresholds chosen for the two algorithms are respectively $\Delta_{\text{GSA}}=10^{-10}$ and $\Delta_{\text{GCD}}=10^{-6}$. Finally, notice that the value $t_{\text{max}}=500$ for the~\acrshort{GSA} has been chosen as a good compromise between the efficiency of the algorithm and the need of actually reaching the true ground state for all the sizes considered.

A few points can be highlighted by looking at~\autoref{tab:GSA_statistics}. First of all, the number of samples $\mathcal{N}^*_s$ on which the~\acrshort{GSA} converged when lowering the field strength $H$ gives a ``measure'' of the~\acrshort{RS} stability of the~\acrshort{BP} fixed point. Indeed, if for $N=10^3$ its decrease with $H$ is rather smooth, when increasing the size $N$ a quite sharp threshold appears, in correspondence of a value of $H$ slightly larger than the actual value~$H_{\text{dAT}}$, that can be evaluated via the~\acrshort{PDA} as in~\autoref{chap:XYinField}:
\begin{equation}
	\qquad H_{\text{dAT}}/J = 1.059(2)
\end{equation}
and that can be recovered here when sending $t_{\text{max}}\to\infty$.

\begin{table}[p]
	\centering
	\caption[Statistics about GSA and GCD on different samples for several sizes and fields]{Statistics about the~\acrshort{GSA} and the~\acrshort{GCD} on $\mathcal{N}_s$ samples for several sizes~$N$ and field strengths $H/J$. $\mathcal{N}^*_s$ is the number of samples on which the~\acrshort{GSA} actually reached convergence with an accuracy $\Delta_{\text{GSA}}=10^{-10}$ within $t_{\text{max}}=500$ iterations. $t^*_{\text{GSA}}$ is the average number of the~\acrshort{GSA} iterations performed on the samples where it actually converged within $t_{\text{max}}$, while $t^*_{\text{GCD}}$ is the average number of the~\acrshort{GCD} iterations performed on all the samples to reach the accuracy $\Delta_{\text{GCD}}=10^{-6}$.}
	\label{tab:GSA_statistics}
	\begin{footnotesize}
	\begin{tabular}{*{13}{c}}
		\toprule
		 & \multicolumn{3}{c}{$N=10^3$} & \multicolumn{3}{c}{$N=10^4$} & \multicolumn{3}{c}{$N=10^5$} & \multicolumn{3}{c}{$N=10^6$}\\
		 & \multicolumn{3}{c}{($\mathcal{N}_s=400$)} & \multicolumn{3}{c}{($\mathcal{N}_s=200$)} & \multicolumn{3}{c}{($\mathcal{N}_s=100$)} & \multicolumn{3}{c}{($\mathcal{N}_s=50$)}\\
		\cmidrule(lr){2-4}\cmidrule(lr){5-7}\cmidrule(lr){8-10}\cmidrule(lr){11-13}
		$H/J$ & $\mathcal{N}^*_s$ & $t^*_{\text{GSA}}$ & $t^*_{\text{GCD}}$ & $\mathcal{N}^*_s$ & $t^*_{\text{GSA}}$ & $t^*_{\text{GCD}}$ & $\mathcal{N}^*_s$ & $t^*_{\text{GSA}}$ & $t^*_{\text{GCD}}$ & $\mathcal{N}^*_s$ & $t^*_{\text{GSA}}$ & $t^*_{\text{GCD}}$\\
		\midrule
		50.0 & 400 & 18 & 5 & 200 & 21 & 5 & 100 & 24 & 5 & 50 & 27 & 5\\
		25.0 & 400 & 21 & 5 & 200 & 25 & 6 & 100 & 29 & 6 & 50 & 33 & 6\\
		10.0 & 400 & 26 & 7 & 200 & 31 & 8 & 100 & 38 & 8 & 50 & 43 & 8\\
		9.00 & 400 & 26 & 8 & 200 & 31 & 8 & 100 & 37 & 8 & 50 & 43 & 9\\
		8.00 & 400 & 26 & 8 & 200 & 33 & 9 & 100 & 39 & 9 & 50 & 45 & 9\\
		7.00 & 400 & 28 & 9 & 200 & 35 & 10 & 100 & 42 & 10 & 50 & 48 & 11\\
		6.00 & 400 & 31 & 10 & 200 & 38 & 12 & 100 & 45 & 13 & 50 & 52 & 14\\
		5.00 & 400 & 33 & 14 & 200 & 41 & 18 & 100 & 49 & 23 & 50 & 60 & 29\\
		4.00 & 400 & 38 & 27 & 200 & 48 & 60 & 100 & 60 & 132 & 50 & 73 & 341\\
		3.00 & 400 & 49 & 46 & 200 & 66 & 103 & 100 & 89 & 212 & 50 & 111 & 600\\
		2.50 & 399 & 62 & 61 & 200 & 87 & 156 & 100 & 118 & 350 & 50 & 150 & 899\\
		2.00 & 389 & 87 & 96 & 199 & 131 & 234 & 99 & 182 & 565 & 50 & 251 & 1443\\
		1.80 & 369 & 104 & 126 & 192 & 151 & 291 & 99 & 224 & 785 & 49 & 324 & 1681\\
		1.60 & 336 & 120 & 124 & 186 & 198 & 305 & 94 & 303 & 768 & 26 & 417 & 1610\\
		1.50 & 310 & 128 & 133 & 181 & 223 & 308 & 72 & 340 & 641 & 16 & 444 & 1959\\
		1.40 & 272 & 139 & 125 & 148 & 247 & 327 & 56 & 385 & 753 & 2 & 459 & 1772\\
		1.30 & 252 & 146 & 135 & 121 & 282 & 318 & 27 & 415 & 833 & 0 & / & 1884\\
		1.25 & 215 & 151 & 149 & 105 & 287 & 348 & 16 & 422 & 921 & 0 & / & 1808\\
		1.20 & 194 & 158 & 144 & 83 & 315 & 330 & 7 & 471 & 780 & 0 & / & 1635\\
		1.15 & 171 & 162 & 159 & 72 & 328 & 348 & 0 & / & 716 & 0 & / & 1882\\
		1.10 & 148 & 165 & 148 & 50 & 318 & 310 & 3 & 455 & 711 & 0 & / & 1803\\
		1.05 & 127 & 162 & 160 & 29 & 354 & 335 & 0 & / & 758 & 0 & / & 1702\\
		1.00 & 108 & 177 & 151 & 20 & 353 & 309 & 0 & / & 771 & 0 & / & 1819\\
		0.90 & 75 & 195 & 151 & 9 & 367 & 293 & 0 & / & 750 & 0 & / & 1590\\
		0.80 & 52 & 233 & 130 & 4 & 388 & 244 & 0 & / & 680 & 0 & / & 1215\\
		0.70 & 28 & 214 & 129 & 2 & 352 & 221 & 0 & / & 459 & 0 & / & 1093\\
		0.60 & 16 & 262 & 144 & 0 & / & 230 & 0 &  / & 485 & 0 & / & 1056\\
		0.50 & 7 & 205 & 176 & 0 & / & 223 & 0 & / & 519 & 0 & / & 1031\\
		\bottomrule
	\end{tabular}
	\end{footnotesize}
\end{table}

Secondly, the average convergence time $t^*_{\text{GSA}}$ increases when approaching the~\acrshort{dAT} point --- as expected for second-order phase transitions ---, ideally diverging in the $t_{\text{max}}\to\infty$ limit. On the other hand, $t^*_{\text{GSA}}$ also increases with the size $N$ at the same value of $H/J$.

Thirdly, also the average time $t^*_{\text{GCD}}$ spent by the~\acrshort{GCD} increases with $N$ at the same~$H$ --- as expected --- and, more interestingly, when approaching the~\acrshort{dAT} point at the same size $N$. This slowing down of the algorithm is a signature of a flatter energy landscape around the minima, namely of a larger fraction of very small eigenvalues of the Hessian. This claim will be confirmed in the next Section, when actually computing the spectral density $\rho(\lambda)$ of $\mathbb{H}$.

\subsection{Quality of the inherent structures reached}

At this point, we may wonder whether the local minimum reached through the~\acrshort{GSA} plus the~\acrshort{GCD} is actually the ground state $\{\theta^*_i\}$ of the energy landscape of the analyzed sample, or otherwise how ``far'' it is from it. In other words, we would like to evaluate the ``quality'' of the \textit{inherent structures}~\cite{Cavagna2009, BaityJesiEtAl2015, BaityJesiParisi2015} reached via this approach.

\begin{figure}[t]
	\centering
	\includegraphics[scale=1]{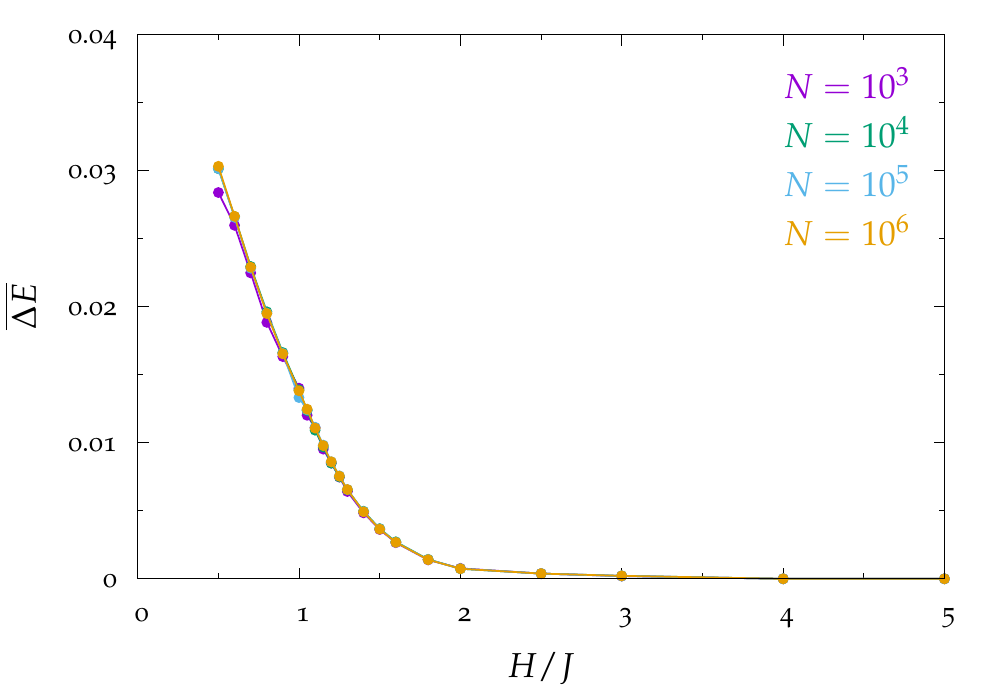}
	\caption[Energy difference in the local minima via different approaches]{Average over the samples for the percentage energy difference $\Delta E \equiv (E-E')/E'$ between the local minimum at energy $E$ reached via the combined algorithm \acrshort{GSA} plus \acrshort{GCD} and the one at energy $E'$ reached just by using \acrshort{GCD}. The two algorithms are supposed to start from the same initial condition for a given sample. It is quite clear how \acrshort{BP}-based approaches can go deeper in the energy landscape when in presence of~\acrshort{RSB} --- end even slightly above the \acrshort{dAT} point --- with respect to relaxation-based ones, so providing ``better'' inherent structures.}
	\label{fig:energy_BPplusMC_vs_onlyMC}
\end{figure}

A quantitative answer can be provided by comparing the energy $E$ of the minimum reached via the combined algorithm introduced above with the energy $E'$ of the minimum reached via the only use of the~\acrshort{GCD}, given the same set of initial conditions for both algorithms. In~\autoref{fig:energy_BPplusMC_vs_onlyMC} we report the corresponding energy difference $\Delta E \equiv (E-E')/E'$, averaged over the $\mathcal{N}_s$ samples for each size~$N$, for different values of the field intensity $H/J$. If for large values of $H$ the two algorithms relax to the same minimum, which is actually unique and hence the sought ground state, when getting closer to the critical point they provide different results. In particular, the~\acrshort{GCD} is just a pure-relaxation algorithm and hence it can not overcome the barriers appearing in the energy landscape when close to the \acrshort{dAT} point. At variance, the~\acrshort{GSA} can do it, with the subsequent use of the~\acrshort{GCD} that just ensures the proper optimization over the real-valued $\theta_i$'s. On average, the purely relaxation approach just stops at some metastable minimum with an energy that is about $1\%$ higher than that of the actual ground state, that at variance is believed to be actually reached by~\acrshort{BP}. Moreover, below $H_{\text{dAT}}$, the difference is even more striking, since the very rugged nature of the energy landscape in such regime: despite neither of the two algorithms reach the true ground state, \acrshort{GSA} can go deeper in the energy landscape by nearly a $2\%-3\%$ with respect to the only \acrshort{GCD}, referring to the range of $H/J$ values in~\autoref{fig:energy_BPplusMC_vs_onlyMC}.

This is an important finding, since in numerical simulations on finite-dimension lattices one typically exploits relaxation algorithms based on the~\acrshort{GCD}, as e.\,g. done in Ref.~\cite{BaityJesiEtAl2015} with Heisenberg spins for $d=3$. Despite being by far faster than~\acrshort{BP}-based algorithms, they most of times get stuck into metastable minima when the energy landscape is rugged, so that the resulting low-temperature statistical properties actually refer to them rather than to the true ground state. At variance, on sparse random graphs \acrshort{BP} allows to exactly reach the global minimum of the energy almost everywhere in the~\acrshort{RS} region, then still providing appreciable results when in presence of rugged landscapes due to the occurrence of~\acrshort{RSB}.

Finally, still looking at the energy $E$ of the minimum $\{\theta^*_i\}$ reached via the~\acrshort{GSA} plus the~\acrshort{GCD}, we can check whether there are substantial differences between the samples on which the~\acrshort{GSA} has converged within the $t_{\text{max}}$ allowed iterations and the other ones. It turns out that the energy distribution is actually the same, hence it is reasonable to claim that no biases are introduced if considering also the latter samples in the forthcoming analysis.

\section{Considerations about the spectral density}
\label{sec:considerations_rhoLambda}

Before actually computing the spectral density of the Hessian matrix, let us make some considerations about what we should expect about it.

\subsection{The fully connected case}

First of all, we can consider the fully connected version of our model. In absence of an external field, the Hessian $\mathbb{H}$ is a dense symmetric matrix with $O(1/\sqrt{N})$ entries, since each coupling $J_{ij}$ is Gaussian distributed with zero mean and variance $1/N$. In the $N\to\infty$ limit, hence, the central limit theorem holds, namely the entries become actually independent Gaussian-distributed random variables, leading to the famous Wigner semicircle law~\cite{Wigner1958, EdwardsJones1976, Book_Mehta2004}
\begin{equation}
	\rho(\lambda)=\frac{1}{2\pi}\sqrt{4-\lambda^2} \quad , \qquad \lambda\in[-2,2]
	\label{eq:Wigner_law}
\end{equation}
The presence of a field of intensity $H$ in the system just shifts rightward the spectral density~\autoref{eq:Wigner_law} of a quantity $O(H)$. Hence, if it is large enough, a gap opens in the spectral density~$\rho(\lambda)$, providing a strictly positive lower band edge $\tilde{\lambda}$.

The Wigner semicircle law is very common, since the Gaussian ensemble of random matrices turns out to correctly describe the statistical properties of a large number of physical phenomena. We redirect the reader to Refs.~\cite{Beenakker1997, Book_Mehta2004, Book_Forrester2010, Book_AkemannEtAl2011} for reviews and further details.

Here, we are particularly interested in its square-root behaviour close to the lower band edge, that holds whether or not a gap is present:
\begin{equation}
	\rho(\lambda) \sim \bigl(\lambda-\tilde{\lambda}\bigr)^{1/2} \quad , \qquad \lambda \gtrsim \tilde{\lambda}
\end{equation}
This square-root behaviour makes the spin glass susceptibility $\chi_{\text{SG}}$ be finite only if $\tilde{\lambda}>0$, namely if the Hessian matrix $\mathbb{H}$ is positive definite. This typically occurs if the field intensity $H$ is large enough. However, if e.\,g. $H$ decreases down to a certain value $H_{\text{dAT}}$, then the gap closes, nonintegrable soft modes appear in the spectral density and $\chi_{\text{SG}}$ eventually diverges, signaling the breaking of replica symmetry.

This is the typical picture for the onset of \acrshort{RSB} in fully connected models~\cite{BrayMoore1979, CugliandoloKurchan1993, CavagnaEtAl1998, Plefka2002, SharmaEtAl2016}, and the square-root behaviour close to the lower band edge seems to be even more general, also occurring for other fully connected models where $\rho(\lambda)$ is no longer given by the (shifted) Wigner semicircle law~\cite{FranzEtAl2015}.

\subsection{A large-field expansion in the sparse case}

If in the large-$N$ limit the Hessian $\mathbb{H}$ exhibits the Wigner semicircular spectral density, it is reasonable to not expect a similar behaviour in the sparse case. Indeed, each row (as well as each column) of $\mathbb{H}$ has just $O(1)$ nonnull elements, and the correlations between the entries are quite strong. Instead of the Wigner law~\autoref{eq:Wigner_law}, the spectral density of the Hessian matrix when $\mathcal{G}$ is a $C$-\acrshort{RRG} with $C=O(1)$ should rather resemble in some way the spectral density of the corresponding adjacency matrix~$\mathbb{A}$
\begin{equation}
	\mathbb{A}_{ij} = 
	\left\{
	\begin{aligned}
		&1 && \quad (i,j) \text{ edge of } \mathcal{G}\\
		&0 && \quad \text{otherwise}
	\end{aligned}
	\right.
\end{equation}
namely the Kesten-McKay law~\cite{Kesten1959, McKay1981}:
\begin{equation}
	\rho(\lambda) = \frac{C\sqrt{4(C-1)-\lambda^2}}{2\pi(C^2-\lambda^2)} \quad , \qquad \lambda\in[-2\sqrt{C-1},2\sqrt{C-1}]
	\label{eq:Kesten_McKay_law}
\end{equation}
or equivalently, if rescaling the entries of $\mathbb{A}$ by $\sqrt{C-1}$
\begin{equation}
	\rho(\lambda) = \frac{1}{2\pi}\sqrt{4-\lambda^2}\cdot\biggl(\frac{C}{C-1}-\lambda^2\frac{1}{C}\biggr)^{-1} \quad , \qquad \lambda\in[-2,2]
	\label{eq:Kesten_McKay_law_resc}
\end{equation}
so to recover the Wigner law~\autoref{eq:Wigner_law} in the large-$C$ limit.

However, the Hessian matrix $\mathbb{H}$ \textit{is not} the adjacency matrix $\mathbb{A}$. Firstly, its diagonal entries are different from zero, depending on both the local external field~$\boldsymbol{H}_i$ and the couplings $J_{ij}$'s with the nearest neighbours; hence, for large enough values of $H$, the spectral density~\autoref{eq:Kesten_McKay_law} should at least be shifted rightward, and also a well pronounced peak should appear at $\lambda \simeq H$. Secondly, offdiagonal entries are no longer all equal to $1$, but they depend on the relative angles $(\theta^*_i-\theta^*_j)$'s; this should imply a further modification of the spectral density~\autoref{eq:Kesten_McKay_law_resc}, most likely on its tails, due to the strong correlations between nearest-neighbour variables.

The different behaviour in the spectral density of sparse random matrices with respect to their fully connected limit is a well known issue in the literature~\cite{RodgersBray1988, BrayRodgers1988, DorogovtsevEtAl2003, Kuhn2008, RogersEtAl2008}. In particular, for what regards the spectral density of the Hessian matrix $\mathbb{H}$, in Ref.~\cite{BaityJesiEtAl2015} the $m=3$ version of our model is studied on a $d=3$ cubic lattice. From such numerical study some relevant features come out, that could be recovered also in the random diluted case. Firstly, a gap in the spectral density is observed for very large values of~$H$, while at a certain value $H=H_{\text{gap}}$ it closes; however, such closure does not correspond to a breaking of the replica symmetry. Secondly, the density of the soft modes in the gapless region does not follow any longer the square-root behaviour $\rho(\lambda) \sim \lambda^{\alpha},\,\alpha=1/2$, rather the data are compatible with the value $\alpha=3/2$. Indeed, no divergence of $\chi_{\text{SG}}$ is implied.

The advantage of working on sparse random graphs rather than in finite dimension is that we can also approach the $H_{\text{dAT}}$ value and then even explore the \acrshort{RSB} region --- whose existence is at variance still debated in finite dimension even for the Ising case~\cite{CharbonneauYaida2017} --- though still using the \acrshort{RS} ansatz. Moreover, we saw that the~\acrshort{GSA} can go deeper in the energy landscape with respect to relaxational MonteCarlo approaches --- that have been used in Ref.~\cite{BaityJesiEtAl2015} ---, actually reaching the true ground state for almost all the samples down to the \acrshort{dAT} critical point.

Moreover, exploiting again the sparsity of the graph, it is possible to suddenly get a reliable approximation of the spectral density of $\mathbb{H}$ in the large-$H$ region. Indeed, given the Hamiltonian $\mathcal{H}$ of Eq.~\autoref{eq:Hamiltonian_XY_sparse_in_Field_chap7}, the ground state $\{\theta^*_i\}$ is given by the extremal condition:
\begin{equation}
	\frac{\partial \mathcal{H}}{\partial \theta_i} = 0 \quad \forall i \qquad \Rightarrow \qquad \sum_{k\in\partial i}J_{ik}\sin{(\theta^*_i-\theta^*_k)} + H\sin{(\theta^*_i-\phi_i)} = 0 \quad \forall i
	\label{eq:Hess_stat}
\end{equation}
that can be perturbatively solved via an expansion in $J/H$. The resulting approximated expression of the ground state is then plugged into the Hessian~\autoref{eq:Hessian_XY}, which is then diagonalized to get $\rho(\lambda)$.

At the zeroth order of the expansion, we have that it is the field direction $\phi_i$ that rules the direction of the $i$-th spin, while its neighbours $\partial i$ are completely neglected:
\begin{equation}
	\theta^*_i = \phi_i
	\label{eq:LargeH_expansion_zeroth}
\end{equation}
At the next order, then, we have a first correction due to the presence of the neighbours:
\begin{equation}
	\theta^*_i = \phi_i + \delta\theta^{(1)}_i
	\label{eq:LargeH_expansion_first}
\end{equation}
with $\delta\theta^{(1)}_i$'s that have to satisfy the stationary condition~\autoref{eq:Hess_stat}, so giving:
\begin{equation}
	\delta\theta^{(1)}_i = -\sum_{k\in\partial i}\frac{J_{ik}}{H}\sin{(\phi_i-\phi_k)}
	\label{eq:LargeH_expansion_first_delta}
\end{equation}
Finally, at the second order, we write:
\begin{equation}
	\theta^*_i = \phi_i + \delta\theta^{(1)}_i + \delta\theta^{(2)}_i
	\label{eq:LargeH_expansion_second}
\end{equation}
and again from stationary condition~\autoref{eq:Hess_stat} we obtain:
\begin{equation}
	\delta\theta^{(2)}_i = -\sum_{k\in\partial i}\frac{J_{ik}}{H}\cos{(\phi_i-\phi_k)}\Bigl(\delta\theta^{(1)}_i-\delta\theta^{(1)}_k\Bigr)
	\label{eq:LargeH_expansion_second_delta}
\end{equation}

\begin{figure}[!t]
	\centering
	\includegraphics[scale=1]{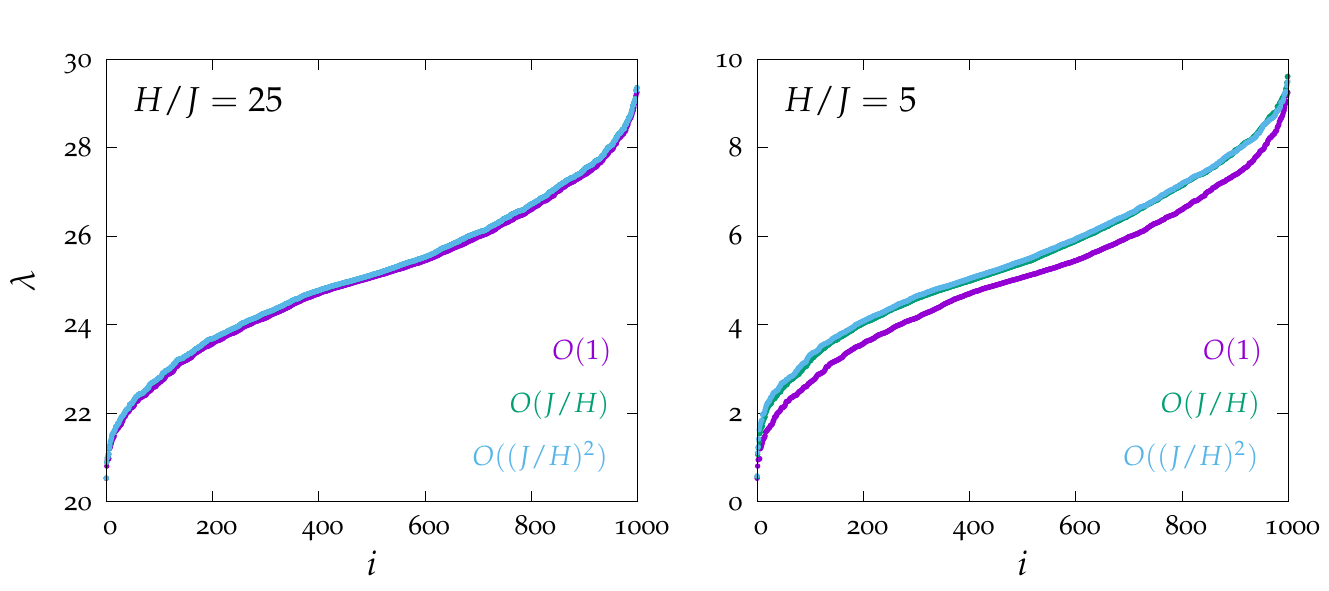}
	\caption[Large-field expansion for the Hessian eigenvalues]{Sorted eigenvalues of the Hessian $\mathbb{H}$ of the spin glass XY model in a random field for a given sample of size $N=10^3$. The underlying topology is as usual a~\acrshort{RRG} of connectivity $C=3$. The ground state $\{\theta^*_i\}$ has been evaluated via the large-field expansion at different perturbative orders in $J/H$. Left panel refers to $H/J=25$, right panel to $H/J=5$.}
	\label{fig:LargeH}
\end{figure}

In order to check the reliability of this approximation, let us focus on a given sample of size $N=10^3$. For each perturbative order in the approximation of the ground state, we evaluate the Hessian and then we compute its eigenvalues, finally reporting them in~\autoref{fig:LargeH} for the field strengths $H/J=25$ (left panel) and $H/J=5$ (right panel). It is evident that for very large values of $H$ the corrections given by the next-to-leading orders are quite negligible, due to the strong aligning effect given by the external field. Instead, when lowering $H$, the effect of neighbours in the evaluation of the true ground state becomes relevant and hence the first order provides a substantial correction with respect to the zeroth order.

If further lowering $H$, a nonnegligible fraction of eigenvalues becomes negative, signaling a no longer positive (semi)definite approximated Hessian matrix. Indeed, in such regime, $J/H$ is of order $1$ and hence the large-field expansion just fails, providing a completely wrong estimation of the actual ground state. Hence, we should rely again on the~\acrshort{GSA} in such regime.

In~\autoref{sec:rhoLambda}, then, we will compare~(see~\autoref{fig:some_rhoLambda_sampleAveraged}) the spectral density~$\rho(\lambda)$ computed via the previous large-field expansion --- averaged over different samples --- and the one given by alternative and more reliable approaches, for several values of $H/J$. In this way, we will actually realize where and how the approach of large-field expansion fails.

\subsection{An analytic argument for the density of soft modes}

Let us now consider a different point of view, in order to analyze the region in which the previous approach fails.

When dealing with systems possessing continuous symmetries, two main kinds of low-energy excitations can be identified: \textit{Goldstone modes}, related to the breaking of the continuous symmetry itself, and \textit{non-Goldstone modes}, which are not produced by the breaking of the continuous symmetry.

Acoustic phonons and spin waves in ordered media belong to the first category and they can be easily described in terms of wave-like equations of motion. This still holds when a weak disorder is introduced in the system, once coarse-grained the system over a scale larger than the typical wavelength. This implies that low-frequency excitations --- namely low-energy ones --- are the least affected by the presence of disorder.

But when the presence of Goldstone modes is ruled out --- as it happens in our case, due to the presence of the random field --- the possible presence of low-energy modes must originate from a different mechanism. Indeed, it is the disorder itself that can produce soft modes in the system, since the ground state strongly depends on the disorder configuration.

In order to justify this claim, let us follow the argument of Refs.~\cite{IlinEtAl1987, Parshin1994, GurarieChalker2003}, considering a one-dimensional picture of the energy landscape, say $U(x)$, which is supposed to be ``smooth enough'':
\begin{equation}
	U(x) = \sum_{n=1}^{\infty}a_n\frac{x^n}{n!}
\end{equation}
and where $U(0)=0$ for simplicity. The concept of quenched disorder translates into the randomness of $a_n$'s coefficients, so that $U(x)$ can be thought as a \textit{random potential}. Then, let $x_0$ be a minimum of $U(x)$, which for the moment is not required to be the global one. The previous expansion can be hence centered around $x_0$, keeping the terms up to the fourth order:
\begin{equation}
	U(x_0) \simeq U(x_0) + \frac{b_2}{2}(x-x_0)^2 + \frac{b_3}{6}(x-x_0)^3 + \frac{b_4}{24}(x-x_0)^4
\end{equation}
so to take into account the possibility of having other minima different from the one in $x_0$. The curvature of the minimum in $x_0$ is given by the $b_2$ coefficient, so that fluctuations around it have a frequency $\omega \sim \sqrt{b_2}$.

For small curvatures $b_2$ --- the ones we are interested in, since we are looking for soft modes --- it can be shown that $\mathbb{P}(b_2) \propto b_2$, once provided that the coefficients~$a_n$'s --- and hence also~$b_n$'s --- are drawn from a smooth distribution with no zeros and no divergences. So the \textit{density of states} $g(\omega)$ goes as:
\begin{equation}
	g(\omega) \sim \omega^{\delta} \quad , \qquad \delta = 3
\end{equation}

At this point, we add the requirement that $x_0$ is not only a local minimum, but also the global one. This additional condition can be ensured by bounding the term involving the third derivative of $U(x)$:
\begin{equation}
	\abs{b_3} < \sqrt{3 b_2 b_4}
\end{equation}
and in turn it further suppresses the probability distribution of small curvatures, $\mathbb{P}(b_2) \propto b_2^{3/2}$, from which in the end:
\begin{equation}
	g(\omega) \sim \omega^{\delta} \quad , \qquad \delta = 4
\end{equation}

Finally moving from the density of states $g(\omega)$ to the spectral density $\rho(\lambda)$:
\begin{equation}
	g(\omega) \sim \omega^\delta \quad \xrightarrow{\,\,\lambda = \omega^2\,\,} \quad \rho(\lambda) \sim \lambda^{\alpha} \qquad , \qquad \alpha = \frac{\delta - 1}{2}
\end{equation}
we get the prediction of the power-law exponent $\alpha$ for the density of soft modes in presence of a disorder-dependent ground state:
\begin{equation}
	\rho(\lambda) \sim \lambda^{\alpha} \quad , \qquad \alpha=\frac{3}{2}
\end{equation}
so providing an analytic argument for the numerical evidence of Ref.~\cite{BaityJesiEtAl2015}.

A few points could be raised against this argument, as already pointed out by Gurarie and Chalker~\cite{GurarieChalker2003}. Firstly, when the system is characterized by a very slow relaxation or even by a breaking of the ergodicity, then it is not usually expected to reach the actual ground state, but just to stop in a quite deep local minimum. However, the previous result regards the presence of other deeper minima in the very surroundings of $x_0$, that is typically ruled out by the \textit{fast regime} of the relaxation. Indeed, close minima are separated by energy barriers that are not likely to be high, so that the system rapidly relaxes toward the deepest minimum in the surroundings. Secondly, what happens when considering more than one dimension, or when moving to sparse random topologies? This argument is quite general and it is expected to hold even in these cases, provided the corresponding localization length of these soft excitations is finite. On the other side, exactly in the infinite-dimension limit it fails, so allowing the recovery of the Wigner distribution~\autoref{eq:Wigner_law}.

\begin{figure}[!t]
	\centering
	\includegraphics[scale=1]{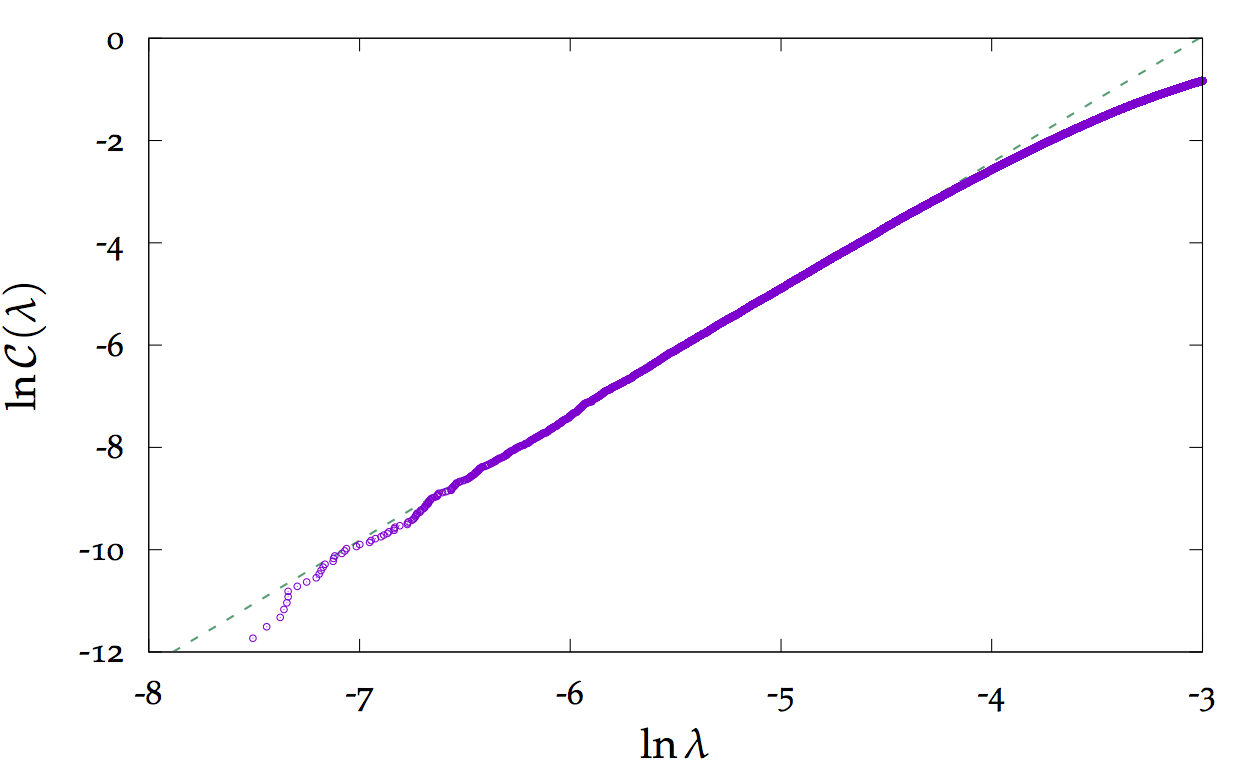}
	\caption[Cumulative distribution of the curvature of the global minimum of a random quartic potential]{Cumulative function $\mathcal{C}(\lambda)$ of the curvature $\lambda$ around the global minimum $x_0$ of $N=10^6$ random quartic polynomials. The straight line fits the data with a slope $\alpha+1 = 2.47(1)$.}
	\label{fig:polin_4grado}
\end{figure}

As a sudden check of this argument, let us compute the probability distribution of the curvature $\lambda$ around the global minimum of $N$ random quartic potentials. The corresponding cumulative function $\mathcal{C}(\lambda)$ computed for $N=10^6$ --- zoomed in the region of the left tail --- is reported in~\autoref{fig:polin_4grado}, together with the straight line that better fits the data in the log-log scale. We find a good agreement with the theoretical prediction:
\begin{equation}
	\alpha+1 = 2.47(1)
\end{equation}

\section{The spectral density of the Hessian}
\label{sec:rhoLambda}

At this point, we can finally diagonalize the Hessian matrices computed so far and extract their spectral density and their eigenvectors, so to characterize the energy landscape.

However, since we are dealing with quite large matrices, we will use different methods in order to compute $\rho(\lambda)$. Then, we will check the reliability of the different approaches by comparing the results for the Hessian matrices of linear size $N=10^3$, namely the smallest ones we computed.

\subsection{Direct diagonalization}

The usual algorithms for direct diagonalization of a $N \times N$ matrix require a computational time which scales as $N^3$. Indeed, they manipulate the matrix irrespective of their structure, and hence there is no difference whether they are dense or sparse. This approach works efficiently only for the smallest sizes we use, namely $N=10^3$ and $N=10^4$, while the direct diagonalization of larger matrices is practically unfeasible.

\begin{figure}[p]
	\centering
	\includegraphics[scale=0.99]{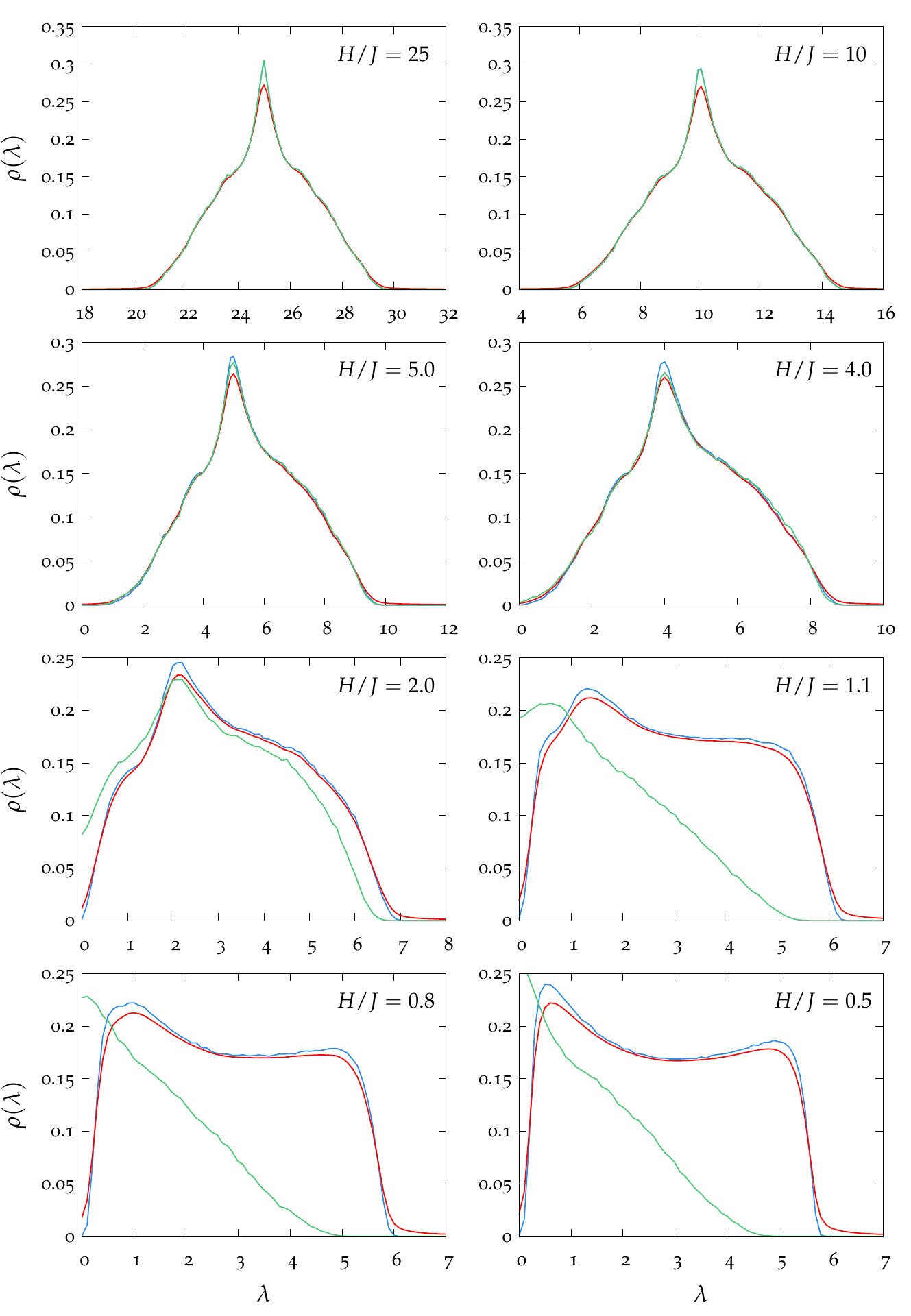}
	\caption[Sample-averaged spectral density of the Hessian for several values of~$H$, computed via different techniques]{Spectral density $\rho(\lambda)$ of the Hessian matrix for several values of the field intensity~$H$, averaged over $\mathcal{N}_s=400$ samples of size $N=10^3$. Blue curves refer to direct diagonalization of the Hessian matrices, red curves are obtained through the resolvent approach, green curves have been computed via the second-order large-field expansion. Notice that the latter approach fails for small values of $H$, providing completely wrong spectral densities having support also on negative values of $\lambda$.}
	\label{fig:some_rhoLambda_sampleAveraged}
\end{figure}

For $N=10^3$ and $N=10^4$ we can still do it through \mbox{\textit{Mathematica}\texttrademark} in a short time~($\sim 0.1\,\mathrm{s}$ for $N=10^3$ and $\sim 100\,\mathrm{s}$ for $N=10^4$), and the resulting spectral density for $N=10^3$ averaged over $\mathcal{N}_s=400$ samples from the $C=3$~\acrshort{RRG} ensemble is reported through blue curves in~\autoref{fig:some_rhoLambda_sampleAveraged} for several values of the field strength $H/J$.

As expected, for large values of $H$ the mean value of the spectral density $\rho(\lambda)$ is centered around $H$ itself, with a width of the spectrum of order $4JC$. This is due to the quasi-diagonal nature of $\mathbb{H}$ for large values of $H$, since the $C$ offdiagonal elements per row are of order $J \ll H$ and hence quite negligible. Of course, a wide gap separates the spectrum from the $\lambda=0$ axis, since even the weakest fluctuation requires a large energy cost. In this regime, the large-field expansion up to the second order in $J/H$ (green curves in the figure) works quite well, being quite indistinguishable from the actual spectral density computed via the direct diagonalization.

When lowering the field strength, then, the gap diminishes and seems to close between $H/J=5$ and $H/J=4$. In order to exactly compute the value $H_{\text{gap}}$ at which it actually happens, we should previously compute the spectral density for larger sizes, so to take into account finite-size effects in the closure of the gap. This task will be accomplished through more efficient approaches.

Notice also that in correspondence of these values of the field strength $H$, the large-field expansion begins to fail, since $J/H$ is no longer a reliable parameter for a perturbative expansion. The inexact location of the ground state for small values of $H$ so implies a Hessian that is no longer positive definite, hence the spectral density computed in this way acquires a support also on negative values of $\lambda$.

\subsection{Arnoldi method}

So far, we have not really exploited the structured nature of the Hessian matrix. In fact, if we take it into account, the spectral density $\rho(\lambda)$ can be computed more easily. One of the most used algorithms to diagonalize sparse matrices is the Arnoldi one~\cite{Arnoldi1951, Book_Lehoucq1998, Software_Arpack}, which relies on the repeated multiplication of a basis of $m \ll N$ vectors by the matrix. In the long-time limit, the ratios between the old and the new norm of the vectors give the $m$ largest eigenvalues, while the directions along which they align represent the corresponding eigenvectors.

If we perform the following tranformation:
\begin{equation}
	\mathbb{H} \to \mathbb{H}' \equiv \lambda_{\text{max}}\mathbb{I}-\mathbb{H}
\end{equation}
where $\lambda_{\text{max}}$ is the largest eigenvalue of $\mathbb{H}$ and $\mathbb{I}$ is the $N \times N$ identity matrix, then the Arnoldi algorithm run on $\mathbb{H}'$ actually provides the $m$ smallest eigenvalues of~$\mathbb{H}$, which are the ones we need for the study of the gap closure.

In this way we compute the $m=50$ smallest eigenvalues of $\mathbb{H}$ for all the samples listed in~\autoref{tab:GSA_statistics}, in a time growing with $N^2$ instead of $N^3$. Notice that, due to the nature of Arnoldi algorithm, in order to compute the~$m$ eigenvalues within a reasonable accuracy, a basis of $M \gg m$ vectors has actually to be used. In our case we found that $M=3m=150$ is enough to recover the $m=50$ smallest eigenvalues computed before by direct diagonalization.

However, even though being very efficient when just dealing with the smallest eigenvalues of $\mathbb{H}$, the Arnoldi method becomes quite useless when dealing with the bulk of the spectral density. So let us move to a further approach, which helps us in this case.

\subsection{The resolvent}

The Hessian matrix $\mathbb{H}$ of each sample can be seen as a single realization of an ensemble of \textit{random matrices} with well specified features. In this case, they are sparse and associated to a treelike structure, in particular the $C=3$ \acrshort{RRG} one. So, usual random matrix techniques can be exploited also in this case~\cite{Book_Mehta2004, Book_AkemannEtAl2011}.

Among all the properties associated with random matrices, the most important one for us is the fact that they possess a well defined average spectral density $\rho(\lambda)$, and hence the spectral density of each realization is just the average one plus the noise due to statistical fluctuations.

The average spectral density $\rho(\lambda)$ can be computed introducing the Green function $\mathcal{R}(\lambda)$ associated to the matrix $\mathbb{H}$, which is mostly known as the \textit{resolvent} of~$\mathbb{H}$:
\begin{equation}
	\mathcal{R}(\lambda)\equiv\frac{1}{\lambda\mathbb{I}-\mathbb{H}}
\end{equation}
So it is a matrix of linear size $N$ as well.

It is clear that when $N$ is of order $10^5$ or $10^6$, as for the largest samples we analyzed, the direct computation of $\mathcal{R}(\lambda)$ by inverting the matrix $\lambda\mathbb{I}-\mathbb{H}$ is unfeasible. So it has to be computed in a different way.

The first computation of $\mathcal{R}(\lambda)$ has been performed in the same spirit of the~\acrshort{PDA} for solving~\acrshort{BP} equations~\cite{AbouChacraEtAl1973}, and actually it is just when the~\acrshort{PDA} has been introduced and exploited for the first time. Each diagonal element $\mathcal{R}_{ii}$ can be computed iteratively by following the Schur complement formula, namely by removing the $i$-th row and column from $\mathbb{H}$, so obtaining:
\begin{equation}
	\mathcal{R}_{ii}(\lambda) = \biggl[\lambda - \mathbb{H}_{ii} - \sum_{j,k \neq i}\mathbb{H}_{ij}\mathcal{R}^{(i)}_{jk}(\lambda)\mathbb{H}_{ki}\biggr]^{-1}
\end{equation}
where $\mathcal{R}^{(i)}$ is the resolvent of the matrix $\mathbb{H}$ without its $i$-th row and column~\cite{AbouChacraEtAl1973, CizeauBouchaud1994, CilibertiEtAl2005, Book_MoroneEtAl2015}. Offdiagonal elements can be computed similarly.

The removal of the $i$-th row and column from the Hessian $\mathbb{H}$ sounds like the actual removal of site $i$ from the graph, and indeed it is so. Hence, in the sparse treelike case, it means that the original matrix $\mathbb{H}$ is divided into independent submatrices, corresponding to the subtrees into which the original graph $\mathcal{G}$ is splitted by the removal of site $i$. So the above recurrency relation directly brings to a set of cavity relations for the resolvent $\mathcal{R}$, where a cavity resolvent $\widetilde{\mathcal{R}}$ can be defined in perfect analogy with the~\acrshort{BP} cavity marginals~\cite{AbouChacraEtAl1973, BiroliEtAl2010, BogomolnyGiraud2013}:
\begin{equation}
	\widetilde{\mathcal{R}}_{i\to j}(\lambda) \equiv \biggl[\lambda-\mathbb{H}_{ii}-\sum_{k\in\partial i\setminus j}\mathbb{H}^2_{ik}\widetilde{\mathcal{R}}_{k\to i}(\lambda)\biggr]^{-1}
	\label{eq:Res_BP_eqs}
\end{equation}
where also the symmetric nature of $\mathbb{H}$ has been exploited. Once reached the fixed point $\widetilde{\mathcal{R}}^*$ of these self-consistency equations, then the proper resolvent can be obtained:
\begin{equation}
	\mathcal{R}_{ii}(\lambda) \equiv \mathcal{R}_i(\lambda) = \biggl[\lambda-\mathbb{H}_{ii}-\sum_{k\in\partial i}\mathbb{H}^2_{ik}\widetilde{\mathcal{R}}^*_{k\to i}(\lambda)\biggr]^{-1}
	\label{eq:Res_sparse}
\end{equation}

These equations can in principle be solved by using the~\acrshort{PDA} as for the proper~\acrshort{BP} equations~\cite{AbouChacraEtAl1973}. However, as already pointed out at the beginning of this Chapter, the Hessian matrix~$\mathbb{H}$ contains long-range correlations that can not be taken into account if using the~\acrshort{PDA} instead of the~\acrshort{GSA}. Consequently, also in this case we have to compute the resolvent $\mathcal{R}$ on a given instance --- namely for a given Hessian matrix~$\mathbb{H}$ --- by following steps analogous to the~\acrshort{GSA} (see pseudocode~\ref{alg:RS_GI_zeroTemp}), then compute the spectral density $\rho(\lambda)$ as it will be explained in a while and only at the end average over the $\mathcal{N}_s$ samples.

There exist two types of solution for the recursive equations~\autoref{eq:Res_sparse}. In the first case, all the eigenvalues~$\lambda$'s live on the real axis and it is so throughout the computation. Consequently, the cavity resolvent $\widetilde{\mathcal{R}}$ is real at each time step and in the end also $\mathcal{R}$~is real as well. In the second case, if a small positive imaginary part~$\epsilon$ is added to $\lambda$'s when initializing the algorithm, then also the cavity resolvent acquires an imaginary part and the same happens to~$\mathcal{R}$. In the end, the limit $\epsilon\to 0$ has to be performed. At this point, two scenarios are possible. In the first one, in the $\epsilon\to 0$ limit the imaginary part of $\mathcal{R}$ goes to zero as well, so that the first kind of solution is actually recovered. In the second one, instead, the imaginary part of $\mathcal{R}$ does not vanish in the $\epsilon\to 0$ limit, and so the corresponding solution is qualitatively different from that of the first type.

These two kinds of solution have a well precise physical meaning, as firstly pointed out in Ref.~\cite{AbouChacraEtAl1973}: a real resolvent corresponds to a set of localized eigenvectors for the matrix~$\mathbb{H}$, while a nonvanishing imaginary part of the resolvent is related to the presence of extended eigenvectors. Indeed, the stability of the completely real solution of~\autoref{eq:Res_sparse} can be exploited in order to detect the localization threshold~\cite{AbouChacraEtAl1973}.

At this point, the spectral density $\rho(\lambda)$ can be finally computed, according to which one of the two cases above occurs. In the localized regime, it turns out that the resolvent develops a singularity when $\lambda$ approaches one of the eigenvalues $\lambda_i$'s of matrix~$\mathbb{H}$. So the spectral density has to be proportional to the probability distribution of the inverse of the resolvent going to zero:
\begin{equation}
	\rho(\lambda) \propto \lim_{\mathcal{R}^{-1}_i(\lambda)\to 0}\mathbb{P}[\mathcal{R}^{-1}_i(\lambda)] = \lim_{\mathcal{R}_i(\lambda)\to \infty}\mathcal{R}^2_i(\lambda)\,\mathbb{P}[\mathcal{R}_i(\lambda)]
\end{equation}
having performed a suitable coordinate transformation. More details about the localized case for sparse treelike matrices can be found in Ref.~\cite{BogomolnyGiraud2013}.

When eigenvectors are extended, instead, the computation of the spectral density is quite simpler. Indeed, it is given by the imaginary part of the resolvent trace~\cite{BrezinEtAl1978, Book_MoroneEtAl2015}:
\begin{equation}
\begin{split}
	\lim_{\epsilon\to 0}\Tr{\bigl[\mathcal{R}(\lambda+i\epsilon)\bigr]} &= \lim_{\epsilon\to 0}\sum_{i=1}^{N}\frac{1}{\lambda+i\epsilon-\lambda_i}\\
	&=N\lim_{\epsilon\to 0}\int \di\lambda'\frac{\rho(\lambda')}{\lambda+i\epsilon-\lambda'}\\
	&=N\Biggl[\dashint\di\lambda'\frac{\rho(\lambda')}{\lambda-\lambda'}-i\pi\rho(\lambda)\Biggr]
\end{split}
\end{equation}
where in the first step we introduced the spectral density as defined in~\autoref{eq:rhoLambda_def}, while in the second step we performed the integration around the singularity in the complex plane. So it finally reads:
\begin{equation}
	\rho(\lambda) = -\frac{1}{\pi N}\lim_{\epsilon\to 0}\Im\Bigl[\Tr{\bigl[\mathcal{R}(\lambda+i\epsilon)\bigr]}\Bigr]
	\label{eq:rhoLambda_through_Res}
\end{equation}
still referring to a given realization of the matrix~$\mathbb{H}$. Then, the usual average over the disorder realization has to be performed.

From the direct diagonalization of the $\mathcal{N}_s$ Hessian matrices of size $N=10^3$, we already know that --- quite surprisingly --- eigenvectors are not ``properly'' localized for any value of the external field strength $H$ (we will discuss this in~\autoref{sec:eigenvectors_Hessian}). So we have to exploit the second strategy, moving on the complex plane. In order to numerically implement it, we define the real and the imaginary parts of the cavity resolvent as:
\begin{equation}
	\Re[\widetilde{R}_{i\to j}(\lambda)] \equiv \widetilde{A}_{i\to j}(\lambda) \qquad , \qquad \Im[\widetilde{R}_{i\to j}(\lambda)] \equiv \widetilde{B}_{i\to j}(\lambda)
\end{equation}
and then we separate the two contributions, rewriting the cavity equations~\autoref{eq:Res_BP_eqs} for the resolvent as:
\begin{subequations}
	\begin{equation}
		\widetilde{\mathcal{A}}_{i\to j}(\lambda) = \frac{\lambda-\mathbb{H}_{ii}-\sum_{k\in\partial i\setminus j}\mathbb{H}^2_{ik}\widetilde{\mathcal{A}}_{k\to i}(\lambda)}{\Bigl[\lambda-\mathbb{H}_{ii}-\sum_{k\in\partial i\setminus j}\mathbb{H}^2_{ik}\widetilde{\mathcal{A}}_{k\to i}(\lambda)\Bigr]^2+\Bigl[\epsilon-\sum_{k\in\partial i\setminus j}\mathbb{H}^2_{ik}\widetilde{\mathcal{B}}_{k\to i}(\lambda)\Bigr]^2}
		\label{eq:Res_BP_eqs_Re}
	\end{equation}
	\begin{equation}
		\widetilde{\mathcal{B}}_{i\to j}(\lambda) = -\frac{\epsilon-\sum_{k\in\partial i\setminus j}\mathbb{H}^2_{ik}\widetilde{\mathcal{B}}_{k\to i}(\lambda)}{\Bigl[\lambda-\mathbb{H}_{ii}-\sum_{k\in\partial i\setminus j}\mathbb{H}^2_{ik}\widetilde{\mathcal{A}}_{k\to i}(\lambda)\Bigr]^2+\Bigl[\epsilon-\sum_{k\in\partial i\setminus j}\mathbb{H}^2_{ik}\widetilde{\mathcal{B}}_{k\to i}(\lambda)\Bigr]^2}
		\label{eq:Res_BP_eqs_Im}
	\end{equation}
	\label{eq:Res_BP_eqs_Re_and_Im}%
\end{subequations}
Analogously, also the proper resolvent $\mathcal{R}$ can be split into a real and an imaginary part, respectively $\mathcal{A}$ and $\mathcal{B}$, computed starting from cavity resolvent $\widetilde{\mathcal{A}}$ and $\widetilde{\mathcal{B}}$:
\begin{subequations}
	\begin{equation}
		\mathcal{A}_i(\lambda) = \frac{\lambda-\mathbb{H}_{ii}-\sum_{k\in\partial i}\mathbb{H}^2_{ik}\widetilde{\mathcal{A}}_{k\to i}(\lambda)}{\Bigl[\lambda-\mathbb{H}_{ii}-\sum_{k\in\partial i}\mathbb{H}^2_{ik}\widetilde{\mathcal{A}}_{k\to i}(\lambda)\Bigr]^2+\Bigl[\epsilon-\sum_{k\in\partial i}\mathbb{H}^2_{ik}\widetilde{\mathcal{B}}_{k\to i}(\lambda)\Bigr]^2}
		\label{eq:Res_sparse_Re}
	\end{equation}
	\begin{equation}
		\mathcal{B}_i(\lambda) = -\frac{\epsilon-\sum_{k\in\partial i}\mathbb{H}^2_{ik}\widetilde{\mathcal{B}}_{k\to i}(\lambda)}{\Bigl[\lambda-\mathbb{H}_{ii}-\sum_{k\in\partial i}\mathbb{H}^2_{ik}\widetilde{\mathcal{A}}_{k\to i}(\lambda)\Bigr]^2+\Bigl[\epsilon-\sum_{k\in\partial i}\mathbb{H}^2_{ik}\widetilde{\mathcal{B}}_{k\to i}(\lambda)\Bigr]^2}
		\label{eq:Res_sparse_Im}
	\end{equation}
	\label{eq:Res_sparse_Re_and_Im}%
\end{subequations}

Now, $\epsilon$ acts as a \textit{regularizer} in the denominator of $\mathcal{R}$ and $\widetilde{R}$, preventing overflows to occur during numerical implementations. However, its presence has the well known effect of broadening the peaks corresponding to each eigenvalue in the spectral density. Since eigenvalues become closer and closer when increasing the linear size $N$, then $\epsilon$ has to scale with some inverse power of $N$ in order to obtain a well defined behaviour in the thermodynamic limit. Indeed, the two limits $\epsilon\to 0$ and $N\to\infty$ are not independent.

In order to find the optimal scaling, we first of all remember that eigenvalues of a random matrix typically repel each other~\cite{Book_Mehta2004}, and hence they are closer in the bulk of the spectral density --- where their typical spacing is of order $1/N$ --- and farther in the tails. This implies that the broadening of the peaks is more evident in the tails, and hence the scaling of $\epsilon$ with $N$ has to refer in particular to these regions.

It is reasonable to assume a power-law growth of the spectrum in the left edge, as it can be observed in~\autoref{fig:some_rhoLambda_sampleAveraged} and from the arguments in~\autoref{sec:considerations_rhoLambda}:
\begin{equation}
	\rho(\lambda) \sim (\lambda-\tilde{\lambda})^{\alpha} \quad , \qquad \lambda \gtrsim \tilde{\lambda}
\end{equation}
where the lower band edge $\tilde{\lambda}$ is $0$ when the spectrum is gapless, while in the gapped region it is of order $H-2JC$. The $\alpha$ exponent will be actually computed in~\autoref{sec:gapless_region} by looking at the smallest eigenvalues of $\mathbb{H}$ obtained via the Arnoldi method. For the moment, relying on the argument by Gurarie and Chalker~\cite{GurarieChalker2003} exposed in~\autoref{sec:considerations_rhoLambda}, we set it to $3/2$.

Since there are exactly $N$ eigenvalues in the spectrum, then the integral of the spectral density between $\tilde{\lambda}$ and the smallest eigenvalue $\lambda_{\text{min}}$ has to be of order~$1/N$:
\begin{equation}
	\int_{\tilde{\lambda}}^{\lambda_{\text{min}}}\di\lambda\,\rho(\lambda) \sim \frac{1}{N}
\end{equation}
Hence, by assuming the above power-law growth, we get:
\begin{equation}
	\frac{1}{\alpha+1}\,\left(\lambda_{\text{min}}-\tilde{\lambda}\right)^{\alpha+1} \sim \frac{1}{N}
\end{equation}
from which we get the expression for the typical spacing $\Delta\lambda$ of eigenvalues in the tails of the spectral density as a function of $N$:
\begin{equation}
	\Delta\lambda \equiv \lambda_{\text{min}} - \tilde{\lambda} \sim \left(\frac{1}{N}\right)^{\frac{1}{\alpha+1}}
\end{equation}
Hence, $\epsilon$ has to scale with the same inverse power of $N$:
\begin{equation}
	\epsilon \sim \left(\frac{1}{N}\right)^{\frac{1}{\alpha+1}}
\end{equation}
i.\,e. $\epsilon \sim N^{-2/5}$, if $\alpha$ is actually equal to $3/2$. Notice, then, that $\Delta\lambda$ is by far smaller in the bulk of the spectrum, so that such scaling of $\epsilon$ can be safely used to reconstruct the whole spectral density $\rho(\lambda)$.

In this way, the resolvent technique allows the computation of the spectral density of $\mathbb{H}$ with an algorithm which scales only linearly with $N$, so providing a huge enhancement with respect to direct diagonalization algorithms. However, we should check how reliable is the spectral density obtained in this way with respect to the ``true'' one computed by direct diagonalization. Red curves in~\autoref{fig:some_rhoLambda_sampleAveraged} refer to $\rho(\lambda)$ computed as in~\autoref{eq:rhoLambda_through_Res} for $N=10^3$ and then averaged over the $\mathcal{N}_s=400$ samples. The two spectral densities are perfectly superimposed almost everywhere and for all the values of $H$ analyzed, so confirming the reliability of the computation through the resolvent. The main disagreement occurs in the tails of the spectrum, and the reason of this is directly related to the presence of the regularizer $\epsilon$ in the cavity equations for the resolvent, which makes the tails fatter. As a consequence, the peaks of the spectral density are slightly lower, due to the normalization constraint.

\begin{figure}[t]
	\centering
	\includegraphics[scale=1]{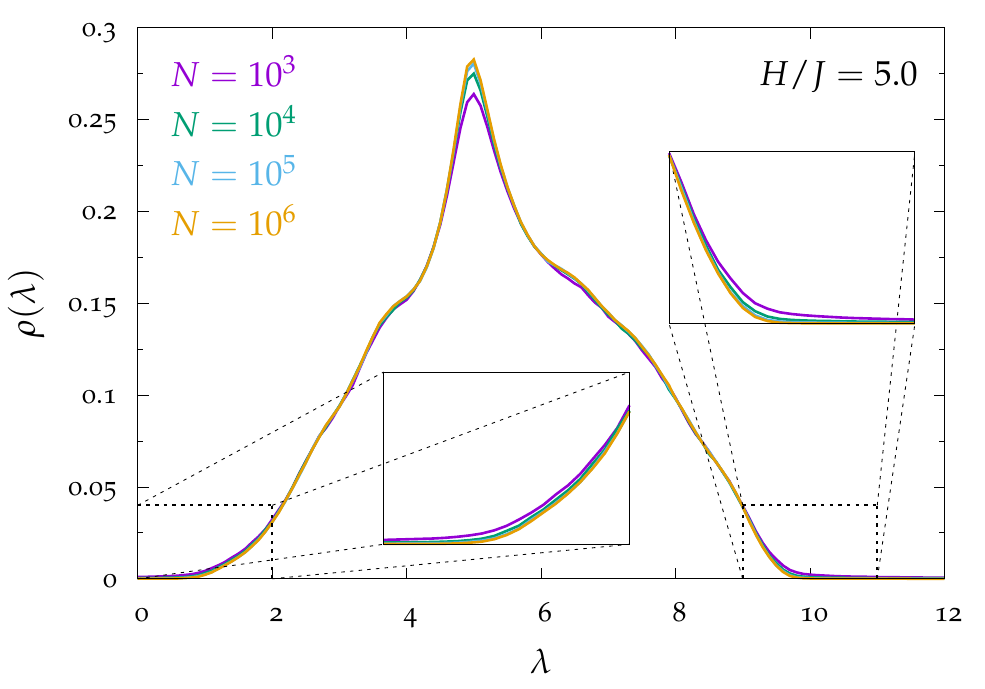}
	\caption[Finite-size effects of the spectral density computed through the resolvent technique]{Finite-size effects of the spectral density of $\mathbb{H}$ computed through the resolvent technique at the field strength $H/J=5.0$, averaged over the $\mathcal{N}_s$ samples for each size~$N$ listed in~\autoref{tab:GSA_statistics}.}
	\label{fig:finite_size_effects_Res}
\end{figure}

Fortunately, the overestimation of the tails of the spectral density is a well known finite-size effect and indeed it vanishes when considering larger and larger sizes~$N$, as shown in~\autoref{fig:finite_size_effects_Res}. So, in conclusion, the resolvent approach turns out to be the most efficient and reliable method for computing the whole spectral density for Hessian matrices of the largest sample we analyzed. However, we would like to stress once more that the spectral density computed via the imaginary part of the resolvent just takes into account the eigenvalues corresponding to extended eigenvectors, while the localized contributions to the spectral density should be computed apart, as previously explained, via the real part of the resolvent.

\section{The gapless region}
\label{sec:gapless_region}

\begin{figure}[t]
	\centering
	\includegraphics[scale=1]{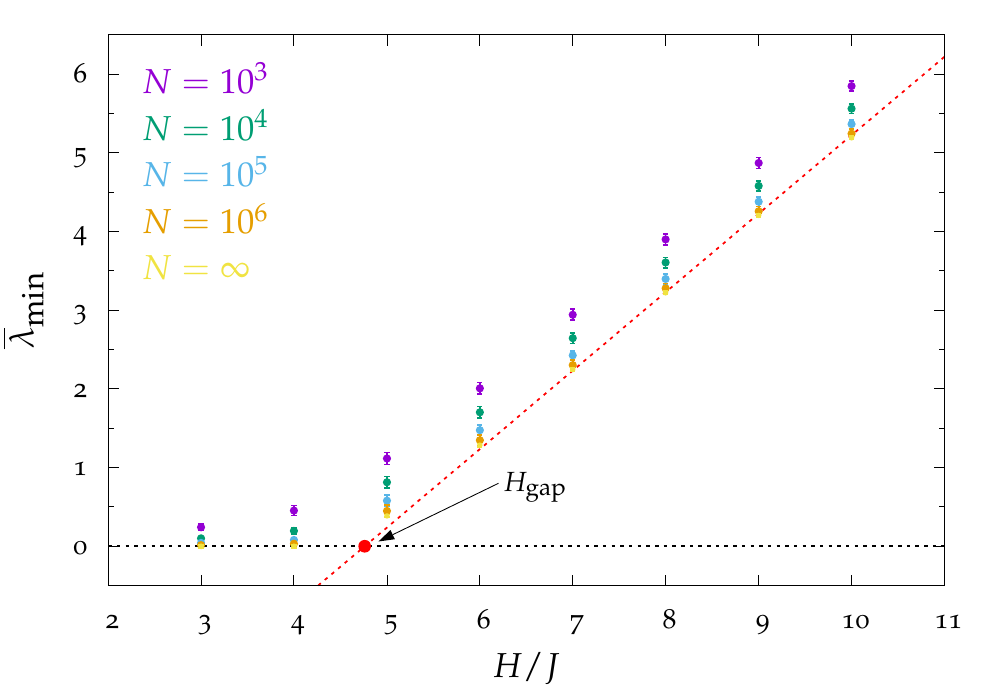}
	\caption[Closure of the gap in the spectral density]{Closure of the gap in the spectral density of the Hessian matrix for the XY model on $C=3$ \acrshort{RRG}. Red straight line corresponds to a linear fit performed over the values of $\overline{\lambda}_{\text{min}}$ extrapolated in the infinite-size limit. Also finite-size values are reported, in order to highlight such finite-size effects. Notice that when too close to the closure of the gap, anomalous finite-size corrections should be taken into account in order to get the correct infinite-size extrapolation.}
	\label{fig:gap_closure}
\end{figure}

Now we can finally compute the value $H_{\text{gap}}$ of the field strength at which the lower band edge $\tilde{\lambda}$ approaches zero. This task can be accomplished in three steps. In the first one, we average the smallest eigenvalue $\lambda_{\text{min}}$ for each value of the field strength $H$ and of the size $N$ over the $\mathcal{N}_s$ samples. Then, we actually estimate the position of the lower band edge in the thermodynamic limit through a linear fit over $\overline{\lambda}_{\text{min}}(H;N)$ vs $N^{-1/(\alpha+1)}$, so exploiting the result obtained above for the typical size of statistical fluctuations of $\lambda_{\text{min}}$. Third, a further linear fit over $\overline{\lambda}_{\text{min}}(H;N=\infty)$ vs $H$ finally gives the sought value $H_{\text{gap}}$, as shown in~\autoref{fig:gap_closure}. Notice that we are still using the value $\alpha=3/2$ provided by the argument in~\autoref{sec:considerations_rhoLambda}. We will check its actual value for the XY model in a while.

So for the $C=3$~\acrshort{RRG} ensemble we have:
\begin{equation}
	H_{\text{gap}}/J = 4.76(5)
\end{equation}
The rather large error in this estimation is systematic and it is due to the infinite-size extrapolation we performed. Indeed, close to $H_{\text{gap}}$, anomalous finite-size corrections on $\overline{\lambda}_{\text{min}}$ could dominate instead of the above scaling $N^{-1/(\alpha+1)}$.

Remarkably, the gap in the spectral density closes in correspondence of a value $H_{\text{gap}}$ of the field strength by far larger than the critical value $H_{\text{dAT}}$ for the~\acrshort{RS} stability:
\begin{equation}
	\qquad H_{\text{dAT}}/J = 1.059(2)
\end{equation}
This means that there is a whole range of $H/J$ values on the $T=0$ axis in which the energy landscape exhibits flat directions, just as found in Ref.~\cite{BaityJesiEtAl2015}.

Actually, the occurrence of soft modes in a disordered system with a disorder-dependent ground state --- though without implying the~\acrshort{RSB} --- is an evidence of the validity of the argument in~\autoref{sec:considerations_rhoLambda} also for the present case under investigation.

In this range of fields between $H_{\text{gap}}$ and $H_{\text{dAT}}$, the shape of the spectrum changes through two main phenomena, clearly visible in~\autoref{fig:some_rhoLambda_sampleAveraged}: \textit{i)} the whole spectrum gets closer to the $\lambda=0$ axis, and \textit{ii)} the central peak due to the contribution of diagonal elements $\mathbb{H}_{ii}$'s lowers and moves toward the left, so that instead of a single peak now we have two peaks separated by a dip, similarly to the Kesten-McKay spectrum~\autoref{eq:Kesten_McKay_law} of the adjacency matrix of a~\acrshort{RRG}~\cite{Kesten1959, McKay1981}. Note that it is also in this range that the large-field expansion fails --- as already said before --- due to the flattening of the energy landscape and hence to the appearance of soft modes, that can not be taken into account by that expansion.

Our interest particularly focuses on the region very close to the $\lambda=0$ axis, where quasi-zero eigenvalues accumulate. Indeed, it is this region that provides the exponent $\alpha$ of the power-law scaling of the spectral density in the gapless regime:
\begin{equation}
	\rho(\lambda) \sim \lambda^{\alpha} \quad , \qquad \lambda \gtrsim 0
	\label{eq:power_law_rhoLambda}
\end{equation}
which in turn is related to several important features of the model: among them, the divergence of the spin glass susceptibility $\chi_{\text{SG}}$. The numerical computation of~$\alpha$ will be performed through different though equivalent approaches.

\subsection{Left tail of the spectrum}

First of all, let us analyze the distribution of the smallest eigenvalues of the Hessian matrices computed by means of the Arnoldi method.

\begin{figure}[!t]
	\centering
	\includegraphics[scale=1]{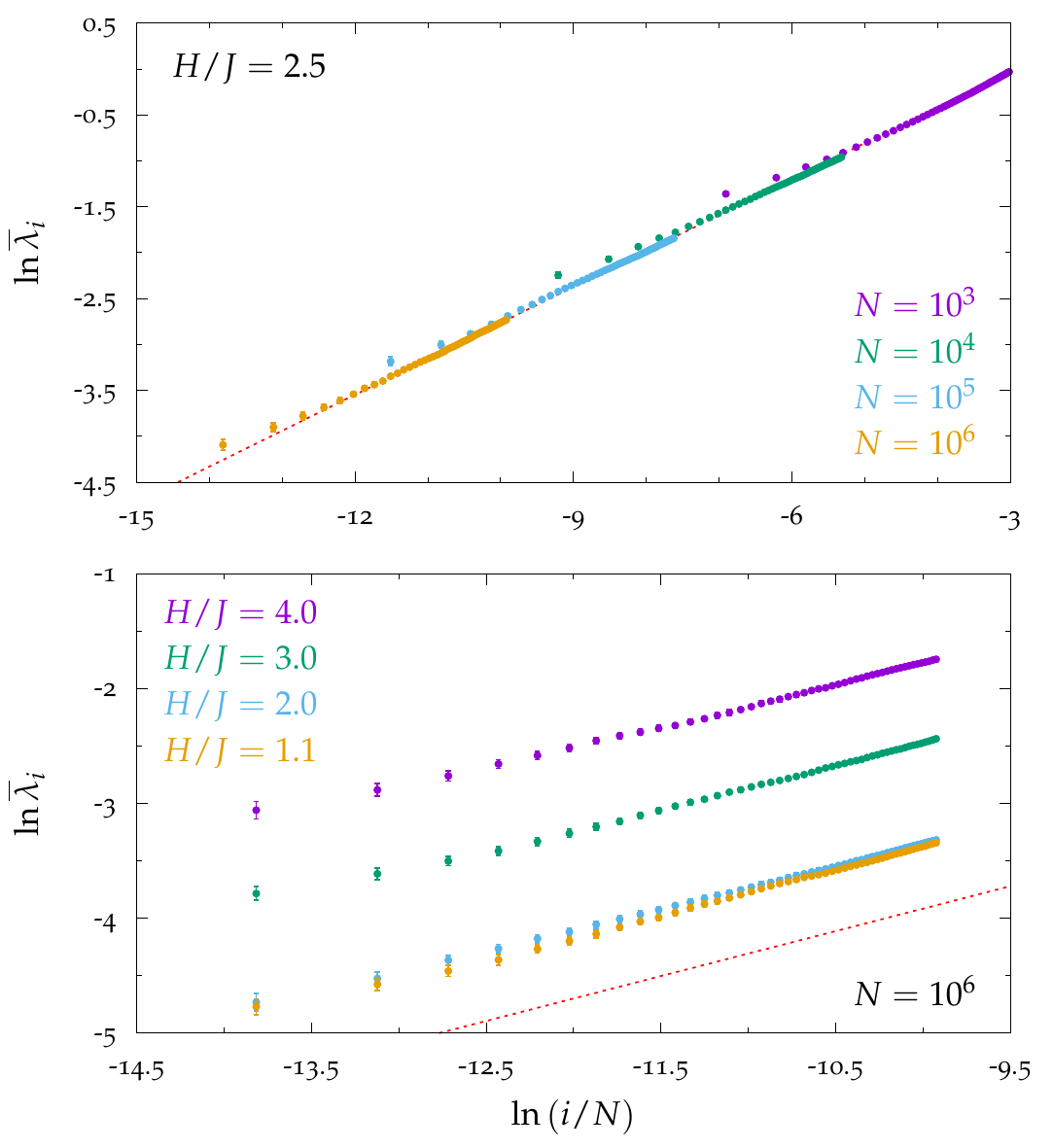}
	\caption[Lower band edge of the spectral density]{\textit{Upper panel}: Logarithm of the sample-averaged $50$ smallest eigenvalues of the Hessian matrix $\mathbb{H}$ at $H=2.5$ (hence in the gapless region) vs the logarithm of the rescaled rank. The slope $1/(\alpha+1)$ of the linear fit (red dashed line) gives a value of~$\alpha$ equal to $1.56(7)$, compatible with the analytic prediction $3/2$. \textit{Lower panel}: again the sample-averaged $50$ smallest eigenvalues of~$\mathbb{H}$ for different field intensities in the gapless region, showing a constant value of~$\alpha$ in the whole region. The red dashed line, of slope $1/(\alpha+1)=2/5$ (i.\,e. $\alpha=3/2$), is just a guide for the eye.}
	\label{fig:smallestLambda}
\end{figure}

Indeed, the $\alpha$ exponent of the power-law growth of $\lambda_i$'s in the left tail of the spectrum can be computed by looking at their arrangement with respect to the rescaled rank $i/N$, as shown in the upper panel of~\autoref{fig:smallestLambda}. In the log-log scale, the averaged eigenvalues $\overline{\lambda}_i$'s vs the rescaled rank $i/N$ arrange in a straight line of slope $1/(\alpha+1)$. Notice, however, that this scaling does not hold for the very first eigenvalues --- due to their larger dependence on finite size --- and furthermore when too close to the bulk, where the power-law trend changes due to the proximity of the first peak of the spectral density.

So a linear fit over the averaged $m=50$ smallest eigenvalues --- just discarding the very first of them --- actually provides an estimation of $\alpha$ at $H=2.5$:
\begin{equation}
	\frac{1}{1+\alpha} = 0.39(1) \qquad \Rightarrow \qquad \alpha = 1.56(7)
\end{equation}
which is compatible with the $3/2$ value obtained through the analytic argument provided in~\autoref{sec:considerations_rhoLambda}, as well as with finite-dimension numerical evidences of Ref.~\cite{BaityJesiEtAl2015}.

Furthermore, this value does not change --- compatibly with the size of the error bars --- for all the values of the field strength between $H_{\text{gap}}$ --- namely where the gap actually closes --- and $H_{\text{dAT}}$ --- i.\,e. where the~\acrshort{RS} ansatz fails --- as shown in the lower panel of~\autoref{fig:smallestLambda} for the size $N=10^6$. It is just the coefficient of~$\lambda^{3/2}$ that changes when lowering the field, so making the whole spectrum getting closer to the $\lambda=0$ axis (as seen in~\autoref{fig:some_rhoLambda_sampleAveraged}).

So we proved the validity of the quite general argument by Gurarie and Chalker~\cite{GurarieChalker2003} also for the spin glass XY model in a field on sparse random graphs, so paving the way to a different interpretation of the \acrshort{RSB} mechanism with respect to the fully connected case.

\subsection{Distribution of the smallest eigenvalue}

From random matrix theory it is well known that the smallest eigenvalue of a gapless positive semi-definite matrix is not actually zero\footnote{The only eigenvalue that can be zero also for finite $N$ is the ``trivial'' one due to the $\mathrm{O}(2)$ rotational symmetry, which is now explicitly broken by the presence of the random field.} for finite values of $N$, since it scales with an inverse power of $N$ itself~\cite{Book_Mehta2004}. Furthermore, in the previous Section we already computed this scaling in the case of a generic lower band edge $\tilde{\lambda}$, which in this case turns out to be zero. So we automatically get the scaling of $\lambda_{\text{min}}$ with~$N$:
\begin{equation}
	\lambda_{\text{min}} \sim \left(\frac{1}{N}\right)^{\frac{1}{\alpha+1}}
\end{equation}
that is indeed what we used above in order to compute $\alpha$ from the left tail of the spectral density.

More interestingly, random matrix theory also provides a description of the sample-to-sample statistics of $\lambda_{\text{min}}$~\cite{Book_Mehta2004, Book_AkemannEtAl2011}. Indeed, eigenvalues of Hessian matrices $\mathbb{H}$'s are lower bounded by $0$ in the gapless region and hence from the Fisher-Tippett-Gnedenko theorem~\cite{FisherTippett1928, Gnedenko1943} the \textit{extreme value statistics} of $\lambda_{\text{min}}$ is found to follow the Weibull distribution:
\begin{equation}
	\mathbb{P}_{\text{Weib}}(\lambda) = \frac{\alpha+1}{\lambda'}\left(\frac{\lambda}{\lambda'}\right)^{\alpha}e^{-(\lambda/\lambda')^{\alpha+1}} \quad , \qquad \lambda \geqslant 0
	\label{eq:P_Weibull}
\end{equation}
where $\alpha$ is exactly the exponent of the spectral density close to the lower band edge, while $\lambda'$ is the ``characteristic scale'' of $\lambda_{\text{min}}$. Its cumulative distribution is instead given by:
\begin{equation}
	\mathcal{C}_{\text{Weib}}(\lambda) = 1 - e^{-(\lambda/\lambda')^{\alpha+1}} \quad , \qquad \lambda \geqslant 0
	\label{eq:C_Weibull}
\end{equation}

\begin{figure}[!t]
	\centering
	\includegraphics[scale=1]{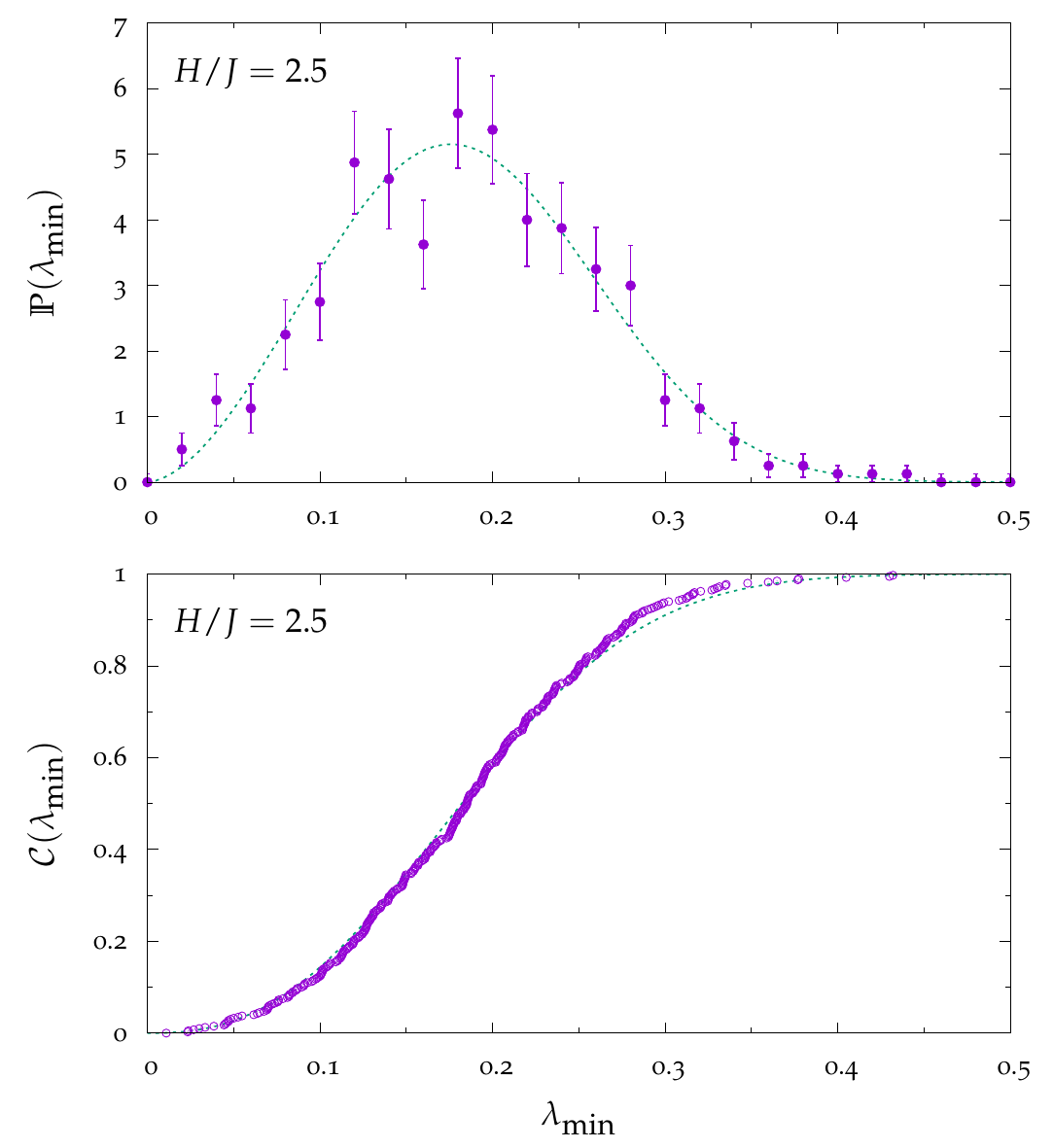}
	\caption[Weibull distribution of the smallest Hessian eigenvalue]{\textit{Upper panel}: Histogram of $\lambda_{\text{min}}$ of $\mathbb{H}$ from the $\mathcal{N}_s$ samples of size $N=10^3$ at $H=2.5$. The green dashed line is the Weibull distribution~\autoref{eq:P_Weibull} of free parameters $\alpha$ and $\lambda'$ that best fits the data, giving $\alpha=1.7(2)$. \textit{Lower panel}: Cumulative distribution of $\lambda_{\text{min}}$ as in the upper panel. The green dashed line is the Weibull cumulative distribution~\autoref{eq:C_Weibull} with $\alpha$ fixed to $3/2$ and $\lambda'$ to be fitted over the data, giving $\lambda'=0.21(1)$.}
	\label{fig:Weibull}
\end{figure}

In the upper panel of~\autoref{fig:Weibull} we plot the histogram of $\lambda_{\text{min}}$ collected over the $\mathcal{N}_s$ samples of size $N=10^3$, fitted by a Weibull distribution~\autoref{eq:P_Weibull} of free parameters $\alpha$ and $\lambda'$. The value obtained for $\alpha$:
\begin{equation}
	\alpha = 1.7(2)
\end{equation}
is again compatible with the value $\alpha=3/2$. Notice that the large error in the estimation of $\alpha$ is due to the binning used in the histogram, in turn due to the relatively small number of samples analyzed ($\mathcal{N}_s=400$ for $N=10^3$).

At this point we can safely set $\alpha$ equal to $3/2$ and then fit the cumulative distribution of $\lambda_{\text{min}}$ at $H/J=2.5$ and $N=10^3$ with the Weibull cumulative distribution~\autoref{eq:C_Weibull}, obtaining the plot in the lower panel of~\autoref{fig:Weibull}, with:
\begin{equation}
	\lambda' = 0.21(1)
\end{equation}

\section{Eigenvectors of the Hessian and delocalization}
\label{sec:eigenvectors_Hessian}

At this point, the picture begins to be clearer. The energy landscape is endowed with few strictly stable minima (or even just one minimum) for large field intensities, $H>H_{\text{gap}}$; in other words, the spectral density $\rho(\lambda)$ of the Hessian matrix is gapped. Then, below this value, it develops soft modes, distributed as $\lambda^{3/2}$ or equivalently as $\omega^4$, predicted by analytic arguments and then confirmed by strong numerical evidences. There is a quite large range of~$H$ values, from $H_{\text{gap}}$ down to $H_{\text{dAT}}$, where the energy landscape exhibits such flat regions but no \acrshort{RSB} occurs. Coherently, no divergence in the spin glass susceptibility $\chi_{\text{SG}}$ is detected, due to the $\alpha=3/2$ exponent in the lower band edge of the spectral density that persists in the whole gapless region. Finally, at $H=H_{\text{dAT}}$, the~\acrshort{RS} solution becomes unstable, signaled by the divergence in the average convergence time of the~\acrshort{GSA}. Still $\chi_{\text{SG}}$ remains finite.

It is clear that the~\acrshort{RSB} occurring at $H_{\text{dAT}}$ is not ruled by the same mechanism of fully connected models, since soft modes are here integrable over the system ($\alpha$~is equal to $3/2$ and not to $1/2$), in turn making the spin glass susceptibility $\chi_{\text{SG}}$ not diverge. On the other hand, in fully connected models the~\acrshort{RSB} is related to the appearance of \textit{extended} soft modes, while here they could be \textit{localized} at $H_{\text{gap}}$ and then become extended only at $H_{\text{dAT}}$, via some kind of \textit{delocalization} transition. So in order to throw more light into this problem, trying to understand how the features of the Hessian at $H=H_{\text{dAT}}$ are linked to the~\acrshort{RSB}, let us look at the eigenvectors associated with these soft modes.

Before going on, let us notice that when two (or more) eigenvalues of a matrix are too close, then the Arnoldi method is not so effective in correctly distinguishing the corresponding eigenvectors. Indeed, they form a nearly degenerate subspace, and hence the number of time steps required for correctly distinguishing them is exponentially large in the spacing of the two eigenvalues. This implies that different algorithms could provide different eigenvectors corresponding to the same eigenvalue. However, we are actually interested in the statistical properties of all the soft modes of the Hessian, so we will see that this degeneration is not a problem in this case.

On finite-dimension lattices, the typical quantity signaling the localization of a generic vector $\boldsymbol{v}$ is the~\acrfull{IPR}, so defined:
\begin{equation}
	\Upsilon \equiv \frac{\sum_{i=1}^{N}v^4_i}{\left(\sum_{i=1}^{N}v^2_i\right)^2}
\end{equation}
where --- when normalizing the variance of $\boldsymbol{v}$ to unity, $\sum_{i=1}^{N}v^2_i=1$ --- a completely localized vector corresponds to a unitary~\acrshort{IPR}:
\begin{equation}
	v_0=1\,\,,\,\,v_i=0 \quad \forall i \neq 0 \qquad \Rightarrow \qquad \Upsilon=1
\end{equation}
while a completely delocalized vector has an~\acrshort{IPR} equal to $1/N$:
\begin{equation}
	v_i=\frac{1}{\sqrt{N}} \quad \forall i \qquad \Rightarrow \qquad \Upsilon=\frac{1}{N}
\end{equation}
at the end going to zero in the thermodynamic limit. So in general the~\acrshort{IPR} measures the fraction of components of $\boldsymbol{v}$ that are ``substantially'' different from zero.

\begin{figure}[!t]
	\centering
	\includegraphics[scale=1]{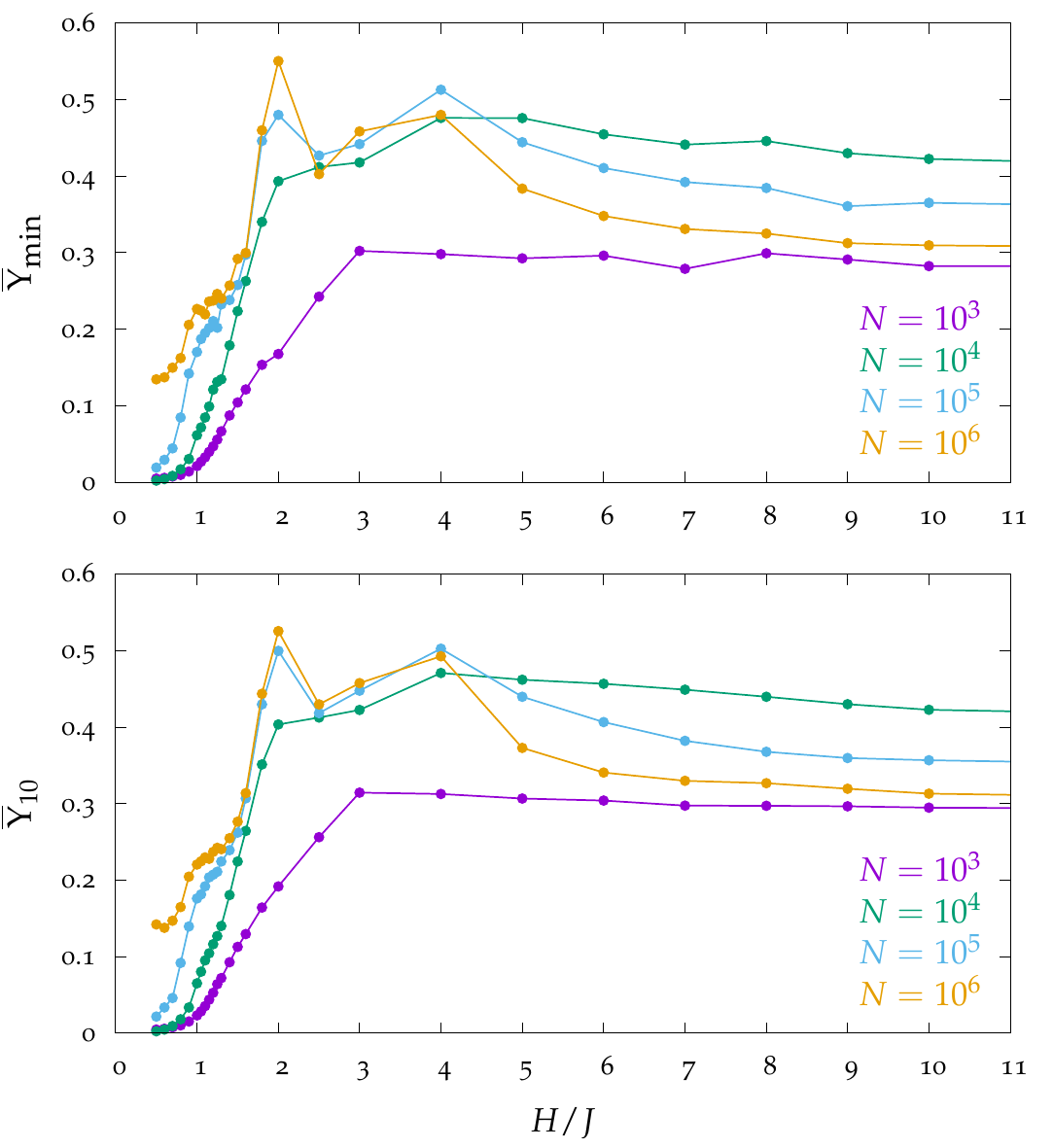}
	\caption[Inverse Participation Ratio of Hessian eigenvectors]{\textit{Upper panel}: The \acrshort{IPR} $\Upsilon$ of $\ket{\lambda}_{\text{min}}$ for each field strength $H$ and graph size $N$, averaged over the $\mathcal{N}_s$ samples. \textit{Lower panel}: Again the~\acrshort{IPR} $\Upsilon$ for each $H$ and $N$, but now averaged on the first $10$ eigenvectors of $\mathbb{H}$.}
	\label{fig:ipr}
\end{figure}

In our case, the simplest idea is to look at the~\acrshort{IPR} of $\ket{\lambda}_{\text{min}}$, which should be the ideal candidate for the supposed delocalization transition. In the upper panel of~\autoref{fig:ipr} we report $\overline{\Upsilon}_{\text{min}}$ --- namely the~\acrshort{IPR} of $\ket{\lambda}_{\text{min}}$ averaged over the $\mathcal{N}_s$ samples --- for each size $N$ when varying the field strength $H$.

For very large values of $H$, $\overline{\lambda}_{\text{min}}$ is by far larger than zero, and the corresponding value of $\overline{\Upsilon}_{\text{min}}$ is around $0.3\,-\,0.4$. Then, when getting closer to $H_{\text{gap}}$, $\overline{\lambda}_{\text{min}}$ goes to zero and $\ket{\lambda}_{\text{min}}$ becomes slightly more localized, with $\overline{\Upsilon}_{\text{min}}$ around $0.4\,-\,0.5$. Finally, when lowering the field strength below $H/J \simeq 2$, we observe a sort of delocalization of~$\ket{\lambda}_{\text{min}}$. Notice that in this region finite-size effects are quite evident, in particular below $H_{\text{dAT}}$. The reason is that for $N=10^3$ the size of typical loops is very small and hence the delocalization of $\ket{\lambda}_{\text{min}}$ is given by the short loops rather than by an actual extended correlation on the treelike topology. Of course, this effect becomes negligible when increasing~$N$, so that the~\acrshort{IPR} does no longer take into account loop effects. Indeed, from the analysis on the dataset at $N=10^6$ it is evident that the delocalization of the softest mode $\ket{\lambda}_{\text{min}}$ does not actually involve the whole system, as we would have expected for a proper delocalization transition with $\overline{\Upsilon}_{\text{min}} \to 0$ when $H \to H_{\text{dAT}}$.

However, as stated before, the smallest eigenvalues of the Hessian matrix $\mathbb{H}$ are nearly degenerate --- in particular in the gapless region and at least for the larger sizes --- and hence we should consider not only the ``softest'' eigenvector $\ket{\lambda}_{\text{min}}$, but the whole subspace spanned by the first soft eigenvectors. So in the lower plot of~\autoref{fig:ipr} we plotted the sampled-averaged \acrshort{IPR} computed on the set of eigenvectors $\{\ket{\lambda}_i\}_{i=1,\dots,10}$ corresponding to the $10$ smallest eigenvalues --- for the same fields and sizes of the upper plot --- finally labeling it as $\overline{\Upsilon}_{10}$:
\begin{equation}
	\overline{\Upsilon}_{10} \equiv \frac{1}{10\,\mathcal{N}_s}\sum_{s=1}^{\mathcal{N}_s}\,\sum_{i=0}^{9}\Upsilon^{(s)}_i
\end{equation}
with the superscript $s$ being the sample index and the subscript $i$ the eigenvector index.

For the largest sizes analyzed, this just results in an increased statistics, since the first eigenvectors are believed to have just the same statistical properties of~$\ket{\lambda}_{\text{min}}$. So $\overline{\Upsilon}_{10}(H)$ has the same qualitative trend of $\overline{\Upsilon}_{\text{min}}(H)$ in the upper panel. Instead, when $N=10^3$, $10$ eigenvectors are too many with respect to the total number of degrees of freedom. Hence, some of them begin to behave as if they were in the bulk of the spectral density, being more extended and hence having a smaller \acrshort{IPR}. This results in a curve $\overline{\Upsilon}_{10}(H)$ which is lower than $\overline{\Upsilon}_{\text{min}}(H)$.

\begin{figure}[p]
	\centering
	\includegraphics[scale=0.99]{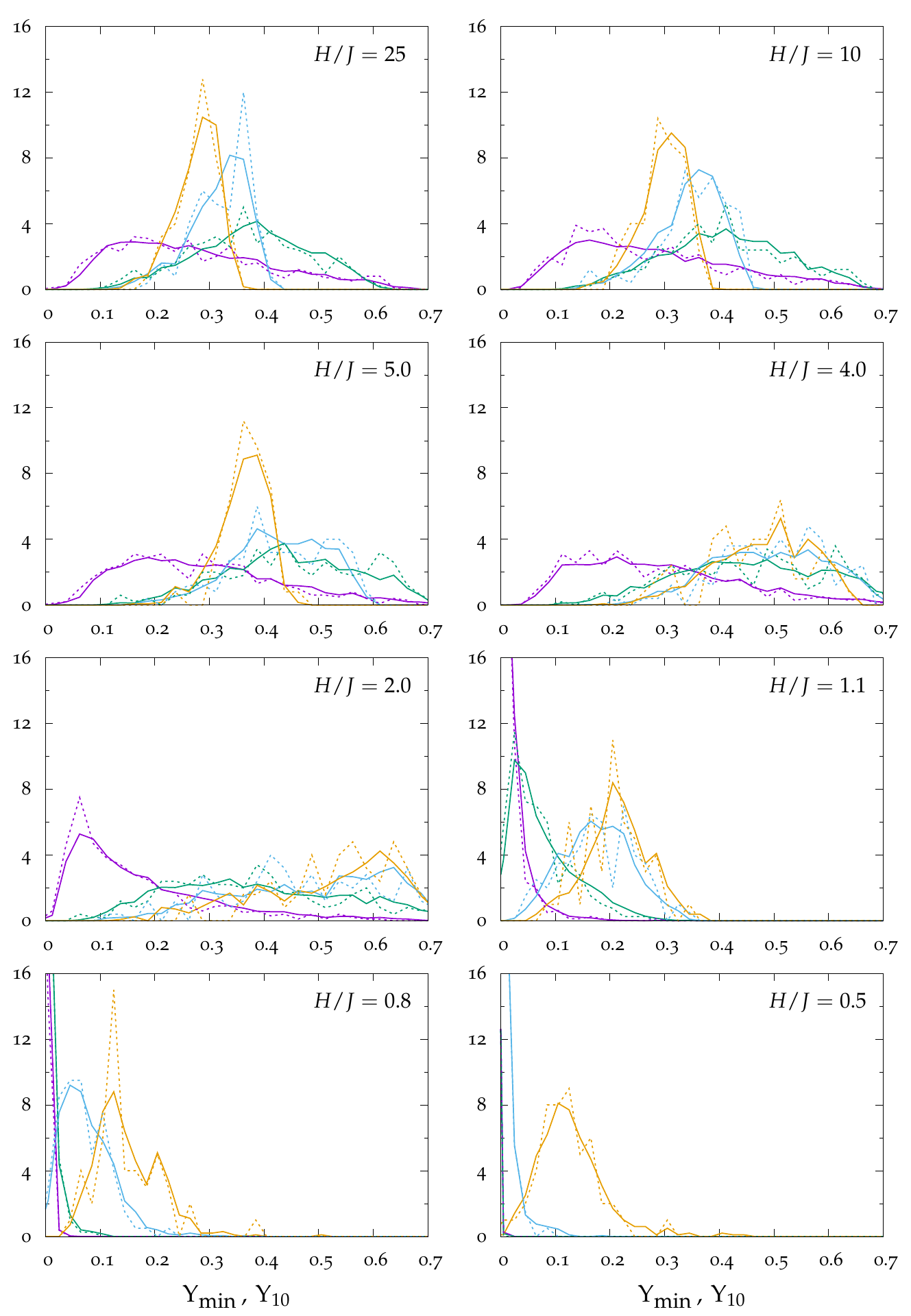}
	\caption[Histograms of the IPR over the samples]{Histograms of $\Upsilon_{\text{min}}$ (dashed lines) and of $\Upsilon_{10} \equiv \{\Upsilon_{i}\}_{i=0,\dots,9}$ (solid lines) over the~$\mathcal{N}_s$ samples for each size $N$ and for some representative values of the field intensity~$H$. Color legend is the same one exploited so far in this Chapter: purple stands for $N=10^3$, green for $10^4$, blue for $10^5$ and yellow for $10^6$.}
	\label{fig:ipr_histo}
\end{figure}

The statistical equivalence between the first $10$ eigenvectors and the very first one --- at least for $N \gtrsim 10^4$ --- can be better appreciated in~\autoref{fig:ipr_histo}, where we plotted the histogram over the samples of $\Upsilon_{\text{min}}$ and of $\Upsilon_{10} \equiv \{\Upsilon_{i}\}_{i=0,\dots,9}$ for some representative values of the field intensity $H$ and for the four sizes $N$ analyzed. Indeed, solid lines (computed on the first $10$ eigenvectors) just appear as a smoother version of the dashed ones (computed on just the first one), evidence of an increased statistics in the former with respect to the latter, without any remarkable difference in their probability distribution.
 
Furthermore, \autoref{fig:ipr_histo} provides even more important information. Indeed, the heterogeneity provided by the sparse topology reflects on the broad probability distribution of $\Upsilon$ over the samples: the~\acrshort{IPR} of the very first eigenvectors is broadly distributed into the allowed range $[0,1]$, so that its average value $\overline{\Upsilon}$ is not enough representative of the entire probability distribution. Moreover, we can also verify that the apparent delocalization of the softest modes for $H \to H_{\text{dAT}}$ at the smaller sizes is just a finite-size effect: the histogram of $\Upsilon$ at $N=10^6$ remains indeed well centered on finite values, without any accumulation on the zero.

\begin{figure}[!t]
	\centering
	\includegraphics[scale=1]{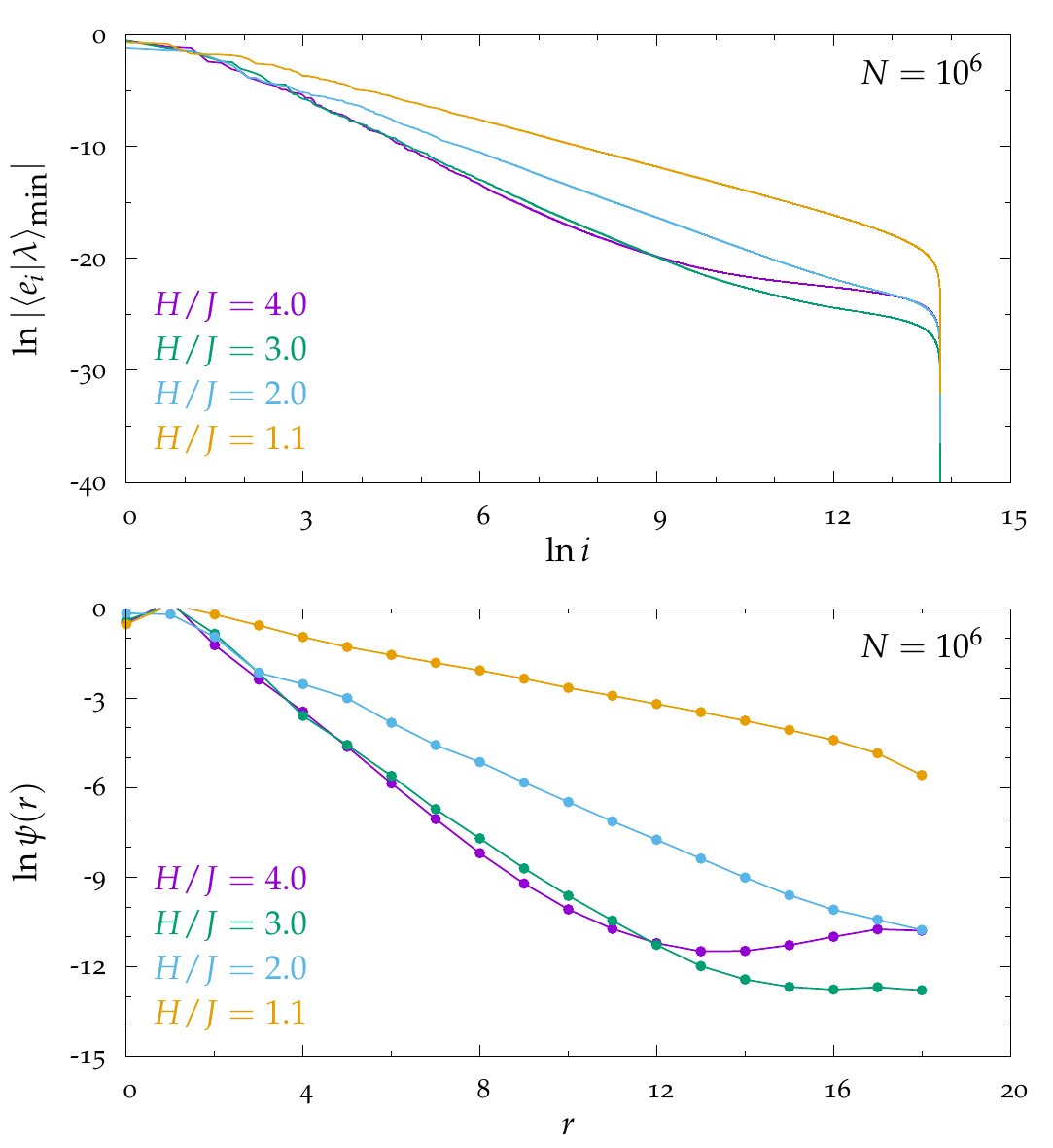}
	\caption[Sorted components of Hessian eigenvectors]{\textit{Upper panel}: The sorted (in absolute value) components $\abs{\braket{e_i|\lambda}_{\text{min}}}$ for a given sample of size $N=10^6$, clearly showing a power-law decay due to the power-law growth of the number of neighbours~$N(r)$ with the distance~$r$ on the graph. \textit{Lower panel}: The previous sorted components of $\ket{\lambda}_{\text{min}}$, now grouped and summed into blocks of size~$N(r)$, with $r$ being the distance from the largest component, seen as the ``origin'' of the softest mode. The resulting ``integrated correlation function'' $\psi(r)$, defined in~\autoref{eq:psi_r}, does now exhibits an exponential decay, but again no evidence of a critical behaviour arise at~$H=H_{\text{dAT}}$ from its behaviour.}
	\label{fig:sorted_eigenvectors}
\end{figure}

The not complete delocalization of $\ket{\lambda}_{\text{min}}$ at $H_{\text{dAT}}$ --- as well as of the others eigenvectors spanning the degenerate soft subspace --- can be better appreciated in~\autoref{fig:sorted_eigenvectors}, where we just focus on a single sample of size $N=10^6$. If we label as $\braket{e_i|\lambda}_{\text{min}}$ the $i$-th component of $\ket{\lambda}_{\text{min}}$, then we can sort them according to their absolute value, obtaining the result in the upper panel of~\autoref{fig:sorted_eigenvectors}. The clear power-law decay is due to the treelike topology of the graph, which makes the number of neighbours $N(r)$ at distance $r$ grow as a power of $r$ itself; e.\,g. for a~\acrshort{RRG} we have
\begin{equation}
	N(r) = C(C-1)^{r-1} \quad , \qquad r \geqslant 1
\end{equation}
That is just why the usual definition of localization of a vector in a finite-dimension lattice is here unsuitable, making almost useless the~\acrshort{IPR} parameter $\Upsilon$.

On the other hand, we can recover the usual exponential decay of correlations out of the critical point when taking into account such power-law growth in the number of neighbours. If we consider the largest component (in absolute value) of $\ket{\lambda}_{\text{min}}$ as the ``origin'' of the soft mode, then it is reasonable to consider the $C$ next components (still sorted according to their absolute value) as referred to the sites of $\mathcal{G}$ at distance $r=1$, then the next $C(C-1)$ components as referred to the sites at distance $r=2$, and so on. In the end, we can group into blocks of size $N(r)$ --- and then sum --- the sorted components of $\ket{\lambda}_{\text{min}}$ believed to refer to the sites at distance $r$ from the origin of the softest mode, respectively,
\begin{equation}
	\psi(r) \equiv \sum_{i=N_{\text{tot}}(r-1)+1}^{N_{\text{tot}}(r)}\abs{\braket{e_i|\lambda}_{\text{min}}}
	\label{eq:psi_r}
\end{equation}
where $N_{\text{tot}}(r)\equiv\sum_{r'=0}^{r}N(r')$ it the \textit{total} number of sites at distance $r'$ up to $r$. In this way we obtain a kind of ``integrated correlation function'' $\psi(r)$, that should now decay exponentially with the distance $r$. It is actually the case, as it can be seen in the lower panel of~\autoref{fig:sorted_eigenvectors}. Unfortunately, in the same panel we also see that the decay rate of $\psi(r)$ does not go to zero when approaching $H_{\text{dAT}}$. The same occurs for the other eigenvectors belonging to the degenerate soft subspace of~$\mathbb{H}$. This seems hence to rule out the hypothesis of a complete delocalization of a single soft mode as the cause of the~\acrshort{RSB}.

At this point, let us go back to the initial task of this Section. We would like to connect the \acrshort{RSB} with some features of the Hessian, and hence we focused on the delocalization of soft modes. Unfortunately, the standard \acrshort{IPR} analysis seems not to provide useful hints in this sense. Indeed, it is true that there is a partial delocalization of the soft modes when approaching $H_{\text{dAT}}$ from above, but $\overline{\Upsilon}(H)$ does not extrapolate to zero at such point. On the other hand, we also showed that the standard~\acrshort{IPR} analysis is not so useful when dealing with treelike topologies, due to the power-law growth of the number of neighbours with the distance on the graph. Notwithstanding then taking it into account, no complete delocalization can be found as well in correspondence of the critical point~$H_{\text{dAT}}$.

The fact that soft modes do not completely delocalize should not surprise too much the reader. Indeed, as pointed out at the beginning of this Chapter, the occurrence of soft modes is directly linked to the strong disorder dependence of the ground state, namely to the well known picture of a rugged energy landscape. So if on one hand the scenario connected to the \acrshort{RSB} does not seem to be any longer the one in which just one soft mode completely delocalizes, on the other hand we could suppose an opposite mechanism: namely, there could be a large number of rather localized soft modes, and it could be their ``cooperation'' that makes the~\acrshort{RS} ansatz fail at $H=H_{\text{dAT}}$.

In order to validate this second scenario, we suggest to adopt a different approach. On a given sample, we first reach the \acrshort{BP} fixed point via the~\acrshort{GSA}. Then, we randomly perturb this fixed point, creating a ``twin population'' of cavity messages. By running again the~\acrshort{GSA} on the second population, we can detect in which region of the graph it converge the most slowly to the previously reached fixed point. Such region should hence be the one related with the smallest $\lambda_{\text{min}}$, while the corresponding eigenvector $\ket{\lambda}_{\text{min}}$ should be given by some kind of ``difference'' between the first fixed point population and the second population still evolving toward the fixed point.

We started this analysis just right now, already obtaining some positive feedback. However it is too early to claim something in this sense and hence we postpone the presentation of these results to future works.

\clearpage{\pagestyle{empty}\cleardoublepage}

\begingroup
	\makeatletter
	\let\ps@plain\ps@empty
	\part{Conclusions}
	\cleardoublepage
\endgroup

\chapter*{Conclusions and outlooks}
\markboth{Conclusions and outlooks}{Conclusions and outlooks}
\addcontentsline{toc}{chapter}{Conclusions and outlooks}
\label{chap:end}
\thispagestyle{empty}

The main aim that pushed us to face this topic was to provide a well established framework for the analysis of vector spin models on disordered systems, both on the analytical and the numerical point of view. It is surprising how on one hand several physical phenomena can be interpreted and explained by means of a vector spin model, while on the other hand very few general results have been carried out so far on them.

This difference becomes even more striking when moving from the universally known mean-field approach provided by fully connected graphs to the less known --- but more useful and rich of insights on the finite dimensional case --- mean-field approach on sparse random graphs. For this reason, we focused on the simplest vector spin model, the XY model, defining it on the sparse topology of random graphs.

In~\autoref{part:Preliminaries} we provided a general introduction to the field of disordered systems and to the tools typically exploited for solving them, namely the belief propagation and the cavity method. A particular attention has been devoted to the motivations that lead us to the use of sparse random graphs rather than fully connected graphs, so to get a description closer to what actually happens on finite dimensional lattices. Then, a brief review over spin glasses has been inserted, in order to make comfortable even those readers not very well experienced in this field. In particular, some key results about fully connected vector spin glasses have been here recapped, so to allow a comparison with the analogous results on the sparse topology obtained in the central part of the thesis.

In~\autoref{part:XYmodelNoField} we mainly focused on the study of the XY model in absence of any external field, in order to provide a well established framework for its analysis on sparse random graphs. All the related results can be also found in Ref.~\cite{LupoRicciTersenghi2017a}. Indeed, in most of the works about the XY model, only the two expectation values $(m_x,m_y)$ of each spin are taken into account, or equivalently the direction $\theta$ in the $xy$ plane. Instead, we tried to convince the reader that the XY model can be fully understood only if the whole probability distribution of $\theta$ is considered, instead of just its mean value. If for the Ising model the two approaches are exactly equivalent, for vector spin models they are not, due to their continuous nature. So moving from $m=1$ to $m>1$ spin components actually implies a passage from a finite dimensional to an infinite dimensional problem, but unfortunately this seems not to be a well recognized issue.

If on one hand this makes analytical computations more involving and numerical simulations more requiring, on the other hand we showed that an efficient and reliable numerical proxy for the XY model is given by the $Q$-state clock model, which actually projects the problem from the infinite dimensional space to a $Q$ dimensional space. We proved that the error committed in the evaluation of physical observables decays exponentially fast in $Q$, so providing a considerable speedup. Moreover, also the universality class of the clock model becomes the same one of the XY model at very small values of $Q$, so further validating the use of the clock model with rather small values of $Q$ --- e.\,g. $Q=32$ or $Q=64$ --- when studying the XY model. Unfortunately, the generalization to the $m>2$ case seems not to be so straightforward, due to the difficulty of discretizing the resulting polar coordinates.

We want to stress two main consequences of the results coming from this analysis. First, it is just the presence of the quenched disorder that enhances the convergence, in a certain sense unfreezing the discrete degrees of freedom of the clock model toward the XY model ones. Indeed, we saw that the most ordered the system --- due to both mostly ferromagnetic couplings and low temperatures --- the slowest the convergence. Second, spin glass XY model is by far more glassy than the Ising model, mainly due to the continuous nature of its spins. In particular, we showed that at zero temperature an infinitesimal quantity of disorder is enough the break the symmetry between replicas, at variance with the Ising case. This could lead to a better explanation of some glassy features experimentally observed, such as memory and rejuvenation, which are indeed reduced by the presence of strong anisotropies.

Once established the methods for the analysis of the XY model on sparse random graphs, in~\autoref{part:XYmodelInField} we moved to the more cumbersome case of the XY model in a field. Indeed, some of the analytical tools exploited before can not be used now, so that most of the results come from numerical --- but still exact --- approaches.

The analysis started by reversing the point of view adopted in~\autoref{part:XYmodelNoField}: the quenched disorder is introduced via an external field randomly oriented on each site, while all interactions are ferromagnetic. Also in this case the XY model surprisingly showed a more pronounced glassiness with respect to the Ising model in a random field. Indeed, in the latter case there is an analytical argument that prevents the occurrence of the replica symmetry breaking, but it does not hold for vector spin models. This ``opportunity'' is actually exploited by the random field XY model, which indeed shows a tiny replica symmetry breaking region in the very low-temperature region. These results can be found in the working paper~\cite{LupoRicciTersenghi2017c}.

Then, we moved to the characterization of the behaviour of the spin glass XY model in a field, in both cases of a homogeneous field and a randomly oriented field. It is crucial the observation that what matters is the direction of the external field, while its modulus is not relevant for the stability of the replica symmetric solution. According to the distributions of the field direction previously enumerated, the breaking of replica symmetry occurs in different ways: via the Gabay--Toulouse (GT) transition when the field is uniform, or via the well known de Almeida--Thouless (dAT) transition when the field direction is randomly distributed. We also linked these two instabilities with the breaking of spin symmetries, identifying them respectively with transverse and longitudinal perturbations of the replica symmetric solution. Finally, the critical behaviour for intermediate distributions of the field direction has been studied, finding that as soon as the perfect alignment of the external field on each site is perturbed, even by a tiny amount, then the resulting critical line is no longer GT-like but dAT-like. Hence, dAT-like instabilities represent the more generic and robust mechanism through which the replica symmetry breaking occurs in a disordered system, with GT-like instabilities relegated to some very peculiar situations. These results can be found in Ref.~\cite{LupoRicciTersenghi2017b}.

Finally, in~\autoref{part:EnergyLandscape} we focused on the zero-temperature limit of the spin glass XY model in a randomly oriented field, which allowed us to explore its energy landscape. Indeed, the continuous nature of the XY spins permits to study the harmonic fluctuations around the ground state via the standard analysis of the corresponding Hessian matrix. Long-range correlations compelled us to study single instances of the problem and then average over them, with a consistent numerical effort.

We proved that the occurrence of soft modes in the energy landscape is not directly related to the breaking of replica symmetry, but it is due to the disorder dependent nature of the ground state. Direct consequence of this dependence is the anomalous density of soft modes in the energy landscape, $\rho(\lambda)\sim\lambda^{3/2}$, that differs from fully connected approaches (which at variance shows a $1/2$ exponent) but matches with other numerical and experimental finite dimensional evidences. Moreover, it is neither just the density of these soft modes that causes the breaking of replica symmetry, nor their delocalization. More likely, it is their combined effects that causes the replica symmetry breaking in the sparse case: soft modes appear inside many but rather localized ``valleys'' in the energy landscape, separated by low energetic barriers. So when perturbed enough from the state toward which it relaxed, the system abandons a valley and moves toward a different one. Even though we have not yet provided a firm proof of this scenario, at the moment it seems to be the most likely one.

Similar studies have been performed out on vector spin glasses in finite dimension via MonteCarlo simulations, but in our case a crucial enhancement is provided by the belief propagation algorithm. Indeed, it allows us to actually reach the true ground state of the energy landscape almost everywhere above the critical point, instead of getting stuck inside some metastable state. Moreover, once reached the fixed point in the space of cavity messages and then perturbed it, the resulting trajectory could provide important hints on the motion of the system in the energy landscape, going beyond the harmonic approximation given by the Hessian. This route is still under analysis, with already some positive results, but it is too early to claim something definitive about it now. Further developments will be provided in a future work~\cite{LupoEtAl2017}.

Even though the analysis of the critical properties of the spin glass XY model is not yet concluded, by means of this thesis we believe to have provided several tiny but important results on the long way toward the full comprehension and explanation of the behaviour of disordered systems with continuous variables. Moreover, there are several analogies between low-energy excitations in vector spin models and the ones in structural glasses, hence our analysis could also provide some insights on the longstanding problem of the comprehension of the glass transition. Indeed, we already showed that our analytically solvable model represents a simple though effective tool to actually reproduce and study the density of states in glassformers. Further connections could finally be established with all those fields in which low-energy modes in a rugged energy landscape play a crucial role, e.\,g. inference, continuous satisfiability problems and machine learning.

\clearpage{\pagestyle{empty}\cleardoublepage}

\appendix

\begingroup
	\makeatletter
	\let\ps@plain\ps@empty
	\part*{Appendices\addcontentsline{toc}{part}{Appendices}}
	\cleardoublepage
\endgroup

\chapter{The BP equations for the XY model}
\label{app:BPeqs_XYmodel}
\thispagestyle{empty}

The aim of this appendix is to make explicit all the manipulations performed on the~\acrshort{BP} equations for the XY model, which have only been touched on in the main text. First of all, we derive \acrshort{BP} equations in the most generic case of factor graphs, which usually turns out to be more useful in numerical simulations. Then, we derive their linearized version, which is needed for the check of the linear stability of \acrshort{BP} fixed points. Of course, also the corresponding expressions in the pairwise case are provided. Finally, we perform the zero-temperature limit on both the full~\acrshort{BP} equations and their linearized version.

\section{The BP equations at finite temperature}

Let us write the most generic Hamiltonian for the $m=2$ case, namely the XY model, which we study throughout this thesis:
\begin{equation}
	\mcH[\{\theta\}]= -\sum_{(i,j)}J_{ij}\,\boldsymbol{\sigma}_i\,\mathbb{U}_{ij}\,\boldsymbol{\sigma}_j-\sum_i\boldsymbol{H}_i\cdot\boldsymbol{\sigma}_i
	\label{eq:Hamiltonian_XY_sparse_mostGeneric_1}
\end{equation}
Notice that the strength and the sign of the interaction between each couple of nearest-neighbour spins, $\boldsymbol{\sigma}_i$ and $\boldsymbol{\sigma}_j$, is ruled by the corresponding coupling constant~$J_{ij}$. Moreover, we also insert a matrix~$\mathbb{U}_{ij}$ that applies a random rotation of an angle~$\omega_{ij}$ to one of the two spins during the interaction. Even though being a little bit redundant, this notation will allow us to easily recover the~\acrshort{BP} equations for each specific shape of interaction.

Due to the normalization constraint, $\norm{\boldsymbol{\sigma}_i}=1$, each spin can be described by a single angle $\theta_i$ belonging to the $[0,2\pi)$ interval. So interactions acquire a cosine-like shape, while field $\boldsymbol{H}_i$ can be decomposed into its modulus $H_i$ and its direction $\phi_i\in[0,2\pi)$:
\begin{equation}
	\mcH[\{\theta\}]= -\sum_{(i,j)}J_{ij}\cos{(\theta_i-\theta_j-\omega_{ij})}-\sum_i H_i\cos{(\theta_i-\phi_i)}
	\label{eq:Hamiltonian_XY_sparse_mostGeneric_2}
\end{equation}
This is the basic Hamiltonian of the XY model that we use throughout the thesis, each time removing or modifying some terms according to the specific model we are going to study.

\subsection{The factor graph case}

Once explicitly written $\mathcal{H}$, it is easy to write the corresponding \acrshort{BP} equations. Even though we are dealing with pairwise interactions between spins, we firstly focus on the~\acrshort{BP} equations written in the factor graph formalism, since it has some advantages in numerical simulations. Then, we will rewrite them in a simpler way by getting rid of $\widehat{\eta}$'s messages, so recovering the most used notation in the pairwise case.

We start from the finite-temperature case. For the variable-to-check edge we have:
\begin{equation}
	\left\{
	\begin{aligned}
	&\eta_{i\to j}(\theta_i) = \frac{1}{\mathcal{Z}_{i\to j}}\,e^{\,\beta H_i\cos{(\theta_i-\phi_i)}}\prod_{k\in\partial i\setminus j}\widehat{\eta}_{k\to i}(\theta_i)\\
	&\mathcal{Z}_{i\to j} = \int \di \theta_i\,e^{\,\beta H_i\cos{(\theta_i-\phi_i)}}\prod_{k\in\partial i\setminus j}\widehat{\eta}_{k\to i}(\theta_i)\\
	\end{aligned}
	\right.
	\label{eq:BP_XY_eta}
\end{equation}
while for the check-to-variable edge we have:
\begin{equation}
	\left\{
	\begin{aligned}
	&\widehat{\eta}_{i\to j}(\theta_j) = \frac{1}{\widehat{\mathcal{Z}}_{i\to j}}\,\int \di \theta_i\,e^{\,\beta J_{ij}\cos{(\theta_i-\theta_j-\omega_{ij})}}\,\eta_{i\to j}(\theta_i)\\
	&\widehat{\mathcal{Z}}_{i\to j} = \int \di \theta_j\,\di \theta_i\,e^{\,\beta J_{ij}\cos{(\theta_i-\theta_j-\omega_{ij})}}\,\eta_{i\to j}(\theta_i)\\
	\end{aligned}
	\right.
	\label{eq:BP_XY_etahat}
\end{equation}

Then, in order to study the stability of their solution --- namely of the \acrshort{BP} fixed points --- the typical approach is to perturb it and check if these perturbations do or do not grow when iterating the~\acrshort{BP} equations. This is nothing but the well known study of the \textit{linear stability} of a fixed point. Being $\eta$'s and $\widehat{\eta}$'s continuous functions as long as temperature is larger than zero, their perturbations have to be continuous functions over the $[0,2\pi)$ interval as well:
\begin{equation}
	\left\{
	\begin{aligned}
		&\eta_{i\to j}(\theta_i)=\eta^*_{i\to j}(\theta_i) \quad &&\to \quad &&\eta_{i\to j}(\theta_i)=\eta^*_{i\to j}(\theta_i)+\delta\eta_{i\to j}(\theta_i)\\
		&\widehat{\eta}_{i\to j}(\theta_j)=\widehat{\eta}^*_{i\to j}(\theta_j) \quad &&\to \quad &&\widehat{\eta}_{i\to j}(\theta_j)=\widehat{\eta}^*_{i\to j}(\theta_j)+\delta\widehat{\eta}_{i\to j}(\theta_j)\\
	\end{aligned}
	\right.
	\label{eq:BP_XY_eta_etahat_pert}
\end{equation}

We start from equations \autoref{eq:BP_XY_eta}, involving variable-to-check edges. When adding the small perturbations to cavity distributions in both sides of the equation --- making explicit also the normalization constant $\mathcal{Z}_{i\to j}$ --- we get:
\begin{equation}
	\eta_{i\to j}(\theta_i) + \delta\eta_{i\to j}(\theta_i) = \frac{e^{\,\beta H_i\cos{(\theta_i-\phi_i)}}\prod_{k\in\partial i\setminus j}\bigl[\widehat{\eta}_{k\to i}(\theta_i)+\delta\widehat{\eta}_{k\to i}(\theta_i)\bigr]}{\int \di \theta_i\,e^{\,\beta H_i\cos{(\theta_i-\phi_i)}}\prod_{k\in\partial i\setminus j}\bigl[\widehat{\eta}_{k\to i}(\theta_i)+\delta\widehat{\eta}_{k\to i}(\theta_i)\bigr]}
	\label{eq:BP_XY_eta_linear}
\end{equation}
Numerator of the right hand side can be rewritten up to the first order as:
\begin{equation}
\begin{split}
	&e^{\,\beta H_i\cos{(\theta_i-\phi_i)}}\prod_{k\in\partial i\setminus j}\bigl[\widehat{\eta}_{k\to i}(\theta_i)+\delta\widehat{\eta}_{k\to i}(\theta_i)\bigr]\\
	&\qquad \simeq e^{\,\beta H_i\cos{(\theta_i-\phi_i)}}\prod_{k\in\partial i\setminus j}\widehat{\eta}_{k\to i}(\theta_i)\\
	&\qquad \qquad + e^{\,\beta H_i\cos{(\theta_i-\phi_i)}}\sum_{k\in\partial i\setminus j}\delta\widehat{\eta}_{k\to i}(\theta_i)\prod_{k'\in\partial i\setminus\{j,\,k\}}\widehat{\eta}_{k'\to i}(\theta_i)
\end{split}
\end{equation}
This expansion can be used also in denominator, leading to:
\begin{equation}
\begin{split}
	&\int \di \theta_i\,e^{\,\beta H_i\cos{(\theta_i-\phi_i)}}\prod_{k\in\partial i\setminus j}\bigl[\widehat{\eta}_{k\to i}(\theta_i)+\delta\widehat{\eta}_{k\to i}(\theta_i)\bigr]\\
	&\qquad \simeq \int \di \theta_i\,e^{\,\beta H_i\cos{(\theta_i-\phi_i)}}\prod_{k\in\partial i\setminus j}\widehat{\eta}_{k\to i}(\theta_i)\\
	&\qquad \qquad + \int \di \theta_i\,e^{\,\beta H_i\cos{(\theta_i-\phi_i)}}\sum_{k\in\partial i\setminus j}\delta\widehat{\eta}_{k\to i}(\theta_i)\prod_{k'\in\partial i\setminus\{j,\,k\}}\widehat{\eta}_{k'\to i}(\theta_i)\\
	&\qquad = \int \di \theta_i\,e^{\,\beta H_i\cos{(\theta_i-\phi_i)}}\prod_{k\in\partial i\setminus j}\widehat{\eta}_{k\to i}(\theta_i)\\
	&\qquad \qquad \times \left[1+\frac{\int \di \theta_i\,e^{\,\beta H_i\cos{(\theta_i-\phi_i)}}\sum_{k\in\partial i\setminus j}\delta\widehat{\eta}_{k\to i}(\theta_i)\prod_{k'\in\partial i\setminus\{j,\,k\}}\widehat{\eta}_{k'\to i}(\theta_i)}{\int \di \theta_i\,e^{\,\beta H_i\cos{(\theta_i-\phi_i)}}\prod_{k\in\partial i\setminus j}\widehat{\eta}_{k\to i}(\theta_i)}\right]\\
	&\qquad \equiv \mathcal{Z}_{i\to j}\,\left(1+\frac{\delta\mathcal{Z}_{i\to j}}{\mathcal{Z}_{i\to j}}\right)
\end{split}
\end{equation}
with $\delta\mathcal{Z}_{i\to j}$ so defined:
\begin{equation}
	\delta\mathcal{Z}_{i\to j} \equiv \int \di \theta_i\,e^{\,\beta H_i\cos{(\theta_i-\phi_i)}}\sum_{k\in\partial i\setminus j}\delta\widehat{\eta}_{k\to i}(\theta_i)\prod_{k'\in\partial i\setminus \{j,\,k\}}\widehat{\eta}_{k'\to i}(\theta_i)
\end{equation}
At this point, the entire fraction in the right hand side of~\autoref{eq:BP_XY_eta_linear} can be expanded up to the first order in perturbations:
\begin{equation}
\begin{split}
	&\frac{e^{\,\beta H_i\cos{(\theta_i-\phi_i)}}\prod_{k\in\partial i\setminus j}\bigl[\widehat{\eta}_{k\to i}(\theta_i)+\delta\widehat{\eta}_{k\to i}(\theta_i)\bigr]}{\int \di \theta_i\,e^{\,\beta H_i\cos{(\theta_i-\phi_i)}}\prod_{k\in\partial i\setminus j}\bigl[\widehat{\eta}_{k\to i}(\theta_i)+\delta\widehat{\eta}_{k\to i}(\theta_i)\bigr]}\\
	&\qquad \simeq \frac{e^{\,\beta H_i\cos{(\theta_i-\phi_i)}}\prod_{k\in\partial i\setminus j}\widehat{\eta}_{k\to i}(\theta_i)}{\mathcal{Z}_{i\to j}}\\
	&\qquad \qquad + \frac{e^{\,\beta H_i\cos{(\theta_i-\phi_i)}}\sum_{k\in\partial i\setminus j}\delta\widehat{\eta}_{k\to i}(\theta_i)\prod_{k'\in\partial i\setminus\{j,\,k\}}\widehat{\eta}_{k'\to i}(\theta_i)}{\mathcal{Z}_{i\to j}}\\
	&\qquad \qquad - \frac{e^{\,\beta H_i\cos{(\theta_i-\phi_i)}}\prod_{k\in\partial i\setminus j}\widehat{\eta}_{k\to i}(\theta_i)}{\mathcal{Z}_{i\to j}}\,\frac{\delta\mathcal{Z}_{i\to j}}{\mathcal{Z}_{i\to j}}\\
	&\qquad = \eta_{i\to j}(\theta_i) + \frac{e^{\,\beta H_i\cos{(\theta_i-\phi_i)}}\sum_{k\in\partial i\setminus j}\delta\widehat{\eta}_{k\to i}(\theta_i)\prod_{k'\in\partial i\setminus\{j,\,k\}}\widehat{\eta}_{k'\to i}(\theta_i)}{\mathcal{Z}_{i\to j}}\\
	&\qquad \qquad - \eta_{i\to j}(\theta_i)\,\frac{\delta\mathcal{Z}_{i\to j}}{\mathcal{Z}_{i\to j}}
\end{split}
\end{equation}
Inserting it back into~\autoref{eq:BP_XY_eta_linear}, in the end we get the set of self-consistency equations for the perturbations of variable-to-check cavity distributions:
\begin{equation}
\begin{split}
	\delta\eta_{i\to j}(\theta_i) = \frac{1}{\mathcal{Z}_{i\to j}}\left[e^{\,\beta H_i\cos{(\theta_i-\phi_i)}}\sum_{k\in\partial i\setminus j}\delta\widehat{\eta}_{k\to i}(\theta_i)\prod_{k'\in\partial i\setminus\{j,\,k\}}\widehat{\eta}_{k'\to i}(\theta_i)-\delta\mathcal{Z}_{i\to j}\,\eta_{i\to j}(\theta_i)\right]
\end{split}
\label{eq:BP_XY_eta_linear_final}
\end{equation}
with the second term in the square brackets that automatically enforces the normalization, namely perturbations must have a zero mean, as expected by definition for a perturbation of a well normalized probability distribution.

We now move to equations~\autoref{eq:BP_XY_etahat}, involving check-to-variable edges:
\begin{equation}
	\left\{
	\begin{aligned}
	&\widehat{\eta}_{i\to j}(\theta_j) = \frac{1}{\widehat{\mathcal{Z}}_{i\to j}}\,\int \di \theta_i\,e^{\,\beta J_{ij}\cos{(\theta_i-\theta_j-\omega_{ij})}}\,\eta_{i\to j}(\theta_i)\\
	&\widehat{\mathcal{Z}}_{i\to j} = \int \di \theta_j\,\di \theta_i\,e^{\,\beta J_{ij}\cos{(\theta_i-\theta_j-\omega_{ij})}}\,\eta_{i\to j}(\theta_i)\\
	\end{aligned}
	\right.
\end{equation}
The addition of small perturbations to cavity distributions in both sides of the equation yields:
\begin{equation}
	\widehat{\eta}_{i\to j}(\theta_j) + \delta\widehat{\eta}_{i\to j}(\theta_j) = \frac{\int \di \theta_i\,e^{\,\beta J_{ij}\cos{(\theta_i-\theta_j-\omega_{ij})}}\bigl[\eta_{i\to j}(\theta_i)+\delta\eta_{i\to j}(\theta_i)\bigr]}{\int \di \theta_j\,\di \theta_i\,e^{\,\beta J_{ij}\cos{(\theta_i-\theta_j-\omega_{ij})}}\bigl[\eta_{i\to j}(\theta_i)+\delta\eta_{i\to j}(\theta_i)\bigr]}
	\label{eq:BP_XY_etahat_linear}
\end{equation}
Numerator can be expanded up to the first order as before:
\begin{equation}
\begin{split}
	&\int \di \theta_i\,e^{\,\beta J_{ij}\cos{(\theta_i-\theta_j-\omega_{ij})}}\bigl[\eta_{i\to j}(\theta_i)+\delta\eta_{i\to j}(\theta_i)\bigr]\\
	&\qquad = \int \di \theta_i\,e^{\,\beta J_{ij}\cos{(\theta_i-\theta_j-\omega_{ij})}}\,\eta_{i\to j}(\theta_i)+\int \di \theta_i\,e^{\,\beta J_{ij}\cos{(\theta_i-\theta_j-\omega_{ij})}}\,\delta\eta_{i\to j}(\theta_i)
\end{split}
\end{equation}
as well as denominator:
\begin{equation}
\begin{split}
	&\int \di \theta_j\,\di \theta_i\,e^{\,\beta J_{ij}\cos{(\theta_i-\theta_j-\omega_{ij})}}\bigl[\eta_{i\to j}(\theta_i)+\delta\eta_{i\to j}(\theta_i)\bigr]\\
	&\qquad = \int \di \theta_j\,\di \theta_i\,e^{\,\beta J_{ij}\cos{(\theta_i-\theta_j-\omega_{ij})}}\,\eta_{i\to j}(\theta_i)+\int \di \theta_j\,\di \theta_i\,e^{\,\beta J_{ij}\cos{(\theta_i-\theta_j-\omega_{ij})}}\,\delta\eta_{i\to j}(\theta_i)\\
	&\qquad = \int \di \theta_j\,\di \theta_i\,e^{\,\beta J_{ij}\cos{(\theta_i-\theta_j-\omega_{ij})}}\,\eta_{i\to j}(\theta_i)\,\left[1+\frac{\int \di \theta_j\,\di \theta_i\,e^{\,\beta J_{ij}\cos{(\theta_i-\theta_j-\omega_{ij})}}\,\delta\eta_{i\to j}(\theta_i)}{\int \di \theta_j\,\di \theta_i\,e^{\,\beta J_{ij}\cos{(\theta_i-\theta_j-\omega_{ij})}}\,\eta_{i\to j}(\theta_i)}\right]\\
	&\qquad \equiv \widehat{\mathcal{Z}}_{i\to j}\,\left(1+\frac{\delta\widehat{\mathcal{Z}}_{i\to j}}{\widehat{\mathcal{Z}}_{i\to j}}\right)
\end{split}
\end{equation}
with $\delta\widehat{\mathcal{Z}}_{i\to j}$ so defined:
\begin{equation}
	\delta\widehat{\mathcal{Z}}_{i\to j} \equiv \int \di \theta_j\,\di \theta_i\,e^{\,\beta J_{ij}\cos{(\theta_i-\theta_j-\omega_{ij})}}\,\delta\eta_{i\to j}(\theta_i)
\end{equation}
As before, the entire fraction in the right hand side of~\autoref{eq:BP_XY_etahat_linear} can be then expanded:
\begin{equation}
\begin{split}
	&\frac{\int \di \theta_i\,e^{\,\beta J_{ij}\cos{(\theta_i-\theta_j-\omega_{ij})}}\bigl[\eta_{i\to j}(\theta_i)+\delta\eta_{i\to j}(\theta_i)\bigr]}{\int \di \theta_j\,\di \theta_i\,e^{\,\beta J_{ij}\cos{(\theta_i-\theta_j-\omega_{ij})}}\bigl[\eta_{i\to j}(\theta_i)+\delta\eta_{i\to j}(\theta_i)\bigr]}\\
	&\qquad \simeq \frac{\int \di \theta_i\,e^{\,\beta J_{ij}\cos{(\theta_i-\theta_j-\omega_{ij})}}\,\eta_{i\to j}(\theta_i)}{\widehat{\mathcal{Z}}_{i\to j}} + \frac{\int \di \theta_i\,e^{\,\beta J_{ij}\cos{(\theta_i-\theta_j-\omega_{ij})}}\,\delta\eta_{i\to j}(\theta_i)}{\widehat{\mathcal{Z}}_{i\to j}}\\
	&\qquad \qquad - \frac{\int \di \theta_i\,e^{\,\beta J_{ij}\cos{(\theta_i-\theta_j-\omega_{ij})}}\,\eta_{i\to j}(\theta_i)}{\widehat{\mathcal{Z}}_{i\to j}}\,\frac{\delta\widehat{\mathcal{Z}}_{i\to j}}{\widehat{\mathcal{Z}}_{i\to j}}\\
	&\qquad = \widehat{\eta}_{i\to j}(\theta_j) + \frac{\int \di \theta_i\,e^{\,\beta J_{ij}\cos{(\theta_i-\theta_j-\omega_{ij})}}\,\delta\eta_{i\to j}(\theta_i)}{\widehat{\mathcal{Z}}_{i\to j}} - \widehat{\eta}_{i\to j}(\theta_i)\,\frac{\delta\widehat{\mathcal{Z}}_{i\to j}}{\widehat{\mathcal{Z}}_{i\to j}}
\end{split}
\end{equation}
Substituting it back into~\autoref{eq:BP_XY_etahat_linear}, also the self-consistency equations for $\delta\widehat{\eta}$'s can be finally obtained:
\begin{equation}
	\delta\widehat{\eta}_{i\to j}(\theta_j) = \frac{1}{\widehat{\mathcal{Z}}_{i\to j}}\,\left[\int \di \theta_i\,e^{\,\beta J_{ij}\cos{(\theta_i-\theta_j-\omega_{ij})}}\,\delta\eta_{i\to j}(\theta_i)-\delta\widehat{\mathcal{Z}}_{i\to j}\,\widehat{\eta}_{i\to j}(\theta_j)\right]
	\label{eq:BP_XY_etahat_linear_final}
\end{equation}

\subsection{The pairwise case}

Though necessary when dealing with many-body interactions, the factor graph formalism is not essential when each check node involves just two variables at each time. Indeed, for the $a$-th interaction involving $\theta_i$ and $\theta_j$ variables, we e.\,g. write $\widehat{\eta}_{i\to j}(\theta_j)$ instead of $\widehat{\eta}_{a\to j}(\theta_j)$. Now, we can get rid of $\widehat{\eta}$'s cavity messages and so recover the pairwise \acrshort{BP} equations that we actually use in the main text, which at variance turn out to be more useful in analytic computations.

In order to do this, it is enough to discard the multiplicative normalization given by $\widehat{\mathcal{Z}}_{i\to j}$ in~\autoref{eq:BP_XY_etahat} and then substitute the resulting expression for $\widehat{\eta}$'s in the right hand side of~\autoref{eq:BP_XY_eta}, so obtaining:
\begin{equation}
	\left\{
	\begin{aligned}
	&\eta_{i\to j}(\theta_i) = \frac{1}{\mathcal{Z}_{i\to j}}\,e^{\,\beta H_i\cos{(\theta_i-\phi_i)}}\prod_{k\in\partial i\setminus j}\int \di \theta_k\,e^{\,\beta J_{ik}\cos{(\theta_i-\theta_k-\omega_{ik})}}\,\eta_{k\to i}(\theta_k)\\
	&\mathcal{Z}_{i\to j} = \int \di \theta_i\,e^{\,\beta H_i\cos{(\theta_i-\phi_i)}}\prod_{k\in\partial i\setminus j}\int \di \theta_k\,e^{\,\beta J_{ik}\cos{(\theta_i-\theta_k-\omega_{ik})}}\,\eta_{k\to i}(\theta_k)\\
	\end{aligned}
	\right.
	\label{eq:BP_XY_eta_pairwise}
\end{equation}

Then, we have also to linearize these equations. Following exactly the same steps of the factor graph case, we obtain:
\begin{equation}
\begin{split}
	\delta\eta_{i\to j}(\theta_i) &= \frac{1}{\mathcal{Z}_{i\to j}}\Biggl[e^{\,\beta H_i\cos{(\theta_i-\phi_i)}}\sum_{k\in\partial i\setminus j}\int \di \theta_k\,e^{\,\beta J_{ik}\cos{(\theta_i-\theta_k-\omega_{ik})}}\,\delta\eta_{k\to i}(\theta_k)\\
	&\qquad \times \prod_{k'\in\partial i\setminus\{j,\,k\}}\int \di \theta_{k'}\,e^{\,\beta J_{ik'}\cos{(\theta_i-\theta_{k'}-\omega_{ik'})}}\,\eta_{k'\to i}(\theta_{k'})-\delta\mathcal{Z}_{i\to j}\,\eta_{i\to j}(\theta_i)\Biggr]
\end{split}
\label{eq:BP_XY_eta_linear_final_pairwise}
\end{equation}
with $\delta\mathcal{Z}_{i\to j}$ so defined:
\begin{equation}
\begin{split}
	\delta\mathcal{Z}_{i\to j} &\equiv \int \di\theta_i\,e^{\,\beta H_i\cos{(\theta_i-\phi_i)}}\sum_{k\in\partial i\setminus j}\int \di \theta_k\,e^{\,\beta J_{ik}\cos{(\theta_i-\theta_k-\omega_{ik})}}\,\delta\eta_{k\to i}(\theta_k)\\
	&\qquad \times \prod_{k'\in\partial i\setminus\{j,\,k\}}\int \di \theta_{k'}\,e^{\,\beta J_{ik'}\cos{(\theta_i-\theta_{k'}-\omega_{ik'})}}\,\eta_{k'\to i}(\theta_{k'})
\end{split}
\end{equation}

\section{The limit of zero temperature}

As long as temperature is larger than zero, each cavity distribution $\eta$ and $\widehat{\eta}$ is positive definite on the whole $[0,2\pi)$ interval. Nevertheless, when $T$ decreases, least probable values of $\theta$ become exponentially suppressed in $\beta$, implying cavity distributions to become Dirac delta functions in the $\beta\to\infty$ limit. However, even in the zero-temperature limit, small fluctuations around the most probable value are allowed for a system with continuous degrees of freedom, meaning that reducing spin marginals to Dirac delta functions implies a loss of precious information about the physics of the model in the limit of very low temperatures. Hence the correct ansatz is to describe the zero-temperature cavity distributions via the large-deviation formalism:
\begin{equation}
	\left\{
	\begin{aligned}
	&\eta_{i\to j}(\theta_i) \equiv e^{\,\beta h_{i\to j}(\theta_i)}\\
	&\widehat{\eta}_{i\to j}(\theta_j) \equiv e^{\,\beta u_{i\to j}(\theta_j)}
	\end{aligned}
	\right.
	\label{eq:eta_etahat_large_dev}
\end{equation}
so to correctly take into account the amplitude of small fluctuations around the most probable values.

By using this notation, the~\acrshort{BP} equations~\autoref{eq:BP_XY_eta} and~\autoref{eq:BP_XY_etahat} can be rewritten in terms of the \emph{cavity fields} $h$'s and the \emph{cavity biases} $u$'s:
\begin{equation}
	\left\{
	\begin{aligned}
	&e^{\,\beta h_{i\to j}(\theta_i)} = \frac{e^{\,\beta\left[H_i\cos{(\theta_i-\phi_i)} + \sum_{k\in\partial i\setminus j} u_{k\to i}(\theta_i)\right]}}{\int \di \theta_i\,e^{\,\beta\left[H_i\cos{(\theta_i-\phi_i)} + \sum_{k\in\partial i\setminus j} u_{k\to i}(\theta_i)\right]}}\\
	&e^{\,\beta u_{i\to j}(\theta_j)} = \frac{\int \di \theta_i\,e^{\,\beta\left[J_{ij}\cos{(\theta_i-\theta_j-\omega_{ij})} + h_{i\to j}(\theta_i)\right]}}{\int \di \theta_j\,\di \theta_i\,e^{\,\beta\left[J_{ij}\cos{(\theta_i-\theta_j-\omega_{ij})} + h_{i\to j}(\theta_i)\right]}}\\
	\end{aligned}
	\right.
	\label{eq:BP_XY_eta_etahat_largeDev}
\end{equation}
Then, evaluating integrals through the saddle-point method when $\beta\to\infty$, the zero-temperature \acrshort{BP} equations can be derived:
\begin{equation}
	\left\{
	\begin{aligned}
	&h_{i\to j}(\theta_i) \cong H_i\cos{(\theta_i-\phi_i)} + \sum_{k\in\partial i\setminus j} u_{k\to i}(\theta_i)\\
	&u_{i\to j}(\theta_j) \cong \max_{\theta_i}\bigl[J_{ij}\cos{(\theta_i-\theta_j-\omega_{ij})} + h_{i\to j}(\theta_i)\bigr]\\
	\end{aligned}
	\right.
	\label{eq:BP_XY_h_u}
\end{equation}
where normalizations are now additive and they are still provided by $\mathcal{Z}_{i\to j}$ and $\widehat{\mathcal{Z}}_{i\to j}$, respectively, evaluated in the $\beta\to\infty$ limit, so yielding:
\begin{equation}
	\max_{\theta_i}h_{i\to j}(\theta_i) = 0 \qquad , \qquad \max_{\theta_j}u_{i\to j}(\theta_j) = 0
\end{equation}
In this way, $h$'s and $u$'s are actually large-deviation functions, being negative semidefinite.

In the pairwise case, the~\acrshort{BP} equations are analogous to the previous ones, with the sole cavity messages $h$'s:
\begin{equation}
	h_{i\to j}(\theta_i) \cong H_i\cos{(\theta_i-\phi_i)} + \sum_{k\in\partial i\setminus j} \max_{\theta_k}\bigl[J_{ik}\cos{(\theta_i-\theta_k-\omega_{ik})} + h_{k\to i}(\theta_k)\bigr]
	\label{eq:BP_XY_h_pairwise}
\end{equation}
with the additive normalization explained before.

\subsection{Linearization at zero temperature}

As in the finite-temperature case, the stability of fixed points of the~\acrshort{BP} equations can be studied by linearization. However, this situation is more cumbersome than before. Indeed, when $T>0$ linear evolution of perturbations in the XY model is given by the application of an infinite-dimensional matrix --- namely of an integral transform --- in which all entries are different from zero. But when $\beta\to\infty$ most of these entries vanish, making perturbations no longer continuous functions of their argument, but singular. For this reason, it is not clear \textit{a priori} if the $\beta\to\infty$ limit and the linearization of \acrshort{BP} equations do commute.

So let us start from \acrshort{BP} equations~\autoref{eq:BP_XY_eta_etahat_largeDev}, where cavity messages have already been written in terms of large deviation functions, and let us perturb them:
\begin{equation}
	\left\{
	\begin{aligned}
	&e^{\,\beta\left[h_{i\to j}(\theta_i)+\delta h_{i\to j}(\theta_i)\right]} = \frac{e^{\,\beta\left[H_i\cos{(\theta_i-\phi_i)} + \sum_{k\in\partial i\setminus j} u_{k\to i}(\theta_i)+\sum_{k\in\partial i\setminus j}\delta u_{k\to i}(\theta_i)\right]}}{\int \di \theta_i\,e^{\,\beta\left[H_i\cos{(\theta_i-\phi_i)} + \sum_{k\in\partial i\setminus j} u_{k\to i}(\theta_i)+\sum_{k\in\partial i\setminus j}\delta u_{k\to i}(\theta_i)\right]}}\\
	&e^{\,\beta\left[u_{i\to j}(\theta_j)+\delta u_{i\to j}(\theta_j)\right]} = \frac{\int \di \theta_i\,e^{\,\beta\left[J_{ij}\cos{(\theta_i-\theta_j-\omega_{ij})} + h_{i\to j}(\theta_i) + \delta h_{i\to j}(\theta_i)\right]}}{\int \di \theta_j\,\di \theta_i\,e^{\,\beta\left[J_{ij}\cos{(\theta_i-\theta_j-\omega_{ij})} + h_{i\to j}(\theta_i) + \delta h_{i\to j}(\theta_i)\right]}}\\
	\end{aligned}
	\right.
\end{equation}
Then, we expand both left and right hand sides up to the first order in perturbations. For the first equation, the variable-to-check one, we get:
\begin{equation}
\begin{split}
	&e^{\,\beta h_{i\to j}(\theta_i)}\bigl[1+\beta\delta h_{i\to j}(\theta_i)\bigr]\\
	&\qquad\qquad =\frac{e^{\,\beta\left[H_i\cos{(\theta_i-\phi_i)} + \sum_{k\in\partial i\setminus j} u_{k\to i}(\theta_i)\right]}\left[1+\beta\sum_{k\in\partial i\setminus j}\delta u_{k\to i}(\theta_i)\right]}{\int \di \theta_i\,e^{\,\beta\left[H_i\cos{(\theta_i-\phi_i)} + \sum_{k\in\partial i\setminus j} u_{k\to i}(\theta_i)\right]}\left[1+\beta\sum_{k\in\partial i\setminus j}\delta u_{k\to i}(\theta_i)\right]}\\
	&\qquad\qquad =\frac{e^{\,\beta\left[H_i\cos{(\theta_i-\phi_i)} + \sum_{k\in\partial i\setminus j} u_{k\to i}(\theta_i)\right]}\left[1+\beta\sum_{k\in\partial i\setminus j}\delta u_{k\to i}(\theta_i)\right]}{\mathcal{Z}_{i\to j}\left(1+\beta\frac{\delta\mathcal{Z}_{i\to j}}{\mathcal{Z}_{i\to j}}\right)}\\
	&\qquad\qquad =\frac{e^{\,\beta\left[H_i\cos{(\theta_i-\phi_i)} + \sum_{k\in\partial i\setminus j} u_{k\to i}(\theta_i)\right]}}{\mathcal{Z}_{i\to j}}\left(1+\beta\sum_{k\in\partial i\setminus j}\delta u_{k\to i}(\theta_i)-\beta\frac{\delta\mathcal{Z}_{i\to j}}{\mathcal{Z}_{i\to j}}\right)
\end{split}
\end{equation}
from which:
\begin{equation}
	\delta h_{i\to j}(\theta_i) = \sum_{k\in\partial i\setminus j}\delta u_{k\to i}(\theta_i)-\sum_{k\in\partial i\setminus j}\delta u_{k\to i}(\theta^*_i)
	\label{eq:BP_XY_h_linear}
\end{equation}
where the normalization constant comes from the evaluation of $\delta\mathcal{Z}_{i\to j}/\mathcal{Z}_{i\to j}$ in the $\beta\to\infty$ limit:
\begin{equation}
	\theta^*_i \equiv \argmax_{\theta_i}{\Bigl[H_i\cos{(\theta_i-\phi_i)} + \sum_{k\in\partial i\setminus j} u_{k\to i}(\theta_i)\Bigr]}
	\label{eq:BP_XY_h_linear_norm}
\end{equation}
and it has the physical meaning of a null perturbation in correspondence of the maximum of the related large-deviation function, i.\,e. on the most probable value in the zero-temperature limit. Note that this normalization for perturbations is different from the finite-temperature one, since we are now dealing with (negative semidefinite) large-deviation functions and no longer in terms of (positive semidefinite) probability distributions.

The second equation, the check-to-variable one, can be evaluated in the same manner:
\begin{equation}
\begin{split}
	&e^{\,\beta u_{i\to j}(\theta_j)}\bigl[1+\beta\delta u_{i\to j}(\theta_j)\bigr]\\
	&\qquad\qquad =\frac{\int \di \theta_i\,e^{\,\beta\left[J_{ij}\cos{(\theta_i-\theta_j-\omega_{ij})} + h_{i\to j}(\theta_i)\right]}\bigl[1 + \beta\delta h_{i\to j}(\theta_i)\bigr]}{\int \di \theta_j\,\di \theta_i\,e^{\,\beta\left[J_{ij}\cos{(\theta_i-\theta_j-\omega_{ij})} + h_{i\to j}(\theta_i)\right]}\bigl[1 + \delta h_{i\to j}(\theta_i)\bigr]}\\
	&\qquad\qquad =\frac{\int \di \theta_i\,e^{\,\beta\left[J_{ij}\cos{(\theta_i-\theta_j-\omega_{ij})} + h_{i\to j}(\theta_i)\right]}\bigl[1 + \beta\delta h_{i\to j}(\theta_i)\bigr]}{\widehat{\mathcal{Z}}_{i\to j}\left(1 + \beta\frac{\delta\widehat{\mathcal{Z}}_{i\to j}}{\widehat{\mathcal{Z}}_{i\to j}}\right)}\\
	&\qquad\qquad =\frac{\int \di \theta_i\,e^{\,\beta\left[J_{ij}\cos{(\theta_i-\theta_j-\omega_{ij})} + h_{i\to j}(\theta_i)\right]}\bigl[1+\beta\delta h_{i\to j}(\theta_i)\bigr]}{\widehat{\mathcal{Z}}_{i\to j}}\left(1 - \beta\frac{\delta\widehat{\mathcal{Z}}_{i\to j}}{\widehat{\mathcal{Z}}_{i\to j}}\right)
\end{split}
\end{equation}
and hence:
\begin{equation}
	\delta u_{i\to j}(\theta_j) = \delta h_{i\to j}(\theta^*_i(\theta_j)) - \delta h_{i\to j}(\theta^*_j)
	\label{eq:BP_XY_u_linear}
\end{equation}
where $\theta^*_i(\theta_j)$ refers to the saddle-point evaluation of the integral in the numerator of the fraction:
\begin{equation}
	\theta^*_i(\theta_j) \equiv \argmax_{\theta_i}\Bigl[J_{ij}\cos{(\theta_i-\theta_j-\omega_{ij})} + h_{i\to j}(\theta_i)\Bigr]
	\label{eq:BP_XY_u_linear_norm_1}
\end{equation}
while $\theta^*_j$ refers to the additive normalization that shifts to zero the perturbation in correspondence of the maximum of the large-deviation function:
\begin{equation}
	\theta^*_j \equiv \argmax_{\theta_j}\Bigl[J_{ij}\cos{(\theta^*_i(\theta_j)-\theta_j-\omega_{ij})} + h_{i\to j}(\theta^*_i(\theta_j))\Bigr]
	\label{eq:BP_XY_u_linear_norm_2}
\end{equation}

Let us now follow the opposite procedure, namely we firstly perform the zero-temperature limit and then we linearize the equations. So starting from the zero-temperature \acrshort{BP} equations~\autoref{eq:BP_XY_h_u} and perturbing cavity fields and biases, we automatically get:
\begin{equation}
	\left\{
	\begin{aligned}
	&\delta h_{i\to j}(\theta_i) \cong \sum_{k\in\partial i\setminus j} \delta u_{k\to i}(\theta_i)\\
	&\delta u_{i\to j}(\theta_j) \cong \delta h_{i\to j}(\theta^*_i(\theta_j))\\
	\end{aligned}
	\right.
\end{equation}
with $\theta^*_i(\theta_j)$ defined exactly as in~\autoref{eq:BP_XY_u_linear_norm_1}. These are just the same expressions obtained via the previous procedure, so the two steps (the $\beta\to\infty$ limit and the linearization) actually commute. However, by using this second method, the proper normalization of perturbations remains hidden, while in the former case it naturally comes out during the saddle-point evaluation of integrals.

Finally, we can write down the linearized zero-temperature \acrshort{BP} equations also for the pairwise case:
\begin{equation}
	\delta h_{i\to j}(\theta_i) = \sum_{k\in\partial i\setminus j} \delta h_{k\to i}(\theta^*_k(\theta_i)) - \sum_{k\in\partial i\setminus j} \delta h_{k\to i}(\theta^*_i)
	\label{eq:BP_XY_h_pairwise_linear}
\end{equation}
where:
\begin{equation}
	\left\{
	\begin{aligned}
		&\theta^*_k(\theta_i) \equiv \argmax_{\theta_k}\Bigl[J_{ik}\cos{(\theta_i-\theta_k-\omega_{ik})} + h_{k\to i}(\theta_k)\Bigr]\\
		&\theta^*_i \equiv \argmax_{\theta_i}{\Biggl[\sum_{k\in\partial i\setminus j}\Bigl[J_{ik}\cos{(\theta^*_k(\theta_i)-\theta_k-\omega_{ik})} + h_{k\to i}(\theta^*_k(\theta_i))\Bigr]\Biggr]}\\
	\end{aligned}
	\right.
	\label{eq:BP_XY_h_pairwise_linear_norm}
\end{equation}
Again, perturbations are normalized in such a way that they vanish in correspondence of the most probable value for their argument, namely in correspondence of the maximum of the related cavity fields $h$'s.

\clearpage{\pagestyle{empty}\cleardoublepage}

\chapter{Fourier expansion of BP self-consistency equations for the XY model}
\label{app:Fourier_coeff_expansion}
\thispagestyle{empty}

In this appendix we analyze the self-consistency equations~\autoref{eq:Fourier_coeff_XY_selfcons} for Fourier coefficients $a$'s and $b$'s in the XY model, in order to infer their scaling just below the critical temperature.

First of all, let's rewrite here the one for $a$'s coefficients:
\begin{equation}
\begin{split}
	a^{(i\to j)}_l & = \frac{2}{\mathcal{Z}_{i\to j}}\int\di\theta\,\cos{(l\theta)}\prod_{k\in\partial i\setminus j}\Biggl\{I_0(\beta J_{ik})\\
	& \qquad +\sum_{p=1}^{\infty}I_p(\beta J_{ik})\left[a^{(k\to i)}_p\cos{(p\theta)}+b^{(k\to i)}_p\sin{(p\theta)}\right]\Biggr\}
\end{split}
\label{eq:a_XY_selfcons}
\end{equation}
being the $b$'s set of equations exactly equivalent.

The idea is to expand right hand side of~\autoref{eq:a_XY_selfcons} in a perturbative way, labeling orders according to the number of coefficients therein. Actually, we will find that the magnitude of coefficients will depend on their Fourier order $l$, but this will be clear only at the end of the expansion.

For this task, we will retain those terms containing up to three coefficients, namely up to the third order in $a$'s and $b$'s.

\section{Numerator expansion}
Let's start our expansion from the numerator of~\autoref{eq:a_XY_selfcons}:
\begin{equation}
\begin{split}
	a^{(i\to j)}_l & = \frac{2}{\mathcal{Z}_{i\to j}}\,\int\di\theta\,\cos{(l\theta)}\prod_{k\in\partial i\setminus j}\Biggl\{I_0(\beta J_{ik})\\
	& \qquad +\sum_{p=1}^{\infty}I_p(\beta J_{ik})\left[a^{(k\to i)}_p\cos{(p\theta)}+b^{(k\to i)}_p\sin{(p\theta)}\right]\Biggr\}\\
	& = \frac{2}{\mathcal{Z}_{i\to j}}\,\prod_{k\in\partial i\setminus j}I_0(\beta J_{ik})\int\di\theta\cos{(l\theta)}\\
	& \qquad \times \prod_{k\in\partial i\setminus j}\left\{1+\sum_{p=1}^{\infty}\frac{I_p(\beta J_{ik})}{I_0(\beta J_{ik})}\left[a^{(k\to i)}_p\cos{(p\theta)}+b^{(k\to i)}_p\sin{(p\theta)}\right]\right\}\\
	& \simeq \frac{2}{\mathcal{Z}_{i\to j}}\,\prod_{k\in\partial i\setminus j}I_0(\beta J_{ik})\int\di\theta\cos{(l\theta)}\left[1+C_1(\theta)+C_2(\theta)+C_3(\theta)\right]
\end{split}
\end{equation}
where $C_1(\theta)$, $C_2(\theta)$ and $C_3(\theta)$ are respectively the terms of first, second and third order in Fourier coefficients $a$'s and $b$'s:
\begin{equation}
\begin{split}
	C_1(\theta) & = \sum_{k\in\partial i\setminus j}\,\sum_{p=1}^{\infty}\frac{I_p(\beta J_{ik})}{I_0(\beta J_{ik})}\left[a^{(k\to i)}_p\cos{(p\theta)}+b^{(k\to i)}_p\sin{(p\theta)}\right]\\
	& \\
	C_2(\theta) & = \sum_{k_1,k_2\in\partial i\setminus j}\,\sum_{p_1,p_2=1}^{\infty}\frac{I_{p_1}(\beta J_{ik_1})I_{p_2}(\beta J_{ik_2})}{I_0(\beta J_{ik_1})I_0(\beta J_{ik_2})}\\
	& \qquad \times \left[a^{(k_1\to i)}_{p_1}\cos{(p_1\theta)}+b^{(k_1\to i)}_{p_1}\sin{(p_1\theta)}\right]\\
	& \qquad \times \left[a^{(k_2\to i)}_{p_2}\cos{(p_2\theta)}+b^{(k_2\to i)}_{p_2}\sin{(p_2\theta)}\right]\\
	& \\
	C_3(\theta) & = \sum_{k_1,k_2,k_3\in\partial i\setminus j}\,\sum_{p_1,p_2,p_3=1}^{\infty}\frac{I_{p_1}(\beta J_{ik_1})I_{p_2}(\beta J_{ik_2})I_{p_3}(\beta J_{ik_3})}{I_0(\beta J_{ik_1})I_0(\beta J_{ik_2})I_0(\beta J_{ik_3})}\\
	& \qquad \times \left[a^{(k_1\to i)}_{p_1}\cos{(p_1\theta)}+b^{(k_1\to i)}_{p_1}\sin{(p_1\theta)}\right]\\
	& \qquad \times \left[a^{(k_2\to i)}_{p_2}\cos{(p_2\theta)}+b^{(k_2\to i)}_{p_2}\sin{(p_2\theta)}\right]\\
	& \qquad \times \left[a^{(k_3\to i)}_{p_3}\cos{(p_3\theta)}+b^{(k_3\to i)}_{p_3}\sin{(p_3\theta)}\right]
\end{split}
\end{equation}
Note that in the sums over multiple $k$'s indexes, they are meant to be different from each other. Indeed, they come from the product over all the neighbours of $i$ but $j$.

At this point we can perform the integration over $\theta$ for each term in the square brackets of~\autoref{app:Fourier_coeff_expansion}. The integration over the zeroth-order term gives zero, so let's move to the first-order term:
\begin{equation}
\begin{split}
	&\int\di\theta\cos{(l\theta)}\,C_1(\theta)\\
	& \qquad = \frac{1}{2}\sum_{k\in\partial i\setminus j}\,\sum_{p=1}^{\infty}\frac{I_p(\beta J_{ik})}{I_0(\beta J_{ik})}\int\di\theta\,\Bigl\{a^{(k\to i)}_p\cos{[(p+l)\theta]}+a^{(k\to i)}_p\cos{[(p-l)\theta]}\\
	& \qquad \qquad +b^{(k\to i)}_p\sin{[(p+l)\theta]}+b^{(k\to i)}_p\sin{[(p-l)\theta]}\Bigr\}\\
	& \qquad = \frac{2\pi}{2}\sum_{k\in\partial i\setminus j}\,\sum_{p=1}^{\infty}\frac{I_p(\beta J_{ik})}{I_0(\beta J_{ik})}\,a^{(k\to i)}_p\,\delta_{p,l}\\
	& \qquad = \frac{2\pi}{2}\sum_{k\in\partial i\setminus j}\,\frac{I_l(\beta J_{ik})}{I_0(\beta J_{ik})}\,a^{(k\to i)}_l
\end{split}
\end{equation}
so recovering the result already obtained in~\autoref{eq:al_linear_numerator} of main text. Let's go on with the second-order term:
\begin{equation}
\begin{split}
	&\int\di\theta\cos{(l\theta)}\,C_2(\theta)\\
	& \qquad = \frac{1}{2}\sum_{k_1,k_2\in\partial i\setminus j}\,\sum_{p_1,p_2=1}^{\infty}\frac{I_{p_1}(\beta J_{ik_1})I_{p_2}(\beta J_{ik_2})}{I_0(\beta J_{ik_1})I_0(\beta J_{ik_2})}\int\di\theta\,\Bigl\{a^{(k_1\to i)}_{p_1}\cos{[(p_1+l)\theta]}\\
	& \qquad \qquad \qquad +a^{(k_1\to i)}_{p_1}\cos{[(p_1-l)\theta]}+b^{(k_1\to i)}_{p_1}\sin{[(p_1+l)\theta]}+b^{(k_1\to i)}_{p_1}\sin{[(p_1-l)\theta]}\Bigr\}\\
	& \qquad \qquad \times\left[a^{(k_2\to i)}_{p_2}\cos{(p_2\theta)}+b^{(k_2\to i)}_{p_2}\sin{(p_2\theta)}\right]
\end{split}
\end{equation}
Since at this point the number of terms grows very rapidly, we immediately get rid of the terms in which there is a product of a cosine times a sine, since they vanish once integrated over $\theta$. We retain only terms with a cosine times a cosine and with a sine times a sine:
\begin{equation}
\begin{split}
	&\int\di\theta\cos{(l\theta)}\,C_2(\theta)\\
	& \qquad = \frac{1}{4}\sum_{k_1,k_2\in\partial i\setminus j}\,\sum_{p_1,p_2=1}^{\infty}\frac{I_{p_1}(\beta J_{ik_1})I_{p_2}(\beta J_{ik_2})}{I_0(\beta J_{ik_1})I_0(\beta J_{ik_2})}\\
	& \qquad \qquad \times \int\di\theta\,\biggl\{a^{(k_1\to i)}_{p_1}a^{(k_2\to i)}_{p_2} \Bigl[\cos{[(p_1+l+p_2)\theta]}+\cos{[(p_1+l-p_2)\theta]}\Bigr]\\
	& \qquad \qquad \qquad +a^{(k_1\to i)}_{p_1}a^{(k_2\to i)}_{p_2} \Bigl[\cos{[(p_1-l+p_2)\theta]}+\cos{[(p_1-l-p_2)\theta]}\Bigr]\\
	& \qquad \qquad \qquad +b^{(k_1\to i)}_{p_1}b^{(k_2\to i)}_{p_2} \Bigl[\cos{[(p_1+l-p_2)\theta]}-\cos{[(p_1+l+p_2)\theta]}\Bigr]\\
	& \qquad \qquad \qquad +b^{(k_1\to i)}_{p_1}b^{(k_2\to i)}_{p_2} \Bigl[\cos{[(p_1-l-p_2)\theta]}-\cos{[(p_1-l+p_2)\theta]}\Bigr]\biggr\}\\
	& \qquad = \frac{2\pi}{4}\sum_{k_1,k_2\in\partial i\setminus j}\,\sum_{p_1,p_2=1}^{\infty}\frac{I_{p_1}(\beta J_{ik_1})I_{p_2}(\beta J_{ik_2})}{I_0(\beta J_{ik_1})I_0(\beta J_{ik_2})}\\
	& \qquad \qquad \times \biggl\{a^{(k_1\to i)}_{p_1}a^{(k_2\to i)}_{p_2}\Bigl[\delta_{p_2-p_1,l}+\delta_{p_1+p_2,l}+\delta_{p_1-p_2,l}\Bigr]\\
	& \qquad \qquad \qquad +b^{(k_1\to i)}_{p_1}b^{(k_2\to i)}_{p_2}\Bigl[\delta_{p_2-p_1,l}+\delta_{p_1-p_2,l}-\delta_{p_1+p_2,l}\Bigr]\biggr\}
\end{split}
\end{equation}
Finally, we analyze the integration of the third-order term:
\begin{equation}
\begin{split}
	&\int\di\theta\cos{(l\theta)}\,C_3(\theta)\\
	& \qquad = \frac{1}{2}\sum_{k_1,k_2,k_3\in\partial i\setminus j}\,\sum_{p_1,p_2,p_3=1}^{\infty}\frac{I_{p_1}(\beta J_{ik_1})I_{p_2}(\beta J_{ik_2})I_{p_3}(\beta J_{ik_3})}{I_0(\beta J_{ik_1})I_0(\beta J_{ik_2})I_0(\beta J_{ik_3})}\\
	& \qquad \qquad \times \int\di\theta\,\Bigl\{a^{(k_1\to i)}_{p_1}\cos{[(p_1+l)\theta]}+a^{(k_1\to i)}_{p_1}\cos{[(p_1-l)\theta]}\\
	& \qquad \qquad \qquad \qquad +b^{(k_1\to i)}_{p_1}\sin{[(p_1+l)\theta]}+b^{(k_1\to i)}_{p_1}\sin{[(p_1-l)\theta]}\Bigr\}\\
	& \qquad \qquad \qquad \times\left[a^{(k_2\to i)}_{p_2}\cos{(p_2\theta)}+b^{(k_2\to i)}_{p_2}\sin{(p_2\theta)}\right]\\
	& \qquad \qquad \qquad \times\left[a^{(k_3\to i)}_{p_3}\cos{(p_3\theta)}+b^{(k_3\to i)}_{p_3}\sin{(p_3\theta)}\right]
\end{split}
\end{equation}
This time we have a further proliferation of terms, so that we just retain only those which gives a non-vanishing contribution once integrated over $\theta$. We get:
\begin{equation}
\begin{split}
	&\int\di\theta\cos{(l\theta)}\,C_3(\theta)\\
	& \qquad = \frac{2\pi}{8}\sum_{k_1,k_2,k_3\in\partial i\setminus j}\,\sum_{p_1,p_2,p_3=1}^{\infty}\frac{I_{p_1}(\beta J_{ik_1})I_{p_2}(\beta J_{ik_2})I_{p_3}(\beta J_{ik_3})}{I_0(\beta J_{ik_1})I_0(\beta J_{ik_2})I_0(\beta J_{ik_3})}\\
	& \qquad \qquad \times \biggl\{a^{(k_1\to i)}_{p_1}a^{(k_2\to i)}_{p_2}a^{(k_3\to i)}_{p_3}\Bigl[\delta_{p_3-p_1-p_2,l}+\delta_{p_2-p_1-p_3,l}+\delta_{p_2+p_3-p_1,l}\\
	& \qquad \qquad \qquad \qquad +\delta_{p_1+p_2+p_3,l}+\delta_{p_1+p_2-p_3,l}+\delta_{p_1-p_2+p_3,l}+\delta_{p_1-p_2-p_3,l}\Bigr]\\
	& \qquad \qquad \qquad +b^{(k_1\to i)}_{p_1}b^{(k_2\to i)}_{p_2}a^{(k_3\to i)}_{p_3}\Bigl[-\delta_{p_3-p_1-p_2,l}+\delta_{p_2-p_1-p_3,l}+\delta_{p_2+p_3-p_1,l}\\
	& \qquad \qquad \qquad \qquad -\delta_{p_1+p_2+p_3,l}-\delta_{p_1+p_2-p_3,l}+\delta_{p_1-p_2+p_3,l}+\delta_{p_1-p_2-p_3,l}\Bigr]\\
	& \qquad \qquad \qquad +b^{(k_1\to i)}_{p_1}a^{(k_2\to i)}_{p_2}b^{(k_3\to i)}_{p_3}\Bigl[\delta_{p_3-p_1-p_2,l}-\delta_{p_2-p_1-p_3,l}+\delta_{p_2+p_3-p_1,l}\\
	& \qquad \qquad \qquad \qquad -\delta_{p_1+p_2+p_3,l}+\delta_{p_1+p_2-p_3,l}-\delta_{p_1-p_2+p_3,l}+\delta_{p_1-p_2-p_3,l}\Bigr]\\
	& \qquad \qquad \qquad +a^{(k_1\to i)}_{p_1}b^{(k_2\to i)}_{p_2}b^{(k_3\to i)}_{p_3}\Bigl[\delta_{p_3-p_1-p_2,l}+\delta_{p_2-p_1-p_3,l}-\delta_{p_2+p_3-p_1,l}\\
	& \qquad \qquad \qquad \qquad -\delta_{p_1+p_2+p_3,l}+\delta_{p_1+p_2-p_3,l}+\delta_{p_1-p_2+p_3,l}-\delta_{p_1-p_2-p_3,l}\Bigr]\biggr\}
\end{split}
\end{equation}

At this point, the three terms can be put again together. We use a compact notation, noticing that: \textit{(i)} for each choice of $p$'s indexes, the related contribution does not vanish only when they algebraically sum to $l$; \textit{ii)} a minus sign appears when there is an even number of $b$'s coefficients and when the corresponding~$p$'s indexes have the same sign in the algebraic sum up to $l$. So we can put a superscript~$(l)$ over the sum symbols in order to enforce the algebraic sum up to $l$ and also a coefficient taking into account that minus sign. In the end, we get:
\begin{equation}
\begin{split}
	a^{(i\to j)}_l & = \frac{2\pi}{\mathcal{Z}_{i\to j}}\prod_{k\in\partial i\setminus j}I_0(\beta J_{ik})\,\Biggl\{\sum_{k\in\partial i\setminus j}\,\frac{I_l(\beta J_{ik})}{I_0(\beta J_{ik})}\,a^{(k\to i)}_l\\
	& \qquad + \frac{1}{2}\sum_{k_1,k_2\in\partial i\setminus j}\,\sum_{p_1,p_2=1}^{\infty}{\vphantom{\sum}}^{\mkern-20mu (l)}\,\frac{I_{p_1}(\beta J_{ik_1})I_{p_2}(\beta J_{ik_2})}{I_0(\beta J_{ik_1})I_0(\beta J_{ik_2})}\\
	& \qquad \qquad \times \Bigl[a^{(k_1\to i)}_{p_1}a^{(k_2\to i)}_{p_2}-\sign{(p_1 p_2)}\,b^{(k_1\to i)}_{p_1}b^{(k_2\to i)}_{p_2}\Bigr]\\
	& \qquad + \frac{1}{4}\sum_{k_1,k_2,k_3\in\partial i\setminus j}\,\sum_{p_1,p_2,p_3=1}^{\infty}{\vphantom{\sum}}^{\mkern-32mu (l)}\,\frac{I_{p_1}(\beta J_{ik_1})I_{p_2}(\beta J_{ik_2})I_{p_3}(\beta J_{ik_3})}{I_0(\beta J_{ik_1})I_0(\beta J_{ik_2})I_0(\beta J_{ik_3})}\\
	& \qquad \qquad \times \Bigl[a^{(k_1\to i)}_{p_1}a^{(k_2\to i)}_{p_2}a^{(k_3\to i)}_{p_3}-\sign{(p_1 p_2)}\,b^{(k_1\to i)}_{p_1}b^{(k_2\to i)}_{p_2}a^{(k_3\to i)}_{p_3}\\
	& \qquad \qquad \qquad -\sign{(p_1 p_3)}\,b^{(k_1\to i)}_{p_1}a^{(k_2\to i)}_{p_2}b^{(k_3\to i)}_{p_3}\\
	& \qquad \qquad \qquad -\sign{(p_2 p_3)}\,a^{(k_1\to i)}_{p_1}b^{(k_2\to i)}_{p_2}b^{(k_3\to i)}_{p_3}\Bigr]\Biggr\}
\end{split}
\end{equation}


\section{Denominator expansion}

Also $\mathcal{Z}_{i\to j}$ has now to be expanded up to the third order in Fourier coefficients. Following the same steps as before, we write:
\begin{equation}
\begin{split}
	\mathcal{Z}_{i\to j} & = \int\di\theta\,\prod_{k\in\partial i\setminus j}\Biggl\{I_0(\beta J_{ik})+\sum_{p=1}^{\infty}I_p(\beta J_{ik})\left[a^{(k\to i)}_p\cos{(p\theta)}+b^{(k\to i)}_p\sin{(p\theta)}\right]\Biggr\}\\
	& = \prod_{k\in\partial i\setminus j}I_0(\beta J_{ik})\\
	& \qquad \times \int\di\theta\,\prod_{k\in\partial i\setminus j}\left\{1+\sum_{p=1}^{\infty}\frac{I_p(\beta J_{ik})}{I_0(\beta J_{ik})}\left[a^{(k\to i)}_p\cos{(p\theta)}+b^{(k\to i)}_p\sin{(p\theta)}\right]\right\}\\
	& \simeq \prod_{k\in\partial i\setminus j}I_0(\beta J_{ik})\int\di\theta\,\left[1+C_1(\theta)+C_2(\theta)+C_3(\theta)\right]
\end{split}
\end{equation}
where $C_1(\theta)$, $C_2(\theta)$ and $C_3(\theta)$ are exactly the same terms defined in the expansion of numerator. As before, we can integrate them separately. 

The integration of zeroth-order term just gives a $2\pi$ factor, while the integration of first-order term gives a vanishing contribution. So let's move directly to the second-order term:
\begin{equation}
\begin{split}
	&\int\di\theta\,C_2(\theta)\\
	& \qquad = \sum_{k_1,k_2\in\partial i\setminus j}\,\sum_{p_1,p_2=1}^{\infty}\frac{I_{p_1}(\beta J_{ik_1})I_{p_2}(\beta J_{ik_2})}{I_0(\beta J_{ik_1})I_0(\beta J_{ik_2})}\\
	& \qquad \qquad \times \int\di\theta\,\left[a^{(k_1\to i)}_{p_1}\cos{(p_1\theta)}+b^{(k_1\to i)}_{p_1}\sin{(p_1\theta)}\right]\\
	& \qquad \qquad \qquad \times \left[a^{(k_2\to i)}_{p_2}\cos{(p_2\theta)}+b^{(k_2\to i)}_{p_2}\sin{(p_2\theta)}\right]\\
	& \qquad = \frac{2\pi}{2}\sum_{k_1,k_2\in\partial i\setminus j}\,\sum_{p_1,p_2=1}^{\infty}\frac{I_{p_1}(\beta J_{ik_1})I_{p_2}(\beta J_{ik_2})}{I_0(\beta J_{ik_1})I_0(\beta J_{ik_2})}\\
	& \qquad \qquad \times \biggl[a^{(k_1\to i)}_{p_1}a^{(k_2\to i)}_{p_2}\delta_{p_1-p_2,0}+b^{(k_1\to i)}_{p_1}b^{(k_2\to i)}_{p_2}\delta_{p_1-p_2,0}\biggr]\\
	& \qquad = \frac{2\pi}{2}\sum_{k_1,k_2\in\partial i\setminus j}\,\sum_{p=1}^{\infty}\,\frac{I_{p}(\beta J_{ik_1})I_{p}(\beta J_{ik_2})}{I_0(\beta J_{ik_1})I_0(\beta J_{ik_2})}\biggl[a^{(k_1\to i)}_p a^{(k_2\to i)}_p+b^{(k_1\to i)}_p b^{(k_2\to i)}_p\biggr]
\end{split}
\end{equation}
where we suddenly discarded those terms which would have given a null contribution. Then, let's integrate also the third-order term:
\begin{equation}
\begin{split}
	&\int\di\theta\,C_3(\theta)\\
	& \qquad = \sum_{k_1,k_2,k_3\in\partial i\setminus j}\,\sum_{p_1,p_2,p_3=1}^{\infty}\frac{I_{p_1}(\beta J_{ik_1})I_{p_2}(\beta J_{ik_2})I_{p_3}(\beta J_{ik_3})}{I_0(\beta J_{ik_1})I_0(\beta J_{ik_2})I_0(\beta J_{ik_3})}\\
	& \qquad \qquad \times \int\di\theta\,\left[a^{(k_1\to i)}_{p_1}\cos{(p_1\theta)}+b^{(k_1\to i)}_{p_1}\sin{(p_1\theta)}\right]\\
	& \qquad \qquad \qquad \times \left[a^{(k_2\to i)}_{p_2}\cos{(p_2\theta)}+b^{(k_2\to i)}_{p_2}\sin{(p_2\theta)}\right]\\
	& \qquad \qquad \qquad \times \left[a^{(k_3\to i)}_{p_3}\cos{(p_3\theta)}+b^{(k_3\to i)}_{p_3}\sin{(p_3\theta)}\right]\\
	& \qquad = \frac{2\pi}{4}\sum_{k_1,k_2,k_3\in\partial i\setminus j}\,\sum_{p_1,p_2,p_3=1}^{\infty}\frac{I_{p_1}(\beta J_{ik_1})I_{p_2}(\beta J_{ik_2})I_{p_3}(\beta J_{ik_3})}{I_0(\beta J_{ik_1})I_0(\beta J_{ik_2})I_0(\beta J_{ik_3})}\\
	& \qquad \qquad \times \biggl\{a^{(k_1\to i)}_{p_1}a^{(k_2\to i)}_{p_2}a^{(k_3\to i)}_{p_3}\Bigl[\delta_{p_3-p_1-p_2,0}+\delta_{p_2-p_1-p_3,0}+\delta_{p_2+p_3-p_1,0}\\
	& \qquad \qquad \qquad \qquad +\delta_{p_1+p_2+p_3,0}+\delta_{p_1+p_2-p_3,0}+\delta_{p_1-p_2+p_3,0}+\delta_{p_1-p_2-p_3,0}\Bigr]\\
	& \qquad \qquad \qquad +b^{(k_1\to i)}_{p_1}b^{(k_2\to i)}_{p_2}a^{(k_3\to i)}_{p_3}\Bigl[-\delta_{p_3-p_1-p_2,0}+\delta_{p_2-p_1-p_3,0}+\delta_{p_2+p_3-p_1,0}\\
	& \qquad \qquad \qquad \qquad -\delta_{p_1+p_2+p_3,0}-\delta_{p_1+p_2-p_3,0}+\delta_{p_1-p_2+p_3,0}+\delta_{p_1-p_2-p_3,0}\Bigr]\\
	& \qquad \qquad \qquad +b^{(k_1\to i)}_{p_1}a^{(k_2\to i)}_{p_2}b^{(k_3\to i)}_{p_3}\Bigl[\delta_{p_3-p_1-p_2,0}-\delta_{p_2-p_1-p_3,0}+\delta_{p_2+p_3-p_1,0}\\
	& \qquad \qquad \qquad \qquad -\delta_{p_1+p_2+p_3,0}+\delta_{p_1+p_2-p_3,0}-\delta_{p_1-p_2+p_3,0}+\delta_{p_1-p_2-p_3,0}\Bigr]\\
	& \qquad \qquad \qquad +a^{(k_1\to i)}_{p_1}b^{(k_2\to i)}_{p_2}b^{(k_3\to i)}_{p_3}\Bigl[\delta_{p_3-p_1-p_2,0}+\delta_{p_2-p_1-p_3,0}-\delta_{p_2+p_3-p_1,0}\\
	& \qquad \qquad \qquad \qquad -\delta_{p_1+p_2+p_3,0}+\delta_{p_1+p_2-p_3,0}+\delta_{p_1-p_2+p_3,0}-\delta_{p_1-p_2-p_3,0}\Bigr]\biggr\}\\
	& \qquad = \frac{2\pi}{4}\sum_{k_1,k_2,k_3\in\partial i\setminus j}\,\sum_{p_1,p_2,p_3=1}^{\infty}{\vphantom{\sum}}^{\mkern-32mu (0)}\,\frac{I_{p_1}(\beta J_{ik_1})I_{p_2}(\beta J_{ik_2})I_{p_3}(\beta J_{ik_3})}{I_0(\beta J_{ik_1})I_0(\beta J_{ik_2})I_0(\beta J_{ik_3})}\\
	& \qquad \qquad \times \Bigl[a^{(k_1\to i)}_{p_1}a^{(k_2\to i)}_{p_2}a^{(k_3\to i)}_{p_3}-\sign{(p_1 p_2)}\,b^{(k_1\to i)}_{p_1}b^{(k_2\to i)}_{p_2}a^{(k_3\to i)}_{p_3}\\
	& \qquad \qquad \qquad -\sign{(p_1 p_3)}\,b^{(k_1\to i)}_{p_1}a^{(k_2\to i)}_{p_2}b^{(k_3\to i)}_{p_3}\\
	& \qquad \qquad \qquad -\sign{(p_2 p_3)}\,a^{(k_1\to i)}_{p_1}b^{(k_2\to i)}_{p_2}b^{(k_3\to i)}_{p_3}\Bigr]\Biggr\}
\end{split}
\end{equation}
where, as before, the superscript~$(0)$ just means that the sum runs over all the possible values of $p_1$, $p_2$ and $p_3$ that algebraically sum up to zero.

Let's put all the terms back into $\mathcal{Z}_{i\to j}$:
\begin{equation}
\begin{split}
	\mathcal{Z}_{i\to j} & = 2\pi\prod_{k\in\partial i\setminus j}I_0(\beta J_{ik})\,\Biggl\{1+\frac{1}{2}\sum_{k_1,k_2\in\partial i\setminus j}\,\sum_{p=1}^{\infty}\,\frac{I_p(\beta J_{ik_1})I_p(\beta J_{ik_2})}{I_0(\beta J_{ik_1})I_0(\beta J_{ik_2})}\\
	& \qquad \qquad \times \Bigl[a^{(k_1\to i)}_p a^{(k_2\to i)}_p+b^{(k_1\to i)}_p b^{(k_2\to i)}_p\Bigr]\\
	& \qquad + \frac{1}{4}\sum_{k_1,k_2,k_3\in\partial i\setminus j}\,\sum_{p_1,p_2,p_3=1}^{\infty}{\vphantom{\sum}}^{\mkern-32mu (0)}\,\frac{I_{p_1}(\beta J_{ik_1})I_{p_2}(\beta J_{ik_2})I_{p_3}(\beta J_{ik_3})}{I_0(\beta J_{ik_1})I_0(\beta J_{ik_2})I_0(\beta J_{ik_3})}\\
	& \qquad \qquad \times \Bigl[a^{(k_1\to i)}_{p_1}a^{(k_2\to i)}_{p_2}a^{(k_3\to i)}_{p_3}-\sign{(p_1 p_2)}\,b^{(k_1\to i)}_{p_1}b^{(k_2\to i)}_{p_2}a^{(k_3\to i)}_{p_3}\\
	& \qquad \qquad \qquad -\sign{(p_1 p_3)}\,b^{(k_1\to i)}_{p_1}a^{(k_2\to i)}_{p_2}b^{(k_3\to i)}_{p_3}\\
	& \qquad \qquad \qquad -\sign{(p_2 p_3)}\,a^{(k_1\to i)}_{p_1}b^{(k_2\to i)}_{p_2}b^{(k_3\to i)}_{p_3}\Bigr]\Biggr\}
\end{split}
\end{equation}

\section{Analysis of the expansion}

Now, if we compare the numerator and denominator expansions for $a^{(i\to j)}_l$, then we can see that both them have the factor $2\pi\prod_{k\in\partial i\setminus j}I_0(\beta J_{ik})$, and hence it can be discarded. Furthermore, we can use the approximation $1/(1+\epsilon)\simeq(1-\epsilon)$ in order to go further in the expansion of $a^{(i\to j)}_l$ up to the third order, so obtaining:
\begin{equation}
\begin{split}
	a^{(i\to j)}_l & = \Biggl\{\sum_{k\in\partial i\setminus j}\,\frac{I_l(\beta J_{ik})}{I_0(\beta J_{ik})}\,a^{(k\to i)}_l+\frac{1}{2}\sum_{k_1,k_2\in\partial i\setminus j}\,\sum_{p_1,p_2=1}^{\infty}{\vphantom{\sum}}^{\mkern-20mu (l)}\,\frac{I_{p_1}(\beta J_{ik_1})I_{p_2}(\beta J_{ik_2})}{I_0(\beta J_{ik_1})I_0(\beta J_{ik_2})}\\
	& \qquad \qquad \qquad \times \Bigl[a^{(k_1\to i)}_{p_1}a^{(k_2\to i)}_{p_2}-\sign{(p_1 p_2)}\,b^{(k_1\to i)}_{p_1}b^{(k_2\to i)}_{p_2}\Bigr]\\
	& \qquad \qquad + \frac{1}{4}\sum_{k_1,k_2,k_3\in\partial i\setminus j}\,\sum_{p_1,p_2,p_3=1}^{\infty}{\vphantom{\sum}}^{\mkern-32mu (l)}\,\frac{I_{p_1}(\beta J_{ik_1})I_{p_2}(\beta J_{ik_2})I_{p_3}(\beta J_{ik_3})}{I_0(\beta J_{ik_1})I_0(\beta J_{ik_2})I_0(\beta J_{ik_3})}\\
	& \qquad \qquad \qquad \times \Bigl[a^{(k_1\to i)}_{p_1}a^{(k_2\to i)}_{p_2}a^{(k_3\to i)}_{p_3}-\sign{(p_1 p_2)}\,b^{(k_1\to i)}_{p_1}b^{(k_2\to i)}_{p_2}a^{(k_3\to i)}_{p_3}\\
	& \qquad \qquad \qquad \qquad -\sign{(p_1 p_3)}\,b^{(k_1\to i)}_{p_1}a^{(k_2\to i)}_{p_2}b^{(k_3\to i)}_{p_3}-\sign{(p_2 p_3)}\,a^{(k_1\to i)}_{p_1}b^{(k_2\to i)}_{p_2}b^{(k_3\to i)}_{p_3}\Bigr]\Biggr\}\\
	& \qquad \times \Biggl\{1-\frac{1}{2}\sum_{k_1,k_2\in\partial i\setminus j}\,\sum_{p=1}^{\infty}\,\frac{I_p(\beta J_{ik_1})I_p(\beta J_{ik_2})}{I_0(\beta J_{ik_1})I_0(\beta J_{ik_2})}\Bigl[a^{(k_1\to i)}_p a^{(k_2\to i)}_p+b^{(k_1\to i)}_p b^{(k_2\to i)}_p\Bigr]\\
	& \qquad \qquad - \frac{1}{4}\sum_{k_1,k_2,k_3\in\partial i\setminus j}\,\sum_{p_1,p_2,p_3=1}^{\infty}{\vphantom{\sum}}^{\mkern-32mu (0)}\,\frac{I_{p_1}(\beta J_{ik_1})I_{p_2}(\beta J_{ik_2})I_{p_3}(\beta J_{ik_3})}{I_0(\beta J_{ik_1})I_0(\beta J_{ik_2})I_0(\beta J_{ik_3})}\\
	& \qquad \qquad \qquad \times \Bigl[a^{(k_1\to i)}_{p_1}a^{(k_2\to i)}_{p_2}a^{(k_3\to i)}_{p_3}-\sign{(p_1 p_2)}\,b^{(k_1\to i)}_{p_1}b^{(k_2\to i)}_{p_2}a^{(k_3\to i)}_{p_3}\\
	& \qquad \qquad \qquad \qquad -\sign{(p_1 p_3)}\,b^{(k_1\to i)}_{p_1}a^{(k_2\to i)}_{p_2}b^{(k_3\to i)}_{p_3}-\sign{(p_2 p_3)}\,a^{(k_1\to i)}_{p_1}b^{(k_2\to i)}_{p_2}b^{(k_3\to i)}_{p_3}\Bigr]\Biggr\}
\end{split}
\end{equation}

Let's analyze order by order the terms which comes out from this product. The first order is simply given by:
\begin{equation}
	\sum_{k\in\partial i\setminus j}\,\frac{I_l(\beta J_{ik})}{I_0(\beta J_{ik})}\,a^{(k\to i)}_l
\end{equation}
and again it just corresponds to the linear analysis performed out in the main text. Then, the second-order term is given by:
\begin{equation}
	\frac{1}{2}\sum_{k_1,k_2\in\partial i\setminus j}\,\sum_{p_1,p_2=1}^{\infty}{\vphantom{\sum}}^{\mkern-20mu (l)}\,\frac{I_{p_1}(\beta J_{ik_1})I_{p_2}(\beta J_{ik_2})}{I_0(\beta J_{ik_1})I_0(\beta J_{ik_2})}\Bigl[a^{(k_1\to i)}_{p_1}a^{(k_2\to i)}_{p_2}-\sign{(p_1 p_2)}\,b^{(k_1\to i)}_{p_1}b^{(k_2\to i)}_{p_2}\Bigr]
\end{equation}
and in the end the third-order term is given by:
\begin{equation}
\begin{split}
	&\frac{1}{4}\sum_{k_1,k_2,k_3\in\partial i\setminus j}\,\sum_{p_1,p_2,p_3=1}^{\infty}{\vphantom{\sum}}^{\mkern-32mu (l)}\,\frac{I_{p_1}(\beta J_{ik_1})I_{p_2}(\beta J_{ik_2})I_{p_3}(\beta J_{ik_3})}{I_0(\beta J_{ik_1})I_0(\beta J_{ik_2})I_0(\beta J_{ik_3})}\\
	& \qquad \qquad \times \Bigl[a^{(k_1\to i)}_{p_1}a^{(k_2\to i)}_{p_2}a^{(k_3\to i)}_{p_3}-\sign{(p_1 p_2)}\,b^{(k_1\to i)}_{p_1}b^{(k_2\to i)}_{p_2}a^{(k_3\to i)}_{p_3}\\
	& \qquad \qquad \qquad -\sign{(p_1 p_3)}\,b^{(k_1\to i)}_{p_1}a^{(k_2\to i)}_{p_2}b^{(k_3\to i)}_{p_3}-\sign{(p_2 p_3)}\,a^{(k_1\to i)}_{p_1}b^{(k_2\to i)}_{p_2}b^{(k_3\to i)}_{p_3}\Bigr]\\
	& \qquad -\sum_{k\in\partial i\setminus j}\,\frac{I_l(\beta J_{ik})}{I_0(\beta J_{ik})}\,a^{(k\to i)}_l\\	
	& \qquad \qquad \times\frac{1}{2}\sum_{k_1,k_2\in\partial i\setminus j}\,\sum_{p=1}^{\infty}\,\frac{I_p(\beta J_{ik_1})I_p(\beta J_{ik_2})}{I_0(\beta J_{ik_1})I_0(\beta J_{ik_2})}\Bigl[a^{(k_1\to i)}_p a^{(k_2\to i)}_p+b^{(k_1\to i)}_p b^{(k_2\to i)}_p\Bigr]
\end{split}
\end{equation}
A careful inspection of these terms tells us that in general none of them is zero. Furthermore, other interesting remarks can be made.
\begin{enumerate}
	\item If there is a ferromagnetic order, then we can always perform a global rotation so that $b_p$'s coefficients become negligible; otherwise, if there is a spin glass order, then it is reasonable that $a_p$'s and $b_p$'s coefficients are of the same order of magnitude; hence, for the following analysis we can focus on those terms only containing $a_p$'s coefficients.
	\item Perturbative expansion for coefficients of order $l>1$ always contains terms like:
	\begin{equation}
		a^{(i\to j)}_l \propto \prod_{k\in\partial i\setminus j} a^{(k\to i)}_{p_k}
	\end{equation}
	with $p_k$'s algebraically summing up to $l$. So we are allowed to make the following \emph{ansatz} on the scaling of coefficients:
	\begin{equation}
		a^{(i\to j)}_l \propto \left(a^{(i\to j)}_1\right)^l
	\end{equation}
	otherwise the perturbative expansion itself would diverge.
\end{enumerate}

\clearpage{\pagestyle{empty}\cleardoublepage}

\backmatter

\cleardoublepage
\phantomsection
\addcontentsline{toc}{chapter}{\bibname}
\sloppy
\printbibliography

\cleardoublepage
\phantomsection
\addcontentsline{toc}{chapter}{\listfigurename}
\listoffigures

\cleardoublepage
\phantomsection
\addcontentsline{toc}{chapter}{\listtablename}
\listoftables

\cleardoublepage
\phantomsection
\addcontentsline{toc}{chapter}{\listalgorithmname}
\listofalgorithms

\cleardoublepage
\phantomsection
\glsnogroupskiptrue 
\setlength{\glslistdottedwidth}{.50\linewidth}
\printglossary[type=\acronymtype,style=listdotted,title=List of Acronyms,toctitle=List of Acronyms]

\glsaddall[types={symbols}]
\cleardoublepage
\phantomsection
\glsnogroupskiptrue 
\setlength{\glslistdottedwidth}{.25\linewidth}
\printglossary[type=symbols,style=symbolstyle,title=List of Symbols,toctitle=List of Symbols]


\end{document}